\documentclass[11pt,a4paper,twoside]{report}
\pdfoutput=1
\usepackage{amsmath,amssymb,mathrsfs,graphicx}
\usepackage[english]{babel}
\usepackage{makeidx}
\newtheorem{mytheo}{Theorem}[section]

\begin{document}


\begin{titlepage}
\begin{center}
\begin{minipage}{0.45\textwidth}
\centering
\includegraphics[height=2.8cm,keepaspectratio]{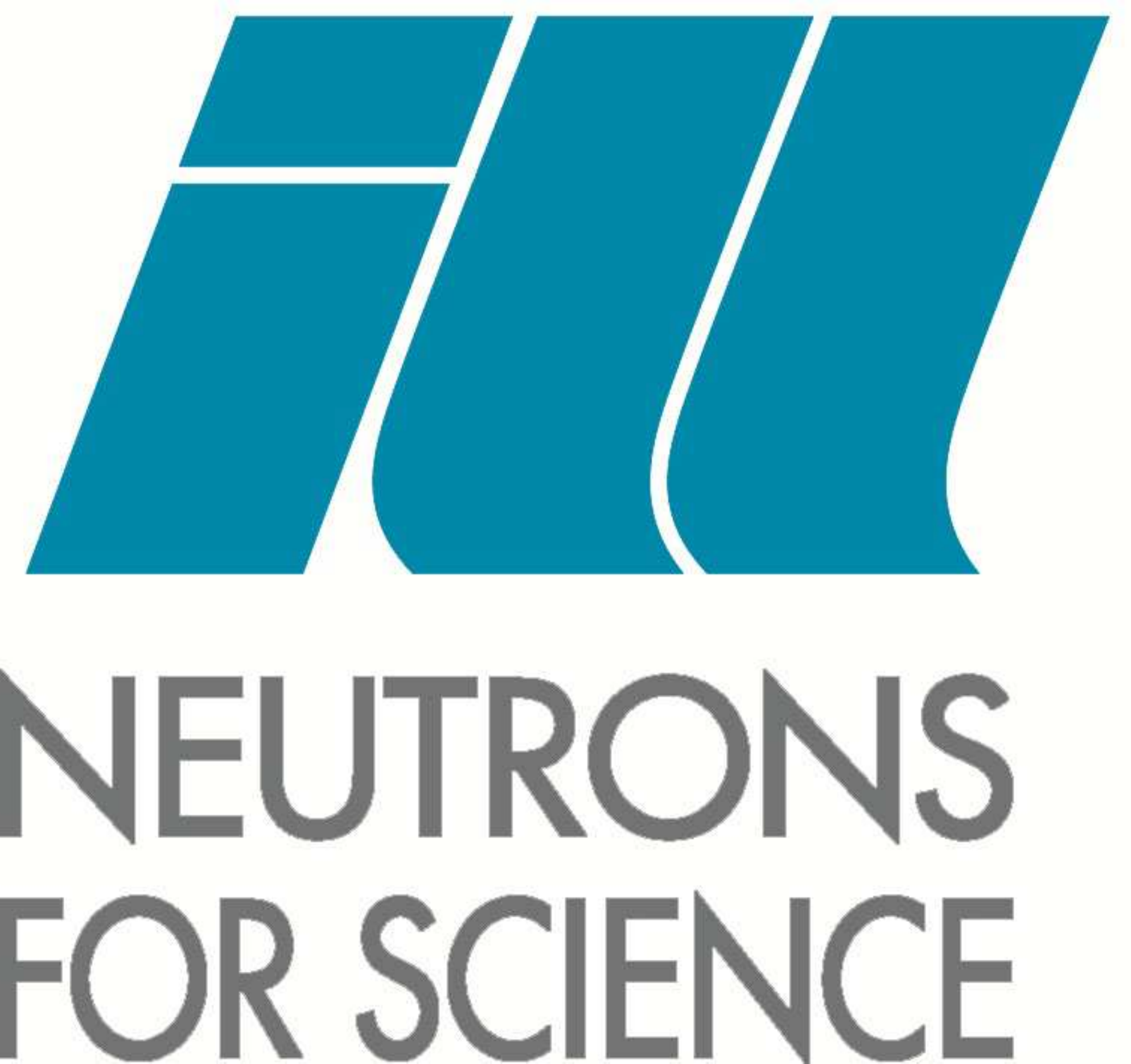}
\\ \vspace{0.2cm}
{\it Institut Laue-Langevin}
\end{minipage} \quad
\begin{minipage}{0.45\textwidth}
\centering
\includegraphics[height=2.8cm,keepaspectratio]{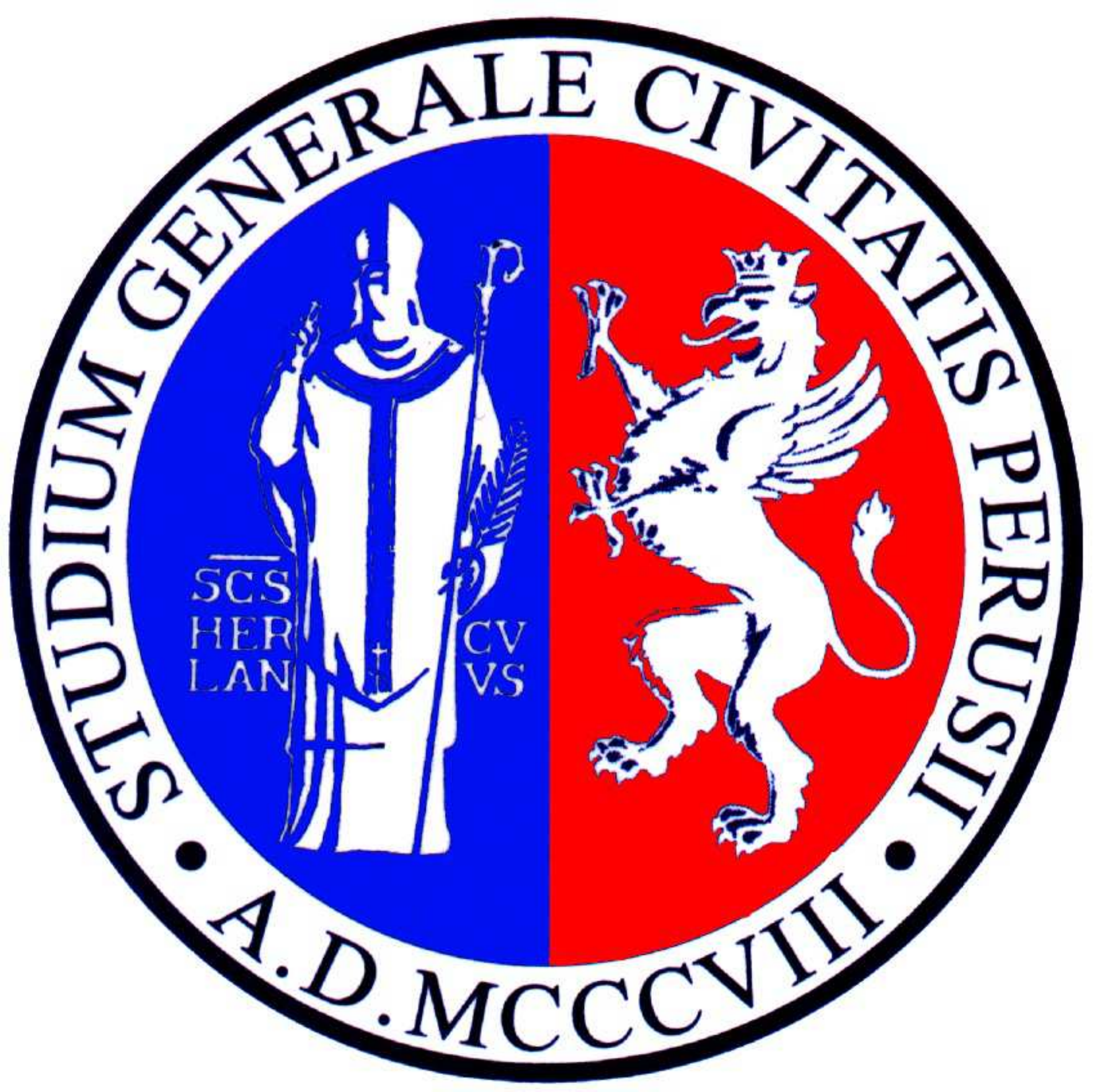}
\\ \vspace{0.2cm}
{\it Universit\`a degli Studi di Perugia}
\end{minipage}
\\ \vspace{2cm}
\textit{\Large Dottorato di Ricerca in Fisica e Tecnologie Fisiche}\\
\vspace{0.5cm}
\textit{\Large XXVI Ciclo}\\
\vspace{2cm}
{\LARGE \bf \bf \sc Boron-10 layers, Neutron Reflectometry}\\ \vspace{0.2cm} {\LARGE \bf \bf \sc and} \\ \vspace{0.5cm}
{\LARGE \bf \bf \sc Thermal Neutron Gaseous Detectors}\\
\vspace{3cm}
\begin{minipage}{0.4\textwidth}
\begin{flushleft} \large
\emph{Author:}\\
Francesco \textsc{Piscitelli}
\end{flushleft}
\end{minipage}
\begin{minipage}{0.4\textwidth}
\begin{flushright} \large
\emph{Supervisor:} \\
Prof. Francesco \textsc{Sacchetti}
\vspace{0.3cm}
\emph{ILL Supervisors:} \\
Dr. Patrick \textsc{Van Esch}\\
Dr. Bruno \textsc{Gu\'erard}\\
\vspace{0.2cm}
\emph{Coordinator:} \\
Prof. Maurizio \textsc{Busso}
\end{flushright}
\end{minipage}
\vfill \rule{70mm}{0.5pt} \\{\large A.A. 2012/2013}
\end{center}
\end{titlepage}
\thispagestyle{empty}
\begin{flushright} {\Large
\emph{To my family.}}
\end{flushright}
\vfill
\begin{flushright} 
{\Large [...] \emph{fatti non foste a viver come bruti, \\ ma per seguir
virtute e canoscenza.}}
\\[0.5cm]
[...] \emph{\footnotesize you were not made to live as brutes, \\
but to follow virtue and knowledge.}
\\[1.5cm]
{\bf \footnotesize Dante Alighieri - Divina Commedia, \\ Inferno
Canto XXVI vv. 119-120}
\end{flushright}
\newpage
\thispagestyle{empty}
{\it \large I would like to thank many people. Each Chapter starts
with a special thank to who was decisive to finalize the work
explained in such Chapter or I had important discussions with or,
more simply, who taught me a lot on a specific subject. I thank my
colleague Anton Khaplanov for what he taught me about the
$\gamma$-ray physics. Philipp Gutfreund, Anton Devishvili, and Boris
Toperverg for what concerns neutron reflectometry: experiments,
theory and data analysis. Federica Sebastiani and Yuri Gerelli for
the discussions we had on neutron reflectometry. Carina H\"{o}glund
for the excellent work done on the boron layers. I want to thank
also Jean-Claude Buffet and Sylvain Cuccaro for the amount of
technical issues they solved for me and for what they taught me
about the mechanics. I want also to thank my colleagues Anton
Khaplanov and Jonathan Correa for the pleasant time spent together
in our office and in the labs.
\\ I wish to every PhD student to have a supervisor as Patrick Van
Esch. A special thank goes to him for the huge amount of things he
explained to me on the neutron scattering. For the help he gave me
to develop the theory of neutron converters and for the time he
spent with me to make this manuscript as it is presented now.
\\ I thank my group leader, Bruno Gu\'{e}rard who believed in the
success of the prototype concept, for having pushed me to construct
the detector. I want to thank Bruno Gu\'{e}rard and Richard
Hall-Wilton for the support they gave in the research of new
technologies in thermal neutron detection.
\\ I thank the SISN (Italian Society of Neutron Spectroscopy)
and all its members, in particular Alessio De Francesco, for what I
learned about neutrons and for all the occasions it gave me.
\bigskip
I would like to thank all the people that, even if they have not
been directly involved in this thesis, I think they have contributed
somehow. \\ The entire ''Italian lunch community'' for the time
spent together eating, drinking and playing sports, in particular
skiing and playing football. I want to thank the ''Tresette''
players for the pleasant time spent after lunch to get back the
concentration to work.
\\ I thank Giuliana Manzin and John Archer because they had care of
me when I came in Grenoble for the first time and because of our
well-established ''eating-friendship''.
\\ I would like to thank ''Los Latinos'' football group for the matches we
played together that really helped me to discharge the stress of
writing a thesis.
\\ I thank my friends, that I left in Italy, to be always interested in
how my life is going.
\bigskip
I thank my family for its strong support. For the moral and concrete
means I received from them over the years.
\\ I would also to thank my family as I conceive it
now including Federica's family for the convivial time spent
together.
\bigskip
I thank Federica Sebastiani to be always present.} \vfill
\begin{flushright}{\Large October 29, 2013}\end{flushright}

\thispagestyle{empty}
\vspace{5cm}
\begin{flushright} {\Large \bf \bf \sc Boron-10 layers, Neutron Reflectometry \\[0.2cm] and \\[0.5cm]
Thermal Neutron Gaseous Detectors} \\ \vfill {\bf Francesco
Piscitelli}
\end{flushright}
\pagenumbering{arabic}
\setcounter{page}{1}
\tableofcontents
\chapter*{Introduction}
\addcontentsline{toc}{chapter}{Introduction}
\section*{Institut Laue-Langevin}
\addcontentsline{toc}{section}{Institut Laue-Langevin} The Institut
Laue-Langevin (ILL) is an international research center at the
leading edge of neutron science and technology. It is situated in
Grenoble, France, among some other important research centers like
C.E.A. \protect\footnote{Commissariat \`a l'\`Energie Atomique} or
E.S.R.F. \index{E.S.R.F.} \protect\footnote{European Synchrotron
Radiation Facility}
\\ ILL was funded and it is managed by the governments of France,
Germany and the United Kingdom, in partnership with $9$ other
European countries \protect\footnote{Italy, Spain, Switzerland,
Austria, the Czech Republic, Sweden, Hungary, Belgium and Poland}.
Every year, more than $1200$ researchers from $30$ countries visit
the ILL. Over $700$ experiments selected by a scientific review
committee are performed annually and research focuses mainly on
fundamental science in a variety of fields; this includes condensed
matter physics, chemistry, biology, nuclear physics and materials
science.
\\The institute operates the most intense neutron source in the World
\protect\footnote{$1.5\cdot10^5$ neutrons per second per $cm^2$},
feeding beams of neutrons to a suite of $40$ high-performance
instruments that are constantly upgraded. From a technical point of
view, neutrons are created by the fission of $^{235}U$ in a nuclear
reactor of about $58\,MW$ of power. Thanks to different procedures,
neutrons with different energies can be produced. They are guided to
the different apparatus in the Guide Halls where the experiments
take place.
\begin{figure}[ht!]
\centering
{\includegraphics[width=12cm,angle=0,keepaspectratio]{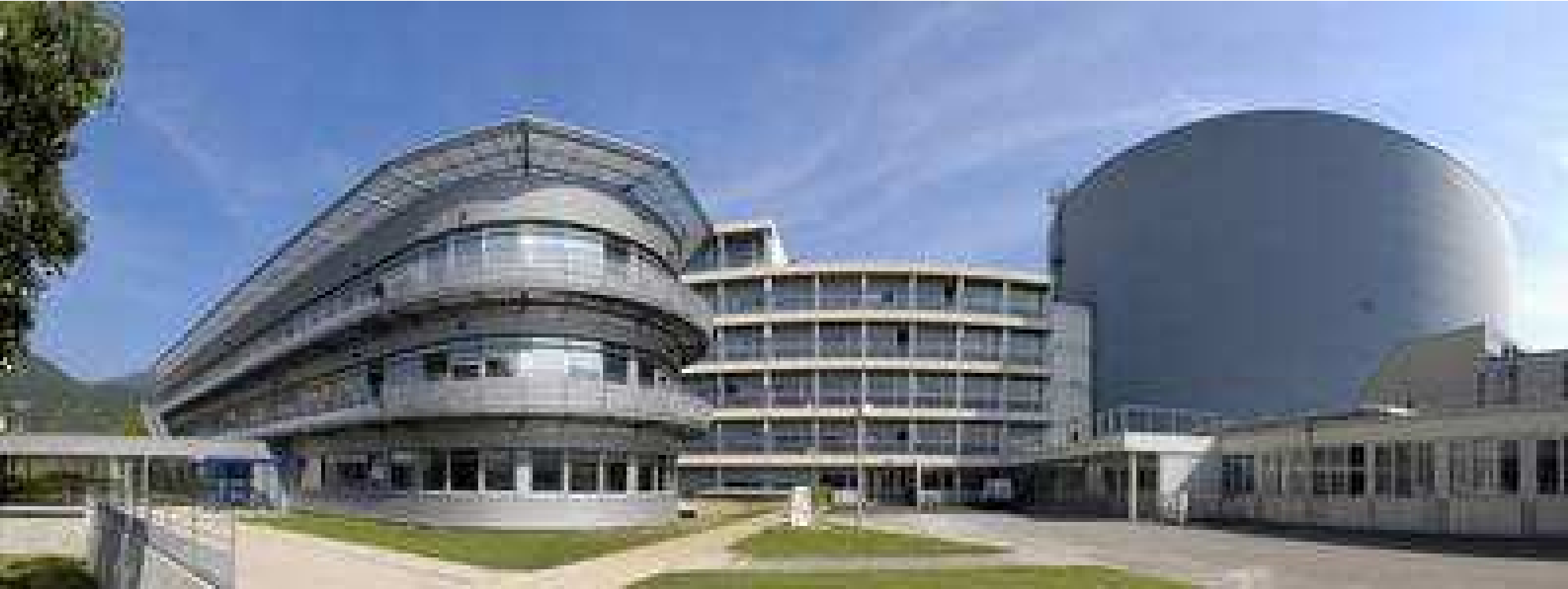}}
\caption{\footnotesize Panoramic view of ILL}
\end{figure}
\bigskip
\newpage
\section*{Outline}
\addcontentsline{toc}{section}{Outline}

By using neutrons we can determine the relative positions and
motions of atoms in a bulk sample of solid or liquid. Neutrons make
us able to look inside the sample with a suitable magnifying glass.
They have no charge, and their electric dipole moment is either zero
or too small to be measured. For these reasons, neutrons can
penetrate matter far better than charged particles and it is also
why they are relatively hard to be detected. Available neutron beams
have inherently low intensities. The combination of weak
interactions and low fluxes make neutron scattering a signal-limited
technique, which is practiced only because it provides information
about the structure of materials that cannot be obtained in simpler,
less expensive ways.

\bigskip

Nowadays neutron facilities are going toward higher fluxes, e.g. the
European Spallation Source (ESS) in Lund (Sweden), and this
translates into a higher demand in the instrument performances:
higher count rate capability, better timing and smaller spatial
resolution are requested amongst others. Moreover, existing
facilities, such as ILL, need also a continuous updating of their
suite of instruments.
\\ Because of its favorable properties, $^3He$
(a rare isotope of $He$) has been the main actor in thermal neutron
detection for years. $^3He$ is produced through nuclear decay of
tritium, a radioactive isotope of hydrogen. By far the most common
source of $^3He$ in the United States is the US nuclear weapons
program, of which it is a byproduct. The federal government produces
tritium for use in nuclear warheads. Over time, tritium decays into
$^3He$ and must be replaced to maintain warhead effectiveness. Until
2001, $^3He$ production by the nuclear weapons program exceeded the
demand, and the program accumulated a stockpile. In the past decade
$^3He$ consumption has risen rapidly. After the terrorist attacks of
September 11, 2001, the federal government began deploying neutron
detectors at the US border to help secure the nation against
smuggled nuclear and radiological material. Thus, starting in about
2001, and more rapidly since about 2005, the stockpile has been
declining.
\\ The World is now experiencing the shortage of $^3He$. This makes
the construction of large area detectors (several squared
meters) not realistic anymore. A way to reduce the $^3He$ demand for
those applications is to move users to alternative technologies.
Some technologies appear promising, though implementation would
likely present technical challenges.
\\ Although scintillators are also widely
employed in neutron detection, they show a higher $\gamma$-ray
sensitivity compared to gaseous detectors that makes their use in
strong backgrounds difficult.
\\ Many research groups in Europe and in the World are exploring
different alternative ways to detect neutrons to assure the future
of the neutron scattering science. They focus mainly on the $^3He$
replacement because this expensive gas, in large quantities, is not
available anymore. Although it is absolutely necessary to replace
$^3He$ for large area applications, this is not the main issue for
what concerns small area detectors ($\sim 1\,m^2$) for which the
research is focused on improving their performances.
\\ There are several aspects that must be investigated in order to validate
those new technologies. E.g. their detection efficiency is one of
the main concerns because it is in principle relatively limited
compared to $^3He$ detectors. The detection of a $\gamma$-ray
instead of a neutron can give rise to misaddressed events. The level
of discrimination between neutrons and background events (e.g.
$\gamma$-rays) a neutron detector can attain is then another key
feature to be studied.

\bigskip

This PhD work was carried out at Institut Laue-Langevin (ILL) in
Grenoble (France) in the Neutron Detector Service group (SDN) which
is mainly in charge of the maintenance of the neutron detectors of
the instruments. This group is also involved in the development of
new technologies for thermal neutron detection.
\\ At ILL we tackled both the problem of $^3He$ replacement for large area
applications and the performance problem for small area detectors.
Both solutions are based on $^{10}B$ layers. $^{10}B$ is about
$20\%$ of the natural abundance of Boron, and thanks to its large
neutron absorbtion cross-section, it is a suitable material to be
employed in neutron detection as a neutron converter. In particular
we used thin layers of magnetron sputtered $^{10}B_4C$ produced by
the Link\"{o}ping University (Sweden).
\\ Although the physical process involved in neutron detection
via layers of solid converter (such as $^{10}B$) is well known,
there a great interest in expanding the theory toward new models and
equations that can help to develop such a technology.
\\ The Multi-Grid gaseous neutron detector was developed at ILL to
face the problem arising for large area applications. We also
implemented the Multi-Blade detector, already introduced at ILL in
2005, but never implemented until 2012, to go beyond the intrinsic
limit in performances of the actual small area detectors. The
Multi-Blade is a small area detector for neutron reflectometry
applications that exploits $^{10}B_4C$-films employed in a
proportional gas chamber. In a $^3He$-based detector the counting
rate and spatial resolution are both limited. The instruments
dedicated to neutron reflectometry studies need detectors of high
spatial resolution ($<1\,mm$) and high counting rate capability.
\\ There is a great interest in expanding the performances of neutron
reflectometry instruments, but due to practical limits in actual
detector resolution and collimation, the technique is probably not
practical. The principal investment is an area detector with
$0.2\,mm$ spatial resolution required in one dimension only.

\bigskip

The Multi-Grid exploits up to $30$ $^{10}B_4C$-layers in a cascade
configuration. For large area applications the main concern about
the $^{10}B$-based technology is the detection efficiency. While in
$^3He$ tubes a $2\,cm$ detection volume assures an efficiency beyond
$70\%$ (at $2.5$\AA); a $30$-layer $^{10}B$ detector is needed to
reach about $50\%$ efficiency for the same neutron wavelength. Since
those layers are arranged in cascade, there is a strong interest in
studying their arrangement in order to improve the efficiency and
the neutron to $\gamma$-ray discrimination.
\\ We elaborated a pure analytical study, proved by experiments,
on the layer arrangement to increase the detector efficiency. We
developed a suite of equations to help the detector construction,
i.e. we derived analytical formulae to optimize the $^{10}B$-coating
thicknesses. Those results can be also applied to other kinds of
solid neutron converter, such as $^6Li$, and not only to
$^{10}B$-films.
\\ This theoretical study has also demonstrated
that the magnetron sputtering is a suitable technique to make
optimized converter layers. We also derived the analytical
expression for the Pulse Height Spectrum (PHS) that helps to predict
the neutron to $\gamma$-ray discrimination.

\bigskip

For a standard neutron detection efficiency, less than $10^{-6}$
$\gamma$-ray sensitivity can be easily achieved in $^3He$ detectors.
The discrimination procedure is also easy to be applied and it is
based on the distinction of the energy deposited in the gas volume.
While there is a good separation in energy between neutron and
$\gamma$-ray events for $^3He$ detectors, this is not the case for
solid-converter-based detectors. We investigated deeply the
$\gamma$-ray sensitivity of $^{10}B$-based detector and we compared
with $^3He$ detectors. We exposed $^{10}B$ and $^3He$ detectors to
the same calibrated $\gamma$-ray background in order to quantify
their sensitivity, when the same energy discrimination method is
used.
\\ Since there is always a certain loss of neutron detection
efficiency for $^{10}B$ detectors, we investigated one more method
to perform the neutron to $\gamma$-ray discrimination for those
detectors.
\\ In order to quantify the $\gamma$-ray background a detector is
exposed to on a real instrument we measured the typical background
on the time-of-flight spectrometer IN6 at ILL.
\\ We elaborated a procedure to measure both the PHS
and the counting curve of $^{10}B$-based detectors free from
$\gamma$-rays that can be compared with the theoretical model we
developed.

\bigskip

The Multi-Blade prototype is a small area detector for neutron
reflectometry applications. It is a Multi Wire Proportional Chamber
(MWPC) operated at atmospheric pressure. The Multi-Blade prototype
uses $^{10}B_4C$ converters at grazing angle with respect to the
incoming neutron beam. The inclined geometry improves the spatial
resolution and the count rate capability of the detector. Moreover,
the use of the $^{10}B_4C$ conversion layer at grazing angle also
increases the detection efficiency. \\ While detection efficiency
increases as the inclination decreases, the reflection of neutrons
at the surface can be an issue. We studied this potential problem by
developing a theoretical model about neutron reflection by strong
absorbing materials such as $^{10}B_4C$.
\\ We characterized $^{10}B_4C$ layers deposited
on several types of substrates by using neutron reflectometry. We
quantified the loss by reflection of such a layer as a function of
the hitting angle and neutron wavelength. We investigated which
properties of the layer and its substrate influence the reflection
that has to be minimized. Our analytical model helped to investigate
the data and to get information about the neutron converter itself.
\\ The Multi-Blade prototype is conceived to be modular in order to be
adaptable to different applications. A significant concern in a
modular design is the uniformity of detector response. Several
effects might contribute to degrade the uniformity and they have to
be taken into account in the detector concept: overlap between
different substrates, coating uniformity, substrate flatness and
parallax errors.
\\ We studied several approaches in the prototype design: number of
converters, read-out system and materials to be used.
\\ We built two versions of the Multi-Blade prototype focusing on
its different issues and features. We measured their detection
efficiency and uniformity on our test beam line. We quantified their
spatial resolution and dead time.
\\ We investigated a different deposition method for the $^{10}B$
converters which is not magnetron sputtering but $^{10}B$ glue-based
painting.

\bigskip

We hope that in our work we have laid a solid theoretical basis,
confirmed by experiments, for the understanding of the main aspects
of solid converter layers employed in neutron detectors. We also
explored practically, by the construction and characterization of
prototypes, a specific type of solid-converter-based neutron
detector, the Multi-Blade, especially suited for application in
neutron reflectometry.

\chapter{Interaction of radiations with matter}\label{chaptintradmatt}
This chapter summarizes the prerequisites needed for the
comprehension of the material aboarded in this manuscript. We
present notions of interactions between particles and matter, and we
introduce some neutron physics. We pay especially attention to the
meaning of coherent and incoherent scattering lengths, and to the
meaning of the imaginary part of the scattering length.
\\ The main sources of the material for this Chapter are:
the books \cite{leo}, \cite{knoll}, \cite{coen}, \cite{schiff},
\cite{squires} and the articles \cite{nelson}, \cite{sears},
\cite{cubitt3}.
\newpage
\section{Definition of cross-section}\label{defcross4}
\begin{figure}[ht!] \centering
\includegraphics[width=10cm,angle=0,keepaspectratio]{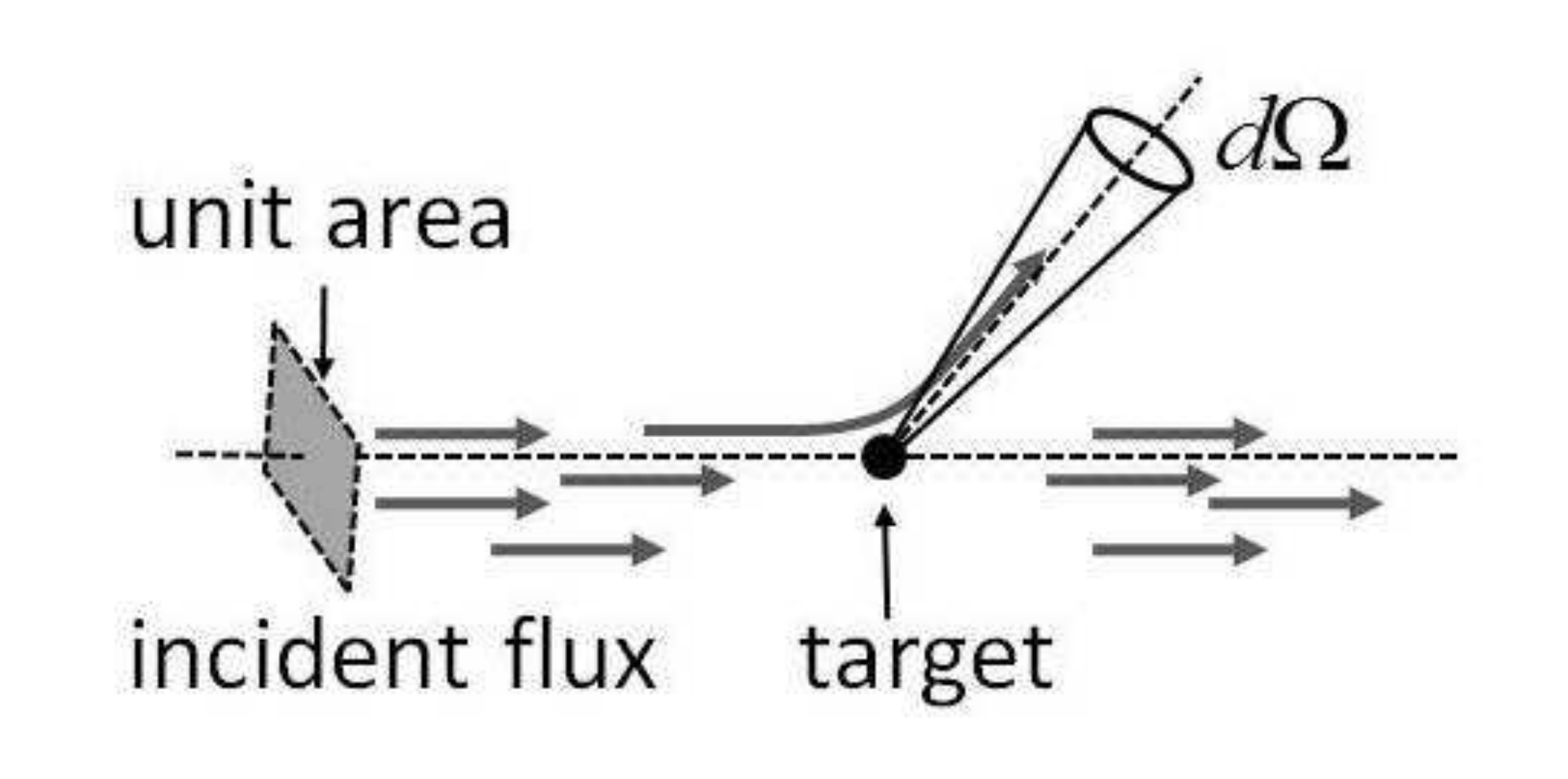}
\caption{\footnotesize Definition of the scattering cross-section
for a single scattering center.}\label{defcrosssect67}
\end{figure}

\begin{figure}[ht!] \centering
\includegraphics[width=10cm,angle=0,keepaspectratio]{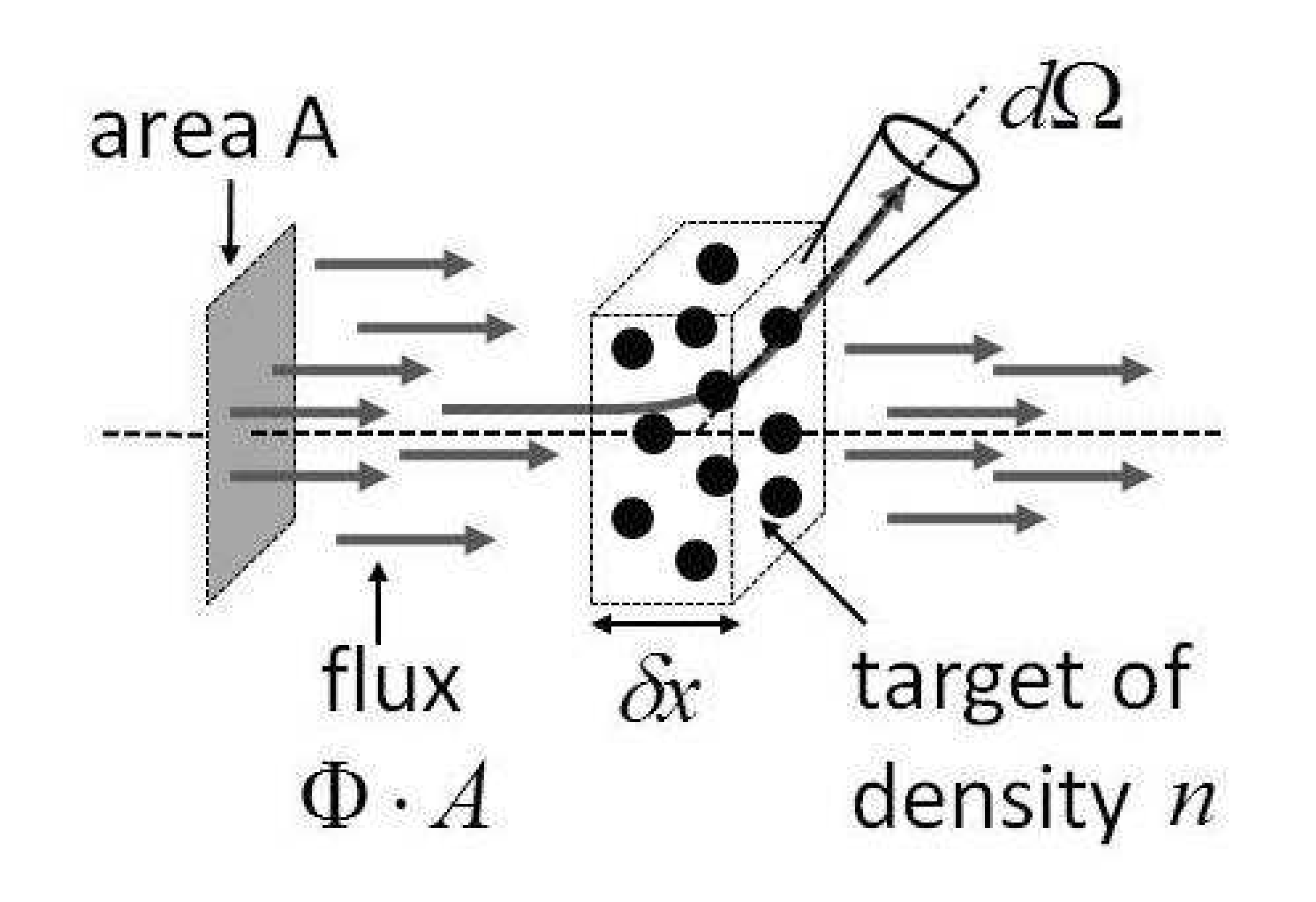}
\caption{\footnotesize Definition of the scattering cross-section
for an extended target.}\label{defcrosssect68}
\end{figure}

The collision or interaction of two particles is generally described
in terms of the \emph{cross-section} \cite{leo}. This quantity
essentially gives a measure of the probability for a reaction to
occur. Consider a beam of particles incident on a target particle
and assume the beam to be broader than the target (see Figure
\ref{defcrosssect67}). Suppose that the particles in the beam are
randomly distributed in space and time. We can define $\Phi$ to be
the flux of incident particles per unit area and per unit time. Now
look at the number of particles scattered into the solid angle $d
\Omega$ per unit time. By \emph{scattering} we mean any reaction in
which the outgoing particle is emitted in the solid angle $\Omega$.
If we average, the number of particles scattered in a solid angle $d
\Omega$ per unit time will tend toward a fixed value $dN_s$. The
\emph{differential cross-section} is then defined as:
\begin{equation}\label{eqha1}
\frac{d\sigma}{d\Omega}\left(E,\Omega
\right)=\frac{1}{\Phi}\cdot\frac{dN_s}{d\Omega}
\end{equation}
that is, $\frac{d\sigma}{d\Omega}$ is the the average fraction of
particles scattered into $d\Omega$ per unit time per unit incident
flux $\Phi$ for $d\Omega$ infinitesimal. Note that because of
$\Phi$, $d\sigma$ has the dimension of an area. We can interpret
$d\sigma$ as the geometric cross sectional area of the target
intercepting the beam. In other words, the fraction of flux incident
on this area will interact with the target and scatter into the
solid angle $d \Omega$ while all those missing $d\sigma$ will not.
\\ In general the differential cross-section of a process varies
with the energy of the reaction and with the angle at which the
particle is scattered. We can calculate a \emph{total
cross-section}, for any scattering whatsoever at an energy $E$, as
the integral of the differential cross-section over all solid angles
as follows:
\begin{equation}\label{eqha2}
\sigma\left(E\right)=\int\frac{d\sigma}{d\Omega}\left(E,\Omega
\right)d\Omega
\end{equation}
Consider now a real target, which is usually a slab of material
containing many scattering centers. We want to know how many
interactions occur on average when that target is exposed to a beam
of incident particles. Assuming that the slab is not too thick so
that the likelihood of interaction is low, the number of centers per
unit perpendicular area which will be seen by the beam is then $n
\cdot \delta x$ where $n$ is the volume density of centers and
$\delta x$ the thickness of the material along the direction of the
beam (see Figure \ref{defcrosssect68}). If $A$ is the perpendicular
area of the target and the beam is broader than the target, the
number of incident particles which are eligible for an interaction
per unit of time is $\Phi \cdot A$. The average number of scattered
particles into $d \Omega$ per unit time is:
\begin{equation}\label{eqha3}
N_s(\Omega)=\Phi \cdot A \cdot n \cdot \delta x
\cdot\frac{d\sigma}{d\Omega}\cdot d\Omega
\end{equation}
The total number of scattered into all angles is similarly:
\begin{equation}\label{eqha4}
N_{tot}=\Phi \cdot A \cdot n \cdot \delta x \cdot\sigma
\end{equation}
In the case the beam is smaller than the target, we need only to set
$A$ equal to the area covered by the beam. We can take another point
of view; that is the probability of an incident particle of the beam
to be scattered. If we divide Equation \ref{eqha4} by the total
number of incident particles per unit time ($\Phi \cdot A$), we have
the probability for the scattering of a single particle in a
thickness $\delta x$:
\begin{equation}\label{eqha5}
P_{\delta x}=n \cdot \sigma \cdot \delta x
\end{equation}
\\ Note that the probability for interaction is proportional to the
distance traveled, $dx$ in first order.
\\ Let us consider now a more general case of any thickness $x$. We
ask what is the probability for a particle not to suffer an
interaction over a distance $x$ traveled in the target. This can be
interpreted as the probability for a particle to \emph{survive} the
interaction process. Let's denote with $P(x)$ the probability of not
having an interaction after a distance $x$ (hence the probability to
survive to the interaction process after a distance $x$ traveled)
and with $w\,dx$ the probability to have an interaction in the
interval $(x,x+dx)$. From Equation \ref{eqha5} we define
$\Sigma=n\cdot \sigma$ the macroscopic cross-section. The
probability of not having an interaction up to $x+dx$ is given by:
\begin{equation}\label{eqha6}
\begin{aligned}
&P(x+dx)=P(x)\cdot \left(1-\Sigma\,dx\right)\quad \Rightarrow \\
&P(x)+\frac{dP}{dx}dx=P(x)-P(x)\,\Sigma\,dx\quad \Rightarrow\\
&\frac{dP(x)}{P(x)}=-\Sigma\,dx\quad \Rightarrow\\
&P(x)=C\cdot e^{-\Sigma\,x}=e^{-\Sigma\,x}\\
\end{aligned}
\end{equation}
where $C$ is an integration constant. Note that $C=1$ because we
require that $P(x=0)=1$. From Equation \ref{eqha6} we can
immediately deduce the probability to have an interaction over a
distance $x$:
\begin{equation}\label{eqha7}
P_{int}(x)=1-P(x)=1-e^{-\Sigma\,x}
\end{equation}
We can define the mean distance $\eta$ traveled by a particle
without interacting; this is known as the \emph{mean free path} that
a particle can travel across the target without suffering any
collision, thus:
\begin{equation}\label{eqha8}
\eta=\frac{\int x P(x) dx}{\int P(x)
dx}=\frac{1}{\Sigma}=\frac{1}{n\cdot \sigma}
\end{equation}
The survival probability of a trajectory of length $x$ becomes:
\begin{equation}\label{eqha11}
P(x)=e^{-\Sigma x}=e^{-\frac{x}{\eta}}
\end{equation}
and the probability of interaction:
\begin{equation}\label{eqha12}
P_{int}(x)=1-e^{-\Sigma x}=1-e^{-\frac{x}{\eta}}
\end{equation}

\section{Charged particles interaction}\label{cpitheo851}
In general two principal features characterize the passage of charge
particles through matter: a loss of energy by the particle and the
deflection of the particle from its incident direction  \cite{leo}.
Mainly these effects are the result of two processes:
\begin{itemize}
    \item electromagnetic interactions with the atomic electrons of the
    material;
    \item elastic scattering from nuclei.
\end{itemize}
These reactions almost occur continuously in matter and it is their
cumulative result which accounts for the principal effects observed.
Other processes include the emission of Cherenkov radiation, nuclear
reaction (this is the case for neutrons) and bremsstrahlung.
\\ It is necessary to separate charged particles into two classes:
electrons and positrons on one side and heavy particles, i.e.,
particles heavier than the electron, on the other.
\subsection{Heavy charged particles}
\begin{figure}[ht!] \centering
\includegraphics[width=5.2cm,angle=0,keepaspectratio]{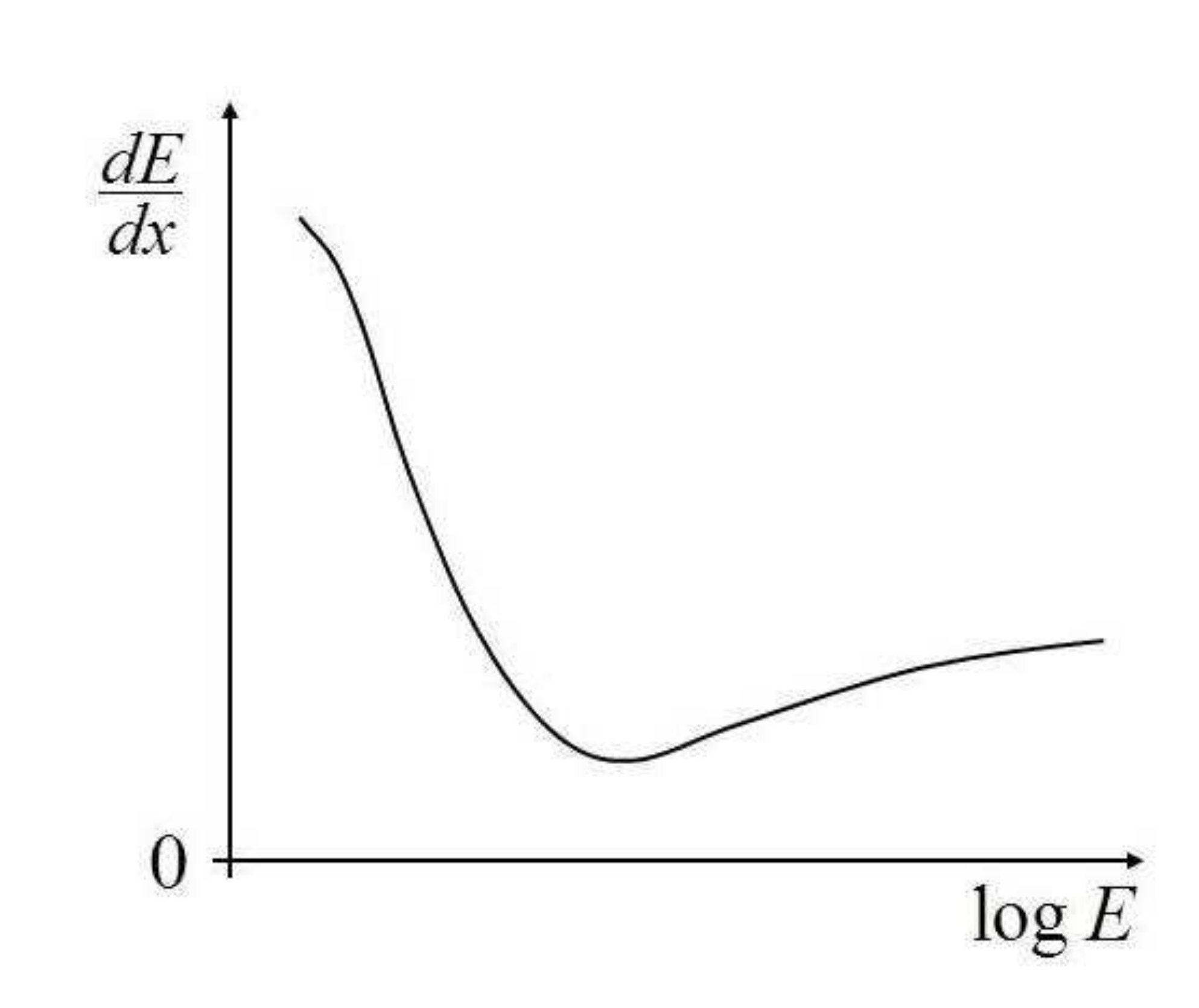}
\includegraphics[width=5.2cm,angle=0,keepaspectratio]{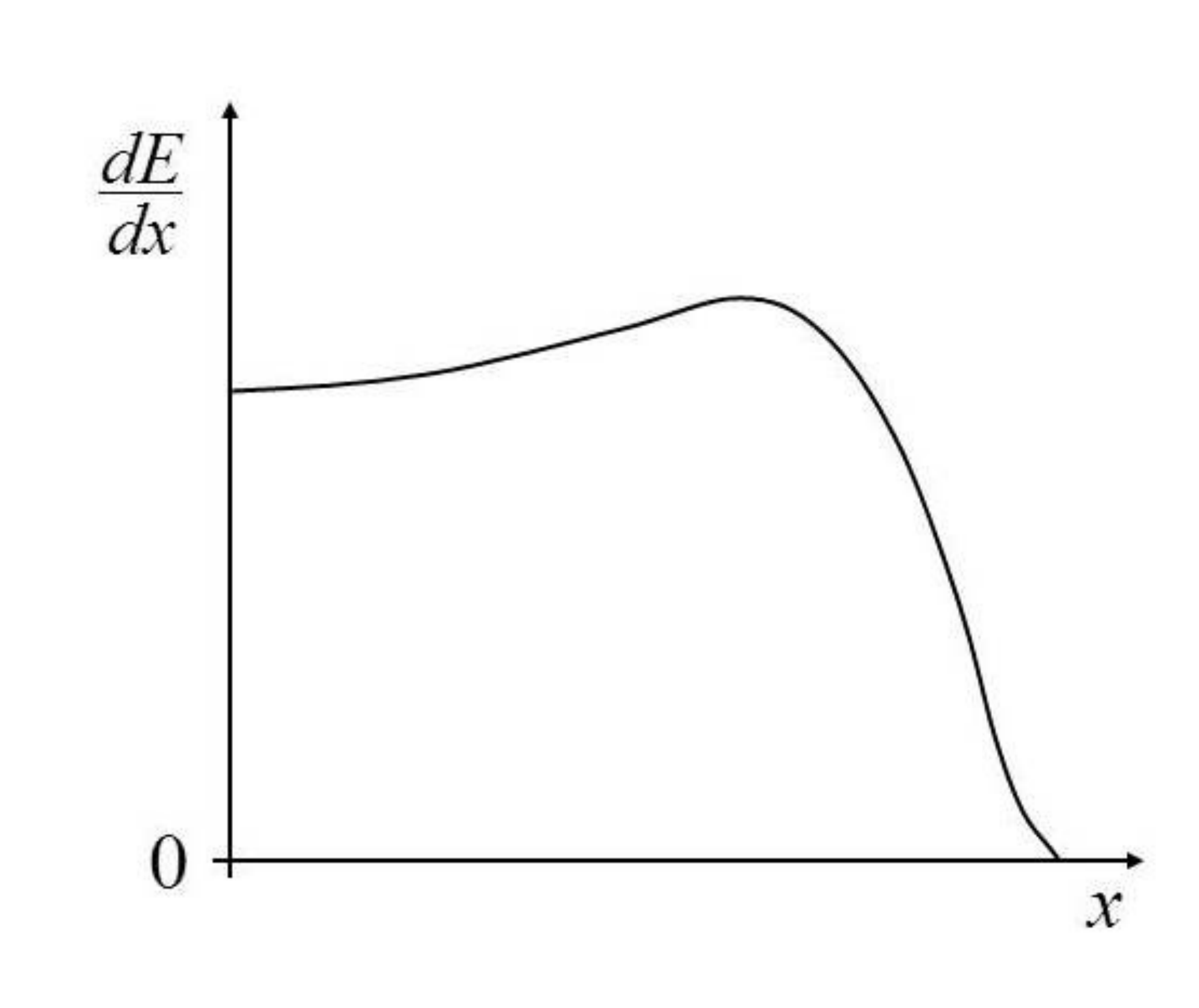}
\includegraphics[width=4.8cm,angle=0,keepaspectratio]{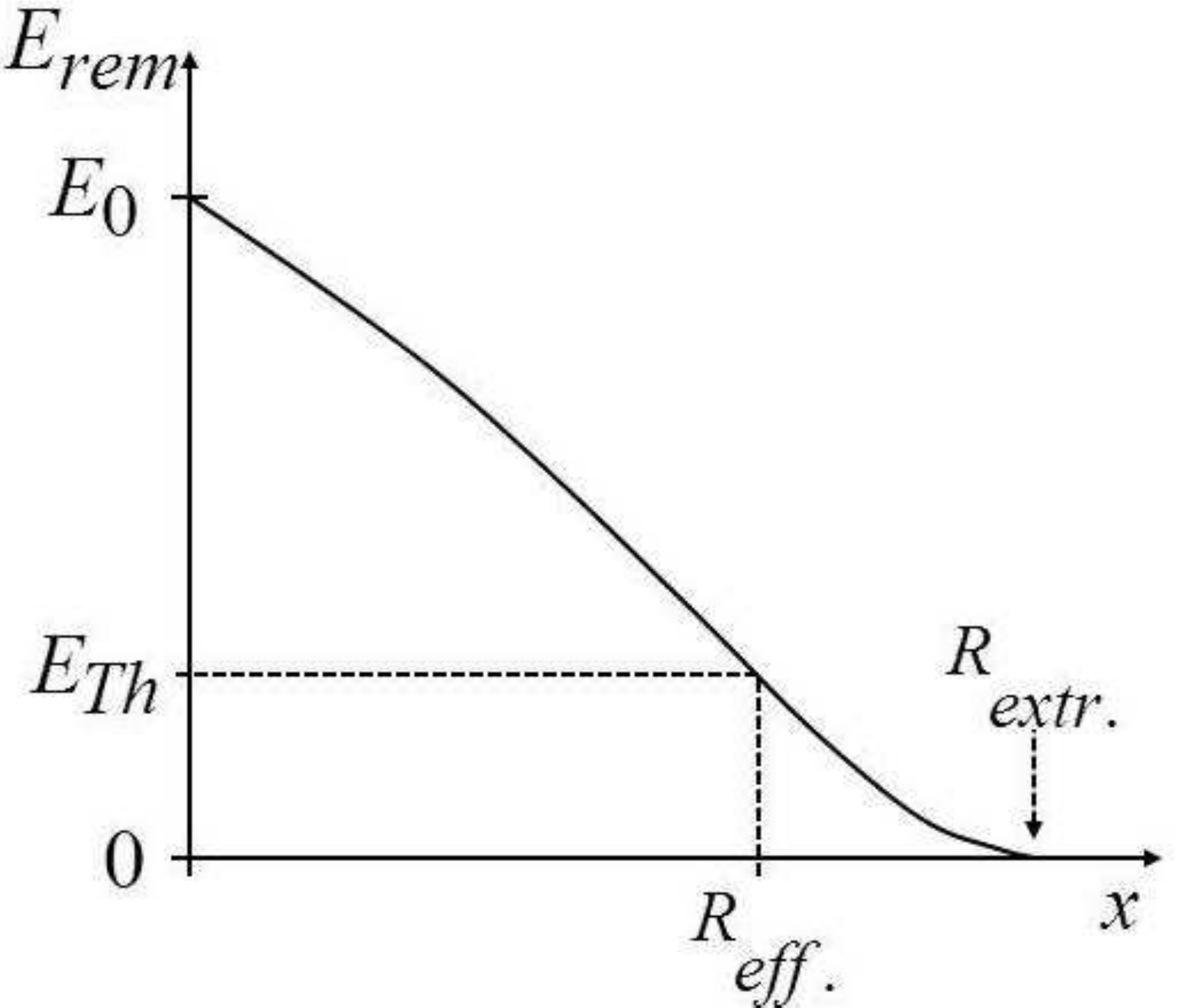}
\caption{\footnotesize The stopping power as a function of $E$
(left) and $x$ (center). Remaining energy $E_{rem}$ as a function of
$x$ (right).}\label{figstopremmip}
\end{figure}

The inelastic collisions with the electrons of the material are
almost solely responsible for the energy loss of heavy particles in
matter. In these collisions energy is transferred from the particle
to the atom causing an ionization or excitation of the latter. The
amount of energy transferred in each collision is generally a very
small fraction of the particle's total kinetic energy; however, in
normally dense matter, the number of collisions per unit path length
is so large, that a substantial cumulative energy loss is observed
in relatively thin layers of material.
\\ Elastic scattering from nuclei also occurs although not as often
as interactions with the bound electrons. In general very little
energy is transferred in these collisions since the masses of the
nuclei of most materials are usually large compared to the incident
particle.
\\ The inelastic collisions are statistical in nature, occurring
with a certain quantum mechanical probability. However, because
their number per macroscopic unit of path length is generally large,
the fluctuations in the total energy loss are small and one can
meaningfully work with the average energy loss per unit path length.
This quantity, often called \emph{stopping power} or
$\frac{dE}{dx}$, was first calculated by Bohr using classical
arguments and later by Bethe and Bloch using quantum mechanics. \\
The classical derivation, that is shown in details in the Appendix
\ref{classstoppow5}, helps to clarify the line of reasoning which
stands behind the result. \\ Even Bohr's classical formula gives a
reasonable description of the energy loss for very heavy particles;
the correct quantum-mechanical calculation leads to the Bethe-Bloch
formula (see Appendix \ref{classstoppow5}).
\\ Software packages are freely available which simulate the energy
loss, from which one can deduce the stopping power. They give a
similar result as the Bethe-Bloch calculation. We use SRIM
\cite{sri}, \cite{sri2} to calculate the stopping power.
\\ It is clear from the behavior in $E$ that as a heavy particle slows
down in matter, its rate of energy loss will change as it loses its
kinetic energy. More energy per unit length will be deposited
towards the end of its path rather than at its beginning. This
effect is more clear in the central plot in Figure
\ref{figstopremmip} which shows the amount of ionization created by
a heavy particle as a function of its position along its
slowing-down path. This is known as a \emph{Bragg} curve, and most
of the energy is deposited close to the end of the trajectory.
\\ If we assume the energy loss to be a continuous function, we can define the
particle \emph{range} which is the average distance a particle can
travel inside a material before stopping. The latter depends on the
type of material, the kind of particle and its energy. In reality,
the energy loss is not continuous but statistical; two identical
particles, with the same initial energy, will not suffer the same
number of collisions and hence the same energy loss. \\ There are
various ways to define the actual range of a particle. In order to
clarify the definition of the range we refer to Figure
\ref{figstopremmip}. The right plot represents the energy a particle
still owns as a function of the distance it has traveled in the
material that can be calculated from the stopping power function
according to:
\begin{equation}\label{eqaa234}
E_{rem}(x) = E_0-\int_0^x  \frac{dE}{d\xi} \, d\xi
\end{equation}
where $E_0$ is the particle initial energy. Note that, in our model,
a particle that has slowed down, below the minimum energy necessary
to create a ion-pair, is considered stopped. As a result we define
\emph{extrapolated range} the average distance a particle can travel
until it carries an energy below the minimum needed to ionize an
atom. In Figure \ref{figstopremmip} the extrapolated range
corresponds to a threshold energy of about $E_{Th}\sim 0$.
\\ An alternative definition of range can be the \emph{effective
range}; this corresponds to the distance a particle on average has
traveled in order to conserve at most the threshold energy $E_{Th}$.
In general this definition is useful when dealing with particle
detectors. Furthermore, if a particle detection system is sensitive
to the particle energies until a minimum detectable threshold (or
LLD - \emph{Low Level Discrimination} \cite{gregor}), it is
meaningful to consider the effective range, that is associated only
to particles that carry a minimum threshold energy necessary to
activate the detector.
\subsection{Electrons and Positrons}
\begin{figure}[ht!] \centering
\includegraphics[width=10cm,angle=0,keepaspectratio]{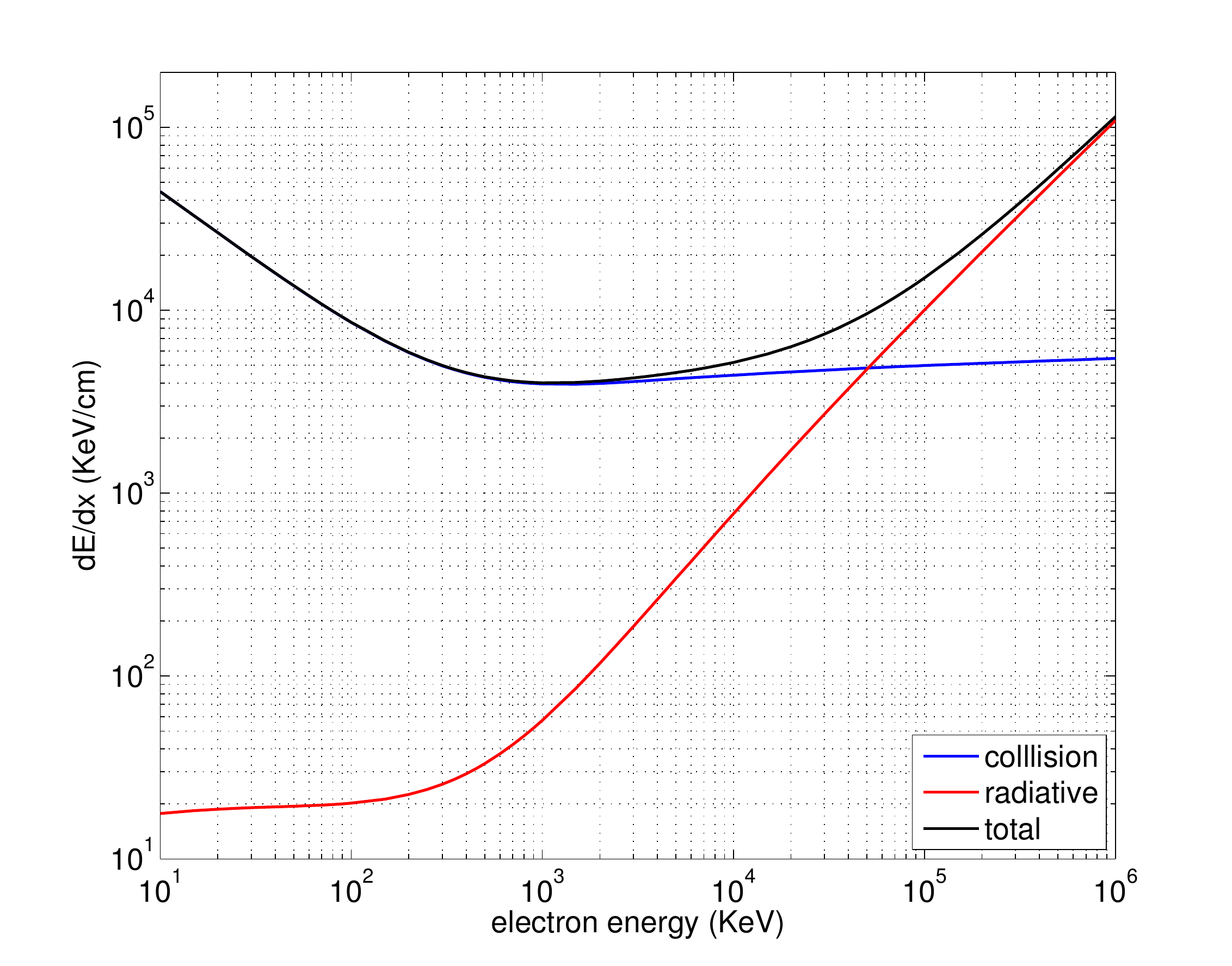}
\caption{\footnotesize The stopping power for electrons in Aluminium
$\rho=2.7g/cm^3$.}\label{fig9iof978jmhs}
\end{figure}

Like heavy charged particles, electrons and positrons also suffer a
collisional energy loss when passing through matter. However,
because of their small mass an additional energy loss mechanism
comes into play: the emission of electromagnetic radiation arising
from scattering in the electric field of a nucleus (bremsstrahlung).
Classically, this can be understood as radiation arising from the
acceleration of the electron as it is deviated from its straight
trajectory by the electrical attraction of the nucleus.
\\ The total energy loss of electrons and positrons, therefore, is composed of two
parts:
\begin{equation}\label{}
\left(\frac{dE}{dx}\right)_{tot}=
\left(\frac{dE}{dx}\right)_{rad}+\left(\frac{dE}{dx}\right)_{coll}
\end{equation}
In Figure \ref{fig9iof978jmhs} the energy loss in radiative and
collisional contributions are plotted for electrons in common
Aluminium ($\rho=2.7\,g/cm^3$). At energies of a few $MeV$ or less,
the radiative loss is still a relatively small factor. However, as
energy increases, the probability of bremsstrahlung rises and it is
comparable to or greater than collision loss.
\\ The Bethe-Bloch formula is still essentially valid, but we have to take into account
two issues. Due to their small mass, the assumption the incident
particle remains undeflected during the collision process is not
valid. Second, for electrons, we are dealing with interactions
between identical particles, and the quantum-mechanical calculation
has to take into account their indistinguishability.
\\ Because of electron's greater susceptibility to multiple
scattering by nuclei, the actual range of electrons is generally
very different from the calculated one obtained from the stopping
power function integration.
\section{Photon interaction}
The behavior of photons in matter is different from that of charged
particles. The probability of single interaction is much lower, but
their effect is much more important. The main interactions of x-rays
and $\gamma$-rays in matter are:
\begin{itemize}
    \item photoelectric effect;
    \item Compton scattering;
    \item pair production.
\end{itemize}
X-rays and $\gamma$-rays are many times more penetrating in matter
than charged particles and a beam of photons is not degraded in
energy as it passes through a thickness of matter but it is only
attenuated in intensity.
\\ The first feature is due to the small cross-section of the three
processes, the second is due to the fact that the three listed
processes remove photons from the beam entirely, either by
absorption or scattering. As a result the photons which pass
straight through are only those which have not suffered any
interactions all. The attenuation suffered by a photon beam can be
expressed by Equation\ref{eqha11}:
\begin{equation}\label{eqrt3}
I(x)=I_0 \, e^{-\mu \, x}
\end{equation}
where $I_0$ is the incident intensity, $x$ is the absorber thickness
and $\mu$ the linear attenuation coefficient.
\\ The linear attenuation coefficient is characteristic of a material
and it is directly related to the total interaction cross-section.
\subsection{Photoelectric effect}
The photoelectric effect involves the absorption of a photon by an
atomic electron with the subsequent ejection of the electron from
the atom. The energy of the outgoing electron is:
\begin{equation}\label{eqrt4}
E_e = \hbar \omega - E_b
\end{equation}
where $\hbar \omega$ is the photon energy and $E_b$ is the electron
binding energy. \\ A free electron can not absorb a photon because
of conservation laws, therefore the photoelectric effect always
occurs on bound electrons with the nucleus absorbing the recoil
momentum.
\\ Figure \ref{fig9654ghg7} shows the linear attenuation coefficient $\mu$
for lead ($\rho=11.34\,g/cm^3$) as a function of the incident photon
energy. The red curve is the contribution given by the photoelectric
effect. In general, rather than the linear attenuation coefficient
$\mu$, one gives the absorption coefficient $\xi$ or the
cross-section $\sigma$; one can be calculated from the other by
using \cite{nelson}:
\begin{equation}\label{}
\mu \left[\frac{1}{cm}\right]= \frac{\sigma \left[cm^2\right] \cdot
N_A \left[\frac{1}{mol}\right] \cdot \rho
\left[\frac{g}{cm^3}\right]}{A \left[\frac{g}{mol}\right]} = \xi
\left[\frac{cm^2}{g}\right] \cdot \rho \left[\frac{g}{cm^3}\right]
\end{equation}
where $\rho$ is the material mass density, $A$ its atomic mass and
$N_A$ is Avogadro's number.
\\Still referring to Figure \ref{fig9654ghg7}, the photoelectric
coefficient value shows discontinuities due to the atomic energy
shells. From the highest energy down to smaller energies, the shells
are called K, L, M, etc. At energies above the highest electron
binding energy of the atom (K-shell), the cross-section is
relatively small but increases as the K-shell energy is approached.
By lowering the energy, the cross-section drops drastically since
the K-electrons are no longer available for the photoelectric
effect. Below this energy, the coefficient $\mu$ rises again and
dips as the L, M, etc. levels, are passed.
\\ In general when a photon knocks-out an electron this leaves a
vacancy in a specific atomic orbital. The higher electrons tend to
relax to the minimum atomic energy with a consequent emission of an
x-ray owning the electron binding energy $E_b$. Since the
probability for an x-ray to undergo the photoelectric effect is even
lager than for a $\gamma$-ray, in most cases it is absorbed again.
Hence, if a $\gamma$-ray energy spectrum is measured through a
scintillator, the energy measured will be the full incoming
$\gamma$-ray energy because the x-ray will be reabsorbed.
\subsection{Compton scattering}
Compton scattering is the scattering of photons on free electrons.
In matter the electrons are bound; however, if the photon energy is
high with respect to the electron binding energy $E_b$, this latter
can be neglected and the electrons can be considered as essentially
free.
\begin{figure}[ht!] \centering
\includegraphics[width=8cm,angle=0,keepaspectratio]{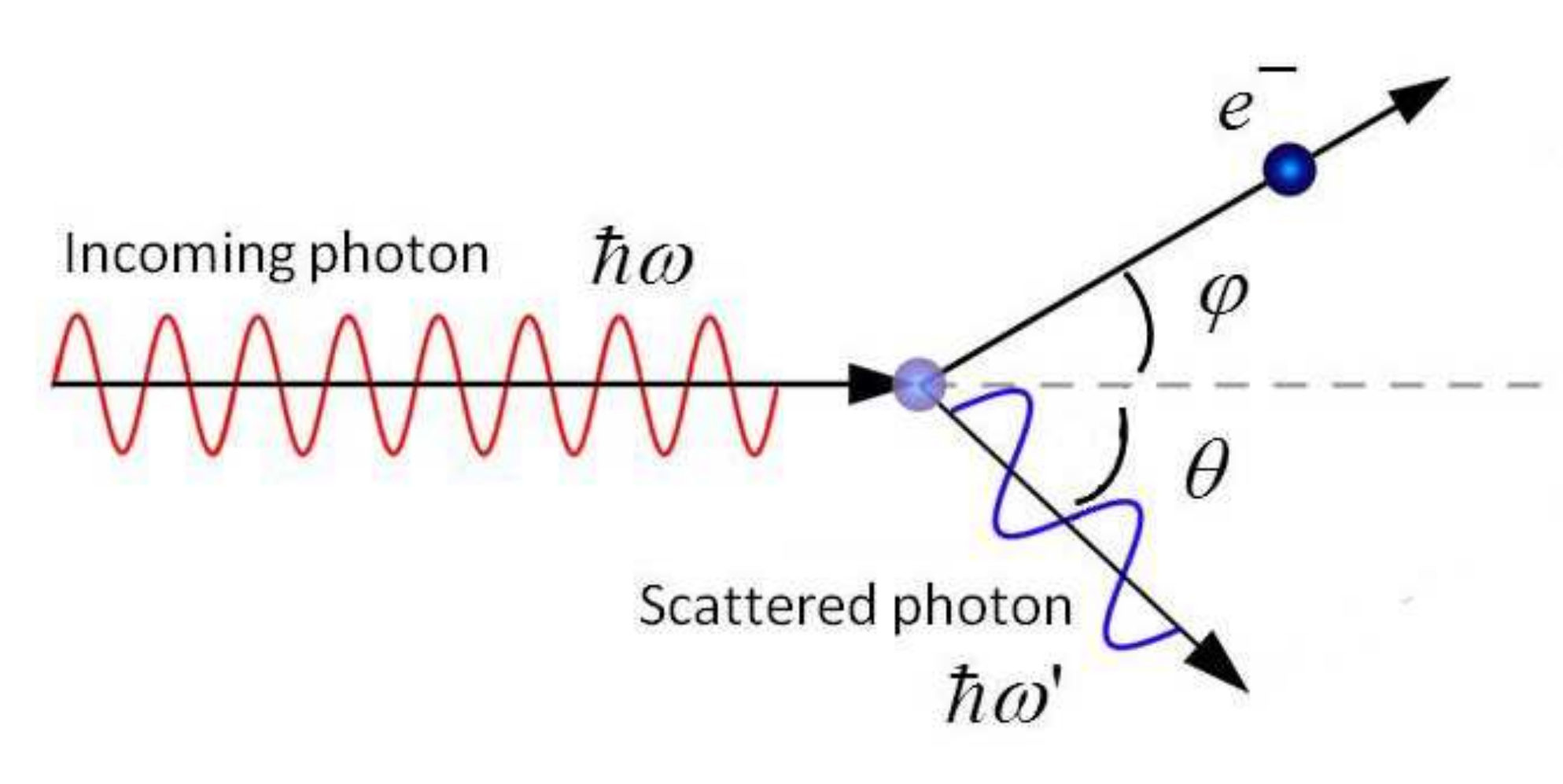}
\caption{\footnotesize Compton scattering
kinematic.}\label{figcompschemkin3}
\end{figure}
\\ Figure \ref{figcompschemkin3} shows the scattering process. By
applying the energy and momentum conservation, the following
relation can be obtained:
\begin{equation}\label{eqrt5}
\hbar\omega'= \frac{\hbar\omega}{1+\gamma\left(1-\cos\theta\right)}
\end{equation}
where $\gamma=\hbar\omega/m_e c^2$. The kinetic energy the knocked
electron gains in the process is:
\begin{equation}\label{eqrt6}
E_e=\hbar\omega-\hbar\omega'=
\hbar\omega\frac{\gamma\left(1-\cos\theta\right)}{1+\gamma\left(1-\cos\theta\right)}
\end{equation}
Figure \ref{fig9654ghg7} shows, in black, the linear absorption
coefficient for Compton scattering in lead as a function of the
incoming photon energy.
\\ Note from Equation \ref{eqrt6} that the maximum energy is
transferred from the photon to the knocked electron when
$\theta=\pi$, that results into
$E_{e(max)}=\hbar\omega\frac{2\gamma}{1+2\gamma}$; this is called
the Compton edge. The latter is always smaller than the energy an
electron can acquire if a photoelectric interaction occurs. As a
result, the continuous spectra due to Compton interactions and the
peak shaped spectrum due to photoelectric interactions will always
be well separated in energy. Figure \ref{figspettornaimio7} shows a
$\gamma$-ray energy spectrum when a NaI scintillator is exposed to
three different $\gamma$-ray sources: a neutron induced source of
$480\,KeV$ photons, a $^{60}Co$ source that emits almost only two
radiations above $1\,MeV$ and a $^{22}Na$ source that emits at
$511\,KeV$ and $1274\,KeV$.
\begin{figure}[ht!] \centering
\includegraphics[width=10cm,angle=0,keepaspectratio]{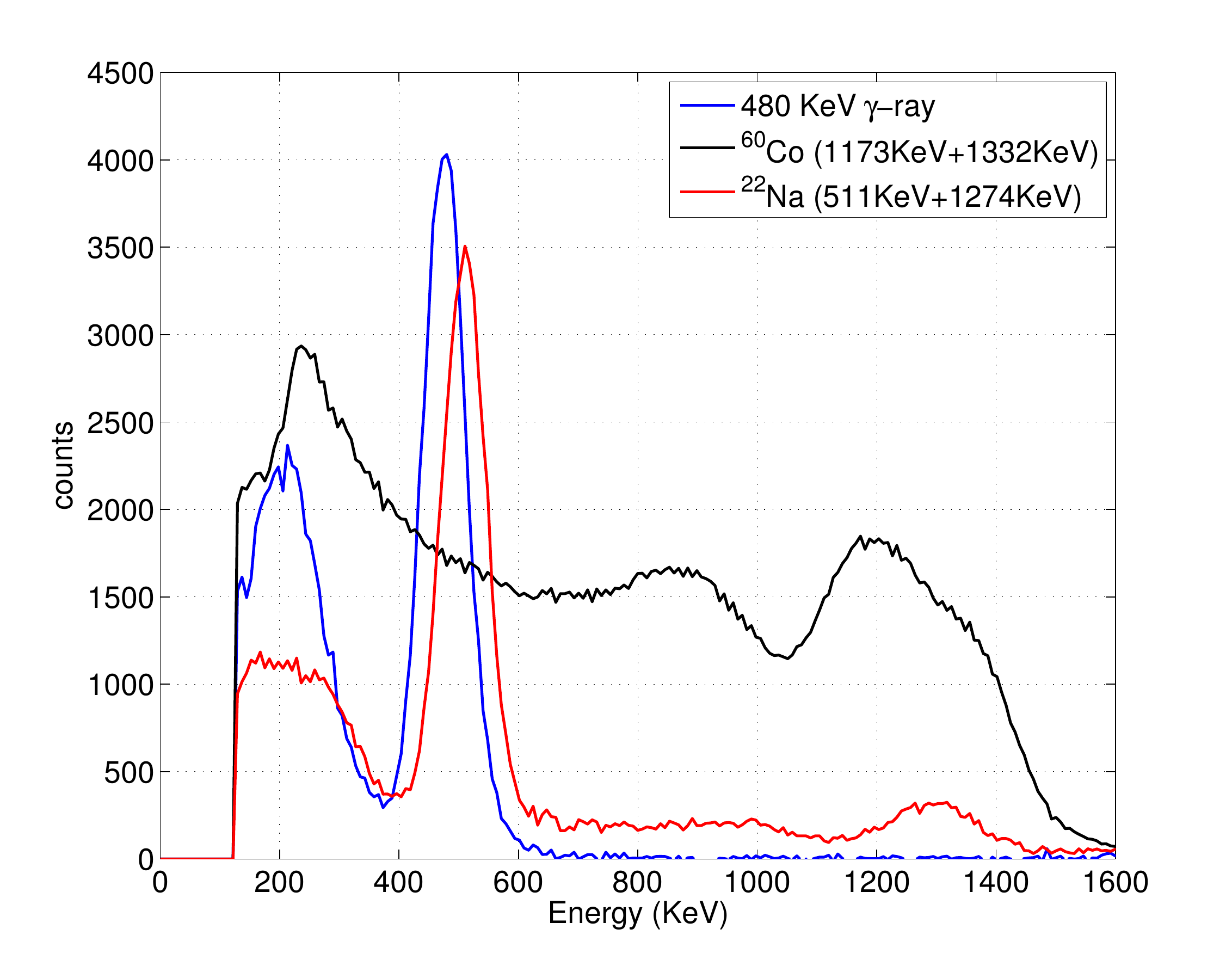}
\caption{\footnotesize Measured $\gamma$-ray spectra with a NaI
scintillator.}\label{figspettornaimio7}
\end{figure}
\\ In the spectra are clearly visible the photo-peaks and the
continuous spectrum extended until the relative Compton edge.
\\ The scintillator energy calibration can be performed by exposing
it to a $\gamma$-ray source. A calibration source should present at
least two photons of well defined and well distinguished energies.
$^{22}Na$ is a suitable calibration source thanks to its $511\,KeV$
and $1274\,KeV$ $\gamma$-rays. The calibration is done by applying a
linear scaling between the two photo-peaks.
\subsection{Pair production}
The process of pair production involves the transformation of a
photon into an electron-positron pair. In order to conserve
momentum, this can only occur in presence of a third body, usually a
nucleus. Moreover, to create the pair, the photon must have at least
the energy of $1022\,KeV$; being $511\,KeV$ the rest mass of an
electron (or positron). In Figure \ref{fig9654ghg7} the pair
production coefficient is plotted in green, and indeed it vanishes
at $1022\,KeV$.
\subsection{Total absorption coefficient and photon attenuation}
\begin{figure}[ht!] \centering
\includegraphics[width=10cm,angle=0,keepaspectratio]{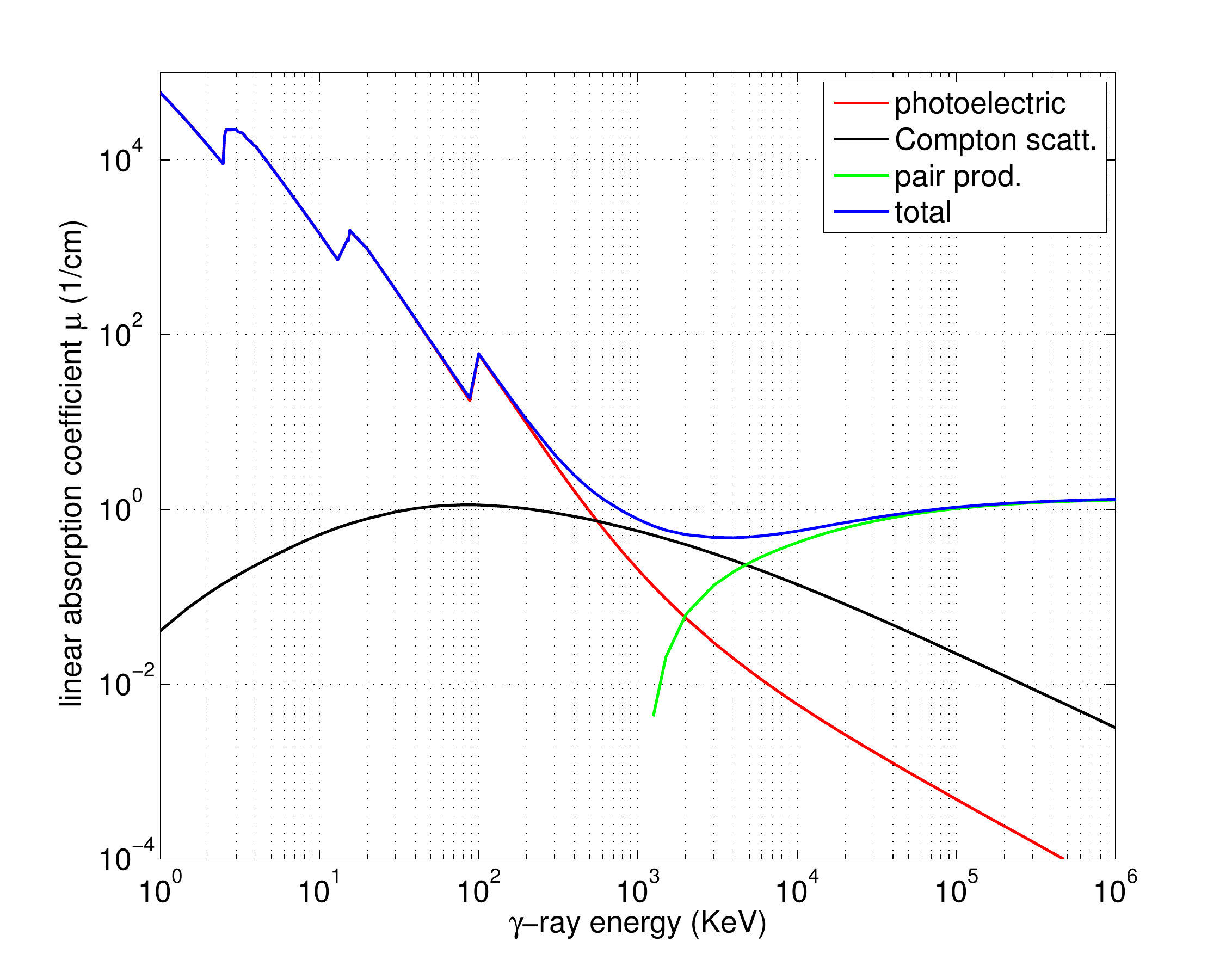}
\caption{\footnotesize The linear absorption coefficient $\mu$ in
lead as a function of the $\gamma$-ray energy for different
processes.}\label{fig9654ghg7}
\end{figure}

The total probability for a photon to interact with matter is the
sum of the individual linear attenuation coefficients (or
cross-sections, or absorption coefficients) we mentioned above.
\begin{equation}\label{}
\mu_{tot}=\mu_{ph.el.}+\mu_{Compt.}+\mu_{pair}
\end{equation}
In Figure \ref{fig9654ghg7} the total linear attenuation coefficient
is plotted in blue.
From Figure \ref{fig9654ghg7} we notice that the three possible
interactions of photons dominate in three different energy regions.
At low energies the photoelectric effect is dominant, Compton
scattering is the main effect at energies around $1\,MeV$, and pair
production prevails at higher energies.
\section{Sources, activity and decay}\label{gsact90}
\subsection{Activity}
\begin{figure}[ht!] \centering
\includegraphics[width=9cm,angle=0,keepaspectratio]{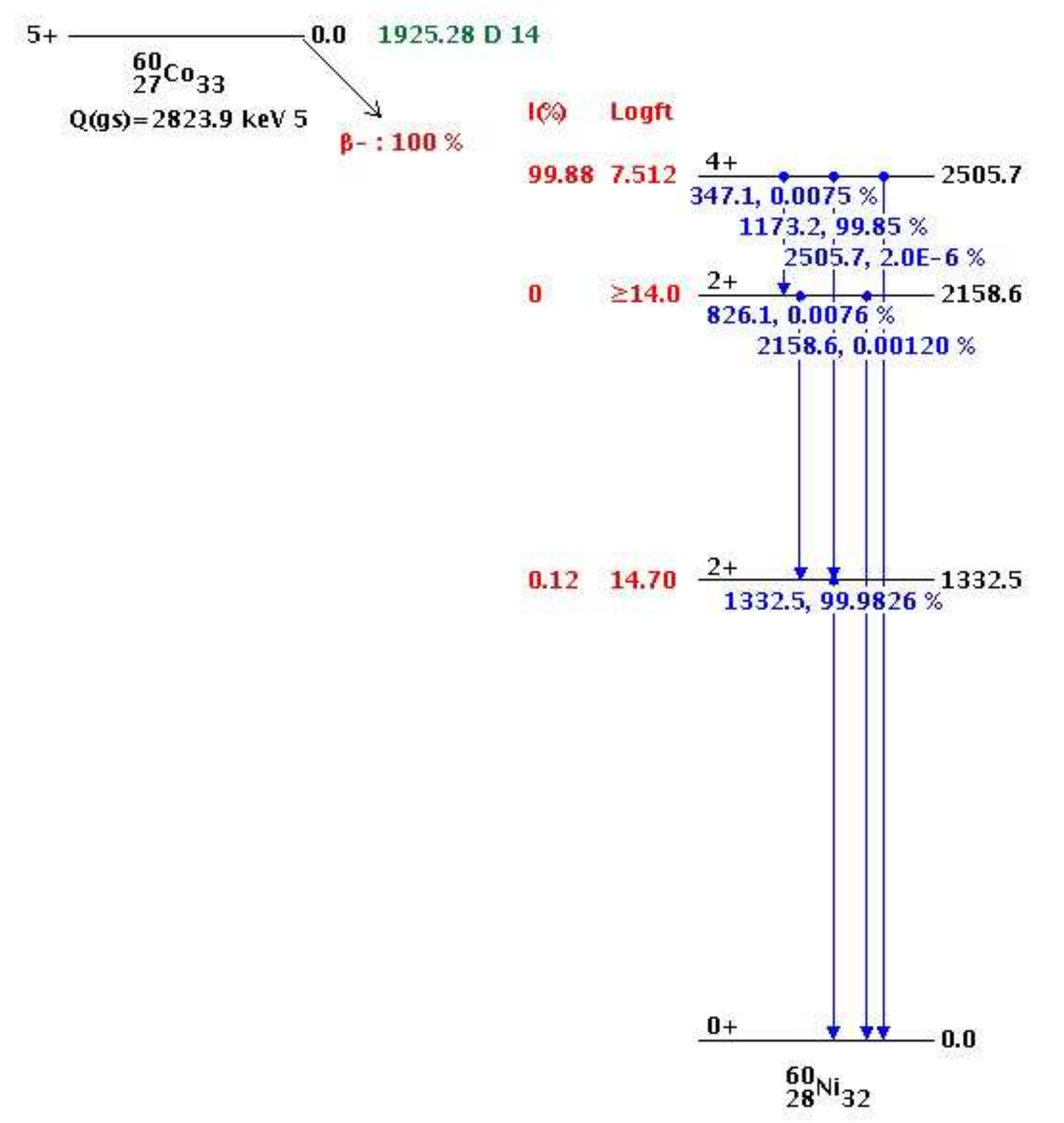}
\caption{\footnotesize $^{60}Co$ decay level
scheme.}\label{figexaco60}
\end{figure}

The activity of a radioisotope source is defined as its rate of
decay where $\lambda_d$, the decay constant, is the probability per
unit time for a nucleus to decay \cite{knoll}:
\begin{equation}\label{eqrt1}
A(t)=\left|\frac{dN(t)}{dt}\right|=\left|-\lambda_d\,N(t)\right|
\quad \Longrightarrow \quad N(t)=N_0 e^{-\lambda_d \, t}
\end{equation}
where $N$ and $N_0$ is the number of radioactive nuclei at the time
$t$ and $t=0$ respectively. Activity can be measured in $Ci$,
defined as $3.7\cdot10^{10}$ disintegrations per second, or in $Bq$
which is its SI equivalent and that is $1$ disintegration per
second. Equivalently the activity can be expressed as a function of
the decay constant $\lambda_d$, the average lifetime
$\tau=1/\lambda_d$ or the half-life $t_{1/2}=\tau\cdot\ln2$.
\\ By knowing the activity at time $t=0$ ($A_0$) it is possible to calculate
the activity at the time $t$ by:
\begin{equation}\label{eqrt2}
A(t)=\lambda_d \cdot N(t)=\lambda_d\cdot N_0 e^{-\lambda_d \,
t}=A_0\cdot e^{-\lambda_d \, t}
\end{equation}
It should be emphasized that activity measures the source
disintegration rate, which is not synonymous with the emission rate
of radiation produced in its decay. Frequently, a given radiation
will be emitted in only a fraction of all the decays, thus a
knowledge of the decay scheme of the particular isotope is necessary
to infer a radiation emission rate from its activity. Moreover, the
decay of a radioisotope may lead to a daughter product whose
activity also contributes to the radiation yield from the source.
\\ Figure \ref{figexaco60} shows the decay scheme for $^{60}Co$, in
green is shown its half life (in days), in black the radioisotopes
and the energy of the levels (in $KeV$) and in blue the
characteristic $\gamma$-ray energies and their probability.
\\ E.g. if we imagine to deal with a source of activity $A=10^4\,Bq$ at
the present time, its emission rate of $\gamma$-rays of energy
$1173.2\,KeV$ is given by $0.9985\cdot10^4\,Bq$.
\subsection{Portable neutron sources}\label{slowdown34ttc4}
The most common portable source of neutrons is obtained by the
bombardment of $Be$ with $\alpha$-particles emitted by an other
element, e.g. $Am$. $\alpha$-particles emitted by the $^{241}Am$
have an energy greater than $5\,MeV$ that is sufficient to overcome
the Coulomb repulsion between the particle and the nucleus
(Beryllium is used because of its low Coulomb force). The reaction
is the following:
\begin{equation}\label{}
\begin{array}{ll}
\alpha + ^9Be \rightarrow & ^{12}C + n
\end{array}
\end{equation}
The resulting neutrons emitted are fast and they have to be slowed
down if thermal neutrons are needed. Most of the $\alpha$-particles
are simply stopped in the target, and only 1 in about $10^4$ reacts
with a $Be$ nucleus. About $70$ neutrons are produced per $MBq$ of
$^{241}Am$. This process to produce neutrons is very inefficient and
an AmBe source has a $\gamma$-ray emission which is orders of
magnitude higher than the neutron yield.
\\ The slowing down of fast neutrons is known as moderation. When a
fast neutron enters into matter it scatters on the nuclei, both
elastically and inelastically, losing energy until it comes into
thermal equilibrium with the surrounding atoms. At this point it
diffuses through matter until it is captured or enters into other
type of nuclear reaction. Elastic scattering is the principal
mechanism of energy loss for fast neutrons. If we consider a single
collision between a neutron, of unity atomic mass, with velocity
$v_0$ and a rest nucleus with an atomic mass $A$; in the
center-of-mass system, the average velocity of the neutron $v_n$
after the collision is:
\begin{equation}\label{}
v_n=\frac{A}{A+1}v_0
\end{equation}
Note that the maximum velocity loss is attained when the neutron
scatter on light nuclei, i.e. protons ($A=1$). Intuitively, the
lighter the nucleus the more recoil energy it absorbs from the
neutron. This implies the slowing down of neutrons is most efficient
when light nuclei are used.
\\ With $^{12}C$ as a moderator of $1\,MeV$ neutrons slowing down to
thermal energies would require about $110$ collisions. For $H$,
instead, only $17$.
\\ When thermal neutron sources are needed, the AmBe core is
surrounded by polyethylene in order to moderate fast neutrons.
\section{Neutron interaction}\label{neuinte45}
\subsection{Neutron interactions with matter}
\begin{figure}[ht!] \centering
\includegraphics[width=9cm,angle=0,keepaspectratio]{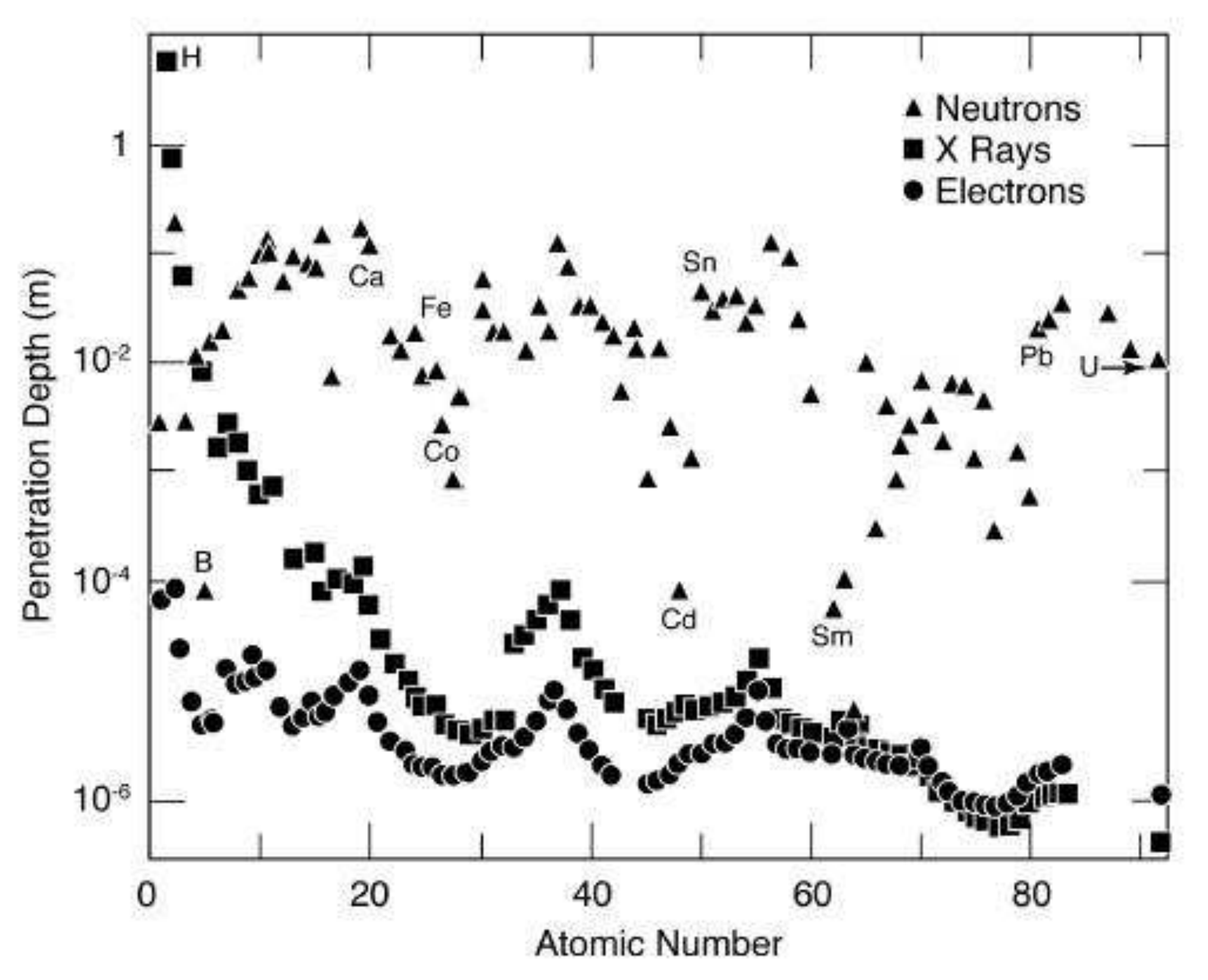}
\caption{\footnotesize The plot shows how deeply a beam of
electrons, x-rays, or thermal neutrons ($1.4$\AA) penetrates a
particular element in its solid or liquid form before the beam
intensity has been reduced by a factor $1/e$.}\label{absne557u}
\end{figure}

Because of their lack of electric charge, neutrons are not subject
to Coulomb forces. Their principal means of interaction is through
the strong force with nuclei. These interactions are much rarer, in
comparison with charged particles, because of the short range of
this force. Purely from a classical point of view, neutrons must
come within $\sim10^{-15}\,m$ of the nucleus before anything can
happen, and since matter is mainly empty space, it is not surprising
that neutrons are very penetrating particles. Figure \ref{absne557u}
shows the penetration depth of a beam of electrons, x-rays, or
thermal neutrons as a function of the atomic number of the element.
The main characteristic of charged particles which distinguishes
them from photons and neutrons is that their penetration into a
material cannot be described by an exponential function. Although
there is a finite probability that a photon or neutron, however low
in energy, can penetrate to a large depth, this is not the case for
a charged particle. There is always a finite depth beyond which a
charged particle will not travel. \\The peculiarity of neutrons to
be weakly absorbed by most materials makes it a powerful probe for
condensed matter research. On the other hand, this is also the
reason why it is rather difficult to build efficient neutron
detectors, particularly if position sensibility is required.
\\The energy $E$ of a neutron, in the non-relativistic limit, can be
described in terms of its wavelength $\lambda$ through the De
Broglie relationship:
\begin{equation}\label{DeBroglie}
\lambda = \frac{2\pi \hbar}{m_n v} \qquad \Rightarrow \qquad E =
\frac{1}{2} m_n v^2 = \frac{\pi \hbar^2}{m_n\lambda^2}
\end{equation}
where $\hbar$ is the Planck's constant and $m_n$ is the mass of the
neutron. In this manuscript we will talk about neutron velocity $v$,
kinetic energy or wavelength equivalently. The energy classification
of neutrons is shown in Table \ref{nenergy45}.
\begin{table}[ht!]
\centering
\begin{tabular}{|l|c|c|c|}
\hline \hline
Energy classification & kinetic energy E (eV) & wavelength (\AA)& velocity (m/s) \\
\hline
ultra cold (UCN)       & $E < 3\cdot10^{-7}$    & $\lambda > 520$                  &  $v < 7.5$\\
very cold  (VCN)       & $3\cdot10^{-7}<E < 5\cdot10^{-5}$    & $520>\lambda > 40$ & $7.5<v<99$\\
cold         & $5\cdot10^{-5}<E < 0.005$    & $40>\lambda > 4$ &$99<v<990$\\
thermal      & $0.005<E<0.5$  & $4>\lambda>0.4$ &$990<v<9900$\\
epithermal   & $0.5<E< 10^3$  & $0.4>\lambda>0.01$&$9900<v<4.4\cdot10^{5}$ \\
intermediate & $10^3<E<10^5$  & $0.01>\lambda>0.001$ &$4.4\cdot10^{5}<v<4.4\cdot10^{6}$\\
fast         & $10^5<E<10^{10}$ & $0.001>\lambda>3\cdot10^{-6}$ & $4.4\cdot10^{6}<v<1.3\cdot10^{9}$\\
\hline \hline
\end{tabular}
\caption{\footnotesize Energy classification of neutrons.}
\label{nenergy45}
\end{table}
\\ Neutron states of motion owe their name to the temperature because they carry
energies that are comparable with the daily life temperatures. E.g.
thermal neutrons are those with energies around $k_BT$ (where $k_B$
is the Boltzmann constant and $T$ is the absolute temperature
corresponding to $20\,^{\circ}C$) or about $25 meV$.
\\ When the neutron interacts with an individual nucleus, it may undergo a variety of nuclear
processes depending on its energy. Among these are:
\begin{itemize}
    \item Elastic scattering from nuclei, i.e. $A(n,n)A$. This is the
    principal mechanism of energy loss for neutrons.
    \item Inelastic scattering, e.g. $A(n,n')A^{*}$. In this
    reaction, the nucleus is left in an excited state which may
    later decay by $\gamma$-ray or some other form of radiative
    emission. In order for the inelastic reaction to occur, the
    neutron must have sufficient energy to excite the nucleus,
    usually order of $1\,MeV$ or more. Below this energy threshold,
    only elastic scattering may occur.
    \item Radiative neutron capture, i.e. $n+(Z,A)\rightarrow \gamma +
    (Z,A+1)$. In general, the cross-section for neutron capture goes
    approximately as $1/v$ with $v$ the neutron velocity. Therefore, absorption
    is most likely at low energies. Depending on the element, there
    may also be resonance peaks superimposed upon the $1/v$
    dependence. At these energies the probability of neutron capture
    is greatly enhanced.
    \item Other nuclear reactions, such as $(n,p)$, $(n,d)$, $(n,\alpha)$,
    $(n,t)$, etc. in which the neutron is captured and charged
    particles are emitted. These generally occur in the $eV$ to $KeV$
    region. Like the radiative capture reaction, the cross-section
    falls as $1/v$. Resonances may also occur depending on the
    element.
    \item Fission, i.e. $(n,f)$. Again this is most likely at thermal
    energies.
    \item High energy hadron shower production. This occurs only for
    very high energy neutrons with $E>100\,MeV$.
\end{itemize}
The neutron has a net charge of zero and a rest mass slightly
greater than that of the proton. $\beta$ decay is therefore possible
according to $ n \rightarrow p + e^- + \bar{\nu_{e}}$. The maximum
electron energy is $781.32\,KeV$ and half-life in free space is
$(11.7 \pm 0.3)$ minutes.
\\ The total probability for a neutron to interact in matter is
given by the sum of the individual cross-sections listed above (as
long as interference effects are not significant), i.e.:
\begin{equation}\label{eqyu2}
\sigma_{tot} = \sum_{i} \sigma_{i}=
\sigma_{elastic}+\sigma_{inelastic}+\sigma_{capture}+\cdots
\end{equation}
If we multiply $\sigma_{tot}$ by the atomic density we obtain the
macroscopic cross-section $\Sigma$ of which the inverse is the mean
free path length $\eta$ as indicated in Equation \ref{eqha11}:
\begin{equation}\label{}
\frac 1 \eta = \Sigma_{tot} = n\cdot \sigma_{tot}= \frac{N_A\cdot
\rho}{A}\sigma_{tot}
\end{equation}
where $\rho$ is the material mass density, $A$ its atomic number and
$N_A$ the Avogadro's number. In general $\Sigma$ is a function of
$\lambda$ because $\sigma$ depends on the neutron energy and it has
units of an inverse length.
\\ In analogy with photons, a beam of neutrons passing through
matter will be exponentially attenuated. The probability for a
neutron of wavelength $\lambda$ to interact with a nucleus of the
matter at depth $x$ in a slab of thickness $dx$, is given by:
\begin{equation}\label{}
K(x,\lambda)\,dx=\Sigma \, e^{-x \Sigma(\lambda)}\,dx
\end{equation}
Consequently, by integration over a finite distance $d$, we obtain
the number of neutrons that have interacted:
\begin{equation}\label{eqsdfgg8}
\frac {N(d)}{N_0} = \int_0^d  dx \, K(x) = \int_0^d dx \, \Sigma
e^{-x \Sigma} = 1 - e^{-d \Sigma}
\end{equation}
with $N_0$ the initial incoming neutron flux. The percentage of
neutrons which pass the layer of thickness $d$ is then $e^{-d
\Sigma}$.
\subsection{Elastic scattering}
\begin{figure}[ht!] \centering
\includegraphics[width=10cm,angle=0,keepaspectratio]{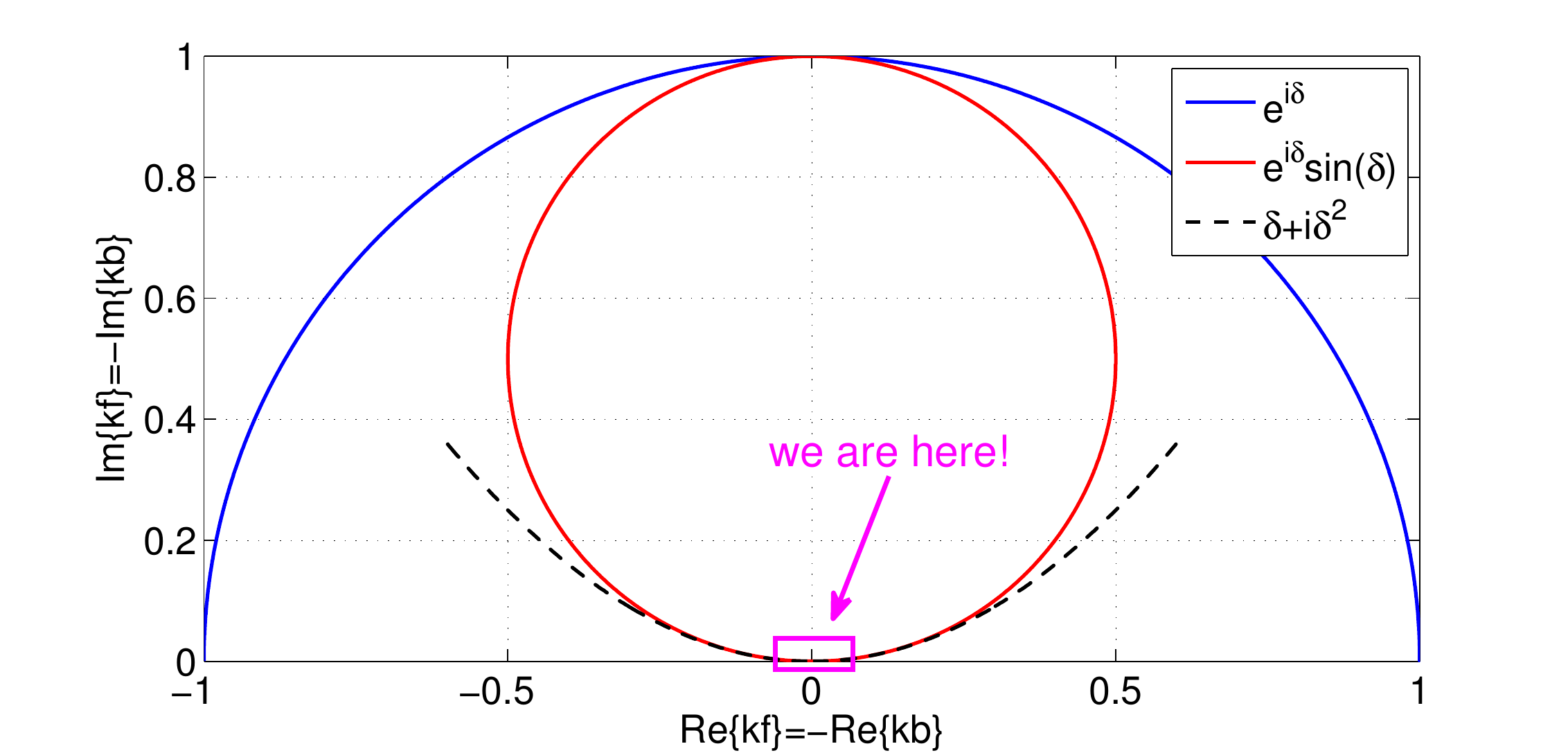}
\caption{\footnotesize The value of the diffusion amplitude $f\cdot
k$ in the complex plane for $\delta\in[0,\pi]$ (in red).
$e^{i\delta}$ in blue and the parabola $\delta + i\,\delta^2$ in
black.}\label{wacgqgqf5}
\end{figure}

We consider now the neutron elastic scattering by a single nucleus
in a fixed position. From general scattering theory \cite{coen},
\cite{schiff} the incoming particle can be described by a plane wave
which interacts, through a potential $V(\bar{r})$, with the nucleus.
The resulting wave-function will be a superposition of the incoming
wave and a spherically diffused wave. This is a solution of the
stationary Schr\"{o}dinger equation.
\begin{equation}\label{eqaghfd8}
\left[-\frac{\hbar^2}{2 m}\nabla^2 + V(\bar{r})\right]\Psi = E \Psi
\end{equation}
In a scattering experiment one wants to determine the probability,
i.e. the cross-section, for the diffusion process to happen. Hence,
one is interested in the asymptotic ($\bar{r}\rightarrow \infty$)
behavior of such a solution:
\begin{equation}\label{eqsdl1}
\Psi_{(\bar{r}\rightarrow \infty)} \sim
e^{ikz}+f(\theta,\phi)\,\frac{e^{ikr}}{r}
\end{equation}
where $f(\theta,\phi)$ is the diffusion amplitude and it depends on
the interaction potential $V(\bar{r})$. One can wonder whether there
are constraints on the form of $f(\theta,\phi)$ for $\Psi$ to be a
solution of the Schr\"{o}dinger equation.
\\ In order to evaluate the cross-section of the process one should
study the diffusion of a wave-packet hitting a potential
$V(\bar{r})$. A simpler way is to calculate the cross-section from
the incident and diffused probability currents:
\begin{equation}\label{eqsdl2}
\bar{J}(\bar{r},t)=-\frac{i \hbar}{2m}\left( \Psi^\ast \frac{d
\Psi}{d\bar{r}}-\Psi \frac{d \Psi^\ast}{d\bar{r}}\right)
\end{equation}
For the incident plane wave and for the scattered wave individually,
it results in:
\begin{equation}\label{eqsdl3}
\bar{J}_i=\frac{\hbar k}{m}\,\hat{u}_z, \qquad \bar{J}_s=\frac{\hbar
k}{m}\frac{1}{r^2}|f(\theta,\phi)|^2 \,\hat{u}_r
\end{equation}
which can be interpreted as the number of particles flowing through
a unity surface per unity time. The scattered wave probability
current, for $\bar{r}\rightarrow \infty$, can be considered only
radial.
\\ The cross-section per unity of solid angle is the ratio between
the number of particles that have been scattered over the number of
incoming particles per unity time over the surface
$d\bar{S}=r^2\,d\Omega$:
\begin{equation}\label{eqsdl4}
d\sigma=\frac{ \frac{\hbar k}{m}\frac{1}{r^2}|f(\theta,\phi)|^2 \,
r^2 d\Omega} {\frac{\hbar k}{m}}=|f(\theta,\phi)|^2 \,d\Omega \quad
\Rightarrow \quad \frac{d\sigma}{d\Omega}=|f(\theta,\phi)|^2
\end{equation}
No assumptions on the potential have been made so far. We consider
now the case of a central potential, $V(\bar{r})=V(r)$. The
hamiltonian commutes with the angular momentum operator, hence there
exist stationary states of well defined energy and angular momentum,
i.e. a common eigenfunction base for both operators. The angular
dependence of these functions will be the spherical harmonics. A
plane wave can be written as a superposition of spherical waves:
\begin{equation}\label{eqsdl5}
e^{ikz} = \sum_{l=0}^{\infty}i^l\sqrt{4\pi \left(
2l+1\right)}j_l(kr)Y_l^0(\theta)
\end{equation}
where the functions $j_l(kr)$ are the spherical Bessel functions and
the $Y_l^0(\theta)$ are the spherical harmonics with $m=0$ because
$\bar{k}$ has been chosen along $z$, hence they do not depend on
$\phi$. For large $r$:
\begin{equation}\label{crgyt564bisfgsve}
j_l(kr)\sim\frac{1}{kr}\sin\left( kr-l\frac{\pi}{2}\right)
\end{equation}
When a plane wave interacts with a central potential, it introduces
a phase shift in the scattered wave amplitudes of each of the
harmonic terms. This can be shown as follows. The asymptotic
behavior of the radial stationary Schr\"{o}dinger equation with a
central potential, assuming $V(r\rightarrow \infty)=0$, is:
\begin{equation}
\left[-\frac{d^2}{dr^2} + k^2\right]u_{k,l}(r) = 0
\end{equation}
with solutions
\begin{equation}\label{crgw5c54w54gcw4g599}
u_{k,l}(r) = Ae^{ikr}+Be^{-ikr}
\end{equation}
for large $r$. This solution contains as well the incoming as the
outgoing particle flux. Unitarity requires $|A|=|B|$, as there is in
a stationary case no destruction or creation of particles within a
large sphere. Hence:
\begin{equation}\label{crgyt564}
u_{k,l}(r) = |A|\left(e^{ikr}e^{i\varphi_A}+
e^{-ikr}e^{i\varphi_B}\right)=C\sin\left(kr-l\frac{\pi}{2}+\delta_l\right)
\end{equation}
with $\delta_l=l\frac{\pi}{2}-\frac{\varphi_B-\varphi_A}{2}$ chosen
to have $\delta_l=0$ when $V(r)=0$, to find the asymptotic behavior
of the spherical Bessel functions of the plane wave expansion as
given in Equation \ref{crgyt564bisfgsve}. Note that $\delta_l$ is a
real quantity.
\\ The Equation \ref{crgyt564} is the solution to the radial part of
Schr\"{o}dinger's equation; the complete asymptotic solution, for
$r\rightarrow \infty$ and considering the spherical waves is:
\begin{equation}\label{crgyt565}
\Phi_{k,l,m}(\bar{r}) = D\,
\frac{\left(e^{-ikr}e^{il\frac{\pi}{2}}-e^{ikr}e^{-il\frac{\pi}{2}}\right)+e^{ikr}e^{-il\frac{\pi}{2}}\cdot\left(1-e^{2i\delta_l}\right)}{2ikr}Y_l^m(\theta,\phi)
\end{equation}
We can interpret the two terms as follows: the incoming wave is a
free particle, as it approaches the region where the potential
increases, it is more and more perturbed by the potential. When it
is scattered, i.e. is the outgoing wave, it has accumulated a phase
shift $2\delta_l$ with respect to the outgoing free wave that it
would be if the potential were $V=0$.
\\ By comparing Equation \ref{eqsdl1} and Equation \ref{crgyt565}, and
using Equation \ref{eqsdl5}, the diffusion amplitude
$f(\theta,\phi)$, in terms of spherical waves, is:
\begin{equation}\label{eqsdl6}
f(\theta) = \frac{1}{k}\sum_{l=0}^{\infty}i^l\sqrt{4\pi \left(
2l+1\right)}e^{i\delta_l}\sin(\delta_l)Y_l^0(\theta)
\end{equation}
The differential scattering cross-section is:
\begin{equation}\label{eqsdl7}
\frac{d\sigma}{d\Omega}=|f(\theta)|^2=\frac{1}{k^2}\left|\sum_{l=0}^{\infty}\sqrt{4\pi
\left( 2l+1\right)}e^{i\delta_l}\sin(\delta_l)Y_l^0(\theta)
\right|^2
\end{equation}
Note that for the wave-function in Equation \ref{eqsdl1} to be a
solution of the Schr\"{o}dinger's equation $f(\theta)$ has to
satisfy Equation \ref{eqsdl6}. Not all the forms for $f$ are
admitted.
\\ We consider now the interaction of neutrons with nuclei.
The nuclear forces which cause the scattering, we recall, have a
range of about $10^{-15}\,m$ while the wavelength of a thermal
neutron is of the order of $10^{-10}\,m$, thus much larger than the
range of those forces. On the scale of a wavelength the potential is
non zero only in a very small region. The potential is central and
can be written as a three-dimensional Dirac's delta of intensity
$a$, which is a real constant:
\begin{equation}\label{eqsdl88}
V(\bar{r})= a\,\delta(\bar{r})
\end{equation}
The multi-pole development of a $\delta$-distribution being limited
to $l=0$, we can consider the only spherical wave to undergo a phase
shift to be the component $Y_0^0(\theta)=\frac{1}{2\sqrt{\pi}}$.
Hence, the scattering amplitude and the cross-section become
(Equations \ref{eqsdl6} and \ref{eqsdl7}):
\begin{equation}\label{eqsdl8}
f = \frac{1}{k}e^{i\delta}\sin(\delta), \qquad
\frac{d\sigma}{d\Omega}=\frac{1}{k^2}\sin^2(\delta)
\end{equation}
The outgoing scattered wave amplitude has to satisfy the Equation
\ref{eqsdl8} with $\delta\in \Re$. Not all the complex plane is
accessible for $f$, but only the circle $e^{i\delta}\sin(\delta)$,
shown in red in Figure \ref{wacgqgqf5}.
\\ In neutron scattering the incoming
neutron can be described as a plane wave and the wave-function of
the scattered neutrons at the point $\bar{r}$ can be written in the
form:
\begin{equation}\label{eqsdl9}
\Psi^{scatt.}_{(\bar{r}\rightarrow \infty)} \sim
e^{ikz}-b\,\frac{e^{ikr}}{r}
\end{equation}
where $r$ is the module of the vector $\bar{r}$ and $b$ is a
constant independent of the polar angles. The minus sign is a
standard convention. If one considers the Fermi pseudo-potential:
\begin{equation}\label{eqyu1}
V(\bar{r})=\frac{2\pi\hbar^2}{m_n}\,b\,\delta(\bar{r})
\end{equation}
and one uses the Born approximation one will find Equation
\ref{eqsdl9} as the solution.
\\ The quantity $b$ is known as the \emph{scattering length} \protect{\footnote{The scattering length relates to a fixed nucleus
and it is known as the bound scattering length. If the nucleus is
free, the scattering must be treated in the center-of-mass system.}}
and depends on the nucleus. Even though in strict potential
scattering, $b$ should be independent of incident neutron energy, in
general scattering theory, it can: this happens when there are
nuclear resonances. The scattering length are experimentally
determined and most $b$ values are of the order of a few $fm$.
\\ The value of $b$ does not only depend on the particular nucleus,
but on the spin state of the nucleus-neutron system. The neutron has
spin $\frac 1 2$. Suppose the nucleus has spin $I$, not zero. Hence
the spin of the system can be either $I+\frac 1 2$ or $I-\frac 1 2$.
Each spin state has its own value of $b$. Every nucleus with
non-zero spin has two values of the scattering length. If the
nucleus spin is zero, the system nucleus-neutron can only have spin
$\frac 1 2$, there is only one value of $b$.
\\ The values for $b$ are determined experimentally, because the
lack of a proper theory. \\ The $b$ is an empirical quantity known
for most nuclei, varying strongly across the periodic table and
often varying sharply between isotopes of the same element. Most
materials have a positive $b$; therefore in a positive potential a
neutron has less kinetic energy and hence a longer wavelength
(opposite to light where the wavelength shortens). This quantity
defines the nature of the neutron-nucleus interaction: whether it is
attractive or repulsive and it also determine the strength of the
interaction. Moreover, this is a specific quantity that strongly
depends on the target nucleus and it can even depend on the neutron
energy, e.g. in the case of $^{113}Cd$ resonances occur as well.
\\ All the neutron scattering theory is based on this simple
assumption that the neutron-nucleus interaction can be considered a
Dirac's delta potential and all the information a scattering
experiment can reveal is all related to the quantity $b$.
\\ By comparing Equations \ref{eqsdl1} and \ref{eqsdl9}, $-b$ is the diffusion amplitude $f$:
\begin{equation}\label{eqsdl10}
f = \frac{1}{k}e^{i\delta}\sin(\delta)= - b \quad \Rightarrow \quad
k\,b = - e^{i\delta}\sin(\delta)
\end{equation}
$b\sim10^{-15}\,m$ and, for thermal neutrons $k=10^{10}\,m^{-1}$,
thus the product $k\,b \sim 10^{-5}$ and
$e^{i\delta}\sin(\delta)<<1$. Expanding Equation \ref{eqsdl10} at
the first order for $\delta<<1$ we obtain:
\begin{equation}\label{eqsdl11}
-k\,b=e^{i\delta}\sin(\delta) = (\cos(\delta)+i\,
\sin(\delta))\cdot\sin(\delta)\sim \delta + i\,\delta^2 \sim
-10^{-5}
\end{equation}
The scattering length $b$ for thermal neutron scattering can be
considered to be a real quantity because its imaginary part is
always at least five orders of magnitude smaller. The values of $b$
make the scattering amplitude $f$ to vary in the small range
$\delta\in[-10^{-5},10^{-5}]$, i.e. $f \in \Re$. In Figure
\ref{wacgqgqf5} is shown the approximated behavior for $b$ in
Equation \ref{eqsdl11} which results in a parabola in the complex
plane ($k\,f=-k\,b$). We see that $b$ is essentially real.
\\ The differential cross-section, Equation \ref{eqsdl8},
becomes:
\begin{equation}\label{eqsdl12}
\frac{d\sigma}{d\Omega}=\frac{\delta^2}{k^2}=b^2
\end{equation}
From which the total scattering cross-section can be derived:
\begin{equation}
\sigma_{s}=4\pi \, b^2
\end{equation}
Let us consider now the scattering by a general system of particles.
Its potential is:
\begin{equation}\label{eqal106}
V=\sum_i V_i\left(\bar{r}-\bar{R_i}\right)=\sum_i
V_i\left(\bar{x_i}\right)
\end{equation}
where $\left(\bar{r}-\bar{R_i}\right)=\bar{x_i}$ and
$V_i\left(\bar{r}-\bar{R_i}\right)$ is the potential the neutron
experiences due to the $i-th$ nucleus.
Explicitly it is:
\begin{equation}
V_i(\bar{x_i})=\frac{2\pi\hbar^2}{m_n}\,b_i\,\delta(\bar{x_i})
\end{equation}
where $b_i$ is the scattering length of the nucleus $i-th$.
\\ One can define the average value of $b$ of the system and the average
value of $b^2$ as:
\begin{equation}
\bar{b} = \sum_i \,\nu_i b_i, \qquad  \overline{b^2} = \sum_i
\,\nu_i b_i^2
\end{equation}
where $\nu_i$ is the frequency with which the value $b_i$ occurs in
the system. Note that the system can be made up of the same element
atoms and the $b_i$ would be different for each nucleus. We recall
the scattering length value depends on the spin states of the
neutron-nucleus system. \\ In a scattering experiment a neutron beam
impinges on a target, which is a large amount of nuclei. Any
combination of spins can occur, i.e. the value of $b_i$ is averaged
over a large number of atoms. On the assumption of no correlation
between the $b$ values of different nuclei:
\begin{equation}\label{eqal506}
\begin{aligned}
\overline{b_i b_j} &= \bar{b}^2 \qquad \mbox{if } i\neq j \\
\overline{b_ib_j} &= \overline{b^2} \qquad \mbox{if } i= j
\end{aligned}
\end{equation}
we can calculate the scattering cross-section of a process averaging
over all the nuclei.
\\ It can be shown that the overall cross-section of the scattering
on such a system consists of two terms: a term depending only on the
correlation of the position of each center with itself (incoherent
part) and a term depending on the correlation of position of pairs
of centers (coherent part) \cite{squires}.
The incoherent part will be proportional to $\sigma_i$ and the
coherent to $\sigma_c$ where we have defined:
\begin{equation}\label{eqal509}
\sigma_c = 4\pi \bar{b}^2, \qquad \sigma_i = 4\pi
\left(\overline{b^2}-\bar{b}^2\right)
\end{equation}
the coherent and incoherent scattering cross-sections.
\\ The physical interpretation is as follows: the actual scattering system has
different scattering lengths associated to different nuclei. The
coherent scattering is the scattering the same system would give if
all the $b$ were equal to $\bar{b}$. The incoherent scattering is
the fluctuation we must add to obtain the scattering due to the
actual system. The latter arises from the random distribution of the
deviations of the scattering lengths from their mean value. As it is
completely random, all interference cancels in this incoherent part.
\\ We define:
\begin{equation}\label{scgcwrg5578967}
b_c^2 = \left|\bar{b}^2\right|, \qquad
b_i^2=\left|\overline{b^2}-\bar{b}^2\right|
\end{equation}
Note that averages taken are defined over the system at hand. In
general the scattering system consists of many spin states. We
recall that the scattering length depends on the coupling of the
neutron and nucleus spins. Let's denote with $I$, the nuclear spin
of the system made up of a single isotope. The resulting spin for
the system neutron-nucleus can be either $I+1/2$ or $I-1/2$. We
associate to the two compositions of spins the scattering length
$b_{+}$ and $b_{-}$. If the system is unpolarized, all the possible
combinations of spins can occur and $b_{+}$ gets a weight of
$\frac{I+1}{2I+1}$ while $b_{-}$ gets a weight of $\frac{I}{2I+1}$
in the statistical ensemble.
\\ If the neutron and the nuclei system are now both polarized the
population over which we average is now defined and if only the
coupling $I+1/2$ occurs, the average would be exactly
$\bar{b}=b_{+}$ with variance
$\left(\overline{b^2}-\bar{b}^2\right)=0$.
\\From the definition of the scattering length comes directly the definition of
the various cross-sections already mentioned in Equation
\ref{eqyu2}.
\\ The actual total scattering cross-section is then given by the
sum of the two contributions.
\begin{equation}
\sigma_{s}=\sigma_c + \sigma_i = 4\pi\overline{b^2}
\end{equation}
If the neutron or the nucleus is unpolarized, the cross-section
becomes:
\begin{equation}
\sigma_{s}=\sigma_{c}+\sigma_{i}=4\pi\left|b_c
\right|^2+4\pi\left|b_i \right|^2
\end{equation}
The scattering cross-section does not depend on the neutron energy
if we have potential scattering ($b$ constant).
\\ The mixture of isotopes or different nuclei, denoted by $j$, is
an additional source of incoherence in the scattering of neutrons.
The average scattering length of such a system is:
\begin{equation}
b=\frac{\sum_j b_j n_j}{\sum_j n_j}
\end{equation}
where $n_j$ is the number density of each isotope or nucleus and
spin state.
\subsection{Absorption}
\begin{figure}[ht!] \centering
\includegraphics[width=7.8cm,angle=0,keepaspectratio]{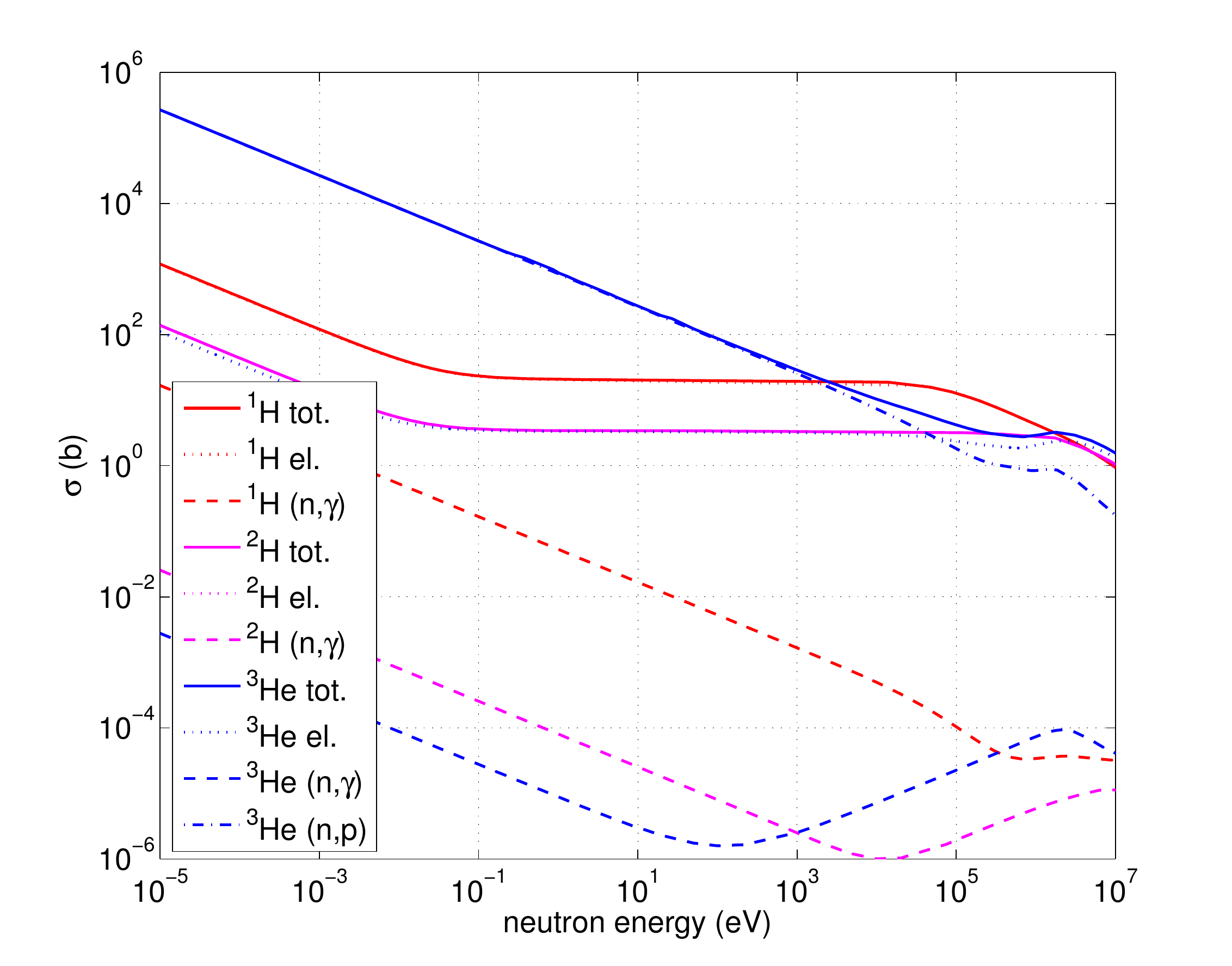}
\includegraphics[width=7.8cm,angle=0,keepaspectratio]{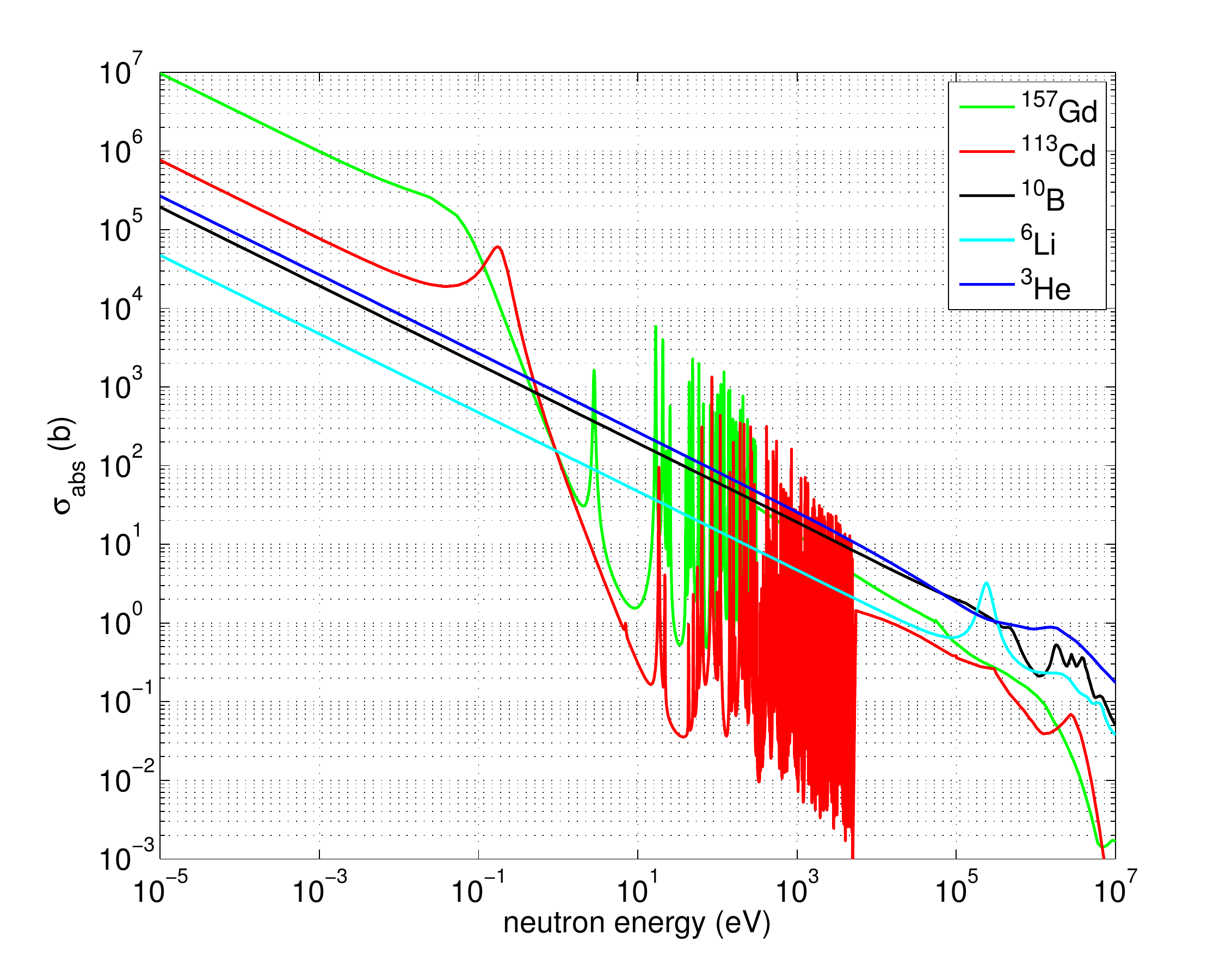}
\caption{\footnotesize Several partial cross-section (expressed in
barn) for $^1H$, $^2H$ and $^3He$ as a function of the neutron
energy (left). Capture cross-sections for $^3He$, $^{10}B$, $^6Li$,
$^{113}Cd$ and $^{157}Gd$ (right).}\label{fgiHHecrosse4}
\end{figure}

Let us consider now the possibility for a neutron to be absorbed.
Let assume the interactions responsible for the absorption to be
invariant under rotation around the origin. As a result, the
scattering amplitude $f$ can still be decomposed in spherical waves.
The phase shift method has to be modified in order to take into
account the absorption. We demonstrated that an incoming plane wave
is shifted by a factor $e^{i\delta}\sin(\delta)=(e^{2i\delta}-1)/2i$
by the potential action. We recall $\delta$ is a real number. Since
$|e^{2i\delta}|=1$, the incoming and outgoing wave amplitudes are
the same: the total probability flux for the plane wave and the
scattered wave is conserved. The total number of particles is
conserved in the scattering process. This was the requirement of
unitarity.
\\ In order to consider absorption this probability is not any more
conserved and we need to add an imaginary part to the phase shift
that results in: $|e^{2i\delta}|<1$. The amplitude of the scattered
wave is smaller than the one of the incoming wave. In Equation
\ref{crgw5c54w54gcw4g599}, $|A|=|B|$ signified that there was no
source or no sink (unitarity) in the large $r$ sphere. Absorption
means that there is a sink, thus $|A|<|B|$: $|\eta|=|A/B|$.
\\ Since $\delta$ is a complex quantity we can denote
$\eta=e^{2i\delta}$ with $|\eta|\leq1$. The asymptotic solutions of
Schr\"{o}dinger's equation (Equation \ref{crgyt565}) become:
\begin{equation}
\Phi_{k,0,0}(r) = A\, \frac{e^{-ikr}e^{il\frac{\pi}{2}}-\eta
e^{ikr}e^{-il\frac{\pi}{2}}}{2ikr}
\end{equation}
where we focus only on the component $l=0$ because, for the
neutron-nucleus interaction, we assume the potential to be central
and a Dirac's delta.
\\ The scattering amplitude and the differential scattering cross-section
(Equation \ref{eqsdl8}) are:
\begin{equation}\label{crgwrtgwgw56954}
f = \frac{1}{k}\frac{\eta-1}{2i}, \qquad
\frac{d\sigma}{d\Omega}=|f|^2=\frac{1}{k^2}\left|\frac{\eta-1}{2i}
\right|^2
\end{equation}
Note that even for a perfect absorbing material, with $\eta=0$,
elastic scattering can and will still occur. The effect is called
\emph{shadow diffusion} and it is a purely quantum effect
\cite{coen}.
\\ The absorption cross-section can be defined in a way similar to the scattering
cross-section, as the number of particles absorbed, i.e. that have
suffered the interaction, over the incident flux per unit time.
\\ To calculate this cross-section we need to quantify the
probability that a particle disappears, $\Delta P$, per unit time
which is given by the difference between the integrals of the
probability current that enters and exits a given volume of the
space, that we take at $r\rightarrow \infty$ to use the asymptotic
solutions of the Schr\"{o}dinger's equation (Equation
\ref{crgyt565}).
\\ By using Equation
\ref{eqsdl2} we calculate the probability current for the asymptotic
solution in Equation \ref{crgyt565}.
\begin{equation}
\bar{J}_{r\rightarrow\infty}=-\frac{\hbar k}{m_n}\frac{\pi}{k^2
r^2}\left(1-|\eta|^2\right)
\end{equation}
We integrate the current to obtain the missing probability:
\begin{equation}
\Delta P=-\int_{r>>1}\bar{J}_{r \rightarrow \infty}r^2\,d\Omega=
\frac{\hbar k}{m_n}\frac{\pi}{k^2}\left(1-|\eta|^2\right)
\end{equation}
The absorption cross-section we normalize to the incident current of
the plane wave (Equation \ref{eqsdl3}):
\begin{equation}\label{csfa8}
\sigma_{abs}=\frac{\pi}{k^2}\left( 1-|\eta|^2\right)
\end{equation}
The absorption cross-section is zero only when
$|\eta|=|e^{2i\delta}|=1$, i.e. when $\delta \in \Re$.
\\ We can still put:
\begin{equation}\label{scrittb78}
b=b'-i\,b''=-f
\end{equation}
We may distinguish three types of nucleus. In the first type $b$ is
complex and varies rapidly with the neutron energy. The scattering
of such nuclei is associated with the resonant absorption and or
scattering of the neutron. In the second type the neutron is not
absorbed, the imaginary part of the scattering length is small and
the scattering length is independent of the neutron energy. For such
a nuclei the scattering length can be considered to be a real
quantity. In the third type there is absorption and scattering but
the complex scattering length is constant.
\\ In most cases the scattering length $b$ can be considered a real
number. We demonstrated that, for thermal neutrons, the product
$\delta \sim k\,b \sim 10^{-5}$ (Equation \ref{eqsdl11}) and the
scattering cross-section is, under these circumstances, given by
Equation \ref{eqsdl12}:
\begin{equation}
\sigma_{s}=\frac{4\pi}{k^2}\sin(\delta)\sim\frac{4\pi}{k^2}\delta^2=4\pi
b^2
\end{equation}
which is independent from $k$, i.e. the neutron energy in as much we
can consider it to be potential scattering.
\\ For absorbing materials we have $Re\{b\}\sim Im\{b\}\sim
10^{-15}\,m$ and $k \sim 10^{10}\,m^{-1}$. Hence:
\begin{equation}
|\eta|^2=|2ikb+1|^2=|2ikb'+2kb''+1|^2=4k^2b'^2+4k^2b''^2+4kb''+1
\sim 4kb''+1
\end{equation}
because we can neglect the quadratic terms which are at least five
order of magnitudes smaller than the first power term. Substituting
in Equation \ref{csfa8}:
\begin{equation}
\sigma_{abs}=\frac{\pi}{k^2}\left(
1-|\eta|^2\right)\sim\frac{\pi}{k^2}4kb''=\frac{4\pi}{k}b''=\frac{4\pi}{k}Im\{b\}
\end{equation}
The absorption cross-section is therefore uniquely given by the
imaginary part of the scattering length. However, the scattering
cross-section depends on both the real and imaginary parts. The
absorption cross-section depends on the neutron energy, it scales as
$1/k\sim 1/v$, the neutron velocity. It should be emphasized that in
the regions where this dependence does not follow the law $1/v$, the
imaginary part of the scattering length $b''$ as well depends on the
neutron energy, which is typically the case for resonance scattering
and absorption.
\\ To be more precise as value for $b''$ one should take the
statistical average over the neutron and nuclear spins. As for
scattering, absorption as well depends on the specific spins
coupling at the moment of the interaction between the neutron and
the nucleus. There are then the coherent and the incoherent
imaginary parts for absorption as well. $Im\{b_c\}$ is the average
of absorption we would obtain on a non-polarized system;
\begin{equation}\label{eqsiabs5}
\sigma_{abs}=\frac{4\pi}{k}Im\{b_c\}
\end{equation}
It is only when the neutron and the nucleus are both polarized that
the imaginary part of the bound incoherent scattering length
$Im\{b_i\}$ contributes to the value of $\sigma_{abs}$.
\\ It should be pointed out that the actual absorption cross-section is defined as follows:
\begin{equation}
\sigma_{a}=\sigma_{capt}+\sigma_{f}
\end{equation}
where $\sigma_{f}$ is the fission cross-section typical for fissile
elements, e.g. $U$. For fissile materials $\sigma_{a}$ is mostly
given by $\sigma_{f}$ and for light atoms by $\sigma_{capt}$. The
absorption cross-section we have discussed so far is more properly
$\sigma_{capt}$. Fissile elements do not show imaginary part in
their scattering lengths because the capture and the fission are two
different physical processes.
\\ Table \ref{crosssect671} shows scattering lengths and partial
cross-sections for several elements, where the absorption
cross-section is tabulated for neutron of $k = 3.49$\AA$^{-1}$
($\lambda=1.8$\AA). In this Table the sign of $b_i$ is given by the
sign of $(b_{+}-b_{-})$, the difference between the spin up
$(I+1/2)$ and the spin down $(I-1/2)$ values for $b$. We defined
$b_i$ somewhat different in Equation \ref{scgcwrg5578967}, as the
standard deviation of the $b$ for a general population of
neutron-target system, and not for a single isotope.
\begin{table}[ht!]
\centering
\begin{tabular}{|l|c|c|c|c|c|c|}
\hline \hline
Isotope & $b_{c} (fm)$ & $b_{i} (fm)$ & $\sigma_{c} (b)$ & $\sigma_{i} (b)$& $\sigma_{s} (b)$ & $\sigma_{a} (b)$\\
\hline
$^{1}H$ & $-3.74$ & $25.27$ & $1.76$ & $80.27$ & $82.03$ & $0.33$ \\
$^{2}H$ & $6.67$  & $4.04$  & $5.59$ & $2.05$  & $7.64$  & $0.00052$ \\
$^{3}He$ & $5.74-1.48i$ & $-2.5+2.57i$ & $4.42$ & $1.6$ & $6$ & $5333$ \\
$^{4}He$ & $3.26$ & $0$ & $1.34$ & $0$ & $1.34$ & $0$ \\
$^{6}Li$ & $2-0.26i$ & $-1.89+0.26i$ & $0.51$ & $0.46$ & $0.97$ & $940$ \\
$^{7}Li$ & $-2.22$ & $-2.49$ & $0.62$ & $0.78$ & $1.40$ & $0.045$ \\
$^{10}B$ & $-0.1-1.07i$ & $-4.7+1.23i$ & $0.14$ & $3$ & $3.1$ & $3835$ \\
$^{11}B$ & $6.65$ & $-1.3$ & $5.56$ & $0.21$ & $5.77$ & $0.0055$ \\
$^{12}C$ & $6.65$ & $0$ & $5.56$ & $0$ & $5.56$ & $0.0035$ \\
$^{16}O$ & $5.80$ & $0$ & $4.23$ & $0$ & $4.23$ & $0.0001$ \\
$^{27}Al$ & $3.45$ & $0.26$ & $1.49$ & $0.0082$ & $1.5$ & $0.231$ \\
$^{28}Si$ & $4.11$ & $0$ & $2.12$ & $0$ & $2.12$ & $0.171$ \\
$^{113}Cd$ & $-8-5.73i$ & $0$ & $12.1$ & $0.3$ & $12.4$ & $20600$ \\
$^{157}Gd$ & $-1.14-71.9i$ & $\pm5-55.8i$ & $650$ & $394$ & $1044$ & $259000$ \\
$^{235}U$ & $10.47$ & $\pm1.3$ & $13.78$ & $0.2$ & $14$ & $680.9$ \\
$^{238}Pu$ & $14.1$ & $0$ & $25$ & $0$ & $25$ & $558$ \\
\hline \hline
\end{tabular}
\caption{\footnotesize Neutron scattering lengths and partial
cross-sections \cite{sears}. Absorption cross-section is tabulated
for $k = 3.49$\AA$^{-1}$ ($\lambda=1.8$\AA).} \label{crosssect671}
\end{table}
\\ We notice that all the isotopes that own an imaginary part in their
scattering length have also a large absorption cross-section.
\\ Figure \ref{fgiHHecrosse4} shows the partial cross-sections for $^1H$, $^2H$ and
$^3He$ as a function of the incoming neutron energy. We recall that
elastic cross-section stands for a processes such as $A(n,n)A$; and
its contribution is implicit in $\sigma_s$. Moreover, processes like
$(n,\gamma)$ or $(n,p)$ can be considered as implicit in
$\sigma_{capt}$.
\\ Since $^3He$ is a strong neutron absorber, its total
cross-section is entirely dominated by $\sigma_{capt}$ and only at
very high energies the elastic scattering prevails. Below $1\,KeV$
the $^3He$ capture cross-section behaves as $1/v$ then the
proportionality is lost. \\ On the other hand, hydrogen is a great
neutron scatterer and its total cross-section is entirely given by
the elastic contribution. Moreover, in a wide range, its
$\sigma_{tot}$ does not depend on energy.
\\In neutron detection, the material with high $\sigma_{abs}$ are the
most interesting because of their property to commute the neutron
into a charged particle that can be detected, this will described
more in details in Chapter \ref{detintrchapt}.
\\ Figure \ref{fgiHHecrosse4} shows the $\sigma_{capt}$ for several
elements employed as neutron converter in neutron detectors.
\\ The $1/v$ dependence of the capture cross-section corresponds to
the imaginary part of the scattering length $b''$ being independent
on the neutron energy. We notice that for $Cd$ or $Gd$ the $1/v$
dependence is broken by nuclear resonances that occur at precise
energies. \\ The most common neutron reactions for neutron converter
elements are listed in Table \ref{eqaa3}. The associated energy that
is liberated at the moment of the absorption is also indicated. At
the moment of the creation of the two fragments, the atomic
electrons are dispersed, thus the two fragments carry a net electric
charge.
\bigskip
\begin{equation}\label{eqaa3}
\begin{array}{ll}
n + ^3He \rightarrow & {^3H} + p + 770 KeV\\
\\
n + ^{10}B \rightarrow & ^7Li + \alpha + 480KeV \gamma$-ray $+ 2300
KeV \,(94\%)\\
\hspace{1.45 cm} \rightarrow & ^7Li + \alpha \hspace{2.82 cm}+ 2780
KeV  \,( 6\%)\\
\\
n + ^6Li\rightarrow & {^3H} + \alpha + 4790 KeV\\
\\
n + ^{157}Gd\rightarrow & Gd^{*}\rightarrow \gamma\hbox{-ray spectrum} \rightarrow \hbox{conversion }e^{-} (< 182 KeV)\\
\\
n + ^{235}U\rightarrow & \hbox{fission fragments } + \sim 80 MeV
\end{array}
\end{equation}
The energies carried by the individual fragments can be calculated
from the energy and momentum conservation. We can neglect the
incoming neutron energy in this calculation because it is several
orders of magnitude smaller than the absorption reaction energy: as
a result the two fragments can be considered emitted back-to-back.
In the thermal range neutrons have energies between $5\,meV$ and
$500\,meV$ to be compared with typical energies of the $MeV$ order
carried by the neutron absorption reaction. Hence the energies
carried by the fragments in Table \ref{eqaa3} are:
\bigskip
\begin{equation}\label{eqaa4}
\begin{array}{ll}
n + ^3He \rightarrow & {^3H}(193\,KeV) + p(577\,KeV)\\
\\
n + ^{10}B \rightarrow & ^7Li(830\,KeV) + \alpha(1470\,KeV) + 480KeV
\gamma$-ray $ \,(94\%)\\
\hspace{1.45 cm} \rightarrow & ^7Li(1010\,KeV) + \alpha(1770\,KeV)
\hspace{2.75 cm} \,( 6\%)\\
\\
n + ^6Li\rightarrow & {^3H}(2740\,KeV) + {\alpha}(2050\,KeV)
\end{array}
\end{equation}
\\ It is important for a neutron converter, used to reveal thermal
neutrons, to have a large absorption cross-section and for its
capture reaction to yield a large amount of energy available to be
detected. This will be treated in details in the next chapter.
\subsection{Reflection of neutrons by interfaces}\label{neutrefltheoint}
\begin{figure}[!ht]
\centering
\includegraphics[width=8cm]{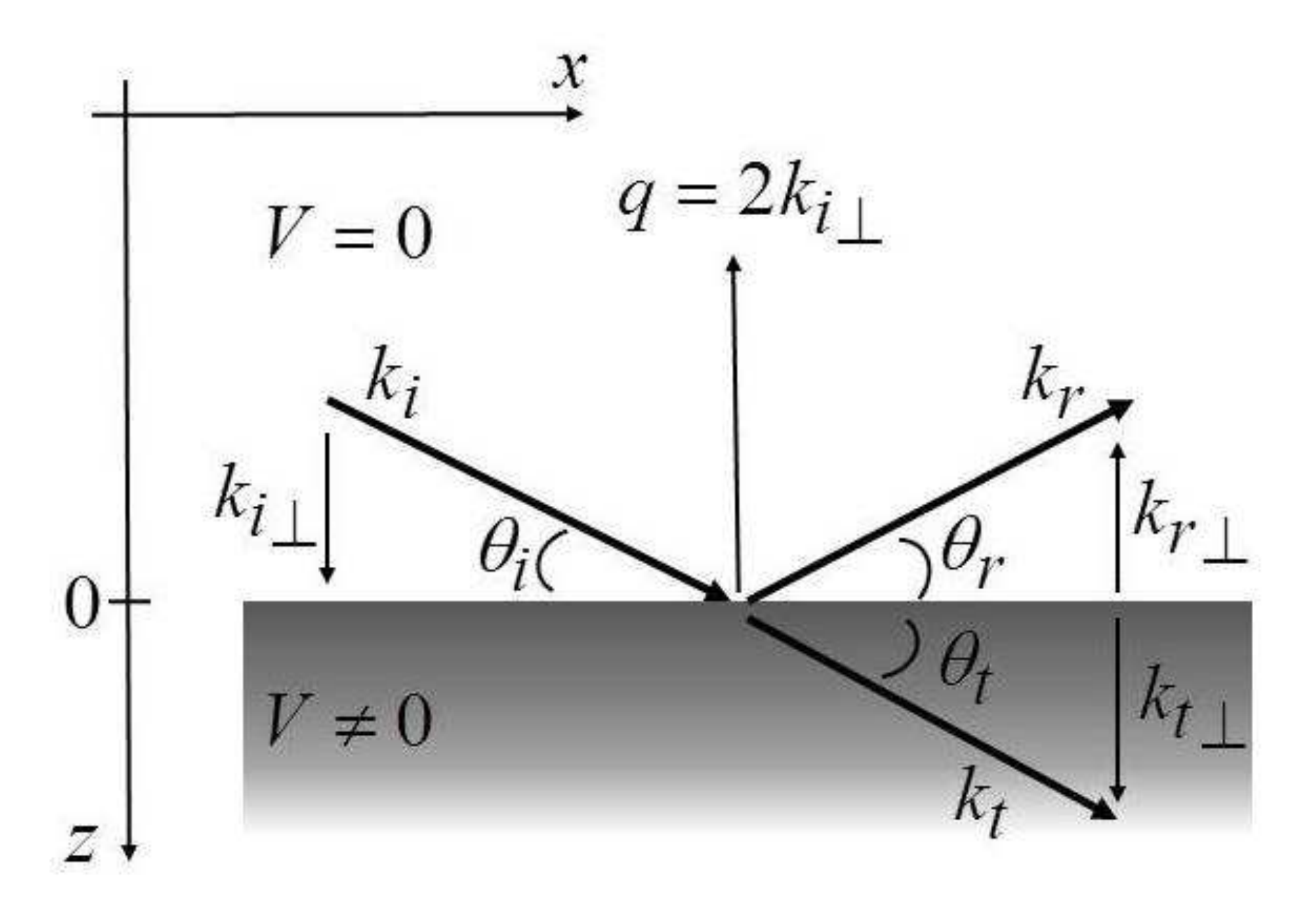}
\caption{\footnotesize Reflection of an incident neutron beam from
an ideally flat interface, $k_i$ and $k_r$ are the incident and
scattered wave vectors, $q$ is the wave vector transfer; and $V$ is
the potential of the semi-infinite substrate.} \label{reflscetch1}
\end{figure}

The reflection of light from surfaces is a well-known phenomenon
caused by the change of refractive index across the interface. The
mirror neutron reflection was demonstrated by Fermi and Zinn in 1944
\cite{fermizinn} and they observed the total reflection of thermal
neutrons below the critical angle. Neutron reflection follows the
same fundamental equations as optical reflectivity but with
different refractive indices. The optical properties of neutron
propagation arise from the fact that quantum-mechanically the
neutron is described by a wave-function. \textbf{The potential in
the Schr\"{o}dinger equation, which is the averaged density of the
scattering lengths of the material, plays the role of a refractive
index.} The neutron refractive index is given by the scattering
length density of its constituent nuclei and the neutron wavelength.
As with light, total reflection may occur when neutrons pass from a
medium of higher refractive index to one of lower refractive index.
\\ Neutron reflection is different from light
reflection because the neutron refractive index of most of materials
is slightly less than that of air or vacuum. As a result total
external reflection is more commonly observed instead of the total
internal reflection experienced with light. The angle where no
neutrons penetrate the surface, hence all of them are reflected, is
called \emph{critical angle}: the reflectivity of neutrons of a
given wavelength from a bulk interface is unity at smaller angles
(ignoring absorption effects) and falls sharply at larger angles. As
with light, interference can occur between waves reflected at the
top and at the bottom of a thin film, which gives rise to
interference fringes in the reflectivity profile \cite{cubitt3}.
\\ Neutron reflection is used as an analytical tool and offers many advantages
over traditional techniques and to x-ray reflection. In particular,
because of the short wavelengths available, it has a resolution of a
fraction of a nanometer, it is nondestructive and it can be applied
to buried interfaces, which are not easily accessible to other
techniques. \\ Neutron reflection is now being used for studies of
surface chemistry (surfactants, polymers, lipids, proteins, and
mixtures adsorbed at liquid/fluid and solid/fluid interfaces),
surface magnetism (ultrathin Fe films, magnetic multilayers,
superconductors) and solid films (Langmuir-Blodgett films, thin
solid films, multilayers, polymer films) \cite{cubitt3}.
\\ Neutron reflection can be described using the Schr\"{o}dinger equation:
\begin{equation}\label{eqaf1}
-\frac{\hbar^2}{2 m_n}\nabla^2 \Psi + V \Psi = E \Psi
\end{equation}
where $V$ is the potential to which the neutron is subject and $E$
its energy. $V$ represents the net effect of the interactions
between the neutron and the scatterers in the medium through which
it moves. We model $V$ as the smeared-out potential of all nuclei,
instead of looking at the individual scattering centers
\cite{opticaligna}. From Equation \ref{eqyu1}, we average out the
Fermi pseudo-potential:
\begin{equation}\label{eqaf2}
V=\frac{2\pi\hbar^2}{m_n} N_b
\end{equation}
where $N_b$ is the \emph{scattering length density} of the medium
the neutron is crossing defined as:
\begin{equation}\label{eqaf3}
N_b=\sum_i b_i n_i
\end{equation}
where $n_i$ is the number of nuclei per unit volume and $b_i$ is the
coherent scattering length of nucleus $i$, because we take the
spin-average (unpolarized beam or sample).
\\ As already mentioned in Section \ref{introchaptheo}, most
materials have a positive scattering length $b$; therefore in a
positive potential a neutron has less kinetic energy and hence a
longer wavelength (light usually behaves in the opposite way).
Moreover, some materials, such as $^{10}B$, present a complex $b$,
of which the imaginary part represents the power for that material
to absorb neutrons. In this Section, we only consider real $b$; the
reflection of neutrons by strong absorbers will be treated in
details in Chapter \ref{chaptreflectometry}.
\\ Referring to Figure \ref{reflscetch1}, we consider
a neutron beam approaching a surface with a bulk potential $V$. The
only potential gradient and hence force is perpendicular to the
surface; this will be not the case when the reflection surface is
not ideally flat and its roughness will play a role in the
reflection process.
\\ Since we are considering a specular reflection only elastic
scattering has to be taken into account. This implies
$k_{r\bot}=k_{i\bot}$.
\\ The solution of the Schr\"{o}dinger equation for this specific
two-dimensional case can be factorized as two plane waves, one
orthogonal to the surface and one parallel:
\begin{equation}\label{}
\begin{aligned}
\Psi(x,z)&=\left(e^{+i\,k_{i\bot}z}+r\,e^{-i\,k_{i\bot}z}\right)\cdot
e^{+i\,k_{i\parallel}x} \qquad & \hbox{if   } z<0\\
Y(x,z)&= t\,e^{+i\,k_{t\bot}z}\cdot e^{+i\,k_{t\parallel}x} \qquad &
\hbox{if   } z>0
\end{aligned}
\end{equation}
where $r$ and $t$ are the probability amplitudes for reflection and
transmission. By substituting the solutions into the Schr\"{o}dinger
equation \ref{eqaf1} we obtain the total energy conservation:
\begin{equation}\label{equaf67bis5}
H\Psi=E\Psi \qquad \Longrightarrow \qquad \frac{\hbar^2}{2
m_n}\left( k_{t\bot}^2+k_{t\parallel}^2\right)=\frac{\hbar^2}{2
m_n}\left( k_{i\bot}^2+k_{i\parallel}^2\right)-V \qquad (z=0)
\end{equation}
Continuity of wavefunctions and their derivatives at boundaries lead
to:
\begin{equation}\label{equaf4bisbis}
\left\{
\begin{matrix}
  \Psi(x,z) = Y(x,z) & \Longrightarrow & \left(1+r \right)e^{+i\,k_{i\parallel}x}=t e^{+i\,k_{t\parallel}x} & \Longrightarrow & 1+r=t\\
  \frac{\partial \Psi(x,z)}{\partial z}=\frac{\partial Y(x,z)}{\partial z} & \Longrightarrow & k_{i\bot}\left(1-r\right)e^{+i\,k_{i\parallel}x}=t\,k_{t\bot}e^{+i\,k_{t\parallel}x} & \Longrightarrow & k_{i\bot}\left(1-r\right)=t\,k_{t\bot}\\
  \frac{\partial \Psi(x,z)}{\partial x}=\frac{\partial Y(x,z)}{\partial x}  & \Longrightarrow & k_{i\parallel}\Psi(x,z) = k_{t\parallel} Y(x,z) & \Longrightarrow & k_{i\parallel} = k_{t\parallel}\\
\end{matrix}
\right.
\end{equation}
The potential the neutrons experience affects only the normal
component of the momentum. Without loss of generality we can deal
with a one-dimensional problem considering only the orthogonal
solution of the Schr\"{o}dinger equation, with
$k_{t\parallel}=k_{i\parallel}$ in Equation \ref{equaf67bis5}:
\begin{equation}\label{equaf4bis}
\begin{aligned}
\Psi_z&=e^{+i\,k_{i\bot}z}+r\,e^{-i\,k_{i\bot}z}  \qquad & \hbox{if   } z<0\\
Y_z&=t\,e^{+i\,k_{t\bot}z}  \qquad & \hbox{if   } z>0\\
\end{aligned}
\end{equation}
In the case of an ideally flat interface only the normal component
of the incoming wave vector $k_i$ is altered by the barrier
potential. It is ''the normal component of the kinetic energy $E_{i
\bot}$'', that determines whether the neutron is totally reflected.
\begin{equation}\label{eqaf4}
E_{i \bot}=\frac{p_{i \bot}^2}{2 m_n}=\frac{\left(\hbar
k_{\bot}\right)^2}{2 m_n}=\frac{\left(\hbar \, k_i \,
\sin(\theta_i)\right)^2}{2 m_n}=\frac{\hbar^2 q^2}{8 m_n}
\end{equation}
where $\theta_i$ is the incoming neutron angle with respect to the
surface. $q=2 k_{i \bot}=2 k_i\sin(\theta_i)$ is the momentum
transfer, as shown in Figure \ref{reflscetch1}. If the normal
component of the kinetic neutron energy does not exceed the barrier
potential ($E_{i \bot}< V$), total reflection occurs and no neutrons
penetrate into the layer. This happens when the momentum transfer
$q$ is smaller than the critical value of wave vector transfer
$q_c$, deduced by the condition $E_{i \bot} = V$ from Equations
\ref{eqaf2} and \ref{eqaf4} we obtain:
\begin{equation}\label{eqaf5}
q_c = \sqrt{16\pi N_b}
\end{equation}
Conservation of momentum implies that $\theta_i=\theta_r$ where
$\theta_r$ is the reflected beam angle; i.e. the reflection is
specular. Off-specular scattering is elastic as well but occurs in
presence of in-plane structures, mixing parallel and normal
components. We do not treat it here.
\\ On the other hand if the perpendicular neutron kinetic energy is $E_{i \bot}>
V$, neutrons can penetrate into the layer, thus reflection is not
total and the neutron can be either reflected or transmitted into
the bulk of the material. As for light the transmitted beam $k_t$
must change direction because its normal component of kinetic energy
is reduced by the potential; i.e. it is refracted. The change in the
normal wave-vector is given by $E_{t \bot}=E_{i \bot}-V$, and it
equals:
\begin{equation}\label{eqaf6}
k_{t \bot}^2 = k_{i \bot}^2 - 4\pi N_b
\end{equation}
The latter relation allows to define the refractive index $n$:
\begin{equation}\label{eqaf7}
n^2 = \frac{k_t^2}{k_i^2}=\frac{k_{i \|}^2+\left(k_{i \bot}^2 - 4\pi
N_b\right)}{k_i^2}=1-\frac{4 \pi N_b}{k_i^2}=1-\frac{\lambda^2
N_b}{\pi}
\end{equation}
where $\lambda$ is the neutron wavelength. For most materials
$N_b<<1$, hence Equation \ref{eqaf7} can be approximated up in the
thermal neutron energy range as $n \approx 1-\frac{\lambda^2
N_b}{2\pi}$. This result confirms that the wavelength change in the
bulk is opposite to that of light (for positive $b$, $n$ is less
than 1). It can be important to notice that materials with constant
$N_b$ are naturally dispersive ($n$ depends on $\lambda$)
\cite{rainbow2}.
\\ Referring to the solution of the Schr\"{o}dinger
equation (Equation \ref{equaf4bis}), if the potential is real and
$E_{i \bot}< V$ (see Equation \ref{eqaf13}), then $k_{t \bot}$ is
imaginary and the solution for $z>0$ in Equation \ref{equaf4bis} is
an evanescent wave (an exponential decay).
\\ From Equation \ref{equaf4bisbis} we can solve the classical Fresnel
coefficients as it is in optics:
\begin{equation}\label{eqaf11}
r=\frac{k_{i\bot}-k_{t\bot}}{k_{i\bot}+k_{t\bot}} \hbox{ \qquad
\qquad} t=\frac{2\,k_{i\bot}}{k_{i\bot}+k_{t\bot}}
\end{equation}
The continuity equation considered for the stationary case
$\frac{\partial P(\bar{r},t)}{\partial t}=0$ where $P(\bar{r},t) =
\left| \Psi(\bar{r},t)\right|^2$, and using unitary (no absorption):
\begin{equation}\label{eqaf11bis}
\frac{\partial P(\bar{r},t)}{\partial
 t}+\nabla \cdot \bar{J}(\bar{r},t)= 0 \qquad \Longrightarrow  \qquad \nabla \cdot \bar{J}(\bar{r},t)= 0
\end{equation}
where $\bar{J}(\bar{r},t)$ is the quantum probability current
defined in Equation \ref{eqsdl2}. Assuming $E_{i \bot}>V$, the
probability for a neutron to be reflected or transmitted into the
layer, given by the ratio of the reflected or transmitted flux over
the incoming flux, is:
\begin{equation}\label{eqaf11bisbis}
R = \frac{k_{i\bot}}{k_{i\bot}}r^2=r^2 \qquad \qquad T =
\frac{k_{t\bot}}{k_{i\bot}}t^2
\end{equation}
By using the Equations \ref{eqaf5}, \ref{eqaf6} and  \ref{eqaf11}
the reflectivity $R$ can be also written as:
\begin{equation}\label{eqaf12}
R=r^2 =\left(\frac{q-\sqrt{q^2-q_c^2}}{q+\sqrt{q^2-q_c^2}} \right)^2
\qquad \quad \hbox{for } \quad q>q_c
\end{equation}
When $q>>q_c$, Equation \ref{eqaf12} at the air-solid interface can
be approximated as $R\approx\frac{16 \pi^2}{q^4}N_b^2$.
\\ $Y_z$, using Equation \ref{eqaf6}, is a real solution when $E_{i \bot}< V$ (or $q< q_c$):
\begin{equation}\label{eqaf13}
Y_z=t\,e^{+i\,k_{t\bot}z}=t\,e^{+i\left(k_{i\bot}^2-4\pi N_b
\right)^{1/2} z}=t\,e^{-\frac{1}{2}\left(q_c^2-q^2 \right)^{1/2} z}
\end{equation}
When the potential barrier is higher than the particle energy normal
to the surface it can still penetrate to a characteristic depth of
$D_{ev}=\left( q_c^2-q^2\right)^{-1/2}$, but there is no quantum
probability current associated to it. This \emph{evanescent} wave
travels along the surface with wave vector $k_{\|}$ and after a very
short time it is ejected out of the bulk in the specular direction.
Taking as example the value of $N_b$ for Si this penetration is on
the order of $100$\AA \, at $q=0$, rising rapidly to infinity at
$q=q_c$. No conservation laws are broken, as the reflectivity is
still unity due to the fact that this wave represents no flux
transmitted into the bulk.
\\ The interface between materials may be rough over a large range of
length scales. A boundary may be smooth but with one material
diffused into the other. It turns out that in both the rough and
diffuse cases the specular reflectivity is reduced as $q^{-4}$. The
resulting density profiles are the same. Equation \ref{eqaf12} in
the case of $q>>q_c$ is affected in the manner:
\begin{equation}\label{eqaf15}
R\approx\left(\frac{16 \pi^2}{q^4}N_b^2\right)\cdot e^{-q^2
\sigma^2} \qquad \quad \hbox{for } \quad q>>q_c
\end{equation}
where $\sigma$ is a characteristic length scale of the layer
imperfection, the surface roughness. In the case of the diffuse
interface the lost intensity given by the factor $e^{-q_z^2
\sigma^2}$ goes into the transmitted beam as there are no potential
gradients in any other direction than normal to the surface. This is
not the case for the rough interface where intensity is lost by
local reflections in directions away from the specular direction or
off-specular scattering. If the in-plane structure is regular as in
an optical grating then the off-specular can be quite dramatic.
\\ The objective of a specular neutron reflection experiment
is to measure the reflectivity as a function of the wave vector
perpendicular to the reflecting surface, $q$. The measurement can be
done by varying either the angle of incidence $\theta$ at constant
wavelength or measuring the time-of-flight, hence varying
wavelength, at constant $\theta$. The corresponding incoming
intensity must also be measured. The reflectivity is simply the
ratio of these two intensities, as a function of $\theta$ or
$\lambda$ which is converted to $q$ by:
\begin{equation}\label{eqaf16}
q=\frac{4\pi}{\lambda}\sin(\theta)
\end{equation}

\chapter{Neutron gaseous detector working principles}\label{detintrchapt}

An overview of the state of the art in neutron detection is
discussed after we are going to describe the working principles of
neutron gaseous detectors.
\\ The main sources of the material for this Chapter are
the books \cite{leo} and \cite{knoll}.

\newpage
\section{Principles of particle detectors}
\subsection{Introduction}
Many types of detector have been developed so far, all are based on
the same fundamental principle: the transfer of part or all of the
radiation energy to the detector matter where it is converted into
some other form more accessible to human perception \cite{leo};
usually an electrical signal. In Chapter \ref{chaptintradmatt} we
discussed the interaction of charged particles and photons with
matter. They transfer their energy to matter through direct
collisions with the atomic electrons, thus excitation or ionization
of the atoms.
\\ The way to produce the electrical signal output is different for
gaseous detectors, scintillators or semiconductors.
\\ Gaseous detectors are based on the direct collection of the
ionization electrons and ions produced in a gas by passing
radiation. During the first half of the $20^{th}$ century ionization
chambers, proportional counters and Geiger counters were developed.
During the late 1960's the multi-wire proportional chamber was
invented.
\\ The scintillation detector makes use of the fact that certain
materials, when struck by a radiation, emit a small flash of light,
i.e. a scintillation. When coupled to an amplifying device such as a
photomultiplier, these scintillations can be converted into
electrical pulses. They detect the passing radiation trough the
indirect process of light detection.
\\ Semiconductors are based on crystalline semiconductors materials,
most notably $Si$ and $Ge$. These detectors are also referred to as
solid-state detectors. The basic operational principle is analogous
to gas ionization devices. Instead of gas, the medium is a
semiconductor. The passage of ionizing radiation creates
electron-hole pairs which are then collected by an electric field.
The advantage of semiconductors is that the average energy required
to create a pair is an order of magnitude smaller than that required
for gas ionization. On the other hand, being crystalline materials,
they also have a greater sensitivity to radiation damage.
\\ The neutron lack of charge and the fact that it is weakly absorbed by most
materials are properties which have contributed to its powerfulness
as a probe for condensed matter research. On the other hand, these
properties make the construction of efficient neutron detectors difficult. \\
Neutral radiation must first undergo some sort of reaction in the
detector producing charged particles, which in turn ionize and
excite the detector atoms. \\ As mentioned in Section
\ref{slowdown34ttc4}, elastic scattering is the principal mechanism
of energy loss for fast neutrons. A fast neutron, hitting a light
atom target, can transfer its energy and generate an energetic
charged particle by recoil. \\ For thermal neutrons the energy is
too low to generate fast charged particles by elastic scattering.
Thermal neutrons are detected indirectly by exploiting a capture
reaction. Those reactions produce either prompt $\gamma$-rays or
heavy charged particles such as protons, tritons, $\alpha$ or
fission fragments. Those secondary radiations have sufficient energy
to be directly detected. The processes used to detect those
secondary particles can be ionization, excitation or scintillation.
\subsection{Gas detectors}
\subsubsection{Charged particles in gas}
\begin{figure}[ht!]
\centering
\includegraphics[width=12cm,angle=0,keepaspectratio]{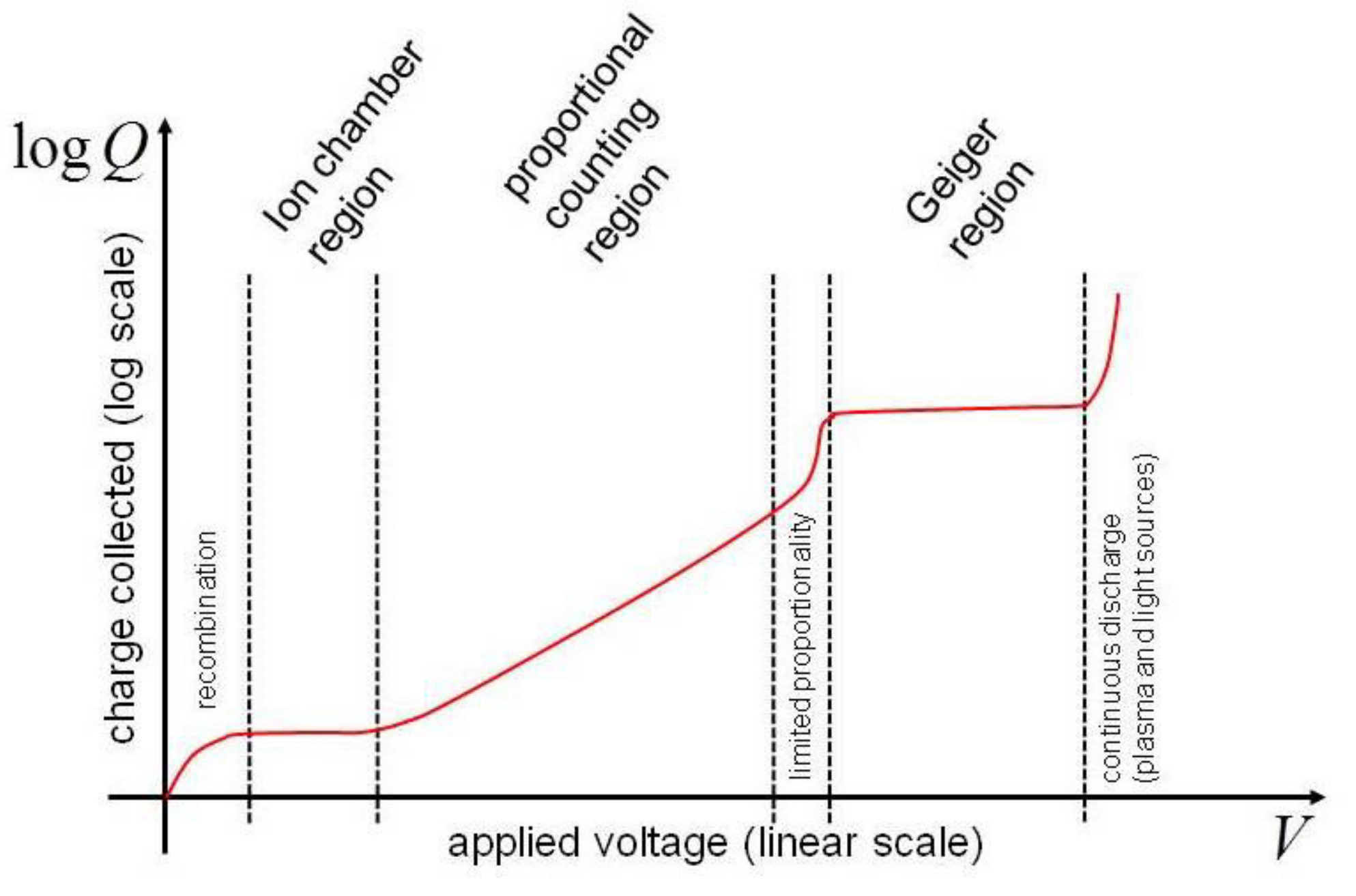}
\caption{\footnotesize Practical gaseous ionization detector
regions.}\label{opermodedetect9}
\end{figure}

A common method to detect a charged particle is based on sensing the
ionization produced when it passes through a gas. As described in
Section \ref{cpitheo851}, the primary interaction of charged
particles is to ionize and excite the gas molecules along their
tracks. After a neutral molecule is ionized, the resulting positive
ion and free electron is called an \emph{ion pair}. The ionization
generated by the heavy charged particles is called \emph{primary
ionization}. The free electrons and ions created may diffuse. Some
of the electrons may also have enough energy to be an ionizing
charged particle themselves and to produce a \emph{secondary
ionization}. We define $n_{pair}$ as the total number of ion pairs
created by the passage of a charged particle, both considering
primary and secondary ionization.
\\ With secondary ionization several different processes are
intended. In this manuscript, we always refer to it as the
ionization induced by electrons that have been created, though
primary ionization, by the passage of a charged particle. In Section
\ref{wefqeu56546546} we will introduce the multiplication process
which is another kind of ionization performed by electrons in strong
electric field; we refer to this ionization as avalanche process.
\\ As already explained in Chapter \ref{chaptintradmatt}, the energy loss of a charged
particle in matter is essentially given by two types of reactions:
ionization and excitation. In Table \ref{tabchaionisdr5} the
excitation and ionization potential for several gases are listed. It
should be emphasized that the actual energy transfer needed, on
average, to create a ion pair is $w_i$. $I_i$ is the ionization
potential. The difference is due to excitation which costs energy to
the particle but which does not create ion pairs. The average energy
lost by a charged particle in the formation of an ion pair is about
$w_i=30\,eV$ and it is not strongly dependent on the type of
molecule or charged particle. The number $n_{pair}$ of ion pairs
created is given by \cite{saul}:
\begin{equation}\label{}
n_{pair} = \frac{\Delta E}{w_i}
\end{equation}
where $\Delta E$ is the charged particle energy loss. The net charge
created by the passage of a charged particle in a medium is directly
proportional to the energy it deposits. E.g. for an
$\alpha$-particle of $1\,MeV$ the process ends up with the creation
of a net charge of a few $fC$.
\begin{table}[ht!]
\centering
\begin{tabular}{|l|c|c|c|}
\hline \hline
gas & $I_{ex} (eV)$ & $I_i (eV)$ & $w_i (eV)$\\
\hline
$H_2$  & $10.8$  & $15.4$  &  $37$\\
$He$   & $19.8$  & $24.5$  &  $41$\\
$N_2$  & $8.1$  & $15.5$  &  $35$\\
$O_2$  & $7.9$  & $12.2$  &  $31$\\
$Ne$   & $16.6$  & $21.6$  &  $36$\\
$Ar$   & $11.6$  & $15.8$  &  $26$\\
$CO_2$ & $5.2$  & $13.7$  &  $33$\\
$CH_4$ & $9.8$ & $13.1$  &  $28$\\
$CF_4$ & $12.5$ & $15.9$  &  $54$\\
\hline \hline
\end{tabular}
\caption{\footnotesize Excitation and ionization potential for
several gases \cite{saul}.} \label{tabchaionisdr5}
\end{table}
\\ The occurrence of the ionizing reactions is statistical in nature,
thus two identical particles will not, in general, produce the same
number of ion pairs. Hence, $n_{pair}$ has to be considered as the
average number of ion pairs created by a ionization process.
\\ While the number of ion pairs created is important for the
efficiency and energy resolution of the detector, it is equally
important that these pairs, or at least the electrons, remain in a
free state long enough to be collected. Two processes contribute to
diminish the net charge created: \emph{recombination} and
\emph{electron attachment} \cite{knoll}. In absence of electric
field, ions and electrons can only diffuse and they will generally
recombine under the force of their electric attraction, emitting a
photon in the process. Electron attachment is the capture of free
electrons by electronegative atoms to form negative ions. The
presence of electronegative gases in the detector severely
diminishes the efficiency of electron collection by trapping the
electrons before they can reach the electrodes. Some well known
electronegative gases are $O_2$, $H_2O$ and $CO_2$. It has been
observed that a small amount of $CO_2$ does not dramatically affect
the detector performances but helps to stop more efficiently the
ionizing particles. In contrast, noble gases $He$, $Ar$, etc. have
negative electron affinities and are suitable for application in
gaseous detectors.
\\ In the absence of electric field, electrons and ions liberated by
the passage of the charged particle diffuse uniformly outward from
their point of creation. It is important to distinguish the words:
\emph{diffusion}, which is the free motion of ions and \emph{drift},
which is the motion due to an external force, e.g. an electric
field.
\\ In the diffusion process the electrons and ions suffer multiple
collisions with the gas molecules and come quickly into thermal
equilibrium with the gas. The charges velocities distribution can be
described by using the Maxwelian distribution.
\\ In Figure \ref{opermodedetect9} we plot the
charge created by ionization, that eventually can entirely be
collected, as a function of the voltage applied to two electrodes
immerse into the gas volume, through which we apply the electric
field. When $V=0$, the electric field between ions and electrons
causes a drift which leads to recombination. Only thermal diffusion
occurs and no net charge can be collected.
\\ As the potential applied to the gaseous detector increases,
several operational modes occur:
\begin{itemize}
    \item ionization mode;
    \item proportional mode;
    \item Geiger mode.
\end{itemize}
Before describing in details the detector operational modes, we are
going to explain how the charge created into the gas is translated
into a readable signal.
\subsubsection{The Shockley-Ramo theorem}
\begin{figure}[ht!]
\centering
\includegraphics[width=8cm,angle=0,keepaspectratio]{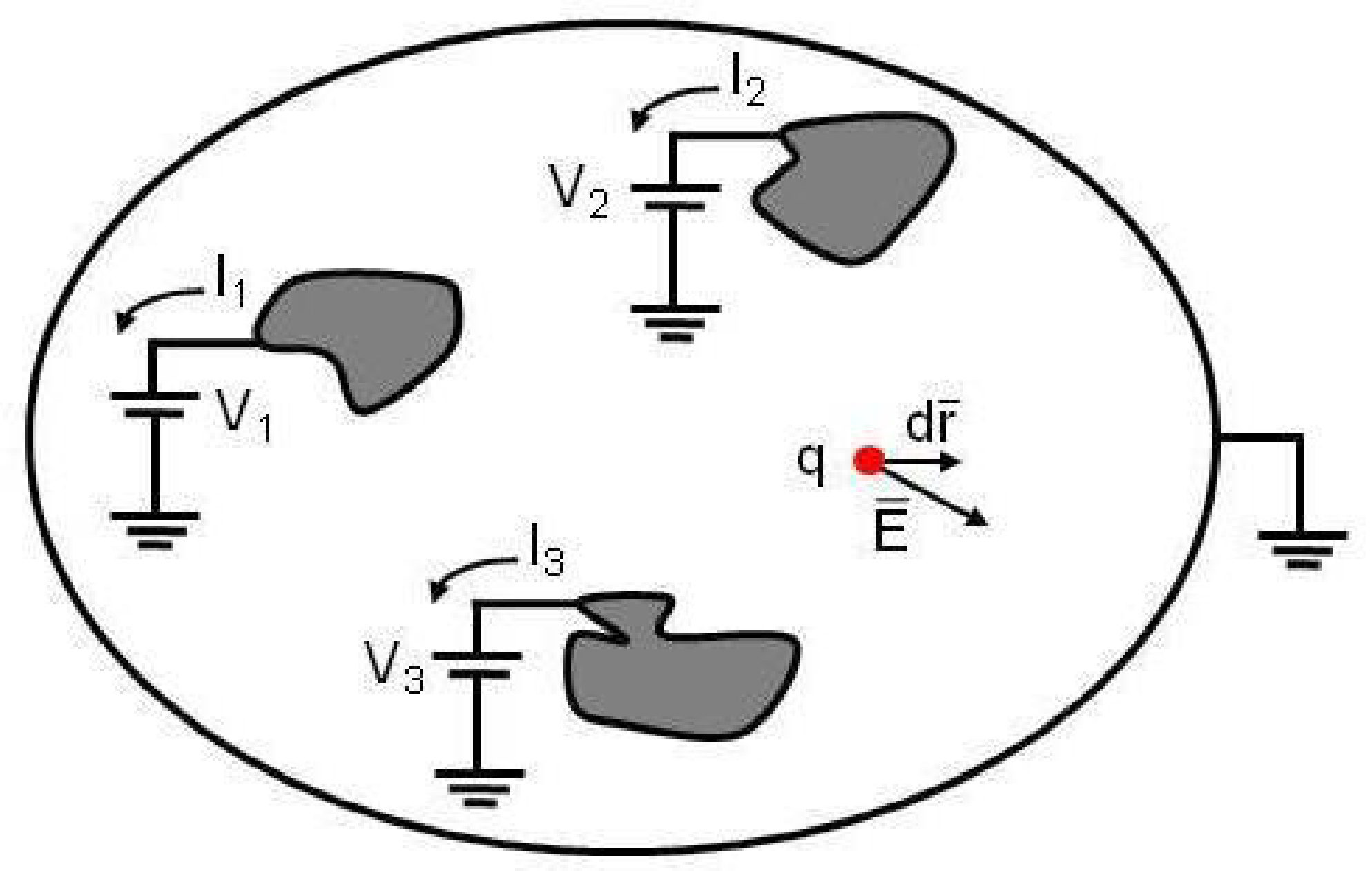}
\caption{\footnotesize Three conductors are polarized through a
generator and a charge $q$ moves along $d\bar{r}$ in the resulting
electric field generated by the
conductors.}\label{fasdcgargr454564gf}
\end{figure}

The signal from a detector is a current signal read from electrodes.
It arises from the \emph{motion} of charge carriers and not from
their physical collection. This statement is valid for gas-filled
detectors as well as for semiconductor detectors. The output pulse
begins to form immediately when the carriers start their motion
toward the electrodes. When the charge carriers deposit their
electrical charge on the electrodes the process is over and no
output signal is induced anymore \cite{knoll}. The time evolution of
the signal is fundamental for understanding the timing properties of
detectors.
\\ The method to calculate the induced charge on electrodes due to the motion of
charges in a detector makes use of the Shockley-Ramo theorem
\cite{shockley}, \cite{ramo}, \cite{zonghe} and the concept of the
\emph{weighting field}.
\\ Referring to Figure \ref{fasdcgargr454564gf}, consider a volume
where there are some, in this case three, conductors. Each conductor
is maintained at his fixed potential $V_i$ through a generator. We
take the volume surrounded by a conductor maintained at ground
potential. Let's imagine a net charge $q$ has been created, e.g.
from the passage of a radiation.
\\ The resulting electric field $\bar{E}$ in a generic point of the space
can be calculated as the superimposition of the single electric
field generated by each conductor.
\begin{equation}
\bar{E} = \sum_i V_i\, \bar{E}_i = V_1\,\bar{E}_1 +V_2\,\bar{E}_2+
V_3\,\bar{E}_3
\end{equation}
where $E_i$ is the weighting field associated to the $i-th$
conductor. It is measured in $1/m$ and can be interpreted as the
electric field that would be generated by the $i-th$ conductor if
its generator is set to $V_i=1V$ and all the other conductors are at
ground potentil. The actual electric field generated by a conductor
can be obtained by setting each potential to zero apart from the one
under interest. E.g. $\bar{E}_i^{real}=V_i\bar{E}_i$.
\\ Let's imagine the charge $q$ to move over an interval $d\bar{r}$. The work
done by the electric field on the charge is:
\begin{equation}\label{aecf44}
dW_q = q\bar{E}\cdot d\bar{r}
\end{equation}
Since the volume is closed, and assuming that the charge $q$ does
not perturb the electric field generated by the conductors, we
assume the conservation of energy in the volume, i.e. $dW_q=dW$.
Where we denote with $dW$ the work done by the sources:
\begin{equation}\label{awectg56}
dW = -\sum_i V_i\,I_i\,dt=-\left( V_1\,I_1+ V_2\,I_2+
V_3\,I_3\right)dt
\end{equation}
By comparing Equations \ref{aecf44} and \ref{awectg56} we obtain:
\begin{equation}\label{ewiof9}
-\left(V_1\,I_1+ V_2\,I_2+ V_3\,I_3\right) = q \bar{E} \cdot
\frac{d\bar{r}}{dt}= q \bar{E} \cdot \bar{v} = q\left(
V_1\,\bar{E}_1 +V_2\,\bar{E}_2+ V_3\,\bar{E}_3 \right)\cdot \bar{v}
\end{equation}
where $\bar{v}$ is the charge velocity.
\\ By differentiating with respect to the $i-th$ potential $V_i$ the Equation \ref{ewiof9}, we
obtain the current $I_i$ that flows through the $i-th$ generator to
compensate the motion of the charge $q$:
\begin{equation}\label{sgsecg7641257}
I_i = -q\bar{E}_i\cdot \bar{v}
\end{equation}
which in its differential form, by using $I=\frac{dQ}{dt}$, is:
\begin{equation}\label{awectg565}
dQ_i = -q\bar{E}_i\cdot \bar{v}\,dt
\end{equation}
$dQ_i$ is the instantaneously induced charge on the $i-th$
conductor. Note that, from Equation \ref{awectg565}, when the charge
$q$ reaches one electrode, i.e. $\bar{v}=0$, there is no more charge
induced $dQ$.
\\ The total amount of charge induced on the $i-th$ conductor is:
\begin{equation}\label{awectg5656}
Q_i = -q\int \bar{E}_i\cdot \bar{v}\,dt
\end{equation}
Let us consider now two plane electrodes facing each other at a
distance $D$. The first is polarized at $V_1=V_0$, i.e. the anode,
and the second at ground, $V_2=0$, i.e. the cathode. In this
configuration the actual electric field, neglecting side effects is
uniform $E=V_0/D$ and orthogonal to the electrodes. We denote its
direction with $x$. The charge $q$ is created at a distance $d$ from
the anode (a distance $D-d$ from the cathode). The weighing filed
associated to the anode is $E_1=1/D$. The current and charge induced
on the first electrode, Equations \ref{sgsecg7641257},
\ref{awectg565} and \ref{awectg5656}, by a negative charge ($q<0$)
going to the anode, is:
\begin{equation}\label{awectg56567}
I_1 = -\frac q D v_x, \qquad dQ_1 = -\frac q D v_x\,dt, \qquad Q_1 =
-q\int \frac 1 D v_x\,dt = \frac q D\int_d^0 dx = \frac q D d
\end{equation}
where $v_x$ is the component of the charge $q$ velocity parallel to
the electric field. A negative charge induces a negative charge on
the anode.
\\ From Equation \ref{awectg56567} we notice that the total induce
charge does not depend on how the charge $q$ moved but only from the
starting and ending points of its path: i.e. the electric field is
conservative. This is not the case for the current which depends on
the charge velocity. The current depends on the path traveled by the
charge and not only on the starting and ending points.
\\ The charge induced on the cathode is the same as the one induced on the anode, except for the sign.
\subsubsection{Signal read-out}
The detector read-out can be essentially carried out in two ways:
\emph{continuous current mode} or \emph{current pulse mode}.
\\ In continuous current mode, the induced current is measured at
the detector electrode. Because in strong irradiation conditions the
detector response time is often long compared with the average time
between events, the effect is to average out many fluctuations and
the output depends on the product of the interaction rate and the
charge per interaction.
\\ In the current pulse mode, the current pulse due to a single event is measured.
The transimpendence amplifier, connected to the electrode, has a
bandwidth corresponding to a time constant $\tau$. $t_c$ is the time
of the charge collection in the gas volume. If $\tau<<t_c$ (short
integration time), the voltage output will follow the instantaneous
value of the current flowing in the detector. On the contrary, if
$\tau>>t_c$ the current flow is integrated over a time $\tau$. The
amplitude of the voltage output will be proportional to the integral
of the current flow, i.e to the charge generated in the detector.
\\ Common values for charge amplifier gains are $G=5\,V/pC$
with integration time of $\tau=2\,\mu s$. A standard charge
amplifier presents a sensitivity of about $10\,fC$, below that
charge level a low-noise amplifier should be used. For example, if a
neutron is converted by $^{235}U$, its fission fragments carry about
$80\,MeV$ that translates into about $3\cdot 10^6$ ion pairs in
$Argon$, i.e. $500\,fC$, which is a sufficient charge to be
amplified electrically. On the other hand, the proton of the
$^3He$-capture reaction carries $577\,KeV$ that turns into a net
charge of about $3\,fC$.
\subsubsection{Ionization chambers}
In the presence of an electric field, the electrons and ions created
by the radiation are accelerated along the field lines toward the
anode and cathode respectively. This acceleration is interrupted by
collisions with the gas molecules which limits the maximum average
velocity which can be attained by the charge carriers. The average
velocity is called \emph{drift velocity} of charges and it is
superimposed on their random diffusion movement. Compared with the
thermal speed, the drift speed of ions is slow, however, for
electrons can be much higher since they are much lighter. \\The
mobility of a charge is defined as:
\begin{equation}\label{ewrth78}
\mu=\frac{u}{E}
\end{equation}
where $u$ is the drift velocity and $E$ is the electric field
strength. For positive ions $u$ is found to depend linearly on the
ratio $E/P$ (with $P$ gas pressure) with relatively large field
values. At constant pressure it implies that mobility is constant
and, for e given $E$, mobility varies as the inverse of the
pressure. The usual mobility of a positive ion in a noble gas is
about $1\,\frac{cm^2}{V\,s}$. \\ The recombination process decreases
the net charge seen by electrodes; it is then necessary to increase
the applied voltage, i.e. the electric field strength, until it is
large enough to avoid recombination. In Figure \ref{opermodedetect9}
when the voltage is sufficient to collect the whole charge a plateau
is attained (ion saturation). A detector operates in ionization mode
when the whole charge created by a ionizing radiation is collected
at the electrodes.
\\ As mentioned in the previous section, neutrons that have been converted by
$^{235}U$ into charged particles lead to induced charge of about
$0.5\,pC$ which is a readable by using a standard amplifier; hence
$^{235}U$ is suitable to build an ionization chamber working in
\emph{pulse mode} because of its reaction energy yield.
\\ Generally ionization chambers work in \emph{continuous current
mode}.
\subsubsection{Proportional counters}\label{wefqeu56546546}
Not all neutron capture reactions can lead to fragments that,
carrying hundreds of $KeV$, produce enough charge to be amplified by
standard amplifiers.
\\ In order to increase the ion pair yield of limited energy
fragments, one can exploit the gas multiplication process. At low
values of the electric field the electrons and ions created simply
drift toward their collection electrodes, as in the case of an
ionization chamber. During the migration ions collide with neutral
gas molecules and, because of their low mobility, they achieve very
little average kinetic energy between collisions. If the electric
field is risen to a sufficient high value, free electrons, on the
other hand, can be accelerated to get enough kinetic energy to
produce further ionizations. Because the average energy of the
electron between collisions increases with the electric field, there
is a threshold value for the field above which this further
ionization occurs. In typical gases, at atmospheric pressure, the
threshold field required is of the order of $10^7\,V/m$.
\\ The liberated electrons are accelerated as well and they
can create additional ionization. The gas multiplication process
therefore takes the form of a cascade, known as \emph{Townsend
avalanche}. The number of electrons per unit path length is governed
by the Townsend equation:
\begin{equation}
\frac{dn}{n}=\alpha\,dx
\end{equation}
where $\alpha$ is the first Townsend coefficient of the gas. Its
value is zero when the electric field is below the threshold value
and generally increases very rapidly as the electric field
increases. In a spatially constant field the electron density grows
exponentially with the distance as the avalanche progresses:
\begin{equation}\label{srgvwec5}
n(x)=n(0)\,e^{\alpha\,x}
\end{equation}
where $n(0)$ is the original charge at the point $x=0$.
\\In the proportional counter, the avalanche terminates when all free
electrons are collected at the anode. As a result, the number of
electrons created by the gas multiplication is proportional to the
number of initial ion pairs created by the incident radiation, i.e.
the net charge created, or read-out, is proportional to the incoming
radiation energy. Referring to Figure \ref{opermodedetect9}, as the
applied voltage rises, the actual electric field increases, hence
the Townsend coefficient increases and the charge created grows
exponentially. \\ This charge amplification reduces the signal to
noise requirement of the amplifiers and significantly improves the
signal-to-noise ratio compared with pulse-type ion chambers.
\\ The formation of an avalanche involves many energetic
electron-atom collisions in which the variety of excited atomic or
molecular states may be formed. The performance of proportional
counters is therefore much more sensitive to the composition of
trace impurities of the fill gas than in the case of ion chambers.
\\ Increasing further the electric field introduces nonlinear
effects. Although free electrons are quickly collected, the positive
ions move much more slowly, and during the time it takes to collect
electrons, they barely move at all. Therefore, each pulse within the
counter creates a cloud of positive ions which is slow to disperse
and represents a space charge that can significantly alter the shape
of the electric field within the detector.
\\ The actual electric field necessary to produce gas multiplication
can be problematic to be realized from a practical point of view of
power supplies or electrical and mechanical constraints. E.g., in
order to increase the electric field up to $2\cdot10^7\,V/m$ in a
planar geometry detector, where the two electrodes are faced one to
the other and they stand at a distance $d_g=10\,mm$, a potential of
$200\,KV$ has to be applied. A way to address this problem is to use
anodes of small radius. The electric field in a cylindrical geometry
is given by:
\begin{equation}\label{vw54434}
\left|E\right| = \frac{V_0}{\left|\bar{r}\right|}\frac{1}{\ln
\left(\frac{r_C}{r_A}\right)}
\end{equation}
where $r_C$ and $r_A$ are the cathode and anode radius respectively;
$V_0$ is the potential difference between them. Note that if $r_C\gg
r_A$, the electric field rises in a region close to the wire, say up
to $\sim5\cdot r_A$. This latter is the so-called
\emph{multiplication region}. The number of pairs created in the
multiplication region rises as we approach the anode and also the
Townsend coefficient increases strongly. Most of pairs are formed
close to the anode surface. The exponential growth (Equation
\ref{srgvwec5}) predicts that half of the charge is created in one
path $\lambda=1/\alpha$. At atmospheric pressure, for $Ar$,
$\lambda=2\,\mu m$. In reality, because of the rising $\alpha$, the
growth is faster than exponential.
\\ As example we take a tube of radius $r_C=5\,mm$ (at ground
potential) in which center is placed an anode wire $r_A=10\,\mu m$.
By only applying $V_0=1000\,V$ we can easily obtain a strong
electric field of $1.6\cdot10^7\,V/m$.
\\ The main contribution to the signal formation on electrodes is due
to positive ions. The multiplication occurs in the multiplication
region, i.e. around the anode wire. Let's take the previous example
of a cylindrical detector. Denote with $q=|n\cdot e|$ the net charge
created by the avalanche process ($n$ is the number of pairs
created) and let us assume for simplicity that it is placed at
$\lambda=2\,\mu m$ from the wire surface, thus at $r=r_A+\lambda$.
\\ By using the Schokley-Ramo theorem, the charge induced on the
anode is given by Equation \ref{awectg5656}:
\begin{equation}\label{awectg5656bis}
Q_{anode} = -q_i\int_{r_1}^{r_2}  \bar{E}_{anode}\cdot \bar{v}\,dt =
-\frac{q_i}{\ln \left(\frac{r_C}{r_A}\right)} \int_{r_1}^{r_2}
\frac{dr}{\left|\bar{r}\right|}=-\frac{q_i}{\ln
\left(\frac{r_C}{r_A}\right)}\ln (r) \mid_{r_1}^{r_2}
\end{equation}
where we use as weighting field $\left|E\right|/V_0$ as defined in
Equation \ref{vw54434}. The charge induced on the cathode, in
presence of only two electrodes, is $Q_{cath.}=-Q_{anode}$. $r_1$
and $r_2$ are the integration limits different for electrons
($q_i=q_e=-q$) and ions ($q_i=q_{ion}=+q$). From the point where the
charge is created, electrons will travel toward the anode
($r_1=r_A+\lambda=12\, \mu m$ and $r_2=r_A=10\,\mu m$) and ions
toward the cathode ($r_1=r_A+\lambda=12\, \mu m$ and
$r_2=r_C=5\,mm$). Hence:
\begin{equation}
\begin{aligned}
Q_{anode}^{e} &= -\frac{q_e}{\ln \left(\frac{r_C}{r_A}\right)}\ln
(r)
\mid_{r_A+\lambda}^{r_A}=  -\frac{q_e}{\ln \left(\frac{r_C}{r_A}\right)}\ln \left(\frac{r_A}{r_A+\lambda}\right)\simeq +0.03\,q_e=-0.03\,q \\
Q_{anode}^{ion} &= -\frac{q_{ion}}{\ln
\left(\frac{r_C}{r_A}\right)}\ln (r) \mid_{r_A+\lambda}^{r_C}=
-\frac{q_{ion}}{\ln \left(\frac{r_C}{r_A}\right)}\ln
\left(\frac{r_C}{r_A+\lambda}\right)\simeq -0.97\,q_{ion}=-0.97\,q
\end{aligned}
\end{equation}
where we denoted with $Q_{anode}^{e}$ and $Q_{anode}^{ion}$ the
charge induced by electrons and ions on the anode. Note that the
ions contribution is larger on the signal formation than the
electrons' one.
\\ Proportional counters are generally operated in \emph{pulse
mode} because they suffer from space charge effects before the count
rate is high enough to have continuous current mode.
\subsubsection{Geiger counters} Referring to Figure
\ref{opermodedetect9}, if the applied voltage is made sufficiently
high, the space charge created by positive ions can become
completely dominant in determining the subsequent history of the
pulse. Under these conditions, the avalanche proceeds until a
sufficient number of positive ions have been created to reduce the
electric field below the point at which additional gas
multiplication can take place. The process is then self-limiting and
will terminate when the same number of positive ions have been
formed regardless of the number of initial ion pairs created by the
incident radiation. Thus, each output pulse from the detector is on
average of the same amplitude and no longer reflects any properties
of the incident radiation. This is called the Geiger-Mueller region
of operation \cite{knoll}.
\\ In a proportional counter each original electron leads to an
avalanche that is basically independent of all other avalanches
formed from other electrons associated with the original ionizing
event. Because all the avalanches are nearly identical, the
collected charge remains proportional to the number of original
electrons. On the other hand, in a Geiger tube, the higher electric
field enhances the intensity of each avalanche; but as well each
avalanche can itself trigger a second avalanche at different
position within the tube and the process becomes rapidly divergent
creating a discharge. In a typical avalanche many excited gas
molecules are formed by electron collisions. Within few $ns$ those
molecules return to their ground state through the emission of a
photon, generally in the UV region. These photons are the key
element in the avalanche chain propagation that makes up the
discharge.
\\ A typical pulse from a Geiger tube represents a large amount
of charge collected, about $10^9$ ion pairs. Therefore, the output
pulse is so intense that in principle no amplifier electronics is
needed; thus a Geiger tube is often an inexpensive choice when a
simple counting system is needed.
\subsubsection{Quencher and stopping gas}
As mention for the Geiger counters, for proportional counters as
well, the gas multiplication process creates many excited molecules
from the collisions of electrons and neutral molecules. These
excited molecules do not contribute directly to the avalanche
process but decay to their ground state through the emission of a
photon, generally an UV-photon. Under certain circumstances these
de-excitation photons could create additional ionization elsewhere
in the fill gas or at the cathode. Although such photon-induced
events are important in the Geiger region of operation, they are
generally undesirable in proportional counters because they can lead
to a loss of proportionality and/or spurious pulses. Furthermore
they can cause avalanches which spread in time and space and,
consequently, they reduce the spatial and time resolutions.
\\ The addition of a small amount of polyatomic gas, such as
methane, to many of the common fill gases, e.g. Argon, will suppress
the photon-induced effects by absorbing the photons in a mode that
does not lead to further ionization. This additional gas is often
called \emph{quench gas}.
\\ Noble gases are commonly used as main gas fill, however for
specific applications the sole use of such a gas can be not enough
to guarantee a certain time or spatial resolution. E.g. neutron
counters are filled with $^3He$, which does not only convert
neutrons but it is as well a suitable gas fill for the detector
operation. Usually to reduce dead time it is recommended to add a
second element, i.e. the \emph{stopping gas} as $CF_4$, to the gas
mixture to increase its global stopping power. In such way the
detector time response improve because ions are stopped on shorter
tracks.
\section{Thermal neutron gas detectors}
\subsection{The capture reaction}
Mechanisms for detecting thermal neutrons in matter are based on
indirect methods \cite{crane}. The process of neutron detection
begins when neutrons, interacting with various nuclei, initiate the
release of one or more charged particles or $\gamma$-rays. The
electrical signal produced by this secondary radiation can then be
processed by the detection system. Thermal neutrons carry a too
small amount of energy to be transferred to a recoil nucleus and to
produce ionization. As introduced in Chapter \ref{chaptintradmatt},
neutrons can cause nuclear reactions. The products from these
reactions, such as protons, alpha particles, $\gamma$-rays, and
fission fragments, can initiate the detection process. Detectors
employing the reaction mechanism can use solid, liquid, or
gas-filled detection media. The choice of reactions is limited and
most of the suitable materials, for their large neutron absorption
cross-section, are listed in Equation \ref{eqaa3}.
\\ The commonest capture process is one which results in the emission
of a prompt $\gamma$-ray. Although these $(n,\gamma)$ reactions
occur for most nuclides, the resulting photons, being uncharged, are
also difficult to detect directly.
\\ Hence, most thermal neutron detectors, with prompt read-out, are
based on the few absorption reaction that result into heavy charged
particles or fission fragments: $^3He$, $^6Li$, $^{10}B$, $^{235}U$,
$Gd$, etc.
\\ In the thermal neutron detection process the information on the
initial neutron energy is completely lost. Once a neutron is
captured, the energy available, to be detected, is the one of the
capture reaction. Thus, in general, neutron detectors provide
information only on the number of neutrons detected and not on their
energy. The only effect of the neutron energy is via the absorption
cross-section which usually follows the $1/v$ law. The latter
influences the probability of the neutron to be detected by the
detector medium.
\\ We focus now on thermal neutron gaseous detectors. To detect a
thermal neutron we need a converter and a stopping gas to be ionized
by the charged particles originated by the neutron capture. Only
fission fragments from $^{235}U$ have enough energy to operate the
detector in ionization mode, usually  $^3He$ and $^{10}B$-based
gaseous detectors are operated in proportional regime. Since $^6Li$
is solid at room temperature and does not produce any $\gamma$-ray
after the neutron capture, is a suitable material to be embedded in
a scintillator.
\\ $^3He$, which is a gas at room temperature, is both a powerful
converter of neutrons and a stopping gas. Its capture reaction
produces a proton and a triton (see Table \ref{eqaa4}). At $1\,bar$
$(\rho=1.3\cdot10^{-4}\,g/cm^3)$ and room temperature the $577\,KeV$
proton is stopped within about $55.3\,mm$; at $10\,bar$ in
$5.5\,mm$. The track length can be too long to efficiently localize
the interaction point of the neutron (see Section
\ref{acgaseg544rt}). Usually a variable amount of a more efficient
stopping gas is added to $^3He$ to reduce the particle traces, e.g.
$CF_4$. In a mixture $^3He/CF_4$ in the ratio $80/20$ at $1\,bar$
the proton track length is reduced to $12.2\,mm$, at $10\,bar$ to
$1.2\,mm$.
\\ In addition to stopping gas can be added a quencher to absorb UV
photons. In neutron detection any quenching gas containing a large
amount of Hydrogen should be avoided because they can cause a strong
neutron scattering.
\\ Another gaseous converter exploited in thermal neutron detection is
$^{10}BF_3$, but for its toxicity its use is limited.
\\ For solid converters, such as $^{10}B$ or $^{10}B_4C$, the
gaseous material acts only as stopping medium. The neutron is
converted in the solid converter and escaping particles ionize the
gas. The description of such a detector will be discussed in details
in Chapter \ref{Chapt1}.
\subsection{Pulse Height Spectrum (PHS) and counting curve
(Plateau)} 
\begin{figure}[ht!]
\centering
\includegraphics[width=7cm,angle=0,keepaspectratio]{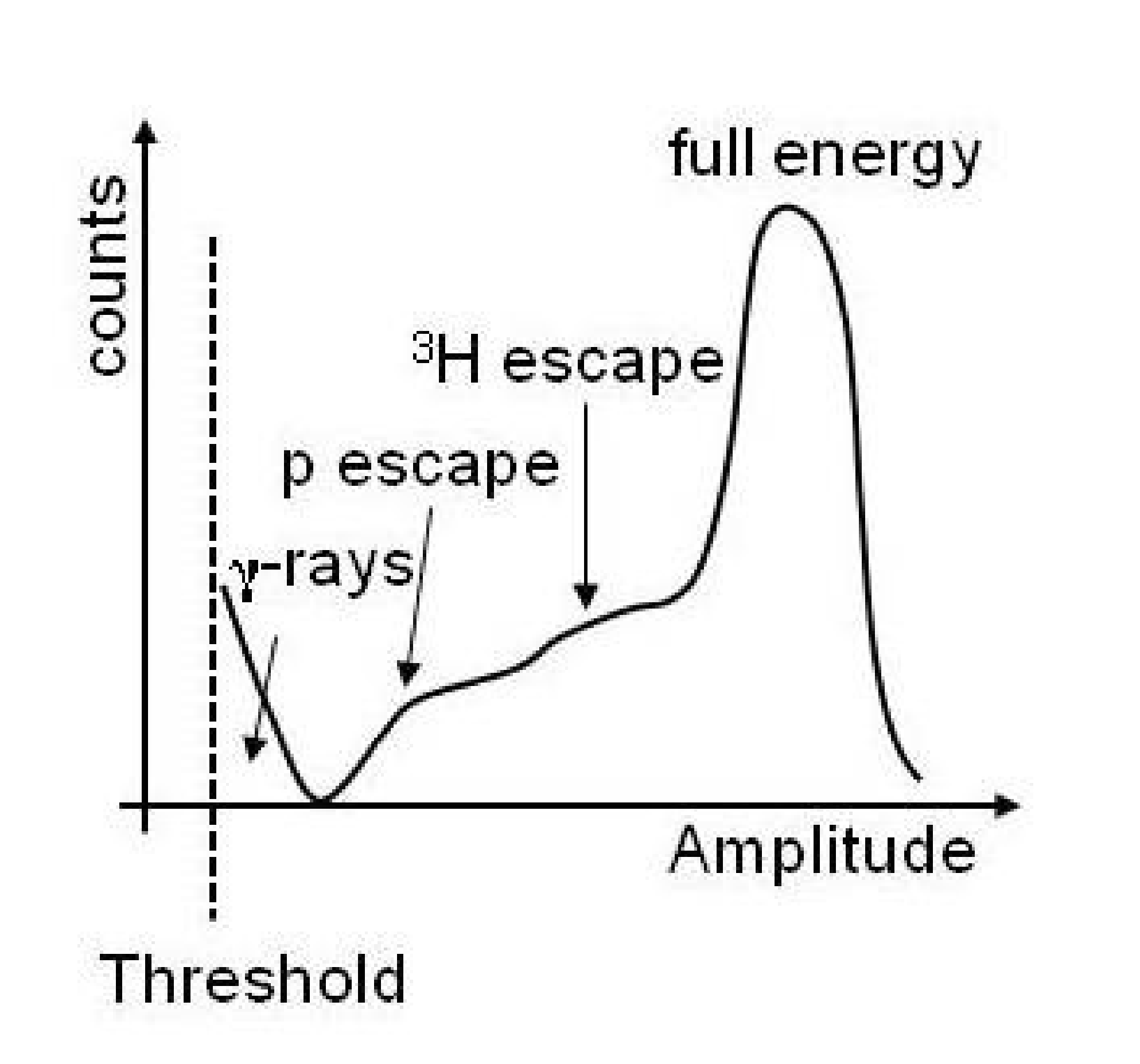}
\caption{\footnotesize Typical PHS of an $^3He$-proportional
tube.}\label{figphsexpschem6}
\end{figure}

\begin{figure}[ht!]
\centering
\includegraphics[width=11cm,angle=0,keepaspectratio]{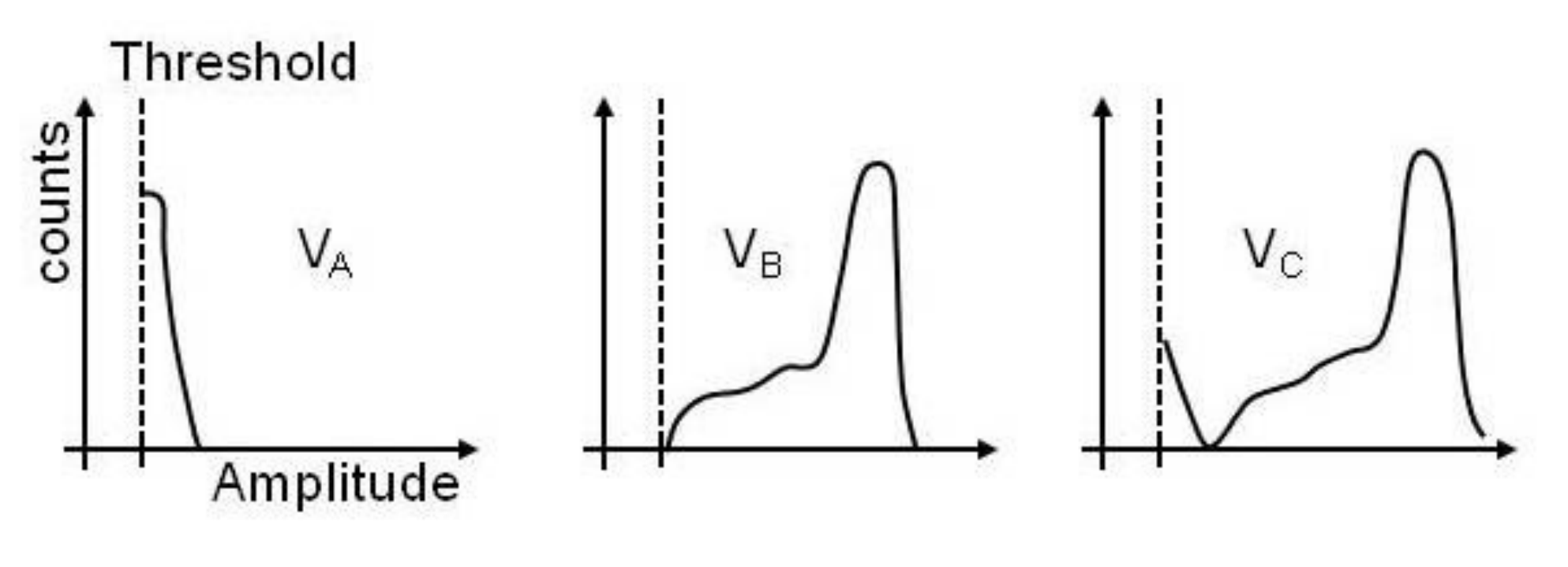}
\caption{\footnotesize Examples of $^3He$-based counter tube PHS as
a function of an increasing operational voltage applied to the
detector.}\label{fig3phsexmpevolt7}
\end{figure}

\begin{figure}[ht!]
\centering
\includegraphics[width=7cm,angle=0,keepaspectratio]{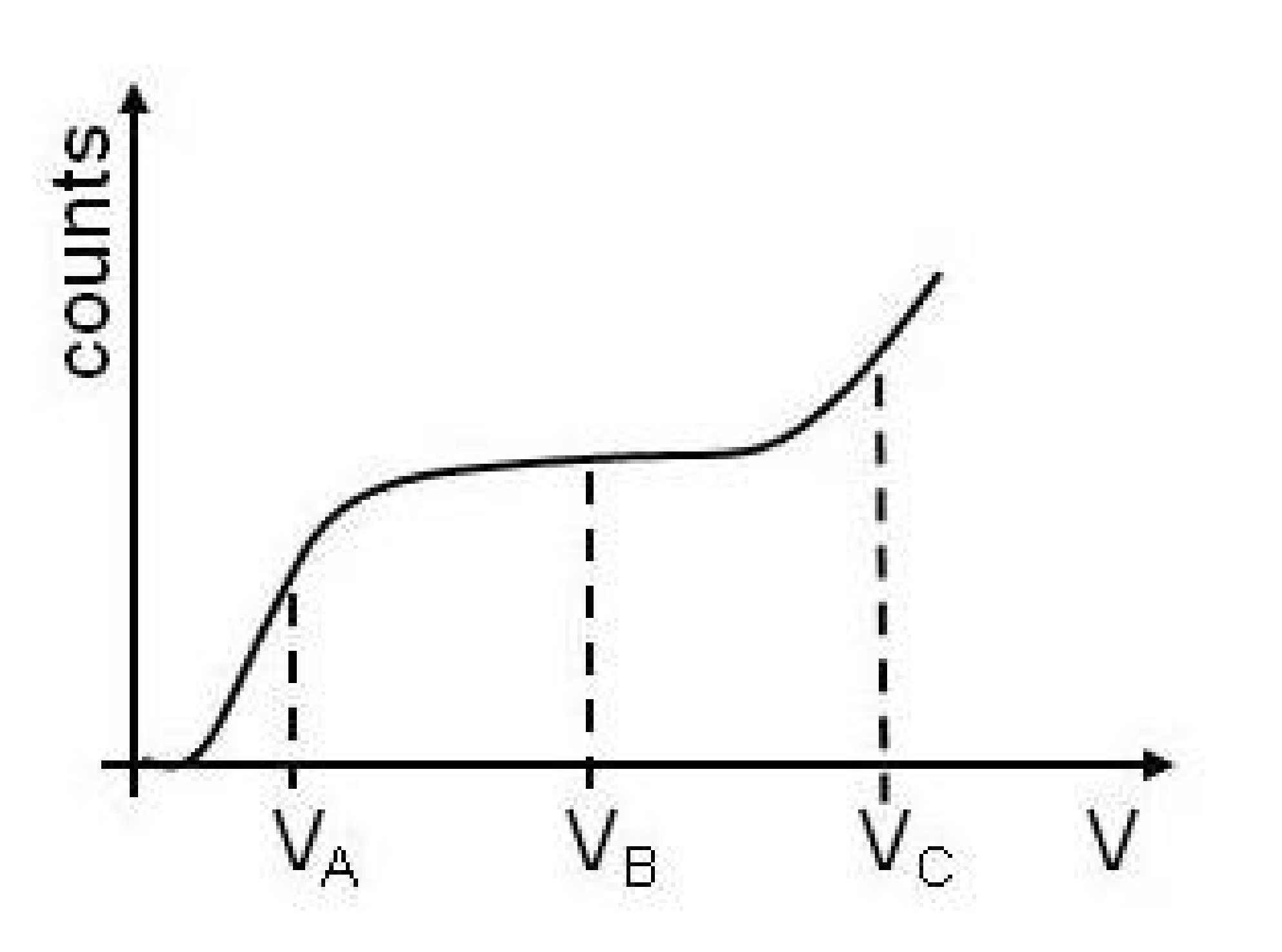}
\caption{\footnotesize Example of counting curve plot (plateau) for
an $^3He$-based counter tube.}\label{figplatexschem23}
\end{figure}

In many applications of radiation detectors, the object
is to measure the energy distribution of the incident radiation.
This is not the case for thermal neutron detectors because the
information on the neutron energy is completely lost in the
conversion process of a neutron into charged particles. The energy
information we can access is always the neutron capture fragment
energy. An important detector aspect is to make it able to
distinguish between real events and background events, e.g. between
neutron events and $\gamma$-rays. As a result, energy resolution is
required on the fragment energies and it is an important feature to
be able to discriminate between particles.
\\ When operating a detector in pulse mode, each individual pulse
amplitude carries the information on the charge generated in the gas
volume, i.e. on the energy deposited. The amplitudes will not all be
the same. The pulse height distribution is a fundamental property of
the detector output that is routinely used to deduce information
about the incident radiation, or, in the case of neutrons, on the
neutron capture reaction. This distribution is more commonly known
as \emph{differential pulse height distribution} and in this
manuscript we will always refer to it as \emph{PHS} or \emph{Pulse
Height Spectrum}. Figure \ref{figphsexpschem6} shows a typical PHS
of a $^3He$-based gaseous detector, in which we recall the capture
reaction yields about $770\,KeV$  of which $577\,KeV$ goes to the
proton and $193\,KeV$ to the triton. When a neutron is converted by
$^3He$ the two fragments immediately release their energy into the
gas volume, consequently an amount of charge proportional to the
total reaction energy will be created. A pulse of amplitude
proportional to the energy will be readable at the detector output.
Let's imagine now that a neutron is converted so close to the
detector wall that one of the two fragments hits the wall without
releasing its energy into the gas. The related output pulse will
have a smaller amplitude. Obviously the energy loss by wall effect
is a continuous process because neutrons can be converted at any
distance from the wall and fragments can be emitted with a random
angle. The effect of the walls is then visible on the PHS as a tail
energy distribution toward smaller energies from the full energy
peak.
\\Generally a neutron detector can be sensitive as well to other
radiations, more commonly $\gamma$-rays. Events generated by
$\gamma$-rays, according to the detector construction, generally
deposit a small amount of energy in the gas volume compared with
capture fragments, hence they are well separated in energy on the
PHS (see Figure \ref{figphsexpschem6}).
\\ A better energy resolution, which is defined as the FWHM (Full Width Half Maximum) of the full energy
peak, would improve the distinction between real events (neutrons)
and spurious events ($\gamma$-rays).
\\ In practice, the PHS is measured by accumulating the pulses using
an integrating amplifier that is characterized by a given electronic
noise. In order to measure properly the PHS a threshold in amplitude
should be set just above the electronic noise to avoid false
triggers.
\\ In the proportional counting region (see Figure
\ref{opermodedetect9}), the charge created in the detector can be
set by changing the operational voltage and it grows exponentially
with it. As a result, in pulse mode, the amplitude of the output
from the detector varies as a function of the operational voltage.
In Figure \ref{fig3phsexmpevolt7} we show three PHS corresponding to
three operational voltages applied, with $V_A<V_B<V_C$. As the
voltage increases, events with smaller energy are amplified over the
noise level threshold of the amplifier and give rise to events on
the PHS. We define the \emph{counting curve} (or \emph{Plateau}) as
the integral of the counts over a given threshold (set by the
electronics used) on the PHS, plotted as a function of the
operational voltage. Its behavior with the voltage $V$ is strictly
related to the PHS shape and an example is shown in Figure
\ref{figplatexschem23}. As the voltage increases, according to the
PHS shape, the counting curve can vary its slope.
\\ Generally, once the plateau is attained, it starts to rise again at
higher voltages because of $\gamma$-ray events.
\\ Once a detector geometry and electronics is defined, a detector
should be operated in a stable regime where a small change in
parameters (voltage, threshold, gas pressure, \dots) does not affect
efficiency. Referring to Figure \ref{figplatexschem23}, the best
operating voltage would be $V_B$. Below this voltage the detector is
not counting the totality of events, hence is less efficient; above
this voltage, we are mixing up neutron events with other events at
low energy.
\\ It should be emphasized that the plateau depends on the detector geometry and read-out system,
e.g. amplifier gain, thus the operational voltage chosen to operate
the detector is related to the detector working configuration.
\subsection{Gas spatial resolution}\label{acgaseg544rt} 
\begin{figure}[ht!] \centering
\includegraphics[width=8cm,angle=0,keepaspectratio]{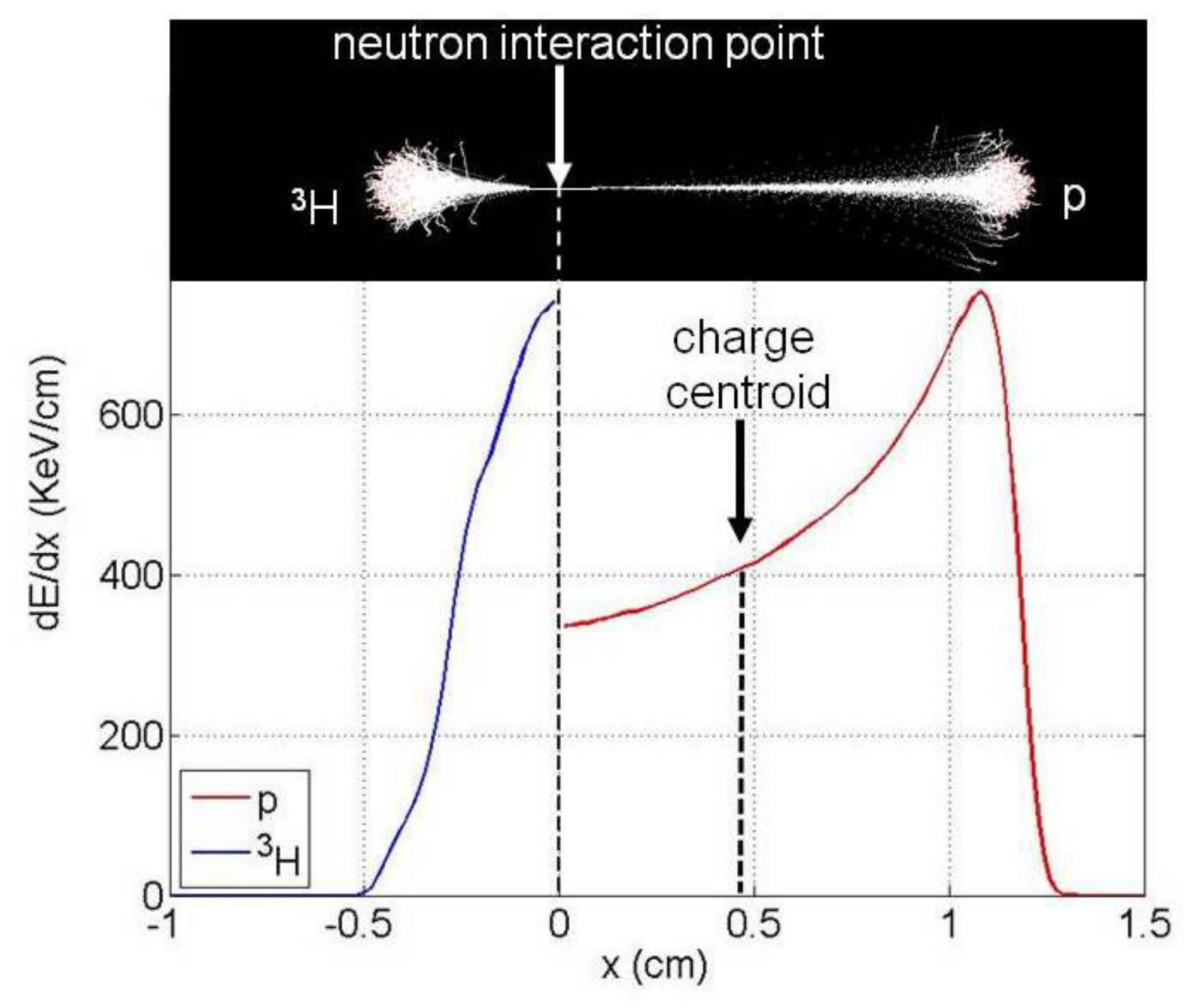}
\caption{\footnotesize Stopping power for the $^3He$ capture
reaction fragments in a gas mixture of $^3He/CF_4$ (80/10)
($\rho=8.8\cdot10^{-4}\,g/cm^3$) at
$1\,bar$.}\label{figcentroid0923mjfg}
\end{figure}

In neutron
scattering science it is often crucial to determine the neutron
impact position on the detector. For such application a PSD
(Position Sensitive Detector) is needed. The detector operational
modes explained are valid for PSD, and an easy way to get the
positional information is to place more anode wires in a gas chamber
(Multi Wire Proportional Chamber - MWPC). Those wires act as
individual detectors hence give the position of the neutron
interaction point. To get the second dimension information a X-Y
coincidence, a delay-line read-out, a charge division read-out, etc.
Those are explained in details in Section \ref{redouchdio98}.
\\ The spatial resolution $\Delta x$ is defined as a distance and represents the ability
for a detector to distinguish between two events that occur at a
distance $\Delta x$, for a given confidence level. A widely used
criterion, to define the spatial resolution, is to give the FWHM
(Full Width Half Maximum) of the image of a point. If the image is a
gaussian, it corresponds to $88\%$ probability to properly
distinguish between two events at that distance.
\\ In the slowing down of a charged
particle the charge is generated all along the particle path, thus
what it is measured at the electrodes is the charge centroid
\cite{convert}. The neutron capture reaction is asymmetric because
of the difference in masses of the two fragments. For $^3He$, most
of the energy will be carried by the proton and it will have a
longer path with respect to the triton. Consequently the charge
centroid does not correspond to the neutron interaction point.
\\ If we use charge centroid read-out the gas spatial resolution is
proportional to the charged particle range in the converter. The
interaction point and the charge centroid approach as the particle
ranges diminish. The higher the gas pressure, the lower are the
particle ranges.
\\ In $^3He$-based neutron detectors the spatial resolution is
limited by mechanical constraints on the vessel containing the
detector which limits the maximal pressure.
\\ For a given neutron interaction point the direction the two
capture fragments are emitted is isotropically distributed. The
detector response, depending on its geometry, will be different
according to the particle ejection direction. What is measured, when
the detector is exposed to a point source, is an average over all
the possibilities that results into a distribution with its own FWHM
that defines the resolution due to the gas.
\\ In Figure \ref{figcentroid0923mjfg} the
stopping power for the the two $^3He$ capture reaction fragments in
$1\,bar$, $T=300\,K$, $^3He/CF_4$ in the ratio $80/20$ is shown. The
resulting gas mixture density is $\rho=8.8\cdot10^{-4}\,g/cm^3$. At
atmospheric pressure in such a gas mixture, the extrapolated range
for the $577\,KeV$ proton is $1.22\,cm$ and for the $193\,KeV$
triton is $0.44\,cm$. The charge centroid is shown in Figure
\ref{figcentroid0923mjfg}, and the error committed between the
actual neutron conversion position and the detector read-out
position is about $0.48\,cm$. This limits the actual spatial
resolution the detector can attain and it can be improved by
increasing the gas pressure. Since the stopping power is directly
proportional to the density, thus the gas pressure, the particle
range is inversely proportional to pressure. At $10\,bar$, in the
same gas mixture just taken as example, the particles ranges are
diminished by a factor $10$. The charge centroid, thus the maximum
spatial resolution achievable, would be around $0.5\,mm$.
\\ It has to be pointed out that modern electronics allows to perform
the reconstruction of the fragment tracks in order to determine the
neutron interaction point. In this case the gas pressure in
$^3He$-based detectors is needed only to increase the detection
efficiency. Pressure remains an issue if the detector is operated in
vacuum. We will not use such electronics.
\subsection{Timing of signals}
The time response of a gaseous detector is related to the time
differences over which the different moving charges generate their
signals on one hand, and the time of the signal development of a single charge on the other hand.
\\ The stopping time of a charge in gases is a few nanoseconds.
Hence, this time is not the bottleneck of the time response of a
detector and it can often be neglected.
\\ What limits the detector time performance is instead the time the
charges, created by the ionization, need to be collected at the
electrodes and the time duration of the signal induction. This
latter depends on the strength of the electric field the detector is
operated at and, according to the geometry, to the mobility of
charges. This latter can be tuned by playing with the gas
composition. Generally, in $^3He$-based neutron detectors the simple
$^3He$ would be enough to assure a correct gas ionization and
collection of charges. On the other hand, a fraction of other gases,
such as $CF_4$, are added to it to form a mixture that increases the
stopping power of the particles, in order to get faster responses
and better resolution.
\\ For example, according to Equation \ref{ewrth78}, the time for a
general ion ($\mu=1\frac{cm^2}{V\,s}$) to drift over $1\,cm$ at
$1\,bar$ in an uniform electric field of $10^5\,V/m$ results in
$t=\frac{1\,cm}{10^3\,cm/s}=1\,ms$. This time for electrons
generally is about three orders of magnitude smaller; i.e. $1\,\mu
s$.
\subsection{Efficiency}
Detection efficiency is defined as the
ratio between the detected events by the detector and the total
amount of incident radiation.
\\ In general, for neutron detectors, the detection efficiency is
related to the absorption probability of a neutron but it is not
strictly the same. The absorption probability increases with the
converter density $\rho$, i.e. pressure for gases, because it
follows the neutron absorption exponential law (see Equation
\ref{eqsdfgg8}) with $\Sigma(\lambda)$ macroscopic cross-section.
Detection efficiency is the number of converted neutrons that give
rise to a signal output over the total. It can happen that not all
the converted neutrons give fragments that release sufficient energy
over the noise level into the gas. Hence some neutrons can be
absorbed without that a signal output is generated above the
threshold. Usually this loss can be neglected and detection
efficiency as well follows the neutron absorption exponential law.
We recall, that since $\sigma_{abs}$ depends on the neutron energy
as $1/v$ the detection efficiency increases with the neutron
wavelength.
\\ By keeping the detector volume constant, a way to
increase the neutron detection efficiency is to increase the
converter gas pressure.
\\ For solid neutron converter based detectors efficiency does not depend in
such a simple way on the neutron wavelength as for gas converters.
\\ A full explanation of that will be given in Chapter \ref{Chapt1}.
\section{Read-out and dead time}\label{redouchdio98}
\subsection{Read-out techniques}
If position sensitivity is required a PDS (Position Sensitive
Detector) is needed. A PSD allows to identify the point of
interaction of the radiation in one direction or in a
two-dimensional plane. The segmentation of the anodes or cathodes in
a gas detector allows to get the positional information. We can
imagine to segment a single cathode plane into strips and to place
many wires close to each other to make an anode plane, i.e. MWPC.
Each wire acts as a single detector. The read-out works in the same
way for cathodes as for anodes. Once an ionization and gas
multiplication occur there will be only a region in a gas chamber
interested by the induction of the signal; there could be one or
more wires where the signal is induced. The \emph{individual
read-out} can be performed by connecting each wire to a single
charge amplifier. To get the second dimension information, a second
wire plane, orthogonal to the first, can be placed in the gas volume
and read-out in the same way. Equivalently, the cathode can be
segmented and read-out by an individual amplifier for each strip.
The number of read-out channels can diverge rapidly for a large
detector. The large amount of electronics required can be expensive
and unwieldy.
\\ In order to reduce the number of read-out channels we discuss here the
delay line and charge division approaches.
\begin{figure}[ht!]
\centering
\includegraphics[width=7cm,angle=0,keepaspectratio]{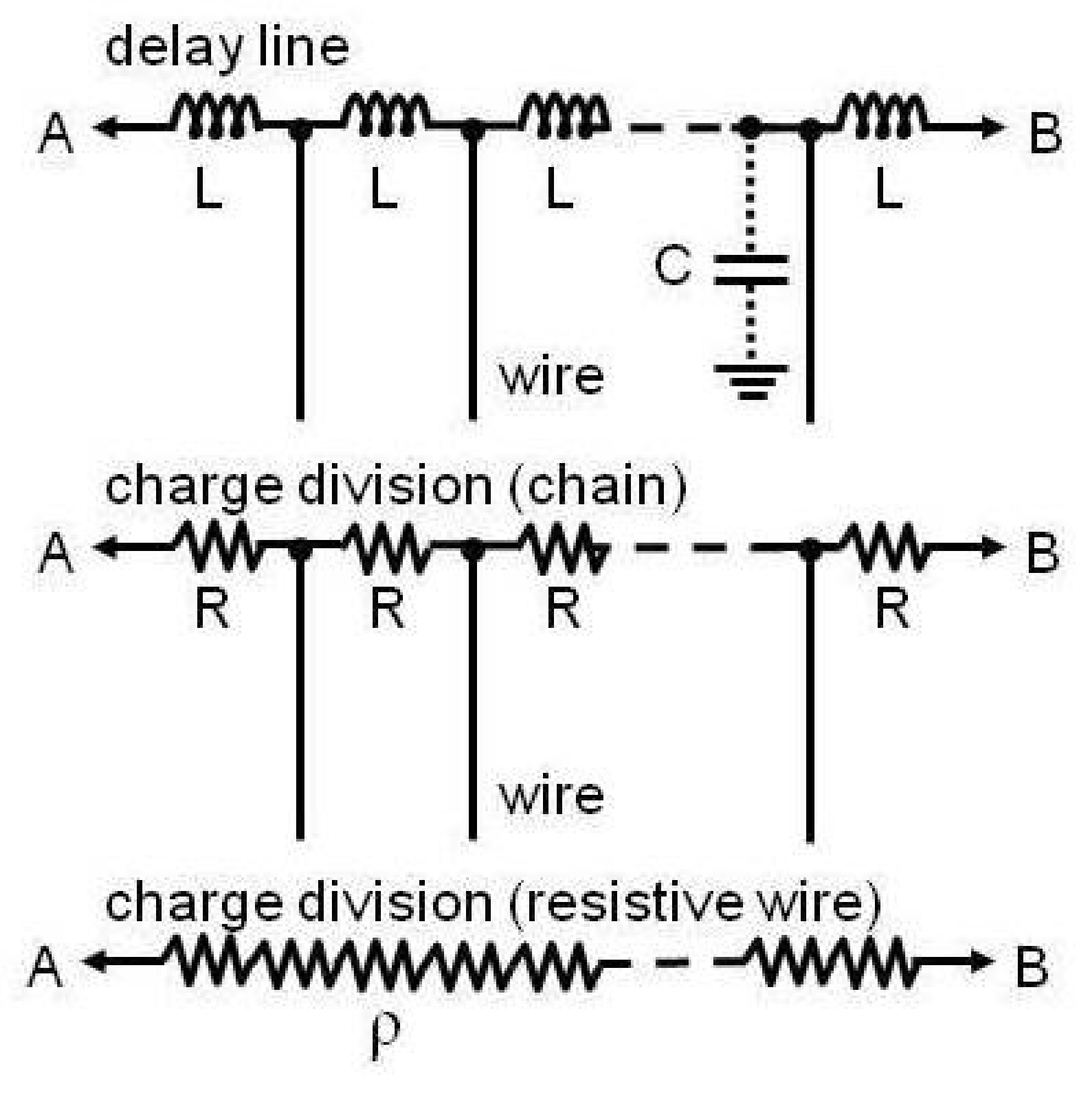}
\caption{\footnotesize Delay line read-out, charge division on
several wires and charge division on a single resistive
wire.}\label{5435y65fyhh}
\end{figure}
\\ Figure \ref{5435y65fyhh} shows the schematic for a delay line and
for charge division. In a delay line $N$ wires (or strips) are
connected in a chain through inductances, say $L$. Each wire has an
intrinsic capacitance to ground ($C$). Usually it is $100\,pF/m$.
The circuit we obtain is a discrete transmission line where signals
travel at the velocity $v=1/\sqrt{L_m\,C_m}$, where $L_m$ and $C_m$
are the inductance and capacitance per unity of length. At the two
outputs, A and B, we measure the difference in the arrival time of
the signals. The position along the delay chain is calculated as:
\begin{equation}
x = \frac{t_A-t_B}{\Delta t_{max}}
\end{equation}
where $\Delta t_{max}$ is the maximum delay possible. The position
$x$ is normalized between $-1,+1$ along the chain. The amplitude of
the two outputs signals, neglecting the resistivity of the chain, is
the same at both ends.
\\ The delay line approach is widely used in radiation detectors
in which the development of the signal is very fast. If the
development of the signal is slow and variable from event to event,
this introduces an error on the determination of the two arrival
times and, consequently, on $x$. This is the case for neutron
detectors where the particle track orientation in the gas introduces
a large jitter on the signals.
\\ Charge division consists of connecting several wires through
resistors or using resistive wires of resistivity $\rho$ (see Figure
\ref{5435y65fyhh}). The charge induced in one wire (for the chain)
or at certain position along the resistive wire, will split
according to the total resistance seen toward the two terminations A
and B. The amplitude of the signals at the two outputs ($Q_A$ and
$Q_B$ proportional to the charge) is decreased proportionally to the
resistance seen from the induction point. The position, normalized
between $[0,1]$, is:
\begin{equation}
x = \frac{Q_A}{Q_A+Q_B}
\end{equation}
where $Q_A+Q_B$ is the total charge induced.
\\ By neglecting any parasite capacitance, the two signals at both
ends arrive at the same time.
\\ By using resistive wires we avoid the use of a second wire plane
in a MWPC because the two-dimensional information is given by the
number of the wire fired and the charge division at its outputs.
\\ A single wire can not be used as a single delay line as for charge
division. The propagation velocity of the signal on the wire is of
the order of the speed of light, and the delay too short to be of
any use.
\\ Johnson noise is inversely proportional to resistance.
While in an ideal delay line there is no problem of noise, in a
resistive chain its is necessary to increase the gas avalanche gain
at which the detector operates in order to increase the signal to
noise ratio. For this reason, neutron detectors read-out in charge
division are operated at higher gain with respect to the same
detector with individual read-out.
\subsection{Dead time}
The \emph{dead time} is the minimum amount of time that must
separate two events in order that they can be recorded as two
separate pulses \cite{knoll}. It can be limited by the processes in
the detector itself or by the associated electronics. Usually the
detector is not susceptible of the incoming radiation in a interval
of a dead time and the information is then lost.
\\ The read-out system also affects the detector dead-time.
\\ A detector can be characterized by a
\emph{paralyzable} or \emph{non-paralyzable} behavior. In a
non-paralyzable detector, an event happening during the dead time of
the previous event, is simply lost. With an increasing event rate
the detector will reach a saturation rate equal to the inverse of
the dead time. In a paralyzable detector, an event happening during
the dead time will not just be missed, but will restart the dead
time. With increasing rate the detector will reach a saturation
point where it will be not capable to record any event at all. A
semi-paralyzable detector exhibits an intermediate behavior, in
which the event arriving during dead time does extend it, but not by
the full amount, resulting in a detection rate that decreases when
the event rate approaches saturation.
\section{The $^3He$ crisis}
\subsection{The $^3He$ shortage}
\begin{figure}[ht!] \centering
\includegraphics[width=7.8cm,angle=0,keepaspectratio]{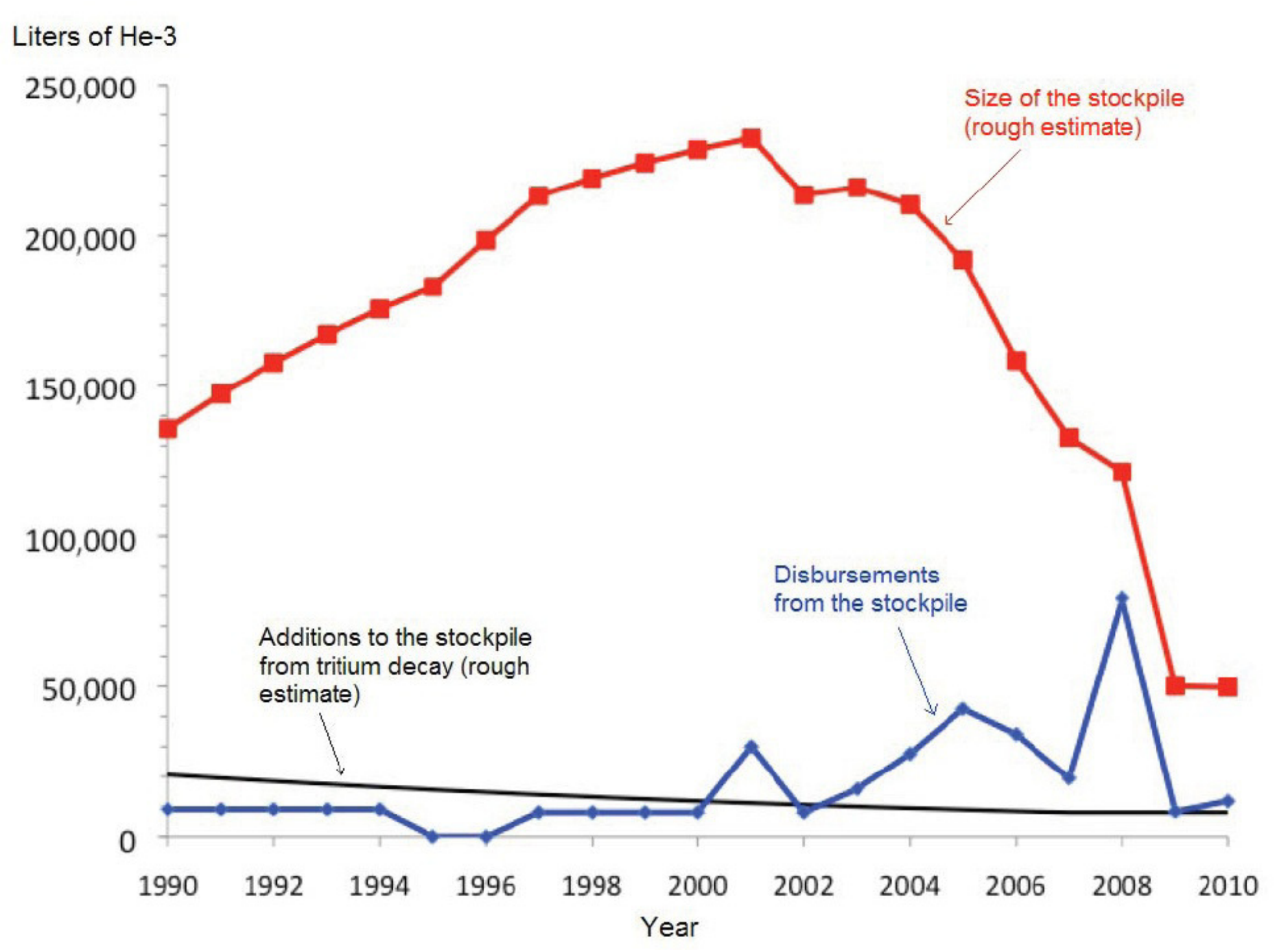}
\includegraphics[width=7.8cm,angle=0,keepaspectratio]{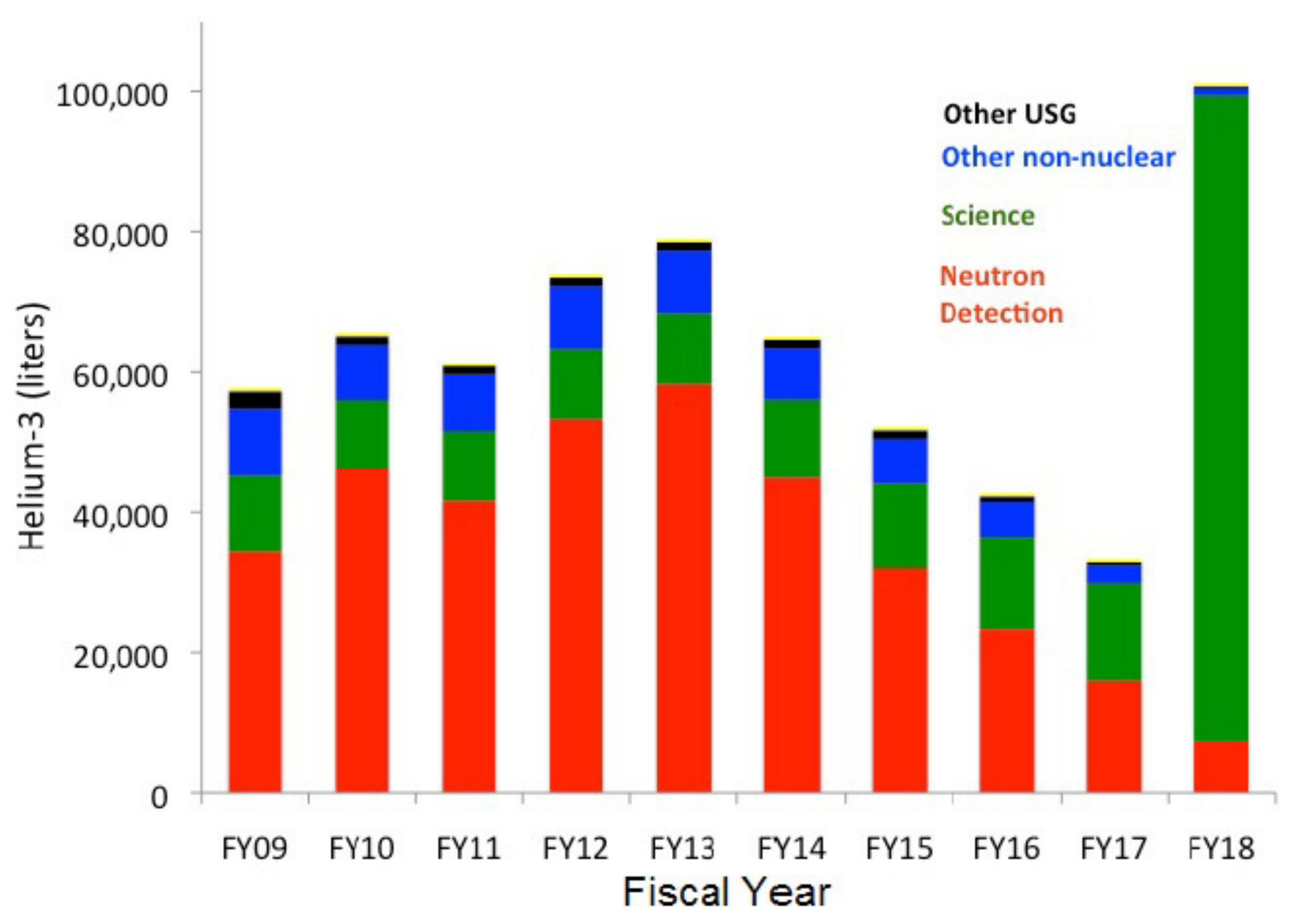}
\caption{\footnotesize Size of the $^3He$ stockpile 1999 - 2010, the
size of the stock in red and the demand in blue (left). Projected
$^3He$ demand 2009 - 2018 (right). Adapted from \cite{shea}.
\label{price1}}
\end{figure}

Scintillators, gaseous detectors and semiconductors are the main
technologies to detect neutrons.
\\ Neutron detection is a key element for applications
in homeland security, industry, and science \cite{shea}. Other uses
are for commercial instruments, dilution refrigerators, for targets
or cooling in nuclear research, and for basic research in condensed
matter physics \cite{kouzes3}.
\\ One of the principal thermal neutron converter materials is $^3He$,
that has been the main actor for years because of its favorable
properties. $^3He$ is a gas with high absorption cross-section and
no electronegativity. $^3He$ is an isotope of helium, an inert,
nontoxic, nonradioactive gas. Most helium is $^4He$. The natural
abundance of $^3He$, as a fraction of all helium, is very small:
only about $1.37$ parts per million. Rather than rely on natural
abundance, one usually manufactures $^3He$ through nuclear decay of
tritium (see Equation \ref{equap1}), a radioactive isotope of
hydrogen. The supply of $^3He$ comes almost entirely from US and
Russia. By far the most common source of $^3He$ in the United States
is the US nuclear weapons program, of which it is a byproduct. The
federal government produces tritium for use in nuclear warheads.
Over time, tritium decays into $^3He$ and must be replaced to
maintain warhead effectiveness. From the perspective of the weapons
program, the extracted $^3He$ is a byproduct of maintaining the
purity of the tritium supply. This means that the tritium needs of
the nuclear weapons program, not demand for $^3He$ itself, determine
the amount of $^3He$ produced.
\begin{equation}\label{equap1}
^3H \, \left( t_{1/2}=12.3y\right)\quad  \rightarrow \quad ^3He +
e^{-} + \bar{\nu_e}
\end{equation}
$^3He$ does not trade in the marketplace as many materials do. It is
accumulated in a stockpile from which supplies are either
transferred directly to other agencies or sold publicly at auction.
Despite declining supply and increasing demand, the auction price of
$^3He$ has been relatively steady, at less than $100$\$ per liter.
\\ Until 2001, $^3He$ production by the nuclear weapons program
exceeded the demand, and the program accumulated a stockpile. In the
past decade $^3He$ consumption has risen rapidly. After the
terrorist attacks of September 11, 2001, the federal government
began deploying neutron detectors at the US border to help secure
the nation against smuggled nuclear and radiological material. Thus
starting in about 2001, and more rapidly since about 2005, the
stockpile has been declining. By 2009, the US government and others
recognized that ongoing demand would soon exceed the remaining
supply.
\\ Nowadays, the world is experiencing a shortage of $^3He$ \cite{shea}.
US federal officials have testified that the shortage is acute and,
unless alternatives are found, will affect federal investments in
homeland security, scientific research, and other areas. Scientists
have expressed concern that the shortage may threaten certain fields
of research.
\\ The $^3He$ stockpile grew from roughly 140000 liters in 1990 to
roughly 235000 liters in 2001. Since 2001, however, $^3He$ demand
has exceeded production. By 2010, the increased demand had reduced
the stockpile to roughly 50000 liters (see Figure \ref{price1}).
\\ The US weapons program currently produces tritium by
irradiating lithium in a light-water nuclear reactor. Before 1988,
the program used heavy-water reactors at the DOE Savannah River Site
in South Carolina. In 1988, the last operating Savannah River Site
reactor, the K reactor, was shut down for safety reasons. For the
next several years, reductions in the nuclear weapons stockpile
meant that tritium recycling met the weapons program's needs without
additional tritium production. Over time, as the tritium produced
before 1988 decayed into $^3He$, the total amount of remaining
tritium decreased. The annual rate of $^3He$ production from the
remaining tritium declined commensurately. The DOE restarted tritium
production for the weapons program in 2003.
\\ Prior to 2001, the demand was approximately 8000 liters per year, which
was less than the new supply from tritium decay. After 2001, the
demand increased, reaching approximately 80000 liters in 2008. The
projections, shown in Figure \ref{price1}, show demand continuing at
above the available new supply for at least the next several years.
These projections contain many variables and are therefore
considerably uncertain.
\\ Given such a large mismatch between supply
and demand, users are likely to seek out alternative technologies,
reschedule planned projects, and make other changes that reduce
demand below what it would be in the absence of a shortage.
\\ One way to address the $^3He$ shortage would be to reduce demand by
moving $^3He$ users to alternative technologies. Some technologies
appear promising, though implementation would likely present
technical challenges. For other applications, alternative
technologies may not currently exist. It is unclear whether federal
agencies and the private sector can reduce demand sufficiently to
match the current $^3He$ supply and still meet priorities for
security, science, and other applications.
\subsection{Alternatives to $^3He$ in neutron detection}
Because of its detection performance, nontoxicity, and ease of use,
$^3He$ has become the material of choice for neutron detection.
Nevertheless, other materials also have a long history of use. With
the current shortage of $^3He$, researchers are reexamining past
alternatives and investigating new ones. Existing alternative
neutron detection technologies have significant drawbacks relative
to $^3He$, such as toxicity or reduced sensitivity. A drop-in
replacement technology does not currently exist. The alternatives
with most short-term promise as $^3He$ replacements are boron
trifluoride, boron-lined tubes, lithium-loaded glass fibers, and
scintillatorcoated plastic fibers. A new scintillating crystal
composed of cesium-lithium-yttrium-chloride (CLYC) also appears
promising. Other materials, less suitable in the short term, show
promise for the long term. Before the $^3He$ shortage became
apparent, most neutron detection research was directed toward
long-term goals such as improving sensitivity, efficiency, and other
capabilities, rather than the short-term goal of matching current
capabilities by alternative means. The neutron detectors used for
homeland security present both an opportunity and a challenge. The
large base of already deployed equipment, if retrofitted with an
alternative technology, could be a substantial source of recycled
$^3He$ for other uses. At the same time, the scale of planned future
deployments presents a potentially large future demand for $^3He$ if
suitable alternatives are not identified. For retrofitting, any
alternative would need to match the dimensions, power requirements,
and other characteristics of the existing technology. For future
deployments, especially beyond the near term, some of these
requirements might be relaxed or altered. In either case, the large
number of systems means that any alternative would need to be
relatively inexpensive. In addition, because of how and where the
equipment is used, any alternative would need to be rugged, safe,
and reliable.
\\ In the last years several efforts have been made to address the $^3He$
shortage. $^{10}B$ technologies have been investigated in gaseous
detectors for neutron science \cite{jonisorma}, \cite{jonitesi1},
\cite{wang1}, \cite{salvat}, \cite{kleinjalousie},
\cite{tsorbatzoglou1} and for homeland security applications
\cite{lacy1}, \cite{lacy2}. $^{10}B$ in solid junction detectors and
GEMs as \cite{kleincascade}, \cite{gregor} and \cite{shoji}. \\ In
this work, we will concentrate on $^{10}B$ solid converters in gas
detectors.

\chapter{Theory of solid neutron converters}\label{Chapt1}

This chapter is born from the collaboration between Patrick Van Esch
and me. I want to really thank him for the precious discussion we
had on the subject \cite{fratheo}. \\ Here are explained the
principles of neutron detection via solid converters and how such
detectors have to be optimized.

\newpage
\section{Introduction}\label{introchaptheo}
\begin{figure}[ht!] \centering
\includegraphics[width=10cm,angle=0,keepaspectratio]{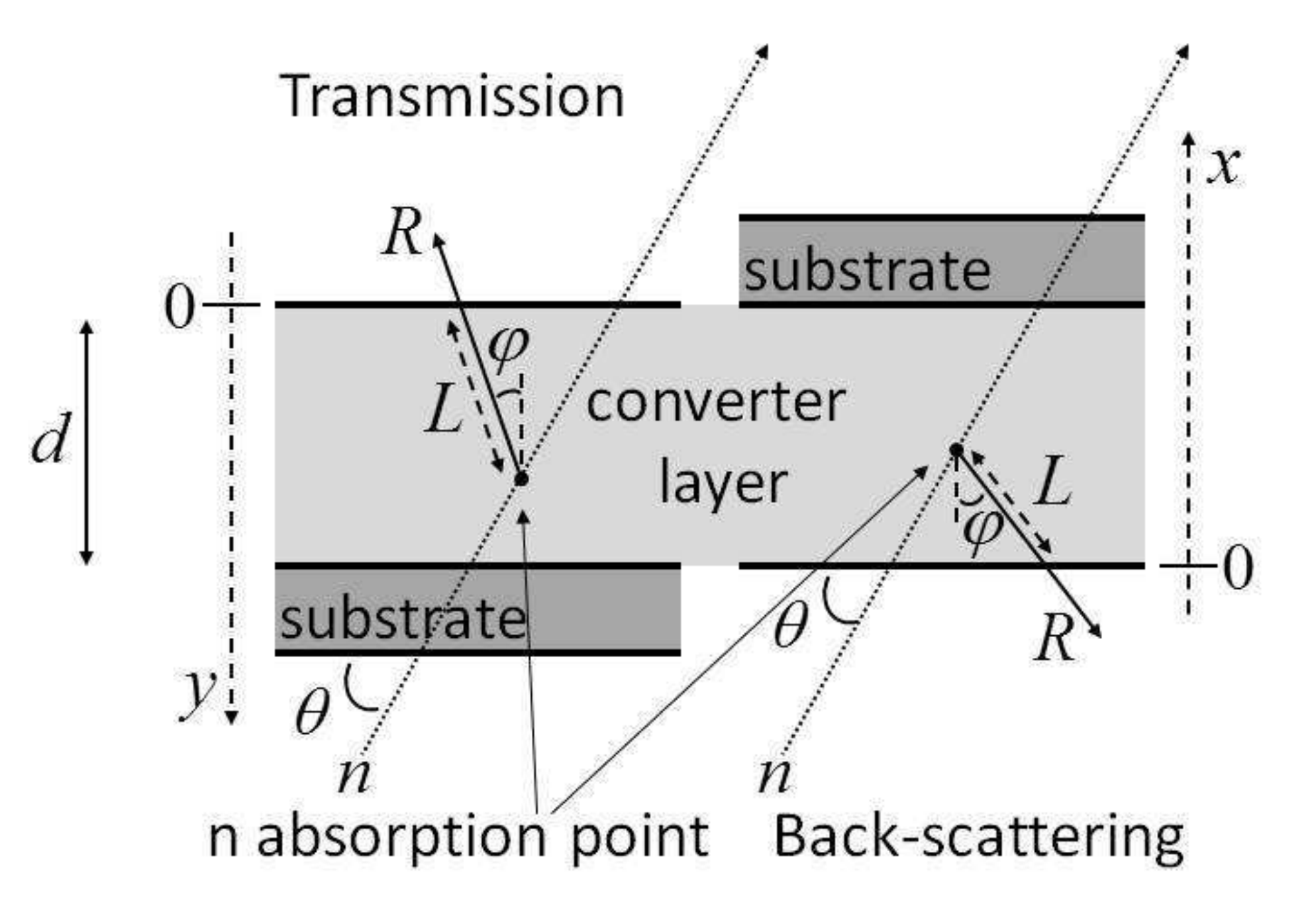}
\caption{\footnotesize Variables definition for a
\emph{back-scattering} and \emph{transmission} layer
calculations.}\label{coorsys}
\end{figure}

\begin{figure}[!ht] \centering
\includegraphics[width=6cm,angle=0,keepaspectratio]{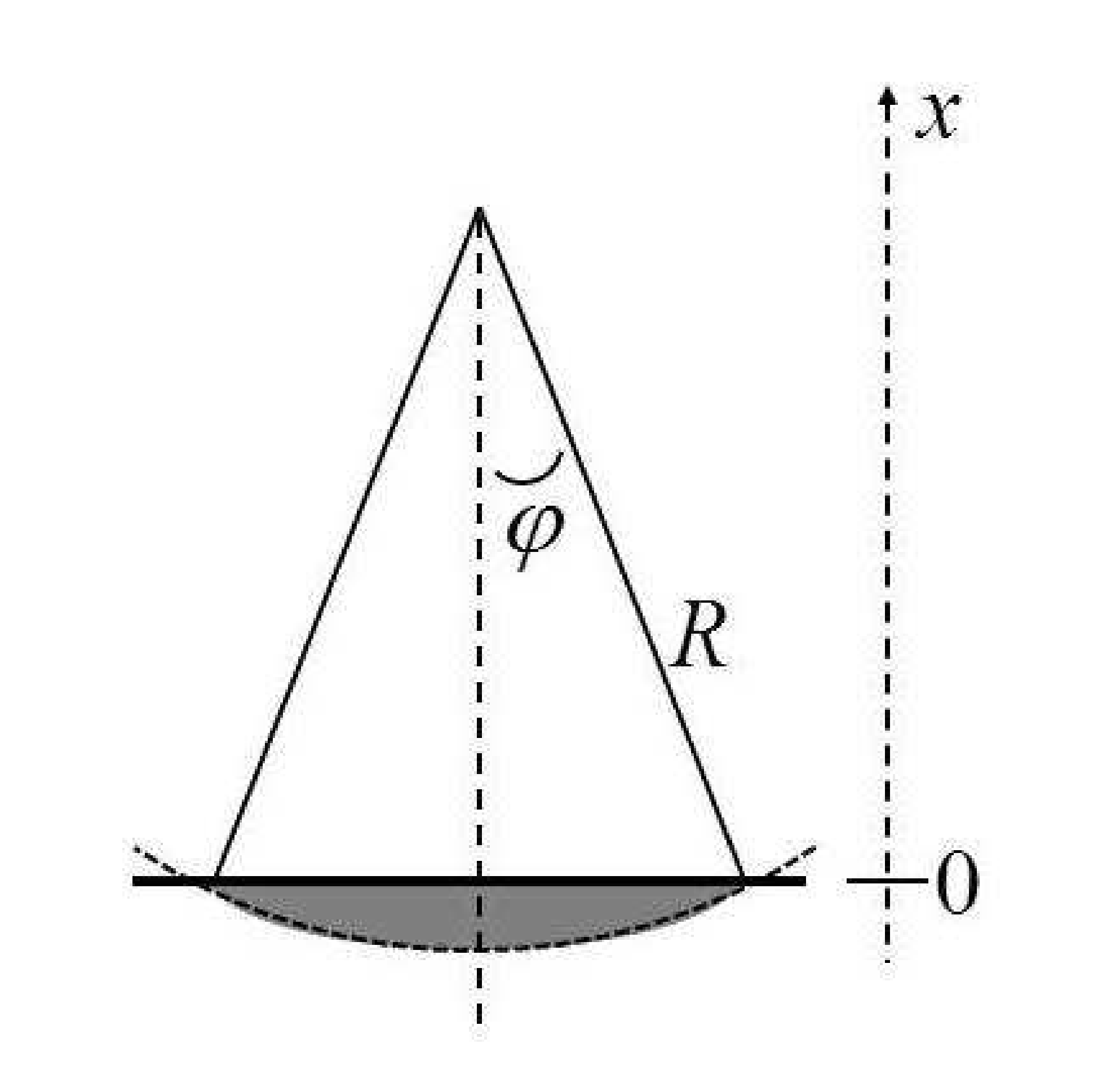}
\caption{\footnotesize Geometric interpretation of Equation
\ref{eqaa5}. \label{spcap1}}
\end{figure}

\begin{figure}[ht!] \centering
\includegraphics[width=7.8cm,angle=0,keepaspectratio]{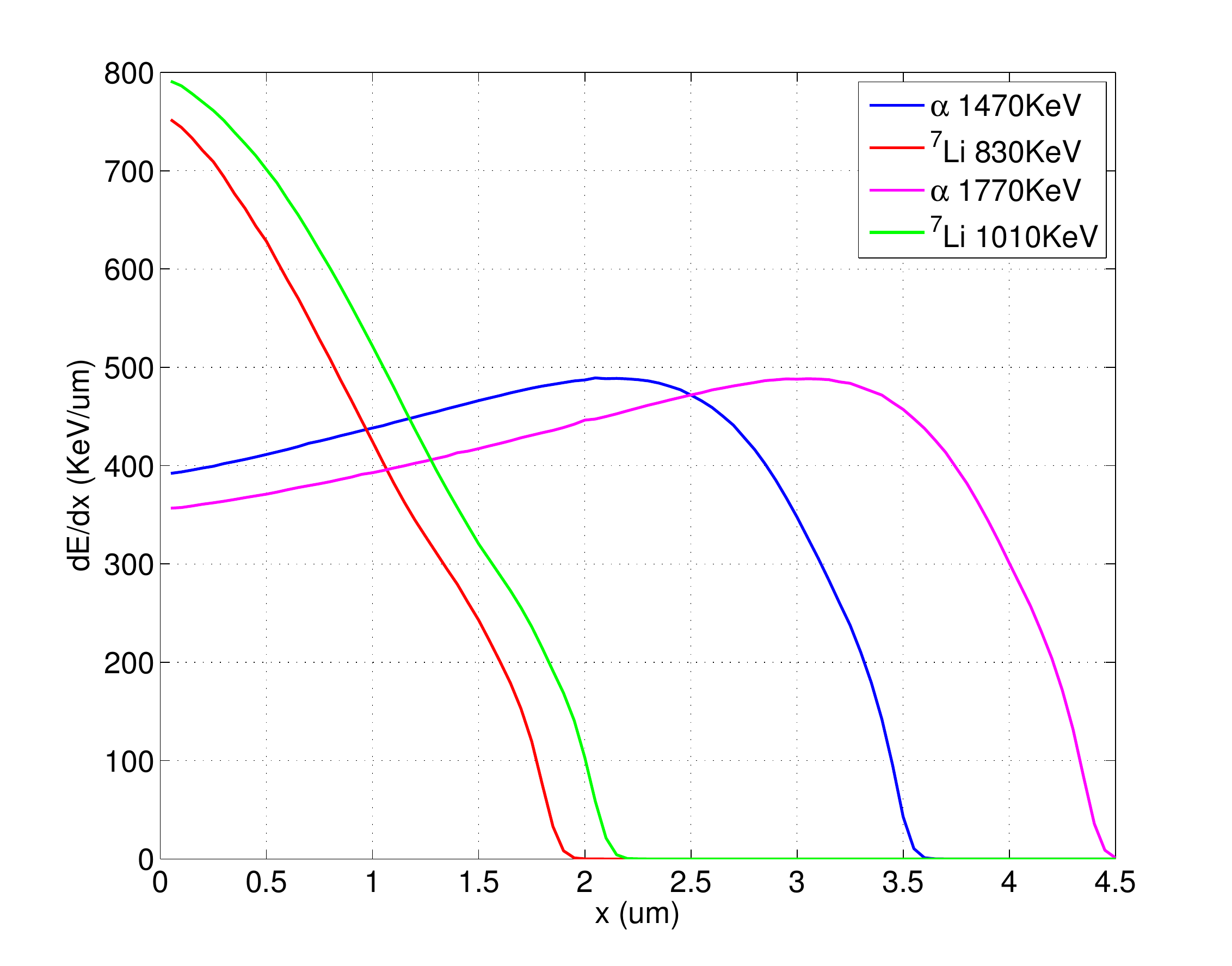}
\includegraphics[width=7.8cm,angle=0,keepaspectratio]{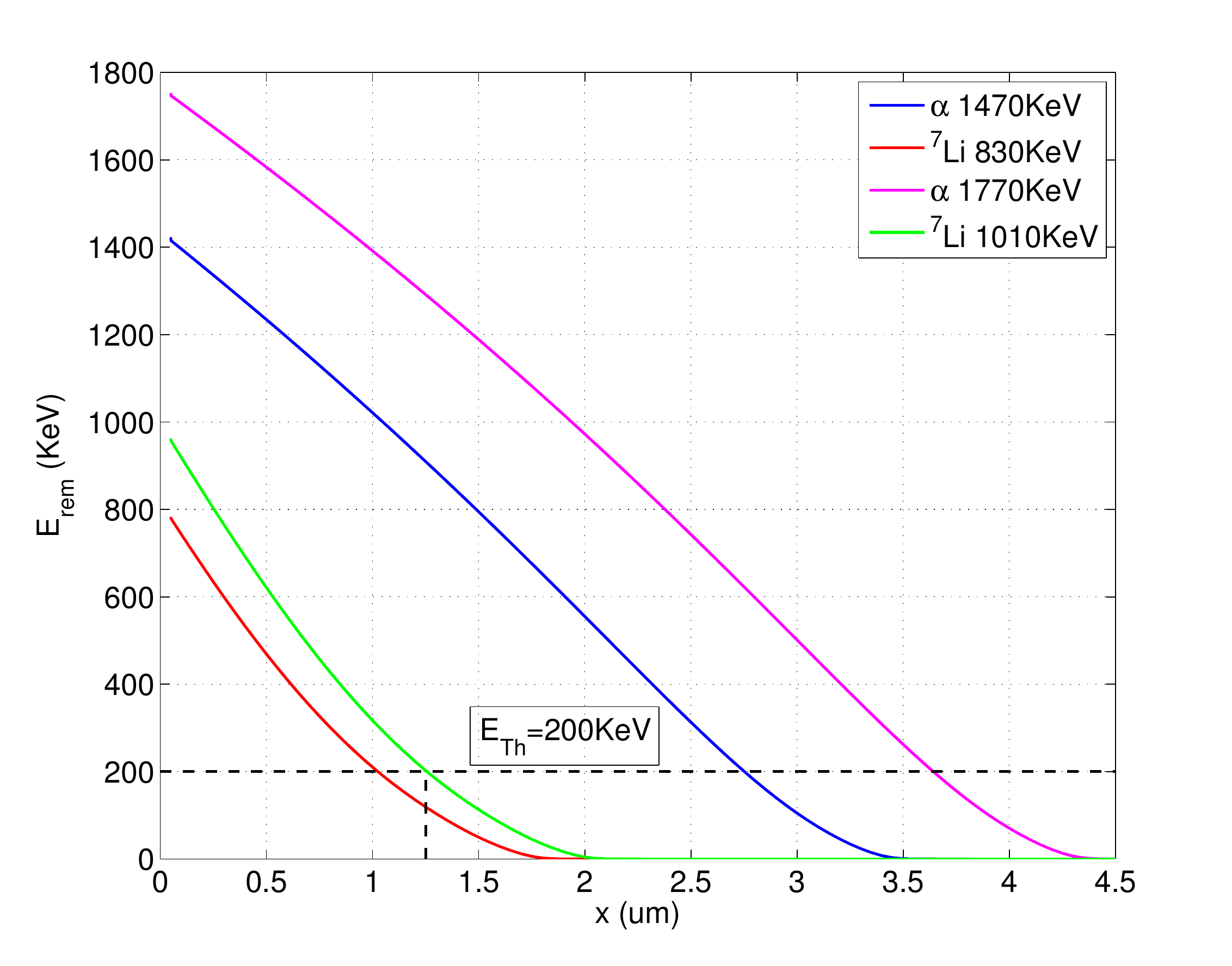}
\caption{\footnotesize Stopping power and remaining energy $E_{rem}$
as a function of the distance traveled $x$ for the four $^{10}B$
neutron capture reaction fragments.}\label{stpow1}
\end{figure}

Using powerful simulation software has the advantage of including
many effects and potentially results in high accuracy. On the other
hand it does not always give the insight an equation can deliver.
\\ The calculations we are going to show originate from the necessity
to understand both the Pulse Height Spectra (PHS) given by solid
neutron converters employed in thermal neutron detectors as in
\cite{jonisorma}, \cite{kleinjalousie}, \cite{lacy1}, and from the
investigation over such a detectors efficiency optimization.
\\ When a neutron is converted in a gaseous medium, such as $^3He$
detector, the neutron capture reaction fragments ionize directly the
gas and the only energy loss is due to the wall effect. As a result,
such detectors show a very good $\gamma$-rays to neutron
discrimination because $\gamma$-rays release only a small part of
their energy in the gas volume and consequently neutron events and
$\gamma$-rays events are easily distinguishable on the PHS.
Moreover, the detector efficiency is essentially given by the
absorption probability.
\\ On the other hand, when
dealing with hybrid detectors, as in \cite{jonisorma},
\cite{bruproceed}, \cite{khaplanov}, where the neutron converter is
solid and the detection region is gaseous, the efficiency
calculation is more complex. Also the $\gamma$-ray to neutron
discrimination for such a detector can be an issue \cite{lacy2},
\cite{athanasiades1}. Indeed, once a neutron is absorbed by the
solid converter, it gives rise to charged fragments which have to
travel across part of the converter layer itself before reaching the
gas volume to originate a detectable signal. As a result, those
fragments can release only a part of their energy in the gas volume.
The neutron PHS can thus have important low energy contributions,
therefore $\gamma$-ray and neutron events are not well separated
just in energy.
\\ In this chapter we want to give an understanding of the important aspects
of the PHS by adopting a simple theoretical model for solid neutron
converters. We will show good agreement of the model with the
measurements obtained with a $^{10}B$-based detector.
\\ The analytical model can help us to optimize the
efficiency for single and multi-layer detector as well as a function
of incidence angle as neutron wavelength distribution.
\\ The model we use is the same as implicitly used in many papers
such as \cite{gregor} or \cite{salvat}. It makes the following
simplifying assumptions:
\begin{itemize}
    \item the tracks of the emitted particles are straight lines
    emitted back-to-back and distributed isotropically;
    \item the energy loss is deterministic and given by the Bragg
    curves without fluctuations;
    \item the energy deposited is proportional to the charge
    collected without fluctuations.
\end{itemize}
We will consider solid neutron converters deposited on an holding
substrate. We consider the substrate and the layer indefinitely
extended in the plane (no lateral border effects). Referring to
Figure \ref{coorsys}, we talk about a \emph{back-scattering} layer
when neutrons are incident from the gas-converter interface and the
escaping particles are emitted backwards into the gas volume; we
call it a \emph{transmission} layer when neutrons are incident from
the substrate-converter interface and the escaping fragments are
emitted in the forward direction in the sensitive volume. We
consider a neutron to be converted at a certain depth ($x$ for
back-scattering or $d-y$ for transmission) in the converter layer
and its conversion yields two charged particles emitted
back-to-back.
\\ For the moment, we consider perpendicular impact on a layer. We
will later see that impact under an angle is easy to implement in
the resulting expressions.
\\ We recall that the probability density per unit of depth for a neutron of wavelength $\lambda$
to be absorbed by a converter at depth $x$ is given by:
\begin{equation}\label{eqaa1}
K(x,\lambda)=\Sigma e^{-x \Sigma(\lambda)}
\end{equation}
where $\Sigma(\lambda)=n\cdot\sigma(\lambda)$ is the macroscopic
absorption cross-section as already demonstrate in Chapter
\ref{chaptintradmatt}; $n$ is the number density of the material and
$\sigma(\lambda)$ the microscopic absorption cross-section
(expressed in barn).
\\ Focusing on the back-scattering case in Figure \ref{coorsys}, where we consider for the moment $\theta=\frac{\pi}{2}$,
the neutron absorption coordinate is denoted by $x$, the escape
probability for one particle, of range $R$, emitted isotropically at
depth $x$, where the neutron is absorbed, is given by:
\begin{equation}\label{eqaa5}
\xi(x)=\begin{cases} \frac 1 2 \left(1-\frac x {R} \right)&\mbox{if  \,} x\leq R\\
0 &\mbox{if \, } x>R
\end{cases}
\end{equation}
As a matter of fact, the ratio $2\pi \cdot \frac x R=2\pi \cdot
\cos{\left(\varphi\right)}$ is the solid angle of unity sphere
coming out of the layer (see Figure \ref{spcap1}).
There are two particles emitted by the neutron reaction and they are
emitted back-to-back. Each time a neutron is absorbed one of the two
fragments is lost because of the holding substrate. The escape
probability in this case is:
\begin{equation}\label{prob2}
\xi(x)=\begin{cases} \frac 1 2 \left(2-\frac x {R_1}-\frac x {R_2}\right) &\mbox{if \, } x\leq R_2<R_1\\
\frac 1 2 \left(1-\frac x {R_1}\right) &\mbox{if \, } R_2 < x \leq R_1\\
0 &\mbox{if \, } R_2 < R_1 < x
\end{cases}
\end{equation}
where $R_1$ and $R_2$ are the two particle ranges, with $R_2< R_1$.
\\ Still referring to Figure \ref{coorsys}, in the case we consider
the {\it Transmission mode} the variable $x$ has to be replaced by
$y$, therefore the neutron is absorbed at depth $d-y$.
\\ As already mentioned in Section \ref{cpitheo851} of Chapter
\ref{chaptintradmatt}, since we are interested in particle
detection, the range defined here is the effective range a particle
has to travel through a material to still conserve an energy
$E_{Th}$ which is the minimum detectable energy. As an example we
take the four fragments originated by the neutron capture in
$^{10}B$ (see Table \ref{eqaa4}). Their stopping powers and
remaining energies $E_{rem}$ (Equation \ref{eqaa234}) are shown in
Figure \ref{stpow1} \cite{sri}; e.g. one of the $^7Li$ particles,
that carries $1010\,KeV$, presents an \emph{extrapolated range} of
about $2\, \mu m$ but if our minimum detectable energy is
$E_{Th}=200\,KeV$ the \emph{effective range} is reduced to about
$1.25\, \mu m$.

\section{Theoretical efficiency calculation}\label{secttheoeff}
\subsection{One-particle model} 
\begin{figure}[!ht] \centering
\includegraphics[width=7.8cm,angle=0,keepaspectratio]{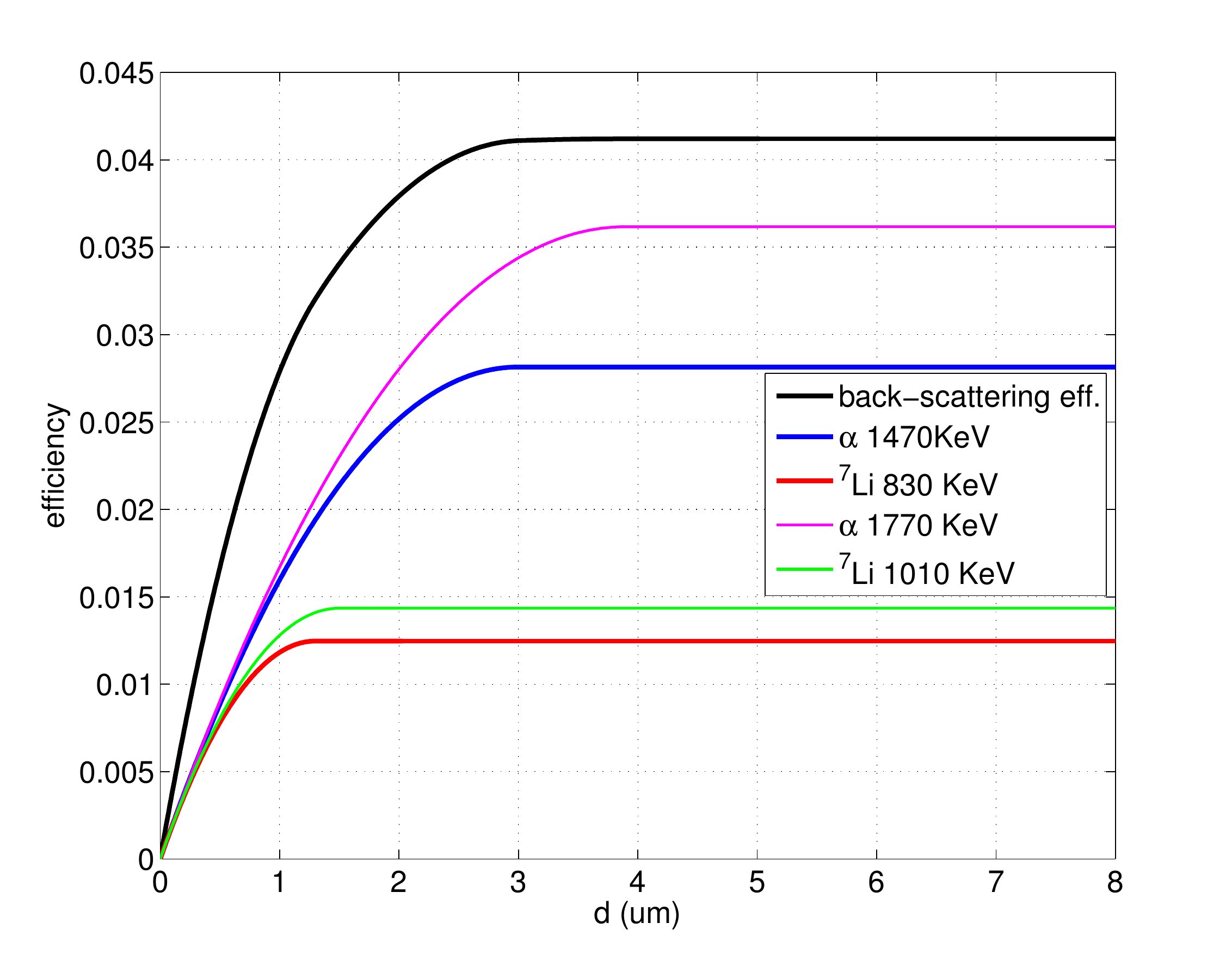}
\includegraphics[width=7.8cm,angle=0,keepaspectratio]{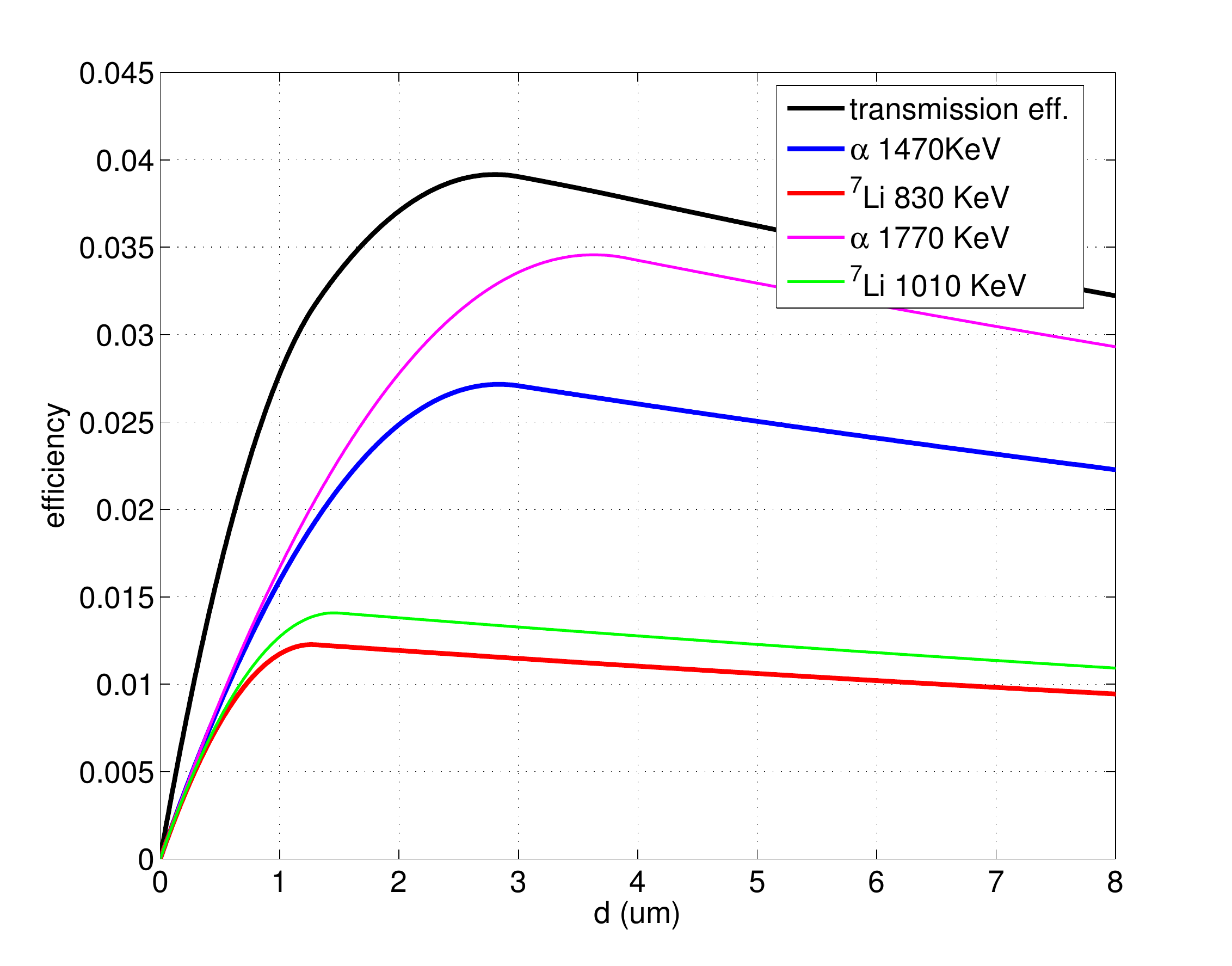}
\caption{\footnotesize Efficiency per particle and for the whole
reaction for a single back-scattering layer of $^{10}B_4C$ (left)
and for a transmission layer (right). $\Sigma=0.04\, \mu m ^{-1}$
and $100\,KeV$ energy threshold applied. \label{effperpart1}}
\end{figure}

\begin{figure}[!ht] \centering
\includegraphics[width=10cm,angle=0,keepaspectratio]{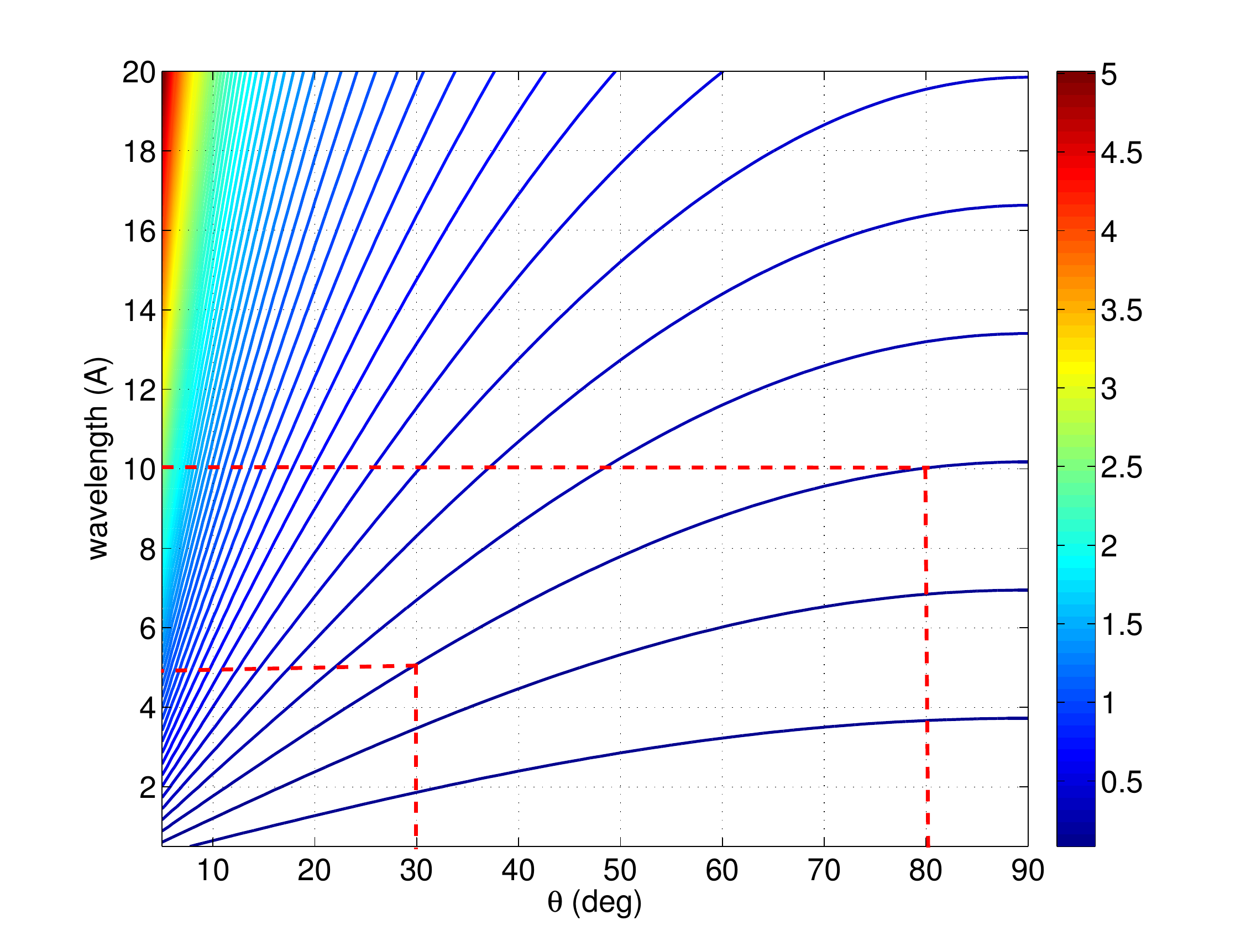}
\caption{\footnotesize Value for the macroscopic neutron
cross-section $\Sigma$ (expressed in $\mu m^{-1}$) for $^{10}B_4C$
($n_{^{10}B}=1.04\cdot10^{23}$ atoms$/cm^3$) as a function of
neutron wavelength ($\lambda$) and neutron incidence angle
($\theta$).}\label{figusigmagross}
\end{figure}

Let's start by considering a
neutron interaction with the material that generates one outgoing
particle isotropically in $4\pi\,sr$. \\Two cases could happen:
either the range of the fragment is longer than the thickness of the
layer or it is shorter. \\In the first case ($d\leq R$) in the
back-scattering case the efficiency is given by:
\begin{equation}\label{BS1}
\varepsilon_{BS}(d)=\int_0^d dx K(x) \xi(x)=\int_0^d dx \Sigma e^{-x
\Sigma} \left(\frac 1 2-\frac x {2R} \right)= \frac 1 2
\left(1-\frac1 {\Sigma R}\right)\left(1-e^{-d\Sigma}\right)+\frac d
{2R}e^{-d\Sigma}
\end{equation}
In the case of transmission:
\begin{equation}\label{T1}
\varepsilon_{T}(d)=\int_0^d dy K(d-y) \xi(y)=\int_0^d dy \Sigma
e^{-(d-y) \Sigma} \left(\frac 1 2-\frac y {2R} \right)= \frac 1 2
\left(1+\frac1 {\Sigma R}\right)\left(1-e^{-d\Sigma}\right)-\frac d
{2R}
\end{equation}
In the second case ($d>R$) the back-scattering efficiency is:
\begin{equation}\label{BS1d}
\varepsilon_{BS}(d)=\int_0^R dx K(x) \xi(x)=\int_0^R dx \Sigma e^{-x
\Sigma} \left(\frac 1 2-\frac x {2R} \right)= \frac 1 2 \left(1-
\frac 1 {\Sigma R} +\frac {e^{-R\Sigma}} {\Sigma R}\right)
\end{equation}
For transmission:
\begin{equation}\label{T1d}
\varepsilon_{T}(d)=\int_0^R dy K(d-y) \xi(y)=\int_0^R dy \Sigma
e^{-(d-y) \Sigma} \left(\frac 1 2-\frac y {2R} \right)= \frac 1
{2\Sigma R} e^{-d\Sigma}  \left(e^{+R\Sigma}-R \Sigma -1\right)
\end{equation}
In Figure \ref{effperpart1} the efficiencies, at $1.8$\AA, for the
four particles of the $^{10}B$ neutron capture reaction are shown.
We took $^{10}B_4C$ of density $\rho=2.24\,g/cm^3$ for which the
four fragments' effective ranges under a threshold of $100\,KeV$
are: $R_{Li(830KeV)}=1.3\, \mu m$, $R_{\alpha(1470KeV)}=3\, \mu m$,
$R_{Li(1010KeV)}=1.5\, \mu m$ and $R_{\alpha(1770KeV)}=3.9\, \mu m$.
The single particle efficiency is plotted for a layer in
back-scattering mode and for one in transmission. The total
efficiency for such a layer is also shown and it is obtained by
adding the four particles efficiencies according to the relative
branching ratio probability.
\\ From Figure \ref{effperpart1} we can see that the maximum efficiency
is attained at different thicknesses for the four fragments, this is
due to the difference in the particle effective ranges. The maximum
efficiency position, considering the whole reaction, lays in between
the four maxima.
\\ A layer operated in back-scattering mode saturates its
efficiency once attained the maximum. This is due to the fact that
adding more material to the layer only gives rise to particles that
can never escape the converter hence they can not deposit energy in
the gas volume. The maximum is reached at the value of the effective
range. On the other hand, the material added to a layer in
transmission mode starts absorbing neutrons without any particle
escaping as in the back-scattering layer but now the actual neutron
flux the layer close to the gas volume can experience is diminished,
consequently the efficiency drops.
\\ In general, the direction of the incoming neutrons could be non-orthogonal
to the converter layer. Referring to Figure \ref{coorsys}, this
angle is indicated by $\theta$ and it is measured starting form the
layer surface. All the formulae deduced for the efficiency are still
valid under an angle. It is sufficient to replace $\Sigma$ by
$\frac{\Sigma}{\sin(\theta)}$. We demonstrate the validity only for
the back-scattering case and $d \leq R$ (Equation \ref{BS1}). By
inclining the layer the effective layer thickness (the path length
of the neutron in the layer) will be $\frac d {\sin(\theta)}$ and
the integration variable $x$ can be modified in the same way to
$z=\frac x {\sin(\theta)}$. However, in $\xi$, the distance is still
the perpendicular distance to the surface, $x= z\,\sin(\theta)$.
\begin{equation}\label{BS1sigmadem}
\begin{aligned}
\varepsilon_{BS}(d,\Sigma,\theta)=&\int_0^{d/{\sin(\theta)}} dz K(z)
\xi(z \cdot \sin(\theta))=\int_0^{d/{\sin(\theta)}} dz \Sigma e^{-z
\Sigma} \left(\frac 1 2-\frac {z \sin(\theta)} {2R} \right)=\\
=&\frac 1 2 \left(1-\frac{\sin(\theta)} {\Sigma
R}\right)\left(1-e^{-\frac{d \Sigma}{\sin(\theta)}}\right)+\frac d
{2R}e^{-\frac{d\Sigma}{\sin(\theta)}}=\varepsilon_{BS}\left(d,\frac{\Sigma}{\sin(\theta)},\theta=90^{\circ}\right)
 \end{aligned}
\end{equation}
Hence:
\begin{equation}\label{eqab2}
\Sigma \rightarrow \frac{\Sigma}{\sin(\theta)}
\end{equation}
This result can be alternatively derived by considering that the
 absorption distance $x$ under an angle $\theta$ is
equivalent to increase it by a factor: $x \rightarrow \frac x
{\sin(\theta)}$.
\\ \textbf{Moreover, this result points out that only the parameter $\Sigma$
plays a role in the efficiency determination, indeed the same value
of $\Sigma$ can be obtained by changing either the inclination of
the layer or the the neutron wavelength or the number density}.
Furthermore, in the cold and thermal neutron energy region the
microscopic neutron absorption cross-section can be approximated by
a linear behavior in $\lambda$, hence:
\begin{equation}\label{eqab2bisimp}
\Sigma(\lambda,\theta)=n \cdot \frac{\sigma(\lambda)}{\sin(\theta)}=
n \cdot \sigma_{1A} \cdot \frac{\lambda}{\sin(\theta)}
\end{equation}
\\ where the constant $\sigma_{1A}$ is the linear
extrapolation of the neutron absorption cross-section from the value
tabulated for $1.8$\AA. In the case of $^{10}B$:
$\sigma_{1A}=\frac{3844}{1.8}[b/$\AA]. As a result the same value
for $\Sigma(\lambda,\theta)$, thus the same efficiency, can be
obtained, for example, at $10$\AA \, under an angle of $80^{\circ}$
or equivalently at $5$\AA \, under an angle of $30^{\circ}$ as shown
in Figure \ref{figusigmagross}.

\subsection{Two-particle model}\label{eqtwopartmodel345} Now we consider two fragments
emitted back to back. The two particles have two different ranges,
and hereafter we consider $R_2<R_1$ without loss of generality.
\\For two particles we have three different cases: the
layer is thinner than both ranges, it is in between or it is thicker
than both of them.
\\Case $d\leq R_2<R_1$ for back-scattering:
\begin{equation}\label{BS21}
\begin{aligned}
\varepsilon_{BS}(d)&=\int_0^d dx K(x) \xi(x)=\int_0^d dx \Sigma
e^{-x \Sigma} \left(1 -\frac x {2R_1}-\frac x {2R_2} \right)=\\&=
\left(1-\frac 1 {2\Sigma R_1}-\frac 1 {2\Sigma R_2}
\right)\left(1-e^{-d \Sigma}\right)+\left(\frac 1 {2R_1}+\frac 1
{2R_2}\right)d e^{-d \Sigma} \end{aligned}
\end{equation}
For transmission:
\begin{equation}\label{T21}
\begin{aligned}
\varepsilon_{T}(d)&=\int_0^d dy K(d-y) \xi(y)=\int_0^d dy \Sigma
e^{-(d-y) \Sigma} \left(1- \frac y {2R_1}-\frac y {2R_2} \right)=
\\&=\left(1+\frac 1 {2\Sigma R_1}+\frac 1 {2\Sigma R_2}
\right)\left(1-e^{-d \Sigma}\right)-\left(\frac 1 {2R_1}+\frac 1
{2R_2}\right)d \end{aligned}
\end{equation}
Case $R_2 < d \leq R_1$ for back-scattering:
\begin{equation}\label{BS22}
\begin{aligned}
\varepsilon_{BS}(d)&=\int_0^{R_2} dx K(x) \xi(x)+\int_{R_2}^d dx
K(x) \xi(x)=\\&=\int_0^{R_2} dx \Sigma e^{-x \Sigma} \left(1 -\frac
x {2R_1}-\frac x {2R_2} \right)+\int_{R_2}^d dx \Sigma e^{-x \Sigma}
\left(\frac 1 2 -\frac x {2R_1}\right)=\\&=\left(1-\frac 1 {2\Sigma
R_1}-\frac 1 {2\Sigma R_2} \right)+\frac{e^{-R_2 \Sigma}}{2R_2
\Sigma}-\frac{e^{-d \Sigma}}{2}\left(1-\frac 1 {R_1 \Sigma }-\frac d
{R_1}\right)
\end{aligned}
\end{equation}
For transmission:
\begin{equation}\label{T22}
\begin{aligned}
\varepsilon_{T}(d)&=\int_0^{R_2} dy K(d-y) \xi(y)+\int_{R_2}^d dy
K(d-y) \xi(y)=\\&=\int_0^{R_2} dy \Sigma e^{-(d-y) \Sigma} \left(1
-\frac y {2R_1}-\frac y {2R_2} \right)+\int_{R_2}^d dy \Sigma
e^{-(d-y) \Sigma} \left(\frac 1 2 -\frac y
{2R_1}\right)=\\&=\frac{e^{(R_2-d) \Sigma}}{2R_2
\Sigma}-\left(1+\frac 1 {2\Sigma R_1}+\frac 1 {2\Sigma R_2}
\right)e^{-d \Sigma}+\frac 1 2 \left(1+\frac 1 {R_1 \Sigma }-\frac d
{R_1}\right)
\end{aligned}
\end{equation}
Case $R_2< R_1< d$ for back-scattering:
\begin{equation}\label{BS23}
\begin{aligned}
\varepsilon_{BS}(d)&=\int_0^{R_2} dx K(x) \xi(x)+\int_{R_2}^{R_1} dx
K(x) \xi(x)=\\&=\int_0^{R_2} dx \Sigma e^{-x \Sigma} \left(1 -\frac
x {2R_1}-\frac x {2R_2} \right)+\int_{R_2}^{R_1} dx \Sigma e^{-x
\Sigma} \left(\frac 1 2 -\frac x {2R_1}\right)=\\&=\left(1-\frac 1
{2\Sigma R_1}-\frac 1 {2\Sigma R_2} \right)+\frac{e^{-R_2
\Sigma}}{2R_2 \Sigma}+\frac{e^{-R_1 \Sigma}}{2R_1 \Sigma}
\end{aligned}
\end{equation}
For transmission:
\begin{equation}\label{T23}
\begin{aligned}
\varepsilon_{T}(d)&=\int_0^{R_2} dy K(d-y) \xi(y)+\int_{R_2}^{R_1}
du K(d-y) \xi(y)=\\&=\int_0^{R_2} dy \Sigma e^{-(d-y) \Sigma}
\left(1 -\frac y {2R_1}-\frac y {2R_2} \right)+\int_{R_2}^{R_1} dy
\Sigma e^{-(d-y) \Sigma} \left(\frac 1 2 -\frac y
{2R_1}\right)=\\&=e^{-d \Sigma}\left(-1-\frac 1 {2\Sigma R_1}-\frac
1 {2\Sigma R_2}+\frac{e^{+R_2 \Sigma}}{2R_2 \Sigma}+\frac{e^{+R_1
\Sigma}}{2R_1 \Sigma} \right)
\end{aligned}
\end{equation}
\newpage
\section{Double layer}\label{Sect2laysub}
\begin{figure}[!ht]
\centering
\includegraphics[width=10cm,angle=0,keepaspectratio]{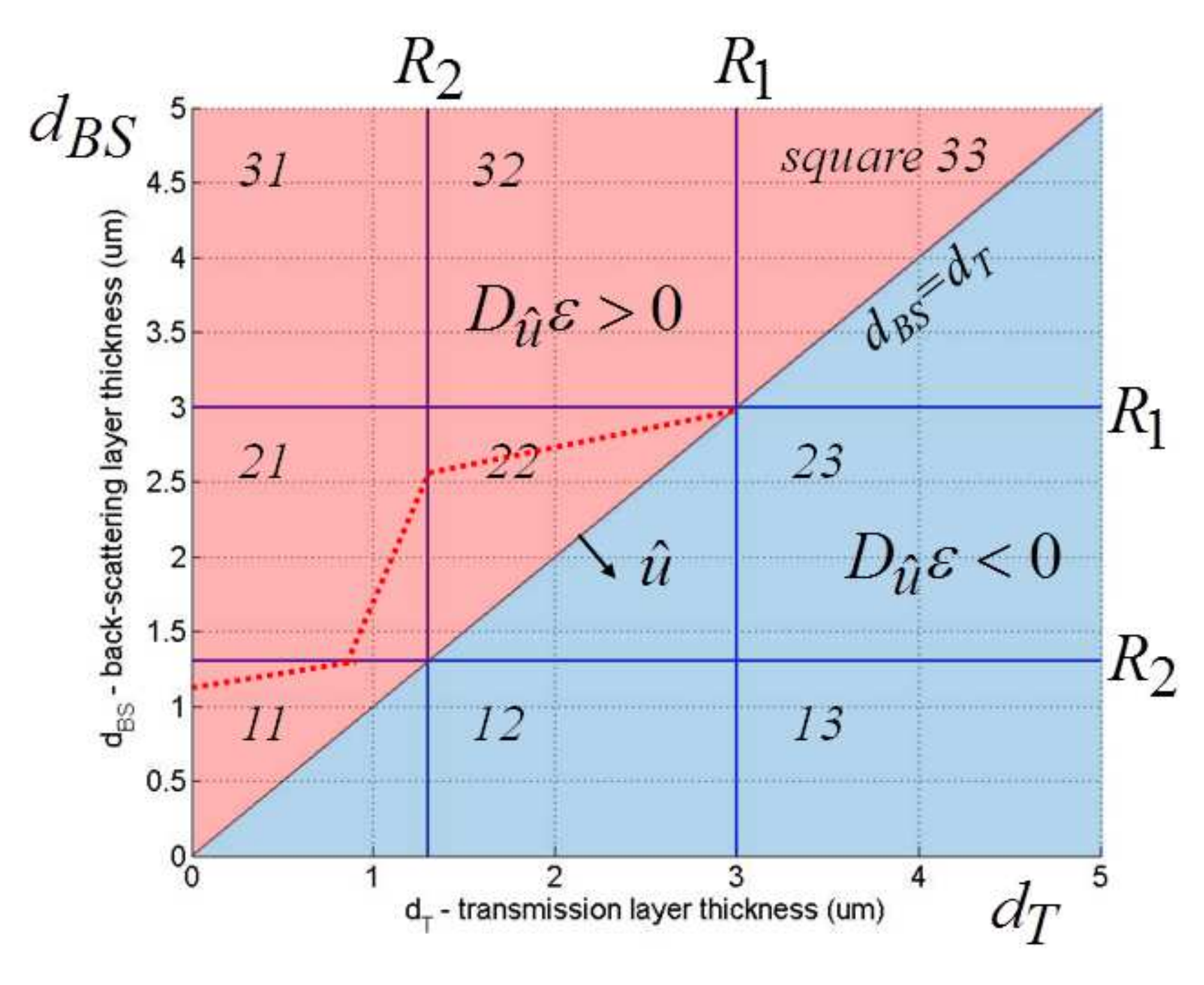}
\caption{\footnotesize Domain for the blade efficiency function. The
domain is divided into 9 partitions according to the neutron capture
fragments ranges. In red and blue are the domain partition where the
efficiency directional derivative along the unity vector $\hat{u}$
is respectively always positive or always negative. The red dotted
line represents the case when the substrate effect is not negligible
and the maximum efficiency can not be attained on the domain
bisector.} \label{fig2laydom}
\end{figure}

We put a double coated blade in a gas detection volume. A
\emph{blade} consists of a substrate holding two converter layers,
one in back-scattering mode and one in transmission mode.
\\ Starting from the analytical formulae derived in Section \ref{secttheoeff}
and \cite{gregor} we are going to derive properties that can help to
optimize the efficiency in the case of a monochromatic neutron beam
and in the case of a distribution of neutron wavelengths. E.g. since
the sputtering technique \cite{carina} coats each side of each
substrate with the same amount of converter material, one can wonder
if this is optimal or if there exists a different optimal thickness
for each side of the substrate to increase the neutron detection
efficiency. The questions are: which are the two coatings
thicknesses that maximize the efficiency of the blade for a given
set of parameters ($\theta$, $\lambda$, \dots)? And in the case the
neutron beam is not monochromatic but it is a distribution of
wavelength?
\\ By denoting with $d_{BS}$ the thickness of the coating for the
back-scattering layer and with $d_{T}$ the transmission layer
thickness, the efficiency of a blade is:
\begin{equation}\label{eqac1}
\varepsilon(d_{BS},d_{T})=\varepsilon_{BS}\left(d_{BS}\right)+e^{-\Sigma
\cdot d_{BS}}\cdot \varepsilon_{T}\left(d_{T}\right)
\end{equation}
where $\varepsilon_{BS}\left(d_{BS}\right)$ and
$\varepsilon_{T}\left(d_{T}\right)$ are the efficiencies for a
single coating calculated as shown in Section \ref{secttheoeff}. The
relation $\nabla \varepsilon(d_{BS},d_{T})=0$ determines the two
optimal layer thicknesses.
\\ In order to keep calculations simple, we consider only two
neutron capture fragments yielded by the reaction. This
approximation will not affect the meaning of the conclusion. In the
case of $^6Li$ Equation \ref{eqac1} is exact, for $^{10}B$ the
expression \ref{eqac1} should ideally be replaced by:
$$\varepsilon(d_{BS},d_{T})=0.94 \cdot\left(
\varepsilon^{0.94}_{BS}\left(d_{BS}\right)+e^{-\Sigma \cdot
d_{BS}}\cdot \varepsilon^{0.94}_{T}\left(d_{T}\right)\right)+0.06
\cdot\left( \varepsilon^{0.06}_{BS}\left(d_{BS}\right)+e^{-\Sigma
\cdot d_{BS}}\cdot \varepsilon^{0.06}_{T}\left(d_{T}\right)\right)$$
where $\varepsilon^{0.94}$ means the efficiency calculated for the
$94\%$ branching ratio reaction with the right effective particle
ranges. We will limit us to the $94\%$ contribution as if it were
$100\%$. As already defined in Section \ref{secttheoeff}, $R_1$ and
$R_2$, with ($R_2<R_1$), are the two ranges of the two neutron
capture fragments. Those regions, delimited by the particle ranges,
in the $\varepsilon(d_{BS},d_{T})$ plot, are marked by the blue
lines in the Figures \ref{fig2lay1} and \ref{fig2lay2}. In case of
$^{10}B_4C$ ($\rho=2.24g/cm^3$), which is the converter used in
\cite{jonisorma}, the two $94\%$ branching ratio reaction particle
ranges are $R_1=3\,\mu m$ ($\alpha$-particle) and $R_2=1.3\,\mu m$
($^7Li$), when a $100\,KeV$ energy threshold is applied (as defined
the minimum detectable energy in \cite{gregor}). If one would like
to use pure $^{10}B$ of density $\rho=2.17g/cm^3$, the two $94\%$
branching ratio reaction particle ranges are $R_1=3.2\,\mu m$
($\alpha$-particle) and $R_2=1.6\,\mu m$ ($^7Li$), when the same
energy threshold is applied. We will take $^{10}B_4C$ as the coating
material in our examples.
\\ As $\varepsilon_{BS}\left(d\right)$ and
$\varepsilon_{T}\left(d\right)$ have different analytical
expressions according to whether $d \leq R_2<R_1$, $R_2<d \leq R_1$
or $R_2<R_1<d$ we need to consider 9 regions to calculate $\nabla
\varepsilon(d_{BS},d_{T})$ as shown in Figure \ref{fig2laydom}. If
we were to include the four different reaction fragments we would
have to consider $25$ domain partitions (see Figure
\ref{fig2laydom}).
We will see later that the most important regions, concerning the
optimization process are the regions \emph{square 11} (where $d_{BS}
\leq R_2<R_1$ and $d_{T} \leq R_2<R_1$) and \emph{square 22} (where
$R_2<d_{T} \leq R_1$, $R_2<d_{BS} \leq R_1$).
\\ In order to consider a non-orthogonal incidence of
neutrons on the layers, it is sufficient to replace $\Sigma$ with
$\frac{\Sigma}{\sin(\theta)}$, where $\theta$ is the angle between
the neutron beam and the layer surface (see Figure \ref{coorsys}).
This is valid for both the single blade case and for a multi-layer
detector. The demonstration was derived in Section
\ref{secttheoeff}.

\subsection{Monochromatic double layer optimization}
In the domain region called \emph{square 11} the efficiency turns
out to be:
\begin{equation}\label{eqac2}
\begin{array}{ll}
\varepsilon_{11}(d_{BS},d_{T})&= \left(1-\frac{1}{2\Sigma
R_1}-\frac{1}{2\Sigma R_2} \right)\left(1-e^{-\Sigma d_{BS}}
\right)+d_{BS} \cdot \left(\frac 1 {2R_1} + \frac 1
{2R_2}\right)e^{-\Sigma d_{BS}}+ \\ & + e^{-\Sigma d_{BS}}
\left(\left(1+\frac{1}{2\Sigma R_1}+\frac{1}{2\Sigma R_2}
\right)\left(1-e^{-\Sigma d_{T}} \right)-\left(\frac 1 {2R_1} +
\frac 1 {2R_2}\right)d_{T} \right)=\\&=A \cdot \left(1-e^{-\Sigma
d_{BS}} \right)+d_{BS} \cdot C \cdot e^{-\Sigma d_{BS}} + e^{-\Sigma
d_{BS}} \left(B \cdot \left(1-e^{-\Sigma d_{T}} \right)-C \cdot
d_{T} \right)
\end{array}
\end{equation}
where we have called:
\begin{equation}\label{eqac2abc}
\begin{aligned}
A&=\left(1-\frac{1}{2\Sigma R_1}-\frac{1}{2\Sigma R_2} \right) \\
B&=\left(1+\frac{1}{2\Sigma R_1}+\frac{1}{2\Sigma R_2} \right)\\
C&=\left(\frac{1}{2 R_1}+\frac{1}{2 R_2} \right)
\end{aligned}
\end{equation}
From Equation \ref{eqac2abc}, we obtain the useful relations:
$\Sigma\left(A-B\right)=-2C$, $\Sigma A=\Sigma-C$ and $\Sigma
B=\Sigma+C$. By calculating $\nabla
\varepsilon_{11}(d_{BS},d_{T})=0$, we obtain the result that $d_{BS}
= d_{T}$ and
\begin{equation}\label{eqac3}
d_{BS} = d_{T} =-\frac{1}{\Sigma}\cdot \ln \left(\frac{C}{\Sigma
B}\right)
\end{equation}
We repeat the procedure for the \emph{square 22} in which the
efficiency is:
\begin{equation}\label{eqac4}
\begin{array}{ll}
\varepsilon_{22}(d_{BS},d_{T})&= \left(1-\frac{1}{2\Sigma
R_1}-\frac{1}{2\Sigma R_2} \right)+\frac{e^{-\Sigma R_2}}{2 \Sigma
R_2}-\frac{e^{-\Sigma d_{BS}}}{2}\cdot \left(1- \frac{1}{\Sigma
R_1}-\frac{d_{BS}}{R_1}\right)+e^{-\Sigma d_{BS}}\cdot\\ & \cdot
\left(-e^{-\Sigma d_{T}}\left(1+\frac{1}{2\Sigma
R_1}+\frac{1}{2\Sigma R_2}\right)+\frac{1}{2}
\left(1+\frac{1}{\Sigma R_1} -
\frac{d_{T}}{R_1}\right)+\frac{e^{-\Sigma \left(d_{T}-R_2
\right)}}{2\Sigma R_2} \right)=\\&= A +\frac{e^{-\Sigma R_2}}{2
\Sigma R_2}-\frac{e^{-\Sigma d_{BS}}}{2}\cdot \left(1-
\frac{1}{\Sigma R_1}-\frac{d_{BS}}{R_1}\right)+e^{-\Sigma
d_{BS}}\cdot\\ & \cdot \left(- B \cdot e^{-\Sigma d_{T}}+\frac{1}{2}
\left(1+\frac{1}{\Sigma R_1} -
\frac{d_{T}}{R_1}\right)+\frac{e^{-\Sigma \left(d_{T}-R_2
\right)}}{2\Sigma R_2} \right)
\end{array}
\end{equation}
We obtain again $d_{BS} = d_{T}$ and
\begin{equation}\label{eqac5}
d_{BS} = d_{T} =-\frac{1}{\Sigma}\cdot \ln \left(\frac{R_2}{R_1}
\left(\frac{1}{2 R_2 \Sigma B - e^{+\Sigma R_2}}\right)\right)
\end{equation}
Naturally each result of Equations \ref{eqac3} and \ref{eqac5} is
useful only if it gives a value that falls inside the region it has
been calculated for.
\\ The points defined by Equations \ref{eqac3} and \ref{eqac5} define
a \emph{maximum} of the efficiency function in the regions either
\emph{square 11} or \emph{square 22} because the Hessian matrix in
those points has a positive determinant and $\frac{\partial^2
\varepsilon}{\partial d_{BS}^2}$ is negative.
\begin{figure}[!ht]
\centering
\includegraphics[width=7.8cm,angle=0,keepaspectratio]{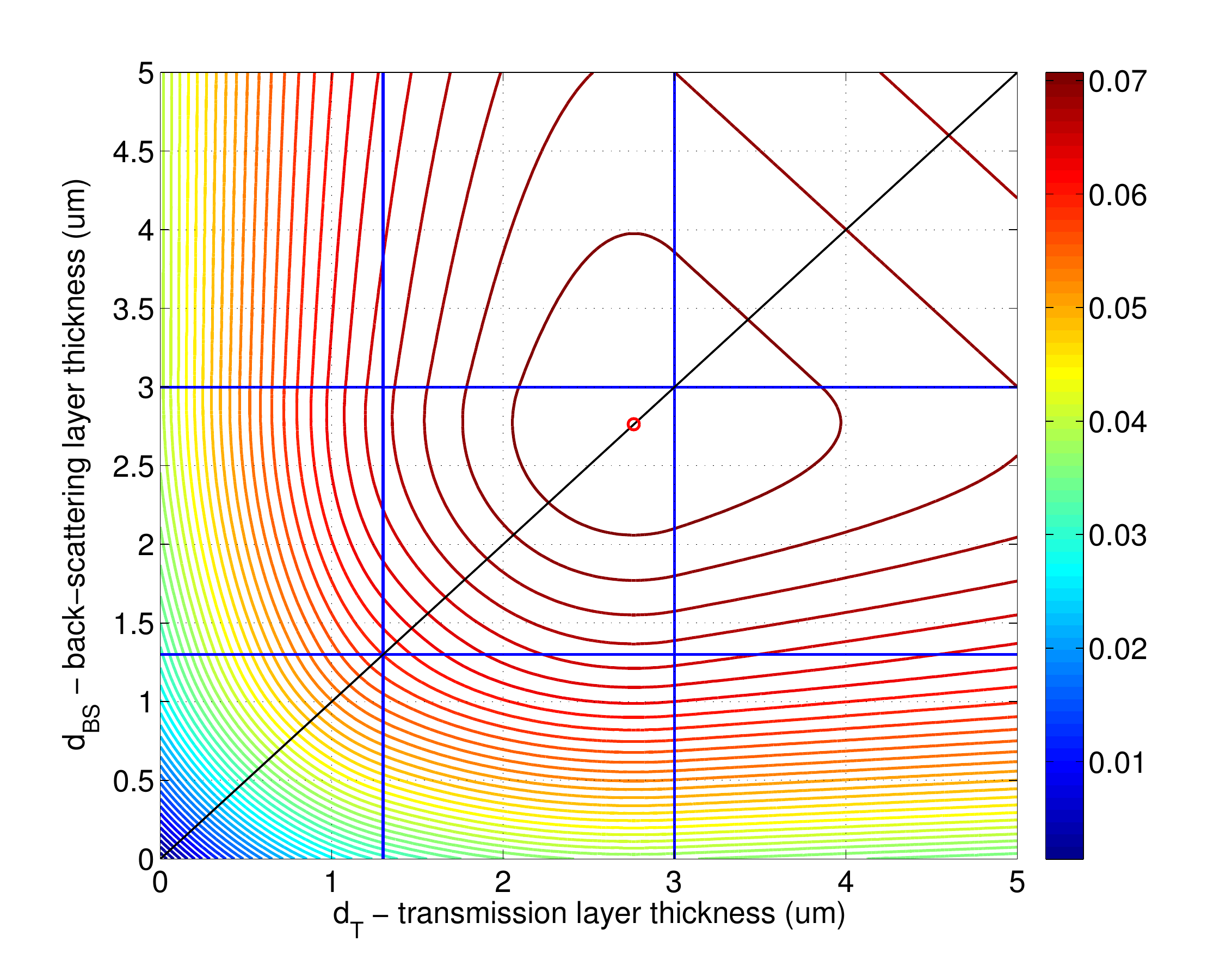}
\includegraphics[width=7.8cm,angle=0,keepaspectratio]{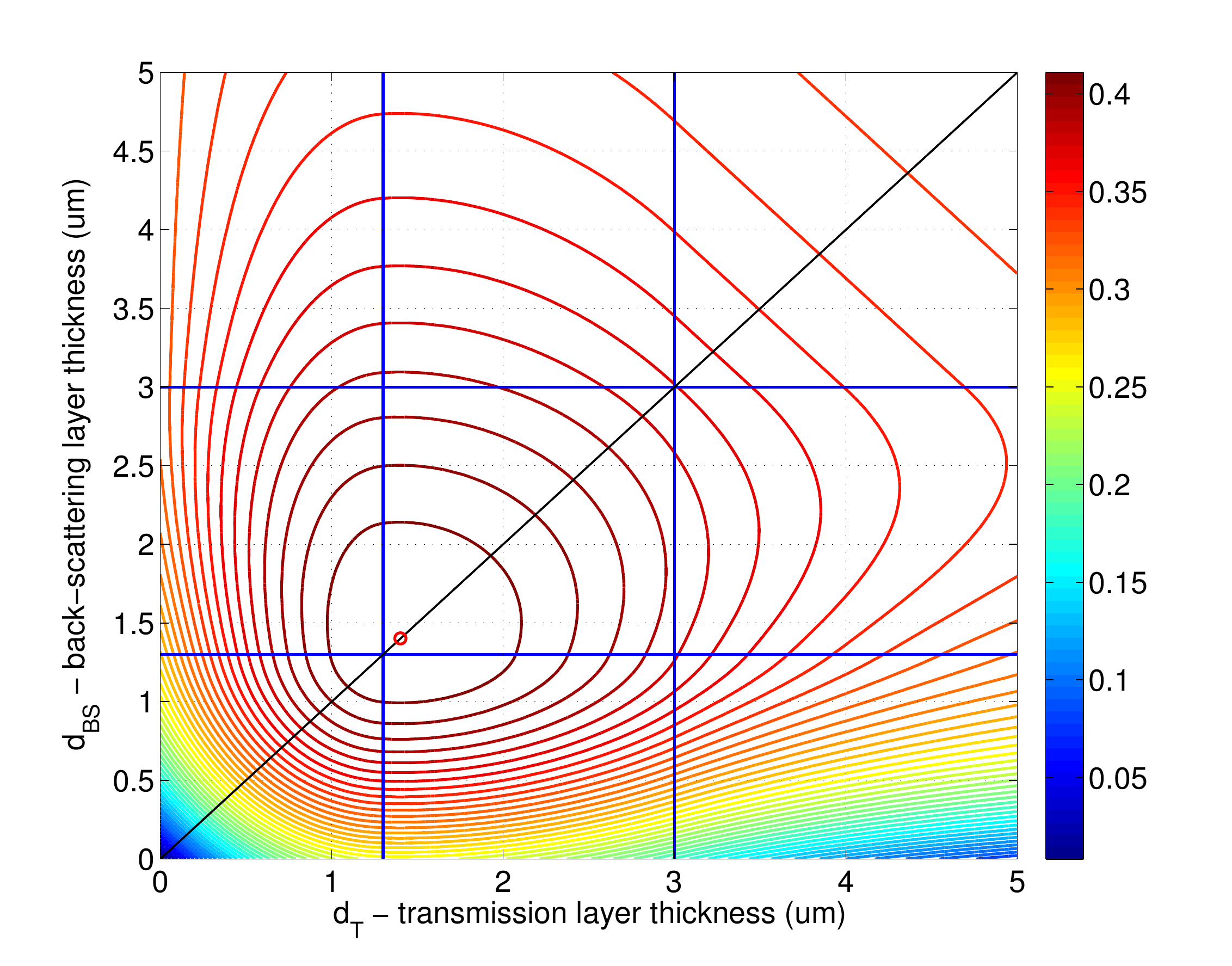}
 \caption{\footnotesize Efficiency plot for a double coated
substrate with $^{10}B_4C$ at $\theta=90{^\circ}$ incidence at
1.8\AA \, (left) and 20\AA \, (right).} \label{fig2lay1}
\end{figure}
\begin{figure}[!ht]
\centering
\includegraphics[width=7.8cm,angle=0,keepaspectratio]{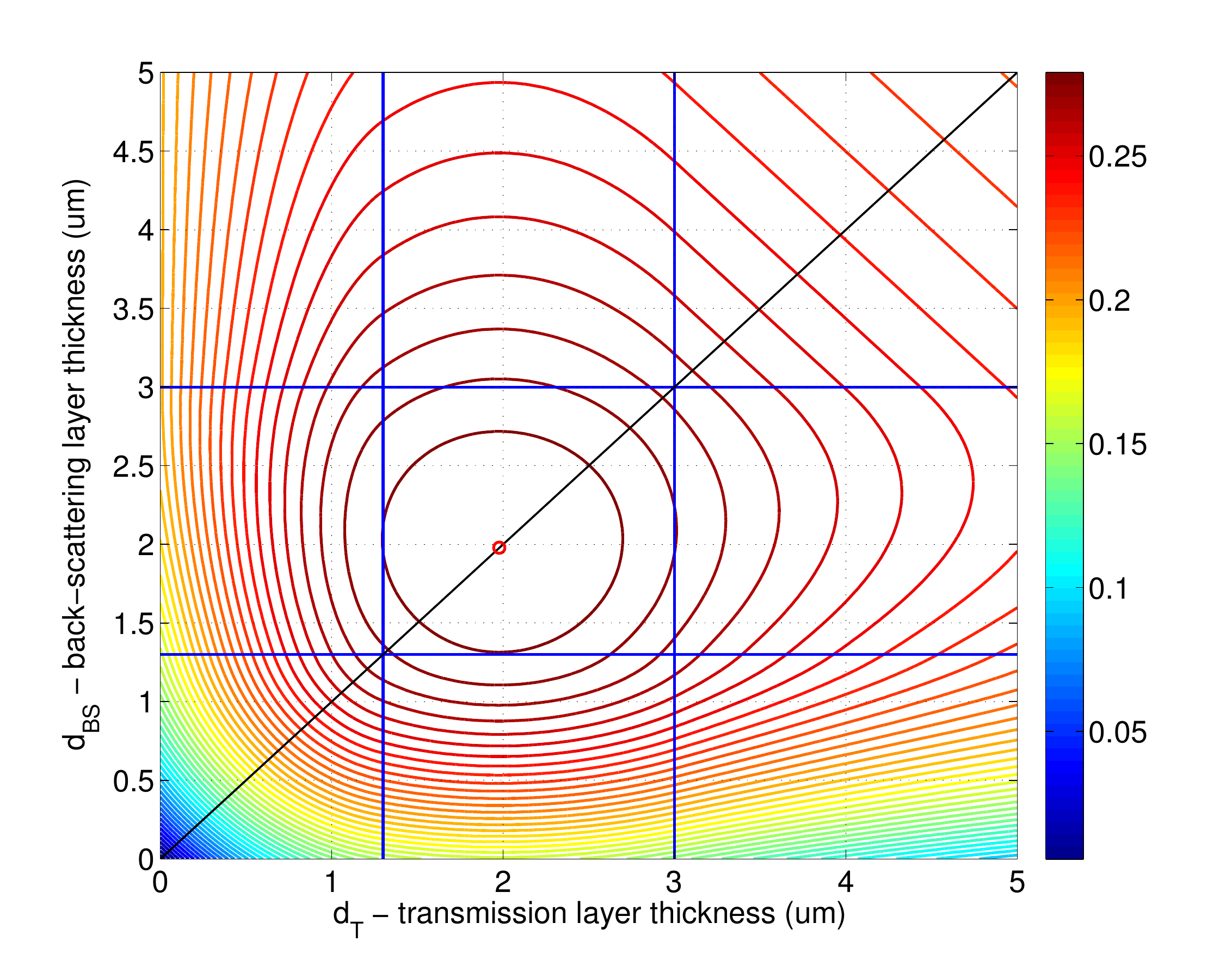}
\includegraphics[width=7.8cm,angle=0,keepaspectratio]{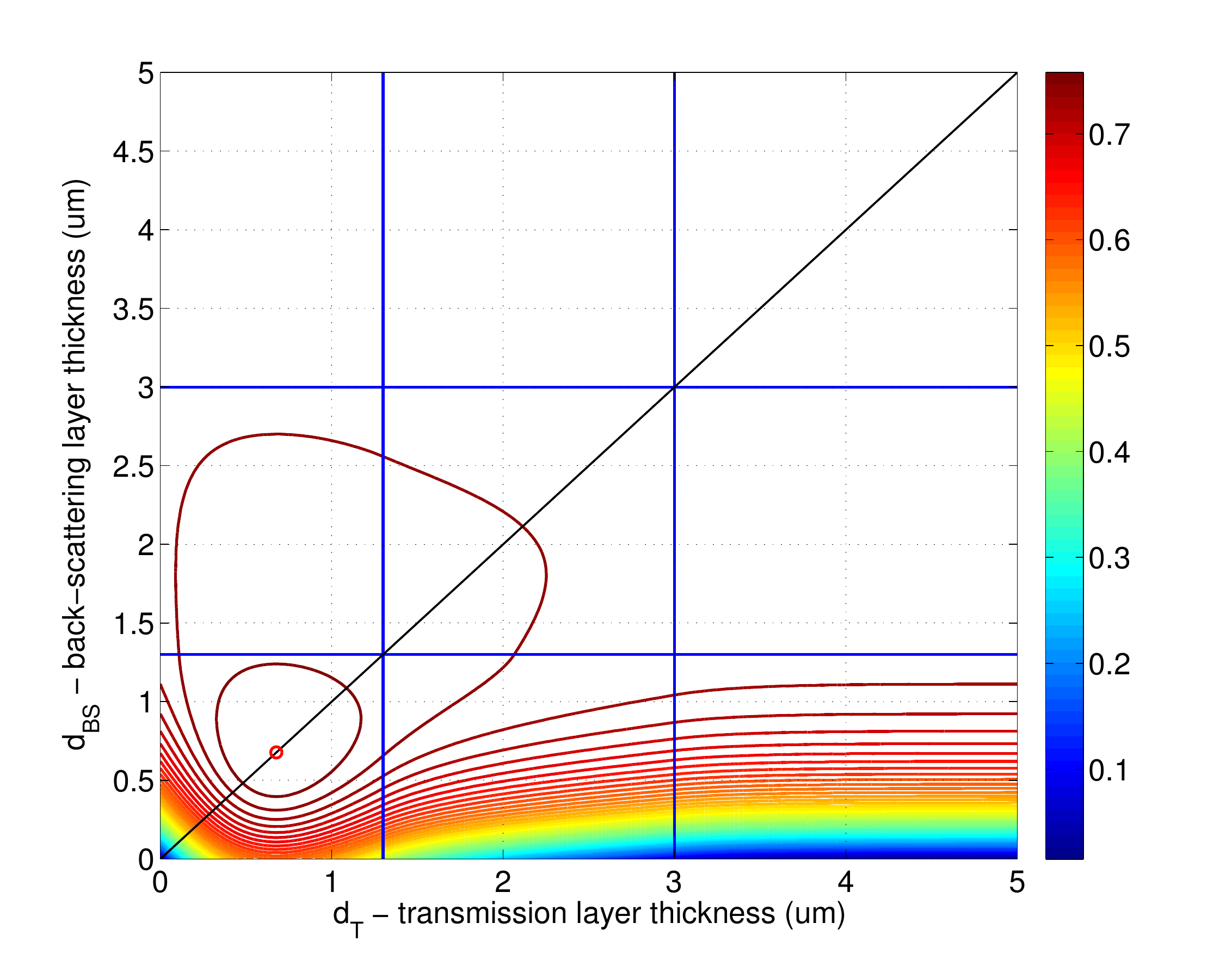}
 \caption{\footnotesize Efficiency plot for a double coated
substrate with $^{10}B_4C$ at $\theta=10{^\circ}$ incidence at
1.8\AA \, (left) and 20\AA \, (right).} \label{fig2lay2}
\end{figure}
It is easy to demonstrate there are no extreme points outside the
domain regions where either $d_{BS} > R_1$ or $d_{T} > R_1$, i.e.
\emph{square $jk$} with $j=3$ or $k=3$ or both. This outcome is also
intuitive. In back-scattering mode when the converter thickness
becomes thicker than the longest particle range ($R_1$) there is no
gain in efficiency by adding more converter material. In the
transmission case, increasing the thickness above $R_1$ will add
material that can only absorb neutrons without any particle
escaping.
\\ For the cases of \emph{squares 12} and
\emph{21}, we obtain that $\nabla \varepsilon(d_{BS},d_{T})=0$ has
no solution; thus the efficiency maximum can never fall in these
domain regions.
\\ In Figures \ref{fig2lay1} and \ref{fig2lay2} the efficiency for
four different cases are plotted. The red circle identifies the
point of maximum efficiency calculated by using Equations
\ref{eqac3} or \ref{eqac5}; it stands out immediately that even
though the efficiency function is not symmetric relatively to the
domain bisector (drawn in black) the maximum nevertheless always
lies on it.
\medskip
\\ \textbf{This is a very important result because the sputtering deposition
method \cite{carina} coats both sides of the substrate with the same
thickness of converter material and it is also suited to make
optimized blades}.
\medskip

\subsection{Effect of the substrate}\label{effsubblad1}
If we consider the neutron loss due to the substrate, the Equation
\ref{eqac1} has to be modified as follows:
\begin{equation}\label{eqac1bis}
\varepsilon(d_{BS},d_{T})=\varepsilon_{BS}\left(d_{BS}\right)+e^{-\Sigma_{sub}
\cdot d_{sub}}\cdot e^{-\Sigma \cdot d_{BS}}\cdot
\varepsilon_{T}\left(d_{T}\right)=\varepsilon_{BS}\left(d_{BS}\right)+\delta(\lambda)
\cdot e^{-\Sigma \cdot d_{BS}}\cdot
\varepsilon_{T}\left(d_{T}\right)
\end{equation}
where $\Sigma_{sub}$ and $d_{sub}$ are the macroscopic cross-section
and the thickness of the substrate. \\ If we optimize the layer
thicknesses, we find the same result of Equations \ref{eqac3} and
\ref{eqac5} for the transmission layer thickness $d_T$ but, on the
other hand, the back-scattering layer thickness does not equal the
transmission layer thickness anymore. It becomes:
\begin{equation}\label{eqau1122}
d_{BS} = \begin{cases} \frac{1}{C} \cdot \left(1-\delta\right)+\delta \cdot d_{T} &\mbox{for \emph{square 11}}\\
R_1 \cdot\left(1-2\delta\right)+\delta \cdot
\left(1+\frac{R_1}{R_2}\right) \cdot d_{T} &\mbox{for \emph{square
21}}\\
R_1 \cdot\left(1-\delta\right)+\delta \cdot d_{T} &\mbox{for
\emph{square 22}}
\end{cases}
\end{equation}
The maximum efficiency can now also lie in \emph{square 21} but not
in \emph{square 12}, because $\delta \in [0,1]$. The dotted line in
\emph{squares 11, 21} and \emph{22}, in Figure \ref{fig2laydom}, are
Equations \ref{eqau1122}. The slope of the line in \emph{squares 11}
and \emph{22} is equal to $\delta(\lambda)<1$. The maximum
efficiency, when $\delta(\lambda)$ is different from 1, now lies on
the dotted lines right above the optimum without substrate effect.
\\ From Equation \ref{eqau1122} we can observe that when $\delta$ is
close to zero, i.e. the substrate is very opaque to neutrons, the
thickness of the back-scattering layer tends to the value $R_1$. On
the other hand, when $\delta$ is close to one, $d_{BS}$ tends to
$d_{T}$. The factor $\delta(\lambda)$ is usually very close to one
for many materials that serve as substrate. We define the relative
variation between $d_{BS}$ and $d_{T}$ as:
\begin{equation}\label{eqau11222}
\Delta_{d} =\left| \frac{d_{T}-d_{BS}}{d_{T}}\right|= \begin{cases} \frac{\left(1-\delta\right)\cdot \left|d_{T}-\frac{1}{C}\right|}{d_T} &\mbox{for \emph{square 11}}\\
\frac{\left|\left(1+\delta+\delta \frac{R_1}{R_2}\right)\cdot
d_{T}-R_1\left(1-2\delta\right)\right|}{d_T} &\mbox{for \emph{square 21}}\\
\frac{\left(1-\delta\right)\cdot \left|d_{T}-R_1\right|}{d_T}
&\mbox{for \emph{square 22}}
\end{cases}
\end{equation}
Still considering the $^{10}B$ neutron capture reaction for the
$94\%$ branching ratio, we list in Table \ref{tabdelta99} the values
for $\delta(\lambda)$ and $\Delta_{d}$ for a $0.5\,mm$ and $3\, mm$
thick Aluminium substrate of density $\rho=2.7\, g/cm^3$ at
$1.8$\AA. In case the substrate is inclined under an angle of
$10^{\circ}$ a substrate of $0.5\,mm$ looks like a substrate of
about $3\, mm$. We consider a neutron to be lost when it is either
scattered or absorbed, therefore, the cross-section used
\cite{sears} is:
$\sigma_{Al}=\sigma_{Al}^{abs}(\lambda)+\sigma_{Al}^{scatt}= 0.2\,b
$(at $ 1.8$\AA )$+1.5\,b=1.7\, b$.
\begin{table}[!ht]
\centering
\begin{tabular}{|c|c|c|}
\hline \hline
$d_{sub}(mm)$ &  $\delta$ (1.8\AA) &$\Delta_{d}$ (1.8\AA) \\
\hline
0.5 &  0.995  & 0.0004 \\
3    &  0.970  & 0.0026  \\
\hline \hline
\end{tabular}
\caption{\footnotesize Neutron loss factor $\delta$ for an Aluminium
substrate and percentage difference between the two coating
thicknesses held by the substrate for $1.8$\AA.} \label{tabdelta99}
\end{table}
\\ Even though the substrate effect can be neglected in most cases
when dealing with a small amount of blades, its effect in a
multi-layer detector can strongly differ from the results obtained
for the ideal case of completely transparent substrate. A further
step is to take its effect into account.

\subsection{Double layer for a distribution of neutron
wavelengths}\label{dlfadnw1} The result of having the same optimal
coating thickness for each side of a blade has been demonstrated for
monochromatic neutrons. We want to prove it now for a more general
case when the neutron beam is a distribution of wavelengths and when
the substrate effect can be neglected ($\delta(\lambda)\approx1$).
\\ We will prove a property that will turn out to be useful. We will
show that the directional derivative of $\varepsilon(d_{BS},d_{T})$,
when descending along the unit vector
$\hat{u}=\frac{1}{\sqrt{2}}\left(1,-1\right)$, is positive up to the
bisector $d_{BS} = d_{T}$ and it changes sign only there. This
vector identifies the orthogonal direction to the bisector (see
Figure \ref{fig2laydom}).
\\ In the \emph{square 11}:
\begin{equation}\label{eqac6}
D_{\hat{u}}\varepsilon(d_{BS},d_{T}) = \hat{u} \times
\nabla\varepsilon(d_{BS},d_{T})  = \frac{C \Sigma}{\sqrt{2}}\cdot
e^{-\Sigma d_{BS}} \left( d_{BS}-d_{T} \right)
\end{equation}
In the \emph{square 22}:
\begin{equation}\label{eqac7}
D_{\hat{u}}\varepsilon(d_{BS},d_{T})=
\frac{\Sigma}{2\sqrt{2}R_1}\cdot e^{-\Sigma d_{BS}} \left(
d_{BS}-d_{T} \right)
\end{equation}
which are both positive above the bisector and negative below. In
the other domain regions the demonstration is equivalent. E.g. in
the \emph{square 12}:
\begin{equation}\label{eqac8}
D_{\hat{u}}\varepsilon(d_{BS},d_{T})= -\frac{\Sigma}{2\sqrt{2}}\cdot
e^{-\Sigma d_{BS}} \left( 1 - \frac{d_{BS}}{R_2}- \frac{d_{BS}}{R_1}
+ \frac{d_{T}}{R_1} \right)
\end{equation}
which is strictly negative in the \emph{square 12} where $d_{BS}
\leq R_2<R_1$ and $ R_2<d_{T} \leq R_1$ except on the bisector where
$d_{BS}=d_T=R_2$. The following theorem is therefore proved.
\begin{mytheo}\label{theo1}
\textit{The efficiency function defined by the Equation \ref{eqac1}
is strictly monotone in the two half-domains identified by the
bisector $d_{BS} = d_{T}$ (see Figure \ref{fig2laydom}).}
\end{mytheo}
In a general instrument design one can be interested in having a
detector response for a whole range of $\lambda$. An elastic
instrument can work a certain time at one wavelength and another
time at another wavelength. In a Time-Of-Flight instrument one can
be interested in having a sensitivity to neutrons of a certain
energy range including or excluding the elastic peak. One can define
a normalized weight function $w\left(\lambda \right)$, such as
$\int_{0}^{+\infty}w\left(\lambda \right)\, d\lambda=1$, that
signifies how much that neutron wavelength is important compared to
others. I.e. the price we want to spend in a neutron scattering
instrument to be able to detect a neutron energy with respect to an
other one. We can also consider an incident beam of neutrons, whose
wavelength distribution is $w\left(\lambda \right)$, and we want to
maximize the efficiency given this distribution.
\\ The efficiency for a blade exposed to a neutron flux which shows
this distribution is:
\begin{equation}\label{eqac9}
\varepsilon_w (d_{BS},d_{T}) = \int_{0}^{+\infty}w\left(\lambda
\right) \varepsilon (d_{BS},d_{T},\lambda) \, d\lambda
\end{equation}
where $\varepsilon (d_{BS},d_{T},\lambda)$ is the efficiency in
Equation \ref{eqac1}.
\\ In order to optimize the efficiency defined by the Equation
\ref{eqac9}, its gradient relative to $d_{BS}$ and $d_{T}$ has to be
calculated:
\begin{equation}\label{eqac10} \nabla\varepsilon_w
(d_{BS},d_{T}) = \nabla\int_{0}^{+\infty}w\left(\lambda \right)
\varepsilon (d_{BS},d_{T},\lambda) \, d\lambda
=\int_{0}^{+\infty}w\left(\lambda \right) \nabla\varepsilon
(d_{BS},d_{T},\lambda) \, d\lambda
\end{equation}
Both gradient components have to cancel out ($\frac{\partial
\varepsilon_w}{\partial d_{BS}} =\frac{\partial
\varepsilon_w}{\partial d_{T}}=0$), this leads to
$D_{\hat{u}}\varepsilon_w=0$. E.g. in \emph{square 11}:
$\int_{0}^{+\infty}w\left(\lambda \right)\left(e^{-\Sigma d_{BS}}
\left(d_T-d_{BS}\right) C \Sigma \right)\, d\lambda=0$. As a result,
in order for the efficiency to attain a maximum, necessarily (but
not sufficiently), its directional derivative along the unity vector
$\hat{u}=\frac{1}{\sqrt{2}}\left(1,-1\right)$:
\begin{equation}\label{eqac12}
D_{\hat{u}}\varepsilon_w(d_{BS},d_{T})=\int_{0}^{+\infty}w\left(\lambda
\right)D_{\hat{u}}\varepsilon(d_{BS}=d_{T},d_{T},\lambda)\, d\lambda
\end{equation}
has to be zero. More explicitly, the condition $\nabla \varepsilon_w
=0$ also implies that $\hat{u} \cdot \nabla \varepsilon_w
=D_{\hat{u}} \varepsilon_w=0$.
\\ For a general family of functions $f(d_{BS},d_{T},\lambda)$ for
which the maximum always lies on the domain bisector it is not true
that the function defined by their positively weighted linear
combination must have the maximum on $d_{BS}=d_{T}$, because in
general $\nabla f (d_{BS},d_{T},\lambda)$ can be positive, null or
negative, thus there are many ways to accomplish $\nabla f_w
(d_{BS},d_{T})=0$. However, thanks to Theorem \ref{theo1},
$D_{\hat{u}}\varepsilon_w=0$ is satisfied only on the bisector.
Below the bisector, $D_{\hat{u}}\varepsilon_w$ is always negative,
as it is a positively weighted integral of strictly negative values;
similarly, above the bisector, $D_{\hat{u}}\varepsilon_w$ is always
positive. The only place where it can be zero, is on the bisector.
Hence, the maximum of $\varepsilon_w$ can only be attained on the
bisector.
\\ The gradient can hence be replaced by
$\frac{\partial }{\partial d_T}$ and the function maximum has to be
searched on the bisector, therefore:
\begin{equation}\label{eqac11} \nabla\varepsilon_w
(d_{BS},d_{T}) =\int_{0}^{+\infty}w\left(\lambda \right)
\frac{\partial}{\partial d_T}\varepsilon
(d_{BS}=d_{T},d_{T},\lambda) \, d\lambda
\end{equation}
\medskip
\\ \textbf{In the end the same layer thickness for both sides
of a blade has to be chosen in order to maximize its efficiency,
even if it is exposed to neutrons belonging to a general wavelength
distribution $w\left(\lambda \right)$}.
\medskip
\\ Equations for the optimal thickness (like in the case of a single wavelength,
Equations \ref{eqac3} and \ref{eqac5}) can be obtained from Equation
\ref{eqac11} once a neutron wavelength distribution has been
established. At first $\frac{\partial}{\partial d_T}\varepsilon
(d_{BS}=d_{T},d_{T},\lambda)$ has to be calculated, furthermore it
has to be integrated over $d\lambda$ before searching for its
solutions. \\ The integration over $\lambda$ can be alternatively
executed in the variable $\Sigma$; indeed $\Sigma$ is just a linear
function in $\lambda$ because $\sigma_{abs}$ is proportional to
$\lambda$ in the thermal neutron region. Moreover, as indicated
previously, $\Sigma$ is also a function of $\theta$ and this is the
only appearance of $\theta$ in the efficiency function.
\textbf{Hence, we can just as well consider a weighting in $\lambda$
and $\theta$ which results in just a weighting function over
$\Sigma$.} In other words, all the results that have been derived
for a wavelength distribution also hold for an angular distribution
or both.
\subsubsection{Flat neutron wavelength distribution example}\label{flatbladedisti3769}
\begin{figure}[!ht]
\centering
\includegraphics[width=7.5cm,angle=0,keepaspectratio]{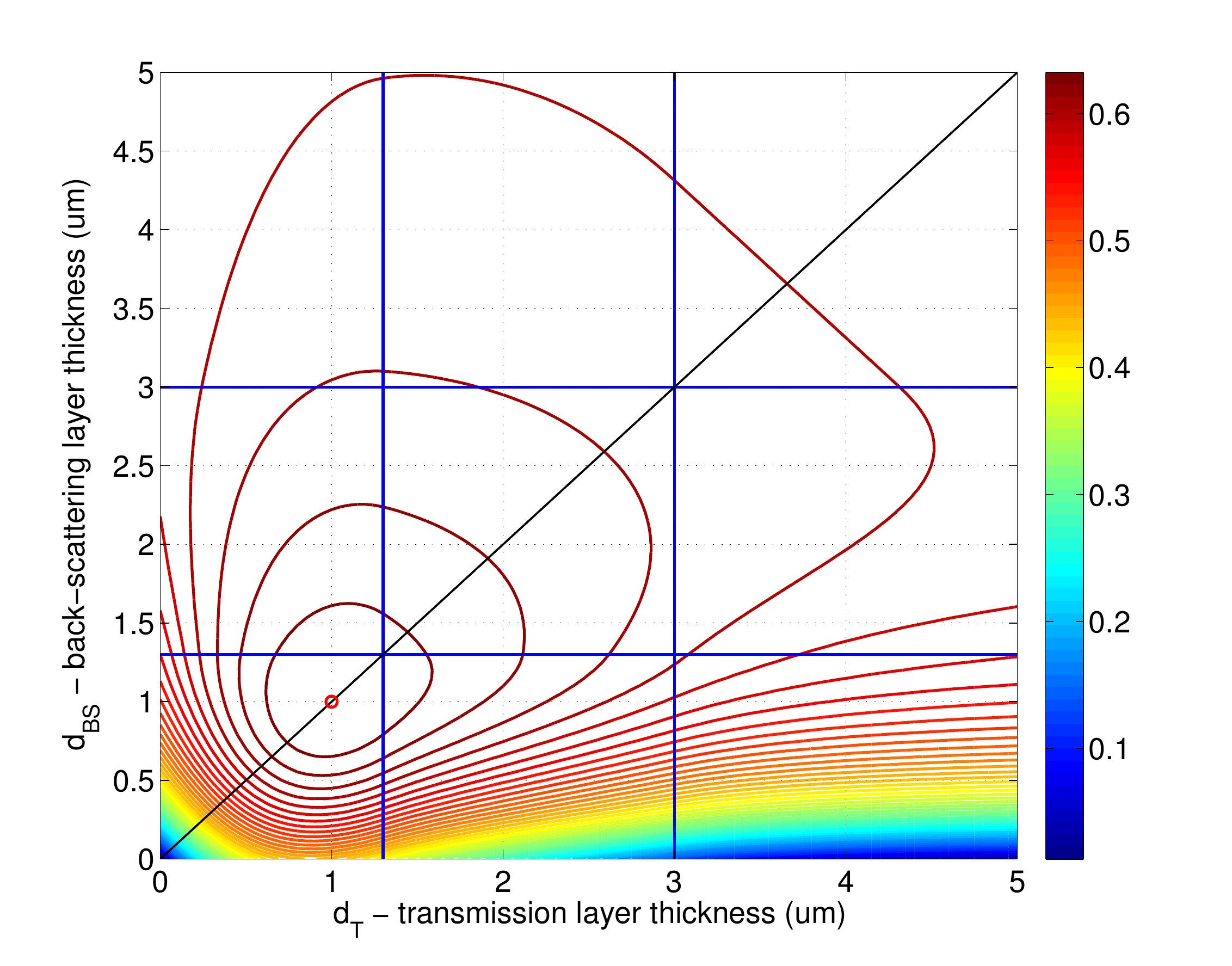}
\includegraphics[width=7.5cm,angle=0,keepaspectratio]{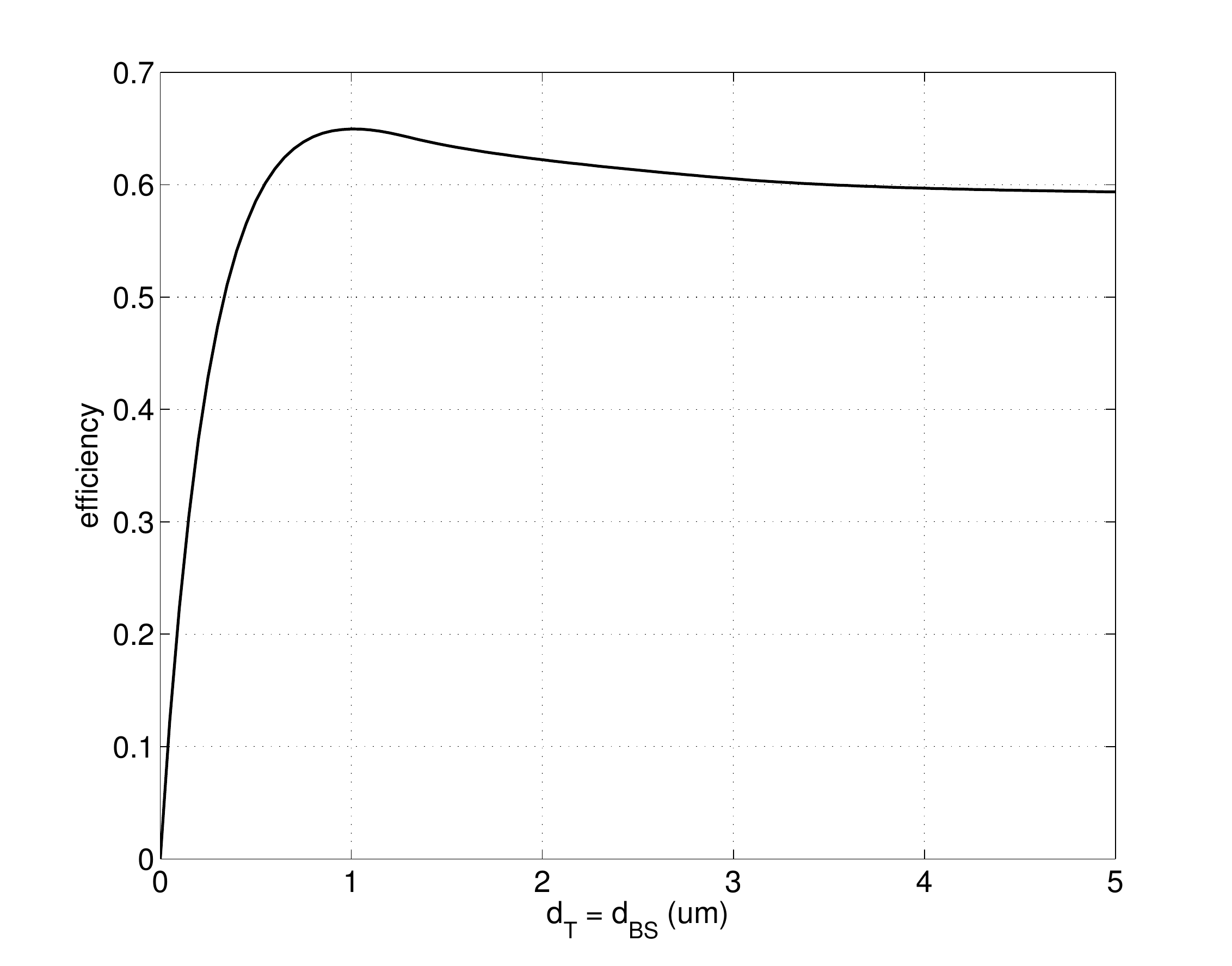}
 \caption{\footnotesize Efficiency plot for a double coated
substrate with $^{10}B_4C$ at $\theta=10{^\circ}$ for a flat
distribution of neutron wavelengths between 1\AA \, and 20\AA \,
(left). Efficiency on the bisector which attains its maximum at $1\,
\mu m$ (right).} \label{fig2lay21umopt}
\end{figure}

As a simple example we take a flat distribution between two
wavelengths $\lambda_1$ and $\lambda_2$ defined as follows:
\begin{equation}\label{eqac13}
w\left(\lambda\right)=\frac{1}{\lambda_2-\lambda_1}
\end{equation}
In \emph{square 11} we obtain:
\begin{equation}\label{eqac14}
\frac{\partial}{\partial d_T}\varepsilon
(d_{BS}=d_{T},d_{T},\lambda)=2e^{-\Sigma d_T}\left( B \Sigma
e^{-\Sigma d_T}-C\right)
\end{equation}
We call $\Sigma_1=\Sigma(\lambda_1)$ and
$\Sigma_2=\Sigma(\lambda_2)$. We recall that A and B are function of
$\Sigma(\lambda)$.
\begin{equation}\label{eqac15}
\begin{aligned}
\nabla\varepsilon_w (d_{BS},d_{T}) &
=\frac{1}{M(\lambda_2-\lambda_1)}\int_{\Sigma_1}^{\Sigma_2}\frac{\partial}{\partial
d_T}\varepsilon \left(d_{BS}=d_{T},d_{T},\Sigma\right) \,
d\Sigma=\\&=\frac{1}{(\Sigma_2-\Sigma_1)}\left[\frac{e^{-2\Sigma
d_T}}{d_T}\left( 2Ce^{+\Sigma d_T}-C-\Sigma-\frac{1}{2d_T}\right)
\right]_{\Sigma_1}^{\Sigma_2}=0
\end{aligned}
\end{equation}
Where the relation $B\Sigma =\Sigma+C$ and
$\Sigma\left(A-B\right)=-2C$ were used.
\\ In the same way the solution in \emph{square
22} can be determined.
\begin{equation}\label{eqac17}
\frac{\partial}{\partial d_T}\varepsilon
(d_{BS}=d_{T},d_{T},\lambda)=e^{-\Sigma d_T}\left( e^{-\Sigma
d_T}\left(2 B \Sigma-\frac{e^{+\Sigma R_2}}{R_2}
\right)-\frac{1}{R_1}\right)
\end{equation}
By integrating we finally obtain:
\begin{equation}\label{eqac16}
\begin{aligned}
\nabla\varepsilon_w (d_{BS},d_{T}) &
=\frac{1}{(\Sigma_2-\Sigma_1)}\left[\frac{e^{-2 \Sigma
d_T}}{d_T}\left( \frac{e^{+\Sigma d_T}}{R_1}-C-\Sigma-\frac{1}{2
d_T}- \frac{d_T \, e^{+\Sigma R_2}}{R_2\left(R_2-2d_T
\right)}\right) \right]_{\Sigma_1}^{\Sigma_2}=0
\end{aligned}
\end{equation}
The solution of Equations \ref{eqac15} and \ref{eqac16} gives the
optimum value for the thickness of the two converter layers in the
region of the domain called \emph{square 11} and \emph{square 22}
respectively for a uniform neutron wavelength distribution between
$\lambda_1$ and $\lambda_2$. E.g. for a uniform neutron wavelength
distribution between $1$\AA \, and $20$\AA \, the optimal thickness
of coatings on both sides of the blade inclined at $10^{\circ}$ is
$1\, \mu m$ (see Figure \ref{fig2lay21umopt}).
\\ Note that in Figure \ref{fig2lay21umopt} the asymptotic efficiency
for thick layers is not much lower than the optimum at $1\,\mu m$.
Thus at an incidence angle of $10^{\circ}$ between $1$\AA \, and
$20$\AA, a very thick piece of $^{10}B_4C$ is almost as efficient as
a double layer.
\newpage
\section{The multi-layer detector design}\label{mgmain4}
\begin{figure}[!ht]
\centering
\includegraphics[width=8cm]{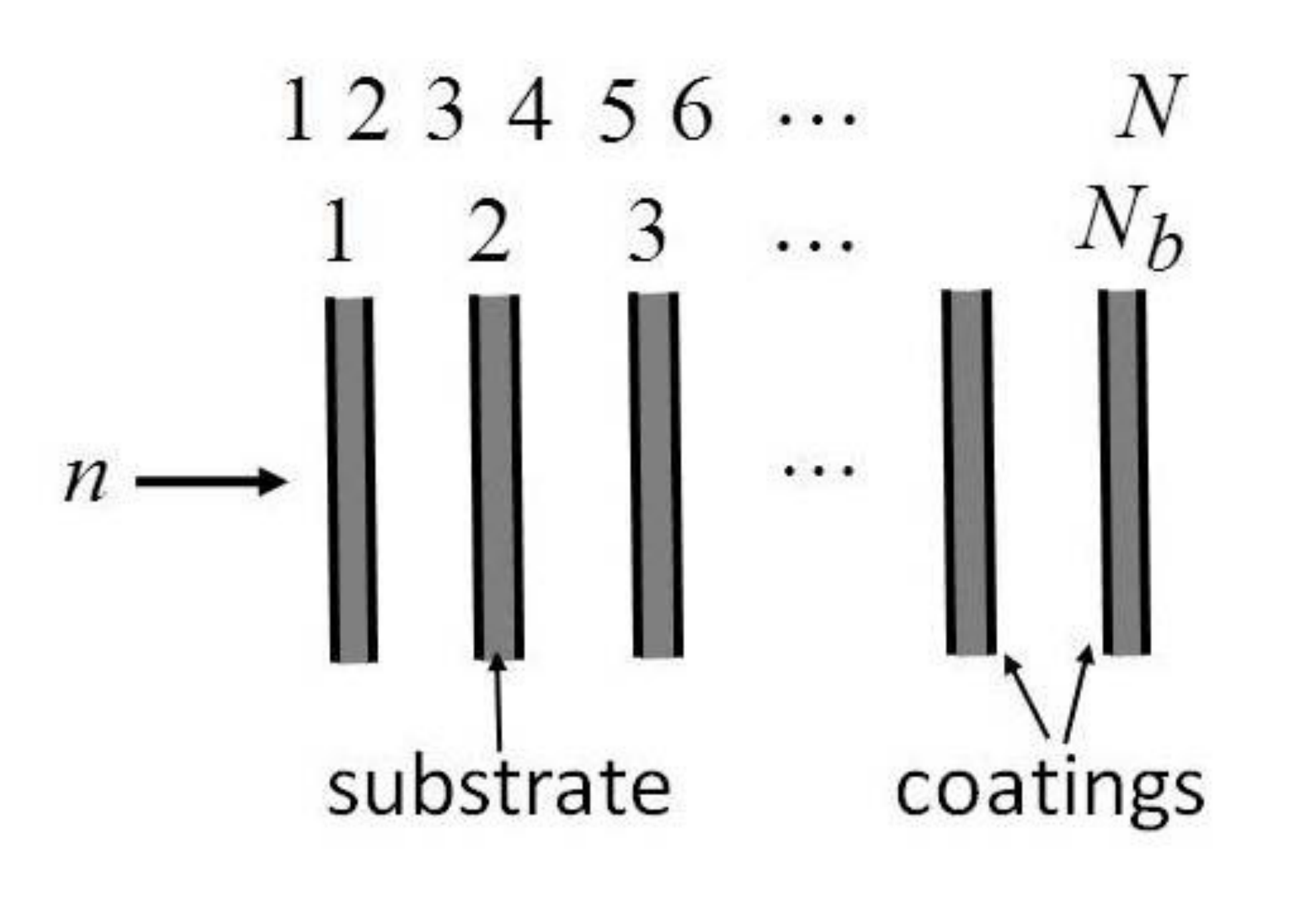}
\caption{\footnotesize A multi-layer detector schematic. $N_b$
blades, holding $N=2\cdot N_b$ converters layers, are placed in
cascade alternate with detection regions.} \label{multigridschet}
\end{figure}
\begin{figure}[!ht]
\centering
\includegraphics[width=14cm]{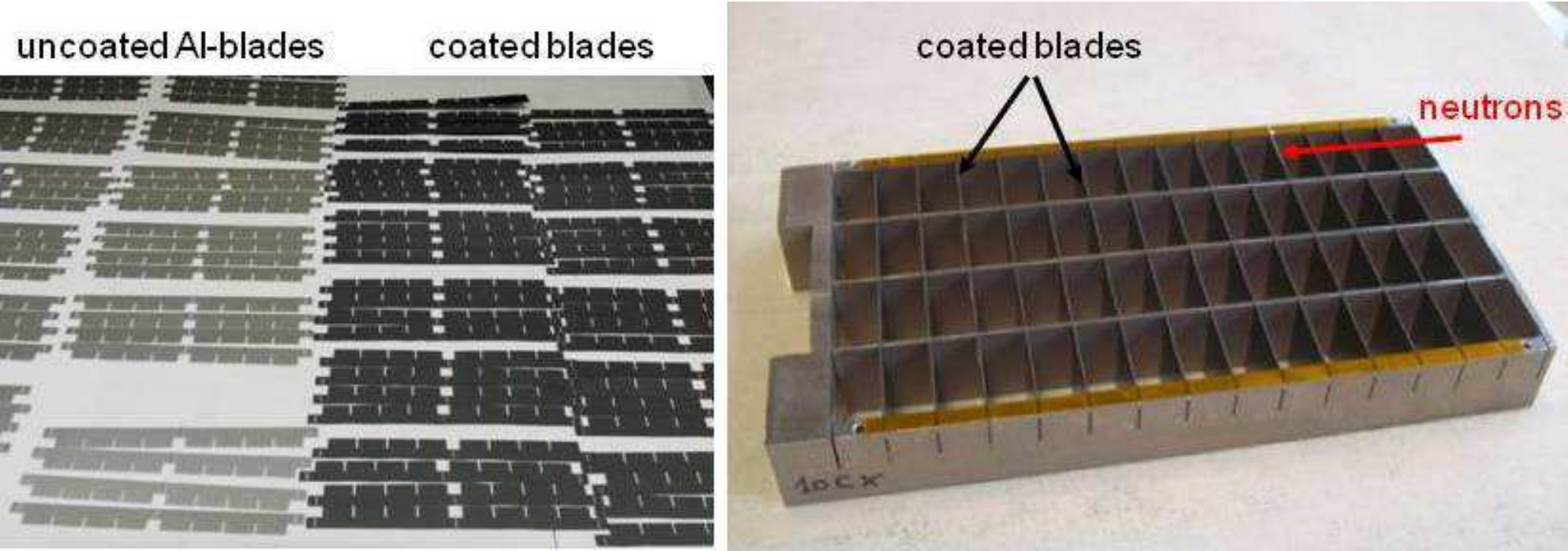}
\caption{\footnotesize The Multi-Grid detector \cite{jonisorma}.
$15$ blades coated with $^{10}B_4C$, are placed in cascade alternate
with gaseous detection regions.} \label{multigridschetdetec}
\end{figure}
In a detector like that presented in \cite{jonisorma} (see Figure
\ref{multigridschetdetec}), \cite{buff2}, or in \cite{wang1}, all
the substrates have the same coating thickness. The actual working
principle of the Multi-Grid will be explained in details in Chapter \ref{Chapt2}. \\
One can ask if it is possible to optimize the coating thicknesses
for each layer individually in order to gain in efficiency. This is
also applicable to neutron detectors which use solid converters
coupled with GEMs \cite{kleincascade}.
\\ In Section \ref{Sect2laysub} we demonstrated that, for a single
substrate holding two converter layers (blade), efficiency is
optimized when both layers have the same thickness such as naturally
happens with a sputtering technique. In general, this property is
not preserved in a multi-layer detector of which we can see the
sketch in Figure \ref{multigridschet}. In a multi-layer detector,
composed by $N$ layers or $N_b=\frac{N}{2}$ blades, the efficiency
can be written as follows:
\begin{equation}\label{eqad1}
\varepsilon_{tot}(N_b) = \varepsilon_1(d_{BS1},d_{T1})+
\sum_{k=2}^{N_b}\varepsilon_1(d_{BSk},d_{Tk})\cdot e^{- \left(
\sum_{j=1}^{\left(k-1\right)} \left(d_{BSj}+d_{Tj}\right)
\right)\cdot \Sigma }
\end{equation}
\\ Where $\varepsilon_1(d_{BSk},d_{Tk})$ represents the efficiency
for a single blade already defined by the Equation \ref{eqac1};
$d_{BSk}$ and $d_{Tk}$ are the coating thicknesses of the $k-th$
blade.
\\ If the detector is assembled with blades with all coatings of the same
thickness, i.e. $d_{BSk}=d_{Tk}=d, \,\, \forall k=1,2,\dots,N_b$,
Equation \ref{eqad1} can be simplified as follows:
\begin{equation}\label{eqad1sthick}
\varepsilon_{tot}(N_b) = \varepsilon_1(d)+
\sum_{k=2}^{N_b}\varepsilon_1(d)\cdot e^{-2 \left(
\sum_{j=1}^{\left(k-1\right)} d \right)\cdot \Sigma }
=\varepsilon_1(d) \cdot \sum_{k=1}^{N_b} e^{-2 \left( k-1 \right) d
\cdot \Sigma }=\varepsilon_1(d) \cdot \frac{1-e^{-2 d \Sigma
N_b}}{1-e^{-2 d \Sigma }}
\end{equation}
Therefore, $\frac{d \varepsilon_{tot}}{dd}=0$ optimizes the
efficiency for one single neutron wavelength for a detector
containing blades of same coating thicknesses:
\begin{equation}\label{eqad1sthickderiv}
\frac{d \varepsilon_{tot}}{dd} = \frac{\varepsilon_1(d) \cdot
2\Sigma}{\left(1-e^{-2 d \Sigma }\right)^2}\left(N e^{-2 d \Sigma
N_b}-e^{-2 d \Sigma}-\left(1-N_b\right)e^{-2 d \Sigma
(N_b+1)}\right)+ \frac{d \varepsilon_1}{dd} \cdot \frac{1-e^{-2 d
\Sigma N_b}}{1-e^{-2 d \Sigma }}=0
\end{equation}
where $\varepsilon_1(d)$ and $\frac{d \varepsilon_1}{dd}$ are
expressed in Equations \ref{eqac2} and \ref{eqac4} where we impose
$d_{BS}=d_T=d$ and Equations \ref{eqac14} and \ref{eqac17}
respectively if we are calculating in the \emph{Square 11} or
\emph{Square 22}.
\\ From Equation \ref{eqad1sthickderiv}, we notice that the maximum
efficiency for a multi-layer detector made up of identical coating
thickness blades, depends explicitly on $N_b$. This means that the
optimal coating thickness for a multi-layer detector depends on how
many blades it is composed of.
\begin{figure}[!ht]
\centering
\includegraphics[width=7.8cm]{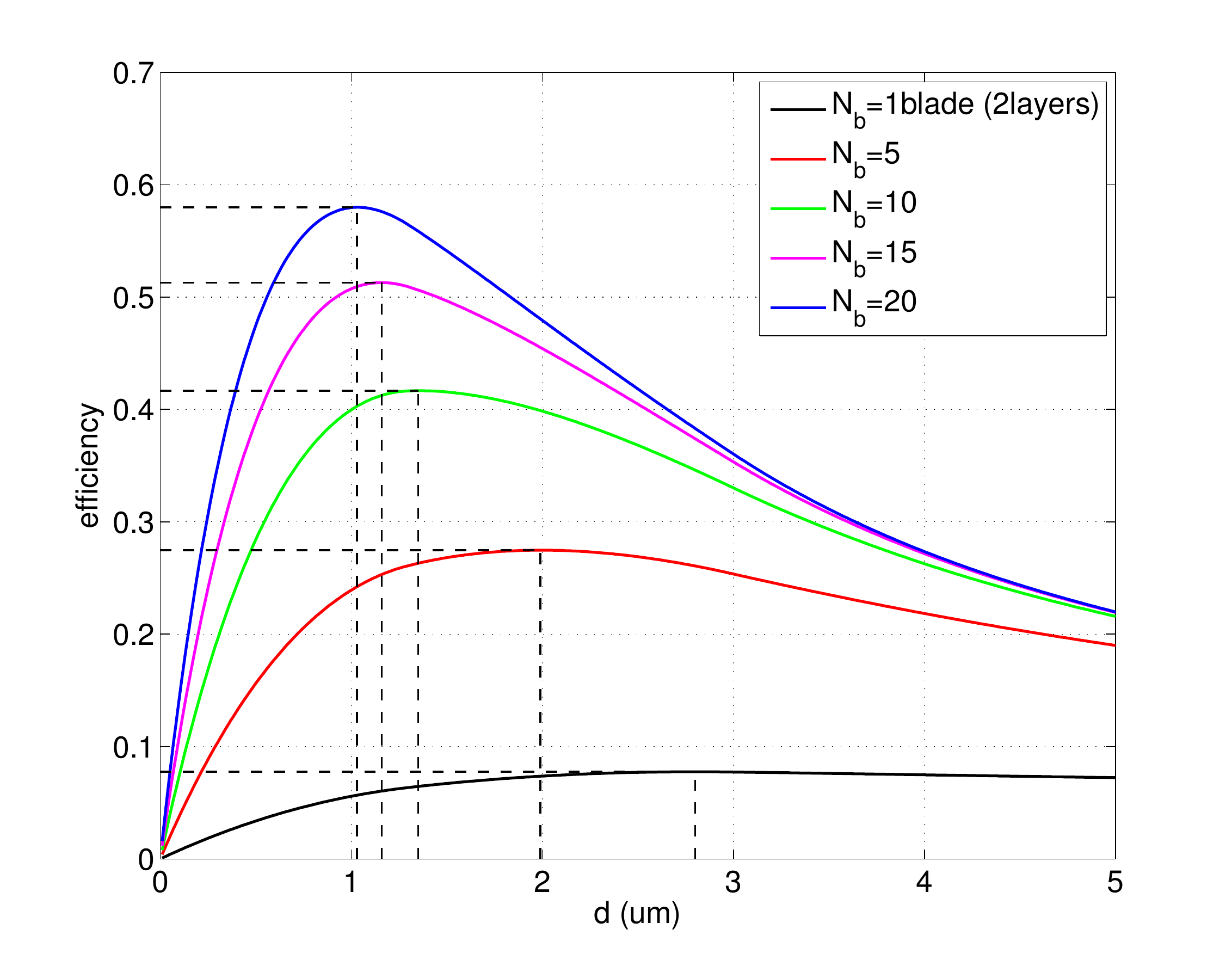}
\includegraphics[width=7.8cm]{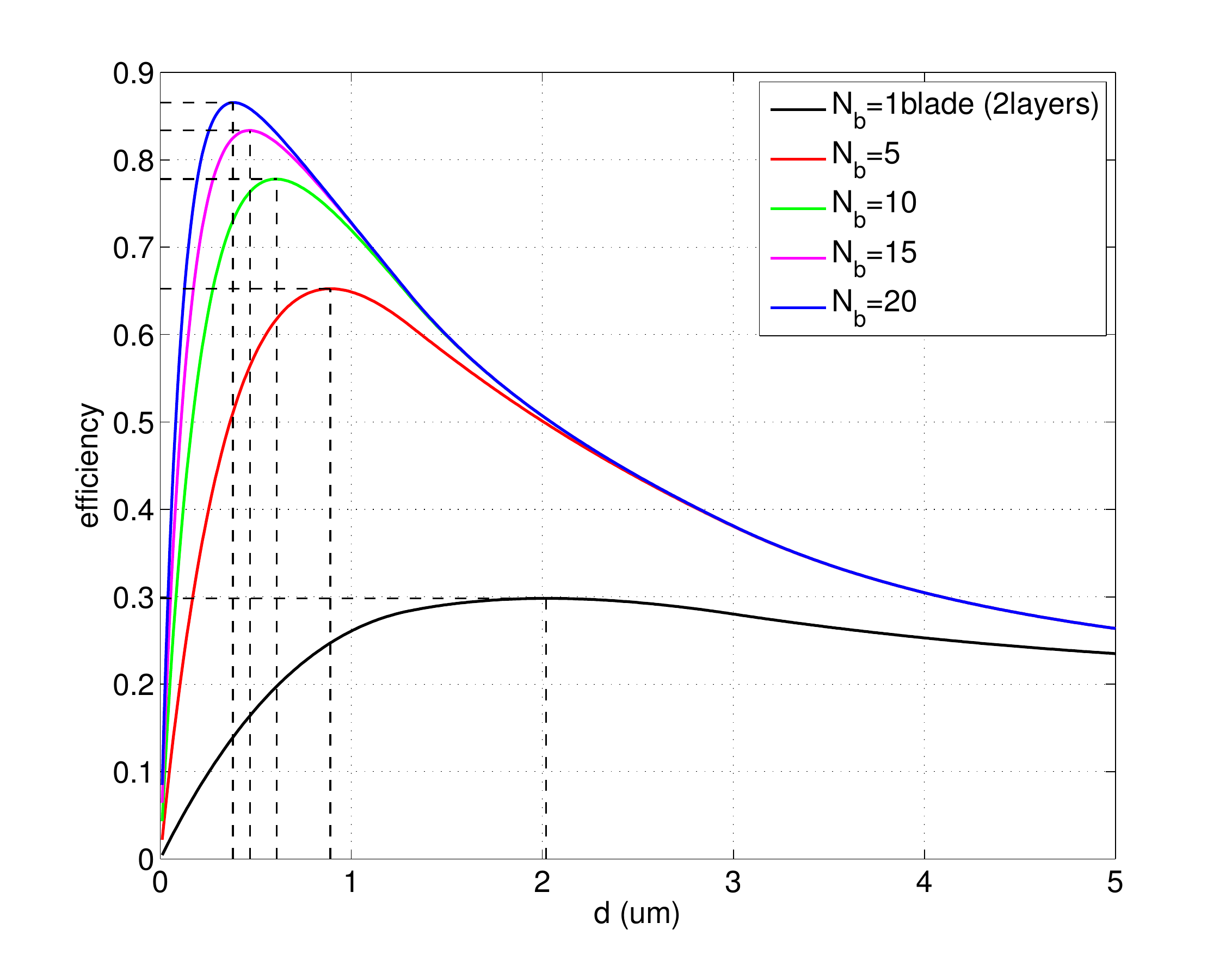}
\caption{\footnotesize Multi-layer detectors containing blades of
same coating thickness optimized for $1.8$\AA \, (left) and for
$10$\AA \, (right).} \label{fighjort457}
\end{figure}
\\ Figure \ref{fighjort457} shows the optimization of several
multi-layer detectors containing 1, 5, 10, 15, 20 blades, done for
both $1.8$\AA \, and for $10$\AA. Note that the longer the neutron
wavelength, the thinner is the thickness of the layers has to be
chosen.
\begin{figure}[!ht]
\centering
\includegraphics[width=7.8cm]{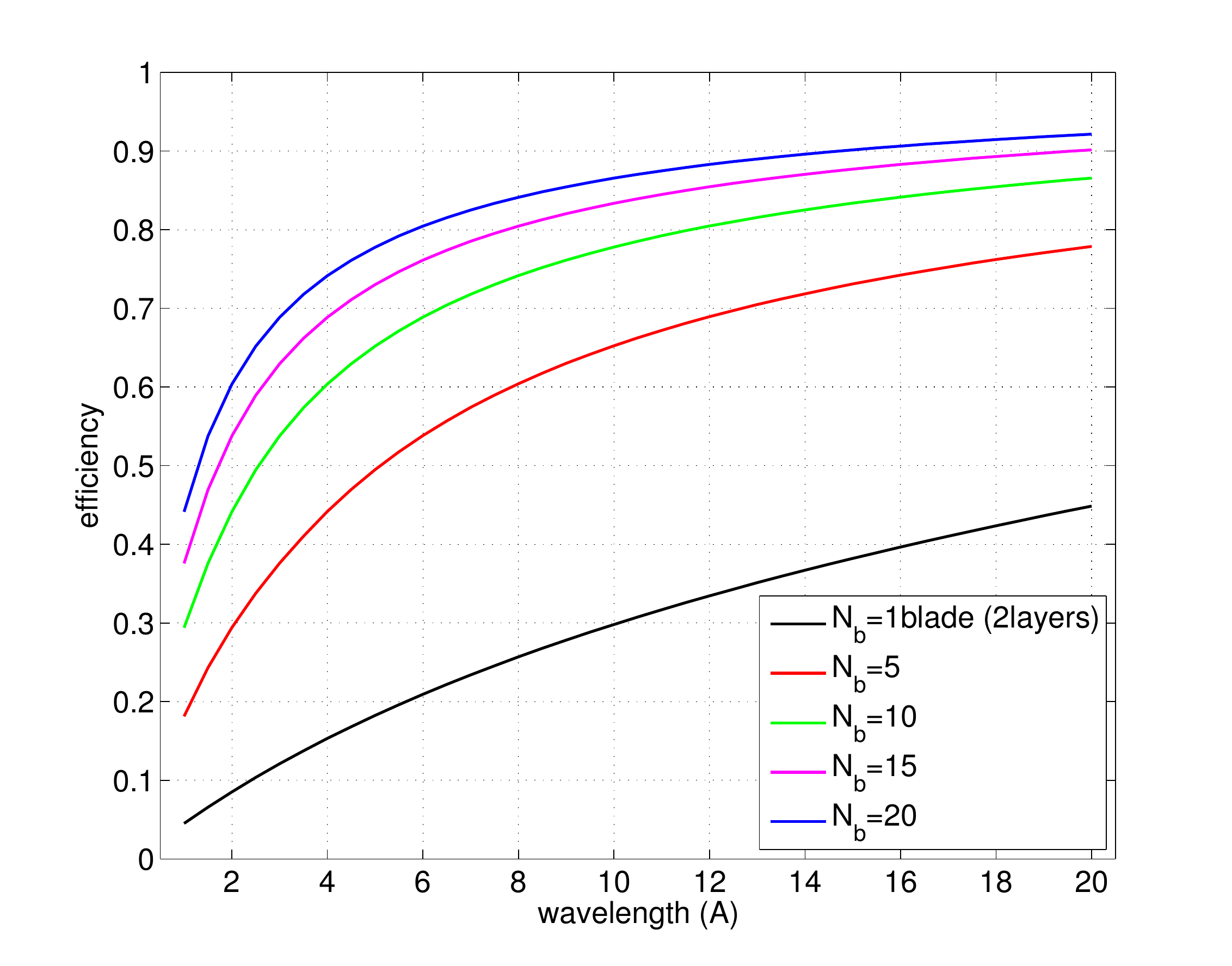}
\includegraphics[width=7.8cm]{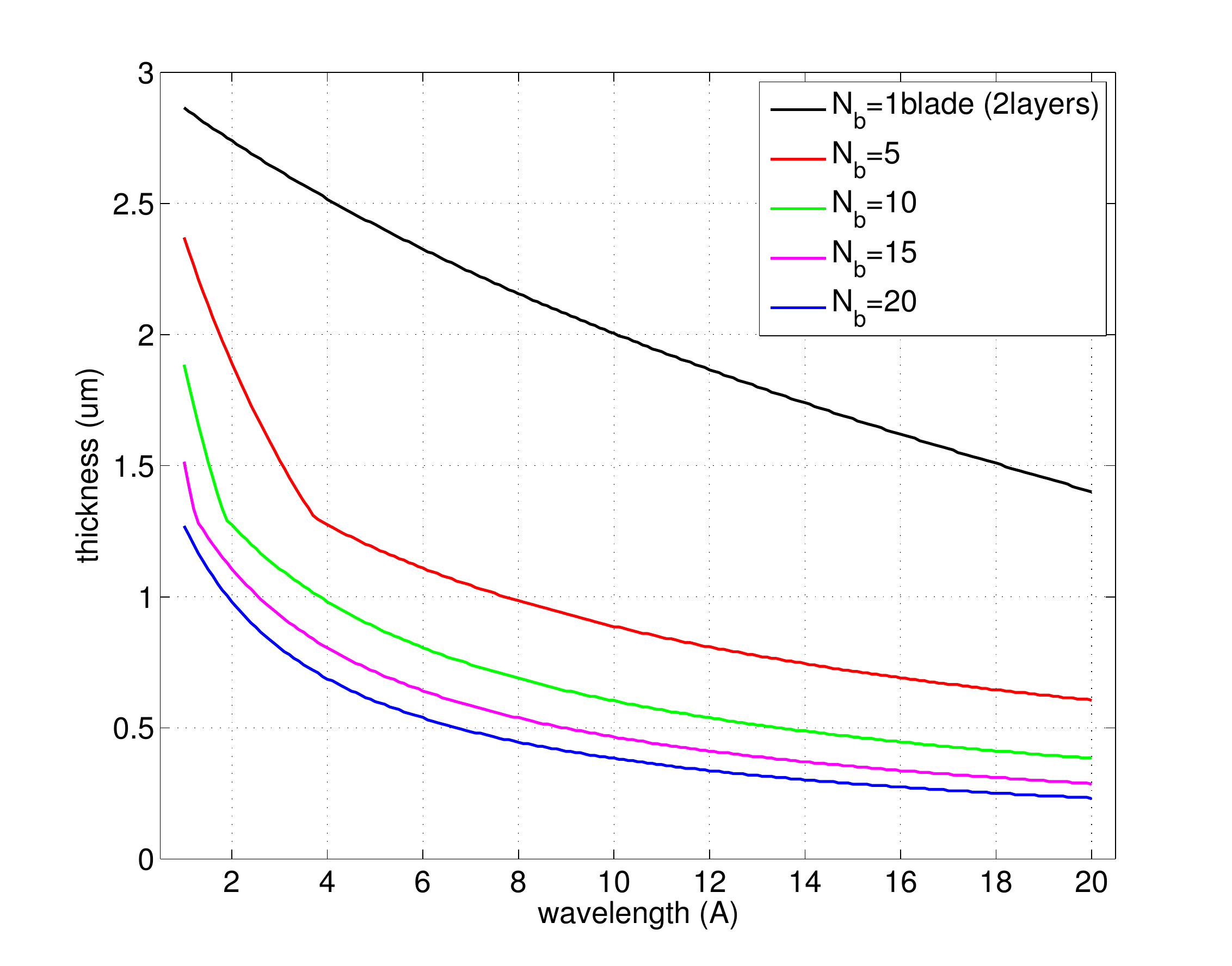}
\caption{\footnotesize Efficiency (left) and optimal thickness of
the identical blades (right) as a function of neutron wavelength for
a 2, 10, 20, 30 and 40 layers multi-layer detector. Solid lines
indicate the optimized efficiency, for each wavelength, for a
detector made up for blades of identical thicknesses.}
\label{2optimizzd567943}
\end{figure}
\\ In Figure \ref{2optimizzd567943} the efficiencies and the
coating thicknesses, given for each blade, of the five multi-layer
detectors already described are plotted as a function of the neutron
wavelength. There, each detector is optimized for each neutron
wavelength and consequently the coating thickness changes with
$\lambda$. The values at $1.8$\AA \, and at $10$\AA \, correspond to
the maxima already shown in Figure \ref{fighjort457}.
\\ We consider now a neutron
wavelength distribution defined by $w\left(\lambda \right)$ (with
$\int_{0}^{+\infty}w\left(\lambda \right)\, d\lambda=1$) as in
Section \ref{dlfadnw1} for a single blade. In this case the
efficiency for such a detector can be written as follows:
\begin{equation}\label{eqad14frtghigxcs}
\varepsilon_{tot}^w (N_b,d) = \int_{0}^{+\infty}w\left(\lambda
\right) \varepsilon_{tot}(N_b,\lambda) \, d\lambda
=\int_{0}^{+\infty}w\left(\lambda \right) \varepsilon_1(d,\lambda)
\cdot \frac{1-e^{-2 d \Sigma(\lambda) N_b}}{1-e^{-2 d
\Sigma(\lambda) }}\, d\lambda
\end{equation}
\\ where $\varepsilon_1(d,\lambda)$ is Equation
\ref{eqac2}. We considered the optimal efficiency of a multi-layer
detector with all-identical layer thicknesses. If we relax this
constraint (that is, if we allow different layers to have different
thicknesses) we can in principle achieve still higher efficiencies.
\bigskip
\\ If individual layer thicknesses have to be optimized for the maximum
efficiency for a multi-layer detector, the approach to follow is
different if we want to optimize for a single neutron wavelength or
for a wavelength distribution. Indeed, for a single lambda, we can
start by optimizing the last layer and, hereafter, we go backward up
to the first layer. If the wavelength if fixed, we are requiring the
layer we are considering to be at its maximum efficiency. On the
other hand, if we deal with a distribution of lambda, the previous
layers to the one we are optimizing will change the distribution it
is receiving. Hence the optimization process has to take into
account the whole detector at once. The Multi-Grid detector
optimization for a given wavelength and for a distribution will be
elucidated.

\subsection{Monochromatic multi-layer detector optimization}\label{monomulayopti}
As already mentioned the way to optimize the layers in multi-layer
detector for a given neutron wavelength is to maximize the last
layer efficiency for this wavelength. Once that is done, we can go
backward until the first layer. Each time we find the optimal
thickness of a layer we fix it and we move on to optimize the
previous one.
\\ By expanding Equation \ref{eqad1} we obtain:
\begin{equation}\label{eqad2}
\begin{aligned}
\varepsilon_{tot}(N_b) & =
\varepsilon_1(d_{BS1},d_{T1})+e^{-(d_{BS1}+d_{T1})\Sigma}\cdot
\varepsilon_1(d_{BS2},d_{T2})+ \\& +e^{-(d_{BS1}+d_{T1})\Sigma}\cdot
e^{-(d_{BS2}+d_{T2})\Sigma} \cdot\varepsilon_1(d_{BS3},d_{T3})+
\dots \\& \dots+e^{-(d_{BS1}+d_{T1})\Sigma}\cdot
e^{-(d_{BS2}+d_{T2})\Sigma}\cdot \dots \cdot
e^{-(d_{BS(N_b-1)}+d_{T(N_b-1)})\Sigma}\cdot
\varepsilon_1(d_{BSN_b},d_{TN_b})=\\&=\varepsilon_1(d_{BS1},d_{T1})+e^{-(d_{BS1}+d_{T1})\Sigma}\cdot\left[
\varepsilon_1(d_{BS2},d_{T2})+e^{-(d_{BS2}+d_{T2})\Sigma}\cdot\right.\\
&\cdot \left[
\varepsilon_1(d_{BS3},d_{T3})+e^{-(d_{BS3}+d_{T3})\Sigma}\cdot \dots
\cdot \left[ \dots \cdot
\left[\varepsilon_1(d_{BS(N_b-1)},d_{T(N_b-1)})+ \right. \right.
\right.
\\& \left. \left. \left. \left.  +e^{-(d_{BS(N_b-1)}+d_{T(N_b-1)})\Sigma}\cdot\varepsilon_1(d_{BSN_b},d_{TN_b})
\right]\dots \right]\right]\right]
\end{aligned}
\end{equation}
Note that the variable $d_{N_b}$ appears only once, in the
efficiency function $\varepsilon_1(d_{BSN_b},d_{TN_b})$. In the case
of a single wavelength, its optimal value can be determined without
taking the other layers into account. We can optimize the detector
starting from the last blade and going backward till the first. Any
change on the previous blades will only affect the \emph{number} of
neutrons that reach the last blade, and we require the last blade to
be as efficient as possible for that kind of neutron. As the layer
thickness optimum of each blade does not depend on the previous
ones, the system of equations is triangular.
\\ This will be not true for the wavelength distribution case and it
will be clarified in Section \ref{monomulti990}. Intuitively, for a
distribution, the gradient of Equation \ref{eqad2} is in addition
integrated over $\lambda$, thus all the blades have to be taken into
account simultaneously in the optimization process.
\\ Looking carefully into the Equation \ref{eqad2}, we can see that
the optimization process involves each time a function of structure:
\begin{equation}\label{eqad3}
f_{k} = \begin{cases}
\varepsilon_1(d_{BSk},d_{Tk})+e^{-(d_{BSk}+d_{Tk})\Sigma}\cdot
\alpha_{k+1} &\mbox{if \,} k < N_b \\ \varepsilon_1(d_{BSk},d_{Tk})
&\mbox{if \, } k = N_b
\end{cases}
\end{equation}
\\ In fact, if the $k-th$ layer it is going to be optimized, all the
following layers up to $N=2\cdot N_b$ have been already optimized,
hence $\alpha_{k+1}$ is a fixed number and represents the cumulative
efficiency of the detector from the blade $(k+1)-th$ to the end. As
a result, the optimization is an iterative process where a function
like \ref{eqad3} has to be optimized each time, taking into account
that $\alpha_{k+1}$ is always a fixed number and not a function of
the thicknesses to be determined, because their optimal values have
been already found.
\\ The two gradient components of Equation \ref{eqad3} are:
\begin{equation}\label{eqad4}
\frac{\partial f_{k}}{\partial d_{BSk}} =
\begin{cases}\frac{\partial}{\partial d_{BSk}}
\varepsilon_1(d_{BSk},d_{Tk})-\Sigma \,
e^{-(d_{BSk}+d_{Tk})\Sigma}\cdot \alpha_{k+1} &\mbox{if \,} k < N_b \\
\frac{\partial}{\partial d_{BSk}}\varepsilon_1(d_{BSk},d_{Tk})
&\mbox{if \, } k = N_b
\end{cases}
\end{equation}
\begin{equation}\label{eqad5}
\frac{\partial f_{k}}{\partial d_{Tk}} = \begin{cases}\frac{\partial
}{\partial d_{Tk}} \varepsilon_1(d_{BSk},d_{Tk})-\Sigma\,
e^{-(d_{BSk}+d_{Tk})\Sigma}\cdot \alpha_{k+1} &\mbox{if \,} k < N_b \\
\frac{\partial}{\partial d_{Tk}}\varepsilon_1(d_{BSk},d_{Tk})
&\mbox{if \, } k = N_b
\end{cases}
\end{equation}
\\ Note that the only difference between Equation \ref{eqad4} and
Equation \ref{eqad5} is the partial derivative variable.
\\ Thus, the condition $\nabla f_k = 0$ turns out to be:
\begin{equation}\label{eqad6}
\begin{cases}\frac{\partial}{\partial d_{BSk}}
\varepsilon_1(d_{BSk},d_{Tk})=\frac{\partial
}{\partial d_{Tk}} \varepsilon_1(d_{BSk},d_{Tk}) \Rightarrow D_{\hat{u}}\varepsilon_1(d_{BSk},d_{Tk})=0 &\mbox{if \,} k < N_b \\
\frac{\partial}{\partial
d_{BSk}}\varepsilon_1(d_{BSk},d_{Tk})=\frac{\partial}{\partial
d_{Tk}}\varepsilon_1(d_{BSk},d_{Tk})=0 &\mbox{if \, } k = N_b
\end{cases}
\end{equation}
\\ The condition in Equation \ref{eqad6} is exactly what
was demonstrated in Section \ref{Sect2laysub}. The Theorem
\ref{theo1} implies that the directional derivative along the unity
vector $\hat{u}=\frac{1}{\sqrt{2}}\left(1,-1\right)$ of the function
$\varepsilon_1(d_{BSk},d_{Tk})$ can only be zero on the bisector
domain, in the case of $k < N_b$. Hence, the maximum efficiency can
only be found, again, on the domain bisector. On the other hand, in
the case of $k=N_b$ the Equation \ref{eqad6} requires the gradient
of the efficiency function to be zero, property which was also
demonstrated in Section \ref{Sect2laysub} (Equations \ref{eqac3} and
\ref{eqac5}) for these kind of functions.
\medskip
\\ \textbf{As a result, even in a Multi-Grid like detector,
that is optimized for a given neutron wavelength with variable layer
thicknesses, it turns out that all the blades have to hold two
layers of the same thickness. On the other hand, thicknesses of
different blades can be distinct.}
\medskip
\\ Thanks to the latter derived property, we can denote with $d_k$
the common thickness of the two layers held by the $k-th$ blade
($d_{BSk}=d_{Tk}=d_k$). The detector efficiency function can be
redraft as follows:
\begin{equation}\label{eqad7}
\begin{aligned}
\varepsilon_{tot}(N,\bar{d}) & =
\varepsilon_1(d_1)+e^{-2d_1\Sigma}\cdot
\varepsilon_1(d_2)+e^{-2d_1\Sigma}\cdot e^{-2d_2\Sigma}
\cdot\varepsilon_1(d_3)+ \dots \\& \dots+e^{-2d_1\Sigma}\cdot
e^{-2d_2\Sigma}\cdot \dots \cdot e^{-2d_{N_b-1}\Sigma}\cdot
\varepsilon_1(d_{N_b})=\\&=\varepsilon_1(d_1)+e^{-2d_1\Sigma}\cdot\left[
\varepsilon_1(d_2)+e^{-2d_2\Sigma}\cdot\left[\varepsilon_1(d_3)+e^{-2d_3\Sigma}\cdot
\right.\right. \dots
\\ & \dots \cdot \left. \left. \left[\varepsilon_1(d_{(N_b-1)})+e^{-2d_{(N_b-1)}\Sigma}\cdot\varepsilon_1(d_{N_b})
\right]\dots \right]\right]=
\\ &= \varepsilon_1(d_{1})+ \sum_{k=2}^{N_b}\varepsilon_1(d_{k})\cdot
e^{-2 \left( \sum_{j=1}^{\left(k-1\right)} d_j \right)\cdot \Sigma }
\end{aligned}
\end{equation}
where $\bar{d}$ is the vector of components $d_k$ for $k=1,2,\dots
,N_b$. The condition $\nabla f_k = 0$ can be simplified by searching
for the maximum on the domain bisector, thus Equations \ref{eqad4}
and \ref{eqad5} turns into:
\begin{equation}\label{eqad8}
\frac{d f_{k}}{d d_k} = \left\{\begin{array}{ll} \frac{d }{d d_{k}}
\varepsilon_1(d_k)-2\Sigma\,
e^{-2 d_k \Sigma}\cdot \alpha_{k+1} &\mbox{if \,} k < N_b \\
\frac{d}{d d_{k}}\varepsilon_1(d_k) &\mbox{if \, } k = N_b\\
\end{array}
\right\}=0
\end{equation}
We recognize into the expression $\frac{d }{d d_{k}}
\varepsilon_1(d_k)$ the derivative in Equations \ref{eqac14} and
\ref{eqac17} that have already been calculated according to the
region domain in Section \ref{Sect2laysub}. In the domain region
called \emph{square 11} as defined in Section \ref{Sect2laysub}, we
obtain:
\begin{equation}\label{eqad9}
\frac{d f_{k}}{d d_k} = \begin{cases}2e^{-\Sigma d_k}\left(
\left(B-\alpha_{k+1}\right) \Sigma e^{-\Sigma
d_k}-C\right)&\mbox{if \,} k < N_b \\
2e^{-\Sigma d_k}\left( B \Sigma e^{-\Sigma d_k}-C\right) &\mbox{if
\, } k = N_b
\end{cases}
\end{equation}
And in the \emph{square 22}:
\begin{equation}\label{eqad10}
\frac{d f_{k}}{d d_k} = \begin{cases}e^{-\Sigma d_k}\left(
e^{-\Sigma d_k}\left(2 \left(B -
\alpha_{k+1}\right)\Sigma-\frac{e^{+\Sigma R_2}}{R_2}
\right)-\frac{1}{R_1}\right)  &\mbox{if \,} k < N_b \\
e^{-\Sigma d_k}\left( e^{-\Sigma d_k}\left(2 B
\Sigma-\frac{e^{+\Sigma R_2}}{R_2} \right)-\frac{1}{R_1}\right)
&\mbox{if \, } k = N_b
\end{cases}
\end{equation}
Where we recall $B=\left(1+\frac{1}{2\Sigma R_1}+\frac{1}{2\Sigma
R_2} \right)$ and $C=\left(\frac{1}{2 R_1}+\frac{1}{2 R_2} \right)$.
\\ Equations \ref{eqad9} and \ref{eqad10} have solutions similar to
Equations \ref{eqac3} and \ref{eqac5} apart from the fact that they
are a recursive form with $\alpha_{k+1}$. In the \emph{square 11}
the solution is:
\begin{equation}\label{eqad11}
d_k^{opt} = \begin{cases} -\frac{1}{\Sigma} \cdot \ln
\left(\frac{C}{\left(B-\alpha_{k+1}\right)\Sigma}\right) &\mbox{if \,} k < N_b \\
-\frac{1}{\Sigma} \cdot \ln \left(\frac{C}{B\Sigma}\right) &\mbox{if
\, } k = N_b
\end{cases}
\end{equation}
In the \emph{square 22}:
\begin{equation}\label{eqad12}
d_k^{opt} = \begin{cases} -\frac{1}{\Sigma} \cdot \ln
\left(\frac{R_2}{R_1}\left(\frac{1}{2 R_2 \Sigma \left(B-\alpha_{k+1}\right)- e^{+\Sigma R_2}}\right)\right) &\mbox{if \,} k < N_b \\
-\frac{1}{\Sigma} \cdot \ln \left(\frac{R_2}{R_1} \left(\frac{1}{2
R_2 \Sigma B - e^{+\Sigma R_2}}\right)\right) &\mbox{if \, } k = N_b
\end{cases}
\end{equation}
\\ As already mentioned in Section \ref{Sect2laysub}, these
solutions are valid if the result they return is included in the
domain region they are defined on.
\\ The optimization method is a recursive procedure that employs the
Equations \ref{eqad11} and \ref{eqad12}; we start from the last
blade, and we find its optimal thickness $d_{N_b}^{opt}$, then we
calculate $\alpha_{N_b}$ as the last layer efficiency using the
optimal thickness found. Now we can calculate $d_{N_b-1}^{opt}$ from
Equations \ref{eqad11} or \ref{eqad12} and $\alpha_{N_b-1}$ and so
on until the first layer. By definition $\alpha_{k+1}$ is the
detector cumulative efficiency starting from the $k+1$ blade to the
end ($N_b$); hence:
\begin{equation}\label{eqad13}
\alpha_{k+1} = \begin{cases}
\varepsilon_1(d_{k+1})+e^{-2d_{k+1}\Sigma}\cdot
\varepsilon_1(d_{k+2})+e^{-2d_{k+1}\Sigma}\cdot e^{-2d_{k+2}\Sigma}
\cdot\varepsilon_1(d_{k+3})+ \dots \\ \dots+e^{-2d_{k+1}\Sigma}\cdot
e^{-2d_{k+2}\Sigma}\cdot \dots \cdot e^{-2d_{N_b-1}\Sigma}\cdot
\varepsilon_1(d_{N_b})=\\=\varepsilon_1(d_{k+1}^{opt})+\sum_{i=k+2}^{N_b}\varepsilon_1(d_{i}^{opt})\cdot
e^{-2 \left(\sum_{j=k+1}^{\left(i-1\right)} d_{j}^{opt} \right)\cdot
\Sigma } &\mbox{if \,} k+1 < N_b
 \\ \hspace{0.43cm} \varepsilon_1(d_{k+1}^{opt}) &\mbox{if \, } k+1 = N_b
\end{cases}
\end{equation}
\subsubsection{Example of application}
\begin{figure}[!ht]
\centering
\includegraphics[width=7.8cm,angle=0,keepaspectratio]{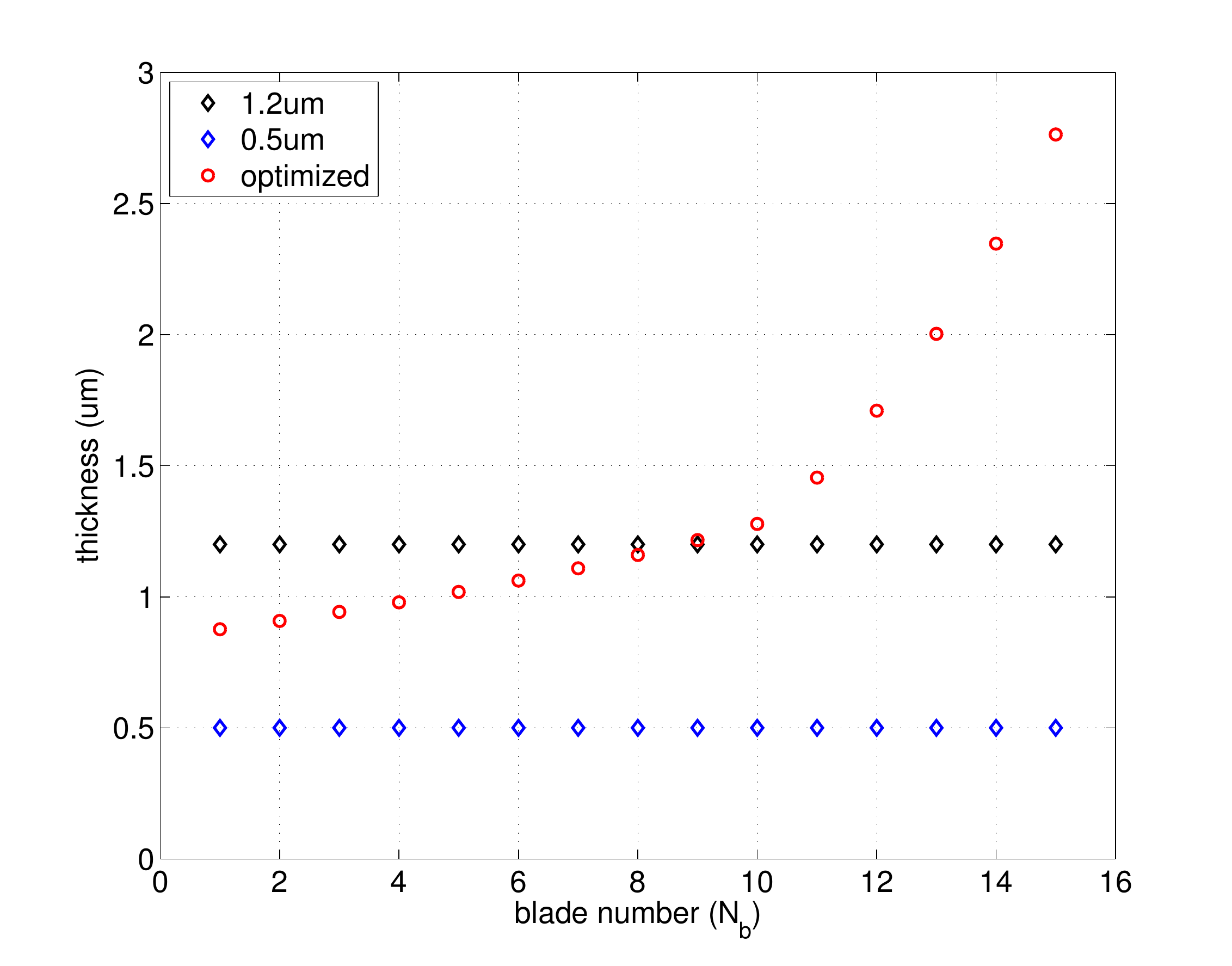}
\includegraphics[width=7.8cm,angle=0,keepaspectratio]{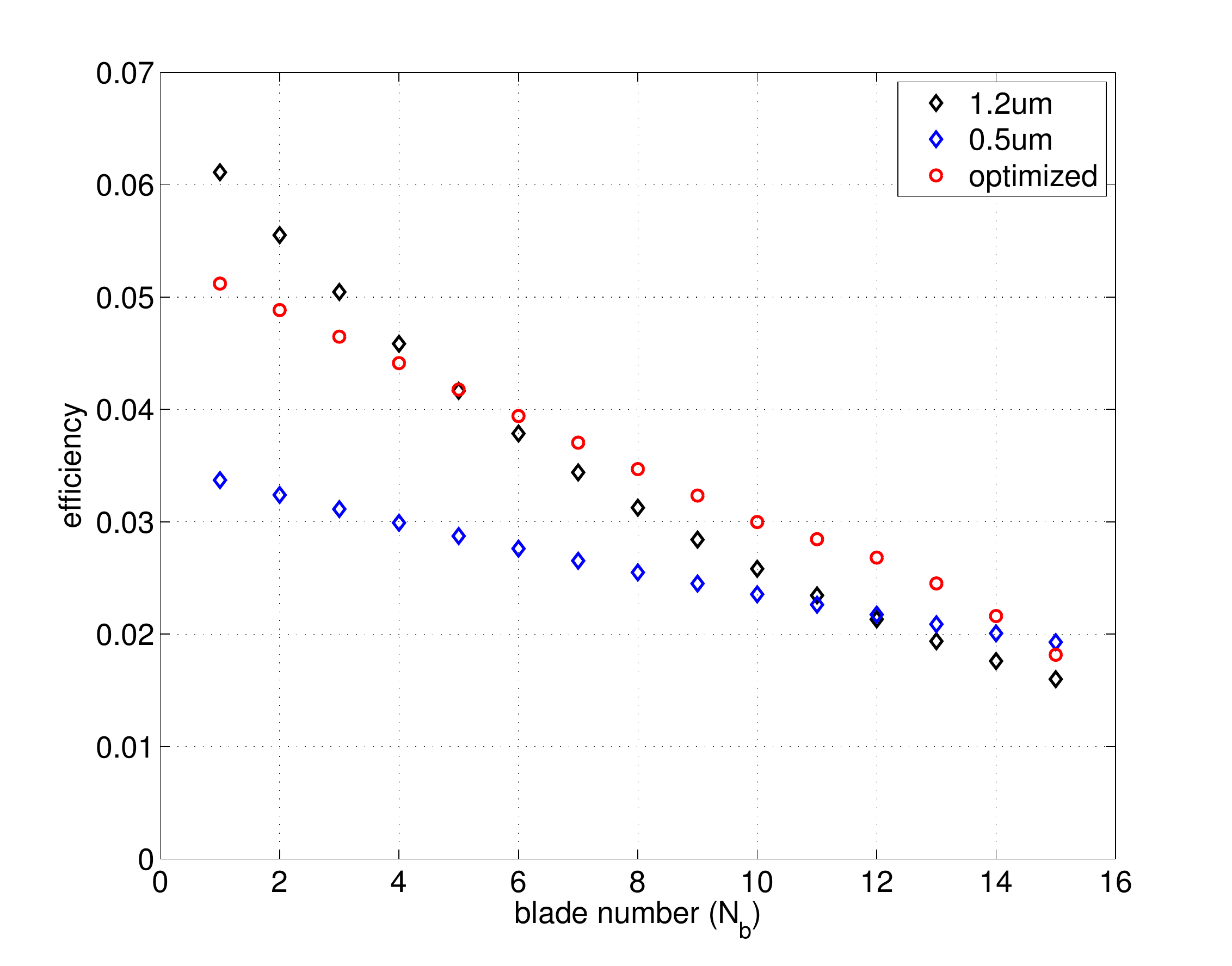}
 \caption{\footnotesize Thicknesses of the blades coatings (left) and their efficiency contribution (right),
 for a detector made up of $15$ identical coating thickness blades
 of $1.2\, \mu m$, $0.5\, \mu m$ and for a detector optimized for $1.8$\AA.} \label{figMG30mon1p8}
\end{figure}

\begin{figure}[!ht]
\centering
\includegraphics[width=7.8cm,angle=0,keepaspectratio]{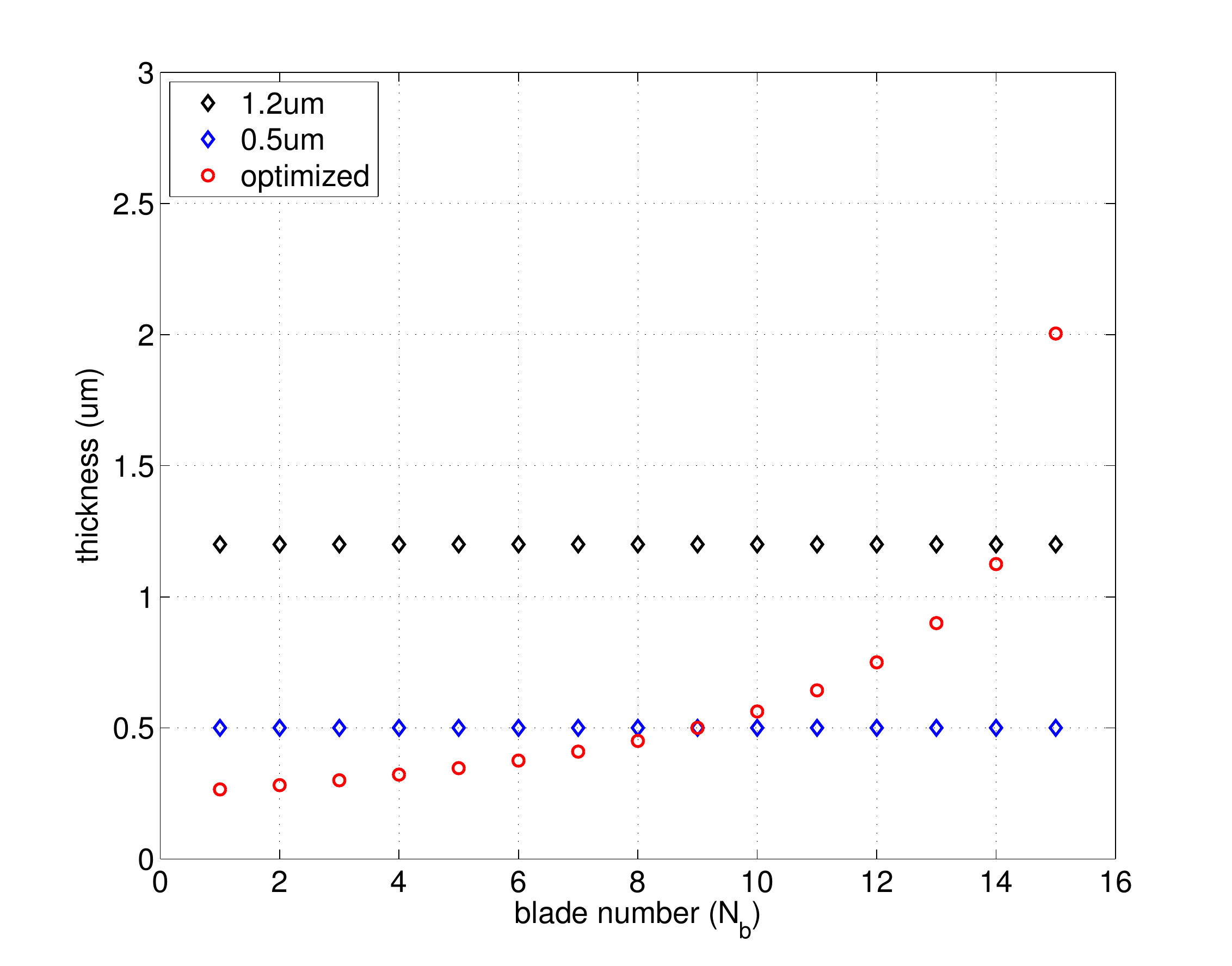}
\includegraphics[width=7.8cm,angle=0,keepaspectratio]{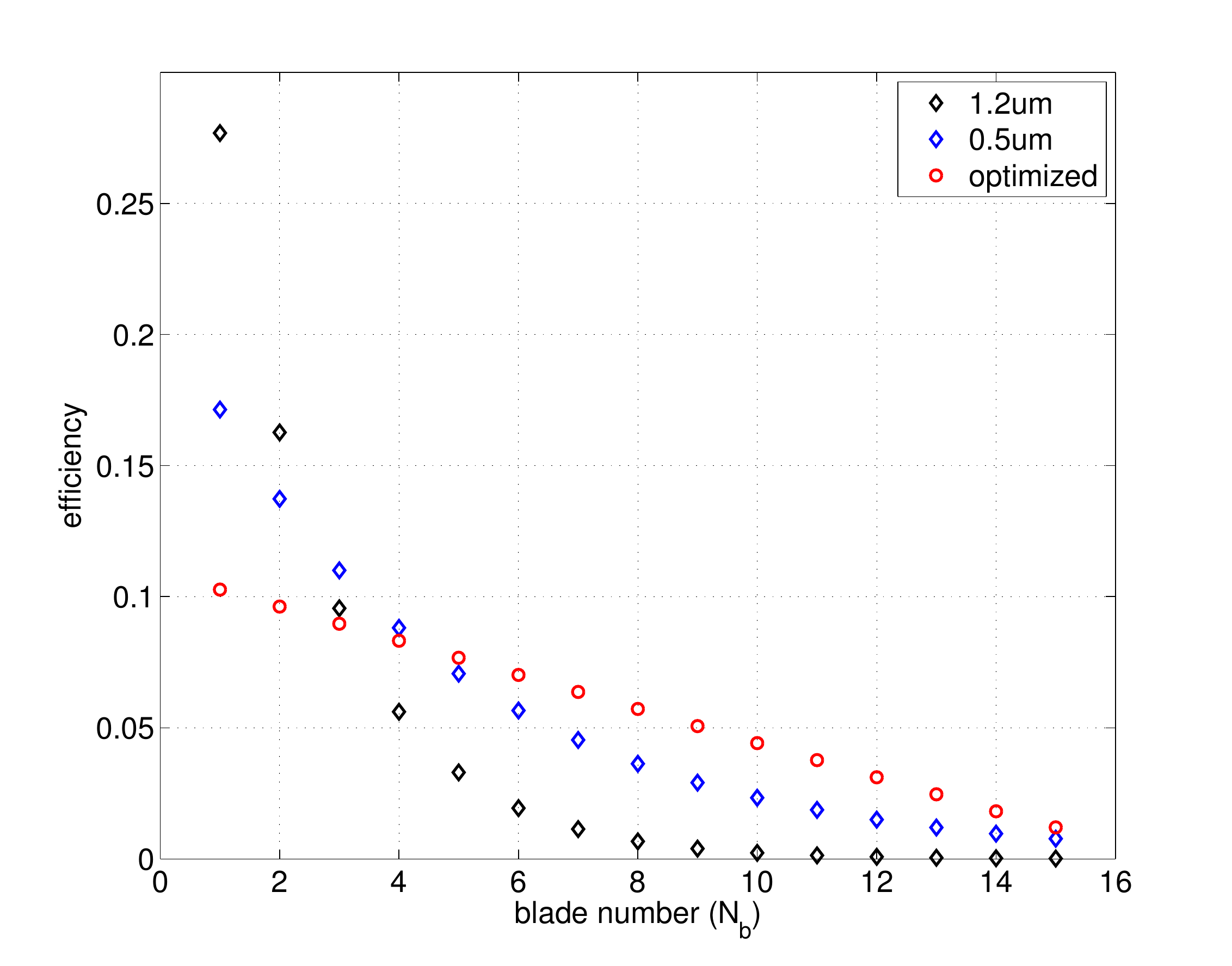}
 \caption{\footnotesize Thicknesses of the blades coatings (left) and their efficiency contribution (right),
 for a detector made up of $15$ identical coating thickness blades
 of $1.2\, \mu m$, $0.5\, \mu m$ and for a detector optimized for $10$\AA.} \label{figMG30mon20}
\end{figure}

\begin{figure}[!ht]
\centering
\includegraphics[width=7.8cm]{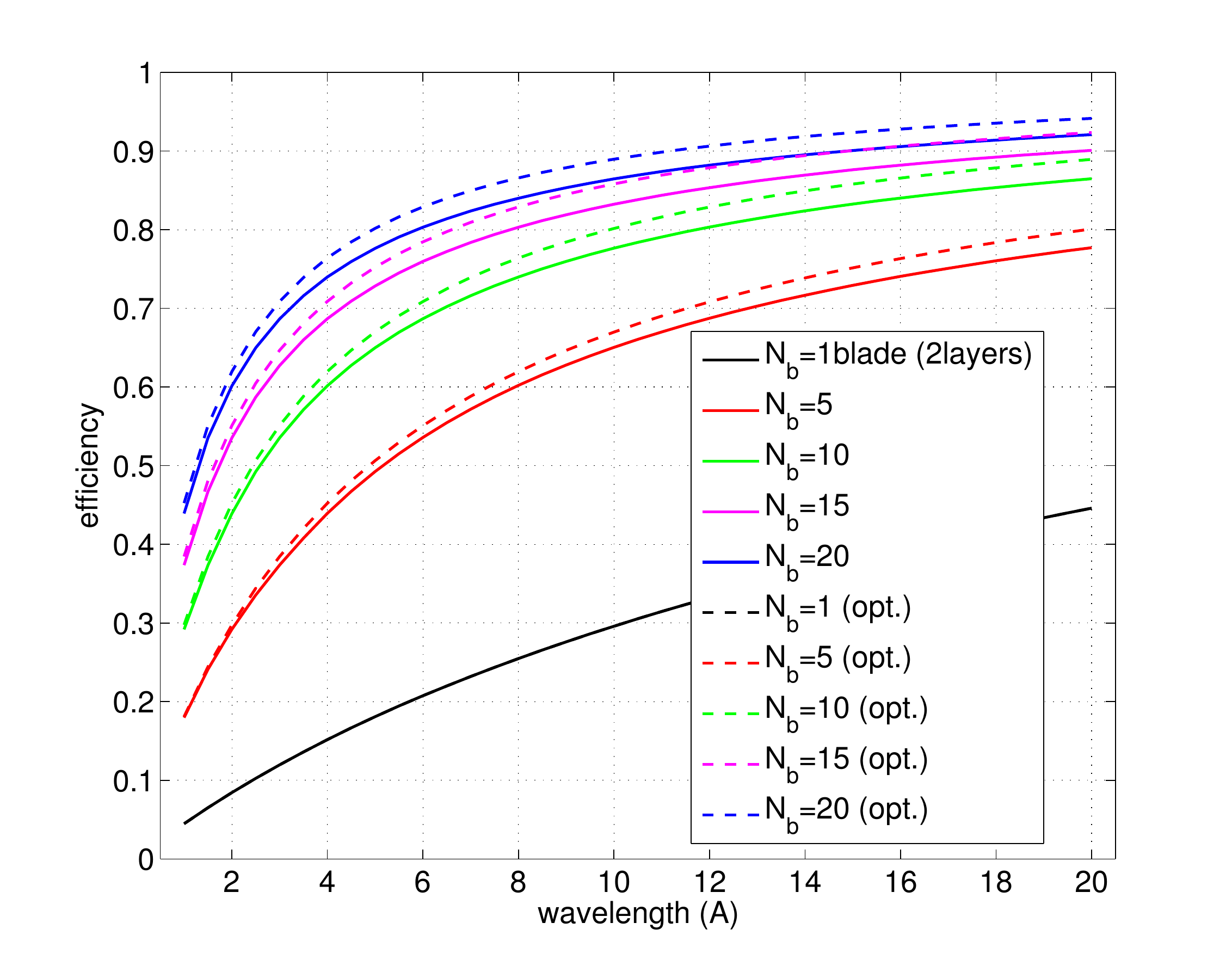}
\includegraphics[width=7.8cm]{figures/theo/MultiGrid/MGoptwaveCFRthicknessEqual}
\caption{\footnotesize Efficiency (left) and optimal thickness of
the identical blades (right) as a function of neutron wavelength for
a 2, 10, 20, 30 and 40 layers multi-layer detector. Solid lines
indicate the optimized efficiency, for each wavelength, for a
detector made up for blades of identical thicknesses; the dashed one
indicate the monochromatic optimization using different thicknesses
inside the detector.} \label{2optimizzd}
\end{figure}

We analyze a detector composed of $30$ successive converter layers
($15$ blades) crossed by the neutron beam at $90^{\circ}$ (like in
Figure \ref{multigridschet}). We consider $^{10}B_4C$ ($\rho=2.24\,
g/cm^3$) as converter; we neglect again the $6\%$ branching ratio of
the $^{10}B$ neutron capture reaction. A $100\,KeV$ energy threshold
is applied and the effective particle range turns out to be $R_1=3\,
\mu m$ ($\alpha$-particle) and $R_2 = 1.3\, \mu m$ ($^7Li$), for the
$94\%$ branching ratio, which we take to be $100\%$.
\\ Figures \ref{figMG30mon1p8} and \ref{figMG30mon20} show the
optimization result for this multi-layer detector; for a
monochromatic neutron beam of $1.8$\AA \, and $10$\AA. On the left,
the optimal thickness given by either Equations \ref{eqad11} or
\ref{eqad12} is plotted in red for each blade; for comparison we use
two similar detectors suitable for short and for long wavelengths in
which the blades are holding $1.2\, \mu m$ and $0.5\,\mu m$ thick
coatings. Those values have been obtained by optimizing the Equation
\ref{eqad1sthick}, the efficiency for a detector holding $15$ blades
of all equal thicknesses for $1.8$\AA \, and for $10$\AA. The
detector with $1.2\, \mu m$ coatings is very close to the one
presented in \cite{jonisorma}. On the right, in Figures
\ref{figMG30mon1p8} and \ref{figMG30mon20}, the efficiency
contribution of each blade is plotted, again for an optimized
detector for $1.8$\AA \, and for an optimization done for $10$\AA.
The expression of the efficiency as a function of the detector depth
is given by Equation \ref{eqad7} for each blade by fixing the index
$k$.
\begin{table}[!ht]
\centering
\begin{tabular}{|c|c|c|c|}
\hline \hline
wavelength (\AA)  & opt. detect. & $0.5\, \mu m$ detect.& $1.2\, \mu m$ detect. \\
\hline
1.8 &  0.525  & 0.388 & 0.510 \\
10  &  0.858  & 0.831 & 0.671 \\
\hline \hline
\end{tabular}
\caption{\footnotesize Efficiency for an optimized multi-layer
detector and for a detector which contains $15$ identical blades of
$1.2\, \mu m$ and $0.5\, \mu m$.} \label{tabeff23}
\end{table}
\\ The whole detector efficiency is given in the end by summing all
the blades' contributions. The whole detector efficiency is
displayed in Table \ref{tabeff23} for the detector of Figures
\ref{figMG30mon1p8} and \ref{figMG30mon20}. By optimizing the
detector for a given neutron wavelength we gain only about $2\%$
efficiency which is equivalent to add a few more layers to the
detectors optimized to hold identical blades.
\\ In Figure \ref{2optimizzd} is shown the
efficiency resulting from the monochromatic optimization process of
the individual blade coatings and the optimization for a detector
containing all identical blades (which thicknesses are shown on the
right for each neutron wavelength), as already shown in Figure
\ref{fighjort457}. Neutrons hit the layers at $90^{\circ}$ and five
cases have been taken into account with an increasing number of
layers. We notice that about for all neutron wavelengths the gain in
optimizing the detector with different blades, let us to gain few
percent in efficiency. The values in Table \ref{tabeff23} are the
values on the pink solid curve and the dashed one at $1.8$\AA \, and
at $10$\AA \, in Figure \ref{2optimizzd}.
\\ Still referring to Figure \ref{2optimizzd}, we notice that a
detector with 15 individually optimized blades (30 layers) has about
the same efficiency (above $10$\AA) as a detector optimized to
contain 20 blades (40 layers) of equal thickness. On the other hand
for short wavelengths the difference is not very significant.
Moreover, there is also a trade off between the constraints of the
detector construction and the complexity of the blade production.

\subsection{Effect of the substrate in a multi-layer detector}
\begin{figure}[!ht]
\centering
\includegraphics[width=7.8cm]{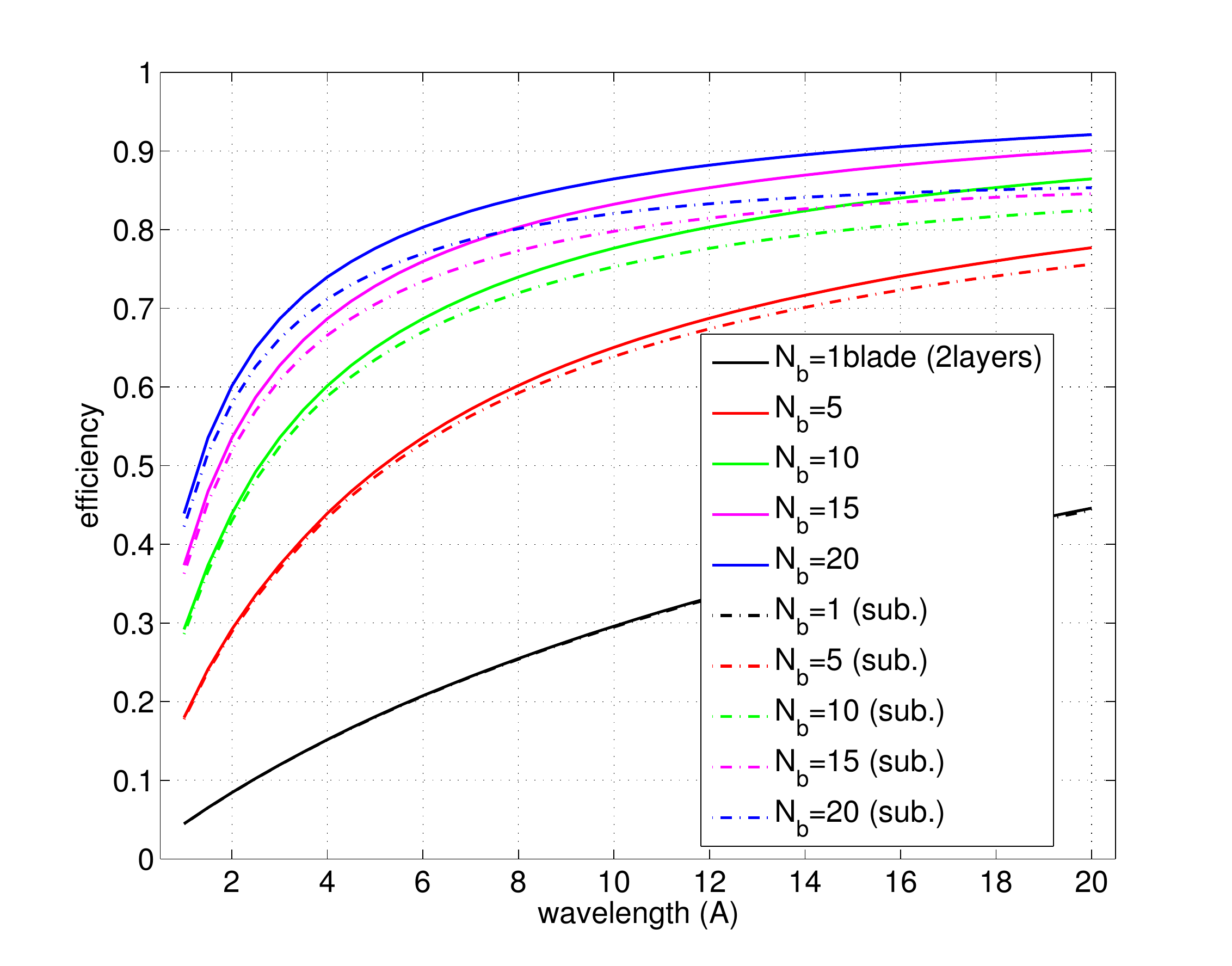}
\includegraphics[width=7.8cm]{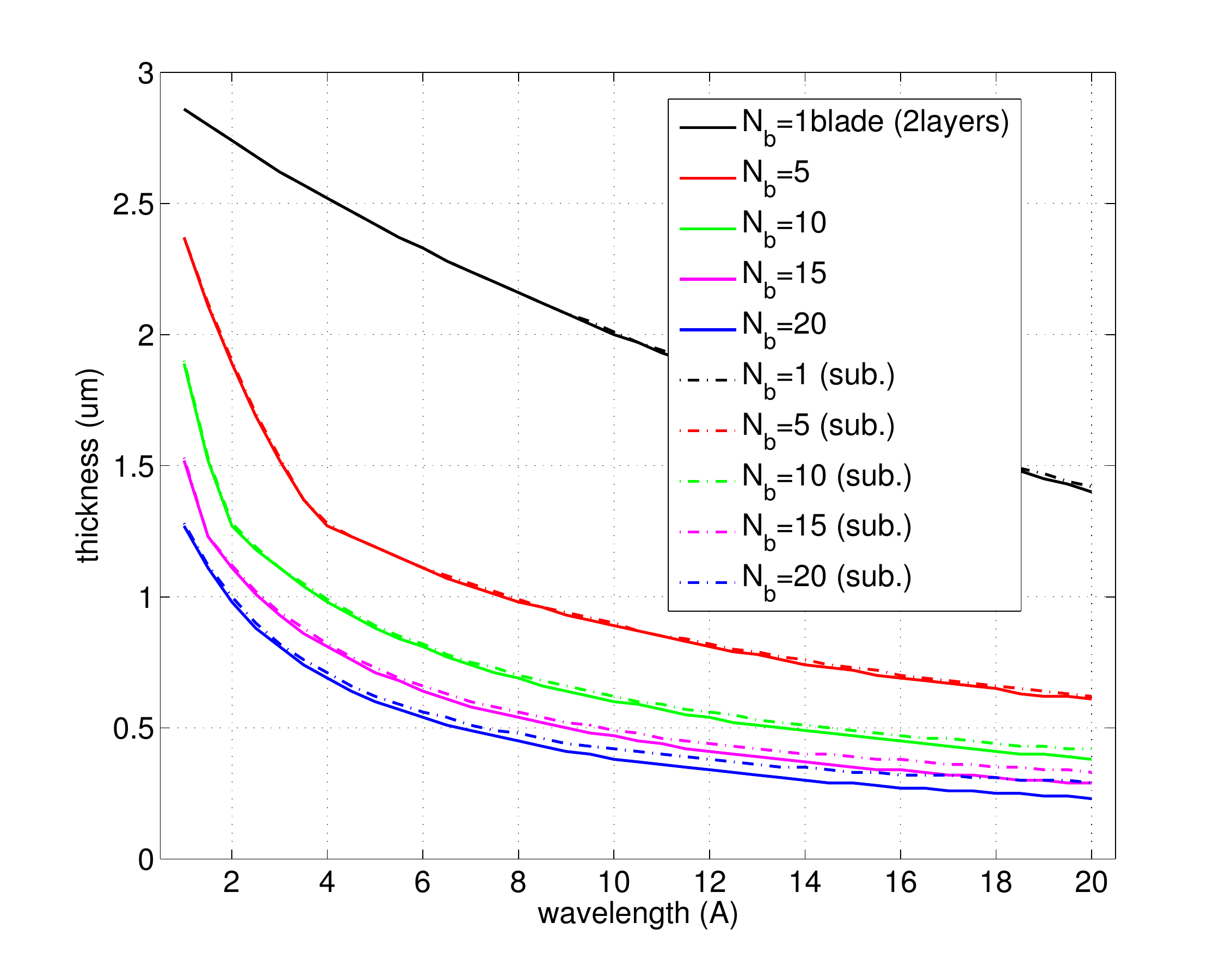}
\caption{\footnotesize Efficiency (left) and optimal thickness of
the identical blades (right) as a function of neutron wavelength for
a 2, 10, 20, 30 and 40 layers multi-layer detector. Solid lines
indicate the optimized efficiency, for each wavelength, for a
detector made up for blades of identical thicknesses; the
dashed-dotted ones indicate the same optimization considering the
substrate.} \label{efsub90876}
\end{figure}

\begin{figure}[!ht]
\centering
\includegraphics[width=10cm]{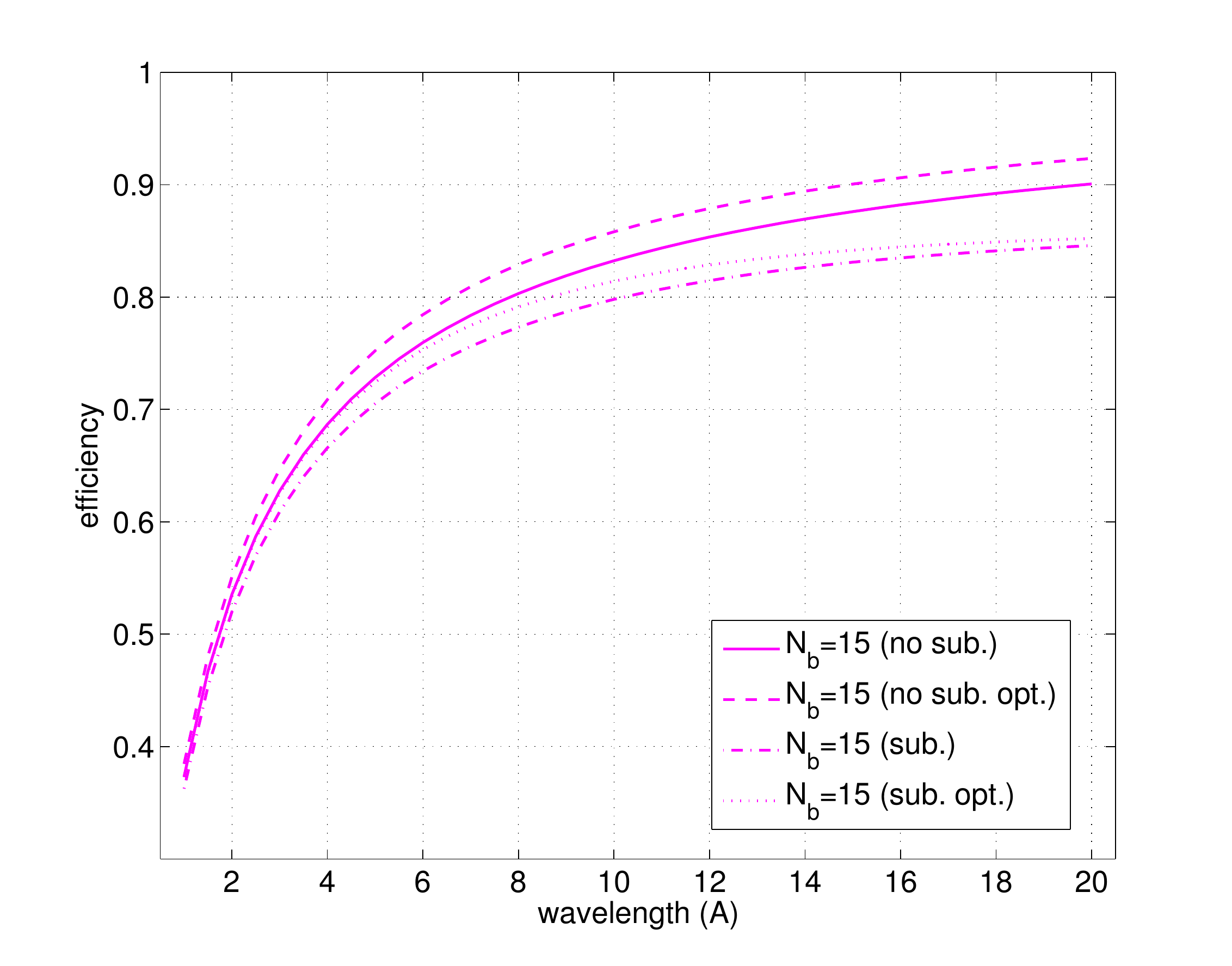}
\caption{\footnotesize Efficiency as a function of neutron
wavelength for a 30-layer detector. Solid lines indicate the
optimized efficiency, for each wavelength, for a detector made up
for blades of identical thicknesses. The dashed one indicates the
monochromatic optimization using different thicknesses inside the
detector. Dashed-dotted and dotted lines are the same as solid and
dashed lines respectively when the substrate is cloistered.}
\label{effsub748952}
\end{figure}

We are going to add the substrate effect in a multi-layer detector;
as we have already treated the case for the single blade in Section
\ref{effsubblad1}. For simplicity we neglect the deviation from the
rule $d_{BS}=d_T$ due to the substrate for each single blade. Hence
we can start from Equation \ref{eqad7}, where each blade has the
same coating thickness for its back-scattering and transmission
layer. The latter in the presence of the substrate becomes:
\begin{equation}
\varepsilon_{tot}^{sub}(N,\bar{d}) = \varepsilon_1^{sub}(d_{1})+
\sum_{k=2}^{N_b}\varepsilon_1^{sub}(d_{k})\cdot e^{-\left( k-1
\right) d_{sub} \cdot \Sigma_{sub} }\cdot e^{-2 \left(
\sum_{j=1}^{\left(k-1\right)} d_j \right)\cdot \Sigma }
\end{equation}
where $\varepsilon_1^{sub}(d)$ is given by Equation \ref{eqac1bis}.
In the case of a detector made up of identical coating thickness
blades we wrote the efficiency as shown in Equation
\ref{eqad1sthick} and considering the substrate it becomes:
\begin{equation}\label{58964l}
\varepsilon_{tot}^{sub}(N_b) =\varepsilon_1^{sub}(d) \cdot
\sum_{k=1}^{N_b} e^{-\left( k-1 \right) d_{sub} \cdot \Sigma_{sub} }
e^{-2 \left( k-1 \right) d \cdot \Sigma }=\varepsilon_1^{sub}(d)
\cdot \frac{1-e^{-\left(2 d \Sigma + d_{sub}
\Sigma_{sub}\right)N_b}}{1-e^{-\left(2 d \Sigma + d_{sub}
\Sigma_{sub}\right) }}
\end{equation}
where, again, $\varepsilon_1^{sub}(d)$ is given by Equation
\ref{eqac1bis}.
\\ As an example we take an Aluminium substrate (density $\rho=2.7\,
g/cm^3$) of $0.5\,mm$ for each blade. We consider a neutron to be
lost when it is either scattered or absorbed, therefore, the
cross-section used is:
$\sigma_{Al}=\sigma_{Al}^{abs}(\lambda)+\sigma_{Al}^{scatt}= 0.2\,b
$(at $ 1.8$\AA )$+1.5\,b=1.7\, b$. Absorption cross-sections at
others neutron wavelengths have been extrapolated linearly in
$\lambda$.
\\ Figure \ref{efsub90876} shows the five detector taken as example
in Figure \ref{2optimizzd} when the detectors are made up for blades
of identical thicknesses considering or not the substrate effect.
The optimization is made for each neutron wavelength separately.
Solid lines represent are the same as in Figure \ref{2optimizzd},
dashed-dotted lines are made for the monochromatic optimization
considering the substrate, i.e. optimizing Equation \ref{58964l}. On
the right in Figure \ref{efsub90876} is shown the optimal thickness
of the identical blades contained in the detector. The effect of the
substrate turns out to change slightly the optimal coating thickness
in order to attain the maximum efficiency.
Furthermore, the higher the number of layers, the bigger is the
deviation between the efficiencies taking into account or not the
effect of the substrate. At long wavelengths, the difference between
a detector composed of 30 or 40 layers becomes smaller taking into
account the substrate.
\\ Moreover, the reasoning explained in Section \ref{monomulayopti}, through
which the efficiency of a multi-layer detector can be optimized by
changing the individual blade coating thicknesses, can be applied
also in presence of the substrate effect. We take a 30-layer
detector that we optimize for 4 different cases. In Figure
\ref{effsub748952} the solid line represents the 30-layer detector
optimized when it contains identical coating thickness blades
without taking the substrate into account, as was the case of solid
line in Figures \ref{2optimizzd} or \ref{efsub90876}. The dashed
line represents the optimization done by changing the coating
thicknesses without substrate effect, as dashed line in Figure
\ref{2optimizzd}. The dashed-dotted line, as in Figure
\ref{efsub90876}, represents the detector efficiency when it is
optimized considering the substrate and it contains identical
blades. The dotted line represents a detector optimized taking the
substrate effect into account allowing the coating thicknesses to
change between different blades.
\\ The substrate effect is to decrease the actual detector
efficiency mainly at higher wavelengths. We can observe from Figure
\ref{effsub748952} that the optimization process helps to gain
efficiency even when substrate plays a role.

\subsection{Multi-layer detector optimization for a distribution of neutron wavelengths}\label{monomulti990}
Let's consider now a multi-layer detector operating on a neutron
wavelength distribution defined by $w\left(\lambda \right)$ and
normalized ($\int_{0}^{+\infty}w\left(\lambda \right)\,
d\lambda=1$). The efficiency for such a detector can be written as
follows:
\begin{equation}\label{eqad14}
\varepsilon_{tot}^w (N_b,\bar{d}_{BS},\bar{d}_{T}) =
\int_{0}^{+\infty}w\left(\lambda \right)
\varepsilon_{tot}(N_b,\lambda) \, d\lambda
\end{equation}
where $\varepsilon_{tot}(N_b,\lambda)$ is the multi-layer detector
efficiency for a single neutron wavelength defined as in Equation
\ref{eqad1}. The efficiency, in this case will be function of
$N=2\cdot N_b$ variables; which can be denoted using the compact
vectorial notation by the two vectors $\bar{d}_{BS}$ and
$\bar{d}_{T}$ of $N_b$ components each. \\ The optimization problem,
in the case of a neutron wavelength distribution, is the
maximization of a $N$-dimensional function. Any change on the
previous blades will change the actual \emph{distribution} of
wavelengths the last blade experiences. The neutron distribution a
blade has to be optimized for depends on all the previous blade
coatings.
\\ Therefore, the $N$-dimensional equation
$\nabla\varepsilon_{tot}^w=0$ has to be solved; explicitly:
\begin{equation}\label{eqad15}
\nabla\varepsilon_{tot}^w = \int_{0}^{+\infty}w\left(\lambda \right)
\nabla\left( \varepsilon_1(d_{BS1},d_{T1})+
\sum_{k=2}^{N_b}\varepsilon_1(d_{BSk},d_{Tk})\cdot e^{- \left(
\sum_{j=1}^{\left(k-1\right)} \left(d_{BSj}+d_{Tj}\right)
\right)\cdot \Sigma }\right) \, d\lambda
\end{equation}
The  $k-th$ $N$-dimensional gradient component for back-scattering
is:
\begin{equation}\label{eqad16}
\frac{\partial \varepsilon_{tot}(N_b,\lambda)}{\partial d_{BSk}} =
\begin{cases}
\frac{\partial \varepsilon_1(d_{BSk},d_{Tk})}{\partial
d_{BSk}}-\Sigma \cdot \sum_{p=\left(k+1\right)}^{N_b}
\varepsilon_1(d_{BSp},d_{Tp}) \cdot e^{- \left(
\sum_{i=1}^{\left(p-1\right)} \left(d_{BSi}+d_{Ti}\right)
\right)\cdot \Sigma }
 \\ \mbox{if \,} k = 1 \\ \\
\frac{\partial \varepsilon_1(d_{BSk},d_{Tk})}{\partial d_{BSk}}\cdot
e^{- \left( \sum_{j=1}^{\left(N_b-1\right)}
\left(d_{BSj}+d_{Tj}\right) \right)\cdot \Sigma }+\\-\Sigma \cdot
\sum_{p=\left(k+1\right)}^{N_b} \varepsilon_1(d_{BSp},d_{Tp}) \cdot
e^{- \left( \sum_{i=1}^{\left(p-1\right)}
\left(d_{BSi}+d_{Ti}\right) \right)\cdot \Sigma } \\ \mbox{if
\, } 1<k<N_b \\
\\ \frac{\partial \varepsilon_1(d_{BSk},d_{Tk})}{\partial d_{BSk}}\cdot
e^{- \left( \sum_{j=1}^{\left(N_b-1\right)}
\left(d_{BSj}+d_{Tj}\right) \right)\cdot \Sigma } \\ \mbox{if \, } k
= N_b
\end{cases}
\end{equation}
Equivalently we find the same expression for the $k-th$ component of
the gradient with respect to the transmission variable; we can
substitute $\partial d_{BSk}$ with $\partial d_{Tk}$ in Equation
\ref{eqad16}.
\\ The condition $\nabla\varepsilon_{tot}^w=0$ implies that for each
$k$ must hold $\frac{\partial
\varepsilon_{tot}(N_b,\lambda)}{\partial d_{BSk}}=0$ and
$\frac{\partial \varepsilon_{tot}(N_b,\lambda)}{\partial d_{Tk}}=0$
at the same time. From Equation \ref{eqad16} and the one for the
transmission variable we finally obtain ($\forall k
=1,2,\dots,N_b$):
\begin{equation}\label{eqad17}
\begin{cases}
\frac{\partial \varepsilon_1(d_{BSk},d_{Tk})}{\partial
d_{BSk}}=\frac{\partial \varepsilon_1(d_{BSk},d_{Tk})}{\partial
d_{Tk}}\Rightarrow D_{\hat{u}}\varepsilon_1(d_{BSk},d_{Tk})=0
&\mbox{if \,} k < N_b \\
\frac{\partial \varepsilon_1(d_{BSk},d_{Tk})}{\partial
d_{BSk}}=\frac{\partial \varepsilon_1(d_{BSk},d_{Tk})}{\partial
d_{Tk}}=0 &\mbox{if \, } k = N_b
\end{cases}
\end{equation}
As for the monochromatic case, the condition in Equation
\ref{eqad17} is exactly what was demonstrated in Section
\ref{Sect2laysub}. The Theorem \ref{theo1} implies that the
directional derivative along the unity vector
$\hat{u}=\frac{1}{\sqrt{2}}\left(1,-1\right)$ of the function
$\varepsilon_1(d_{BSk},d_{Tk})$ can only be zero on the domain
bisector, in the case of $k < N_b$. Hence, the maximum efficiency
can only be found, again, on the bisector. On the other hand, in the
case of $k=N_b$ the Equation \ref{eqad17} requires the gradient of
the efficiency function to be zero, property which was also
demonstrated for these kind of functions in Section
\ref{Sect2laysub} (Equations \ref{eqac3} and \ref{eqac5}).
\medskip
\\ \textbf{In a Multi-Grid like detector, which has to
be optimized for any distribution of neutron wavelengths, all the
blades have to hold two layers of the same thickness. Naturally,
thicknesses of different blades can be distinct.}
\medskip
In this case it is not possible to start the optimization from the
last blade because the thicknesses of the previous layers will
affect the neutron wavelength distribution reaching the deeper
laying blades. We have in this case to optimize an $N_b$-dimensional
function at once. Therefore, the $N_b$-dimensional equation
$\nabla\varepsilon_{tot}^w=0$ has to be solved.
\\ Thanks to the property just derived, we can denote with $d_k$
the common thickness of the two layers held by the $k-th$ blade
($d_{BSk}=d_{Tk}=d_k$). In the detector efficiency function,
expressed by Equation \ref{eqad14}, we can substitute
$\varepsilon_{tot}(N_b,\lambda)$ with its simpler expression shown
in Equation \ref{eqad7}.
\\ For the optimization process the $k-th$ component of the gradient
$\nabla_k\varepsilon_{tot}^w$ can be replaced with $\frac{\partial
\varepsilon_{tot}^w}{\partial d_{k}}$; because now $N_b=\frac N 2$
unknown, instead of $N$, have to be found.
\begin{equation}\label{eqad18}
\nabla_k\varepsilon_{tot}^w = \int_{0}^{+\infty}w\left(\lambda
\right) \frac{\partial \varepsilon_{tot}}{\partial d_{k}} \,
d\lambda= \int_{0}^{+\infty}w\left(\lambda \right) \frac{\partial
}{\partial d_{k}} \left( \varepsilon_1(d_{1})+
\sum_{k=2}^{N_b}\varepsilon_1(d_{k})\cdot e^{-2 \left(
\sum_{j=1}^{\left(k-1\right)} d_j \right)\cdot \Sigma }\right) \,
d\lambda
\end{equation}
$\frac{\partial \varepsilon_{tot}}{\partial d_{k}}$ is an expression
like Equation \ref{eqad16} provided that we impose $d_{BSk}=d_{Tk}
\, \forall k=1,2,\dots,N_b$.
\\ Finally, in order to optimize a detector for a given neutron
wavelength distribution, the following system of $N_b$ equations in
$N_b$ unknown ($d_k$) has to be solved:
\begin{equation}\label{eqad19}
\begin{cases}
\int_{0}^{+\infty}w\left(\lambda \right)\left[ \frac{\partial
\varepsilon_1(d_k)}{\partial d_k}-2\Sigma \cdot \sum_{p=2}^{N_b}
\varepsilon_1(d_p) \cdot e^{- 2\left( \sum_{i=1}^{\left(p-1\right)}
d_i \right)\cdot \Sigma }\right]  \, d\lambda=0
 &\mbox{if \,} k = 1 \\ \\
\int_{0}^{+\infty}w\left(\lambda \right)\left[ \frac{\partial
\varepsilon_1(d_k)}{\partial d_k}\cdot e^{- 2\left(
\sum_{j=1}^{\left(N_b-1\right)} d_j \right)\cdot \Sigma
}\right.+\\-\left.2\Sigma \cdot \sum_{p=\left(k+1\right)}^{N_b}
\varepsilon_1(d_p) \cdot e^{- 2\left( \sum_{i=1}^{\left(p-1\right)}
d_i \right)\cdot \Sigma }\right] \, d\lambda=0
 &\mbox{if \, } 1<k<N_b \\ \\
\int_{0}^{+\infty}w\left(\lambda \right)\left[\frac{\partial
\varepsilon_1(d_k)}{\partial d_k}\cdot e^{- 2\left(
\sum_{j=1}^{\left(N_b-1\right)} d_j \right)\cdot \Sigma }\right] \,
d\lambda=0
 &\mbox{if \, } k = N_b
\end{cases}
\end{equation}
\\ We recall that $\varepsilon_1(d_k)$ and $\Sigma$ are function of
$\lambda$ and $\varepsilon_1(d_k)$ is the blade efficiency defined
in Equations \ref{eqac2} and \ref{eqac4}; its derivative
$\frac{\partial \varepsilon_1(d_k)}{\partial d_k}$ was already
calculated in the Equations \ref{eqac14} and \ref{eqac17} (Section
\ref{Sect2laysub}).
\\ The system of equations \ref{eqad19} can be easily solved
numerically. \\ Comparing this result with the one found in the
monochromatic case in Section \ref{monomulti990}, where the solution
could be found iteratively starting from the last blade, we have now
a system of $N_b$ equations in $N_b$ unknown. In Section
\ref{monomulti990}, the system of equations \ref{eqad8} turned out
to be upper triangular.
\\ This is not the case for the distribution case in which the gradient of Equation
\ref{eqad7} is in addition integrated over $\lambda$, thus all the
blades have to be taken into account at once in the optimization
process. The optimization problem will be the maximization of a
$N$-dimensional function.
\\ To better understand the meaning of that, we should
figure out how the optimization process works. For a single
wavelength, each layer efficiency has to maximize for that well
defined neutron energy; all the previous layers only affect the
number of neutrons that can reach the deeper blades. We require the
last blade to be as efficient as possible on that kind of neutron.
On the other hand, in the case of a wavelength distribution, any
change on the previous blades will change the actual distribution
the last blade experiences. There can be neutrons of a certain
energy that can not get to a layer it was optimized for. Thus, the
neutron distribution a blade has to be optimized for depends on all
the previous blades coatings. In this case, the matrix does not end
up to be triangular.

\subsubsection{Flat neutron wavelength distribution example}
\begin{figure}[!ht]
\centering
\includegraphics[width=7.8cm,angle=0,keepaspectratio]{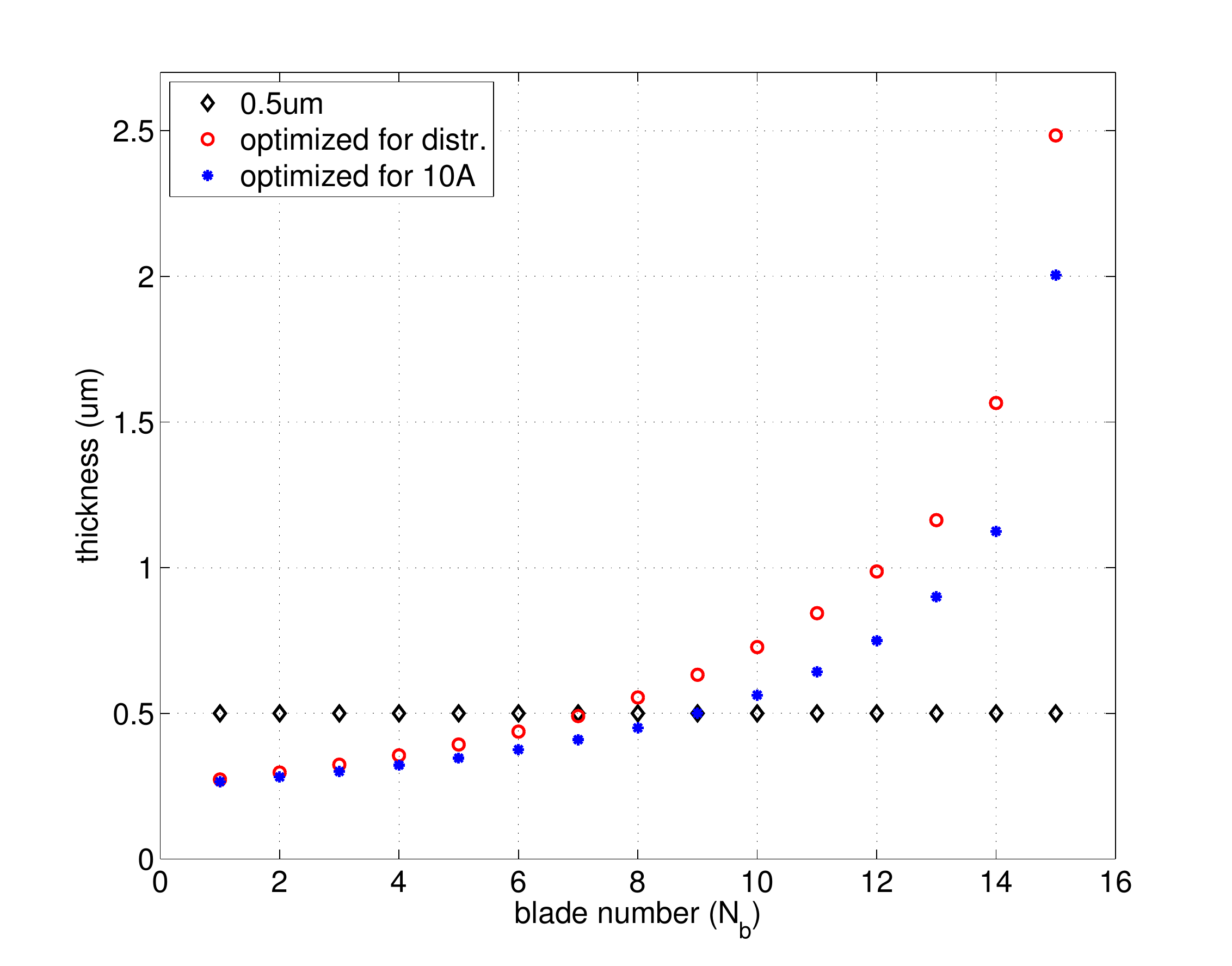}
\includegraphics[width=7.8cm,angle=0,keepaspectratio]{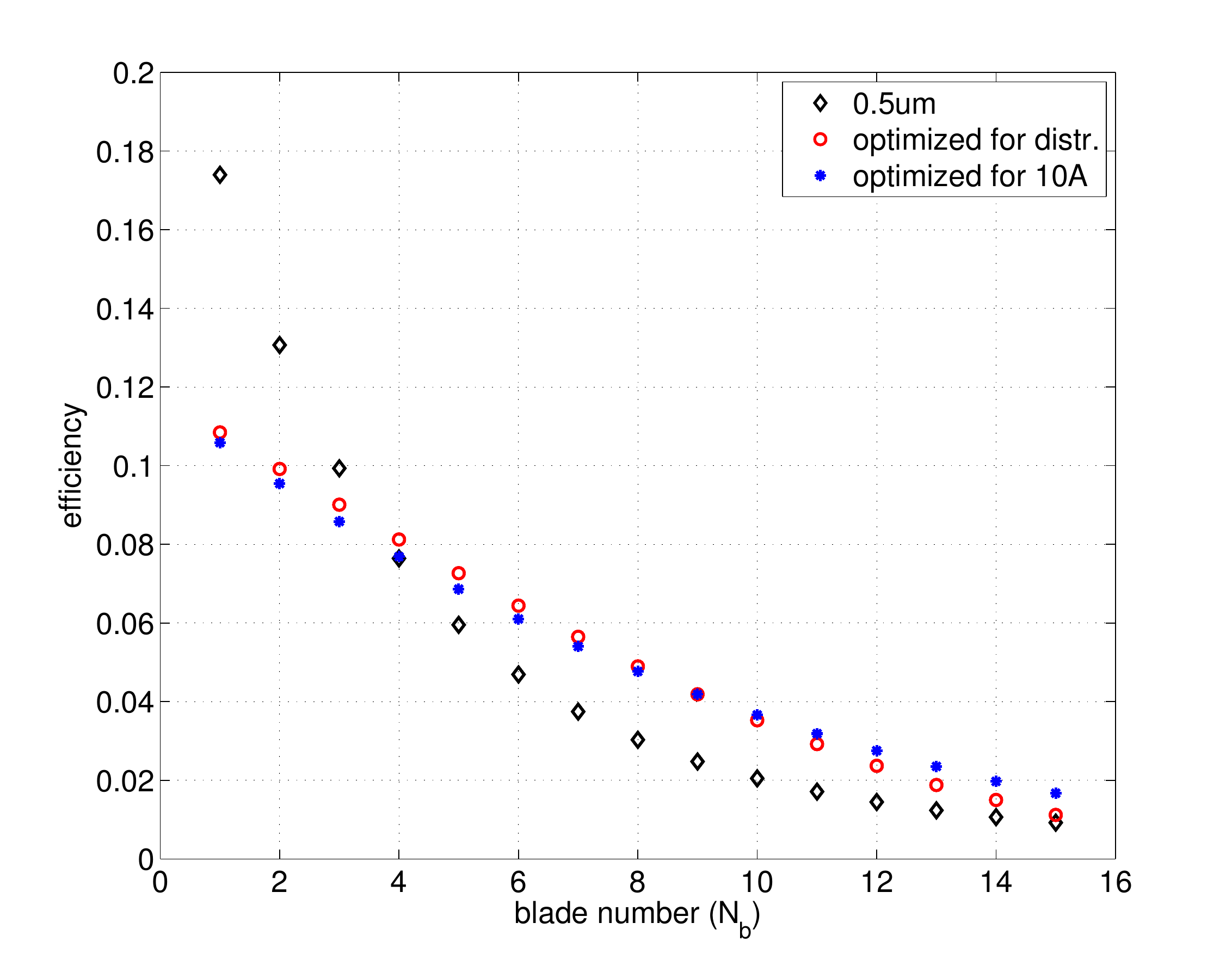}
 \caption{\footnotesize Thicknesses of the blades coatings (left) and their efficiency contribution (right), for a detector made
 up of $15$ identical coating thickness blades of $0.5\, \mu m$, for a detector optimized for the flat distribution
 of wavelengths and for a detector optimized for $10$\AA.} \label{figMG30flat102}
\end{figure}

We take a flat distribution
$w\left(\lambda\right)=\frac{1}{\lambda_2-\lambda_1}$ between the
two wavelengths $\lambda_1=1$\AA \, and $\lambda_2=20$\AA \, as in
Section \ref{Sect2laysub} for the single blade case. In Figure
\ref{figMG30flat102} the thicknesses of each of the blade coatings
and each blade efficiency contribution for a $30$-layer detector are
shown. Three detectors are compared, the one of simplest
construction is a detector holding $15$ identical blades of $0.5\,
\mu m$ coating thickness, the second is a detector optimized
according to Equation \ref{eqad19} for that specific flat
distribution and the last is a detector that has been optimized for
a single neutron wavelength of $10$\AA \, conforming to Equations
\ref{eqad11}, \ref{eqad12} and \ref{eqad13}. The fact to have a
contribution of wavelengths shorter than $10$\AA \, in the case of
the red line makes the coating thicknesses larger compared to the
blue curve.
\\ As a result, frontal layers are slightly more efficient for the
distribution optimized detector than for the one optimized for
$10$\AA; on the other hand, deep layers lose efficiency.
\begin{figure}[!ht]
\centering
\includegraphics[width=7.8cm,angle=0,keepaspectratio]{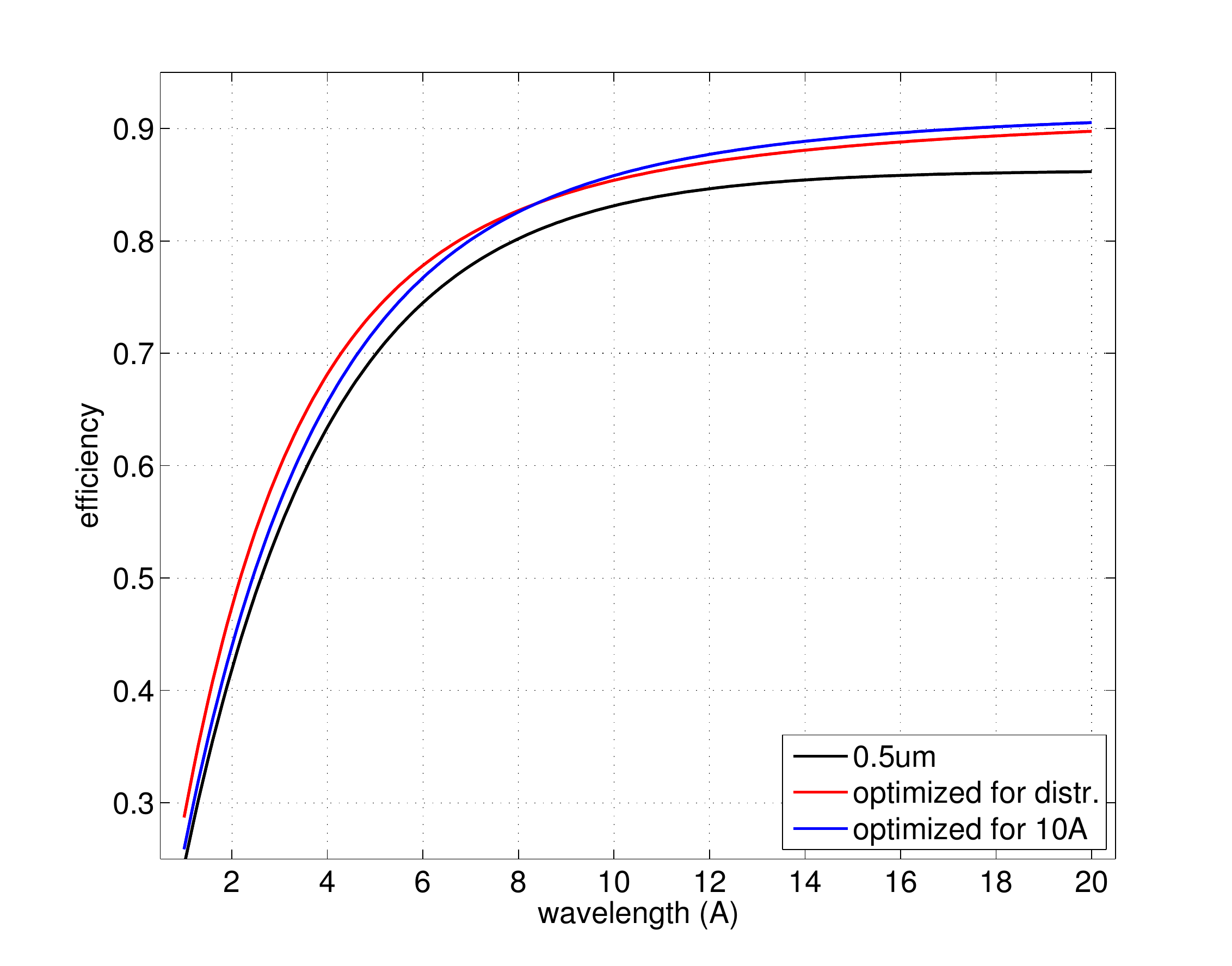}
\includegraphics[width=7.8cm,angle=0,keepaspectratio]{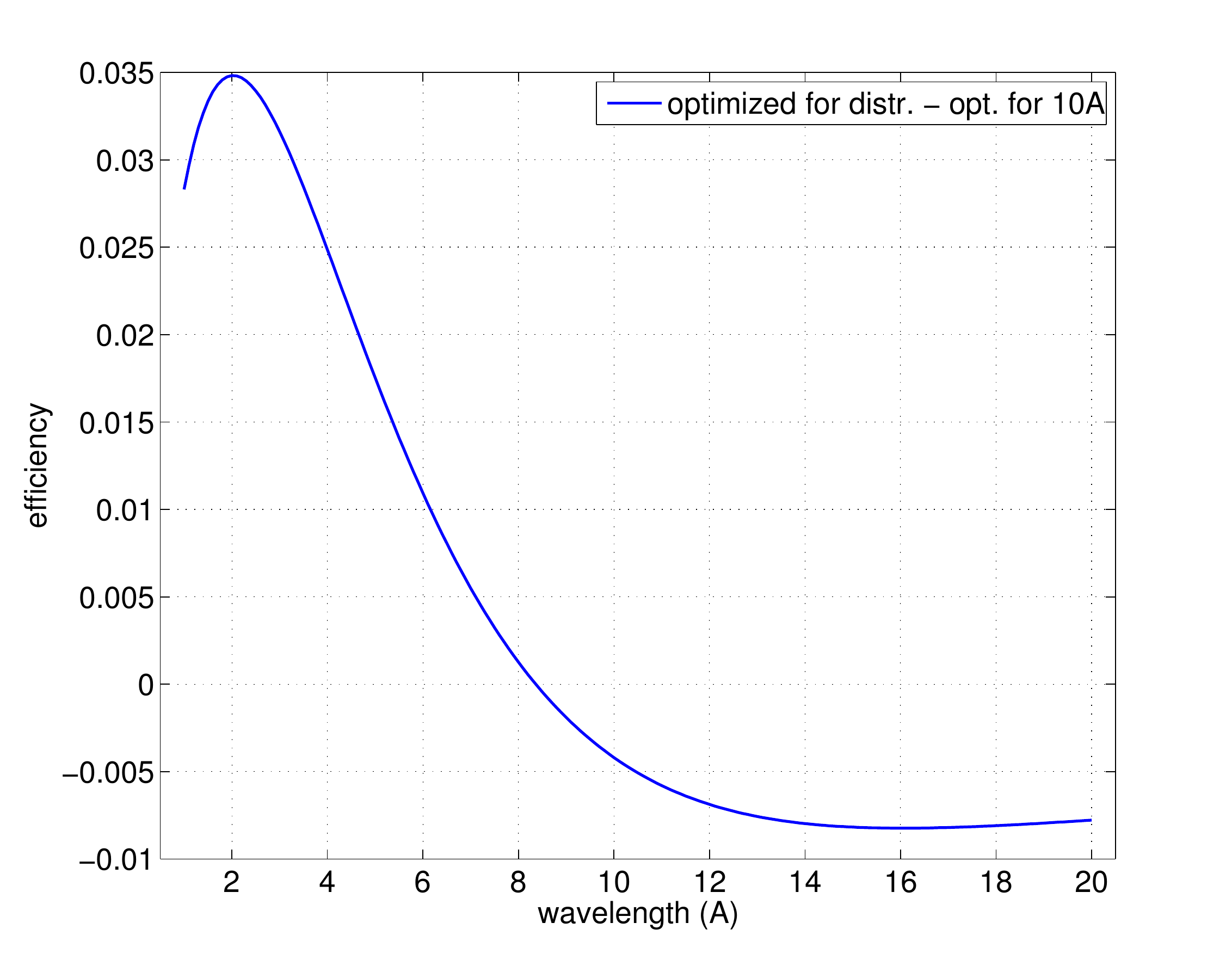}
 \caption{\footnotesize Efficiency as a function of neutron wavelength (left) for a detector made
 up of $15$ identical coating thickness blades of $0.5\, \mu m$, for a detector optimized for the flat distribution
 of wavelengths and for a detector optimized for $10$\AA. Difference between the efficiencies for a detector optimized
 for a flat distribution and for $10$\AA \, as a function of neutron wavelength (right).} \label{figMG30flat103}
\end{figure}
Figure \ref{figMG30flat103} shows the three detector efficiencies as
a function of neutron wavelength. By comparing red and blue lines,
of which the difference is plotted on the right plot, the optimized
detector gains efficiency on shorter wavelengths but loses on
longer. Moreover, we notice that the optimization process explained
in this section let to gain at most $3.5\%$ at short wavelengths
while losing less than $1\%$ on longer ones. The weighted efficiency
over $w\left(\lambda\right)$ is shown in Table \ref{tabeff569}.
\begin{table}[!ht]
\centering
\begin{tabular}{|c|c|c|}
\hline \hline
 opt. detect. & opt. detect. for 10\AA & $0.5\, \mu m$ detect. \\
\hline
0.796 & 0.793 & 0.764 \\
\hline \hline
\end{tabular}
\caption{\footnotesize Averaged efficiency over the flat neutron
wavelength distribution ($1$\AA-$20$\AA) for a detector which
contains $15$ identical blades of $0.5\, \mu m$, for an optimized
multi-layer detector for that specific flat distribution and for a
detector optimized for $10$\AA. (Energy threshold $100\,KeV$
applied).} \label{tabeff569}
\end{table}
We can conclude that if we are interested in optimizing a detector
in a given interval of wavelengths without any preference to any
specific neutron energy; optimizing according to Equation
\ref{eqad19} does not give a big improvement in the average
efficiency compared to optimizing for the neutron wavelength
distribution barycenter (about $10$\AA).
\\ Although the averaged efficiency for the optimized detector
in the neutron wavelength range differs from the one optimized for
$10$\AA \, only by $0.3\%$ one can be interested to have a better
efficiency for shorter wavelengths rather than for longer. It is in
this case that the optimization process can play a significant role.
On that purpose let's move to the following example.

\subsubsection{Hyperbolic neutron wavelength distribution example}
\begin{figure}[!ht]
\centering
\includegraphics[width=7.8cm,angle=0,keepaspectratio]{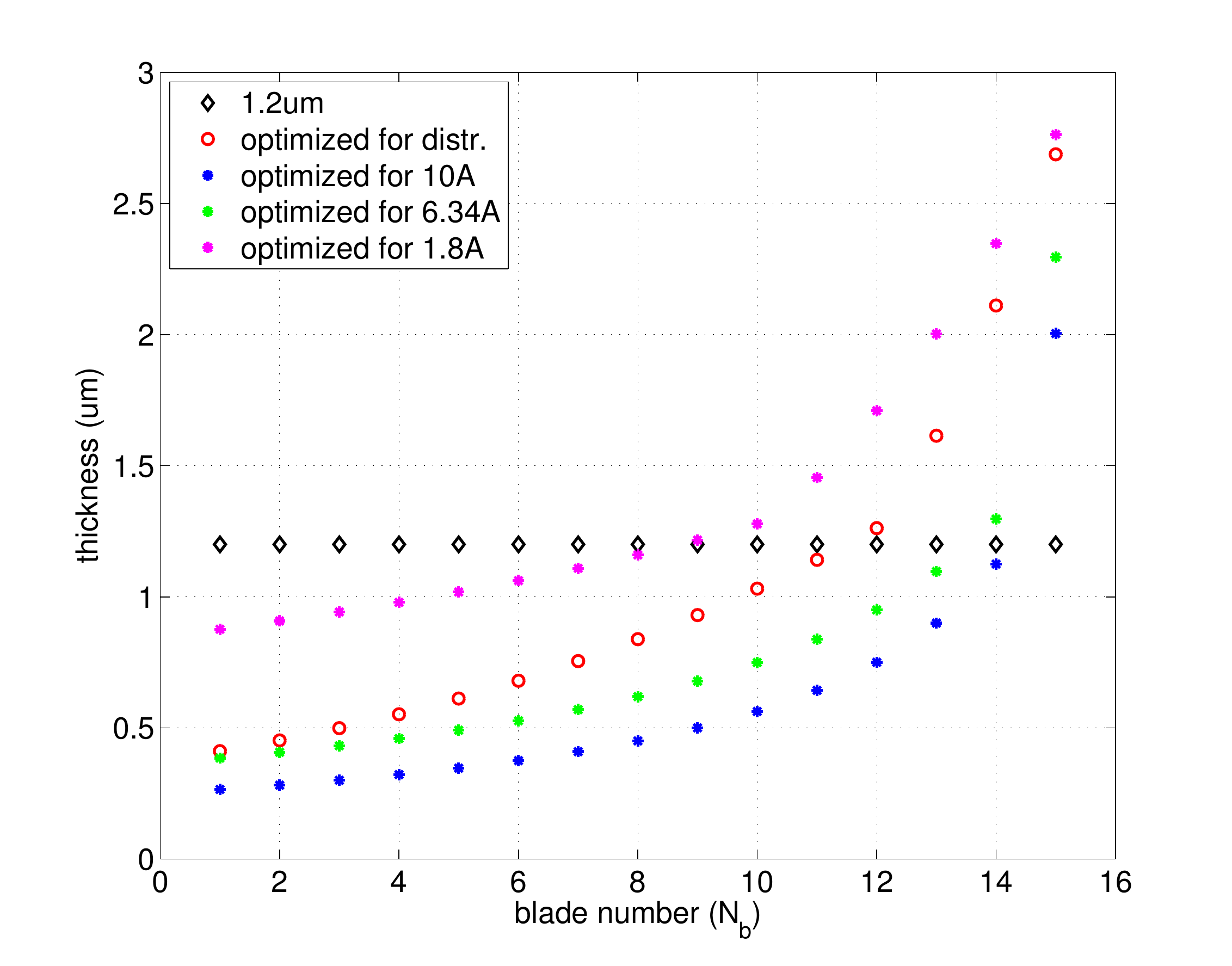}
\includegraphics[width=7.8cm,angle=0,keepaspectratio]{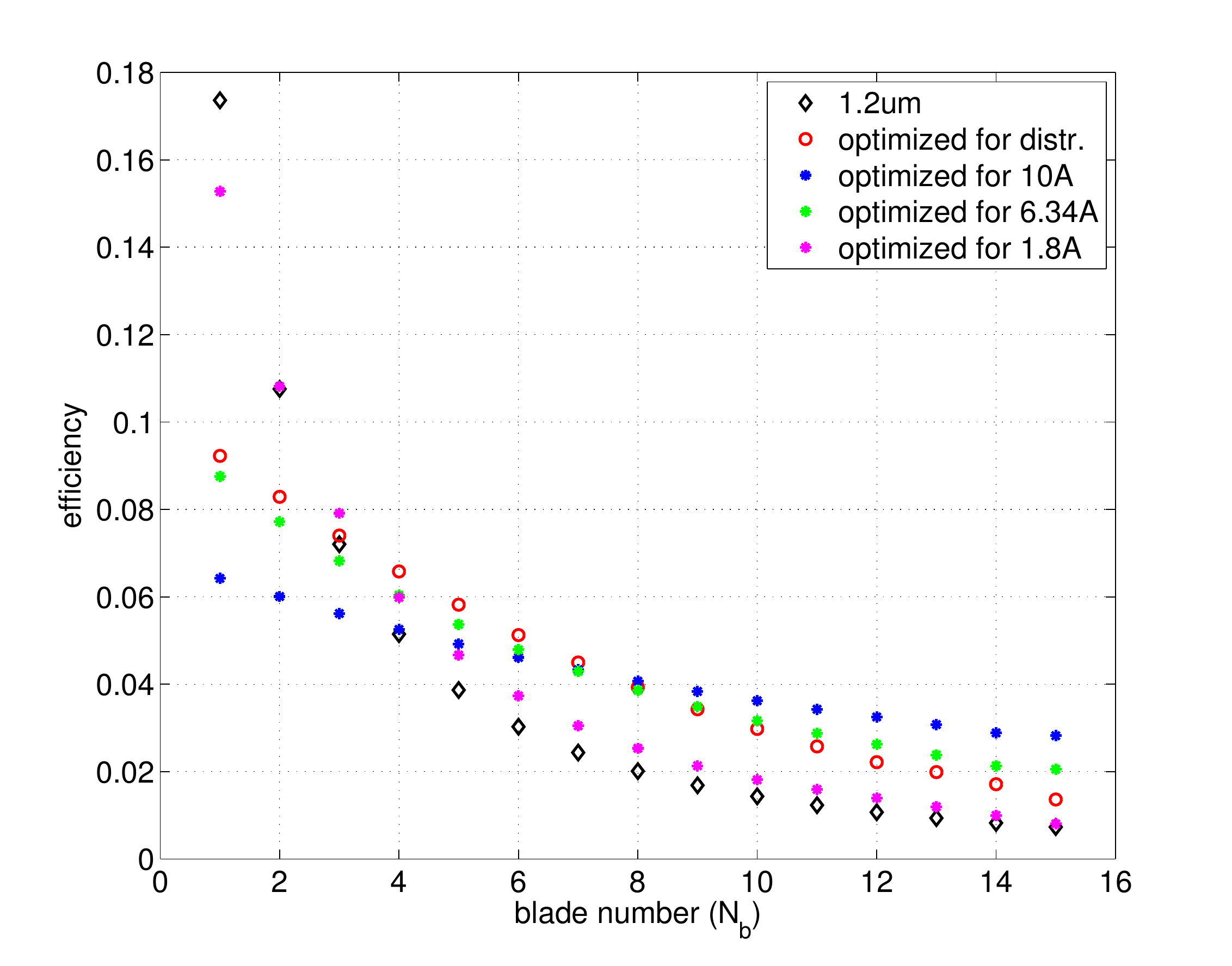}
 \caption{\footnotesize Thicknesses of the blades coatings (left) and their efficiency contribution (right), for a detector made
 up of $15$ identical coating thickness blades of $1.2\, \mu m$, for a detector optimized for an hyperbolic distribution
 of wavelengths and for a detector optimized for $10$\AA, $6.34$\AA \, and for for $1.8$\AA.} \label{figMG30iperb102}
\end{figure}

We consider a hyperbolic neutron wavelength distribution between
$\lambda_1=1$\AA \, and $\lambda_2=20$\AA.
\begin{equation}\label{eqad20}
w\left(\lambda\right)=\frac{1}{
\ln\left(\frac{\lambda_2}{\lambda_1}\right)} \cdot \frac{1}{\lambda}
\end{equation}
\\ This optimization aims for giving equal importance to bins on a logarithmic
wavelength scale. The barycenter of the wavelength distribution
corresponds to
$\int_{\lambda_1}^{\lambda2}w\left(\lambda\right)\lambda \,
d\lambda=6.34$\AA.
\\ In Figure \ref{figMG30iperb102} are shown the
thicknesses of each blade coatings and the efficiency as a function
of the depth direction in the detector for a $30$-layer detector.
Five detectors are compared, the one of $1.2\, \mu m$ coating
thickness, a detector optimized according to Equation \ref{eqad19}
for that specific hyperbolic distribution, a detector that has been
optimized for a single neutron wavelength of $10$\AA, $1.8$\AA, and
the barycenter of the distribution.
\begin{figure}[!ht]
\centering
\includegraphics[width=7.8cm,angle=0,keepaspectratio]{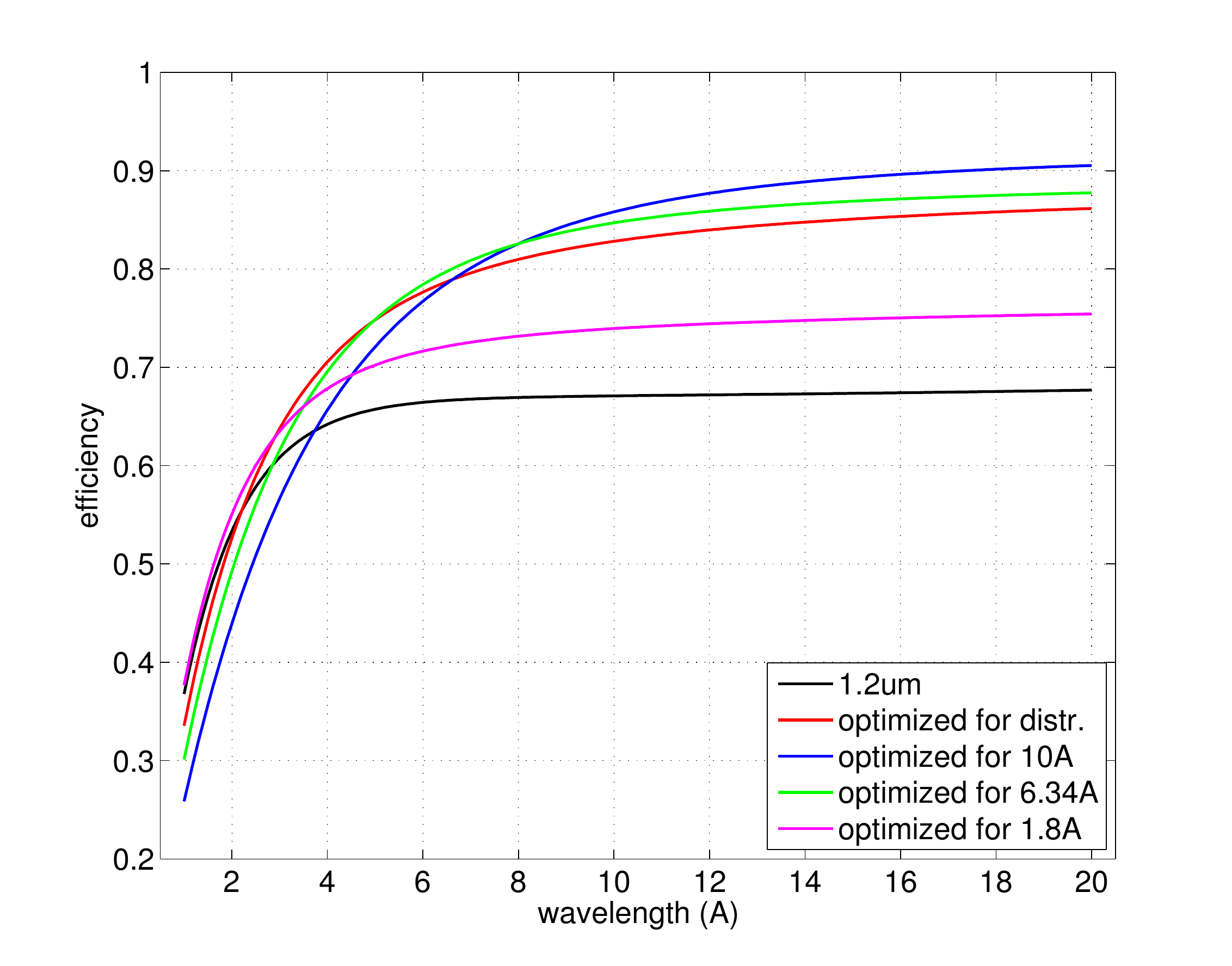}
\includegraphics[width=7.8cm,angle=0,keepaspectratio]{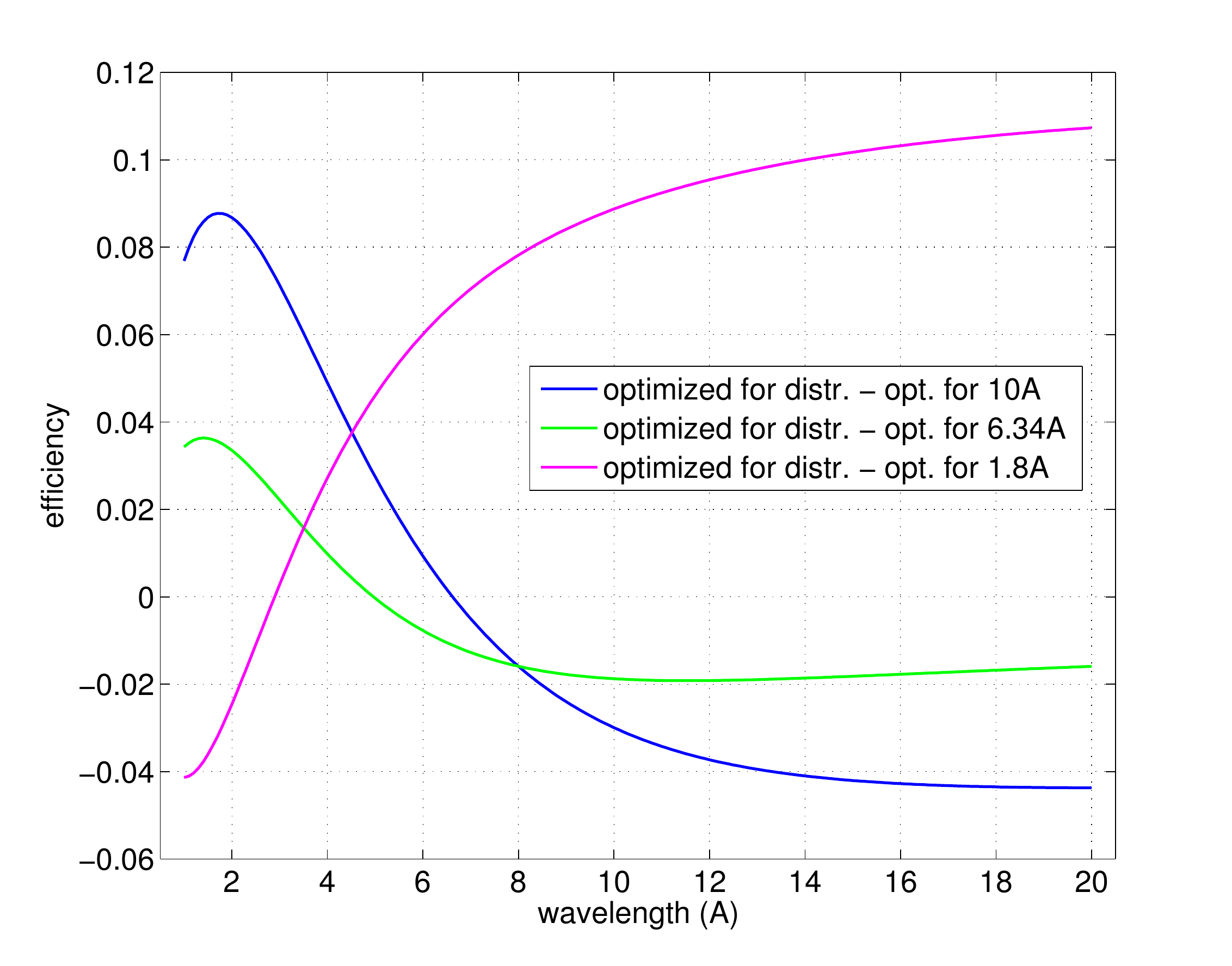}
 \caption{\footnotesize Efficiency as a function of neutron wavelength (left) for a detector made
 up of $15$ identical coating thickness blades of $1.2\, \mu m$, for a detector optimized for an hyperbolic distribution
 of wavelengths and for a detector optimized for $10$\AA \, and for $1.8$\AA. Difference between the efficiencies for a detector optimized
 for a flat distribution and for $10$\AA, $6.34$\AA \, and for $1.8$\AA \, as a function of neutron wavelength (right).} \label{figMG30iperb103}
\end{figure}
\begin{table}[!ht]
\centering
\begin{tabular}{|c|c|c|c|c|}
\hline \hline
opt. detect. & opt. $10$\AA & opt. $6.34$\AA  & opt. $1.8$\AA & $1.2\, \mu m$ detect.\\
\hline
0.671 & 0.641 & 0.664 & 0.639 & 0.597 \\
\hline \hline
\end{tabular}
\caption{\footnotesize Averaged efficiency over the hyperbolic
distribution defined in Equation \ref{eqad20} for a detector which
contains $15$ identical blades of $1.2\, \mu m$, for an optimized
multi-layer detector for that specific distribution and for a
detector optimized for $10$\AA, $6.34$\AA \, and for $1.8$\AA.
(Energy threshold $100\,KeV$ applied).} \label{tabeff570}
\end{table}
\\ By only comparing the averaged efficiencies, shown in Table
\ref{tabeff570}, the distribution optimized detector shows only a
gain of at most about $3\%$ with respect to the detectors optimized
for the distribution barycenter or other wavelengths. It seems that
there is not a big improvement in the detector efficiency over the
full neutron energy range.
\\ Figure \ref{figMG30iperb103} shows the five detector efficiencies as a
function of wavelength and their difference on the right plot. By
comparing the red (distribution optimized detector) and the green
(barycenter optimized detector) lines, of which the difference is
plotted in green on the right plot, we notice that the detector
optimized for such a distribution gains about $4\%$ efficiency at
short wavelengths and loses about $2\%$ at high wavelengths. With
respect to a detector conceived for higher wavelengths (blue curve),
i.e. $10$\AA, the distribution optimized one gains about $9\%$ at
short wavelengths and loses about $4\%$ at high wavelengths. A
detector conceived for short wavelengths, such as the one
represented by the pink line ($1.8$\AA), has an opposite behavior
instead. The distribution optimized detector gains efficiency for
long wavelengths reaching about $11\%$.
\\ Even if the the optimization procedure, explained in this section,
shows that there is not a notable improvement over the full range of
neutron wavelength, it can lead to a significant efficiency
improvement in certain neutron wavelength ranges. As in the case of
a flat distribution, a detector optimized for a distribution
according to Equations \ref{eqad19}, does not show significant
improvement in performances with respect to a detector just
optimized for its barycenter.

\section{Why Boron Carbide?}
We are going to explain here why $^{10}B_4C$ is a suitable material
for neutron detection in solid converter gaseous detectors.
\\ A good solid converter material that can be employed in a thermal
neutron detector should own features like sufficient electrical
conductivity, high neutron absorption cross-section, low density, no
toxicity and easy manipulation. We compare two neutron solid
converters $^{10}B$ and $^{6}Li$ and their compounds $^{10}B_4C$ and
$^{6}LiF$; their characteristics are listed in Table
\ref{tabreswhyb8}. We list their mass density, microscopic and
macroscopic absorption cross-section $\Sigma$, $\eta=1/\Sigma$ is
the mean free path, and the electrical resistivity.
\\ A good converter material
should not be too resistive because, in general, it acts as a
cathode in the gaseous detector. In order for the electric field to
stay constant in time it should evacuate the charges in a reasonable
time. \\ While $^{10}B$, $^{10}B_4C$ and $^{6}LiF$ do not present
any strong reactivity, pure $^{6}Li$ reacts with water easily. It
has to be manipulated and operated in a controlled atmosphere.
\\ Among the features listed in Table \ref{tabreswhyb8}, a converter material
should offer a high density to maximize the number of neutron
conversions per unit volume but, on the other hand, a low density to
let the neutron capture reaction fragments escape easily from the
layer to produce a detectable charge in the gas volume. While the
microscopic absorption cross-section only plays a role in the
neutron capture process; the mass density influences both the
capture and the fragment escape processes. Hence, $\Sigma$, or
equivalently $\eta$, represents the capture power of a material
given its density and its absorption cross-section. The shorter
$\eta$, the higher is the probability for a neutron to be captured.
The energy the fragments own contributes to the escape probability,
i.e. in their ranges. For that purpose $^{6}Li$ fragments carrying
$4790\,KeV$ are more probable to escape.
\begin{table}[!ht]
\centering
\begin{tabular}{|r|c|c|c|c|c|}
\hline \hline
material & $\rho (g/cm^3)$ & $\Sigma (1/\mu m) \,at \, 1.8$\AA & $\eta (\mu m) \,at \,1.8$\AA & $\sigma_{abs} (b) \,at \, 1.8$\AA & $\rho_e (\Omega \cdot m)$  \\
\hline
   $^{nat}Li$  &  $0.53$  & $3.3\cdot10^{-4}$ &$3030$ &$940$ & $10^{-7}$ \\
pure $^{6}Li$  &  $0.46$  & $4.4\cdot10^{-3}$ &$227$ &$940$ & $10^{-7}$ \\
   $^{nat}B$   &  $2.35$  & $1.0\cdot10^{-2}$ &$100$ &$3835$ & $10^{6}$ \\
pure $^{10}B$  &  $2.17$  & $5.0\cdot10^{-2}$ &$20$ &$3835$ & $10^{6}$\\
 $^{nat}LiF$   &  $2.64$  & $4.4\cdot10^{-4}$ &$2273$ &$940$ & $10^{-8}$ \\
$96\%$ enriched $^{6}LiF$  &  $2.54$ & $5.8\cdot10^{-3}$ &$172$ &$940$ & $10^{-8}$ \\
 $^{nat}B_4C$   &  $2.52$  &  $8.4\cdot10^{-3}$ &$119$ &$3835$ & $10^{-3}$ \\
$98\%$ enriched $^{10}B_4C$  & $2.37$  & $4.2\cdot10^{-2}$ &$24$ &$3835$ & $10^{-3}$ \\
\hline
glass & & & & & $>10^{9}$\\
Si & & & & &$1$\\
graphite & & & & & $10^{-5}$\\
Cu    & & & & & $10^{-8}$\\
 \hline \hline
\end{tabular}
\caption{\footnotesize Physical features of common neutron thermal
neutron converter. $\eta=1/\Sigma$ is the mean free path a neutron
can travel across the material before being absorbed. $\sigma_{abs}$
is shown for the active nuclide; i.e. $^6Li$ or $^{10}B$. Natural
compounds refer to $7.5\%$ of $^6Li$ and $92.5\%$ of $^7Li$ for
Lithium and $19.4\%$ of $^{10}B$ and $80.6\%$ of $^{11}B$ for
Boron.} \label{tabreswhyb8}
\end{table}
\\ Moreover, mass density plays in the escape process. In Table
\ref{tabreswhyb09} the ranges for the $^{10}B$ and $^{6}Li$ reaction
fragments (see Table \ref{eqaa4}) are listed for an energy threshold
of $100\,KeV$.
\begin{table}[!ht]
\centering
\begin{tabular}{|r|c|c|c|c|}
\hline \hline
material & $R (\mu m) $ & $R (\mu m) $ & & \\
& $\alpha (2050\,KeV)$ & $^3H (2740\,KeV)$ & & \\
\hline
  pure $^{6}Li$                &  $21$  & $132$ & & \\
$96\%$ enriched $^{6}LiF$  &  $5.2$ & $32.8$ & & \\
 \hline \hline
& $R (\mu m)$ & $R (\mu m)$ & $R (\mu m)$ & $R (\mu m)$\\
& $\alpha (1470\,KeV)$ & $^7Li (830\,KeV)$ & $\alpha (1770\,KeV)$ & $^7Li (1010\,KeV)$\\
\hline
  pure $^{10}B$              &  $3.3$  & $1.5$ & $4.1$ & $1.8$\\
$98\%$ enriched $^{10}B_4C$  &  $2.9$  & $1.2$ & $3.7$ & $1.5$\\
 \hline \hline
\end{tabular}
\caption{\footnotesize Ranges of the capture reaction fragments in
the material. An energy threshold of $100\,KeV$ is applied.}
\label{tabreswhyb09}
\end{table}
\\ It is true that pure $^{10}B$ has a slightly higher boron atom
density and the fragments have somewhat longer ranges which makes
the potential efficiency of a pure $^{10}B$ coated detector in
principle higher. However, given that pure $^{10}B$ is about eight
orders of magnitudes more electrically resistive than $^{10}B_4C$,
therefore one opted in \cite{jonisorma} to use $^{10}B_4C$.
\\ Because the fragments in $^{6}Li$ have a very long
range, although $\eta$ is shorter than the one in $^{6}LiF$, it
could be a very powerful converter material but hygroscopicity makes
it not a very good candidate.
\begin{figure}[!ht]
\centering
\includegraphics[width=7.8cm,angle=0,keepaspectratio]{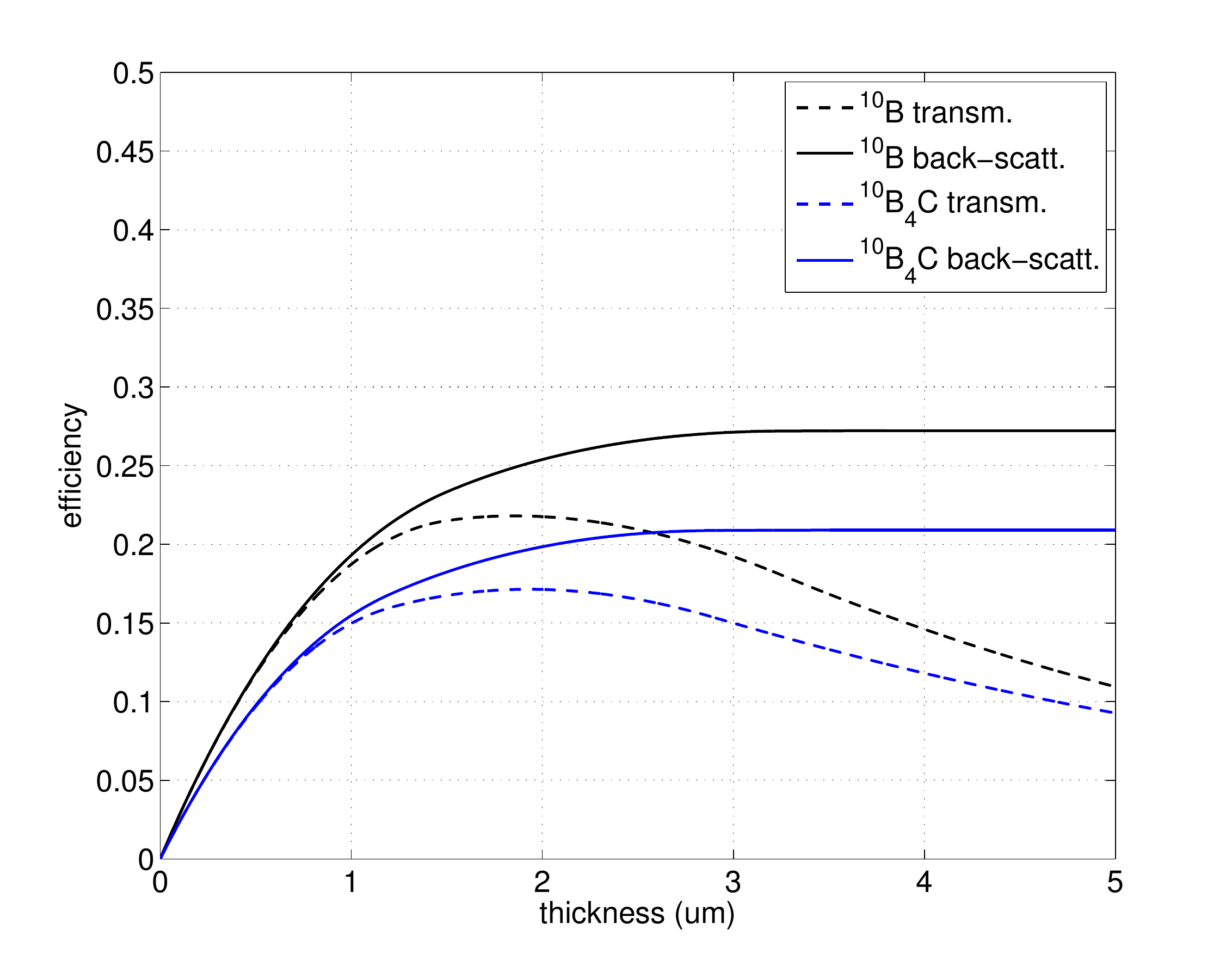}
\includegraphics[width=7.8cm,angle=0,keepaspectratio]{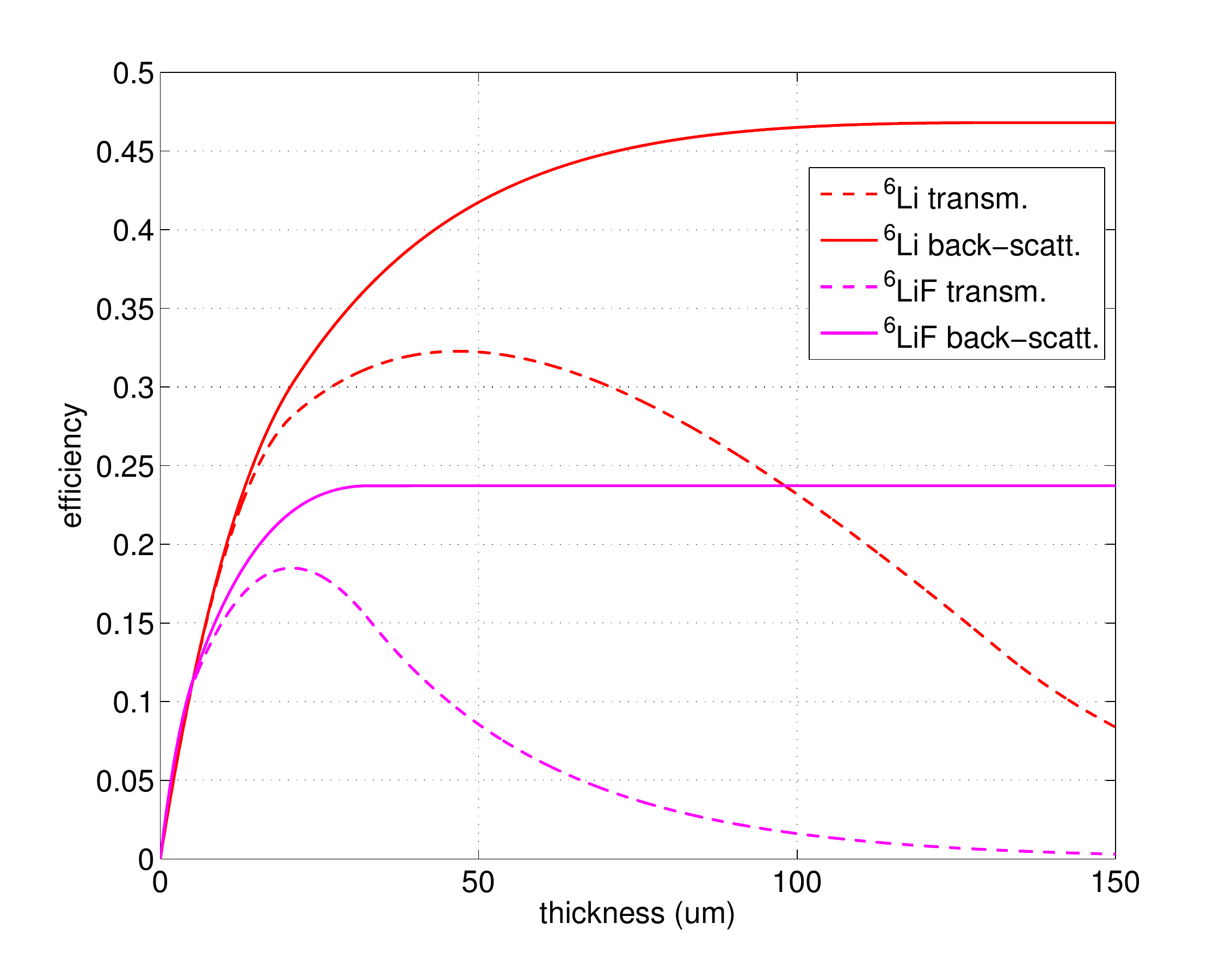}
 \caption{\footnotesize Comparison between the efficiencies of pure $^{10}B$, enriched $^{10}B_4C$, pure $^{6}Li$
 and enriched $^{6}LiF$ plotted as a function of the layer
 thickness. Both efficiencies at $1.8$\AA \, for a single layer in transmission mode
 and in back-scattering mode at $10^{\circ}$ incidence angle are
 plotted. An energy threshold of $100\,KeV$ is applied.} \label{figcompsinglelayersBLiF}
\end{figure}
\\ We compare now $^{10}B_4C$ and $^{6}LiF$. Ranges in $^{6}LiF$
are longer, i.e. have higher escape efficiency; on the contrary, it
is $^{10}B_4C$ that has a higher neutron conversion power because of
its shorter mean free path $\eta$. From only these consideration it
is not easy to figure out which one exhibits the highest detection
efficiency.
\\ A qualitative parameter to evaluate the goodness of a solid
neutron converter is:
\begin{equation}
\chi(\lambda) = \langle R \rangle \cdot \Sigma(\lambda)=\frac{
\langle R \rangle}{\eta(\lambda)}
\end{equation}
where as $\langle R \rangle$ we refer to the average particle range
for a given capture reaction. $\chi$ is a dimensionless number that
takes into account both the power of a material to convert neutrons
of a given wavelength and the ease for the produced particle to
escape. The higher $\chi$, the better is the converter.
$\chi$-values are tabulated in Table \ref{sce6e}.
\begin{table}[!ht]
\centering
\begin{tabular}{|r|c|c|}
\hline \hline
material & $\langle R \rangle (\mu m)$ & $\chi \,at \,1.8$\AA \\
\hline
pure $^{6}Li$  &  $76.5$  & $0.34$  \\
pure $^{10}B$  &  $2.43$  & $0.12$ \\
$96\%$ enriched $^{6}LiF$    &  $19$ & $0.11$ \\
$98\%$ enriched $^{10}B_4C$  & $2.08$  & $0.09$ \\
 \hline \hline
\end{tabular}
\caption{\footnotesize Average range for different neutron
converters and the corresponding $\chi$ calculated for $1.8$\AA. An
energy threshold of $100\,KeV$ is applied.} \label{sce6e}
\end{table}
\\ Figure \ref{figcompsinglelayersBLiF} shows the detection
efficiency for a single converter layer either in transmission mode
or in back-scattering plotted as a function of the film thickness.
We chose $1.8$\AA \, as neutron wavelength and an angle of incidence
of neutrons of $10^{\circ}$. An energy threshold of $100\,KeV$ is
applied. $^{10}B_4C$ and $^{6}LiF$, in back-scattering mode, attain
their maximum efficiency, about $21\%$ and $24\%$ respectively, for
a film of thickness $d=3\, \mu m$ and $d=30\, \mu m$ respectively.
Even $^{6}LiF$ shows a slightly higher efficiency, the fabrication
of layer of $30\,\mu m$ can be an issue in terms of costs and
deposition time unless a single material block is used.
\begin{figure}[!ht]
\centering
\includegraphics[width=7.8cm,angle=0,keepaspectratio]{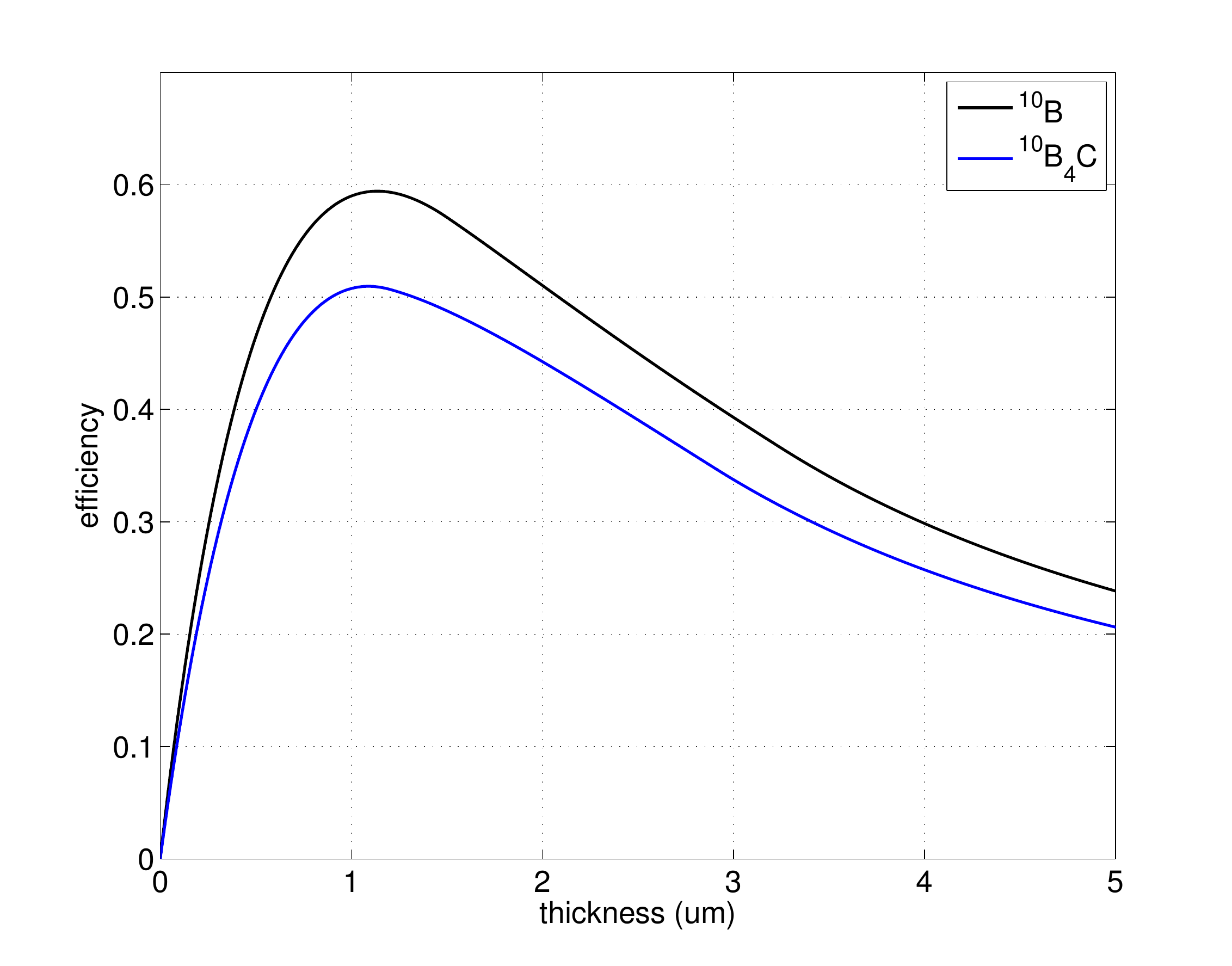}
\includegraphics[width=7.8cm,angle=0,keepaspectratio]{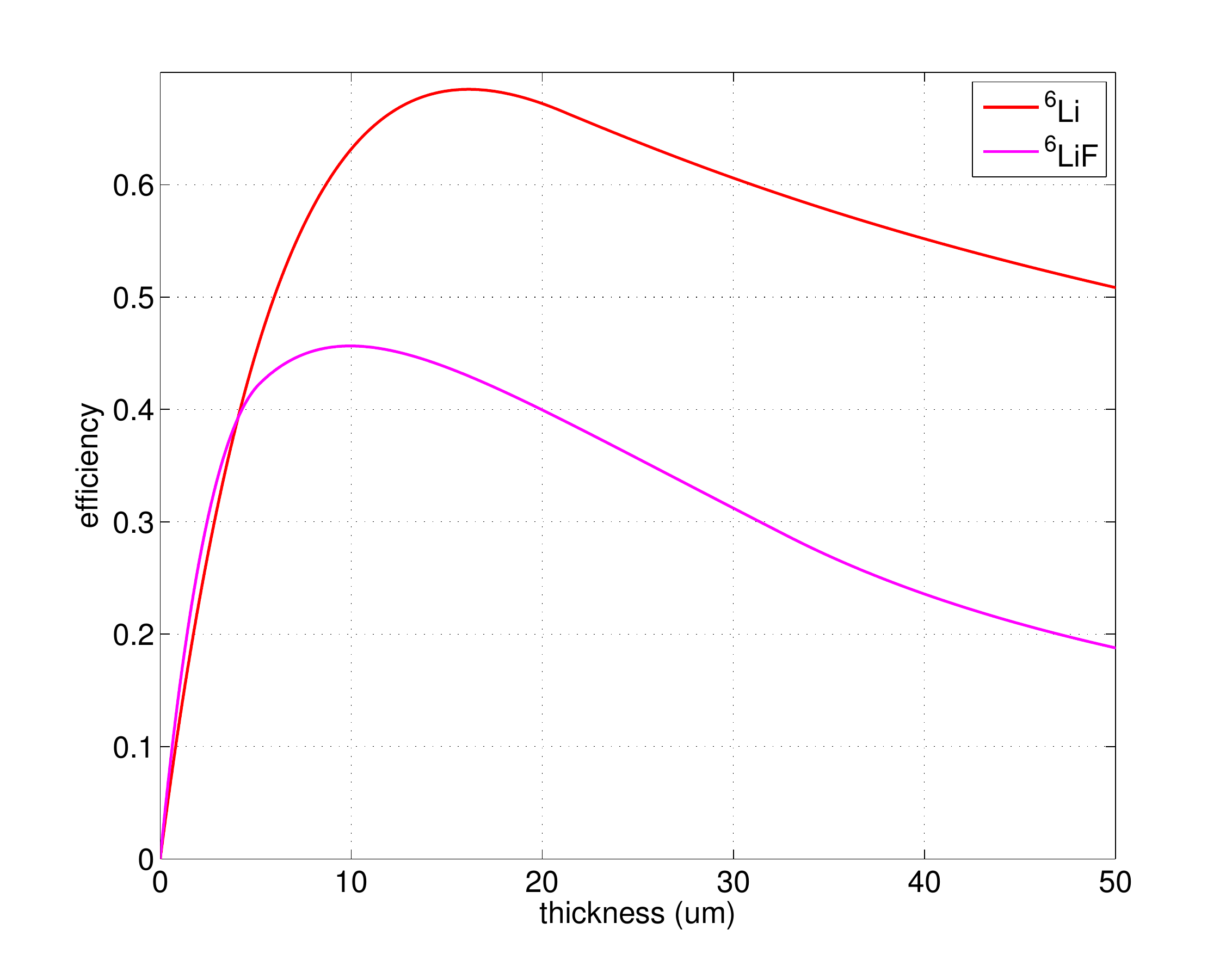}
 \caption{\footnotesize Comparison between the efficiencies of pure $^{10}B$, enriched $^{10}B_4C$, pure $^{6}Li$
 and enriched $^{6}LiF$ plotted as a function of the single layer
 thickness. Efficiencies are calculated at $1.8$\AA \, for a multi-layer detector made up of 30 layers ($N_b=15$ blades)
 at $90^{\circ}$ incidence angle. An energy threshold of $100\,KeV$ is applied.}
\label{figcompsinglelayersBLiF330}
\end{figure}
\\ Figure \ref{figcompsinglelayersBLiF330} shows the efficiency for
a  multi-layer detector composed of $N_b=15$ blades ($30$ layers) as
a function of the single film thickness. \\ This time $^{10}B_4C$
presents a higher efficiency (about $50\%$) with respect to
$^{6}LiF$ (about $45\%$). This makes it the best candidate for a
multi-grid like detector \cite{jonisorma}.
\\ An alternative to exploit the high escape probability of $^{6}Li$
fragments, is to coat a  $^{6}Li$ layer with a very thin layer of
$^{10}B_4C$ that acts as a protection and an additional converter
material. The latter should be optimized to let $^{6}Li$ particles
escape. Moreover, one can make self-supporting layers out of
$^{6}Li$ to be coated with $^{10}B_4C$; it will increase
significantly the neutron detection efficiency.
\section{Theoretical Pulse Height Spectrum calculation}\label{SectThPHSCalc}
The physical model taken into account in \cite{gregor} and in
\cite{salvat} can be used as well to derive the analytical formula
for the Pulse Height Spectra (PHS). A similar work was done in
\cite{salvat} (see Appendix \ref{connfsalvat}) where only Monte
Carlo solutions were shown; here we want to use analytic methods to
understand the structure of the PHS.
\\ We make the approximation mentioned in the Introduction \ref{introchaptheo} and we assume
either a simplified stopping power function (see Section
\ref{strapp33}) or one simulated with SRIM \cite{sri} for the
neutron capture fragments.
\\ Referring to Figure \ref{coorsys}, we calculate the probability
for a particle emitted from the neutron conversion point at certain
depth ($x$ for back-scattering or $d-y$ for transmission) to travel
exactly a distance $L$ on a straight line towards the escape
surface. This distance $L$ is related to the charged particle
remaining energy through the primitive function of the stopping
power. The final electric signal will be proportional to the charge
created by these charged fragments and, thus, to the energy they own
after escaping the layer.
\\ We will demonstrate that even under strong approximations of the
stopping power function the model still predicts quite well the
important physical features of the PHS.
\subsection{Back-scattering mode}\label{backscatt678}
The probability for a neutron to be captured at depth
$\left(x,x+dx\right)$ in the converter layer and for the capture
reaction fragment (emitted isotropically in $4\pi\,sr$) to be
emitted with an angle $\varphi = \arccos(u)$ (between
$\left(u,u+du\right)$) is:
\begin{equation}\label{eqae1}
p(x,u)dx \, du=\begin{cases} \frac 1 2 \cdot \Sigma e^{-\Sigma \cdot x} dx \, du &\mbox{if  \,} x\leq d\\
0 &\mbox{if \, } x > d
\end{cases}
\end{equation}
where $\Sigma=n\cdot\sigma$ with $n$ number density of the
conversion layer and $\sigma$ the neutron absorption cross section;
$d$ is the layer thickness. The factor $\frac 1 2$ takes into
account that half of the time the particle travels toward the layer
substrate and it is therefore lost. \\The fragment will travel a
distance $L$ across the converter layer if $L=\frac{x}{u}$ and it is
formally expressed by a Dirac delta function:
\begin{equation}\label{eqae2}
\delta\left(\frac{x}{u}-L\right)= \frac{x}{L^2} \cdot
\delta\left(u-\frac{x}{L}\right)
\end{equation}
By using the following delta function property $\delta(g(\xi))=
\sum_i \frac{\delta\left(\xi-\xi_i\right)}{\mid g'(\xi_i)\mid} $
where $\xi_i$ are the solutions of $g(\xi)=0$.
\\ The probability for a particle to travel a distance $\left(L,L+dL\right)$ across
the layer is given by:
\begin{equation}\label{eqae3}
\begin{aligned}
P(L)\,dL&= \int_0^d \int_0^1 \delta\left(\frac{x}{u}-L\right) p(x,u)
\, dx du = \frac{\Sigma}{2 L^2} \int_0^d dx \, x e^{-\Sigma \cdot x}
\int_0^1 du \, \delta\left(u-\frac{x}{L}\right)= \\
&= \frac{\Sigma}{2 L^2} \int_0^d dx \, x e^{-\Sigma \cdot x}
\left(H(x)-H(x-L)\right)=\begin{cases} \frac{1}{2 L^2} \left(
\frac{1}{\Sigma}-(\frac{1}{\Sigma}+L)e^{-\Sigma \cdot
L}\right)\,dL &\mbox{if  \,} L \leq d\\
\frac{1}{2 L^2} \left(
\frac{1}{\Sigma}-(\frac{1}{\Sigma}+d)e^{-\Sigma \cdot d}\right)\,dL
&\mbox{if \, } L > d
\end{cases}
\end{aligned}
\end{equation}
where $H$ is the Heaviside step function.
\\ It is sufficient to replace $\Sigma$ with $\frac{\Sigma}{\sin(\theta)}$ if neutrons
hit the layer under the angle $\theta$ with respect to the surface
(see Figure \ref{coorsys}). In the PHS calculation $p(x,u)$ has to
be changed as follows:
\begin{equation}\label{eqae4}
p(x,u,\theta)dx \, du=\begin{cases} \frac 1 2 \cdot \Sigma e^{-\Sigma \cdot \frac{x}{\sin{(\theta)}}} \frac{dx}{\sin{(\theta)}} \, du &\mbox{if  \,} x\leq d\\
0 &\mbox{if \, } x > d
\end{cases}
\end{equation}
The demonstration is identical to the one shown in Section
\ref{secttheoeff} for the efficiency calculation.
\\ Referring to Figure \ref{figusigmagross} in Section \ref{secttheoeff}, this
result means that the same PHS can be obtained, for example, at
$10$\AA \, under an angle of $80^{\circ}$ or equivalently at $5$\AA
\, under an angle of $30^{\circ}$. If, for example, one is
interested in measuring some PHS for a given neutron incidence angle
and only a monochromatic beam is available, from Figure
\ref{figusigmagross} it is possible to get the effect of having a
different wavelength by changing the inclination instead. Every PHS
measured on each equipotential line in Figure \ref{figusigmagross}
is the same.
\\ If $E(L)$ is the remaining energy of a particle that has traveled a distance
$L$ into the layer, $\frac{dE(L)}{dL}$ is the stopping power or
equivalently the Jacobian of the coordinate transformation between
$L$ and $E$.
\\ Once $P(L)dL$ is known we can calculate $Q(E)dE$, therefore:
\begin{equation}\label{eqae5}
P(L)dL=P(L(E))\cdot \left|\frac{dL}{dE}\right| \cdot dE =
P(L(E))\cdot \frac{1}{\left|\frac{dE}{dL}\right|} \cdot dE
\end{equation}
Therefore:
\begin{equation}\label{eqae6}
Q(E)dE= P(L(E))\cdot \frac{1}{\left|\frac{dE}{dL}\right|} \cdot dE
\end{equation}
where $Q(E)dE$ is the probability that an incident neutron will give
rise to a release of an energy $\left(E,E+dE\right)$ in the gas
volume; hence it is the analytical expression for the PHS.
\subsubsection{PHS calculation using SRIM output files for Stopping Power}\label{SRIMsp33}
\begin{figure}[!ht]
\centering
\includegraphics[width=10cm]{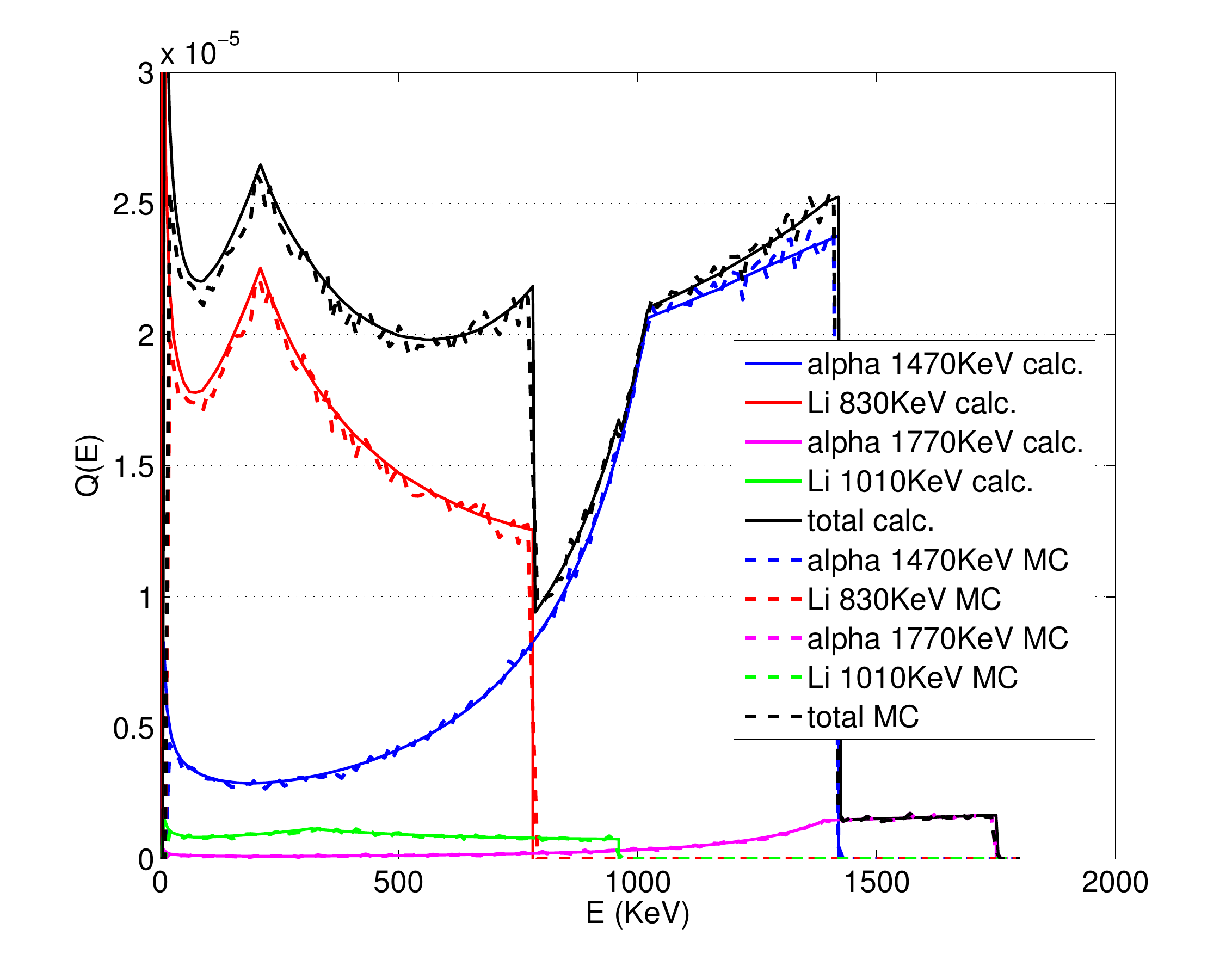}
\caption{\footnotesize PHS calculated and MC simulated for $1\,\mu
m$ back-scattering layer at $1.8$\AA \,and $90^{\circ}$ incidence.}
\label{phscfrMC}
\end{figure}

\begin{figure}[!ht]
\centering
\includegraphics[width=10cm]{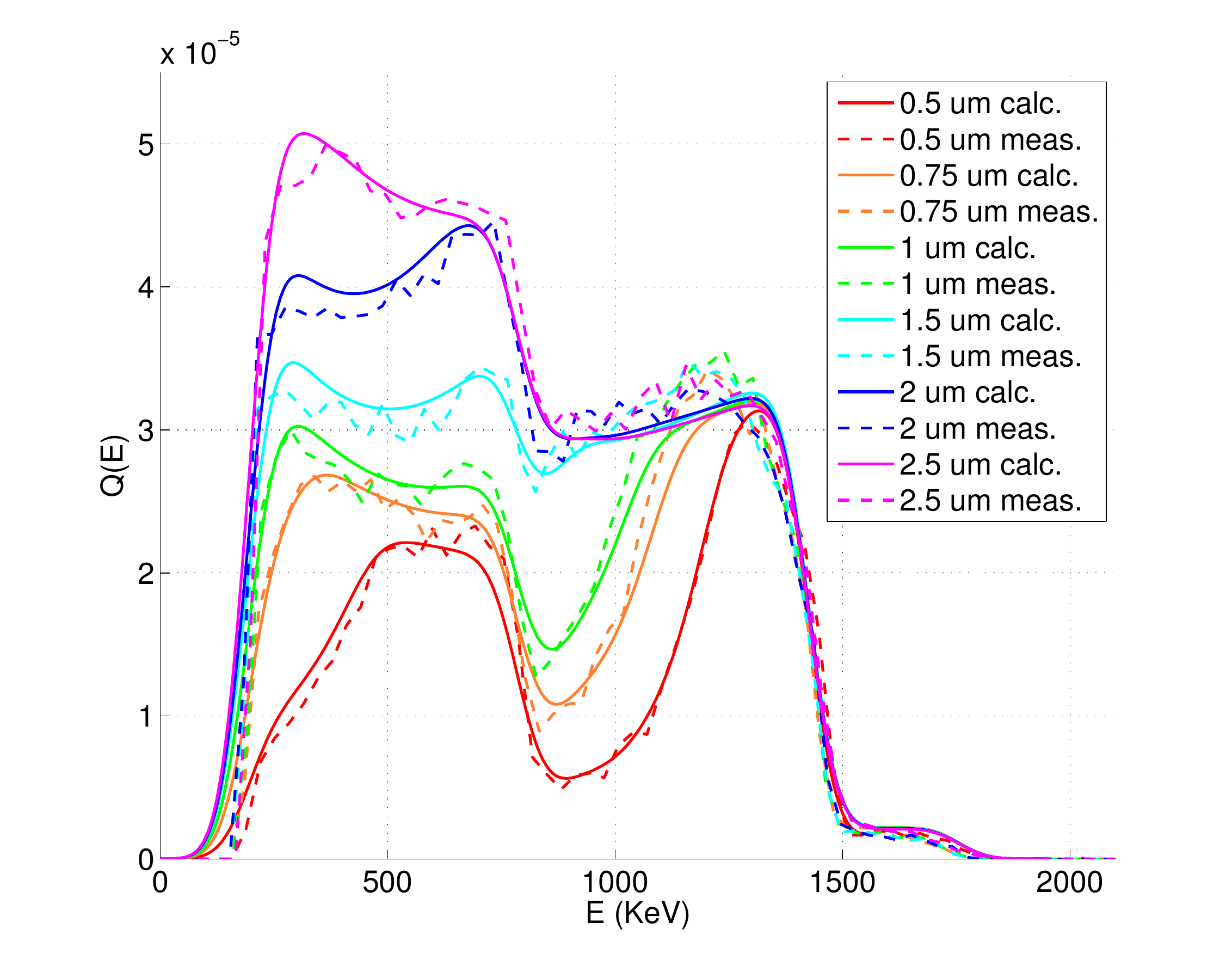}
\caption{\footnotesize Comparison between a PHS calculated and one
measured at ILL-CT2 on a $2.5$\AA \, neutron beam using a Multi-Grid
like detector \cite{jonisorma} where were mounted blades of
different thicknesses.} \label{phscfrMeas}
\end{figure}

We take the case of the $^{10}B$ reaction as example, however
results can be applied to any solid neutron converter. We recall the
energies carried for the $94\%$ branching ratio is $E_0=1470KeV$ for
the $\alpha$-particle and $E_0=830KeV$ for the $^7Li$; for the $6\%$
branching ratio, $E_0=1770KeV$ for the $\alpha$-particle and
$E_0=1010KeV$ for the $^7Li$. Referring to Equation \ref{eqae6}, the
stopping power $\frac{dE}{dL}$ used here was simulated with SRIM
\cite{sri} (see Figure \ref{stpowapprox45}) and $L(E)$ obtained by
numerical inversion of the stopping power primitive function, i.e.
the remaining energy inverse function.
\\ The full PHS that takes into account the full process can be obtained
by adding the four PHS in the case of $^{10}B$ according to the
branching ratio probability:
\begin{equation}\label{eqae7}
\begin{array}{ll}
Q_{tot}(E)dE &= \left(0.94 \cdot \left(Q_{\alpha \,
1470KeV}(E)+Q_{^7Li \,830KeV}(E)\right)\right.+
\\ & +0.06 \cdot \left.\left(Q_{\alpha \,1770KeV}(E)+Q_{^7Li \,1010KeV}(E)\right)\right)\cdot dE
\end{array}
\end{equation}
Consequently the efficiency for a single layer can be calculated by:
\begin{equation}\label{eqae8}
\varepsilon\left(E_{Th}\right)= \int_{E_{Th}}^{+\infty} Q_{tot}(E)dE
\end{equation}
where $E_{Th}$ is the energy threshold applied to cut the PHS. This
result is fully in agreement with what can be calculated by using
the Equations derived in Section \ref{secttheoeff}.
\\ In order to confirm our derived formulae a Monte Carlo simulator
has been developed taking into account the same physical model
exploited in this section. A random number generator simulates the
probability for a neutron to be absorbed a certain depth in the
conversion layer and with a random emission angle for the fragment.
Using the SRIM files, we calculate the remaining energy after a
straight path inside the layer for the fragment in question; this is
the energy released into the gas volume. \\ Figure \ref{phscfrMC}
shows the result of Equation \ref{eqae7} for the four single
particles and the total PHS compared with the Monte Carlo PHS, for
$1\,\mu m$ single back-scattering layer at $1.8$\AA \, and
$90^{\circ}$ incidence.
\\ In order to check in practice the formulae derived, a direct
measure on the neutron beam is necessary. A Multi-Grid-like detector
\cite{jonisorma} was used to collect the data we are going to show
here. The data was collected on CT1 (Canal Technique 1) at the ILL
where a monochromatic neutron beam of $2.5$\AA \, is available. This
particular detector has the peculiarity that in each of its frames
blades of different thickness coating were mounted; as a result the
simultaneous PHS measurement for different layer thicknesses has
been possible. The blades are made up of an Aluminium substrate of
$0.5\,mm$ thickness coated \cite{carina} on both sides by an
enriched $^{10}B_4C$ layer. Thicknesses available in the detector
were: $0.50\,\mu m$, $0.75\,\mu m$, $1\,\mu m$, $1.5\,\mu m$,
$2\,\mu m$ and $2.5\,\mu m$.
\\ In our calculation we are not taking into account several processes,
such as wall effects, gas amplification and fluctuations, space
charge effects, electronic noise, etc. but only the neutron
conversion and the fragment escape. Moreover, while the calculation
has an infinite energy precision, this is not the case on a direct
measurement because many processes give a finite energy resolution.
\\ In order to be able to compare calculations and measurements,
after the PHS were calculated for the thicknesses listed above, they
were convoluted with a gaussian filter of $\sigma=10\,KeV$. The
measured PHS were normalized to the maximum energy yield
($1770\,KeV$). An energy threshold of $180\,KeV$ was applied to the
calculation to cut the spectrum at low energies at the same level
the measured PHS was collected.
\\ We compare calculated and measured PHS in Figure \ref{phscfrMeas}; we can
conclude that the model gives realistic results in sufficient
agreement with the experimental ones, to be able to describe its
main features.

\subsubsection{PHS calculation using a strong approximation}\label{strapp33}
\begin{figure}[!ht]
\centering
\includegraphics[width=7.8cm,angle=0,keepaspectratio]{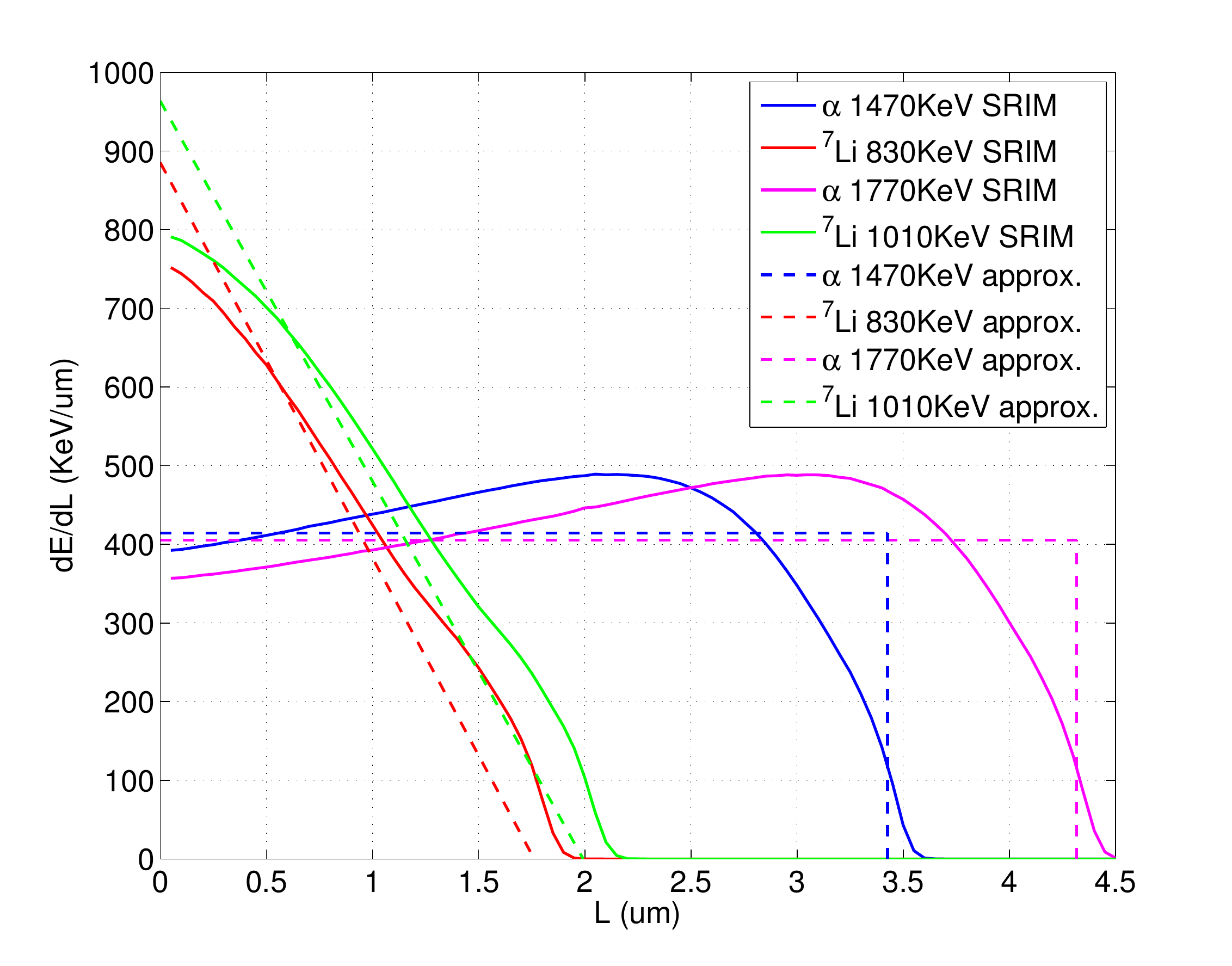}
\includegraphics[width=7.8cm,angle=0,keepaspectratio]{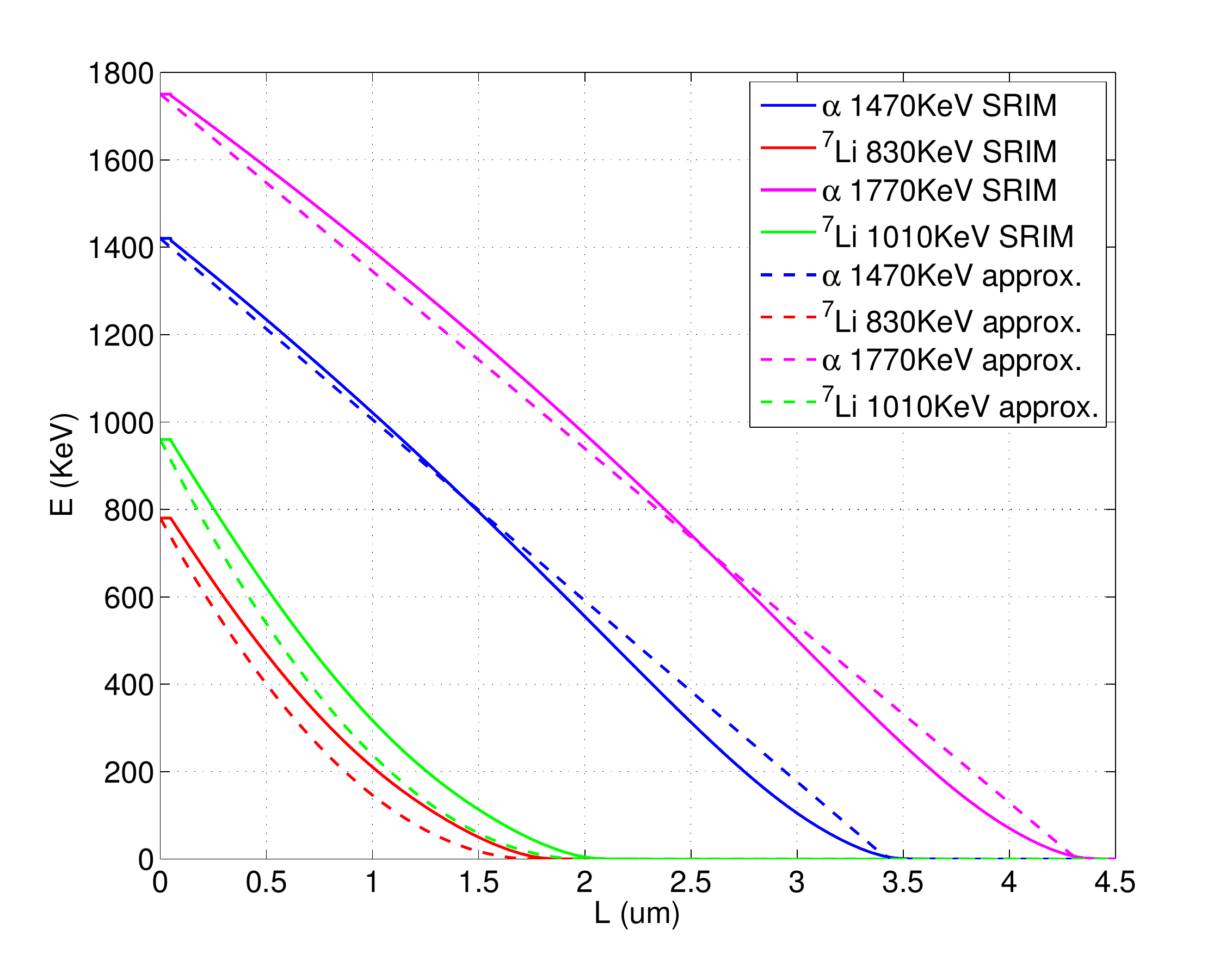}
 \caption{\footnotesize Stopping power and its primitive $E(L)$ for $^{10}B$-reaction fragments,
 solid curves are for functions obtained from SRIM, dashed lines are the approximated behaviors in the
 Equations \ref{eqae9} and \ref{eqae10}.} \label{stpowapprox45}
\end{figure}

\begin{figure}[!ht]
\centering
\includegraphics[width=7.8cm,angle=0,keepaspectratio]{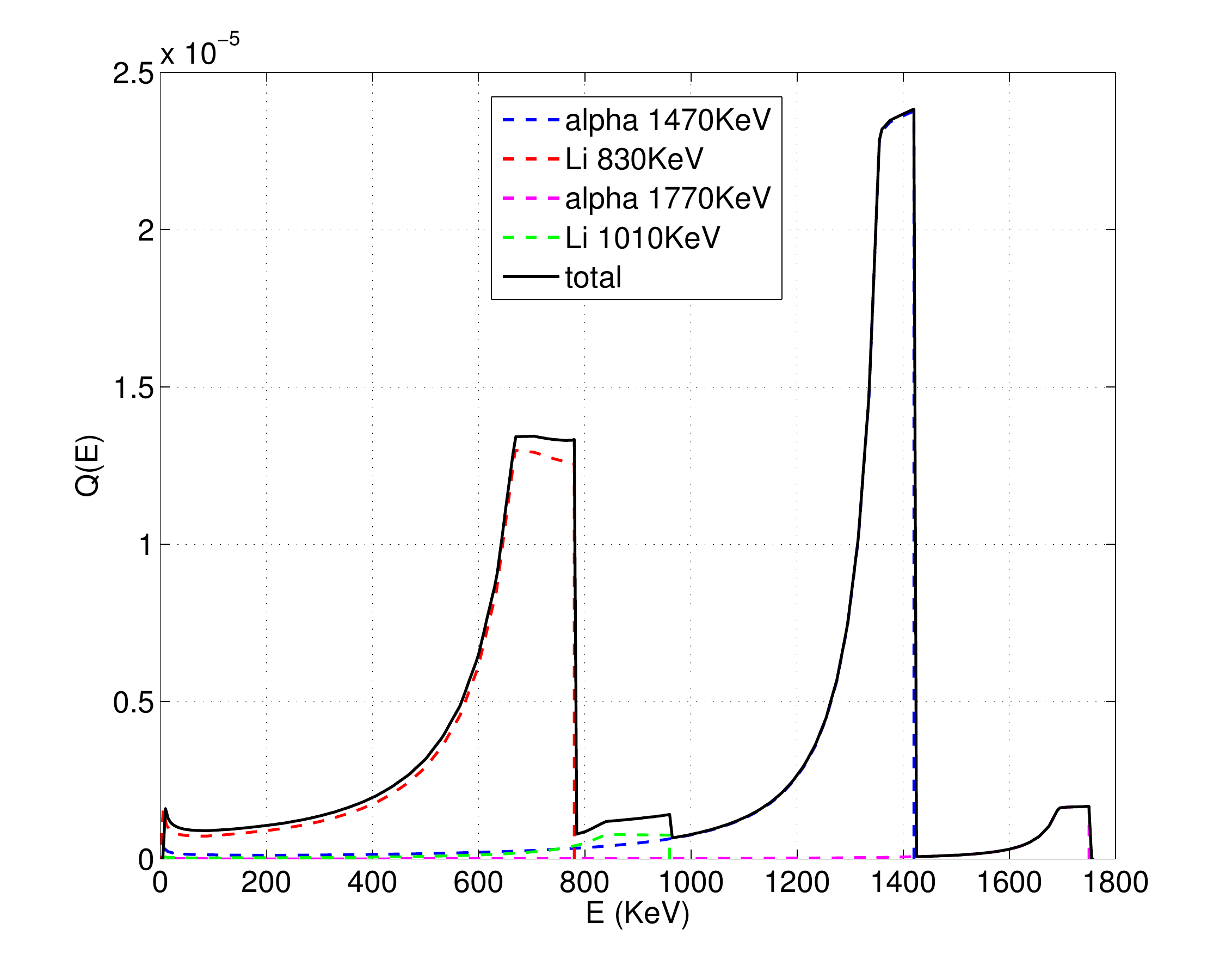}
\includegraphics[width=7.8cm,angle=0,keepaspectratio]{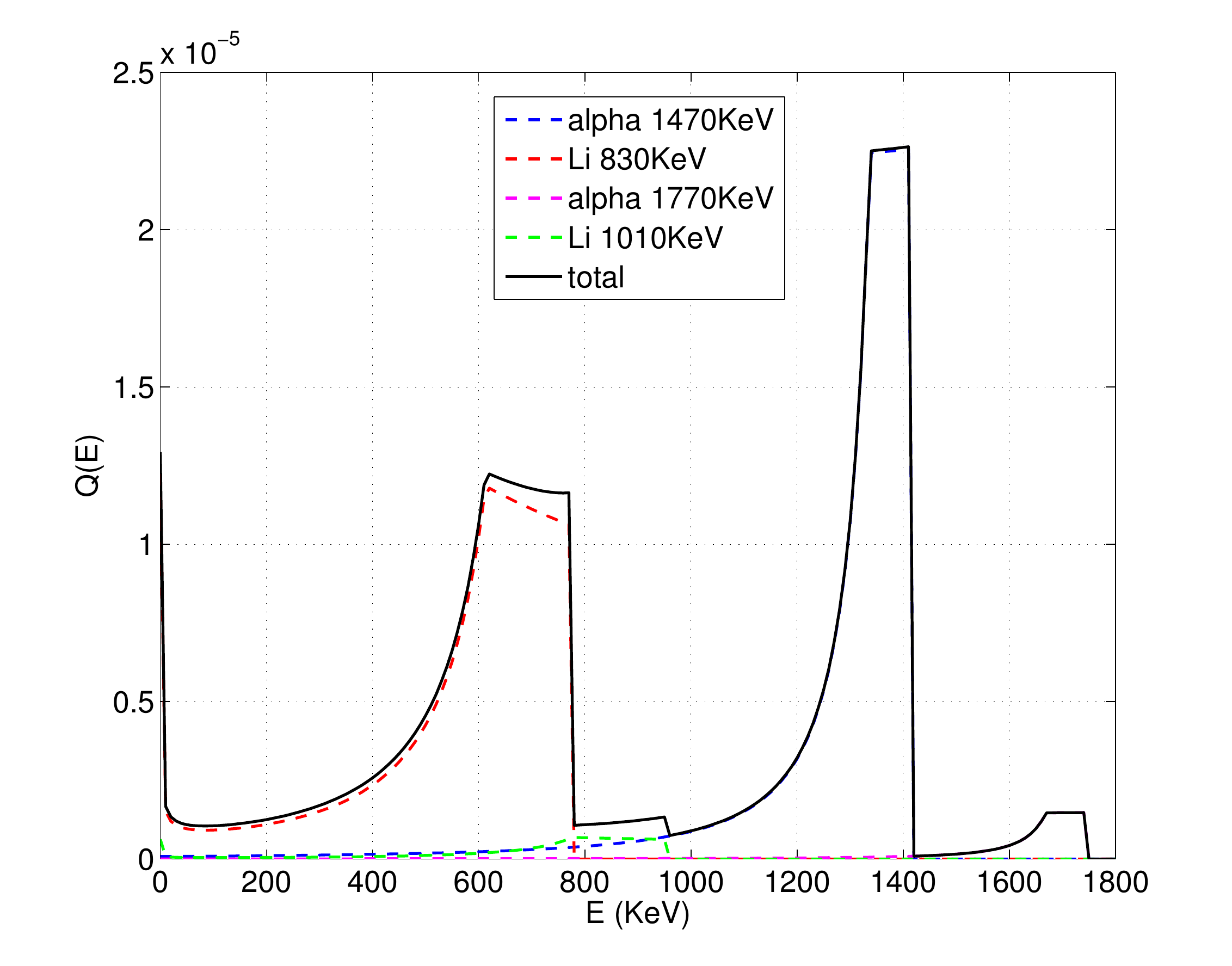}
 \caption{\footnotesize Calculated PHS using SRIM (left) and
 approximated (right) stopping power functions for a single back-scattering layer of $0.2\, \mu m$
 for $1.8$\AA \, and $90^{\circ}$ incidence.} \label{figPHSST401}
\end{figure}

\begin{figure}[!ht]
\centering
\includegraphics[width=7.8cm,angle=0,keepaspectratio]{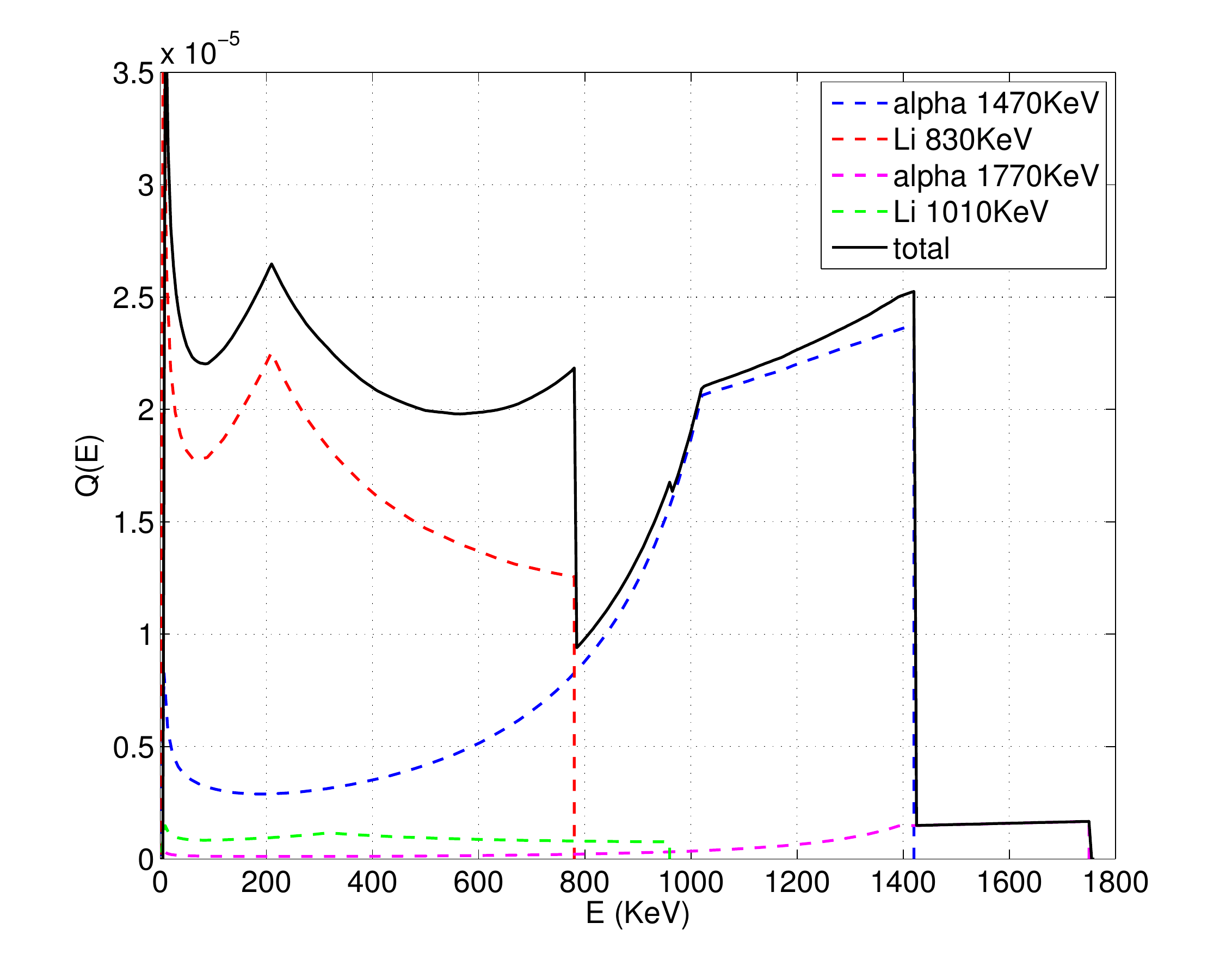}
\includegraphics[width=7.8cm,angle=0,keepaspectratio]{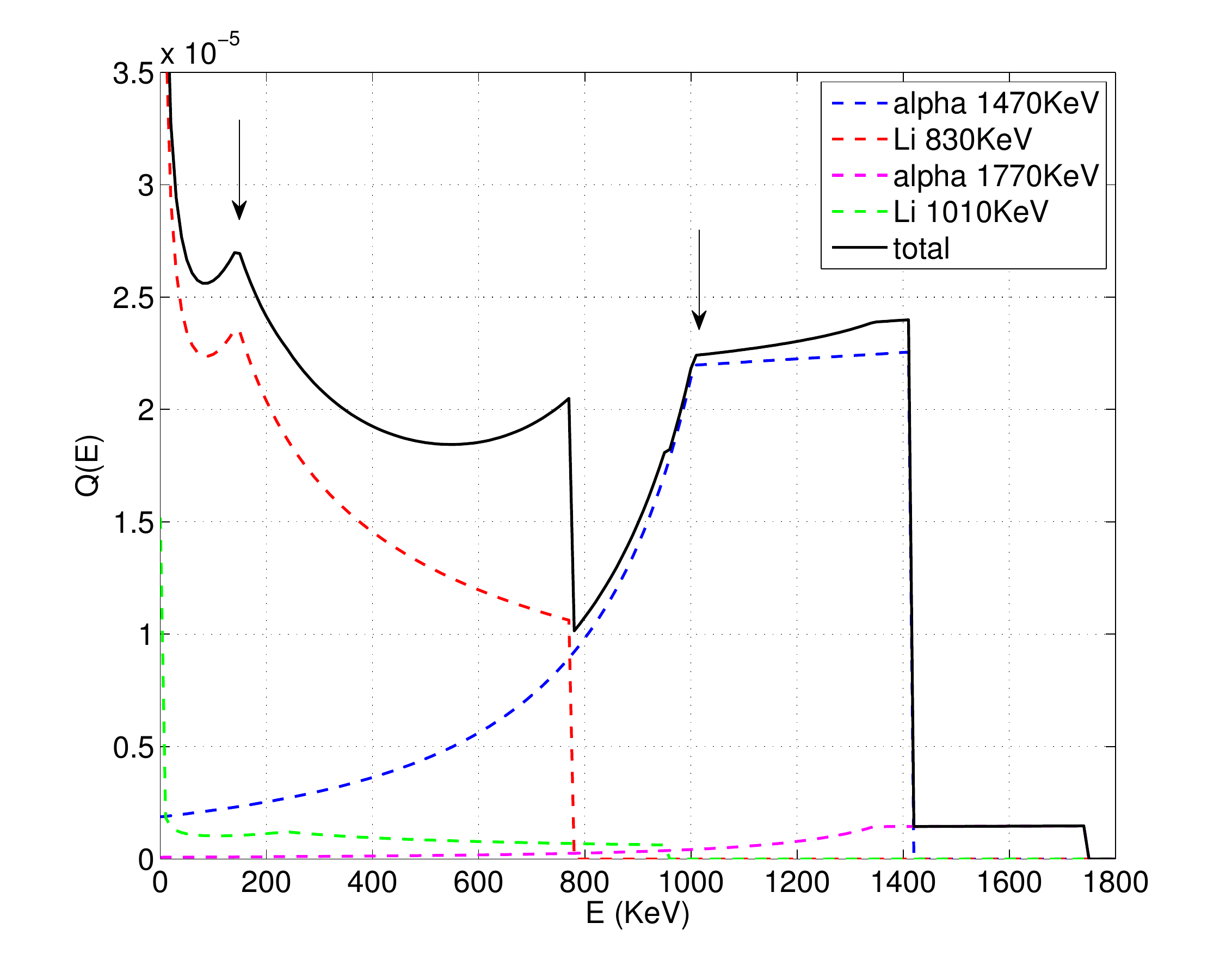}
 \caption{\footnotesize Calculated PHS using SRIM (left) and
 approximated (right) stopping power functions for a single back-scattering layer of $1\, \mu m$
 for $1.8$\AA \, and $90^{\circ}$ incidence.} \label{figPHSST402}
\end{figure}

\begin{figure}[!ht]
\centering
\includegraphics[width=7.8cm,angle=0,keepaspectratio]{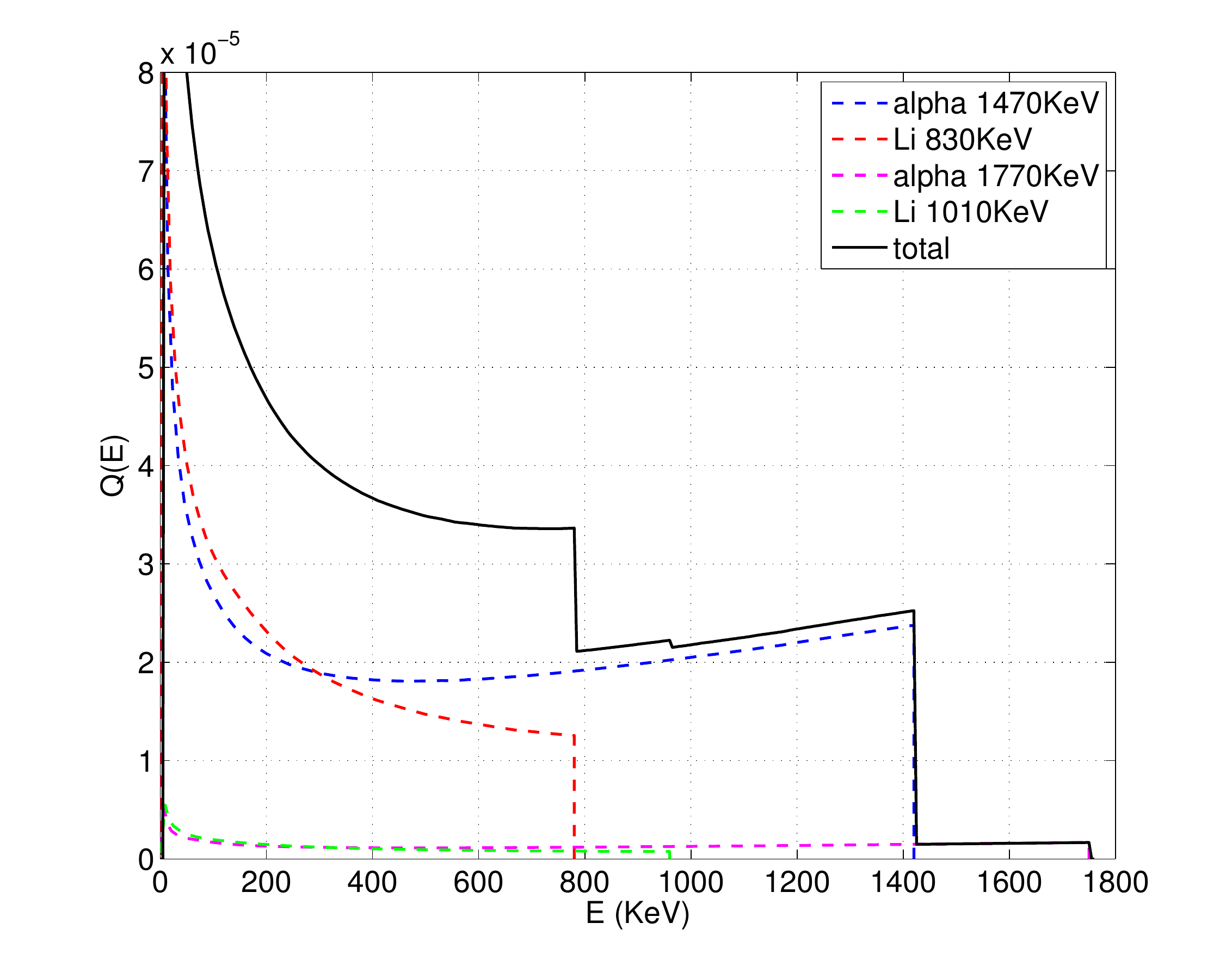}
\includegraphics[width=7.8cm,angle=0,keepaspectratio]{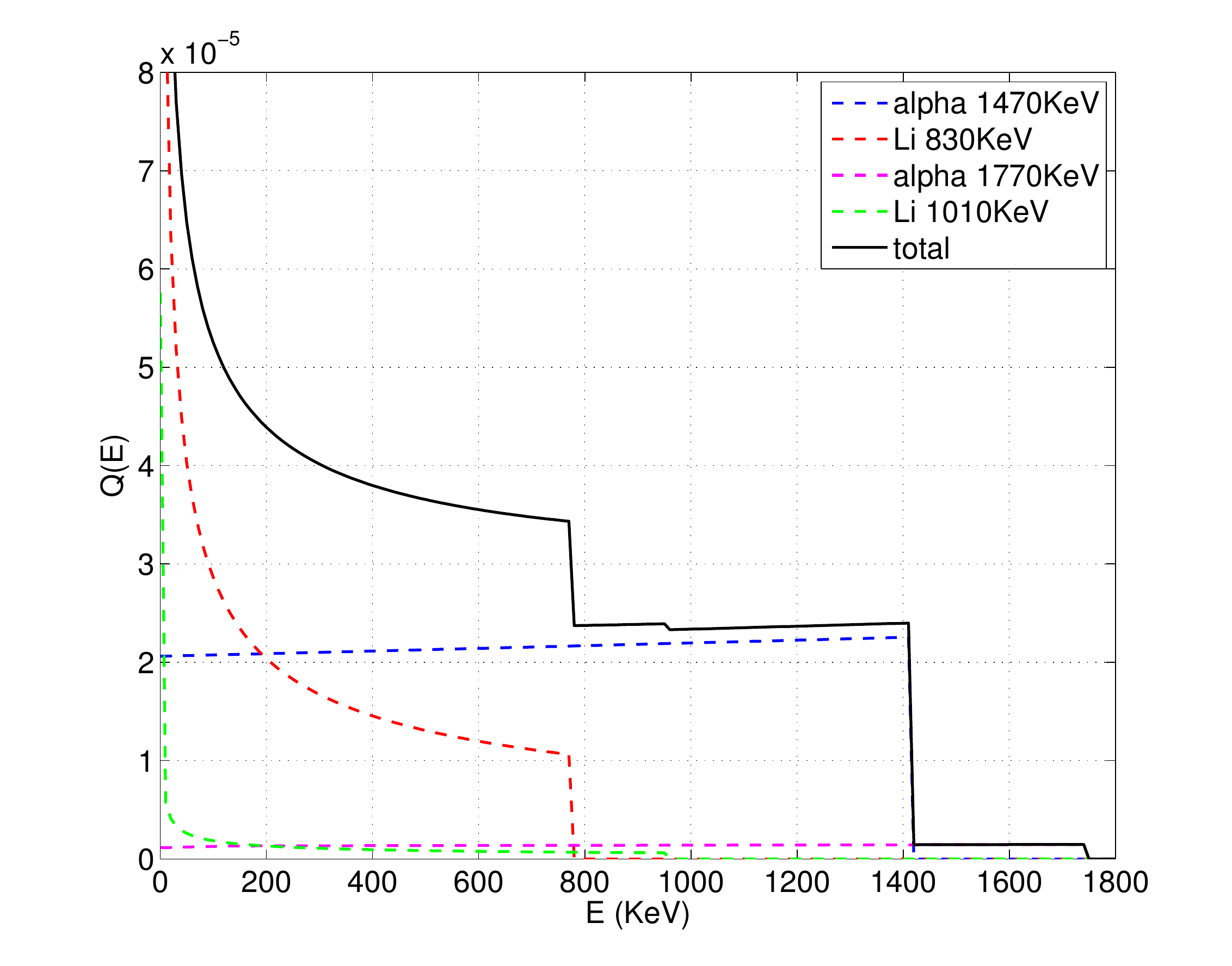}
 \caption{\footnotesize Calculated PHS using SRIM (left) and
 approximated (right) stopping power functions for a single back-scattering layer of $4\, \mu m$
 for $1.8$\AA \, and $90^{\circ}$ incidence.} \label{figPHSST403}
\end{figure}

\begin{figure}[!ht]
\centering
\includegraphics[width=7cm,angle=0,keepaspectratio]{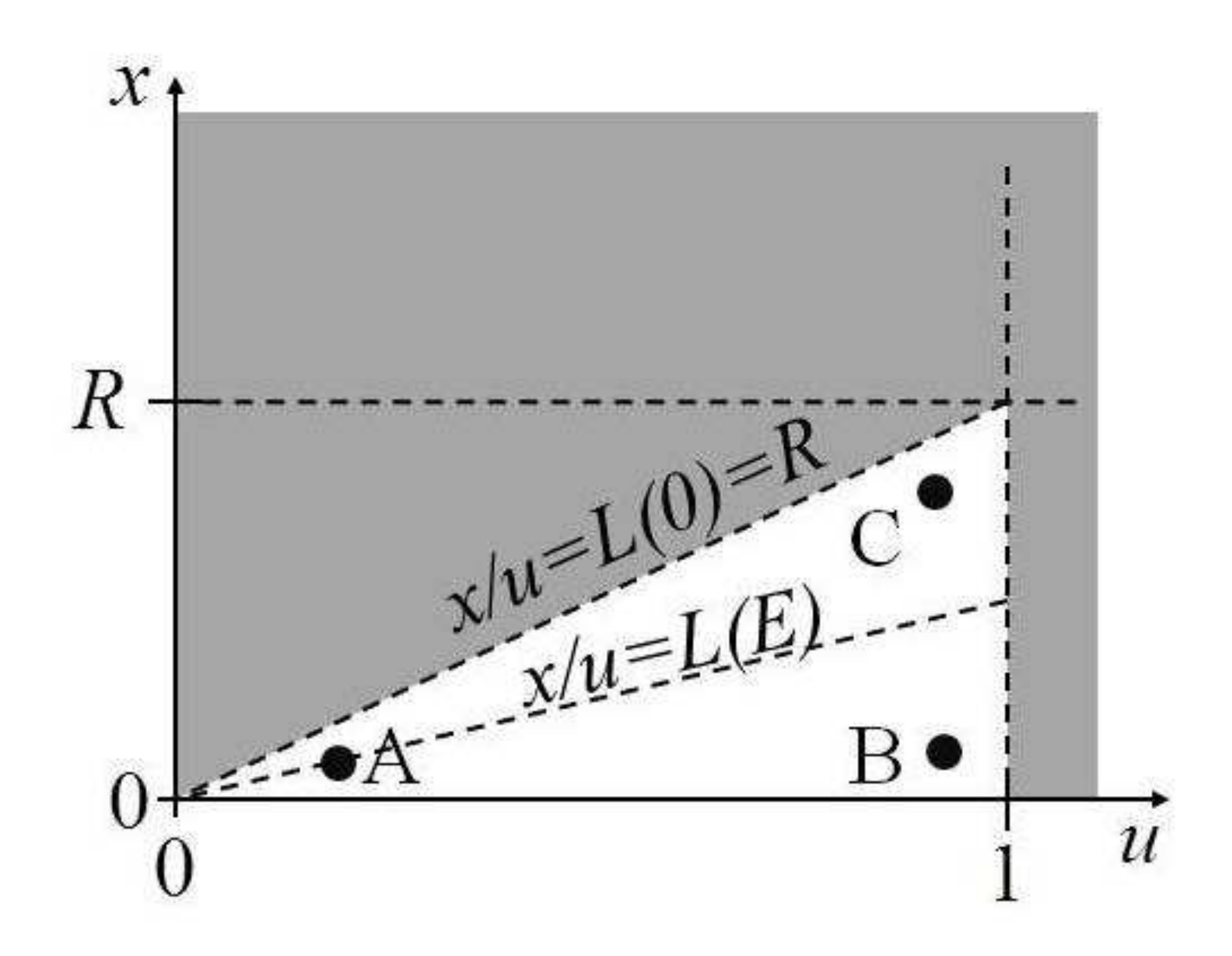}
\includegraphics[width=7cm,angle=0,keepaspectratio]{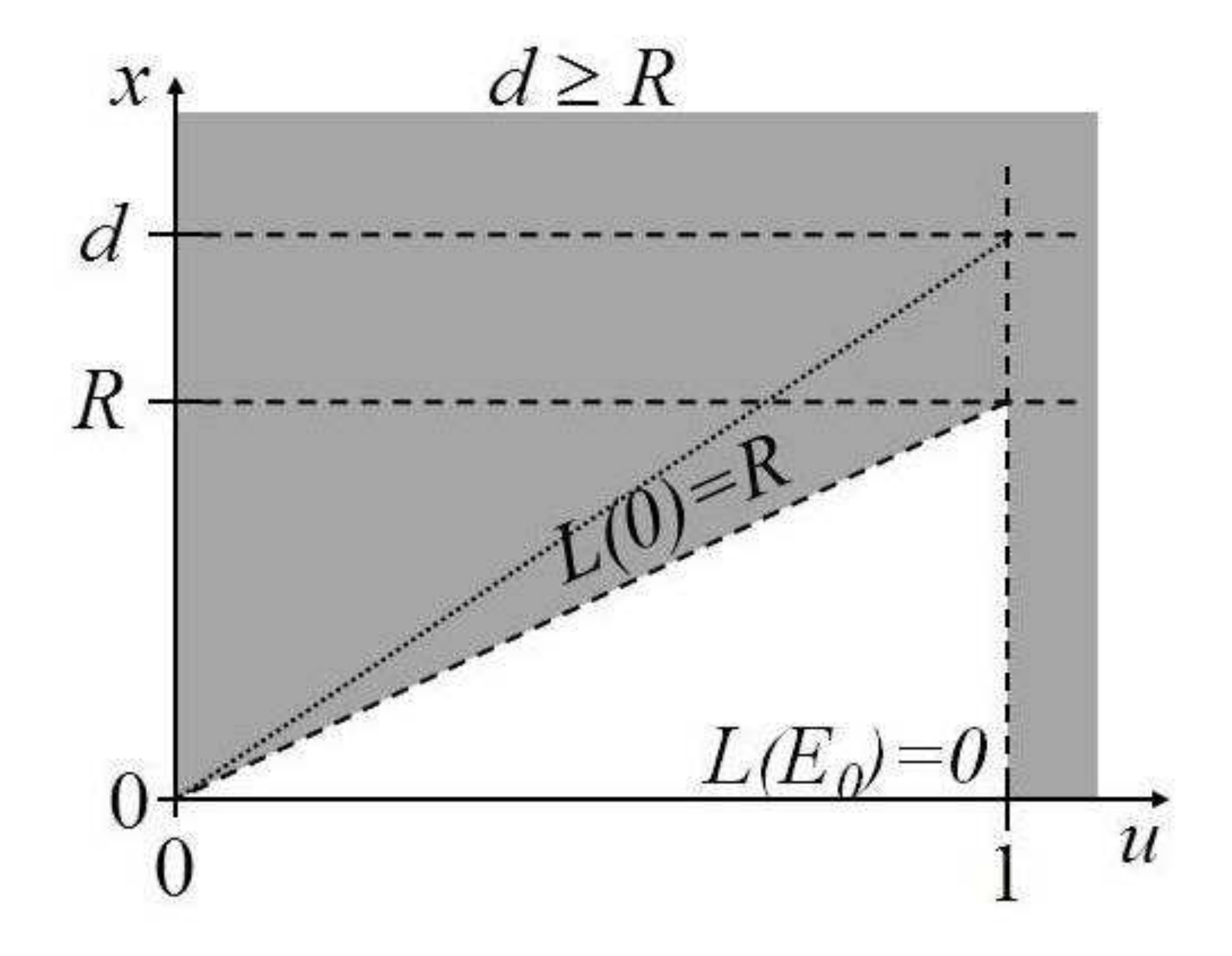}
 \caption{\footnotesize PHS \emph{variable space} and PHS \emph{variable space} when $d \geq R$.} \label{vs34561}
\end{figure}

A fully analytical result that does not appeal to experimental or
SRIM-calculated stopping power functions can be useful to understand
the PHS structure and to determine its properties.
\\ The stopping power functions $\frac{dE}{dL}$ can be
approximated by a constant in the case of an $\alpha$-particle and
with a linear dependency in $L$ for a $^7Li$-ion. As a result the
energy dependency as a function of the traveled distance $L$ is
given by:
\begin{equation}\label{eqae9}
E_{\alpha}(L)=\begin{cases} -\frac{E_0}{R} \left( L-R\right) &\mbox{if  \,} L\leq R\\
0 &\mbox{if \, } L > R
\end{cases}
\end{equation}
And equivalently for the $^7Li$-fragment:
\begin{equation}\label{eqae10}
E_{Li}(L)=\begin{cases} \frac{E_0}{R^2}\left( L-R\right)^2 &\mbox{if  \,} L\leq R\\
0 &\mbox{if \, } L > R
\end{cases}
\end{equation}
\\ Where $R$ is the particle range and $E_0$ its initial energy.
\\ In Figure \ref{stpowapprox45} are shown the stopping power
functions $\frac{dE}{dL}$ for $^{10}B$-reaction fragments and their
primitive $E(L)$, in the case of using SRIM (solid lines) and in the
case we use the expression displayed in the Equations \ref{eqae9}
and \ref{eqae10} (dashed lines).
By substituting Expressions \ref{eqae9} and \ref{eqae10} into the
Equation \ref{eqae6} we obtain a fully analytical formula for the
PHS. It has to be pointed out that each relation, valid for $L \leq
R$, is valid in the range $E \leq E_0$. Hence Equations \ref{eqae11}
and \ref{eqae12} hold for $E \leq E_0$. The two formulations in
Equation \ref{eqae3} for $L\leq d$ and $L > d$, translate in two
different analytical expressions for $Q(E)$ for $E<E^{*}$ and for
$E\geq E^{*}$, with $d=L(E^{*})$. For the $\alpha$-particle:
\begin{equation}\label{eqae11}
\begin{aligned}
Q(E)\,dE=\begin{cases} \frac{1}{2E_0R\left(1-\frac{E}{E_0}\right)^2}
\left( \frac{1}{\Sigma}-(\frac{1}{\Sigma}+d)e^{-\Sigma
d}\right)\,dE & \mbox{if  \,} E < E_0\left(1-\frac d R\right)\\
\frac{1}{2E_0R\left(1-\frac{E}{E_0}\right)^2} \left(
\frac{1}{\Sigma}-\left(\frac{1}{\Sigma}+R\left(1-\frac{E}{E_0}\right)\right)
\cdot e^{-\Sigma R\left(1-\frac{E}{E_0}\right)}\right)\,dE &
\mbox{if \, } E \geq E_0\left(1-\frac d R\right)
\end{cases}
\end{aligned}
\end{equation}
\\ Where the relation $E^{*} = E_0\left(1-\frac d R\right)$ is derived from $d=L(E^{*})$.
\\ For the $^7Li$:
\begin{equation}\label{eqae12}
\begin{aligned}
Q(E)\,dE=\begin{cases}
\frac{1}{4E_0R\sqrt{\frac{E}{E_0}}\left(1-\sqrt{\frac{E}{E_0}}\right)^2}
\left( \frac{1}{\Sigma}-(\frac{1}{\Sigma}+d)e^{-\Sigma \cdot
d}\right)\,dE &\mbox{if  \,} E < E_0\left(1-\frac d R\right)^2\\
\frac{1}{4E_0R\sqrt{\frac{E}{E_0}}\left(1-\sqrt{\frac{E}{E_0}}\right)^2}
\left(\frac{1}{\Sigma}-\left(\frac{1}{\Sigma}+R\left(1-\sqrt{\frac{E}{E_0}}\right)\right)\right.
\cdot \\
\left.\cdot e^{-\Sigma
R\left(1-\sqrt{\frac{E}{E_0}}\right)}\right)\,dE &\mbox{if \, } E
\geq E_0\left(1-\frac d R\right)^2
\end{cases}
\end{aligned}
\end{equation}
\\ Where, again, the relation $E^{*} = E_0\left(1-\frac d R\right)^2$ is
derived from the condition $d=L(E^{*})$.
\\ Figures \ref{figPHSST401}, \ref{figPHSST402} and
\ref{figPHSST403} show the calculated PHS obtained by using the SRIM
stopping power functions and the approximated one displayed in the
Expression \ref{eqae9} and \ref{eqae10} for $0.2\, \mu m$, $1\, \mu
m$ and $4\, \mu m$ respectively, when neutrons hit at $90^{\circ}$
the surface and their wavelength is $1.8$\AA. They show similar
shapes that differ in some points; e.g. focusing on the $1470\,KeV$
$\alpha$-particle, the fact that the approximated $E(L)$ function
(see Figure \ref{stpowapprox45}) differs from the SRIM one at high
$L$ leads to a disappearance of the PHS rise at low energies; it is
clearly visible in Figure \ref{figPHSST403}. We see that as $d$
increases what looked like a single peak splits into two peaks.
While one peak stays constant at the highest fragment energy $E_0$
the second one moves toward lower energies when the layer thickness
increases. This is important when trying to improve the neutron to
gamma-rays discrimination by creating a valley that separates them
in amplitude.
\\ In order to understand the PHS structure, we define the PHS
\emph{variable space}: on the abscissa axis is plotted $u =
\cos(\varphi)$, where $\varphi$ is the angle the fragment has been
emitted under, and, on the ordinates axis, is plotted the neutron
absorption depth $x$. $u \in \left[0,1 \right]$; $x \in \left[0,d
\right]$ if $d<R$ or $x \in \left[0,R \right]$ if $d \geq R$ because
a neutron can only be converted inside the layer and, on the other
hand, if a neutron is converted too deep into the layer, i.e. $x>R$
no fragments can escape whatever the emission angle would be. In
Figure \ref{vs34561}, on its left, the variable space is shown; an
event in the A-position would be a fragment that was generated by a
neutron converted at the surface of the layer and escape the layer
at grazing angle. An event in the position B represents a fragment
that escapes orthogonally to the surface and its neutron was
converted at the surface. An event in C means a neutron converted
deep into the layer with an orthogonal escaping fragment. The
straight lines $x=L(E) \cdot u$ characterize the events with
identical escape energy $E$, that contribute to the same bin in the
PHS. The straight line characterized by $x=R \cdot u$ is the horizon
for the particles that can escape the layer and release some energy
in the gas volume. To be more precise events that give rise to the
zero energy part of the PHS lie exactly on the line $\frac x u
=L(E=0)=R$ because they have traveled exactly a distance $R$ in the
converter material. On the other hand, events that yield almost the
full particle energy $E_0$, will lie on the line identified by
$\frac x u=L(E=E_0)=0$.
\\ The events that generate the PHS have access to a
region, on the variable space, identified by a triangle below the
straight line $x=R \cdot u$ (see Figure \ref{vs34561}).
\begin{figure}[!ht]
\centering
\includegraphics[width=7cm,angle=0,keepaspectratio]{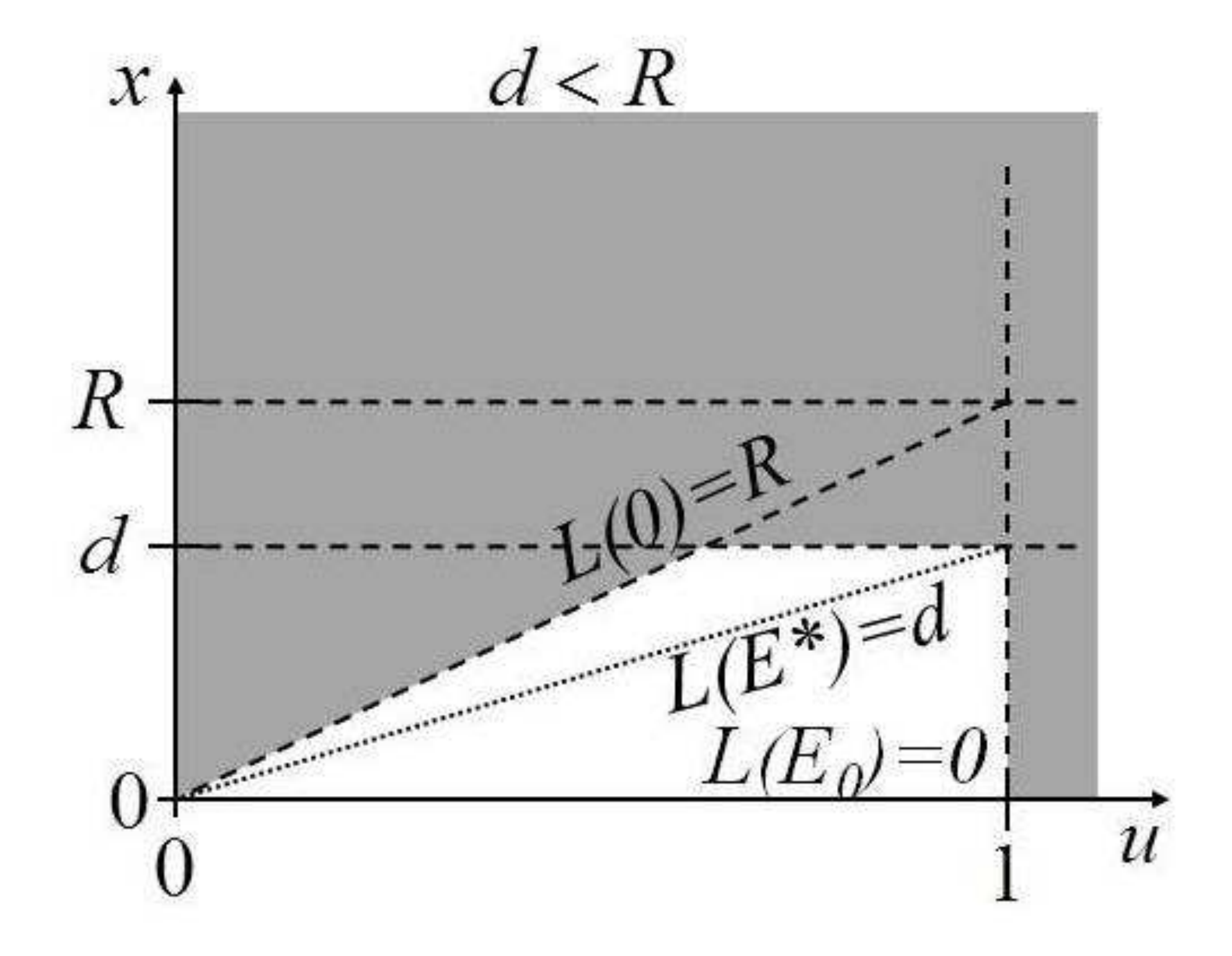}
\includegraphics[width=7cm,angle=0,keepaspectratio]{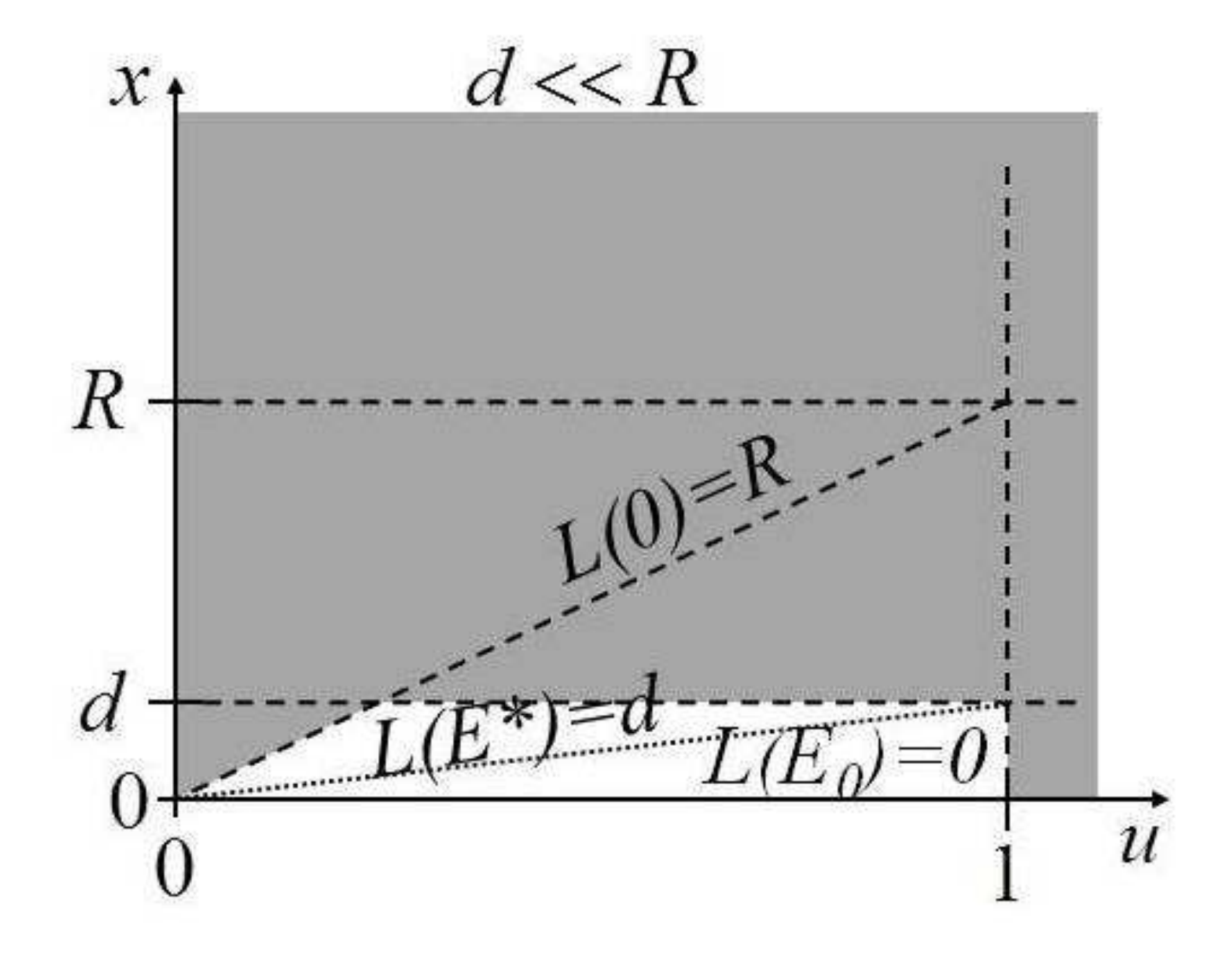}
 \caption{\footnotesize PHS \emph{variable space} for $d < R$ and $d << R$.} \label{vs34562}
\end{figure}
\\ If $d > R$ the variable $x$ can explore the interval
$x \in \left[0,R \right]$. This is the case of the PHS in Figure
\ref{figPHSST403}, where $d=4\,\mu m$, $R_{Li(830KeV)}=1.7\, \mu m$
and $R_{\alpha(1470KeV)}=3.4\, \mu m$. We take the two particles of
the $94\%$ branching ratio of $^{10}B$ reaction as example.
\\ If $d<R$, the variable $x$ can explore the interval $x \in
\left[0,d \right]$ (see Figure \ref{vs34562}), thus the domain is
now a trapezoid. The events near the line $\frac x u =L(E^{*})=d$,
which is the switching condition found in the Equations \ref{eqae11}
and \ref{eqae12}, give rise to a peak because this line has the
\emph{maximum length available}. Thus, we expect a peak in the PHS
around $E^{*}$. This is shown in Figure \ref{figPHSST402} where
$d=1\,\mu m$ and, again, $R_{Li(830KeV)}=1.7\, \mu m$ and
$R_{\alpha(1470KeV)}=3.4\, \mu m$, the peaks that originate from the
condition $\frac x u =L(E^{*})=d$ for the two particles are
indicated by the arrows.
\\ If $d\geq R$, the peak occurs for $L(E^{*})=R$, that is, zero
energy. This is problematic for $\gamma$-ray to neutron
discrimination.
\\ If $d<<R$, the variable space is compressed and the
straight lines identified by $\frac x u =L(E^{*})=d$ and $\frac x u
=L(E)=0$ become more and more similar. The two peaks approach and,
the more $d$ is negligible compared to $R$, the more the two peaks
appears as one single peak (see Figure \ref{figPHSST401}).
\\ If we want to avoid a strong presence of neutrons in the low
energy range of the PHS, where we know the $\gamma$-rays
contamination is strong, it is important to try to get the second
peak higher than the energy threshold ($E_{Th}$). This implies that
the thickness $d$ of any layer in the detector should obey
$d<L(E_{Th})$ for the $L$ corresponding to the particle with the
smallest range. This can be a contradictory requirement with
efficiency optimization in which case a compromise between
$\gamma$-rays rejection and efficiency has to be found.

\subsection{Transmission mode}
Equations for transmission mode can be calculated in the same way
they have been determined for back-scattering mode by substituting
$x$ with $d-y$ in the Expression \ref{eqae1} and $x$ with $y$ in the
expression $\delta\left(\frac x u -L\right)$. As a result we obtain:
\begin{equation}\label{eqae13}
P(L)\,dL = \begin{cases} \frac{1}{2 L^2}
\left(\frac{1}{\Sigma}e^{-\Sigma \cdot
d}+(L-\frac{1}{\Sigma})e^{-\Sigma \cdot (d-L)}\right)\,dL &\mbox{if
\,} L \leq d\\ \frac{1}{2 L^2} \left(\frac{1}{\Sigma}e^{-\Sigma
\cdot d}+(d-\frac{1}{\Sigma})\right)\,dL &\mbox{if \, } L > d
\end{cases}
\end{equation}
Hence, $Q(E)dE$ can be calculated as shown already in Section
\ref{backscatt678}. \\ However the same conclusions can be drawn
concerning the qualitative aspects of the PHS especially the
position of the two peaks.

\chapter{Converters at grazing angles}\label{chaptreflectometry}
The ideas in this Chapter were born in October 2012, when a
discussion between Philipp Gutfreund and me led us to the question:
do the boron converters we are using in detectors reflect neutrons
destroying any possibility to increase the layers' efficiency? A
complete investigation has been made in the next 6 months and it has
led to the discovery of the limits of the solid converter technology
used at grazing angle. I want to thank Carina H\"{o}glund for the
samples. I want to thank Philipp Gutfreund for the time he let me
use the D17 reflectometer at ILL, I want to thank my colleague Anton
Khaplanov, for the discussion and the support in this study. At
last, I want to thank Anton Devishvili, Andrew Dennison and Boris
Toperverg for the time spent on the reflectometer SuperAdam at ILL
and for their important suggestions.

\newpage
\section{Introduction}
\begin{figure}[!ht]
\centering
\includegraphics[width=10cm,angle=0,keepaspectratio]{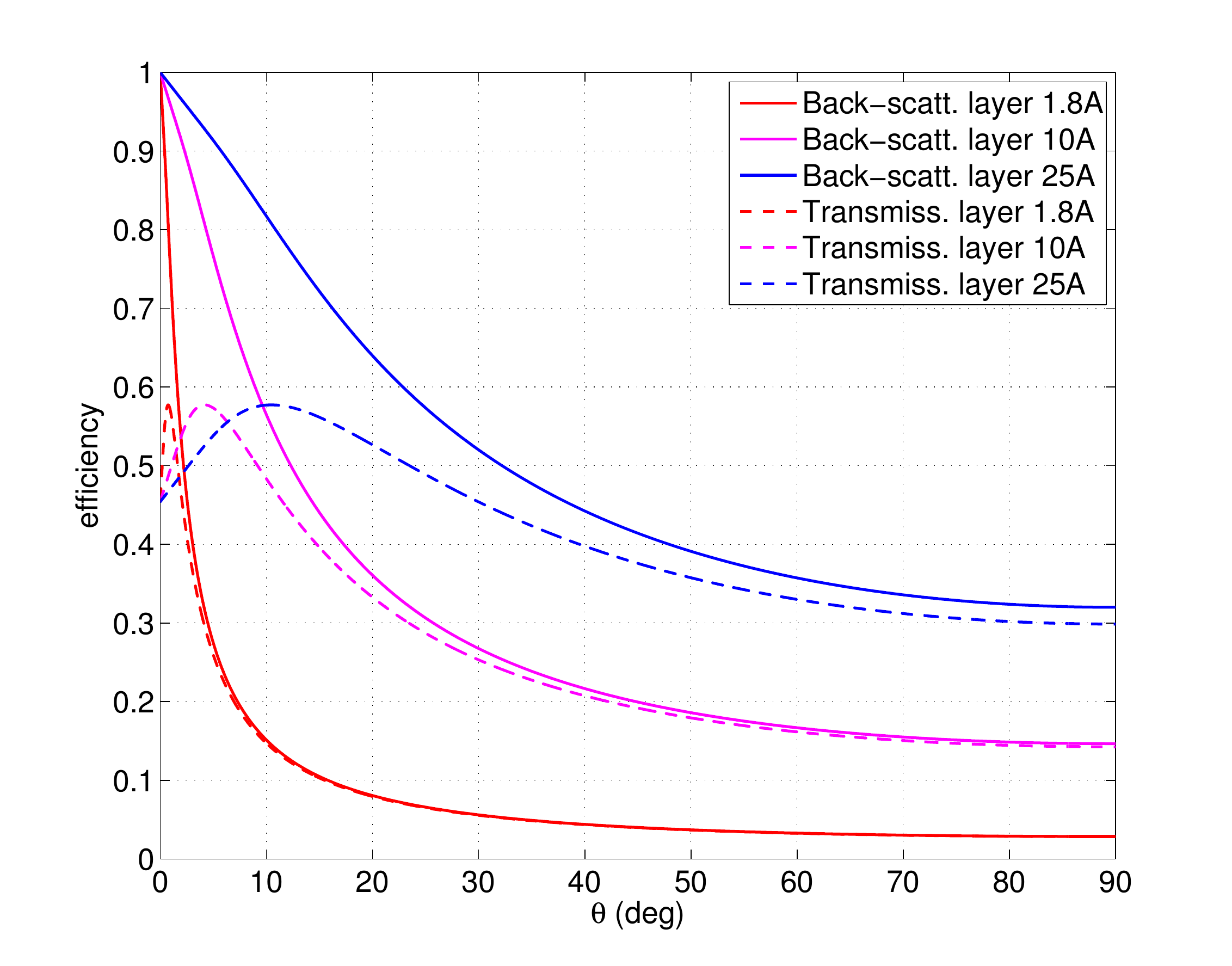}
\caption{\footnotesize Efficiency for a single layer $1\,\mu m$
thick $^{10}B_4C$ as a function of the neutron incidence angle.
Energy threshold of $100\,KeV$ is applied.}
\label{figexpeffanglesi47}
\end{figure}

In Chapter \ref{Chapt1} we developed an analytical model to
calculate the efficiency of a solid converter as a function of
several parameters. We have already demonstrated that the two
parameters $\theta$, the angle under which the neutrons hit the
converter layer, and $\lambda$, their wavelength, enter in the
calculations as a single parameter $\Sigma$ (Equation
\ref{eqab2bisimp}). This means that same features on the PHS or in
the efficiency can be found either by acting on the neutron
wavelength or on the angle $\theta$.
\\ We can, for instance, fix the $\lambda$ and vary the
angle. From the derivation of the physics of neutron conversion
layer explained in Section \ref{eqtwopartmodel345} we can observe
that, as $\theta$ decreases, the efficiency increases. Figure
\ref{figexpeffanglesi47} shows the efficiency for a single
back-scattering layer and for a single transmission layer as a
function of $\theta$ for two neutron wavelengths: $1.8\,$\AA,
$10\,$\AA \, and $25\,$\AA. These neutron wavelengths were chosen to
cover the usual neutron wavelength range used in a neutron
reflectometer instrument. The layers are $1\,\mu m$ thick and
consist of $100\%$ enriched $^{10}B_4C$ ($\rho = 2.24g/cm^3$). An
energy threshold of $100\,KeV$ is applied.
\\ The efficiency for a back-scattering layer according to the theory can, in principle,
reach $1$ for a sufficiently small angle. When $\theta$ decreases
(but not for $\theta=0$) not only the effective layer thickness
increases and consequently the absorption raises, but the source
points where the charged particle are generated when a neutron is
captured, are closer to the surface; therefore they have more
chances to escape the layer.
\\ On the other hand, for a
transmission layer, as the angle decreases also the effective layer
thickness increases starting absorbing neutrons without any increase
in efficiency. In Figure \ref{figexpeffanglesi47}, the transmission
layer efficiency attains a maximum at low angle, and then starts
decreasing again.
\\ These calculations however do not take into account the potential
reflection of the neutron by the conversion layer.
\\ It is important to consider neutron reflection in the efficiency
calculation when a detector involves converters at grazing angle. \\
Layer roughness has to be taken into account because it can affect
the neutron reflection.
\section{Reflection of neutrons by absorbing materials}\label{absrelfec6}
\begin{figure}[!ht]
\centering
\includegraphics[width=7.8cm,angle=0,keepaspectratio]{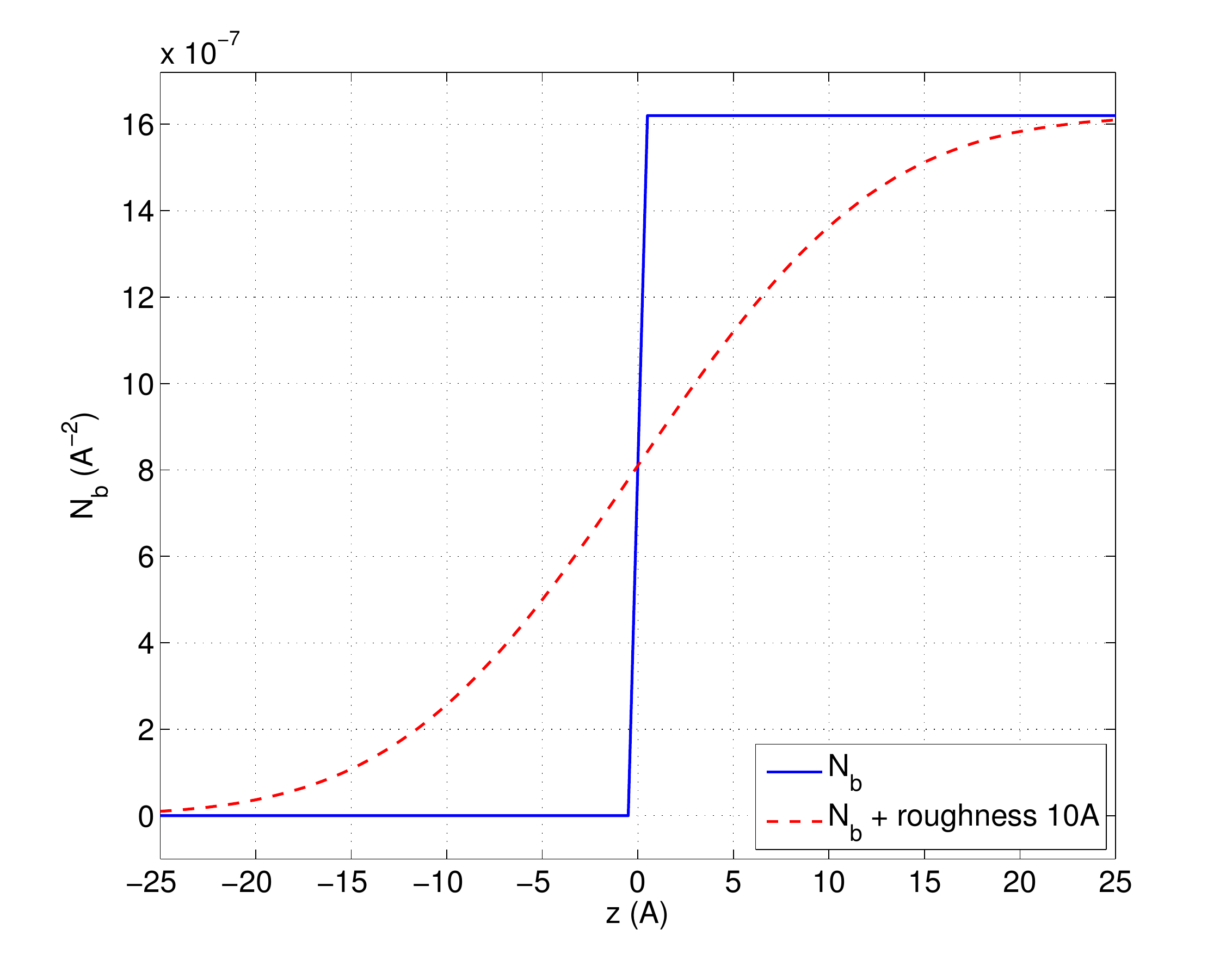}
\includegraphics[width=7.8cm,angle=0,keepaspectratio]{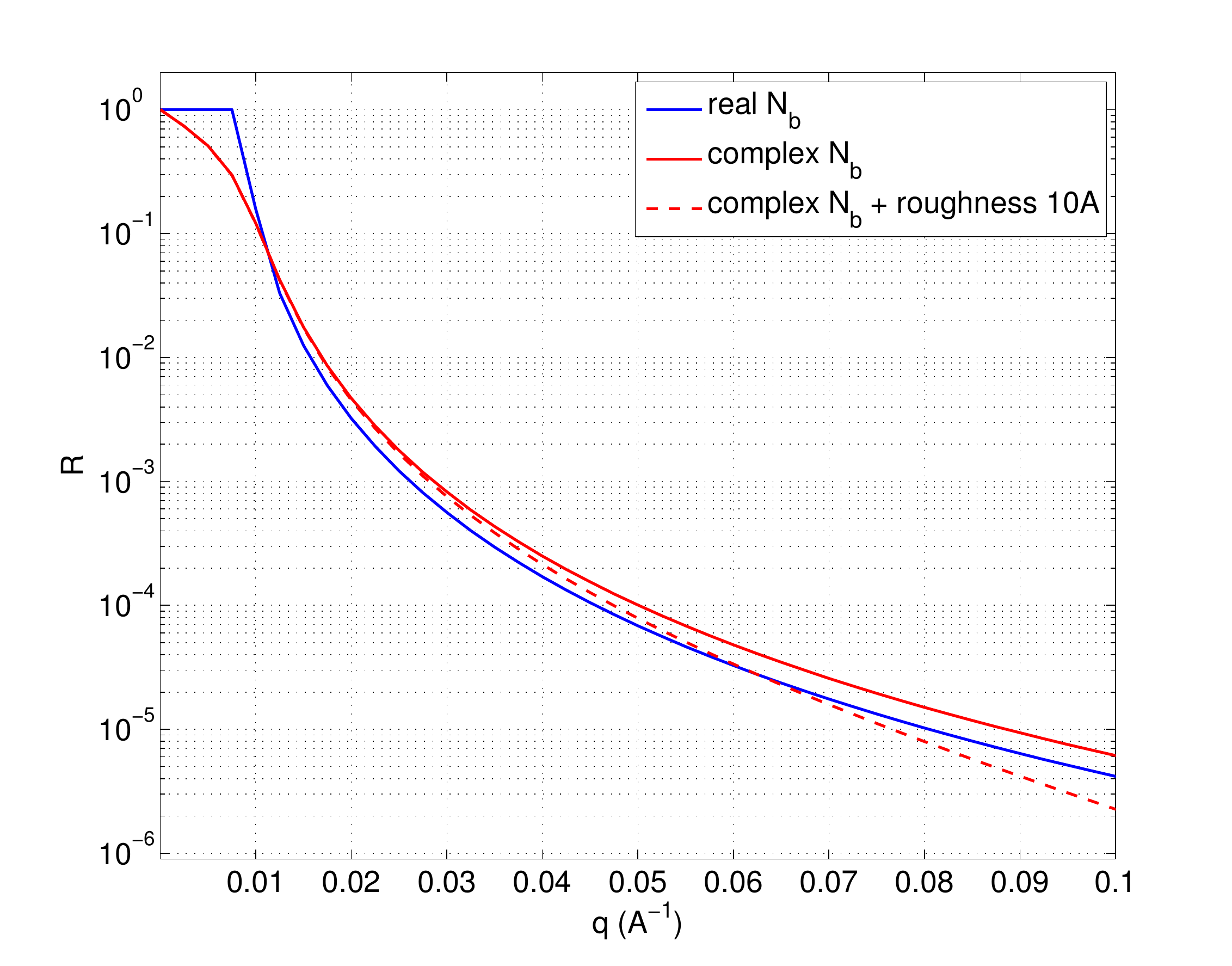}
\caption{\footnotesize Scattering Length Density ($N_b$) for
$^{10}B_4C$ omitting or not the interface roughness (left) and the
corresponding reflectivities (right).} \label{rifexa985}
\end{figure}

\begin{figure}[!ht]
\centering
\includegraphics[width=8cm,keepaspectratio]{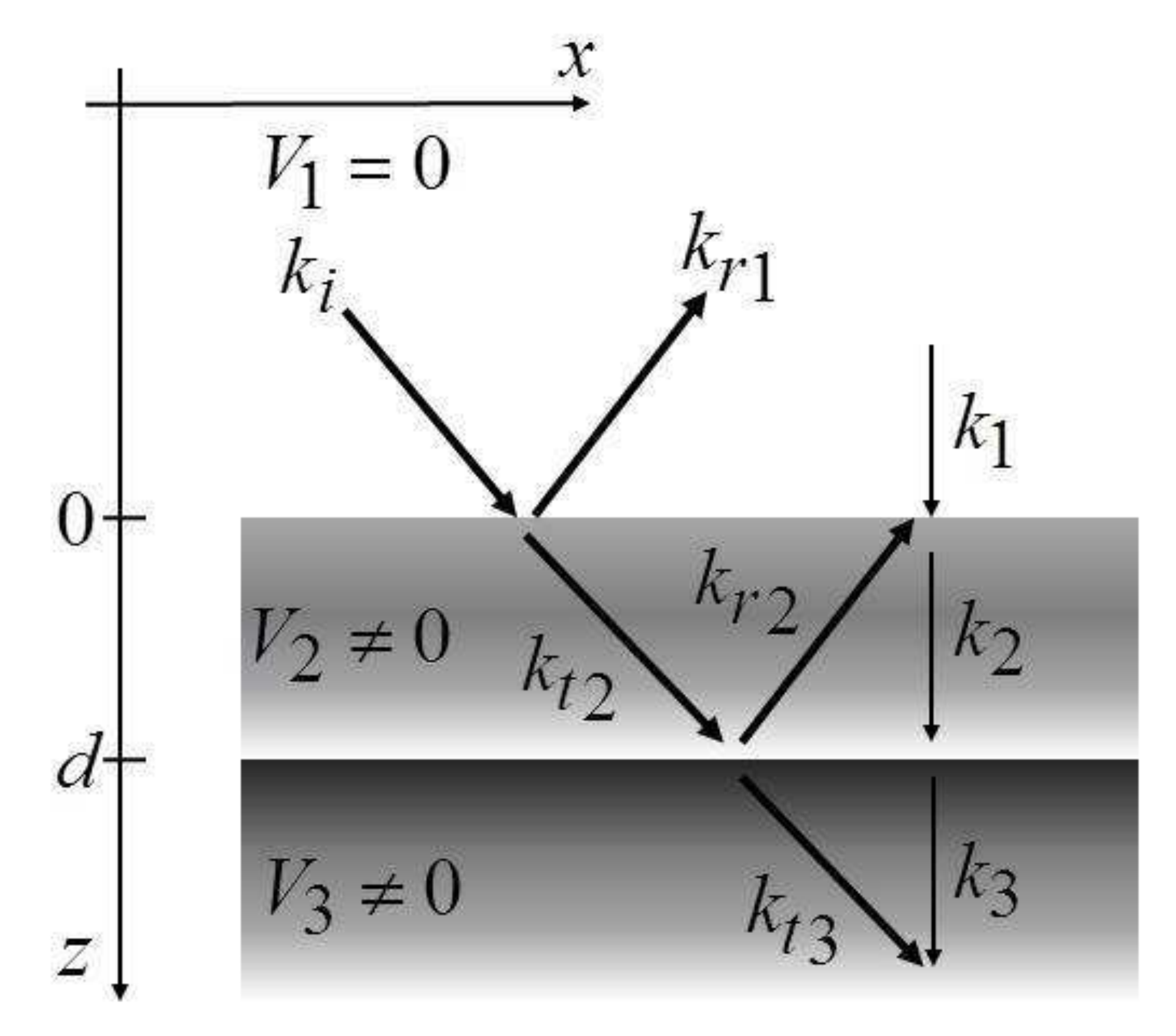}
\caption{\footnotesize Reflection of an incident neutron beam from
an ideally flat interface, $k_1$, $k_2$ and $k_3$ are the normal
component of the wave-vectors in the three regions.}
\label{reflscetch134}
\end{figure}

\begin{figure}[!ht]
\centering
\includegraphics[width=7.8cm,keepaspectratio]{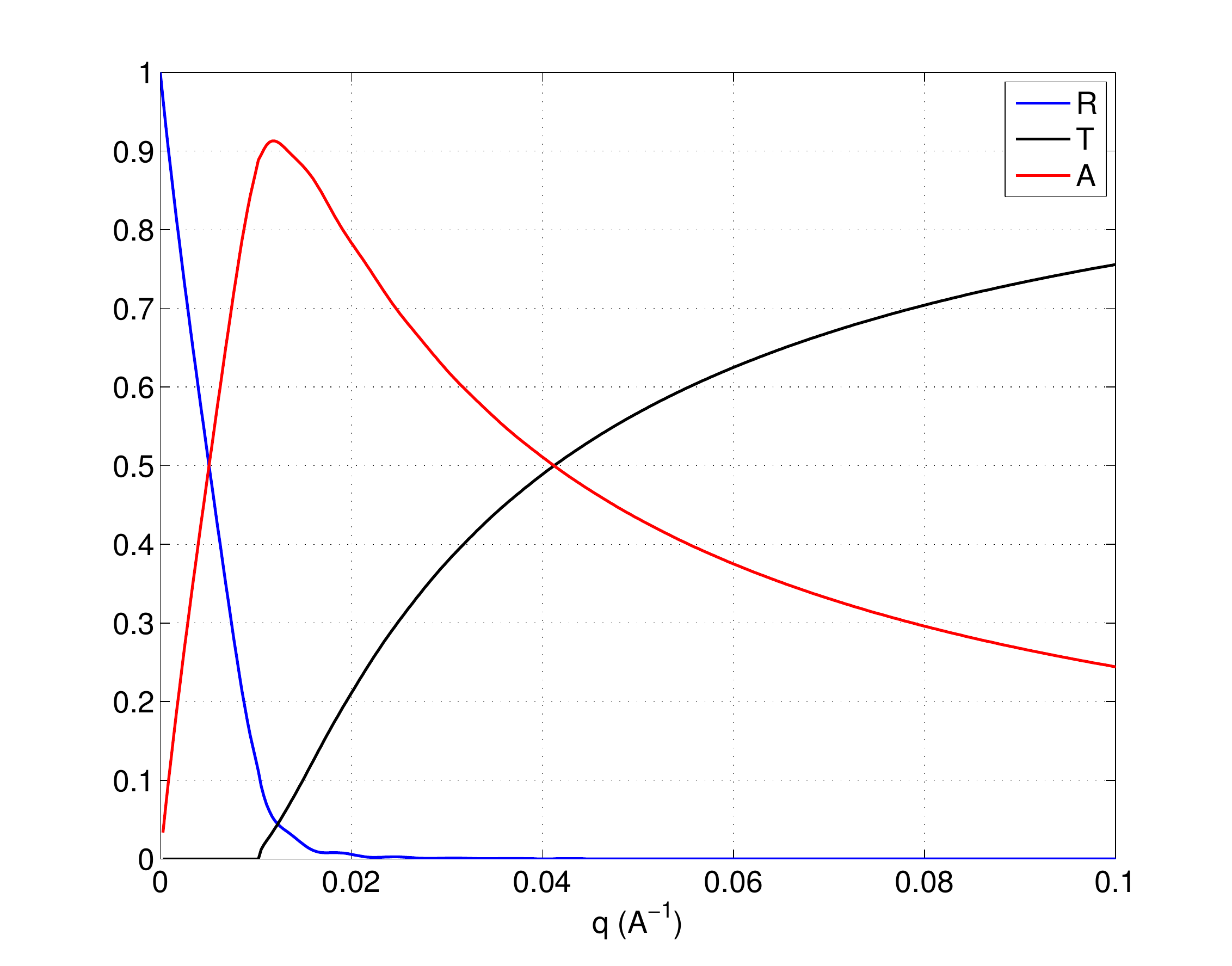}
\includegraphics[width=7.8cm,keepaspectratio]{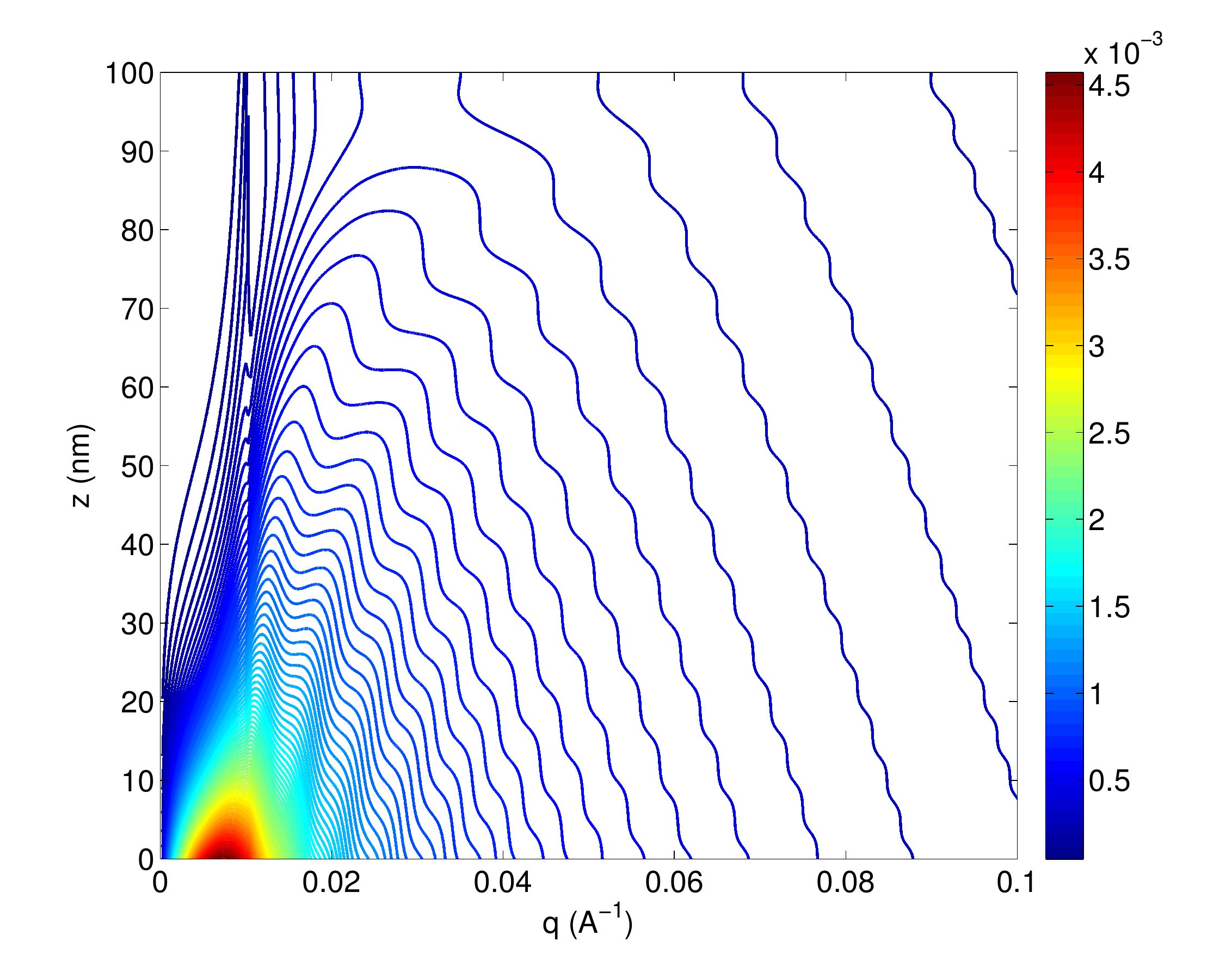}
\caption{\footnotesize Reflectivity, Transmission and Absorption
from a $d=100 \, nm$ $^{10}B_4C$ layer deposited on Si (left),
probability for a neutron to be absorbed in the layer as a function
of $z$ and $q$ (right).} \label{borplo1}
\end{figure}

Materials such as $Cd$ and $Gd$ or $^{10}B$, which are very strong
absorbers of neutrons, can still have significant reflectivities. As
shown in Section \ref{neuinte45}, the scattering and the absorbtion
cross-sections depend both on the complex phase shift $\eta$. For
absorbing materials $|\eta|<1$. Referring to Equation
\ref{crgwrtgwgw56954}, even for a perfect absorber ($\eta=0$) there
is a contribution to the scattering cross-section. This effect is
what is \emph{shadow diffusion} \cite{coen}.
\\ We discussed in Section \ref{neutrefltheoint} the principles of
reflection of neutrons at interfaces. When dealing with neutron
absorbers the theory describing the physical process of reflection
has to be modified to take into account, not only the possibility
for a neutron to be scattered, but also its absorption by nuclei. As
introduced in Chapter \ref{chaptintradmatt}, the scattering length
of a nucleus is, in general, a complex quantity. Its real and
imaginary parts can be associated to the scattering process but only
its imaginary part to the absorption (see Equation \ref{scrittb78}).
\begin{equation}
b_{tot}=b_{coh}=b'_{coh}+i\, b''_{coh}
\end{equation}
where we take only the coherent scattering length for both real and
imaginary parts because we suppose either the sample or the neutrons
to be unpolarized (Equation \ref{csfa8}), and we will average over
the bulk to obtain an effective potential description.
\\ The scalar potential $V$ (Equation \ref{eqaf2}) in the Schr\"{o}dinger equation will
contain the contribution given by the absorption:
\begin{equation}\label{}
V=\frac{2\pi\hbar^2}{m_n}\left( N_b^{real} + i \, N_b^{im}\right)
\end{equation}
The potential the neutron experiences at the interface is now
complex. To model absorption we are violating unitarity in the
Schr\"{o}dinger equation: it is a trick that works. \\ The solutions
of the Schr\"{o}dinger equation, with a complex potential, can still
be written as shown in Section \ref{neutrefltheoint}. The
wave-vectors will be complex quantities. Referring to Equation
\ref{eqaf6}, the change in the normal wave-vector has an imaginary
part given by the complex potential that results into an
exponentially reduced amplitude of the wave-function \cite{hayter}.
Note that if $N_b$ is purely imaginary (a perfectly absorbent
material) it still gives a contribution to the change in the normal
wave-vector (the square root of a purely imaginary number does have
a real part). \\ With absorption, even in the total reflection
regime ($q<q_c$), a neutron wave (this time not evanescent, see
Equation \ref{eqaf13}) can penetrate inside the layer reducing the
total reflection amplitude to less than 1. \\ In the absorbing
regime the characteristic depth a neutron can penetrate the layer
before being absorbed is given by:
\begin{equation}
D=\frac{1}{Im\left\{k_{t \bot}\right\}}
\end{equation}
As example we take enriched $^{10}B_4C$ ($\rho=2.24\,g/cm^3$) to
which corresponds a scattering length density of $N_b = \left(1.62-i
\, 1.11\right)\cdot 10^{-6} \, $\AA$^{-2}$. For simplicity we
imagine a bulk of $^{10}B_4C$ of constant $N_b$. In Figure
\ref{rifexa985} the scattering length density profiles at the
interface air/$^{10}B_4C$ are shown on the left and the
corresponding reflectivities are shown on the right.
\\ By neglecting the imaginary part of the $N_b$, reflectivity
is $1$ when $q<q_c$, and it behaves like $\propto 1/q^4$ for large
$q$. When $N_b$ is, on the other hand, complex the reflectivity
below $q_c$ is reduced. \\ For those materials having a negligible
absorption cross section, absorption does not significantly reduce
the reflectivity.
\\ Figure \ref{rifexa985} shows also what happens to the reflectivity
when a roughness of $\sigma=10$\AA \, is considered at the
air/$^{10}B_4C$ interface: intensity drops at high $q$. Note that
the contribution given by the absorption affects the entire
$q$-range while roughness has an influence only at high $q$.
\\ The reflectivity profile can still be measured arbitrary in ToF or in
monochromatic mode without affecting the results (if we can assume
constant imaginary scattering length density, meaning there are no
absorption resonances). We observe from Equation \ref{equaf4bisbis}
that only the normal component of the wave-vector $k_{i\bot}$ is
affected by the potential, for both complex and real cases.
Moreover, excluding resonances, this potential does not depend on
the neutron wavelength (Equations \ref{eqaf2} and \ref{eqaf3}),
because $b_{abs}$ does not depend on $\lambda$. The reflectivity
then only depends on $\theta$ and $\lambda$ through $k_{i\bot}=\frac
q 2$, so whatever method (ToF or monochromatic) is used to get a
value for the measured reflectivity the result is the same for same
$q$, even if the material is a strong absorber. Neutron
reflectometry at grazing angle by using thermal neutrons gives the
same reflectivity profile as UCN (Ultra Cold Neutron) reflectometry
at large angles if the resulting $k_{i\bot}$ is kept the same for
both techniques.
\\ The continuity equation \ref{eqaf11bis} in the region where the
material is an absorber has to be generalized as \cite{schiff}:
\begin{equation}\label{equaf20bis}
\frac{\partial P(\bar{r},t)}{\partial
 t}+\nabla \cdot J(\bar{r},t)= - \frac{2}{\hbar} P(\bar{r},t) Im\left\{V\right\} \qquad \Longrightarrow  \qquad \nabla \cdot J(\bar{r},t)= - \frac{4 \pi \hbar}{m_n} P(\bar{r},t) N_b^{im}
\end{equation}
assuming stationarity: $\frac{\partial P(\bar{r},t)}{\partial t}=0$.
The probability for a certain number of neutrons to be absorbed,
$A$, is given by the integral over the entire volume of absorbing
material, which in a one-dimensional case, reduces to:
\begin{equation}\label{}
A = \frac{m_n}{\hbar k_1}\int_0^d \nabla \cdot J(z,t) \, dz = -
\frac{4 \pi}{k_1} \int_0^d P(z,t) N_b^{im} \, dz = - \frac{4
\pi}{k_1} \int_0^d \left| Y_z\right|^2 N_b^{im} \,dz
\end{equation}
where $d$ is the thickness of the absorbing layer and $N_b^{im}$ is
the imaginary part of the scattering length density of the absorbing
medium.
\\ Since we want to measure reflectivity of absorbers employed in
neutron detectors; they are generally thin layers deposited on
holding substrates.
\\ Let us consider a finite thickness as in \cite{hai}, $d$, of
absorbing material deposited on a substrate, e.g. Silicon or
Aluminium. Excluding resonances, the solutions of the
Schr\"{o}dinger equation can be written as shown in Section
\ref{neutrefltheoint}, and we can focus only on the normal
components of the wave-functions:
\begin{equation}\label{equaf20}
\begin{aligned}
\Psi_z&=e^{+i\,k_1z}+r_1\,e^{-i\,k_1z} \qquad & \hbox{if   } z<0\\
Y_z&=t_2\,e^{+i\,k_2z} + r_2\,e^{-i\,k_2z}\qquad & \hbox{if   } 0<z<d\\
\Phi_z&=t_3\,e^{+i\,k_3z}\qquad & \hbox{if   } z>d\\
\end{aligned}
\end{equation}
where we have called $k_{n\bot}=k_n$ with $n=1,2,3$, the normal
component of the wave-vectors in the three regions defined by the
potentials $V_n$ (see Figure \ref{reflscetch134}).
\\ By imposing conservation of energy at the interfaces, as already
shown in Section \ref{neutrefltheoint} (Equation \ref{eqaf6}), we
obtain:
\begin{equation}
\begin{aligned}
k_1&=\frac{1}{2}q\\
k_2^2&=k_1^2-4\pi N_b^{(2)}\\
k_3^2&=k_1^2-4\pi N_b^{(3)}\\
\end{aligned}
\end{equation}
where $N_b^{(2)}$ and $N_b^{(3)}$ are the SLD (Scattering Length
Density) of the absorbing layer and the substrate respectively.
\\Consequently, by imposing the continuity of the
wave-function and its derivative at the two boundaries ($z=0$ and
$z=d$), we obtain:
\begin{equation}\label{equaf21}
\begin{aligned}
r_1&=t_2+r_2-1 \hbox{\qquad \qquad \,\,\,}
r_2=\frac{\alpha_{12}\beta_{23}\delta}{1+\beta_{12}\beta_{23}\delta}
\\t_2&=\frac{\alpha_{12}}{1+\beta_{12}\beta_{23}\delta} \hbox{\qquad \qquad}
t_3=e^{-i\left(k_3+k_2\right)d}\cdot\frac{\left(1+\beta_{23}\right)\alpha_{12}\delta}{1+\beta_{12}\beta_{23}\delta}
\end{aligned}
\end{equation}
where we have defined the Fresnel transmission coefficients as
$\alpha_{ij}=\frac{2k_i}{k_i+k_j}$, the reflection coefficients
$\beta_{ij}=\frac{k_i-k_j}{k_i+k_j}$ and $\delta=e^{+2ik_2d}$.
\\ The reflection and transmission probabilities are given by:
\begin{equation}
R = \frac{\left|J_r\right|}{\left|J_i\right|}, \qquad  \qquad T =
\frac{\left|J_t\right|}{\left|J_i\right|}
\end{equation}
with
\begin{equation}
\left|J_i\right|=\frac{\hbar\,k_1}{m_n}, \qquad
\left|J_r\right|=\frac{\hbar\,k_1}{m_n}\left(r_1 \cdot
r_1^\ast\right), \qquad
\left|J_t\right|=\frac{\hbar\,k_3}{m_n}\left(t_3 \cdot
t_3^\ast\right)
\end{equation}
\\ Outside the absorbing regions the Equations \ref{eqaf11bisbis} are
still valid. Referring to Figure \ref{reflscetch134} if the first
medium is air, the second is an absorber ($V_2$ is complex) and the
third is a substrate such as Silicon ($V_3$ is real); the measured
reflectivity, the transmission inside the substrate and the
absorption in the layer are:
\begin{equation}\label{eqaf22}
\begin{aligned}
R&=r_1 \cdot r_1^\ast\\
T&=\frac{k_3}{k_1}\left(t_3 \cdot t_3^\ast\right)\\
A&=1-R-T=\frac{1}{k_1}\int_0^d \nabla \cdot J_2(z,t) \, dz = -
\frac{4 \pi}{k_1} \int_0^d |Y_z|^2 N_b^{im} \, dz
\end{aligned}
\end{equation}
where $J_2$ is the current probability calculated for $Y_z$.
\\ As example we take the same absorber as in Figure
\ref{rifexa985}; in this case the $^{10}B_4C$ is $d=100 \, nm$ thick
and it is deposited on Si ($N_b = 2.14\cdot10^{-6}\,$\AA$^{-2}$). In
Figure \ref{borplo1}, on the left we show reflectivity, transmission
and absorption, as calculated in Equation \ref{eqaf22} as a function
of $q$. On the right we show the probability for a neutron, carrying
a given $q$, to be absorbed at certain depth in the layer, i.e. the
quantity $- \frac{4 \pi}{k_1} |Y_z|^2 N_b^{im}$. We notice that
absorption increases in proximity of the critical edge ($q_c$) that
is, in this case about $q_c =0.01\,$\AA$^{-1}$.

\section{Corrected efficiency for reflection}\label{correffforreflect6}
\begin{figure}[!ht]
\centering
\includegraphics[width=7.8cm,angle=0,keepaspectratio]{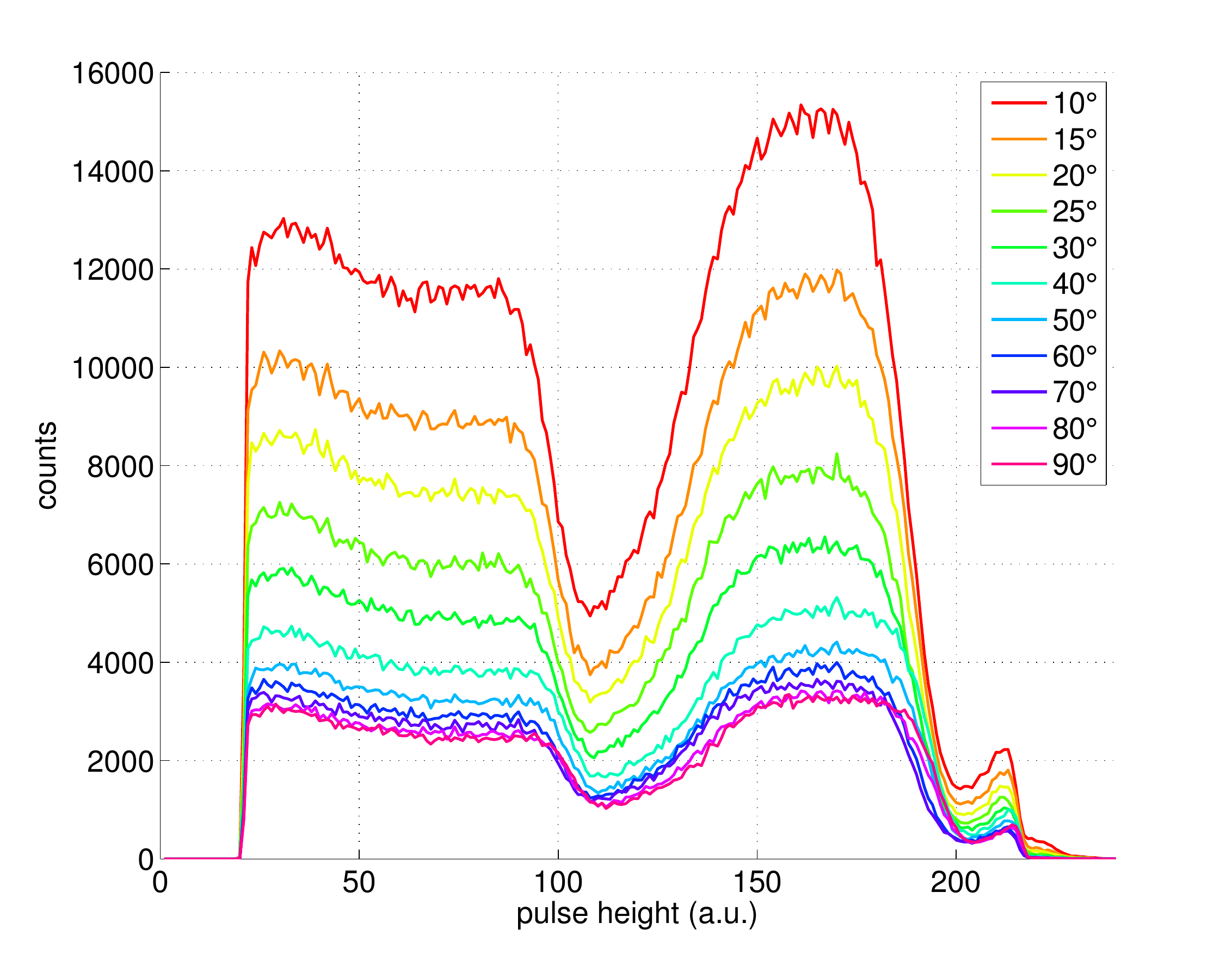}
\includegraphics[width=7.8cm,angle=0,keepaspectratio]{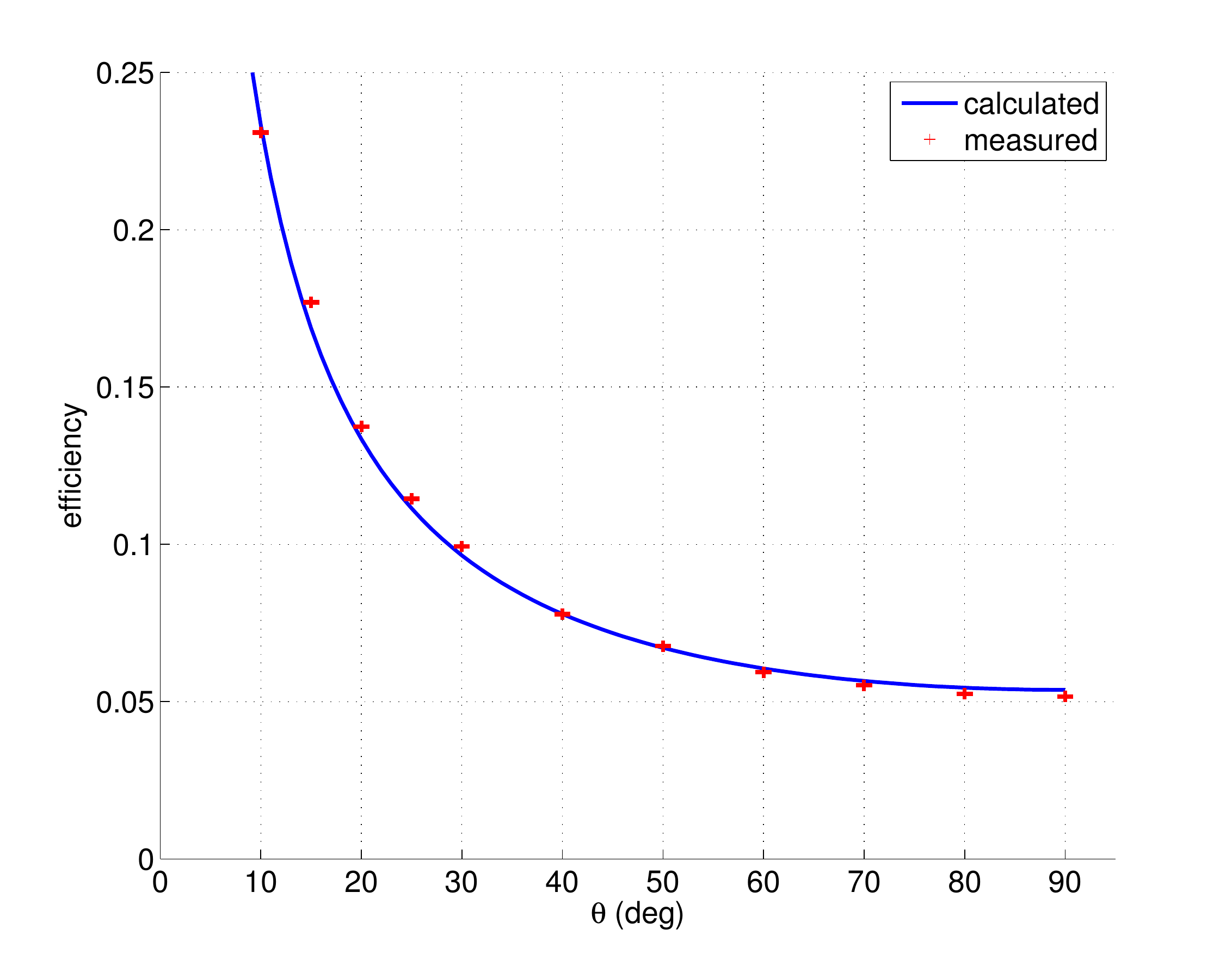}
\caption{\footnotesize PHS and efficiency measured at $2.5\,$\AA \,
for a $1\,\mu m$ $^{10}B_4C$ back-scattering layer operated in a
MWPC.} \label{figPHSexamplmeas0936}
\end{figure}

\begin{figure}[!ht]
\centering
\includegraphics[width=7.8cm,angle=0,keepaspectratio]{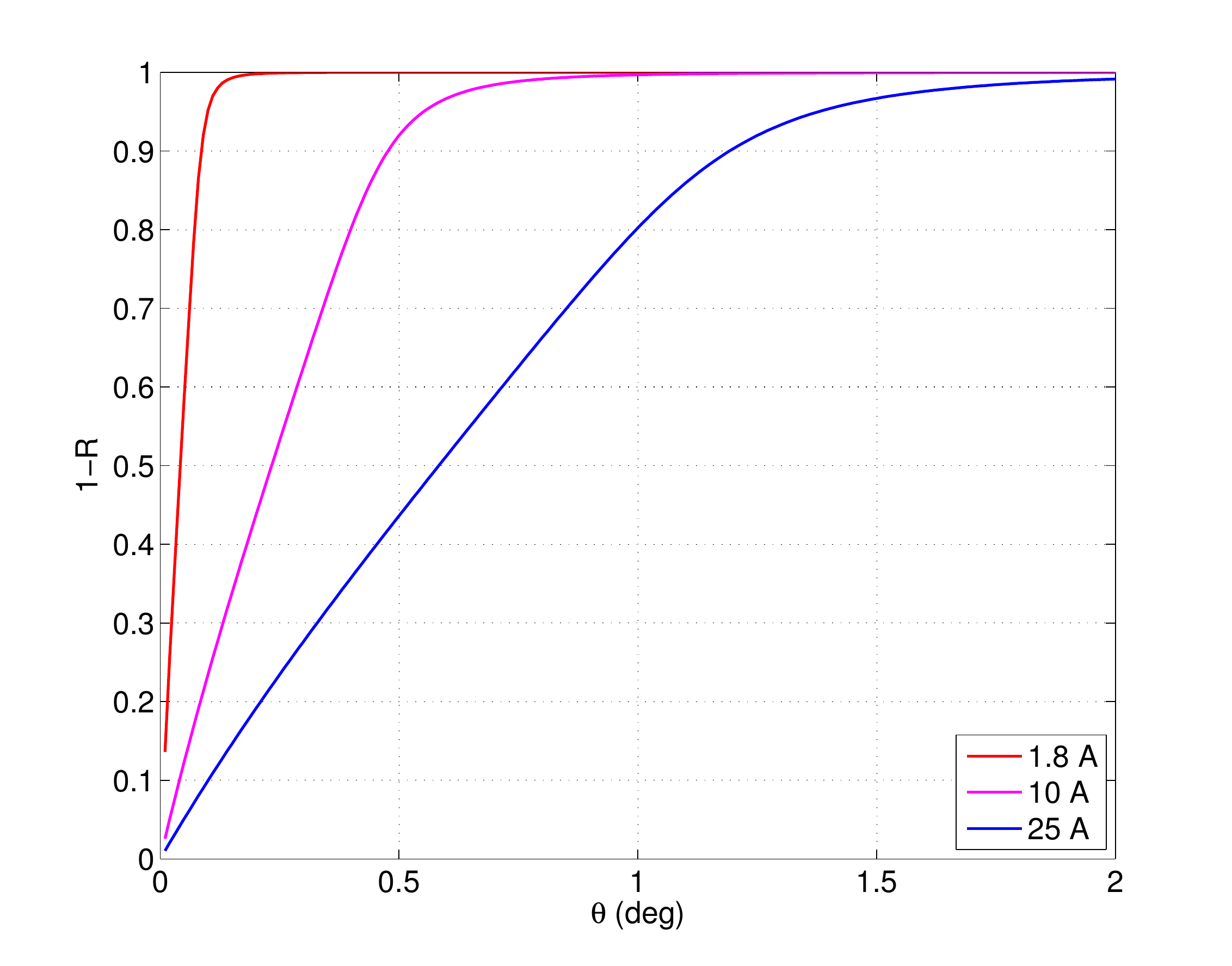}
\includegraphics[width=7.8cm,angle=0,keepaspectratio]{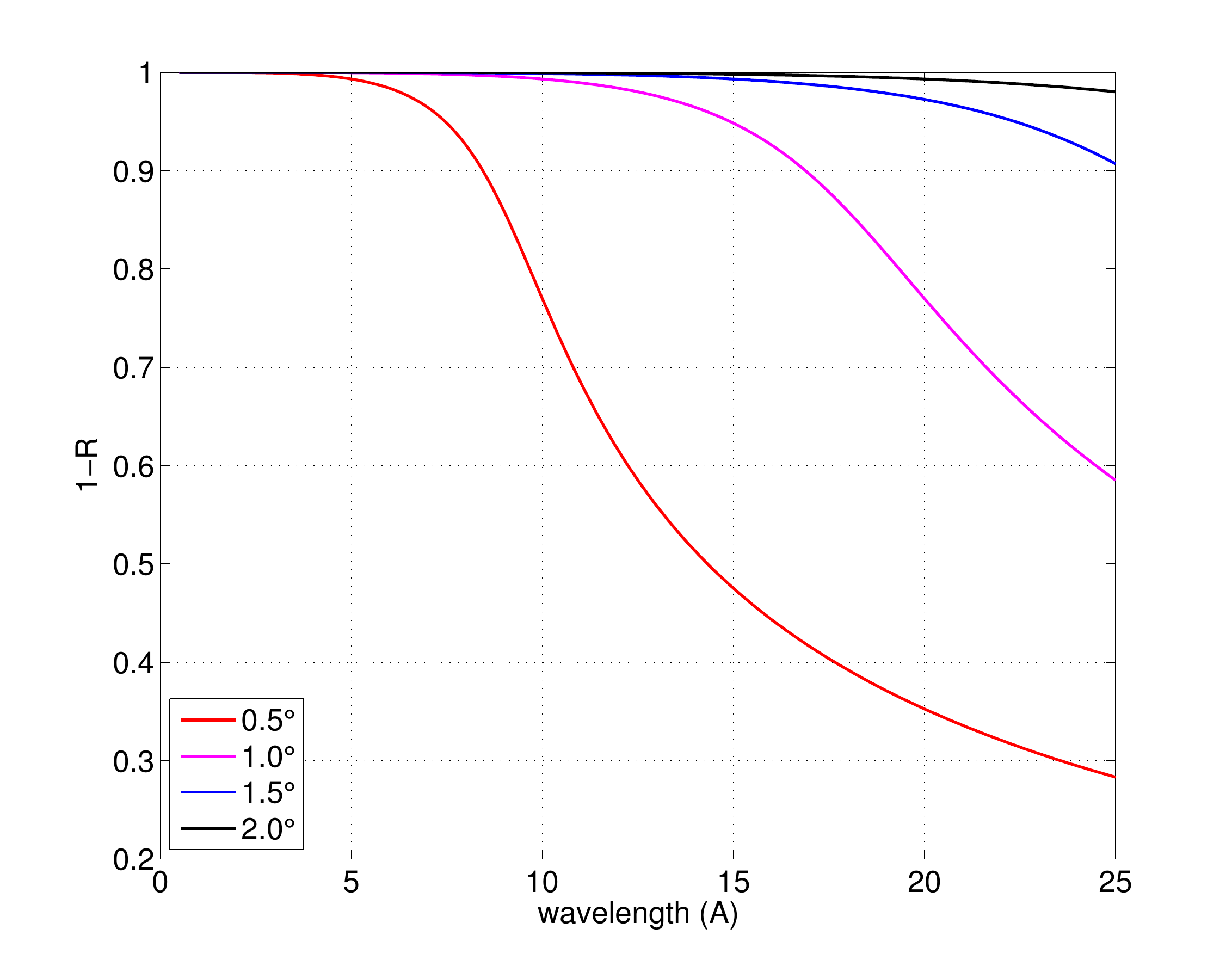}
 \caption{\footnotesize Factor $1-R$ for a $1\,\mu
m$ $^{10}B_4C$ layer ($N_b=1.6-1.11\,i$) as a function of $\theta$
(left) and as a function of neutron wavelength (right).}
\label{fir0945}
\end{figure}

\begin{figure}[!ht]
\centering
\includegraphics[width=10cm,keepaspectratio]{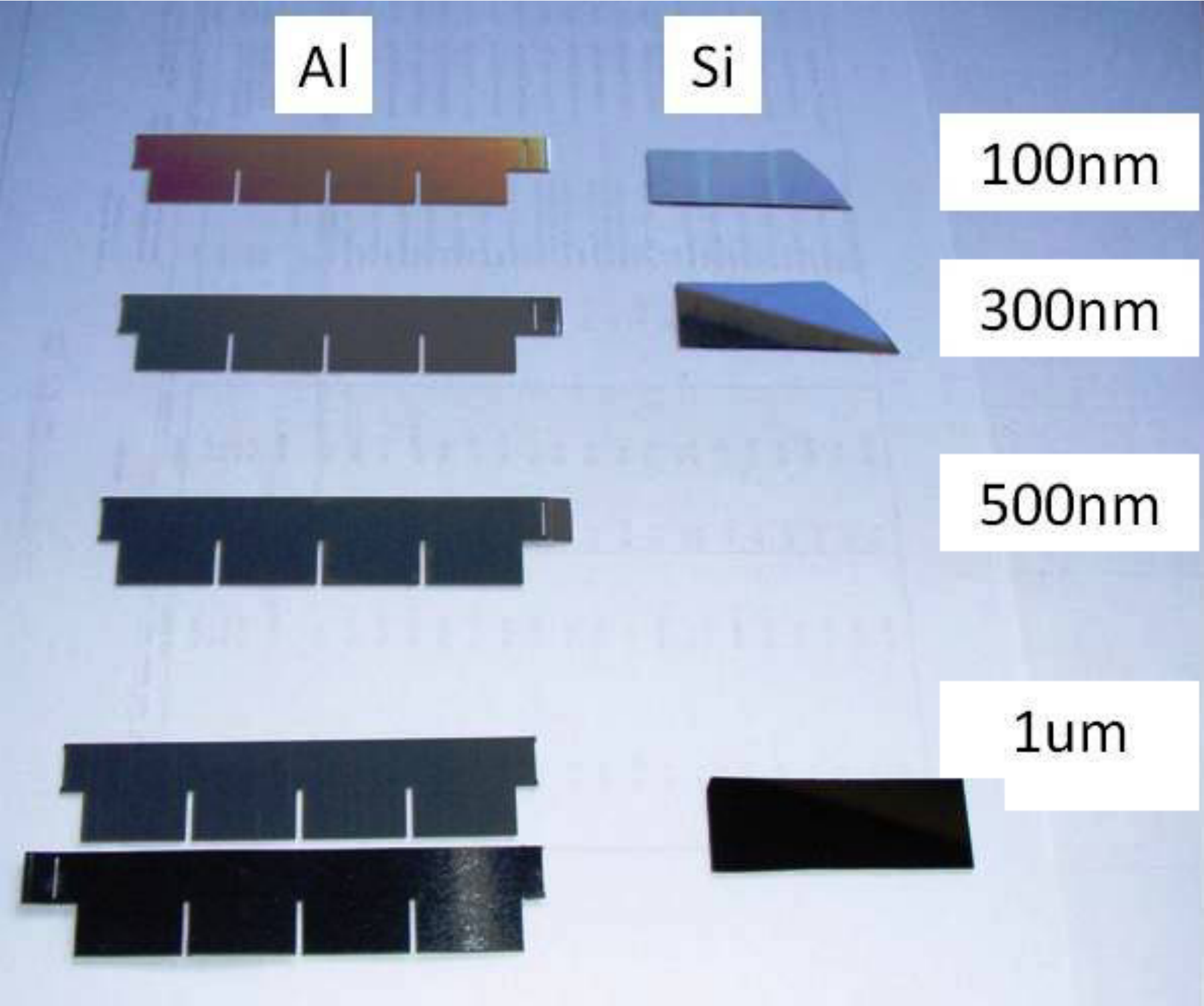}
\caption{\footnotesize Samples of $Al$ and $Si$ coated with
different thicknesses of $^{10}B_4C$ layer \cite{carina}.}
\label{samfig907854}
\end{figure}

The neutron efficiency for a single layer of $1\,\mu m$ $^{10}B_4C$
has been measured as a function of the neutron incidence angle
$\theta$ for a given neutron wavelength of $2.5\,$\AA \, on our
neutron test beam line CT2 at ILL.
\\ The converter layer was mounted in a MWPC of about $100\,cm^2$
operated at $1bar$ of $CF_4$. After calibration and Plateau
measurement we chose $1300V$ as the working voltage. A neutron beam
of $2\times3 \, mm^2$ was focused on the detector and a PHS was
measured as a function of the incidence angle. In Figure
\ref{figPHSexamplmeas0936} the measured PHS for a single
back-scattering layer is shown.
\\ The calculation assumes no reflection. The increase in detection efficiency can be
calculated according to Equation \ref{eqae7} in Section
\ref{SRIMsp33}. The resulting measured and calculated neutron
detection efficiency is shown in Figure \ref{figPHSexamplmeas0936}.
\\ When we deal with detectors that involve converters at grazing
angle, the formulae derived in Chapter \ref{Chapt1} for the
detection efficiency have to be replaced by:
\begin{equation}\label{eqwty7}
\varepsilon_R\left(d,\theta,\lambda\right)=\left(1-R(\theta,\lambda)\right)\cdot
\varepsilon\left(d,\theta,\lambda\right)
\end{equation}
where $R$ is the neutron reflectivity that becomes significant at
cold neutron wavelength or at very small angle. In general for
$\theta>2^{\circ}$ neutron reflection can be neglected ($R\approx
0$) in the thermal-cold neutron wavelength range. As we will
discuss, there is a way to reduce the reflectivity factor ($R$) at
smaller than $2^{\circ}$ angles, by acting on the layer surface
roughness.
\\ In Figure \ref{fir0945} we plot the calculated factors $(1-R(\theta,\lambda))$ according to
the Equations \ref{eqaf12} in Section \ref{neutrefltheoint} in
Chapter \ref{chaptintradmatt} given that
$q=\frac{4\pi}{\lambda}\sin(\theta)$. We take as example a $1\,\mu
m$ $^{10}B_4C$ layer, that we can consider bulk, of Scattering
Length Density $N_b=(1.6-1.11\,i)\cdot10^{-6}\,$\AA$^{-2}$
($\rho=2.24g/cm^3$).
\\ We notice that for application that involve cold neutron
wavelength such as $25\,$\AA \, the detection efficiency is limited
to be at most $60\%$ and not $1$ if the detector employs $1^{\circ}$
as converting angle.
\\ In order to study experimentally the neutron reflectivity of
thin-films neutron converters, two sets of data have been recorded.
The first set of data was taken using D17 \cite{cubittD17} at ILL
which is a ToF reflectometer to preliminary quantify the actual
reflectivity of the coatings. A second experiment has been performed
on SuperAdam \cite{superadam} at ILL, which is a monochromatic
reflectometer ($\lambda=4.4\,$\AA) in a more complete setup. The two
experiments allowed to compare the two techniques (ToF and
monochromatic) in addition to give information on neutron converter
reflectivity. The comparison let us to verify the theoretical
predictions in Section \ref{neutrefltheoint}.
\\ The coatings were fabricated in Link\"{o}ping University by the Thin Film Physics Division
\cite{carina}, \cite{carinatesi} by sputtering technique (PVD). The
$^{10}B_4C$ layers have been deposited either on several $Al$ alloys
or on $Si$ wafers for preliminary characterizations. Examples of
those samples are shown in Figure \ref{samfig907854}. The $Si$
samples look shiny while the higher roughness of the $Al$ alloy
makes the $Al$ samples matte. Moreover, we had two different
$Al$-alloys for the $1\,\mu m$ sample with different layer roughness
(see Figure \ref{samfig907854}).
\\ Physical vapor deposition (PVD) is a technique for thin film
synthesis under vacuum conditions, where a solid or liquid
deposition material is vaporized and its condensation on a substrate
forms a film. Chemical reactions are usually absent in the gas
phase, due to the small probability for collisions of the vapor
species under the applied pressures. The usual physical mechanisms
for source atoms to enter the gas phase are high temperature
evaporation or sputtering. $^{10}B_4C$ films were grown in high
vacuum chambers by reactive magnetron sputter deposition.
\\ To minimize the amount of impurities in the films it is necessary to have good vacuum
conditions in the deposition chamber. All films we discuss here are
deposited at base pressures of $0.1\,mPa$ \cite{carinatesi}.
\\ Ion beam analysis is used for determination of the concentration of specific elements in a
sample. Thickness, compositional gradients and depth positions of
different elements can also be determined. Elastic recoil detection
analysis (ERDA) was used. This technique is based on elastic
scattering of incoming ions with the target atoms. In ERDA the
energy of the knocked out target atom is detected. ERDA is good for
depth profiling and analysis of elements with a mass smaller than
the mass of the incoming heavy ions. In the present setup the
relative concentrations of element was measured at $\sim2\%$ in ERDA
\cite{carinatesi},\cite{carina}.
\begin{figure}[!ht]
\centering
\includegraphics[width=8cm,angle=0,keepaspectratio]{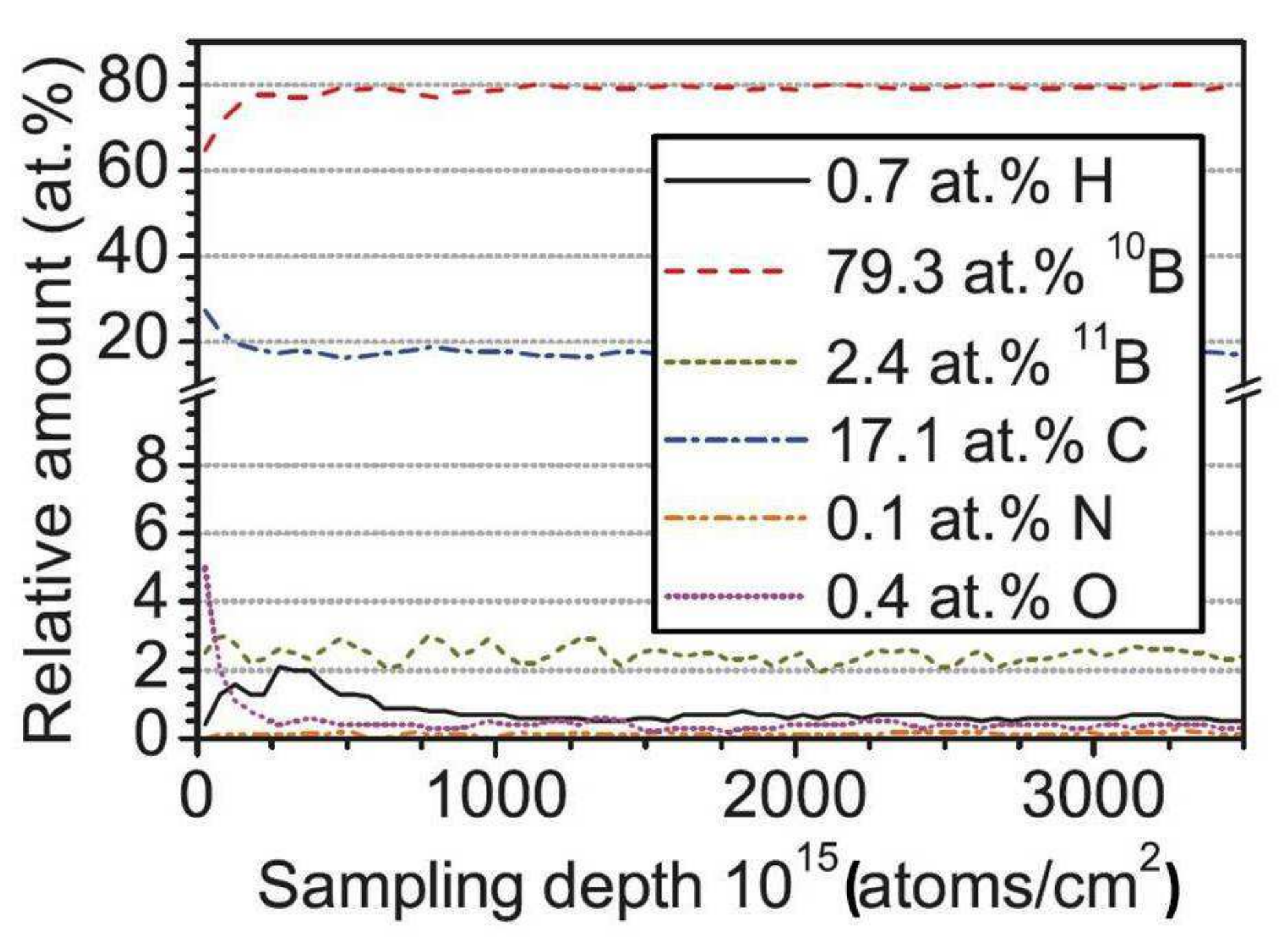}
\includegraphics[width=7cm,angle=0,keepaspectratio]{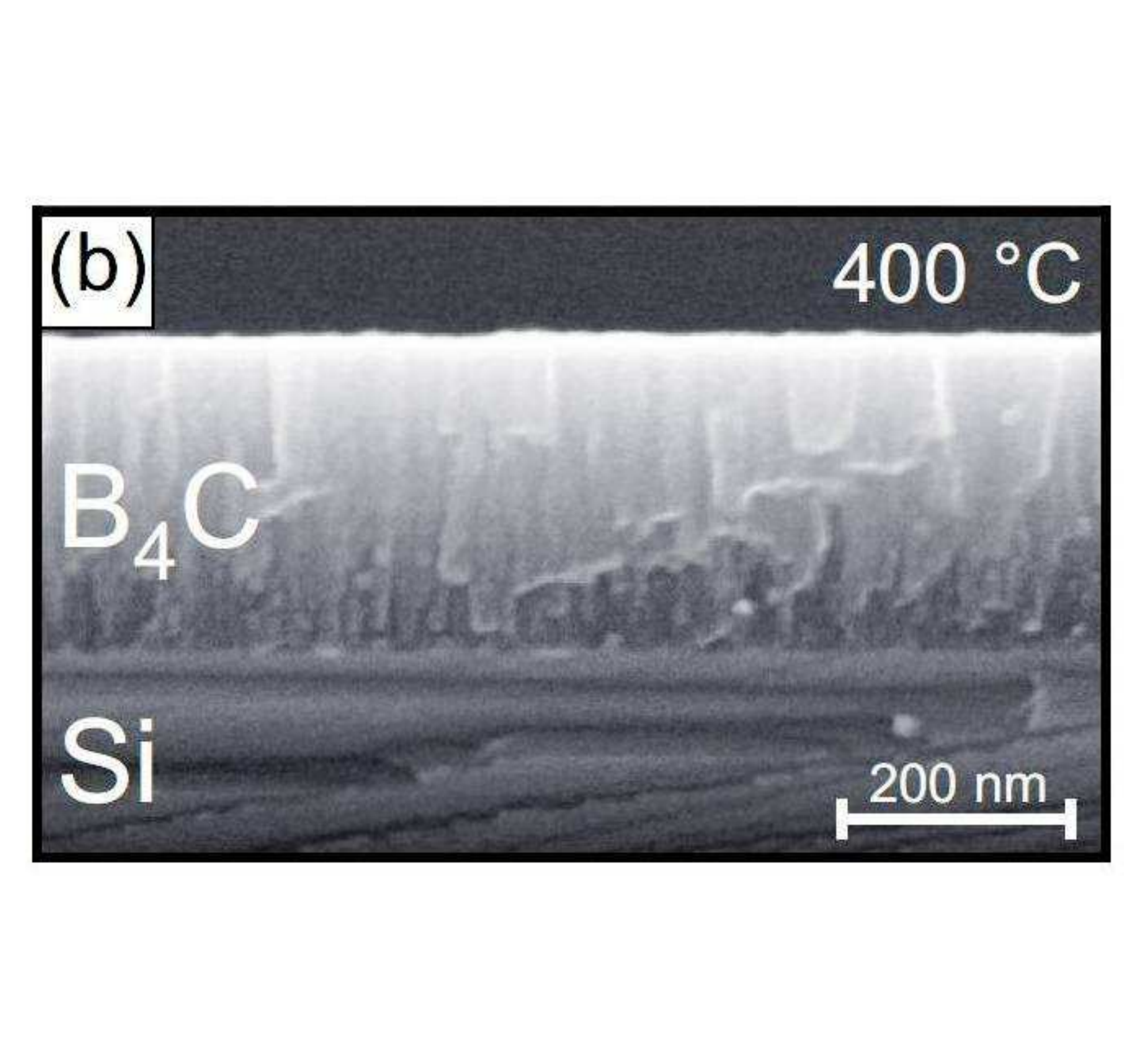}
 \caption{\footnotesize $^{10}B_4C$ film composition and a SEM image of a $^{nat}B_4C$ layer on $Si$. The x-axis correspond to about
 $270\,nm$ of sampling depth \cite{carina}.}
\label{carites2}
\end{figure}
\\ In Figure \ref{carites2} the relative amount of elements in a typical $^{10}B_4C$
film is plotted as a function of the layer depth. The conversion
between the x-axis and the actual depth is given by:
\begin{equation}\label{eqkk1}
d = \frac{x\cdot M_r}{\rho \cdot N_A}
\end{equation}
where $M_r$ is the molar mass of the $^{10}B_4C$ layer, $\rho$ its
density and $N_A$ is the Avogadro's number. Referring to Figure
\ref{carites2} and using the Equation \ref{eqkk1}, we are sampling
about $270\,nm$ on the full scale of x-axis.
\\ According to Equation \ref{eqaf3} and considering the
composition given in Figure \ref{carites2}, the scattering length
density of such a layer is: $N_b=\sum_i b_i
n_i=(1.6-1.11\,i)\cdot10^{-6}\,$\AA$^{-2}$ ($\rho=2.24g/cm^3$). We
use as scattering length density for $Si$ the standard value of
$N_b=2.14\cdot10^{-6}\,$\AA$^{-2}$ and for $Al$
$N_b=2.07\cdot10^{-6}\,$\AA$^{-2}$.
\\ The first experiment has been performed on the D17 instrument \cite{cubittD17}, a Time-of-Flight reflectometer at
ILL, on $1\,\mu m$ samples deposited on both $Si$ and $Al$ (see
Figure \ref{samfig907854}). Reflectivity profiles have been measured
for three angles $\theta=0.5^{\circ},1^{\circ},2^{\circ}$ in
ToF-mode between $\lambda=2$\AA \, and $\lambda=25$\AA. The
resulting range in momentum transfer is from $q=0.005$\AA$^{-1}$ up
to $q=0.2$\AA$^{-1}$ with a resolution of $\Delta q=0.05$\AA$^{-1}$
in the worst case. The reflected intensity (and the direct beam) in
ToF can be measured without scanning in angle but just acquiring the
neutron wavelength spectrum at once. The reflectivity is calculated
as the ratio between the reflected and the direct wavelength
spectra.
\\ In ToF the background is uncorrelated with the instrument timing
and it can be evaluated by looking at a region of the detector where
we are sure there is no reflection and checking that it has no time
structure. This background has been subtracted to the reflected and
the direct beam spectra.
\\ No specular reflection has been observed on all the $Al$-samples, on
the other hand $Si$ has a strong reflectivity and it is shown in
Figure \ref{fgori890} as a function of $q$ (pink curve).
\begin{figure}[!ht]
\centering
\includegraphics[width=10cm,keepaspectratio]{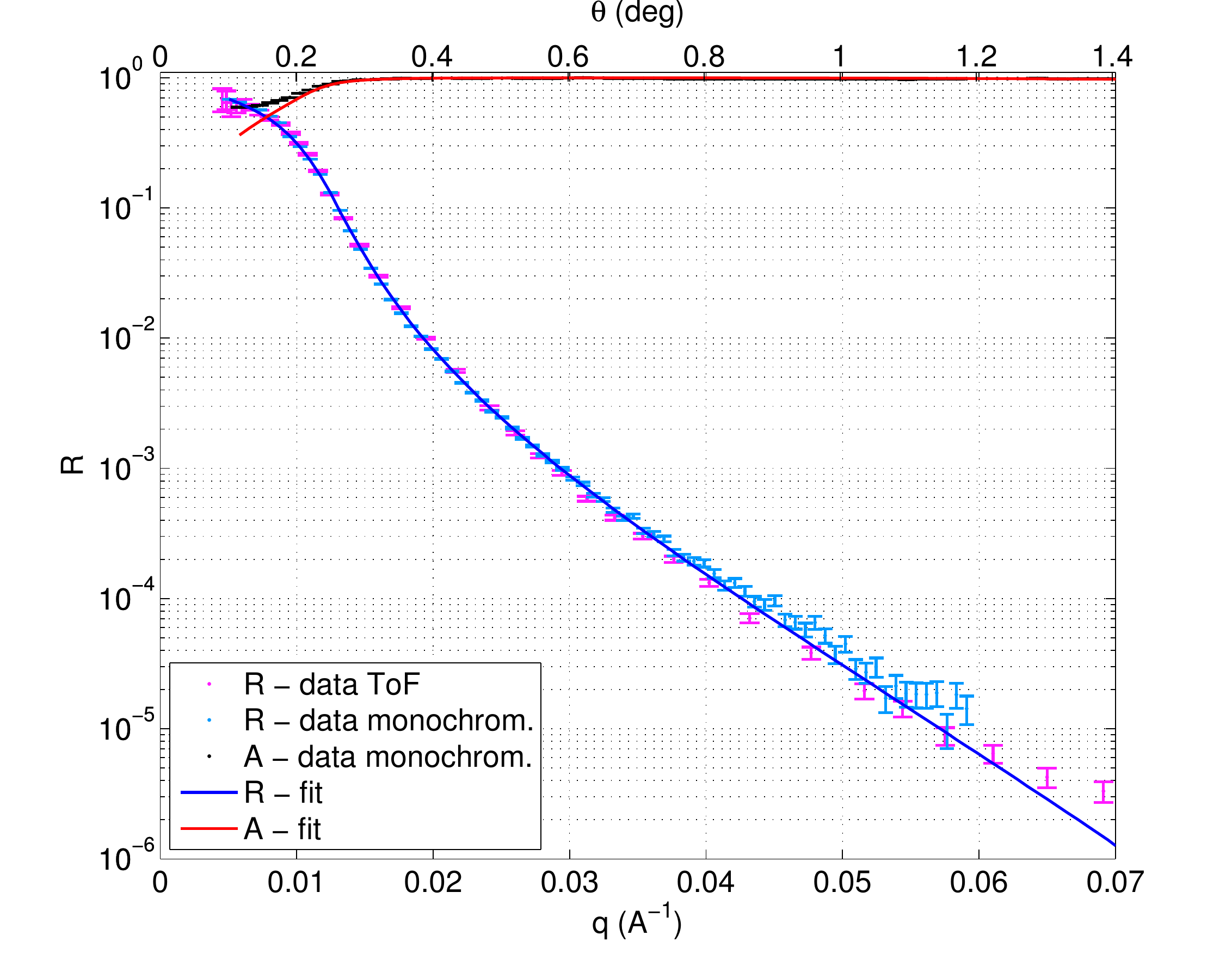}
\caption{\footnotesize Measured reflectivity for the $1\,\mu m$
$^{10}B_4C$ sample on $Si$ as a function of $q$; pink curve has been
obtained from a ToF measurement and cyan from monochromatic angular
scan. Absorption was measured as well from the $\gamma$-ray yield.}
\label{fgori890}
\end{figure}
\\ We repeated the neutron reflectivity experiment by using the
SuperAdam instrument \cite{superadam} in monochromatic mode at
$\lambda=4.4$\AA. A scan in angle has been performed to get the
reflectivity profile in $q$. We measured in addition to the $1\,\mu
m$ samples, one extra sample of $100\,nm$ deposited on a $Si$
substrate.
\\ Not only neutron reflectivity has been measured but, neutron absorption has been measured
as well, thanks to the $\gamma$-ray yield of $^{10}B$. We recall
that in the $94\%$ of the cases when a neutron is captured by a
$^{10}B$-atom a $478\,KeV$ $\gamma$-ray is produced and emitted
isotropically. In Figure \ref{figexpsupadam9046} is shown a sketch
of the experiment performed on SuperAdam. A Germanium detector was
placed close to the sample in order to maximize the solid angle
without affecting the neutron beam. Taking into account both the
Ge-detector efficiency and the solid angle, we estimate the whole
efficiency for the $478\,KeV$ $\gamma$-ray photo-peak detection to
be around $5\%$. The Ge-detector has been calibrated in energy by
using a $^{22}Na$ source.
\begin{figure}[!ht] \centering
\includegraphics[width=10cm,keepaspectratio]{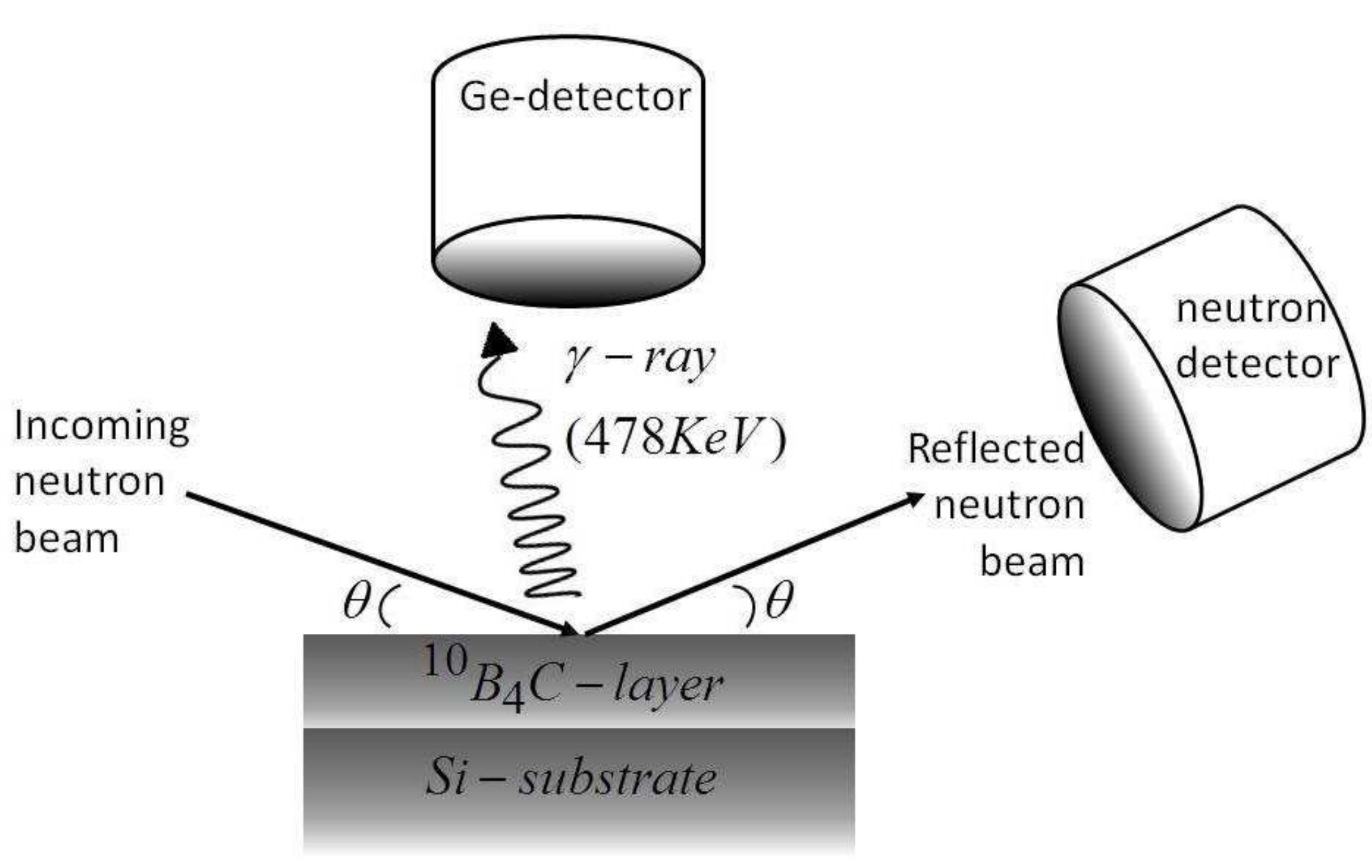}
\caption{\footnotesize Schematic of the experiment on SuperAdam.}
\label{figexpsupadam9046}
\end{figure}
\\ For a given sample we record for each angle during the scan a spectrum for the
Ge-detector and a neutron detector image of the reflected neutrons
knowing the actual normalization given by the direct beam. \\ Each
point of the absorption curve has been obtained by fitting the
Ge-detector spectrum, around the $478\,KeV$ $\gamma$-ray photo-peak,
with a model that includes a linear background. The latter is
subtracted from the actual number of counts and it takes into
account a Compton background deriving from other $\gamma$-ray
energies.
\\ For small $q$ (small $\theta$), there is an overlap between the
direct beam and the reflection images. The raw image is projected
over the $x$-axis and fitted. Reflected intensities have been fitted
using a double gaussian in order to decouple the reflected beam from
the direct beam. Figure \ref{acr578g4} shows an example of this fit
for a given angle. The raw image of the detector shows the two
peaks: on the left, it is the reflected beam and, on the right, the
direct beam through the sample. The number of counts in the
reflected peak is then obtained by the fitting parameters: $a$ the
gaussian amplitude, $m$ its center and its standard deviation
$\sigma$, through $\sqrt{2\pi}\,a\sigma$.
\begin{figure}[!ht]
\centering
\includegraphics[width=7.8cm,angle=0,keepaspectratio]{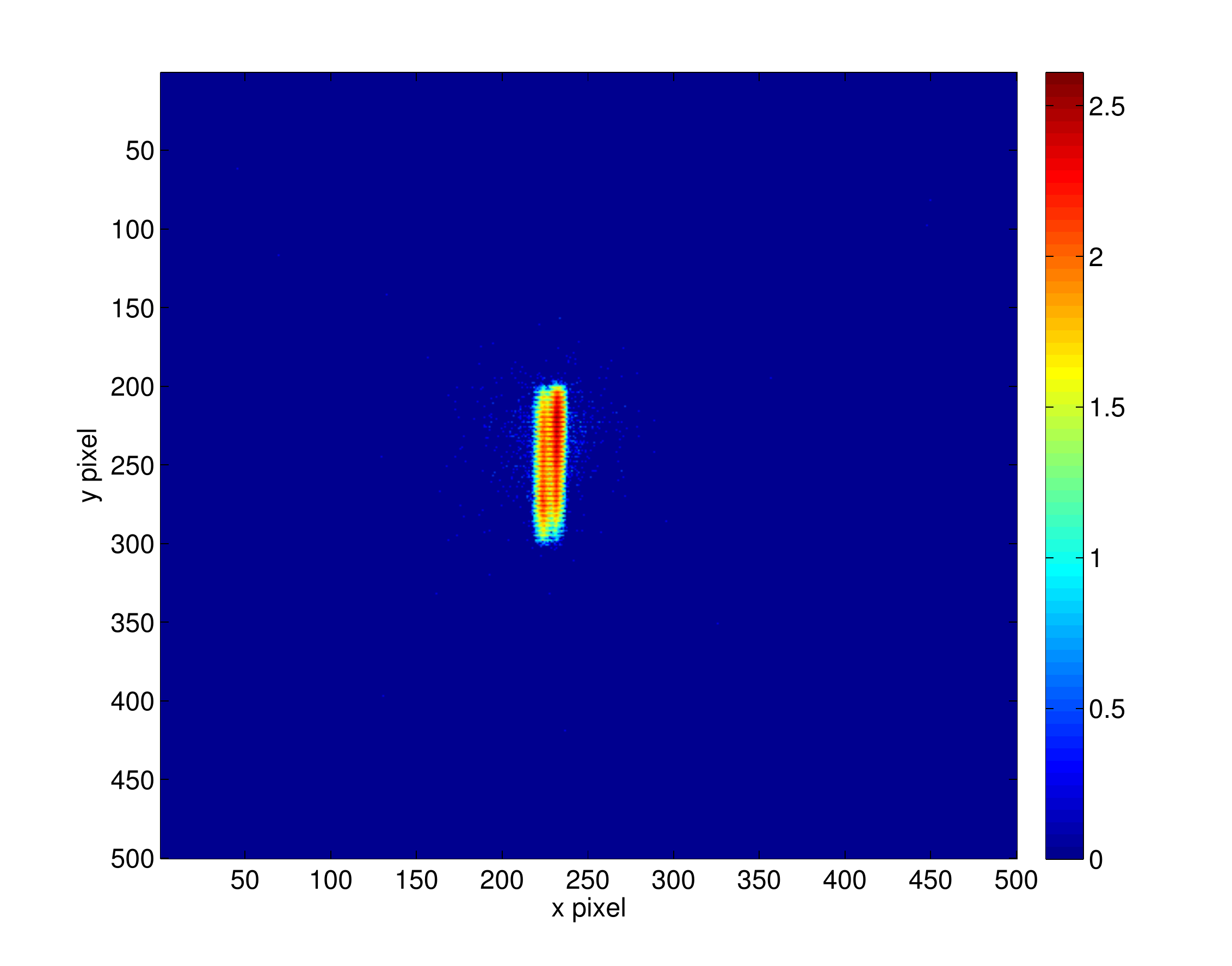}
\includegraphics[width=7.8cm,angle=0,keepaspectratio]{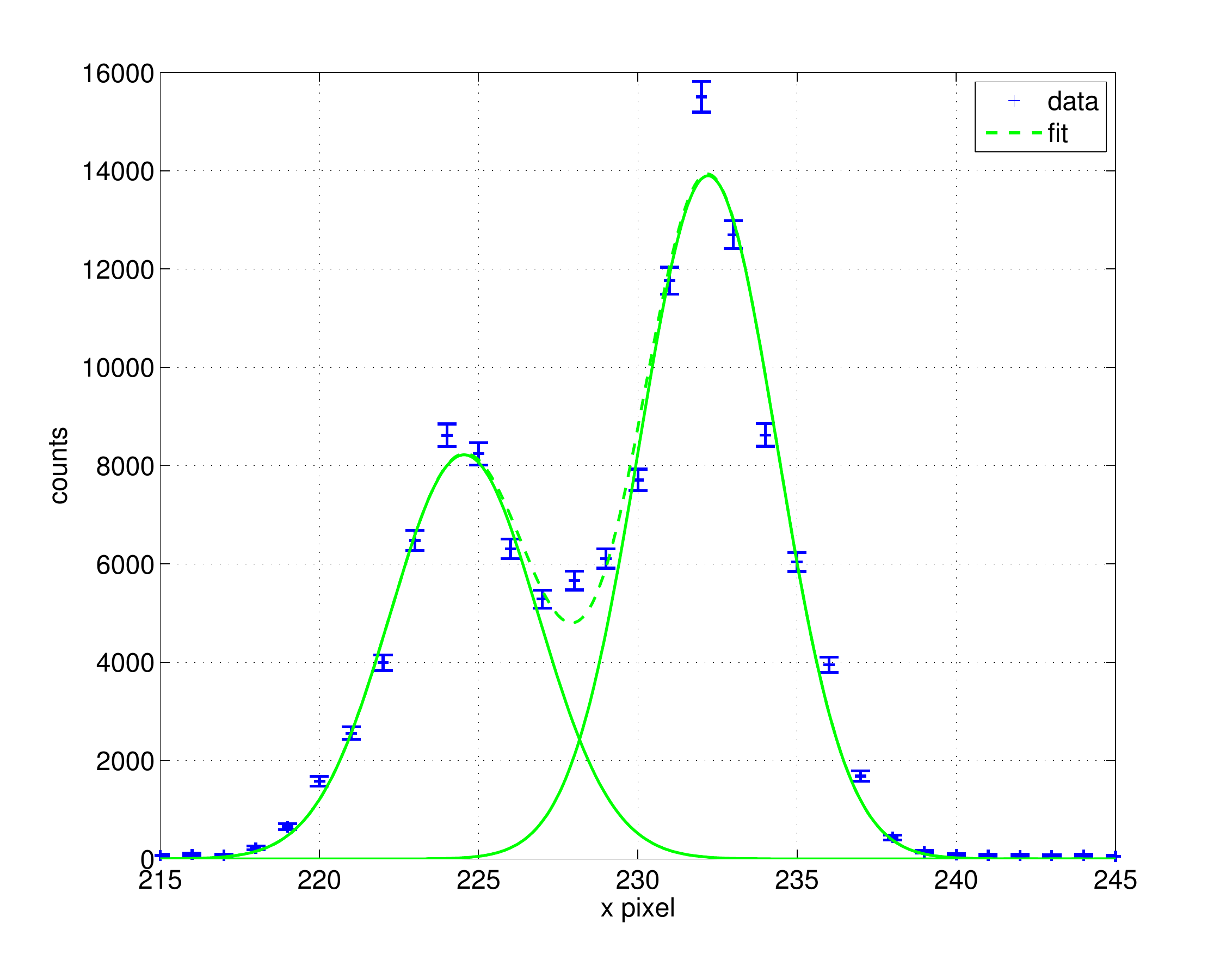}
 \caption{\footnotesize Raw image of the SuperAdam detector (left), projeceted intensities over
 the $x$-axis and the double gaussian fit (right).} \label{acr578g4}
\end{figure}
\\ At higher $q$, higher $\theta$, the direct beam is far
from the reflected beam on the detector. Usually the background is
not symmetric under the reflected peak but it is more intense on the
side of the direct beam. In order to subtract this contribution from
the reflected intensity, it has been fitted with a single gaussian
and a linear background fit. The intensity is again obtained from
the fit parameters.
\\ On a standard reflectivity measurement the angular scan starts from
zero and rises up to one or few degrees; since the sample length is
not infinite there will be a certain point in the scan when the size
of our beam coincides with the actual projected size of the sample,
this is the so-called over-illumination angle ($\theta_{over}$).
Therefore the raw intensity of the reflection rises until
$\theta_{over}$ and then behaves as an absolute reflectivity
profile. Hence a data correction has to be taken into account in
order to transform the intensity of the reflection into
reflectivity. Moreover, on SuperAdam the angles are known with an
error of $0.2\,mrad$; thus a slight shift in the position of the
angle is accepted and also the quantity $\theta_{shift}$ is a free
fitting parameter.
\\ The model considered to fit the data is the one explained in
in Section \ref{absrelfec6} where a $^{10}B_4C$ layer is placed on a
$Si$ substrate. This model is described in details in Appendix
\ref{modelreflect}.
\\ The fitting routine is a standard reduced chi-square minimization
that takes into account both absorption and reflectivity profiles at
the same time. The measurements contain about 150 points for each
curve and $\theta_{over}$, $\theta_{shift}$, $^{10}B_4C$ scattering
length density (real and imaginary parts), layer roughness
$\sigma_r$, thickness, and Ge-detector efficiency are the free
fitting parameters.
\begin{figure}[!ht]
\centering
\includegraphics[width=10cm,keepaspectratio]{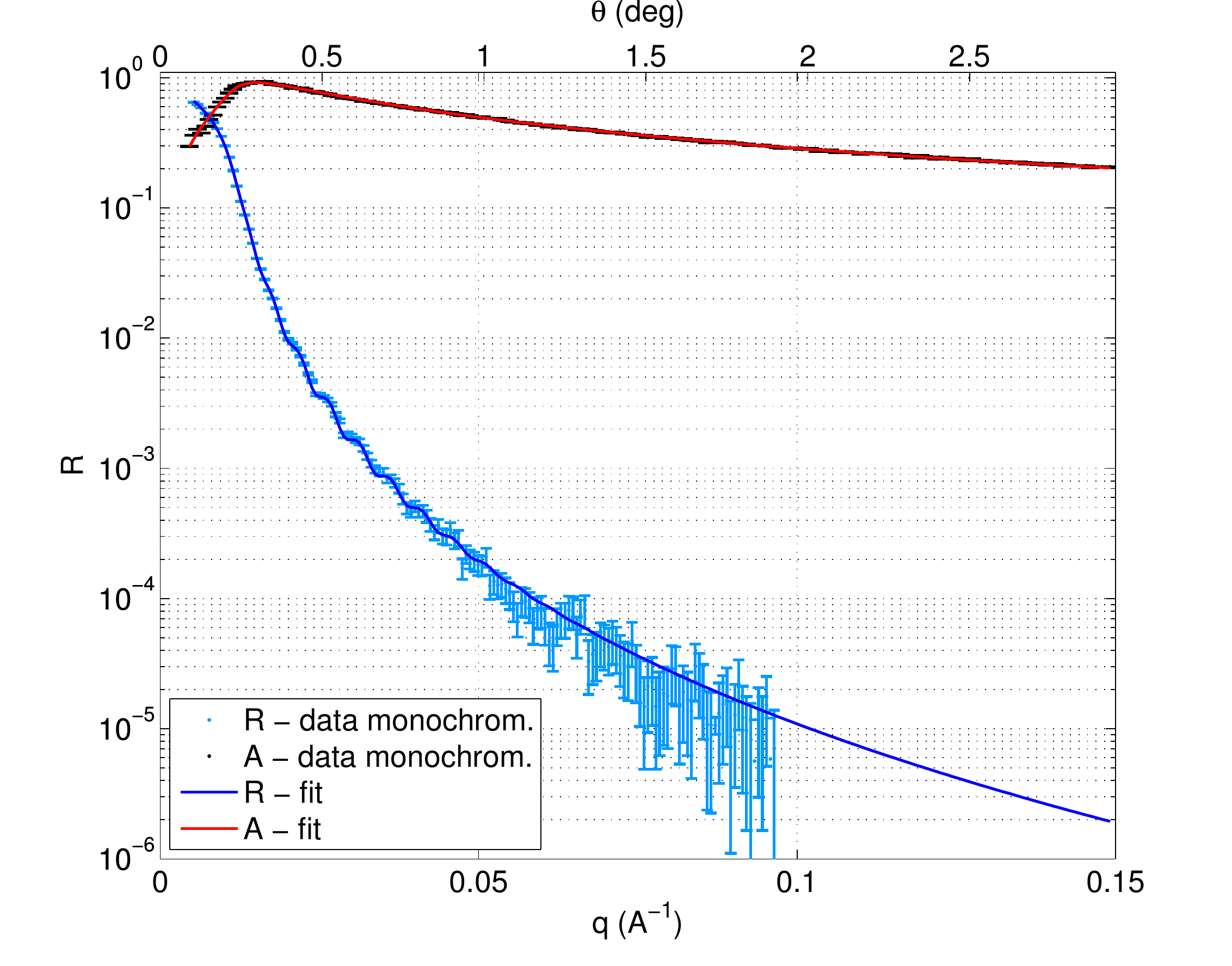}
\caption{\footnotesize Measured reflectivity for the $100\,n m$
$^{10}B_4C$ sample on $Si$ as a function of $q$. Absorption was
measured from the $\gamma$-ray yield.} \label{fgori895667}
\end{figure}
\\ Figure \ref{fgori895667} shows the reflectivity and absorption
profiles for the $100\,n m$ $^{10}B_4C$ sample on $Si$. In Table
\ref{tabd5680232} the fitting parameters obtained are listed.
\\ We suppose the fitting parameters to be independent of each other.
We fix the complete set of parameters as obtained by the chi-square
minimization. We modify one parameter at once until the chi-square
increases by $1$. We take this as the associated error in the
parameter estimation.
\\ Referring to Figure \ref{fgori890}, we can observe that the
reflectivity as measured in a ToF instrument or using angular scan
in monochromatic mode is the same apart from a background tail at
higher $q$ values. This is the experimental confirmation of what
mentioned in Section \ref{neutrefltheoint}, only the normal
component of the wave-vector is affected by the layer independent of
what $\theta$ and $\lambda$ combination is used, and even in the
case of strong absorbing layers.
\\ No layer thickness value has been found for the thicker film
because it can be considered as bulk. Because of absorbtion not a
single neutron can reach the substrate and can be reflected toward
the detector. The thinner layer thickness parameter is estimated to
be $(121\pm2)nm$ thick with a reduced $\chi^2$ of $1.2$. The fitting
routine on the $1\,\mu m$ sample ends up with a worse $\chi^2$ of
about $20$. Referring to Figure \ref{fgori890}, the absorption
measured at low angles is higher than foreseen by the model because
of the presence of more converter material exposed to the beam and
that is not taken into account by the model used to fit the data.
The presence of this additional $^{10}B_4C$, on the substrate back
face and edge, is due to the deposition method.
\begin{table}[!ht]
\centering
\begin{tabular}{|c|c|c|c|c|c|}
\hline \hline
sample & d(nm) & $\theta_{over}(mrad)$ & $\theta_{shift}(mrad)$ & Ge eff.($\%$) & $\chi^2$ \\
\hline
$1\,\mu m$ $Si$ & bulk & $14.0\pm0.1$  & $0.22\pm0.02$ & $4.91\pm0.04$ & $20.7$  \\
$100\,nm$ $Si$  & $121\pm2$  & $18.7\pm0.5$  & $0.74\pm0.04$ & $5.19\pm0.08$ & $1.2$  \\
\hline \hline
\end{tabular}
\\[0.3cm]
\begin{tabular}{|c|c|c|}
\hline \hline
sample & $N_b$(\AA$^{-2}$) & $\sigma_r$(\AA)  \\
\hline
$1\,\mu m$ $Si$ &$((2.48\pm0.02)-(1.01\pm0.02)i)\cdot10^{-6}$ & $38\pm3$ \\
$100\,nm$ $Si$  & $((2.50\pm0.05)-(1.11\pm0.02)i)\cdot10^{-6}$ & $45\pm9$ \\
\hline \hline
\end{tabular}
\caption{\footnotesize Parameters found by the model fitting
routine.} \label{tabd5680232}
\end{table}
\\ Expected SLD is $N_b=(1.6-1.11\,i)\cdot10^{-6}\,$\AA$^{-2}$, while the results for
both samples is about $N_b=(2.5-1.1\,i)\cdot10^{-6}\,$\AA$^{-2}$.
Note the significant difference in the real part and the good
agreement for the imaginary part.
\\ By neglecting minor contaminants in the sputtered layer,
the imaginary part of the SLD is given entirely by $^{10}B$ while
the real part is determined by $^{10}B$, $^{11}B$ and $^{12}C$ (see
Table \ref{crosssect671}). The imaginary part of the fitted SLD
corresponds to the calculated value based on the ERDA measurement
which implies that the $^{10}B$ number density is in agreement and
equals $n_{(^{10}B)}=1\cdot10^{23}\frac{1}{cm^3}$ with an error of
$6\%$.
\\ Given that the scattering length of $^{11}B$ and $^{12}C$ are both
equal to $6.65\,fm$, their ratio only matters in the mass density.
The sum of their densities is determined by the real part of the SLD
and the calculated density of $^{10}B$, itself determined by the
imaginary part. Respecting the error bars on the SLD given by the
fit, this results in a restriction on the interval of the $^{10}B$
fraction and the mass density. They are listed in Table
\ref{tabres56436sevy}.
\begin{table}[!ht]
\centering
\begin{tabular}{|c|c|c|c|c|c|}
\hline \hline
 $\rho\,(g/cm^3)$ & $\%\,^{10}B$ & $\%\,^{11}B+\%\,^{12}C$ \\
\hline
 $[2.34,2.46]$ & $[70,73.5]$ & $[30,26.5]$ \\
\hline \hline
\end{tabular}
\caption{\footnotesize Ranges of density and composition compatible
with the fitted SLD.} \label{tabres56436sevy}
\end{table}
\\ The measured ERDA values are outside the given intervals, and imply higher
levels of non $^{10}B$ atoms. We do not explain the difference.
\\ We only considered a uniform layer consisting of $^{10}B$,
$^{11}B$ and $^{12}C$ in our model. It is true that the ERDA
analysis, in Figure \ref{carites2}, shows a thin oxygen rich layer
of about $20\,nm$ at the surface and some presence of hydrogen both
on the few percent level. A priori such a thin layer should give
interference fringes in the reflectivity profile around
$q=0.016\,$\AA$^{-1}$ and $q=0.047\,$\AA$^{-1}$ which are totally
absent in the data shown in Figure \ref{fgori890}. This is why we
did not include such an extra thin layer in the model.
\\ Values picked within these intervals and inspired by the ERDA
measurement are shown in Table \ref{scge56}.
\begin{table}[!ht]
\centering
\begin{tabular}{|c|c|c|c|c|c|}
\hline \hline
 & $\rho\,(g/cm^3)$ & $\%\,^{10}B$ & $\%\,^{11}B$ & $\%\,^{12}C$ & others\\
\hline
ERDA           & $2.24$ & $79$ & $2.4$ & $17$ & 1.6 \\
Reflectometry  & $2.40$ & $71.5$ & $2.5$   & $26$ & 0\\
\hline \hline
\end{tabular}
\caption{\footnotesize Density and composition of the sputtered
layers.} \label{scge56}
\end{table}
\\ The samples of $1\,\mu m$ deposited on $Al$ have been also measured,
but no specular reflection has been observed while absorption is
comparable to what was observed on the $Si$-sample.
\\ Off-specular reflection is not observed in any sample ($Si$ or $Al$),
it is always below the background level.
\\ In order to diminish the reflection effect in a detector, it is
sufficient to have a rough surface ($>100\,nm$). This can be of
importance if one wants to build a detector based on micro-strips
and solid converters. Operated at small angle, the absorber
deposition on glass could not have a large enough roughness to avoid
significant reflection.
\\ It has to be pointed out that an excessive roughness will also degrade
the efficiency. When the roughness becomes comparable to the
fragments path lengths in the converter ($\sim1\,\mu m$ for
$^{10}B_4C$) the surface can not be considered flat anymore. The
theory of the efficiency for converters under an angle was drawn
assuming the layer to be flat. This flatness is essential for the
neutron to encounter a lot of $^{10}B$, and the conversion fragments
to be close to the surface to be able to escape. If the roughness
starts to be comparable to the conversion fragments ranges, this
assumption does not work anymore. There is a drop in the expected
efficiency.
\\ We repeat, in Figure \ref{ertdb87}, the ToF measured reflectivities
as a function of of the neutron wavelength $\lambda$ (while in
Figure \ref{fgori890} they are shown as a function of $q$). In
Figure \ref{ertdb87} we notice that if we use a converter at
$1^{\circ}$ (red curve) about $30\%$ or more of the neutrons are
reflected, thus not converted, for wavelengths larger than $20$\AA.
\begin{figure}[!ht]
\centering
\includegraphics[width=10cm,keepaspectratio]{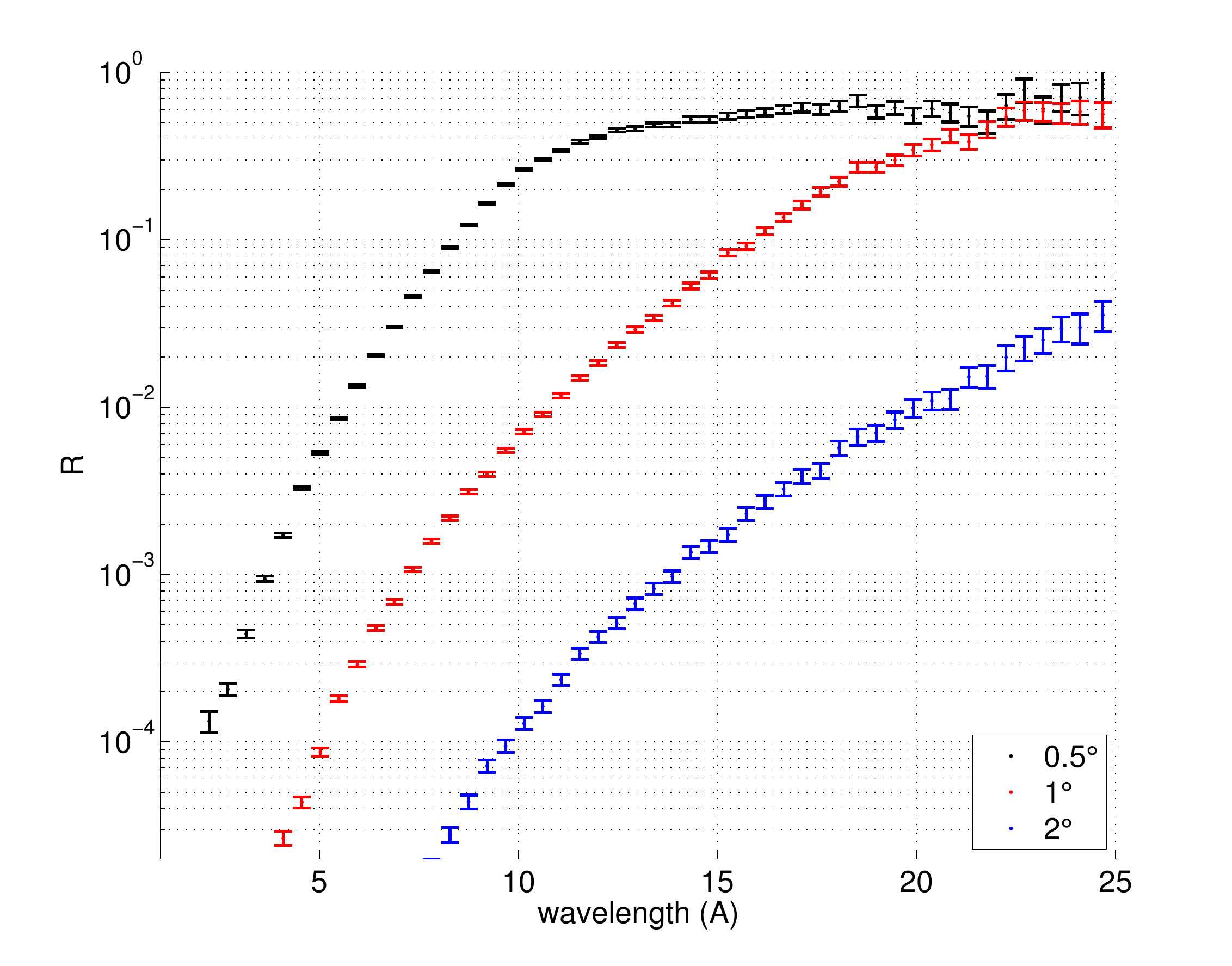}
\caption{\footnotesize Measured reflectivity of $1\,\mu m$
$^{10}B_4C$ deposited on $Si$ as a function of $\lambda$ for 3
different angles.} \label{ertdb87}
\end{figure}
\\ According to Equation \ref{eqwty7} we can calculate the corrected
efficiency for reflection. The plot in Figure
\ref{figexpeffanglesi47} has to be modified as shown in Figure
\ref{figexpeffanglesi4767}. We calculate the corrected efficiency,
for an energy threshold of $100\,KeV$, for a $1\,\mu m$ thick
$^{10}B_4C$ single back-scattering layer for three neutron
wavelengths: $1.8\,$\AA, $10\,$\AA \, and $25\,$\AA. According to
what we found from the fit and summarized in Table
\ref{tabd5680232}, the layer roughness used is about $40\,$\AA \,
and the scattering length density is
$N_b=(2.5-1.1\,i)\cdot10^{-6}\,$\AA$^{-2}$.
\\ We recall that, if the layer has a higher
roughness, the effect of the reflection decreases. Aluminium is a
suitable material to avoid neutron reflection instead of Silicon.
\begin{figure}[!ht]
\centering
\includegraphics[width=10cm,angle=0,keepaspectratio]{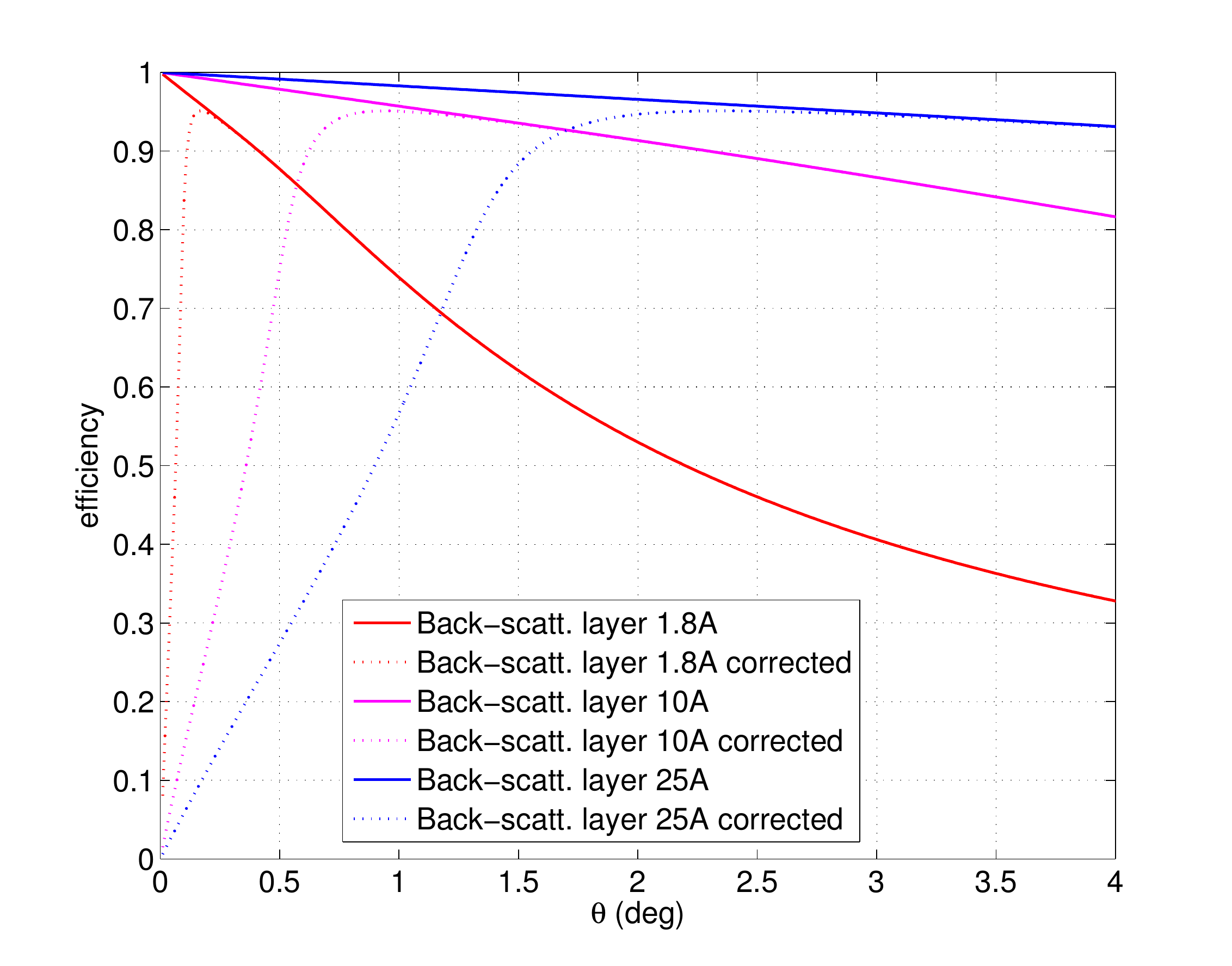}
\caption{\footnotesize Corrected and not corrected efficiency for
reflection calculated for a single layer $1\,\mu m$ thick
$^{10}B_4C$ back-scattering layer at $1.8\,$\AA, $10\,$\AA \, and
$25\,$\AA \, as a function of the neutron incidence angle. Energy
threshold applied: $100\,KeV$.} \label{figexpeffanglesi4767}
\end{figure}
\\ We observed a Doppler effect on the $478\,KeV$ $\gamma$-ray photo-peak due to the sample
orientation. When a neutron is captured by $^{10}B$, it is the
$^7Li$ fragment that, in the $94\%$ branching ratio, goes in an
excited state that emits the $\gamma$-ray. If $^7Li$ is emitted
toward the substrate it is stopped within about $120\,fs$, to be
compared with the $^7Li^{*}$ half life of $t_{1/2}=73\,fs$ before it
goes to its ground state emitting the photon. A significant fraction
of the $^7Li$ fragments will hence emit its photon when it is at
rest. In a thin layer, if $^7Li$ is emitted toward the surface it
travels in air for a much longer time. Then the $\gamma$-ray is also
emitted when it is in motion. In a thick converter the volume to
surface ratio is such that the fragment is almost always stopped,
and a fraction of the photons are emitted when stopped. Referring to
Figure \ref{awcf43gv}, in the case A, a thick block of $B_4C$ rubber
was placed in front of the Ge-detector. The spectrum measured is
symmetric in energy. In Figure \ref{awcf43gv}, the peak at
$511\,KeV$ gives an indication of the Ge-detector energy resolution.
All the measurements have been normalized to the intensity of the
$478\,KeV$ $\gamma$-ray photo-peak to compare the peak shape.
\\ When one of our samples is exposed to the beam, case B or C, the
layer is thin enough to assure the fragments to always escape the
layer. On average half of the times the $^7Li$ emits at rest and
half of the times when travels either away from (B)
(\emph{red-shift}) or toward (C) (\emph{blue-shift}) the
Ge-detector. Consequently the $478\,KeV$ $\gamma$-ray photo-peak
results asymmetric due to the Doppler shift.
\begin{figure}[!ht] \centering
\includegraphics[width=7.8cm,keepaspectratio]{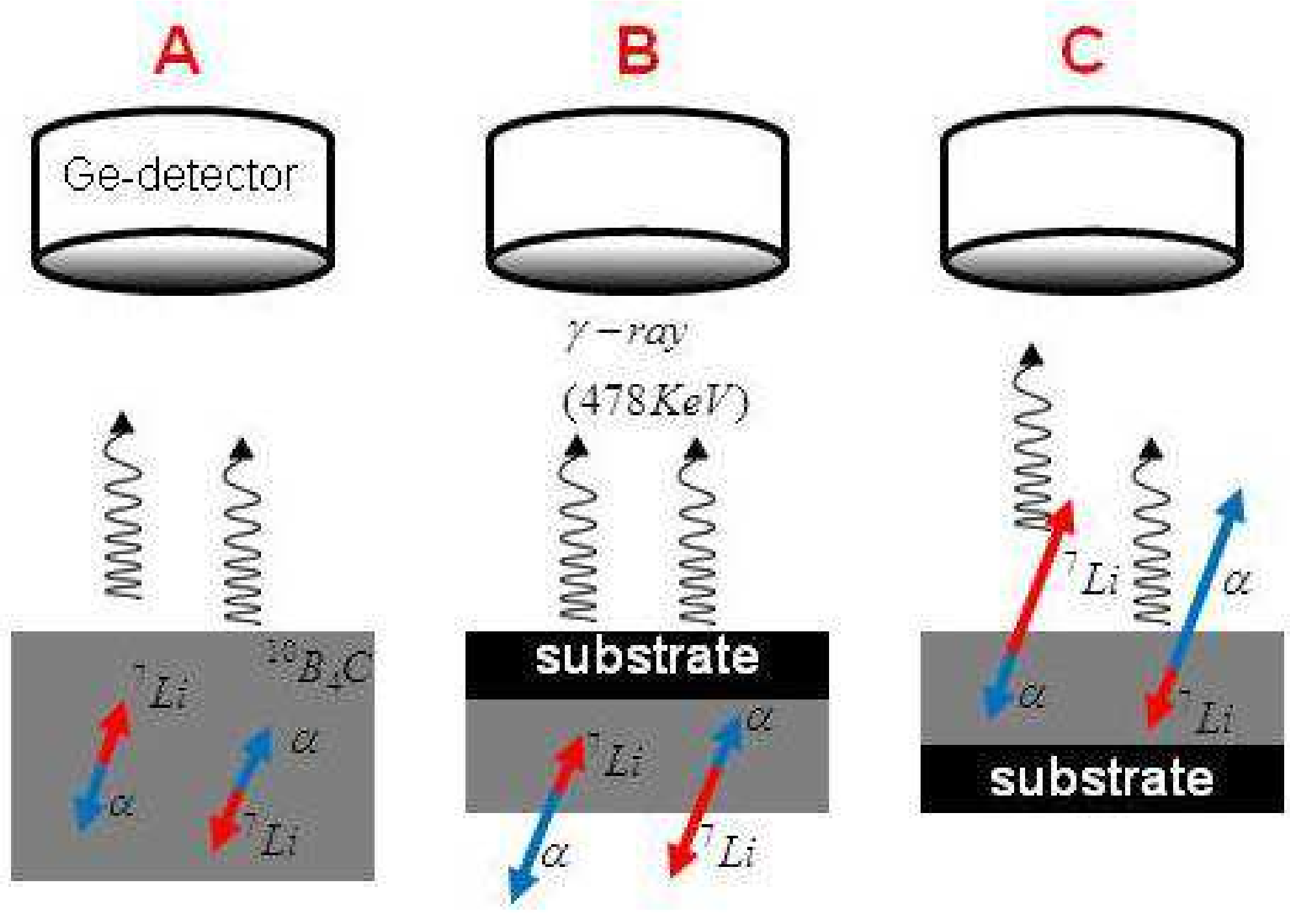}
\includegraphics[width=7.8cm,keepaspectratio]{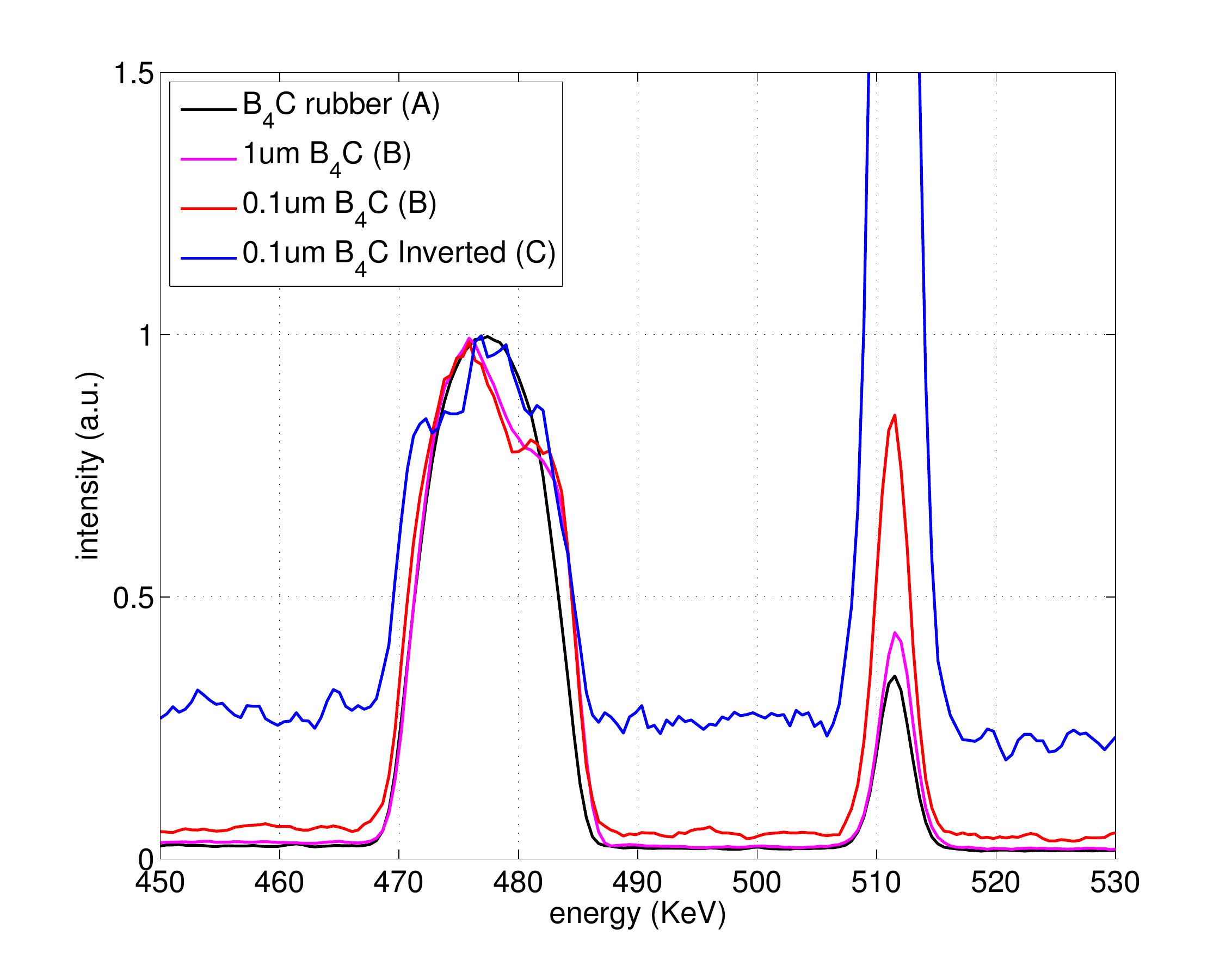}
\caption{\footnotesize Setups used to measure the Doppler effect in
the $478\,KeV$ $\gamma$-ray energy (left). Ge-detector spectrum:
photo-peaks of $478\,KeV$ and $511\,KeV$ $\gamma$-rays (right).}
\label{awcf43gv}
\end{figure}

\chapter{Gamma-ray sensitivity of neutron detectors based on solid converters}\label{Chapt2}

I want to thank Anton Khaplanov who taught me a lot about
$\gamma$-rays. The work presented in this Chapter makes reference to
the work in \cite{khap} and in \cite{in6procc}.

\newpage
\section{Introduction}
Any neutron detector is also a $\gamma$-ray detector. The aim of
this Chapter is to individuate the characteristics that
differentiate neutrons and background non-neutron events, e.g.
$\gamma$-rays. In some cases it is possible to determine the origin
of the electric signal, in other cases the two signals show the same
time structure and same charge yield; hence it is not possible to
distinguish them. \\ Different kinds of radiation can hit the
detector and generate background, but $\gamma$-rays are the most
common in neutron facilities. It is really likely for a neutron to
excite any nucleus with the subsequent emission of $\gamma$-rays.
Thus, a $\gamma$-ray background is always present in neutron
facilities. Moreover at ILL, which is a reactor source, there is an
important contribution that comes directly from the core.
\begin{figure}[h] \centering
\includegraphics[width=7.8cm,angle=0,keepaspectratio]{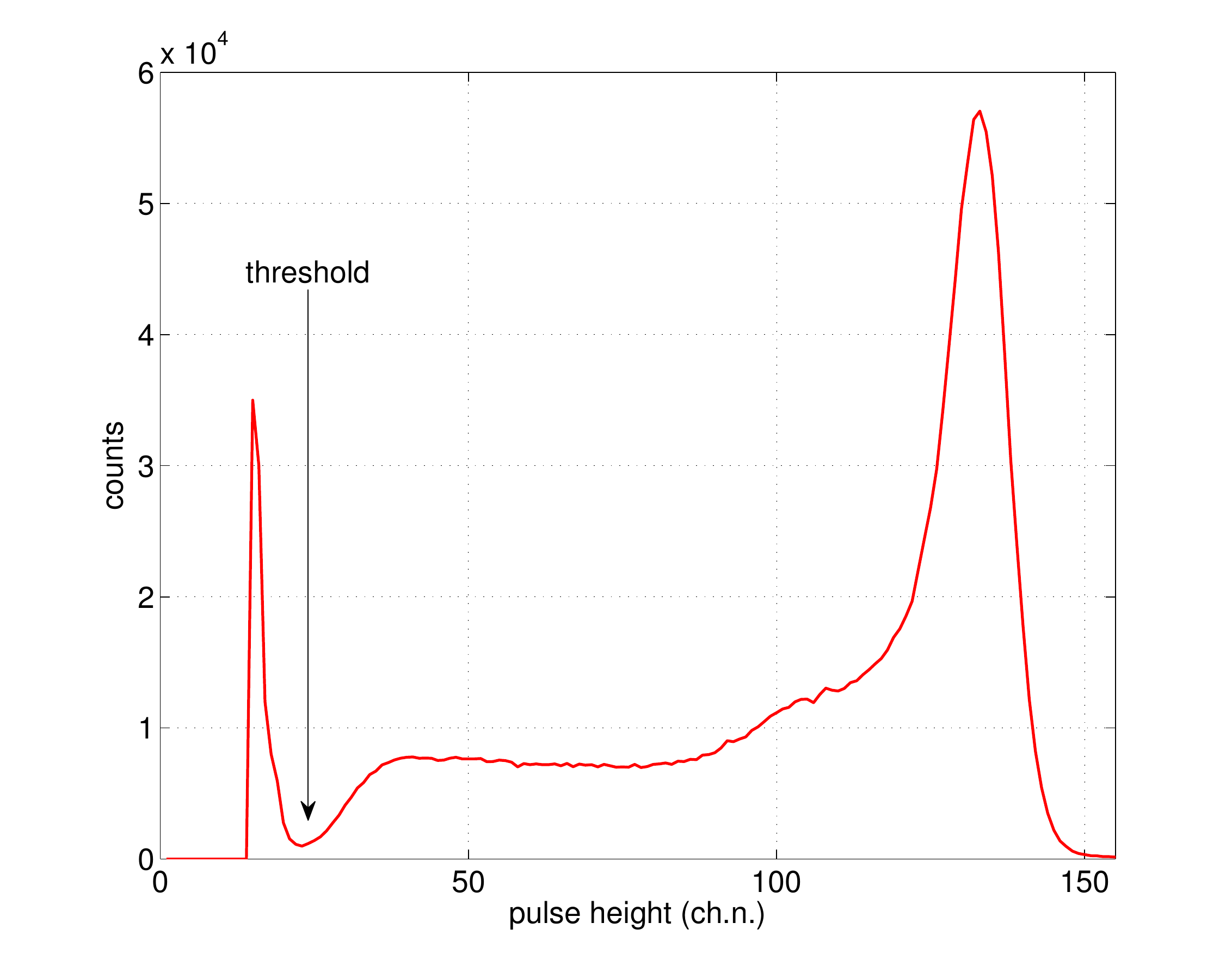}
\includegraphics[width=7.8cm,angle=0,keepaspectratio]{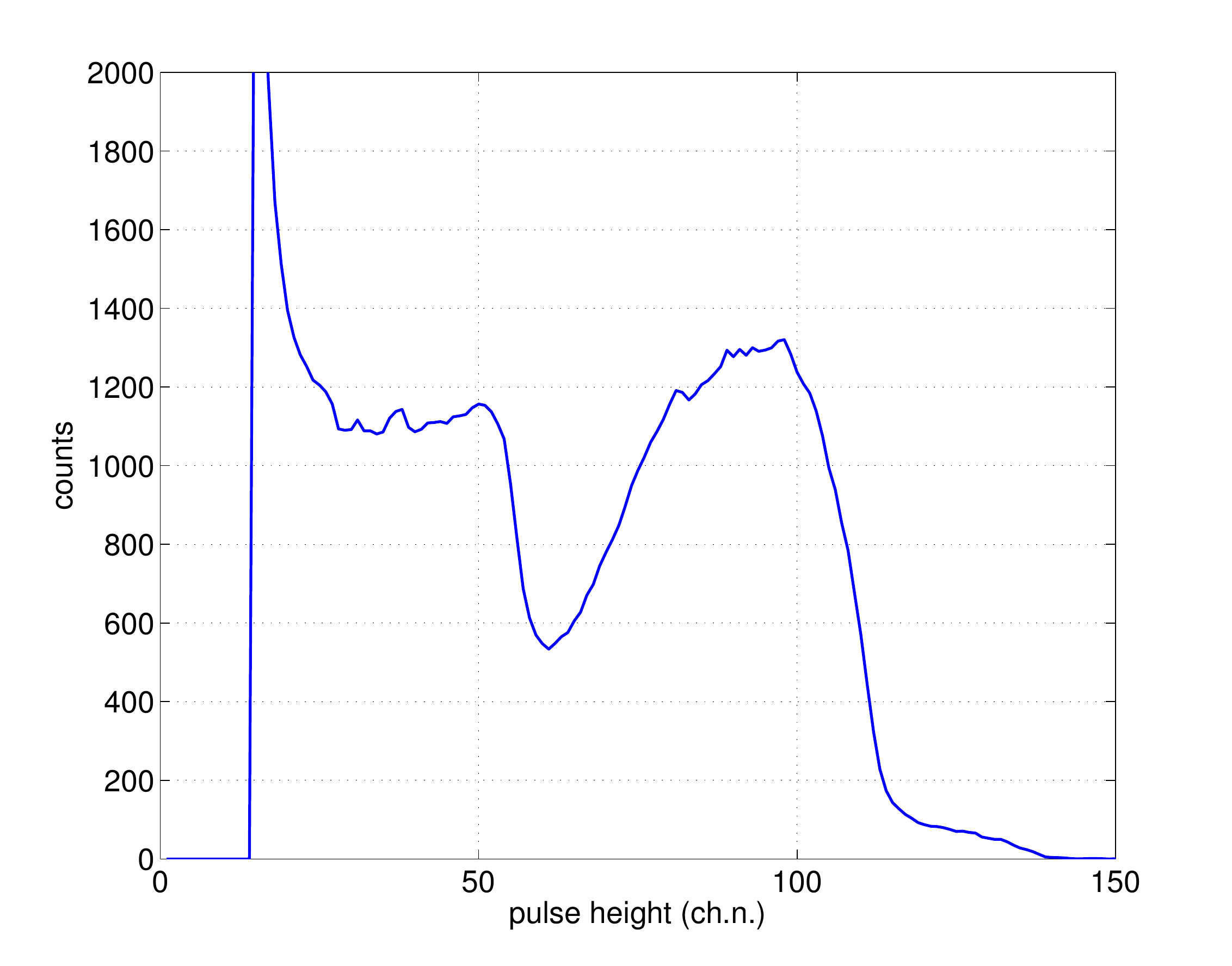}
\caption{\footnotesize $^3He$ PHS (left) and $^{10}B$ PHS (right).
\label{hephs3}}
\end{figure}
\\ In $^3He$-based neutron detectors an efficient neutron to
$\gamma$-ray discrimination can be easily done by applying an energy
threshold. Figure \ref{hephs3} shows two PHS, one for $^3He$ and one
for $^{10}B$. The standard discrimination of $\gamma$-ray and
neutron events is by applying a threshold in the PHS. When a neutron
is captured by $^3He$, the two fragments ionize directly the gas.
Apart from the wall effect, they can always deposit their entire
energy in the gas ($770\,KeV$), and even with the wall effect, a
minimum energy is always deposited, equal to the triton energy. \\
Generally $\gamma$-rays can interact either with the wall of the
detector or directly with the gas. Every time an electron is
generated, by photoelectric absorption or Compton scattering, it
ionizes the gas and gives rise to a net charge. As it will be shown
in more details in Section \ref{geant4faerg}, $\gamma$-rays deposit,
on average, a smaller amount of energy in the gas than neutron
capture fragments. Thus they appear at the lowest energies on the
PHS. Of course there are interactions that give rise to higher
energy events, but those are less likely. In the $^3He$ PHS in
Figure \ref{hephs3}, we notice how $\gamma$-rays originates the low
energy rising on the PHS and how those events are well distinguished
from neutron events. A simple amplitude threshold can be applied to
discriminate against $\gamma$-rays. Naturally, there are some events
that mix up with neutron events but their contribution for most
applications can be neglected. The quantification of the
misaddressed events using the energy threshold method to
discriminate against background for $^3He$ and $^{10}B$ will be
discussed in this chapter.
\\ $^{10}B$ has in principle a better signal-to-noise ratio than $^3He$ because
its energy yield is larger ($2300\,KeV$). A $^{10}BF_3$-based
detector would behave as an $^3He$-based due to its conversion in
the gaseous phase. The resulting PHS would have a wider valley
between neutron and background events.
\\ This is not the case for solid converters, e.g. $^{10}B_4C$.
\\ In Figure \ref{hephs3} a $^{10}B$ PHS is also shown.
The $^{10}B$ does not show a clear difference in the signal
amplitude between a neutron event or a $\gamma$-ray. When a neutron
is captured in solid $^{10}B$ it generates the capture fragments
inside the layer. In order to reach the gas volume they have to
travel toward the escaping surface, hence the energy deposition in
the gas is a continuum down to zero or the minimum detectable energy
(LDD \cite{gregor} or $E_{Th}$). The rise at lower energy is then a
mixture of neutron and background events. It is possible to diminish
the neutron event contribution to the low energy part of the PHS. In
Chapter \ref{Chapt1} we notice that, given the neutron wavelength,
playing with the layer thickness can improve the neutron to
$\gamma$-rays energy separation, but this will affect the layer
efficiency. Let us consider a Multi-Grid \cite{jonisorma} like
detector. The layer thickness has to be optimized to improve the
detector efficiency. A very thin layer shows a better neutron to
background separation, but several layers have to be added in order
to keep the detector efficiency constant. This will increase the
material inside the detector that can cause extra $\gamma$-ray
interactions. On the other hand, thicker layers have a worse energy
separation but the detector is more compact. For a single converter
at grazing angle, the efficiency is maximized, in back-scattering,
when it is very thick. This translates into no energy separation
between neutrons and background on the PHS. Of course, one can still
apply an energy threshold to eliminate $\gamma$-ray events, but
contrary to a $^3He$-based detector this will now also lower the
efficiency of the neutron detector. Hence, if one could find an
efficient method to address the $\gamma$-ray discrimination, this
would be of great interest. In this chapter we explore this
possibility.
\section{The Multi-Grid detector}
\begin{figure}[!ht] \centering
\includegraphics[width=14cm,angle=0,keepaspectratio]{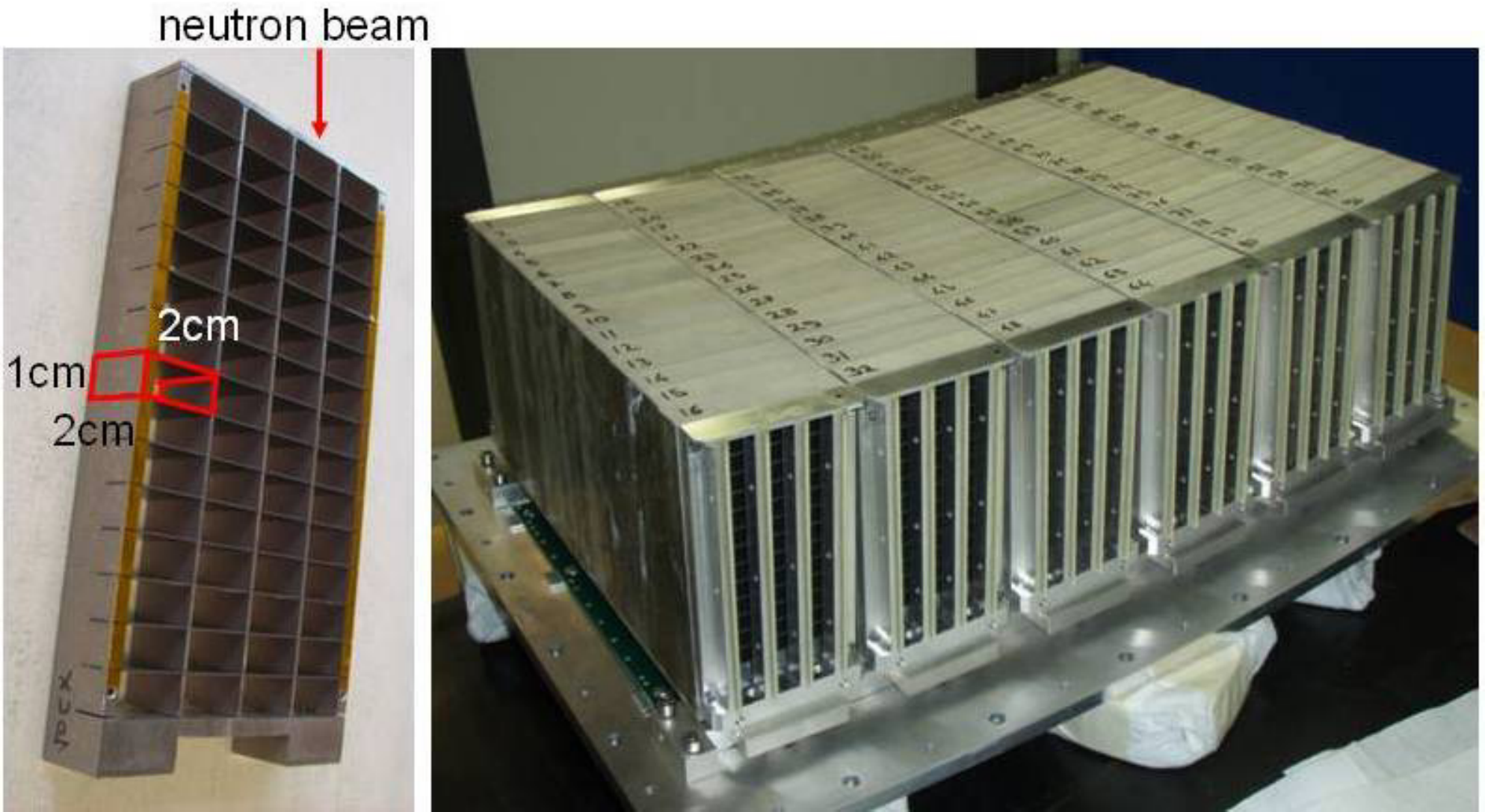}
\caption{\footnotesize A grid containing 15 blades coated with
$^{10}B_4C$ (left). The Multi-Grid detector \cite{jonisorma} tested
on IN6.} \label{MBIN656}
\end{figure}

\begin{figure}[!ht]
\centering
\includegraphics[width=7.8cm,angle=0,keepaspectratio]{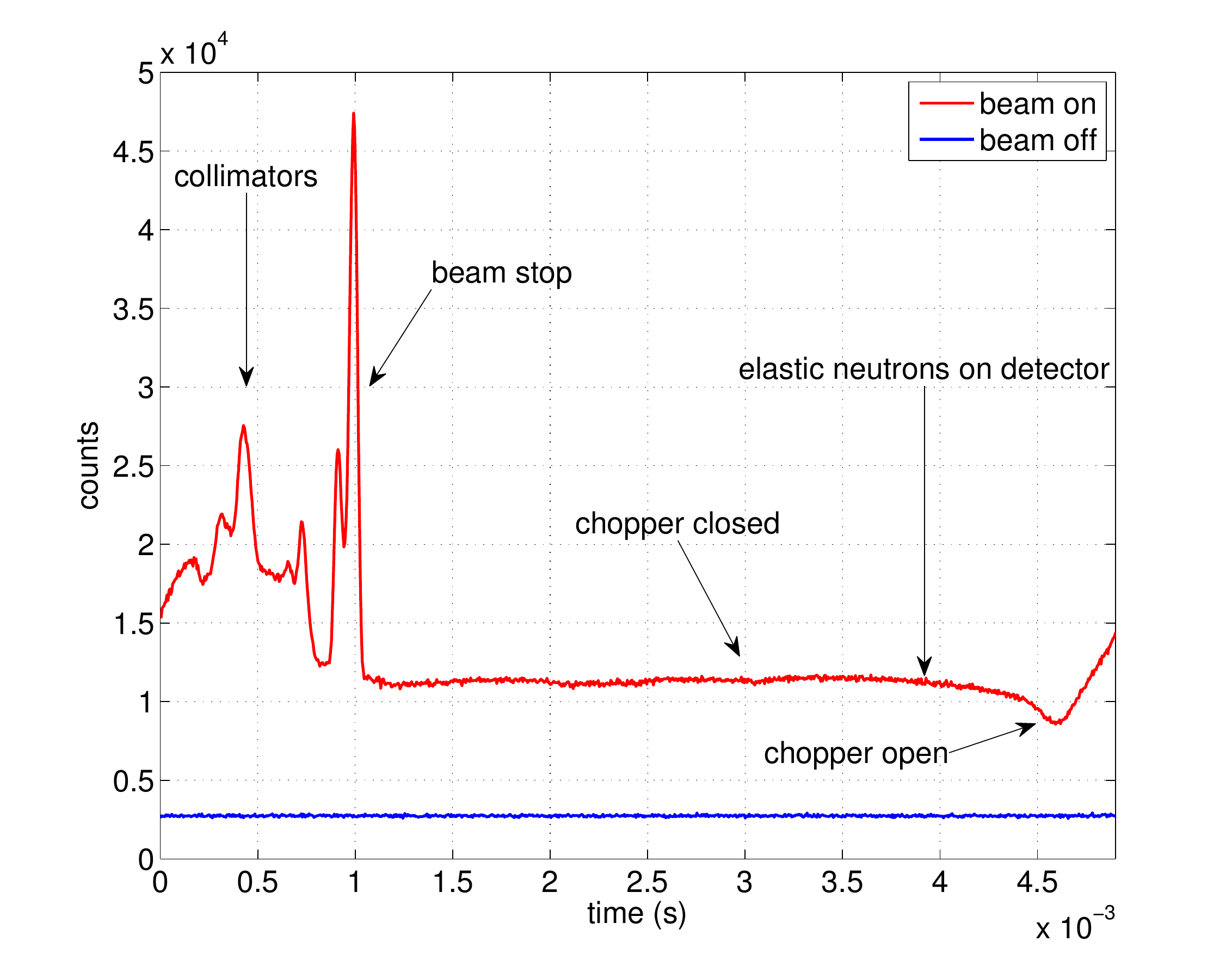}
\includegraphics[width=7.8cm,angle=0,keepaspectratio]{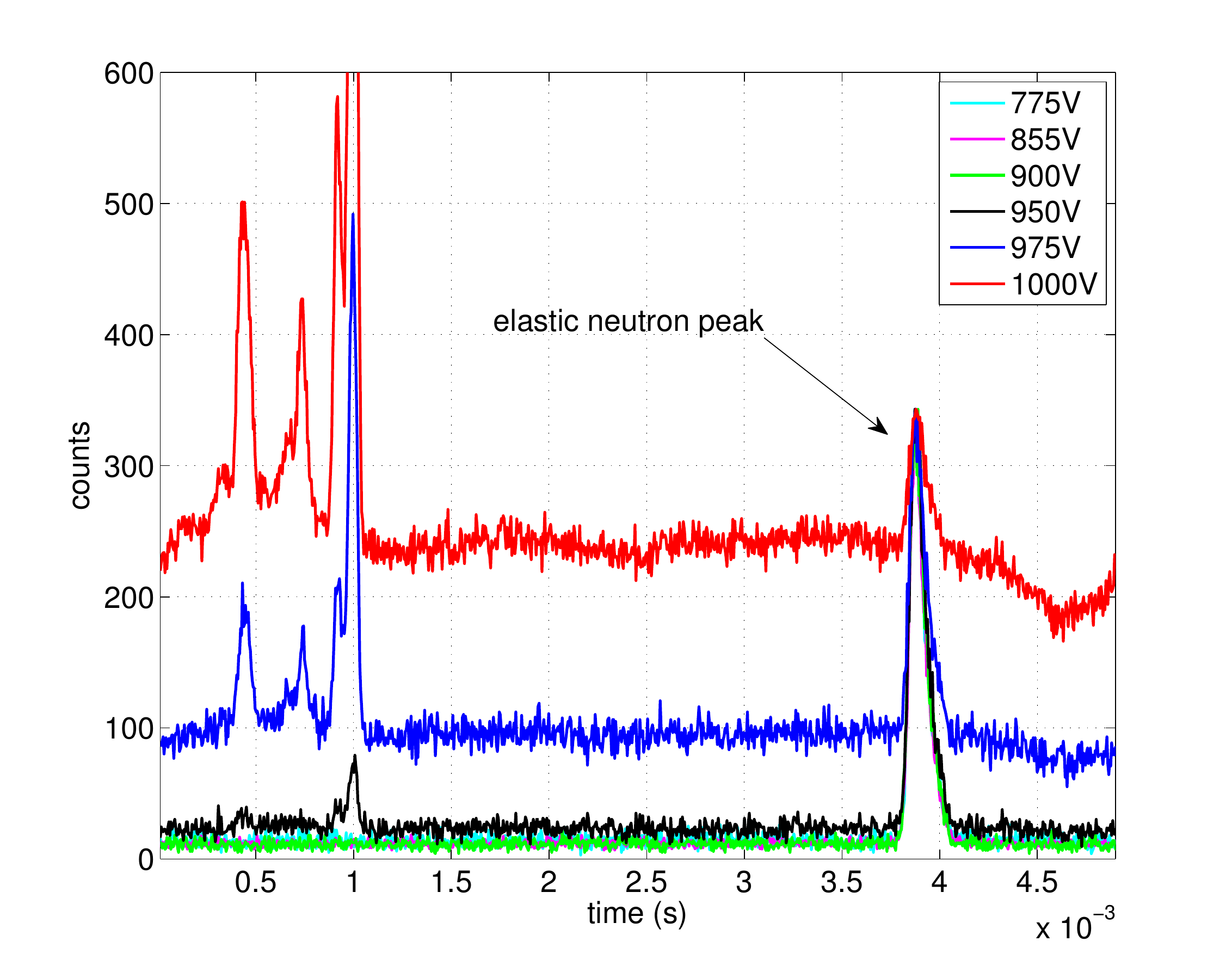}
\caption{\footnotesize A NaI spectrum of the $\gamma$-ray background
on IN6 at ILL temporized with the instrument timing (left). The
Multi-Grid spectrum for several bias voltages. Courtesy of A.
Khaplanov.} \label{IN6spett45g}
\end{figure}

The Multi-Grid detector is a prototype study over the alternatives
to $^3He$ for large area detectors ($>30\,m^2$) for ToF instruments.
\\ The Multi-Grid detector \cite{jonisorma}, \cite{jonitesi1}, is a solid neutron
converter based gaseous detector. Neutron are converted by several
$^{10}B_4C$ layers placed in cascade and alternated with gaseous
detection regions. Neutrons hit each converter at normal incidence.
It contains 15 blades, i.e. 30 converter layers. Each blade is a
$0.5\,mm$ Aluminium substrate coated on both sides with $1\,\mu m$
$^{10}B_4C$ \cite{carina}. In order to get the two-dimensional event
localization the Multi-Grid is segmented into grids. Figure
\ref{MBIN656} shows a grid equipped with 15 blades and the full
mounted detector composed by $96$ grids stacked to form 6
independent modules of 16 grids each. A stack of 16 grids is a
column; 6 columns were placed one after the other in order to cover
about $0.15\,m^2$ (see Figure \ref{MBIN656}).
\\ Each grid acts as a cathode. They are insulated and read-out
independently of each other. Each grid contains 15 blades in order
to form $15 \times 4 = 60$ voxels of $2\times 2 \times 1\,cm^3$
sides. Once the grids are stacked, they form 60 rectangular shaped
tubes. An anode wire is placed all along them. A voxel is identified
by the coincidence between a wire signal and a grid signal. This
detector allows the three-dimensional localization of the neutron
event with the voxel size spatial resolution.
\\ Thanks to its 30 converter layers it shows an efficiency above $50\%$
at $2.5$\AA. The multi-layer detector efficiency optimization was
discussed in details in Chapter \ref{Chapt1}.
\\ To reduce the number of read-out channels the wires were connected
by resistors allowing a charge division read-out. The grids were
connected together row by row. The detector is operated at
atmospheric pressure. $Ar/CO_2$ ($90/10$) was flushed continuously
in the detector.
\\ The resulting detector was installed on IN6 at ILL to be tested and
compared with $^3He$ detectors. The detector replaces 25 tubes of
the standard compliance of IN6 \cite{in6procc}.
\\ A NaI scintillator was placed inside the chamber of the IN6 detector in order to
measure the $\gamma$-ray background. The scintillator energy
calibration was performed by using a $^{22}Na$ source as explained
in Chapter \ref{chaptintradmatt} where we discussed the photon
interactions. IN6 is ToF instrument, a chopper spectrometer more
precisely, where a pulsed monochromatic neutron beam is obtained by
a sequence of choppers that allow to determine the Time-of-Flight of
neutrons, i.e. their energy. Neutrons arriving at the detector show
a time structure. Those scattered elastically in the sample form a
distinct peak in the time spectrum, while those scattered
inelastically arrive earlier or later than the elastic peak
depending on the energy transfer to or from the neutron. A large
$\gamma$-ray background is also generated by the instrument itself
by the surrounding equipment. The background originating in the
instrument also shows a time structure. Figure \ref{IN6spett45g}
shows, on the left, the integral over all the photon energies of the
NaI detector spectrum, recorded both when the neutron beam is opened
and when the IN6 beam is shut off.
\\ While the $\gamma$-ray background with the beam off is
constant, we observe its time structure when the beam is on. Any
interaction of the neutron beam with the instrument materials
generates background. In Figure \ref{IN6spett45g} we can distinguish
when neutrons cross the collimators and the beam stop.
\\ On the right plot in Figure \ref{IN6spett45g} is shown the same
time spectrum as for the NaI scintillator but for the Multi-Grid
prototype for several bias voltage. The spectra have been normalized
to the elastic neutron peak. We notice that as we increase the bias
voltage in the detector the $\gamma$-ray background becomes more and
more important with respect to the neutron contribution. The
background shows the same time structure as for the NaI
scintillator. From this measurement we understand that there exists
an optimal operational voltage to get the best neutron to background
discrimination. The best signal-to-noise ratio is obtained by
setting the bias voltage to $900\,V$.
\section{GEANT4 simulations of interactions}\label{geant4faerg}
\begin{figure}[!ht]
\centering
\includegraphics[width=7.8cm,angle=0,keepaspectratio]{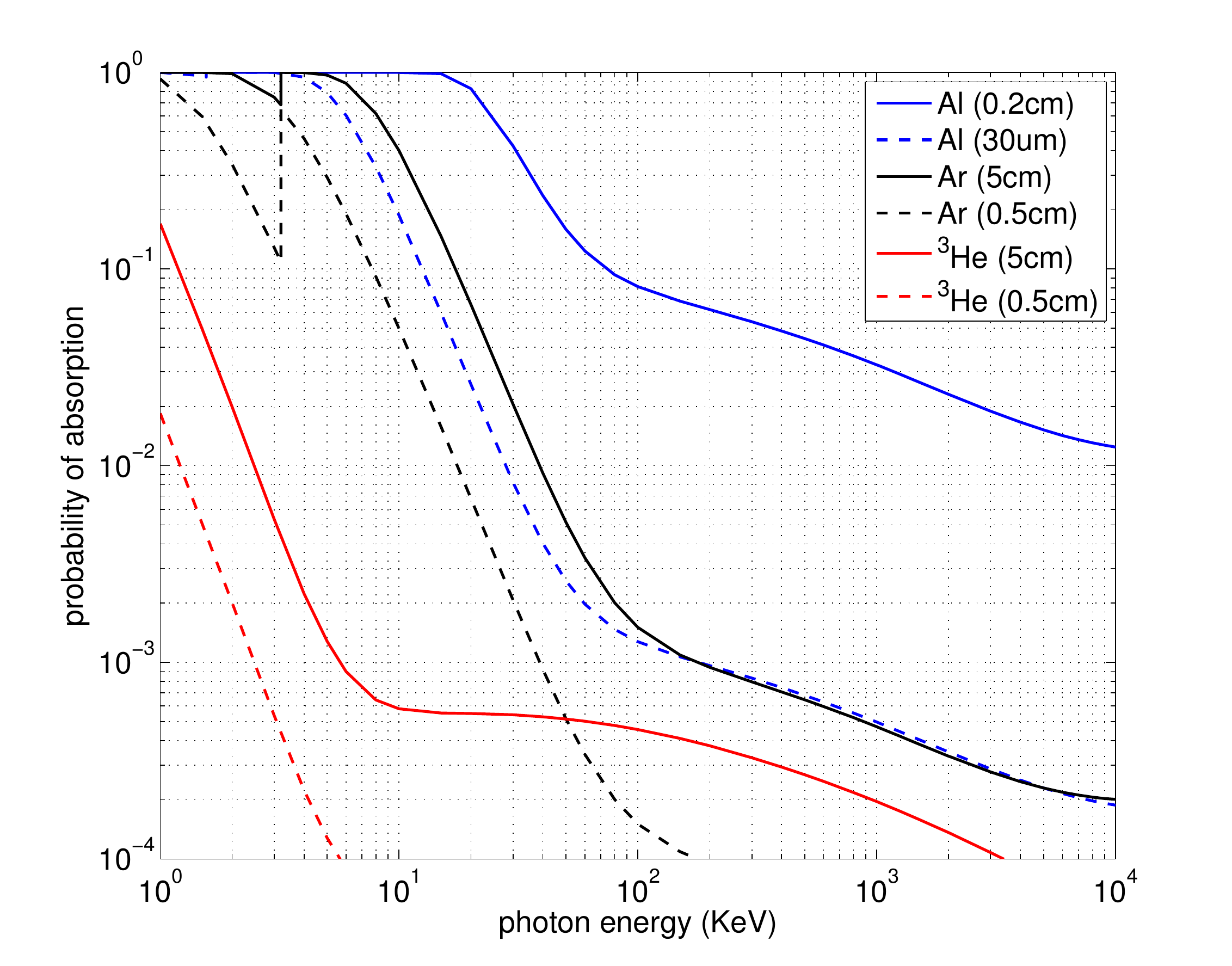}
\includegraphics[width=7.8cm,angle=0,keepaspectratio]{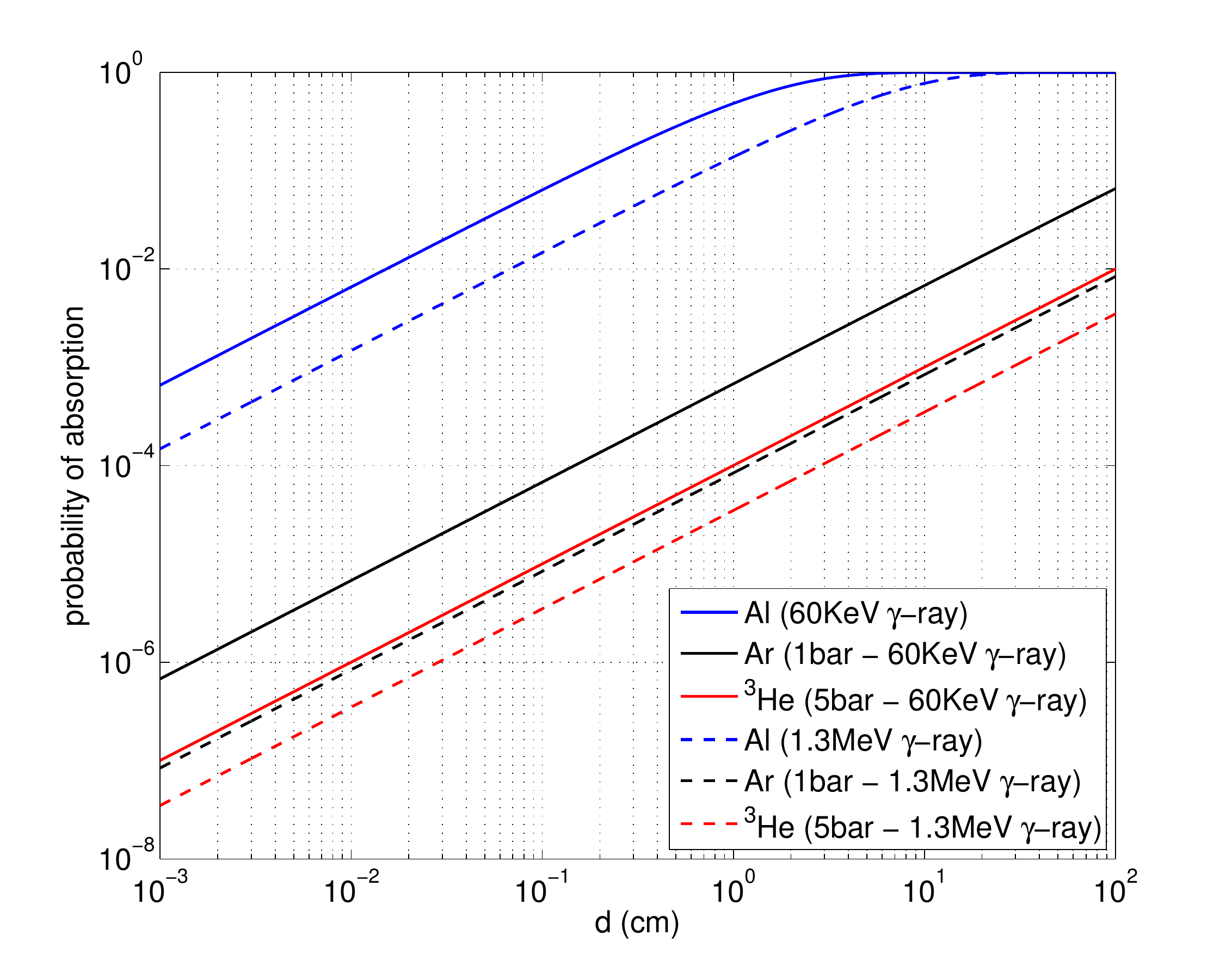}
\caption{\footnotesize Probability of photon interaction with $Al$,
$Ar$ and $^3He$ as a function of the photon energy (left) and of the
traveled distance in matter (right).} \label{figprobintegamdetec67}
\end{figure}

\begin{figure}[!ht]
\centering
\includegraphics[width=10cm,angle=0,keepaspectratio]{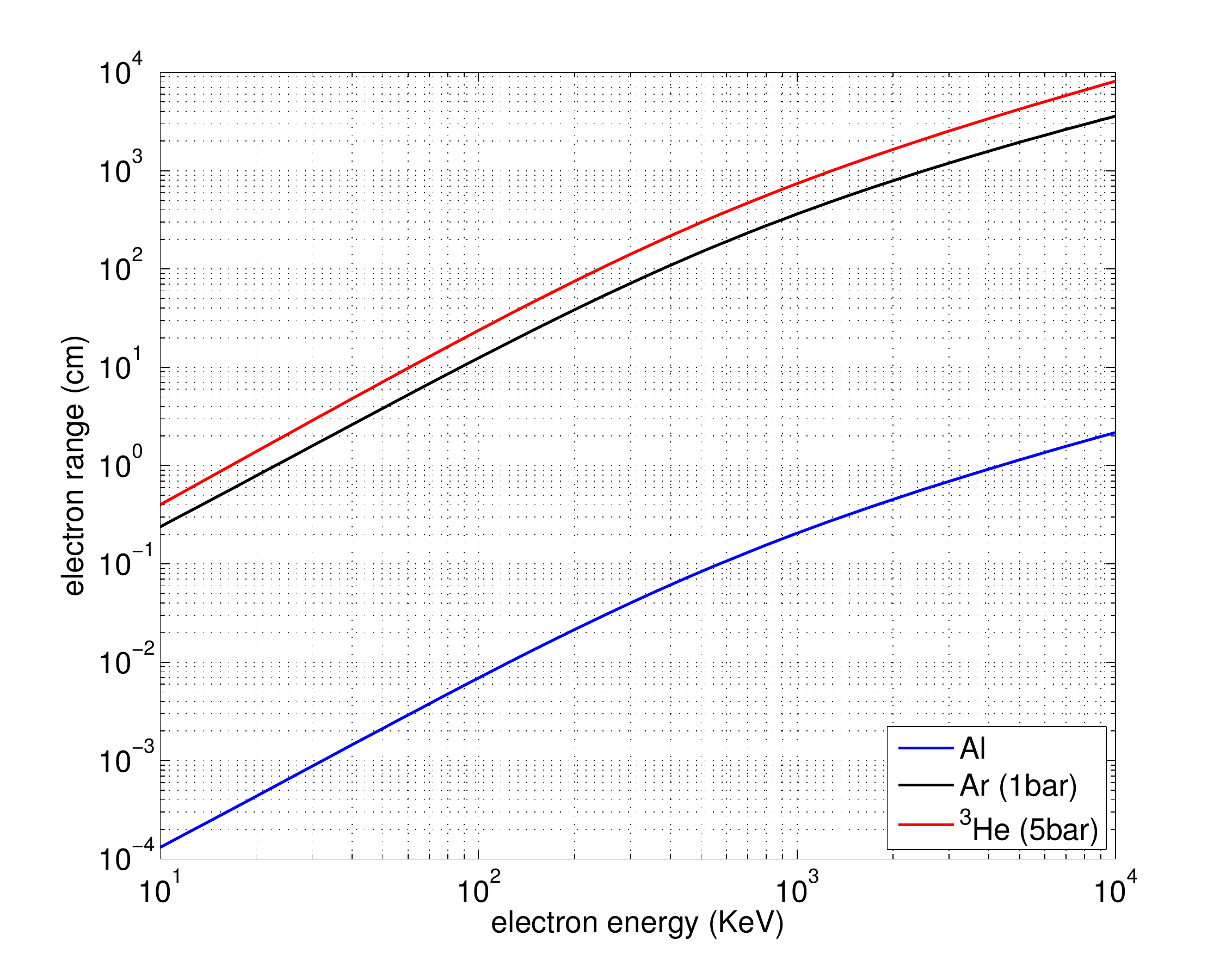}
\caption{\footnotesize Electron range in $Al$, $Ar$ and $^3He$ as a
function of its energy.} \label{figprobintegamdetec67elect}
\end{figure}

\begin{figure}[!ht]
\centering
\includegraphics[width=10cm,angle=0,keepaspectratio]{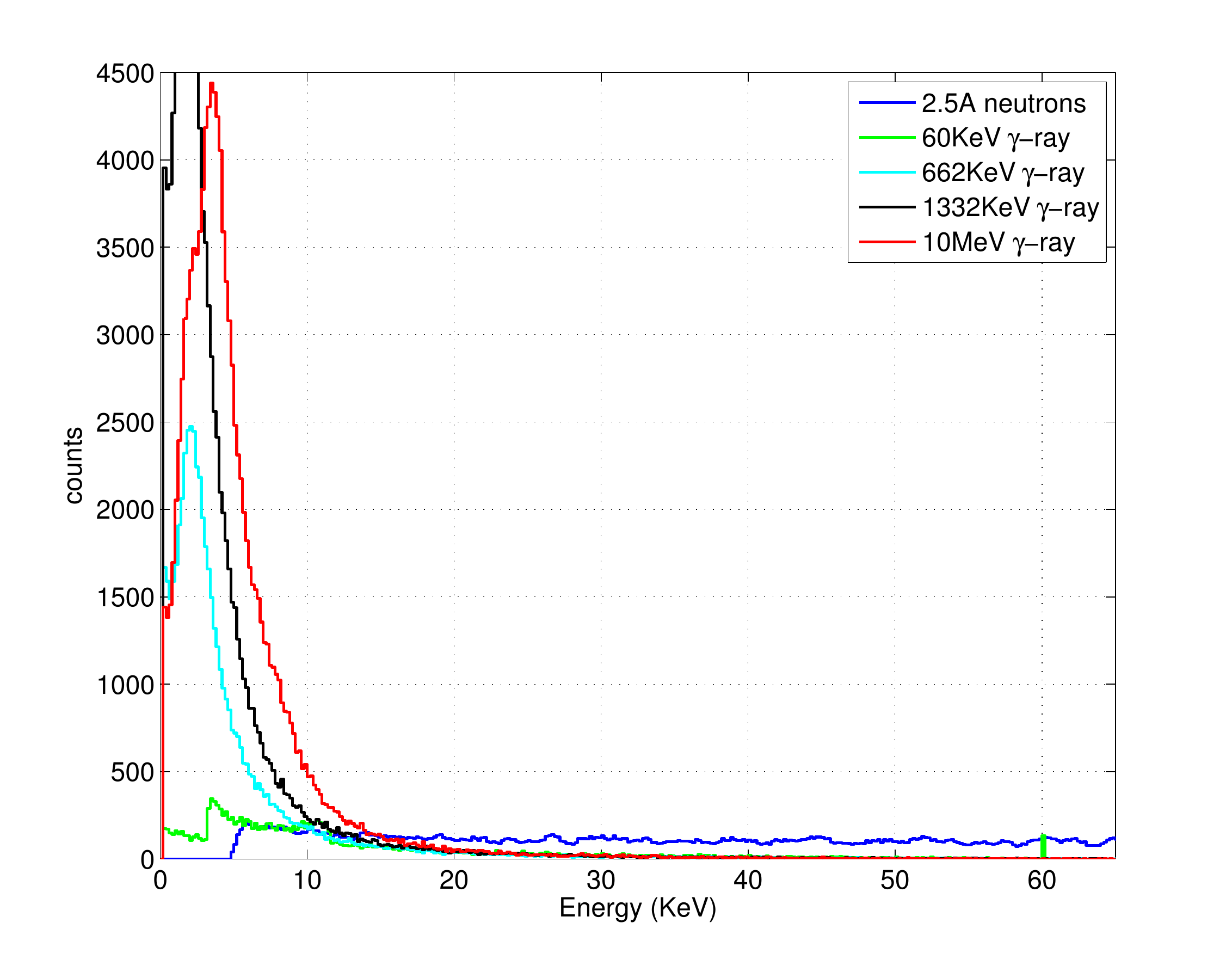}
\caption{\footnotesize GEANT4 simulation of several energies photons
interacting with a Multi-Grid \cite{jonisorma} $Al$-voxel filled
with $Ar/Co_2$ ($90/10$) at $1\,bar$.} \label{geantsimugifjh45654u}
\end{figure}

The energy threshold, applied on the PHS for a gaseous detector, is
an efficient and simple method to discriminate between neutrons and
background events. We mainly focus here on $\gamma$-rays as the
background principally present in a neutron facility. The energy
threshold method is efficient because of the low probability a
$\gamma$-ray has to deposit its entire energy in the gas volume
whatever energy it carries. A detector can be mainly decoupled in
its solid part and its gas from the point of view of a photon.
\\ Figure \ref{figprobintegamdetec67} shows the probability of
interaction of photons with $Al$ ($\rho=2.7\,g/cm^3$), $Ar$ and
$^3He$ at room temperature and at two different pressures. The
interaction probability is calculated from Equation \ref{eqrt3}
using the total attenuation coefficient considering Compton
scattering, photo-electric absorption and pair production. On the
left plot the interaction probability is plotted as a function of
the photon energy and while the amount of material is fixed and vice
versa on the right plot. The material thicknesses have been chosen
according to commonly used in detectors and the photon energies
according to widely employed sources, i.e. $^{241}Am$ and $^{60}Co$.
\\ The probability for a photon to interact with the solid is higher than
the probability of interaction with the gaseous part, for light
gases. If a $\gamma$-ray interacts with the solid, the generated
electron, either by Compton scattering or photo-electric effect, has
to escape the solid in order to deposit its energy in the gas and to
give rise to an electric signal. Intuitively, it is really unlikely
an electron can deposit its entire energy in the gas, and it only
happens if it was generated at the very surface of the solid facing
the gas volume. On the other hand, if a photon has less chances to
interact with the gas, it is also true that it can directly release
its energy in the gas to generate a signal.
\\ Figure \ref{figprobintegamdetec67elect} shows the electron range
in $Al$, $Ar$ and $^3He$ as a function of its energy. An electron
can carry the full photon energy or less, either if it was generated
by photo-electric absorption or by Compton scattering.
\\ In a standard $^3He$ tube or in the Multi-Grid detector \cite{jonisorma} the gas
volume makes at most a few $cm$. Hence, if an electron carries a
limited energy, e.g. $60\,KeV$, it will be stopped by the gas in few
$cm$, thus it is likely to deposit its entire energy. On the other
hand, if it carries higher energy, it can never deposit its full
energy in the gas but will hit the wall of the vessel.
\\ A simulation is needed to fully describe the physical process
that gives rise to the low energy tail on the PHS and allows to find
the energy threshold to be suited. A GEANT4 \cite{ge4} simulation
has been developed. A volume of $2\times2\times1\,cm^3$ is filled
with $Ar/Co_2$ ($90/10$) at $1\,bar$. The volume is a voxel of the
Multi-Grid detector \cite{jonisorma}. The gas volume is surrounded
by $1\,mm$ thick $Al$ and an $Al$-window of $1\,mm$ is placed in
front of the the detector entrance. Two $1\,\mu m$ $^{10}B_4C$
layers are at the front and a the bottom of the voxel.
\\ A $2.5$\AA \, neutron beam is simulated to be compared with
spread photon beams. Four energies have been simulated, $60\,KeV$,
$662\,KeV$, $1332\,KeV$ and $10\,MeV$, in order to evaluate a wide
energy range. These values have been chosen because the $\gamma$-ray
sensitivity measurement has been performed using $\gamma$-ray
sources. $60\,KeV$ emission represents an $^{241}Am$ source,
$662\,KeV$ a $^{137}Cs$ source and $1332\,KeV$ a $^{60}Co$ source
(see Table \ref{newe56nergyqwefxqw45}). The value $10\,MeV$ has been
chosen to represent the $Cd$ neutron induced emission, which is
widely extended in energy up to $10\,MeV$. The energy deposited in
the gas is recorded and the Figure \ref{geantsimugifjh45654u} shows
the corresponding PHS. In the case of a neutron capture a $478\,KeV$
$\gamma$-ray is also emitted and it can consequently deposit its
energy into the gas volume. The electron produced by this
interaction always releases a small fraction of its energy before
hitting the detector wall. A peak at small energies is visible on
the PHS.
\\ As expected, only a small amount of the photon energy
is deposited in the gas volume independently of the its initial
energy.
\\ Since the wide energy spectrum photons only generates a low
energy tail on the PHS, the energy threshold method to discriminate
against background remains a good method as well for solid converter
based neutron detector, e.g. $^{10}B$. Even if the $^{10}B$ PHS is
extended continuously down to low energy, one can obtain a good
$\gamma$-ray rejection by losing only a few neutron events. \\ In
the following sections the $\gamma$-ray sensitivity is quantified
for the threshold method and alternatives methods, e.g. signal shape
analysis, have been investigated to try to improve the
discrimination quality.
\section{$\gamma$-ray sensitivity measurements}
We present here several measurements on the $\gamma$-ray sensitivity
of both $^{10}B_4C$ and $^3He$-based neutron detectors. We define
the detector sensitivity to $\gamma$-rays as its efficiency in
counting background events in precise conditions. E.g. one can be
interested in comparing the neutron detection efficiency with
respect to the efficiency of measuring $\gamma$-rays in the same
conditions. Since the background a detector is exposed can be high
with respect to neutrons coming from the main beam, a very low
$\gamma$-ray sensitivity is required; e.g. below $10^{-6}$.
\\ The neutron to background contribution has been decoupled in the
PHS and in the Plateau measurements in order to validate the
simulations.
\\ The absolute efficiency for neutrons and for $\gamma$-rays has
been measured for both a $^{10}B_4C$ and $^3He$-based neutron
detector.
\subsection{$^{10}B_4C$-based detector PHS}\label{labusltyim4}
\begin{figure}[!ht]
\centering
\includegraphics[width=7.8cm,angle=0,keepaspectratio]{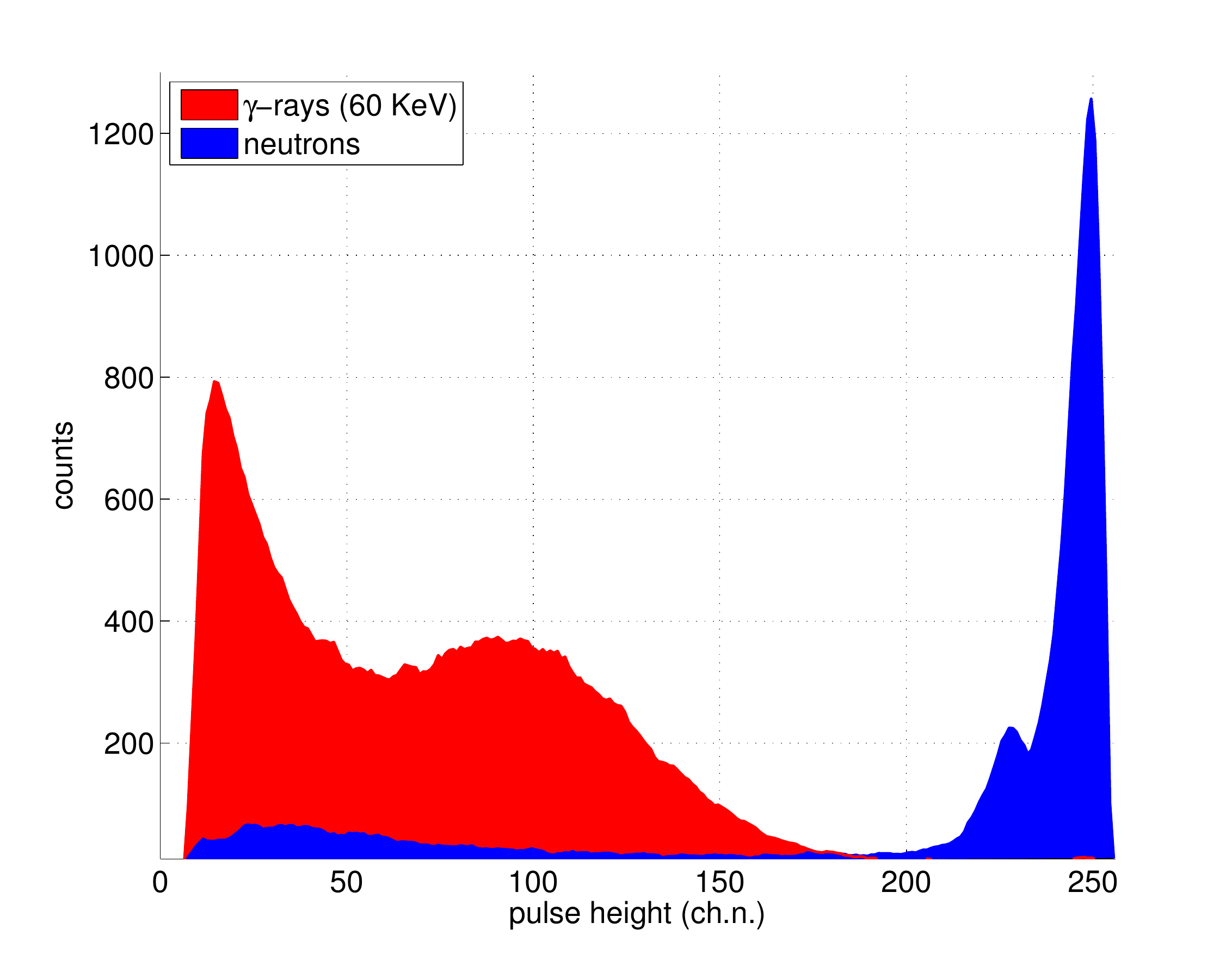}
\includegraphics[width=7.8cm,angle=0,keepaspectratio]{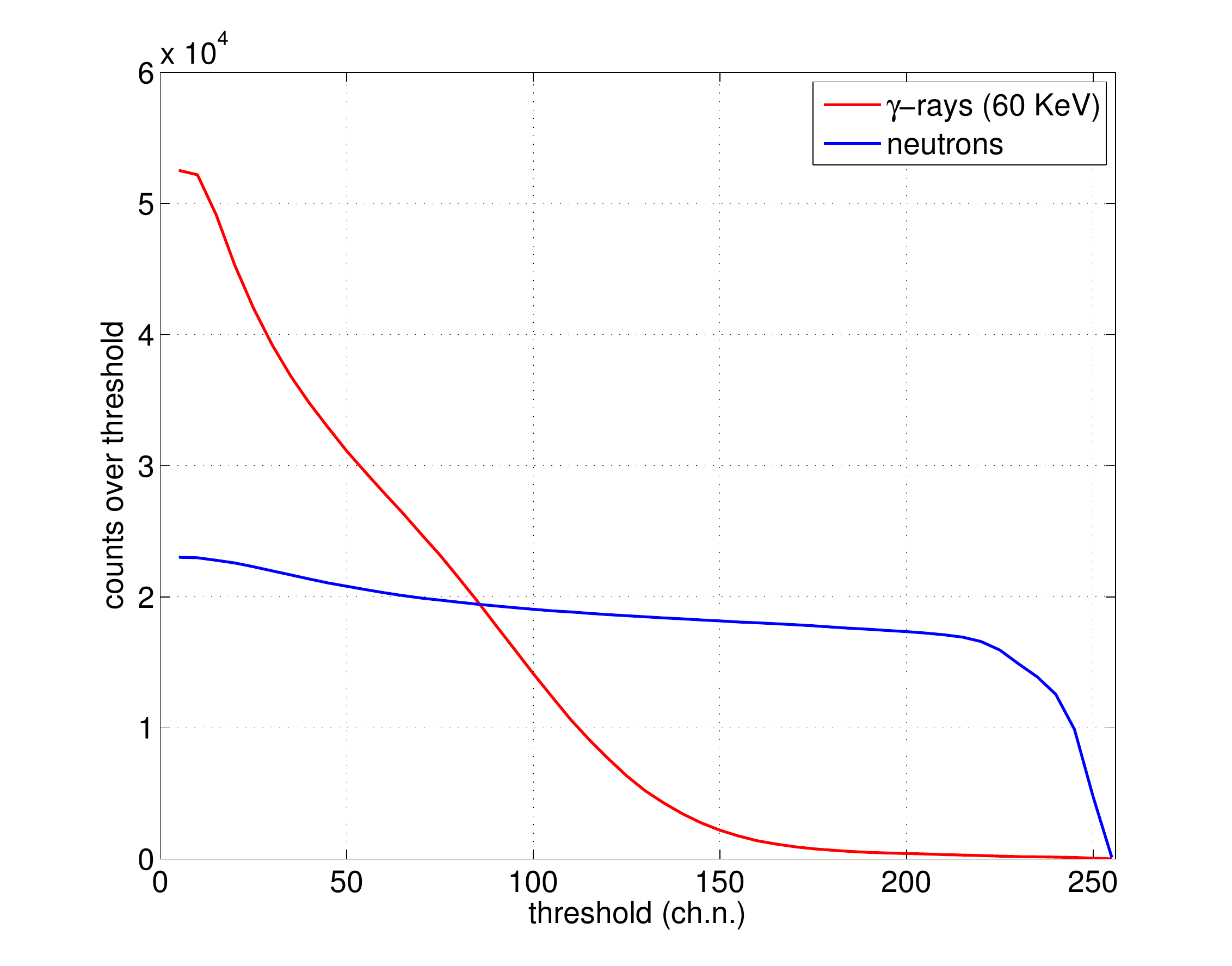}
\caption{\footnotesize PHS measured for neutrons and $\gamma$-rays
with the Multi-Grid detector (left). The evolution of the number of
counts over a threshold as a function of the threshold (right).}
\label{strepcrossefg78}
\end{figure}

\begin{figure}[!ht]
\centering
\includegraphics[width=15cm,angle=0,keepaspectratio]{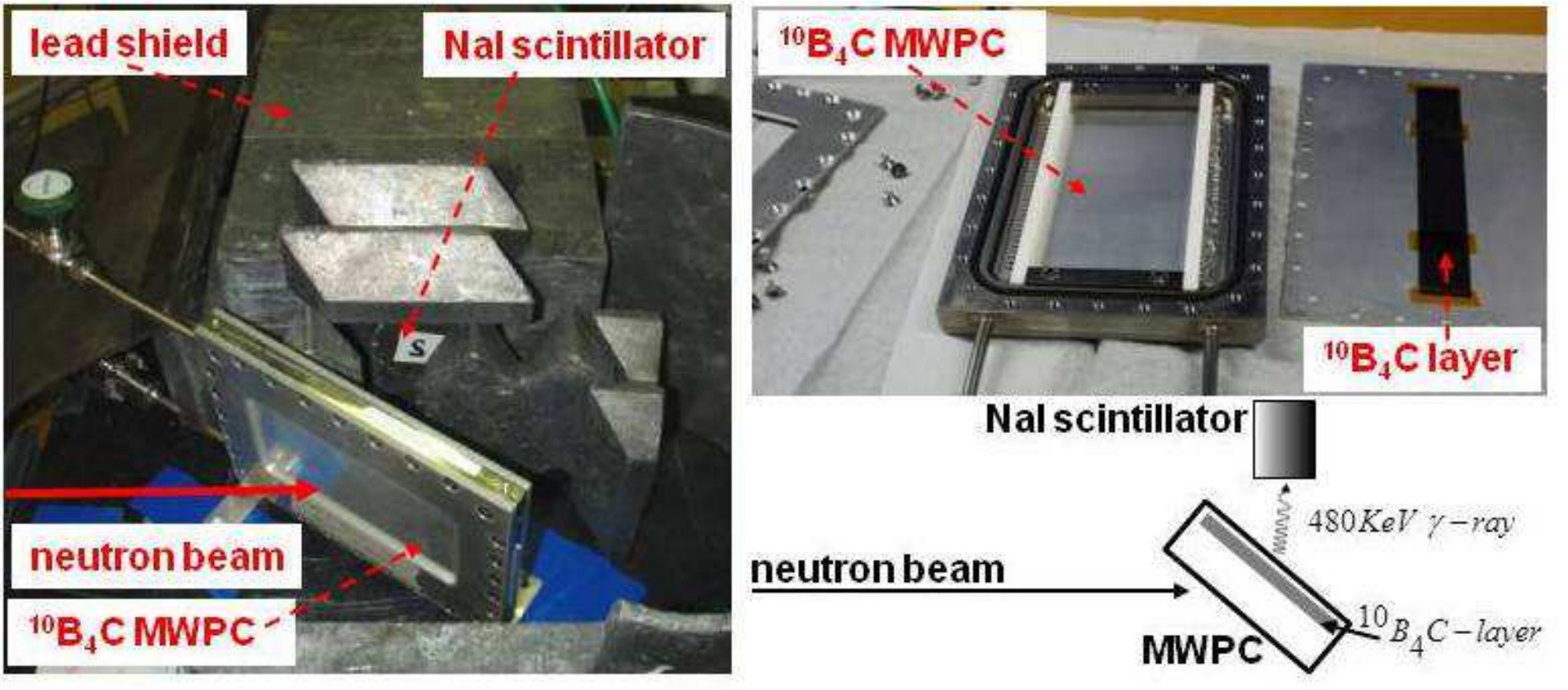}
\caption{\footnotesize Setup used to measure the $^{10}B_4C$ PHS in
coincidence with the emission of the $^{10}B$ neutron capture
$\gamma$-ray.} \label{setupgammacoince5}
\end{figure}

The measurement of the actual contribution given by neutrons and by
$\gamma$-rays to the PHS allows to validate the GEANT4 simulations
and prove the background event contribution mainly concerns the low
energy region of the PHS.
\\ To decouple the neutron and the background contributions to the
PHS one can measure the detector output, screening either one or the
other radiation. This measurement was performed on CT2 at ILL with a
$2.5$\AA \, neutron beam and with an AmBe source. The $^{241}Am$
available (see Table \ref{newe56nergyqwefxqw45}) has a too weak
activity to perform a measurement in a reasonable time given the low
detector sensitivity. Thus we opt for the AmBe of
$3.7\cdot10^{9}\,Bq$ used as $\gamma$-ray source. In case of an AmBe
source, the $\gamma$-ray counts greatly exceed neutron counts.
\\ The PHS was measured with the Multi-Grid detector
\cite{jonisorma} in one of its voxels. To obtain the neutron
contribution, the region of interest of the Multi-Grid was covered
by a $5\,cm$ lead shield in order to screen the $\gamma$-rays. To
obtain the $\gamma$-ray contribution, which is mostly given at
$60\,KeV$ because of the $^{241}Am$, the AmBe source was shielded
with polyethylene and $^{10}B_4C$ sheets to thermalize and stop
neutrons. Since the background contribution is mainly given at low
energy, we operated the detector at high gain in order to rise low
energy events above the threshold. Moreover, the charge amplifier
used has a limited dynamic range, hence high energy events are
saturated.
\\ Figure \ref{strepcrossefg78} shows the measured PHS and the integral
of the number of counts which cross a given value of the threshold,
both for neutrons and $\gamma$-rays. The neutron spectrum is
saturated because of the neutron capture fragment higher energy
yield. The long low energy tail of neutrons is extended down to zero
inside the background. $\gamma$-ray contribution is extended under
the high energy region of the spectrum.
\\ The number of neutron counts as a function of the threshold is
almost constant; in increasing the detector gain there is no actual
raise of the neutron events. If a threshold of $200\,ch.n.$ or
$100\,ch.n.$ is chosen, there is not much to gain in neutron
efficiency. On the other hand, the behavior for background events is
exponential. If a slightly different threshold is chosen there is a
huge counting difference. A trade-off has to be found to maximize
the signal to noise ratio of the detector. \\ The argument chosen is
here arbitrary because the actual neutron and $\gamma$-ray flux is
unknown. We will quantify accurately the detector sensitivity in
Section \ref{gamsens99}.
\\ In Section \ref{correffforreflect6} of Chapter \ref{chaptreflectometry}, we used the $480\,KeV$
$\gamma$-ray emitted by the $^{10}B$ neutron capture reaction (see
Table \ref{eqaa3}) to measure neutron absorption in the layer
together with reflection.
\\ Similarly, in order to decouple properly the two contributions to the PHS given
by neutrons and $\gamma$-rays, we exploit again the emission of this
$480\,KeV$ photon. To measure the $480\,KeV$ $\gamma$-ray we used a
$NaI$ scintillator. Its energy calibration can be obtained as
explained in Chapter \ref{chaptintradmatt}.
\\ Every time a $480\,KeV$ $\gamma$-ray is measured this is the
signature that there was a neutron converted in the $^{10}B_4C$
layer or in other $B_4C$ (shielding, etc.).
\\ Figure \ref{setupgammacoince5} shows the MWPC detector used
and the setup of the experiment. The detector is operated at
atmospheric pressure of $CF_4$. An $Al$-blade, coated with $1\,\mu
m$ $^{10}B_4C$, was placed in a $20\times8\,cm^2$ MWPC. The
converter layer is facing the anode plane. All the anodes are
connected together to a single charge amplifier. The bias voltage is
applied through a decoupling capacitor and it is $1000\,V$.
\\ The wide spectrum $\gamma$-ray background had to be shielded by a
lead housing surrounding the NaI detector. Since most of the neutron
shielding at ILL is made out of $Cd$ or $^{10}B_4C$ sheets, the
contamination of our measurement involves as well the $480\,KeV$
region. It was really crucial to set up the shielding properly.
\\ The MWPC and the NaI detectors were readout by the same data acquisition
system and events were assigned a timestamp. This makes possible to
examine both singles and coincident spectra in the same measurement.
The total rate in the MWPC was $2.1\,KHz$ (neutrons and background
$\gamma$-rays) and the rate of the NaI with the gate around the
$480\,KeV$ photo-peak was $97\,Hz$. This rate is mostly given by the
environment. Because of the small fraction of solid angle covered,
only a few of those counts come from the MWPC. One coincidence is
obtained by looking at an event in the $480\,KeV$ photo-peak by the
NaI scintillator when the MWPC gives a signal over the threshold set
over the electronic noise level. The coincidence rate is $8.3\,Hz$.
The gate around the photo-peak was chosen to decrease the
possibility to have a random coincidence with a background photon of
different energy from $480\,KeV$.
\\ The $^{10}B_4C$ detector was operated at a standard gain ($HV=1000\,V$)
and at a very high gain in order to make it more sensitive to
photons.
\begin{figure}[!ht]
\centering
\includegraphics[width=7.8cm,angle=0,keepaspectratio]{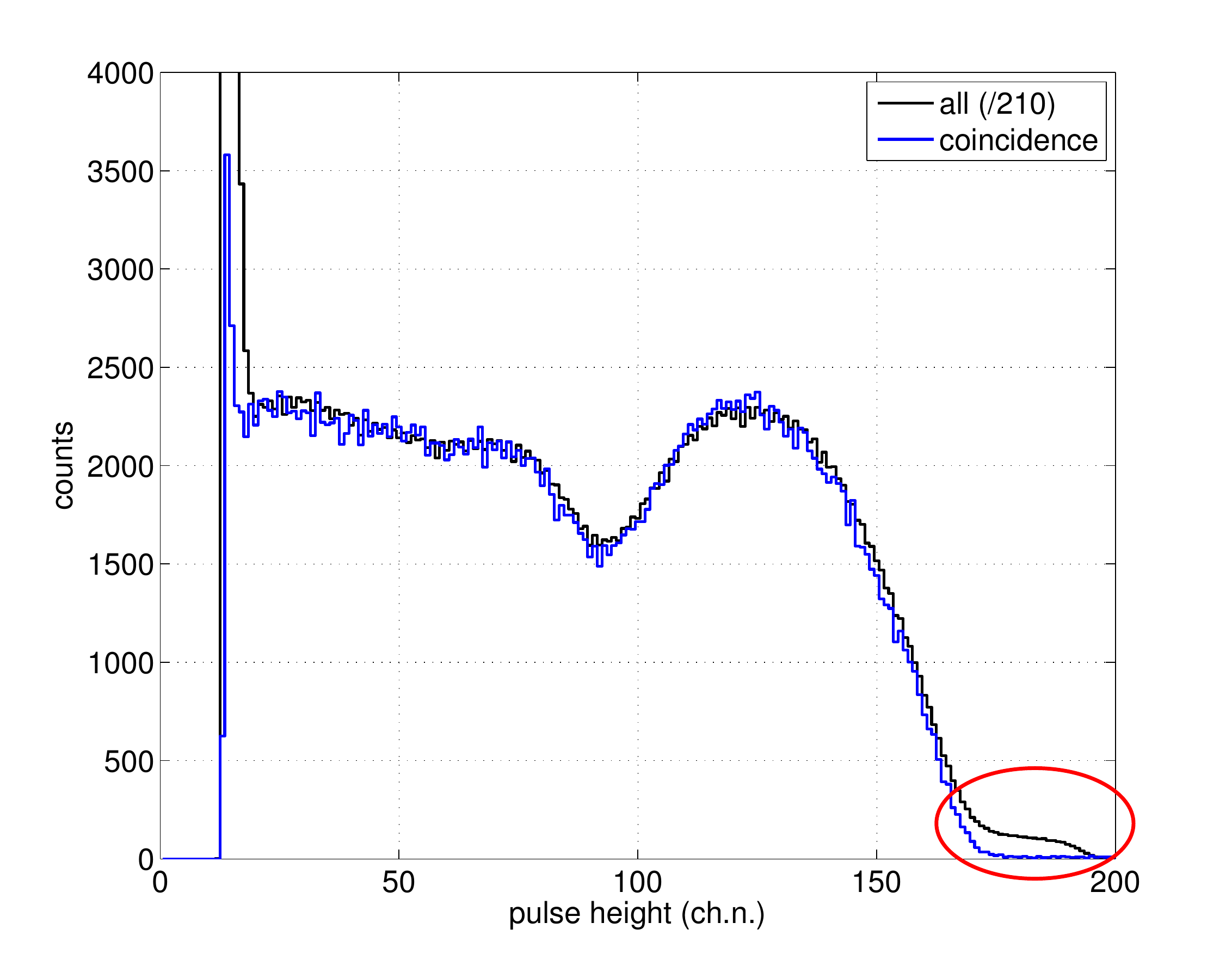}
\includegraphics[width=7.8cm,angle=0,keepaspectratio]{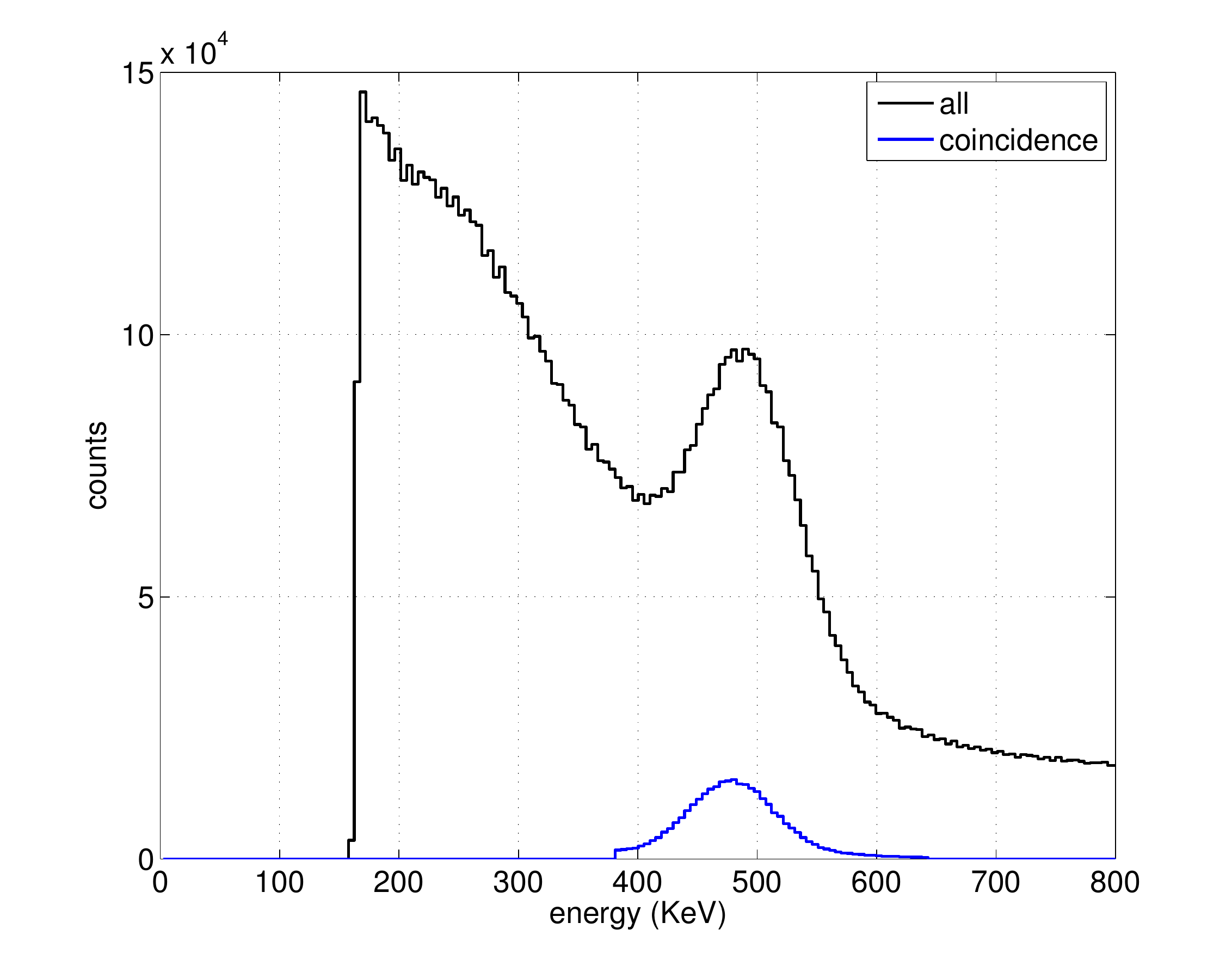}
\caption{\footnotesize $^{10}B_4C$ detector PHS (left) and NaI
scintillator energy spectrum (right) with and without $480\,KeV$
photon coincidence.} \label{figcoincf89sjksncoje}
\end{figure}
\\ Selecting the $480\,KeV$ peak in the NaI in coincidence with a
signal in the MWPC allows to identify true neutron conversion
events, even those whose energies normally make them
indistinguishable from $\gamma$-rays. These are shown in blue in
Figure \ref{figcoincf89sjksncoje}. The probability to detect a
$480\,KeV$ photon emitted from the boron layer is of course much
smaller than one. However, since we know that for large energies
essentially only neutrons contribute to the spectrum, scaling allows
to match the coincident spectrum with the total spectrum. The best
scaling factor turns out to be $210$. The difference then
corresponds to the spectrum due to all non-neutron events or
neutrons converted in the $6\%$ branching ratio of the $^{10}B$
capture reaction. In Figure \ref{figcoincf89sjksncoje}, the
contribution given by the $6\%$ branching ratio $\alpha$-particle
vanishes after coincidence.
\\ The measurement was repeated at higher gain to increase the detector
sensitivity to $\gamma$-rays. The neutron spectrum extends to much
larger energies and only its lower part can be measured without
saturation of the amplifier in these conditions. The measurement was
operated on the neutron beam including or not a $\gamma$-ray source
of high intensity, i.e. the AmBe source.
\\ Figure \ref{figcoincf89sjksncojeew5} shows the results.
Most of this difference spectrum vanishes when the AmBe source is
removed, which confirms that it corresponds to the $\gamma$-ray
spectrum of $Am$: predominantly $60\,KeV$.
\begin{figure}[!ht]
\centering
\includegraphics[width=10cm,angle=0,keepaspectratio]{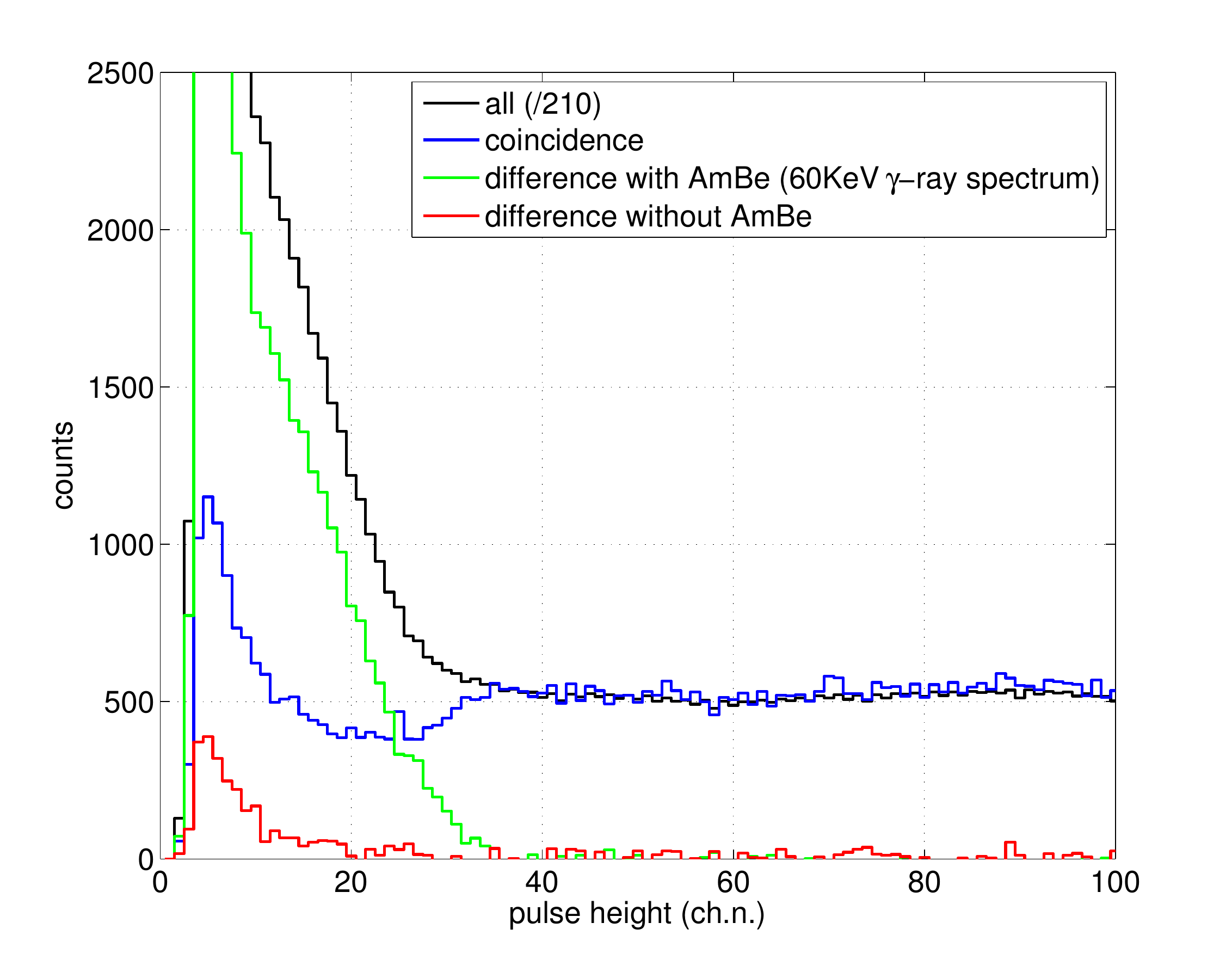}
\caption{\footnotesize $^{10}B_4C$ detector PHS (lower energy tail)
measured in coincidence with the $480\,KeV$ photon with and without
an AmBe source.} \label{figcoincf89sjksncojeew5}
\end{figure}
\\ We confirm the background energy spectrum mostly involves the low
energy region.
\\ The energy spectrum of neutron conversion fragments that reach the
gas has no lower limit for a solid film detector. Therefore any
lower level threshold will reject some neutron events. A minimum
required threshold is determined by the end-point of the
$\gamma$-ray spectrum.
\subsection{$^{10}B_4C$-based detector Plateau}
\begin{figure}[!ht]
\centering
\includegraphics[width=10cm,angle=0,keepaspectratio]{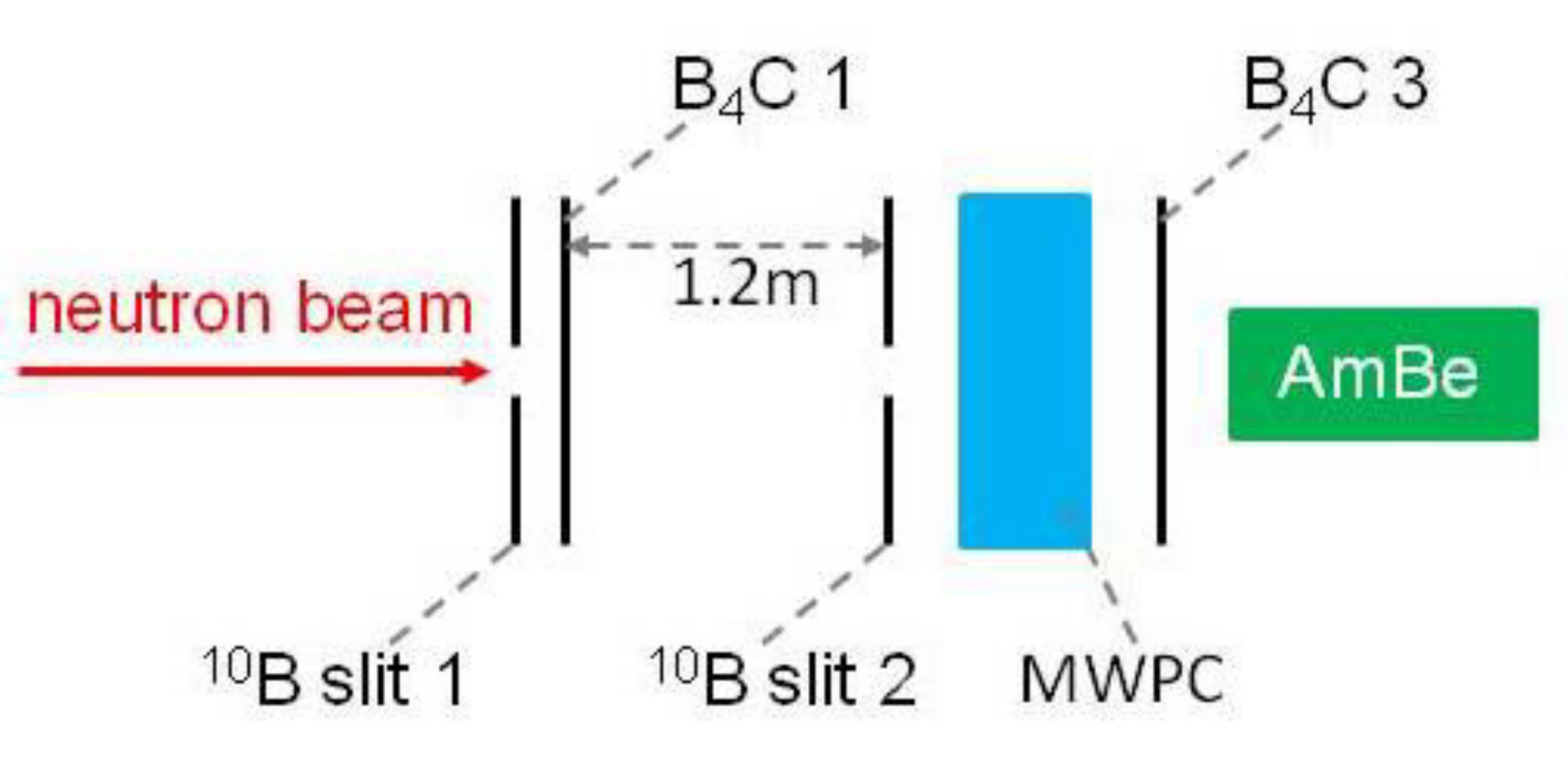}
\caption{\footnotesize Setup used to measure the $^{10}B_4C$
detector plateau.} \label{setuppe67}
\end{figure}

\begin{figure}[!ht]
\centering
\includegraphics[width=10cm,angle=0,keepaspectratio]{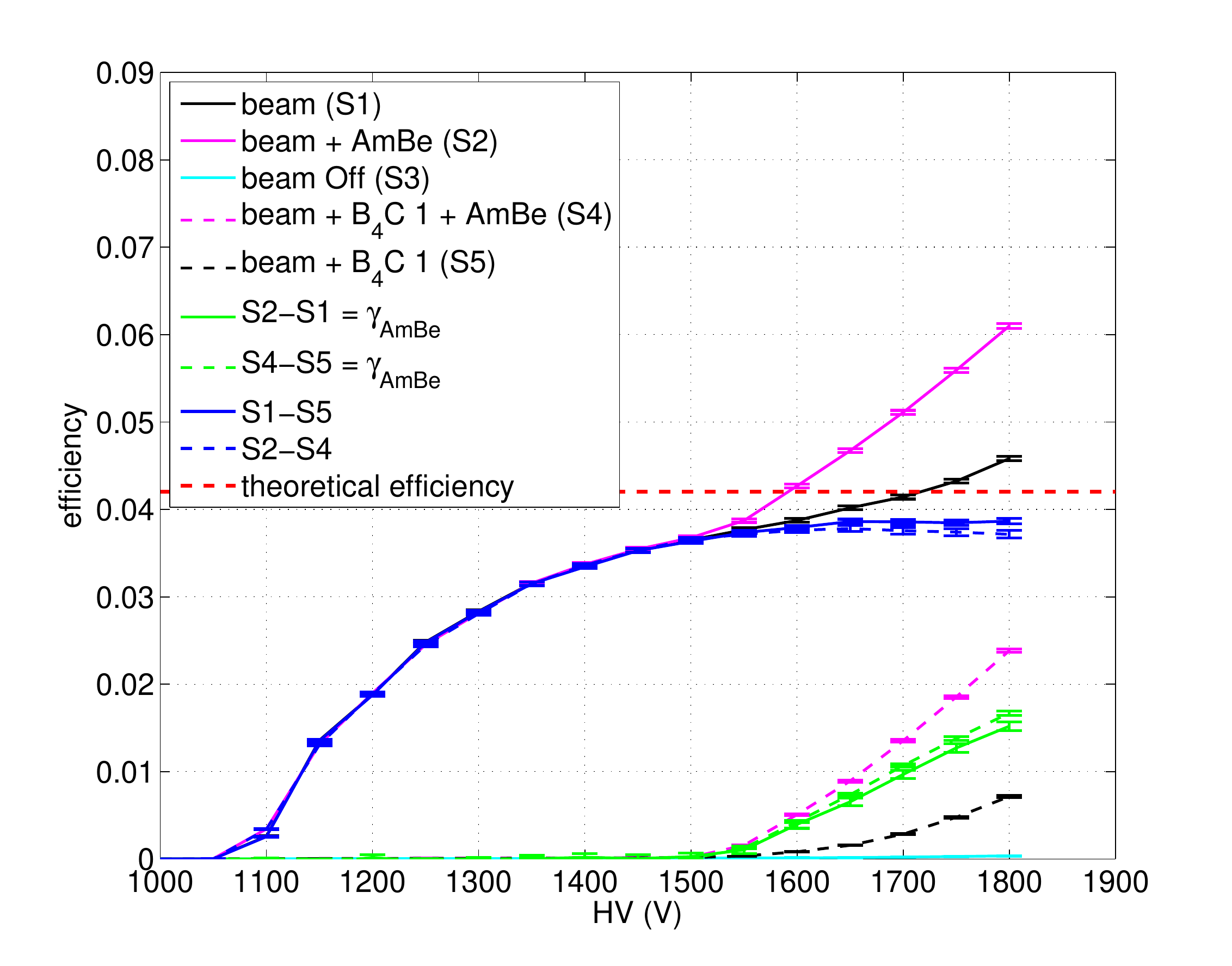}
\caption{\footnotesize Plateau measured in the several configuration
listed in Table \ref{tabset1}.} \label{palte90dfjfurn489}
\end{figure}

The fraction of neutrons rejected due to the overlap of the
background and neutron spectra can be estimated in a counting curve
measurement. In order to measure the background and the neutron
components to the plateau, we have performed a set of measurements.
The measurement has been repeated in several configurations in order
to be able to subtract the single components.
\\ The plateau was measured on CT1 at ILL, with the MWPC containing a single
$^{10}B_4C$ layer shown in Figure \ref{setupgammacoince5} and
already used for the neutron to $480\,KeV$ $\gamma$-ray coincidence
in Section \ref{labusltyim4}.
\\ Figure \ref{setuppe67} shows the complete setup used to perform
the measurements. A neutron beam ($2.5$\AA) was collimated through
two $^{10}B$ slits at $1.2\,m$ distance, to form a $5\times3\,mm^2$
footprint on the detector. An AmBe source, used as $\gamma$-ray
source, was placed on the opposite window of the MWPC. The neutrons
emitted were shielded by a polyethylene and a $B_4C$ sheet ($B_4C
\,\,3$ in Figure \ref{setuppe67}). We can chose to add a second
$B_4C$ sheet ($B_4C \,\,1$ in Figure \ref{setuppe67}) in order to
stop the neutron beam and then measure the background without
affecting the setup unless for the more $\gamma$-rays produced by
the sheet. We can consider those $\gamma$-rays to be negligible
because of the solid angle at $1.2\,m$. Moreover, the measurement
was repeated with the $B_4C$ sheet after the second collimation slit
and only a slight difference was observed at very high gain.
\begin{table}[!ht]
\begin{center}
\begin{tabular}{|c||c|c|c|c|c|c|c|}
  \hline
  \hline
  Setup & beam & $B_4C$ 1 & $B_4C$ 3 & AmBe & Slit 1 & Slit 2\\
  \hline
  $S1$ & x &   &    x &   & x & x \\
  $S2$ & x &   &    x & x & x & x \\
  $S3$ &   &   &    x &   & x & x \\
  $S4$ & x & x &    x & x & x & x \\
  $S5$ & x & x &    x &   & x & x \\
  \hline
  \hline
\end{tabular}
\caption{\footnotesize Setups used in the measurements (the x means
that a specific element was used at the moment of the measurement).}
\label{tabset1}
\end{center}
\end{table}
\\ We performed measurements of the plateau in 5 different conditions. They are
listed in Table \ref{tabset1}. $S1$ is the measurement of the
neutron and the background contributions. $S2$ is as $S1$ with the
addition of the AmBe background. $S3$ allows to measure the
background without the beam. $S5$ is the background measurement with
the beam on, and $S4$ with the additional AmBe contribution. All
parts of the setup produced background that has to be subtracted.
The collimation slits produce $\gamma$-rays when exposed to
neutrons. We list here the single contribution to the counting
curve:
\begin{equation}
\begin{aligned}
  S1 & = n & + S3 & + \Gamma  & + \gamma_{slit1}  & + \gamma_{slit2} &                     &        + \gamma_{B_4C 3}       \\
  S5 & =   &  S3  & + \Gamma  & + \gamma_{slit1}  &                      & + \gamma_{B_4C 1} &
\end{aligned}
\end{equation}
where $n$ is the pure neutron count, $\Gamma$ the unknown background
coming from the environment due to the presence of the beam, e.g.
$\gamma_{slit1}$ the $480\,KeV$ $\gamma$-rays coming from the
$^{10}B$-slit 1. Moreover, $S2 = S1 + \gamma_{AmBe}$, $S4 = S5 +
\gamma_{AmBe}$. Where $\gamma_{AmBe}$ are the $\gamma$-rays emitted
by the AmBe, mostly $60\,KeV$.
\\ The neutron beam was calibrated using the hexagonal detector
according to the procedure in Appendix \ref{apphexmeasexplan}. The
neutron counting at the detector position was $(31850 \pm 20)\,Hz$.
Knowing the neutron flux allows to normalize the counting curve to
get the actual detection efficiency. The expected theoretical
efficiency for a $1\,\mu m$ layer hit at $90^{\circ}$ in
back-scattering with no energy threshold $(E_{Th}=0)$ is $4.2\%$
(red dashed curve in Figure \ref{palte90dfjfurn489}).
\\ Figure \ref{palte90dfjfurn489} shows the single measurements and
the plateaux obtained by substraction. The plateau obtained in $S2$
shows a higher $\gamma$-ray rising at high voltage with respect to
$S1$. \\ The AmBe source contribution can be highlighted by
subtracting either $S2-S1$ or $S4-S5$. \\ The neutron contribution
can be obtained by:
\begin{equation}
S1-S5 = S2-S4 = n + \gamma_{slit2} + \gamma_{B_4C 3} -
\gamma_{B_4C1}
\end{equation}
apart from the background produced by the second collimation slit
($\gamma_{slit2}$) and if we assume $\gamma_{B_4C 1}\simeq
\gamma_{B_4C 3}$. As already mentioned we moved the $B_4C 1$ sheet
after the second collimation slit and no appreciable difference is
observed in the plateau.
\\ We expect the pure neutron plateau to saturate to the
theoretical efficiency.
\\ Setting the high voltage at the point just before the rise of the
$\gamma$-rays detection ($1500\,V$) results in approximately $6\%$
fewer neutron counts compared to the maximum of the plateau. Note
however, that a certain threshold level is required also to reject
electronic noise. The loss of $6\%$ compares to all neutron events
where any finite part of the energy is deposited in the gas, not to
those events that can be detected over the electronic noise level.
The equivalent energy threshold of the noise level becomes smaller
and smaller when the voltage and hence the gas gain rises. The
plateau should approach the theoretical efficiency at high voltage.
\subsection{$^{10}B_4C$ and $^3He$-based detectors $\gamma$-ray
sensitivity}\label{gamsens99} We quantify in this Section the actual
sensitivity of both $^{10}B_4C$ and $^3He$-based detectors to
$\gamma$-rays. The measurements have been performed in a wide range
of $\gamma$-ray energy. The efficiency to detect $\gamma$-rays of a
detector can be measured by using calibrated sources. By knowing the
source activity and the solid angle subtended by the detector, the
actual photon flux can be calculated. The efficiency is the ratio
between the number of detection events and the incoming photon flux
\cite{kouzes1}, \cite{kouzes2}.
\\ Four $\gamma$-ray sources have been used and their activity, main
photons emitted with their relative intensity are listed in Table
\ref{newe56nergyqwefxqw45}. Those four sources allow to explore the
energy range from $x$-rays up to $1\,MeV$. We can consider
$^{241}Am$, $^{137}Cs$ and $^{60}Co$ three distinguished energy
ranges from low energy up to above $1\,MeV$; while $^{133}Ba$ shows
a low-medium energy range emission.
\begin{table}[!ht]
\centering
\begin{tabular}{|l|c|c|c|}
\hline \hline
source & photon energy (KeV) & intensity (\%)\\
\hline
$^{133}Ba$       & $4$    & $14.7$  \\
$(A = 1.85\cdot 10^5 \,Bq)$       & $30.6$ & $31$    \\
$$               & $30.9$    & $57$  \\
$$               & $34.9$    & $15.4$  \\
$$               & $36$    & $3$  \\
$$               & $53$    & $2$  \\
$$               & $79$    & $2.6$  \\
$$               & $81$    & $32.9$  \\
$$               & $276$    & $7$  \\
$$               & $303$    & $18$  \\
$$               & $356$    & $62$  \\
$$               & $383$    & $9$  \\
\hline
$^{241}Am$       & $13.9$    & $14.3$  \\
$(A = 3.5\cdot 10^5 \,Bq)$       & $26$ & $2.3$    \\
$$               & $60$    & $35.9$  \\
\hline
$^{137}Cs$       & $31$    & $2$  \\
$(A = 2\cdot 10^5 \,Bq)$       & $32$ & $3.8$    \\
$$               & $662$    & $85$  \\
\hline
$^{60}Co$       & $1173$    & $99.85$  \\
$(A = 2.31\cdot 10^4 \,Bq)$       & $1332$ & $99.98$    \\
\hline \hline
\end{tabular}
\caption{\footnotesize $\gamma$-ray sources activity, main photon
emitted and their relative intensity.} \label{newe56nergyqwefxqw45}
\end{table}
\\ We compare two detectors: the $^{10}B_4C$-based Multi-Grid \cite{jonisorma} and
the $^3He$-based hexagonal detector used in Appendix
\ref{apphexmeasexplan} to quantify the neutron flux. The latter is
filled with $3\,bar$ of $^3He$ and $1.5\,bar$ of $CF_4$. Multi-Grid
was filled with $1\,bar$ of $CF_4$. The hexagonal detector is
composed of 37 hexagonal tubes with a $7\,mm$ diameter arranged in a
honeycomb formation.
\\Efficiencies or sensitivities to $\gamma$-rays are
defined as the probability for a photon incident on a detector
element (such as a tube) to result in an event confused with a
neutron detection event. Plateau measurements have been used for
this since in order to measure a pulse height spectrum, either the
threshold needs to be set extremely low, or a high bias voltage
needs to be used. At a typical gas amplification used in neutron
detection, the threshold cannot be set low enough to study
$\gamma$-ray signals due to electronic noise. Increasing the bias
voltage results in a high gas gain but in turn leads to poor energy
resolution.
\\ Each detector was exposed to the sources and the actual photon
flux was determined by calculating the solid angle the detector was
subtended \cite{schroer1}, \cite{schroer2}, \cite{schroer}.
\\ The normalized neutron plateau at $2.5$\AA \, for both detectors was measured for
comparison. It was determined by knowing the actual neutron flux
(see Appendix \ref{apphexmeasexplan}).Figure
\ref{paltcomparet84jmfu38jeda} shows the results; where the
efficiency is given per tube.
\begin{figure}[!ht]
\centering
\includegraphics[width=7.8cm,angle=0,keepaspectratio]{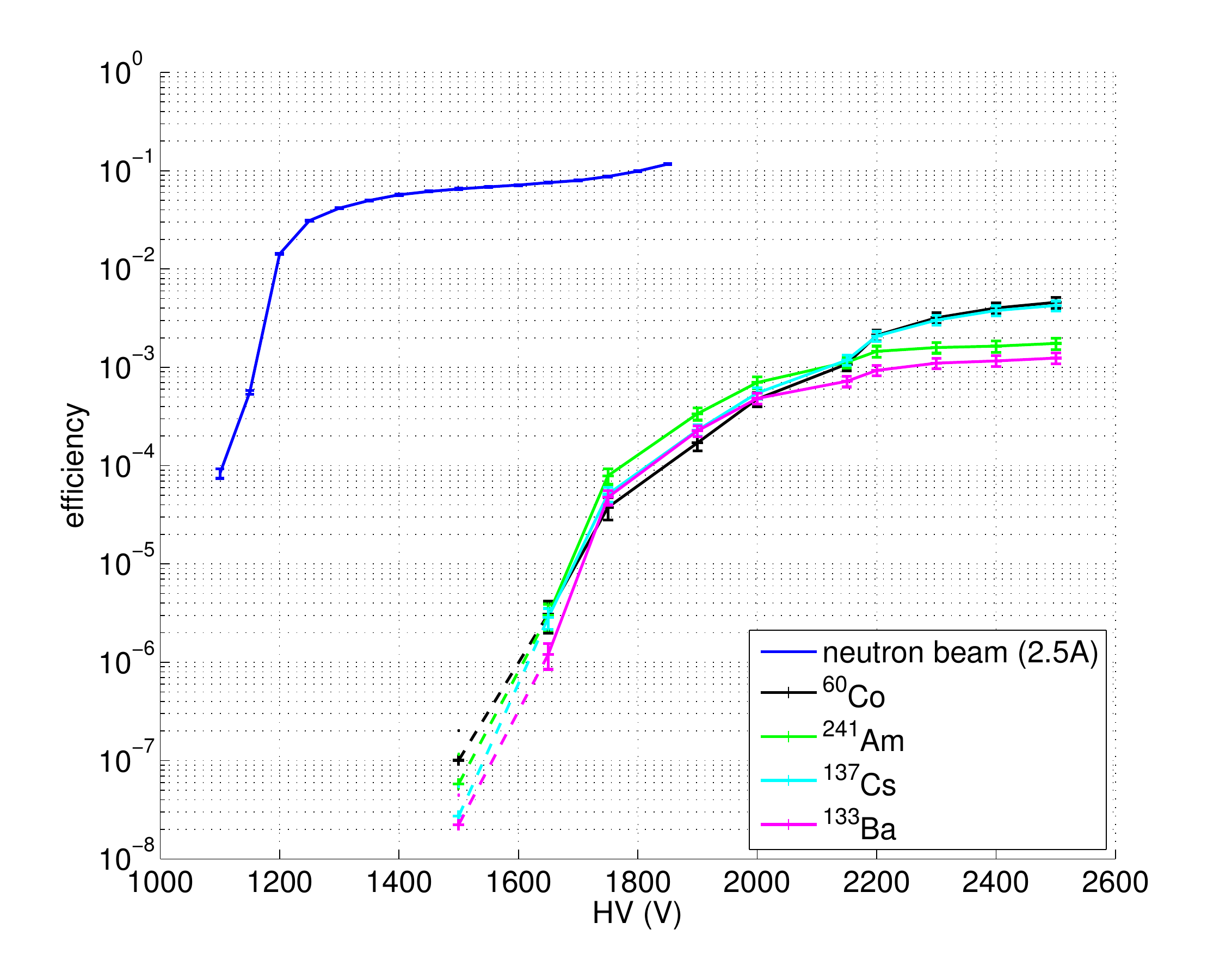}
\includegraphics[width=7.8cm,angle=0,keepaspectratio]{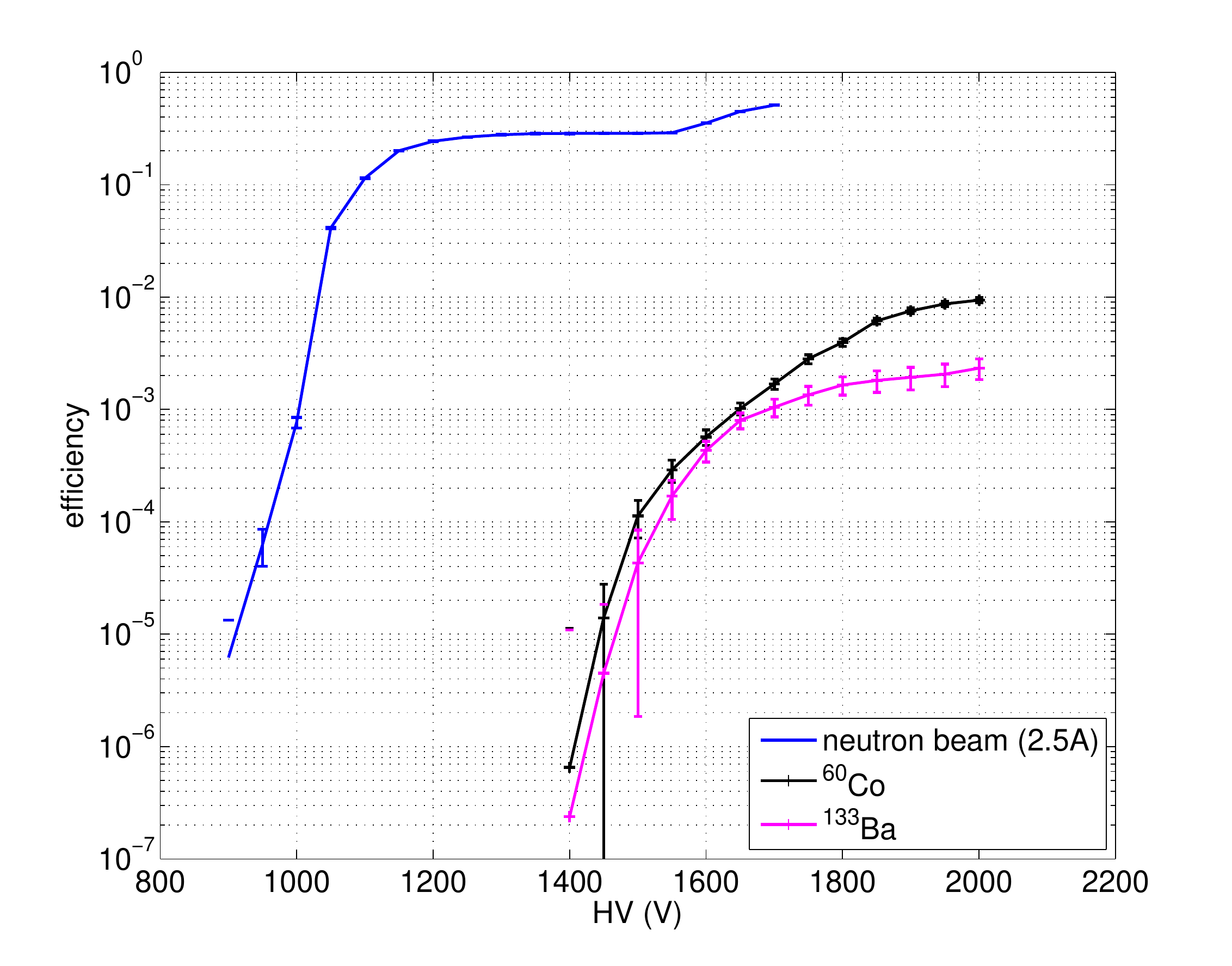}
\caption{\footnotesize Plateau measurements with the Multi-Grid
$^{10}B$ detector (left) and a Multi-Tube $^3He$ detector (right)
with neutrons ($2.5$\AA) and $\gamma$-ray sources. Detection
efficiency per tube is shown. The nominal operating voltages are
$1600\,V$ and $1400\,V$ respectively.}
\label{paltcomparet84jmfu38jeda}
\end{figure}
\\ The disconnected points at the left ends of the curves are upper
limits, sensitivity can be lower than the plotted points, where no
statistically significant counts could be detected over background.
\\ The $^{133}Ba$ and $^{241}Am$ source can be
considered as low energy sources, whereas $^{137}Cs$ and $^{60}Co$
as high energy sources. Note that (see Figure
\ref{paltcomparet84jmfu38jeda}) the difference between low and high
energy photons is only visible at high gain for both detectors. As
the bias voltage increases the count rate due to the $^{60}Co$
source exceeds that of $^{133}Ba$ since a larger number of
interactions, primarily in the solid elements of the detector,
contributes with higher energy electrons.
\\In a realistic configuration where a gas detector is set up to detect
neutrons, the energy threshold should be set so that the signals
from photons are rejected while those from neutrons are not. No
sharp cut-off for the highest pulse height resulting from a specific
primary particle energy exists due to the statistical nature of gas
amplification and charge transport. The final contamination due to
$\gamma$-ray signals will be due to those events where a
$\gamma$-ray signal including statistical fluctuation is over the
threshold. This is most likely to be due to a low-energy
$\gamma$-ray (since the specific energy loss is then high), such as
$60\,KeV$. At the first glance this fact is encouraging since it is
much easier to shield low-energy photons. Note however, that
interactions of high-energy photons as well as nuclear reactions,
such as neutron capture or decay of activated materials, often
result in emission of $x$-rays and internal conversion electrons
which have just the energy that is most likely to contribute to
background. It is therefore important to carefully consider the
external radiation environment as well as the secondary sources that
may exist in the immediate vicinity of the detectors.
\\ The discrimination between neutron and photon signals presents a
challenge in many types of neutron detectors. A high level of
discrimination can be reached with the conventional $^3He$ tube.
Concerning the $^{10}B$ thin film detectors, we have found that the
$\gamma$-ray rejection need not be lower in these detectors than in
$^3He$ tubes, if we allow for a small loss in neutron detection
efficiency. Less than $<10^{-6}$ is easily reached: one needs to
lower the efficiency only by about $0.5\%$.

\subsection{Pulse shape analysis for neutron to $\gamma$-ray discrimination}
In this Section we want to investigate the possibility to achieve a
greater neutron to $\gamma$-ray discrimination than just a PHS
threshold by also processing the signals shapes.
\\ While several methods have been developed in order to discriminate
between neutrons and $\gamma$-rays for neutron scintillators
\cite{spowart}, \cite{morris}; it is not the case for neutron
gaseous detectors. In a scintillator the time structure of the light
output follows two completely different behaviors for neutrons and
for $\gamma$-rays. The technique works well by virtue of the fact
that for gamma initiated scintillations there is much more fast
decay constant light output than slower decay constant light output
as compared to the same relative intensities in light output from
neutron induced scintillations \cite{morris}.
\\ In a gaseous detector the main difference between neutron and
$\gamma$-ray signals is the space charge density created. While a
neutron originates heavy particles which ionize the gas, a
$\gamma$-ray produces a light electron. The way a particle or an
electron ionizes the gas is different for two reasons: the electron
generally carries less energy than the particles and its energy loss
behaves differently. However, the tracks in gas, for electrons and
capture fragments, can be both oriented randomly; this makes it
difficult to find a discrimination criterion that is valid for the
multiplicity of cases that can occur.
\\ To study the shape of the output signals we use the same MWPC
used previously and shown in Figure \ref{setupgammacoince5}. We
operate the detector at a high gain in order to have an intense
$\gamma$-rays sensitivity: i.e. $1750V$ (see Figure
\ref{palte90dfjfurn489}). This voltage corresponds to a region of
limited proportionality of the gas amplification which is in between
the proportional and Geiger detector operational modes. For such a
reason in Figures \ref{1usS1}, \ref{1usS2}, \ref{1usS3} and in
Figures \ref{3nsS1}, \ref{3nsS2} and \ref{3nsS3} in the PHS we can
not recognize the energy carried by the fragments.
\\ Since we do not know a priori the signal time structure, two
amplifiers have been chosen to perform the analysis: a fast one with
$3\, ns$ integration time and a slower one of $1\, \mu s$.
\\ We place a $5\,cm$ lead shield before the detector
in order to decrease the $\gamma$-ray background originated from the
collimation slits. We record $20000$ signal traces each time. The
signal processing was then done off-line. \\ We use in three
different configurations, listed in Table \ref{wertw456h5}: with and
without the neutron beam and with and without the additional
background originated by the AmBe source.
\begin{figure}[!ht]
\centering
\includegraphics[width=10cm,angle=0,keepaspectratio]{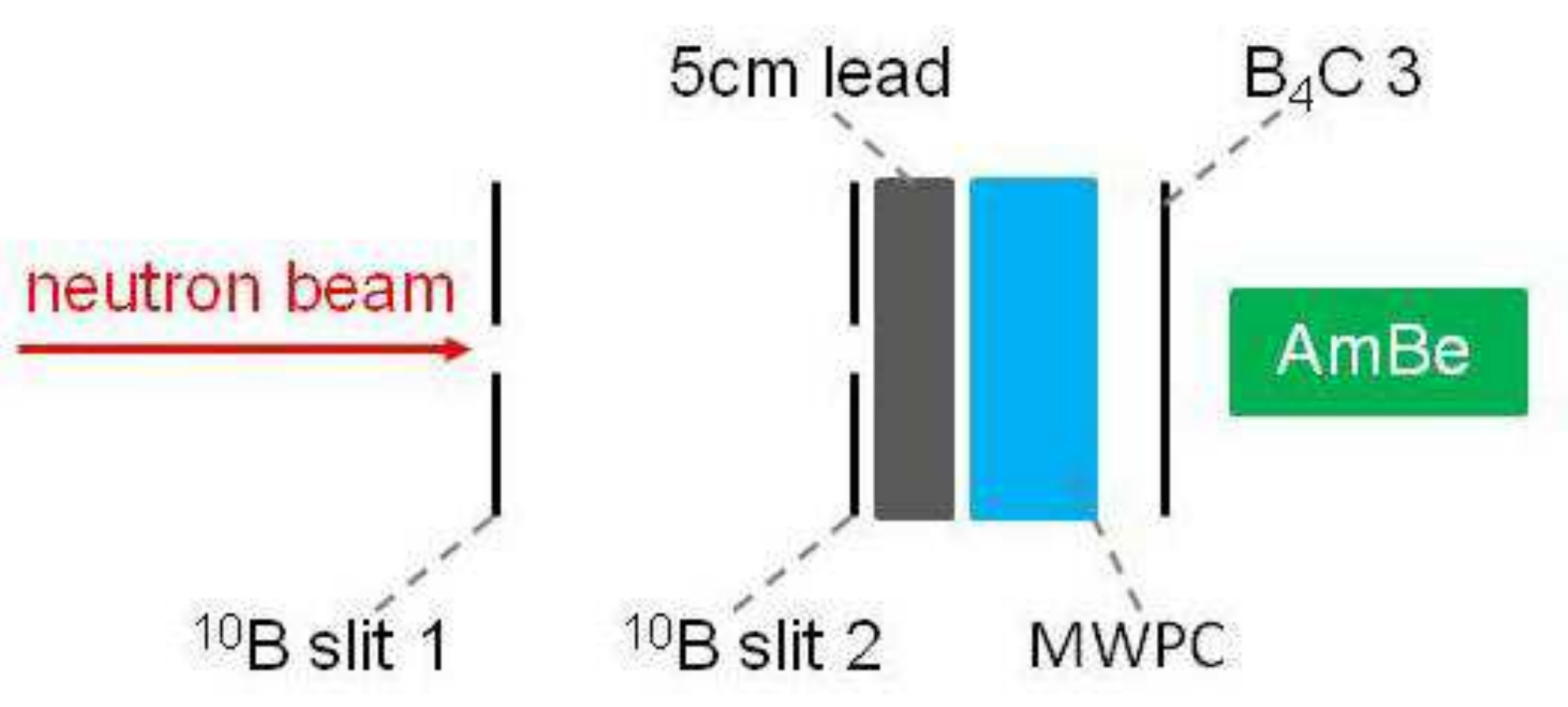}
\caption{\footnotesize Setup used to perform the signal shape
analysis.} \label{setSSA}
\end{figure}
\begin{table}[!ht]
\begin{center}
\begin{tabular}{|c||c|c|c|c|c|c|}
  \hline
  \hline
  Setup & beam & Lead & $B_4C$ 3 & AmBe & Slit 1 & Slit 2\\
  \hline
  $S1$ & x & x &   &   & x & x \\
  $S2$ &   & x & x & x & x & x \\
  $S3$ & x & x & x & x & x & x \\
  \hline
  \hline
\end{tabular}
\caption{\footnotesize Setups used in the measurements (the x means
that a specific element was used at the moment of the measurement).}
\end{center}
\end{table}\label{wertw456h5}
\\ We can assume that $S1$ is a measurement of mostly
neutrons, $S2$ of mostly the $\gamma$-rays from the AmBe source and
$S3$ of both contributions.
\\ The off-line analysis was performed by looking at the signal
structure in terms of charge yield and time-over-threshold (TOT).
For what concerns the $1\, \mu s$ amplifier, the signal amplitude is
proportional to the charge created in the gas volume because its
integration time is longer than the physical charge collection in
the detector. On the other hand, for the $3\, ns$ amplifier the
charge is given by the integral of the signal over time. The
time-over-threshold is the time a signal stays above a given
threshold. The latter was set at the lowest value possible according
to the amplifiers noise level ($3\,mV$ for both).
\\ In each figure that follows, is shown a charge spectrum on the top-left
corner, a TOT spectrum on the top-right corner. A charge versus TOT
is shown in each figure in the bottom plots: one scattered and one
of intensity normalized to the total number of events.
\\ Figures \ref{1usS1}, \ref{1usS2} and \ref{1usS3} show the signal
analysis for the three sets of measurements for the slow amplifier.
\begin{figure}[!ht]
\centering
\includegraphics[width=12cm,angle=0,keepaspectratio]{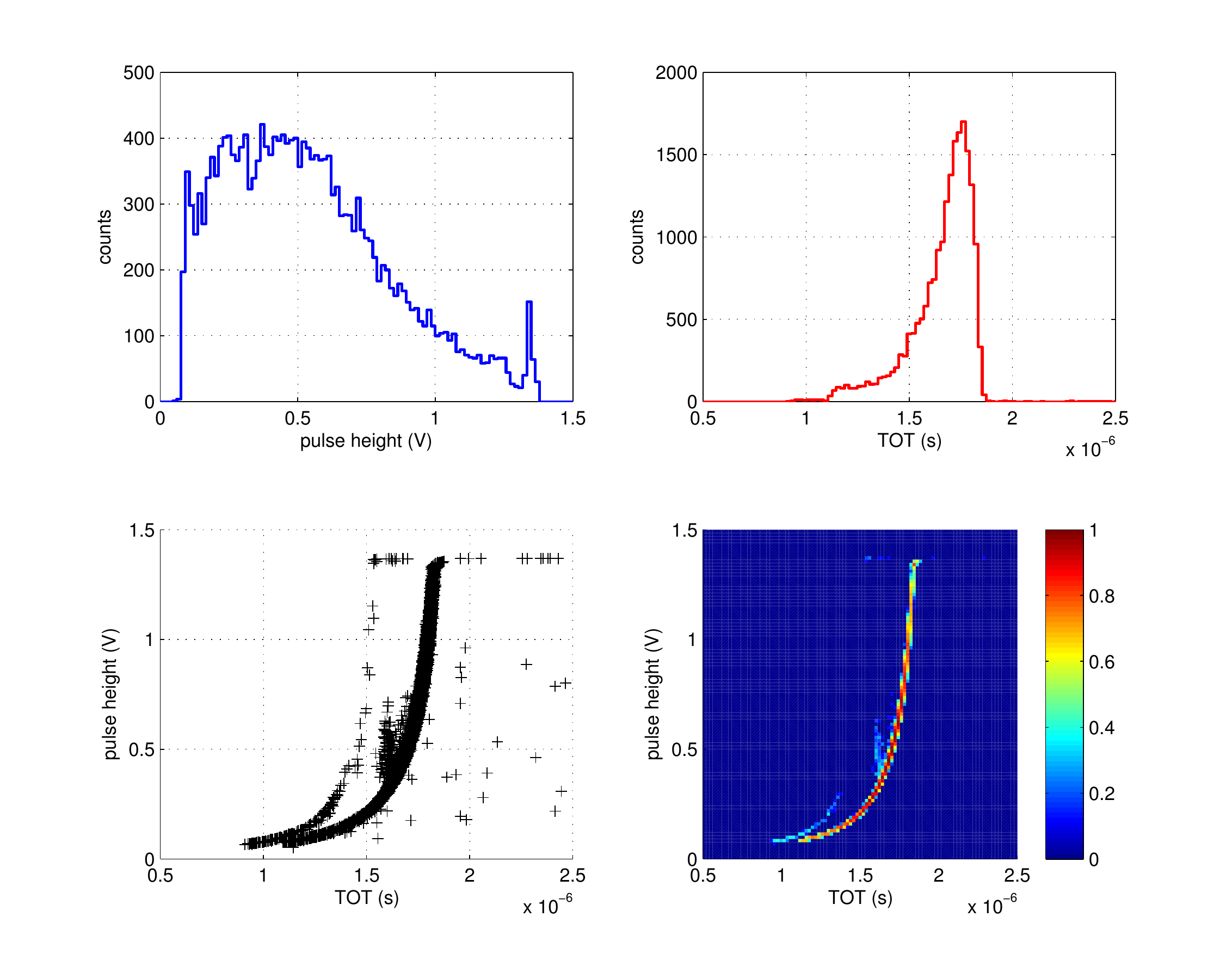}
\caption{\footnotesize Measurement of $S1$ (beam only) - amplifier
$1\, \mu s$. \label{1usS1}}
\end{figure}
\begin{figure}[!ht]
\centering
\includegraphics[width=12cm,angle=0,keepaspectratio]{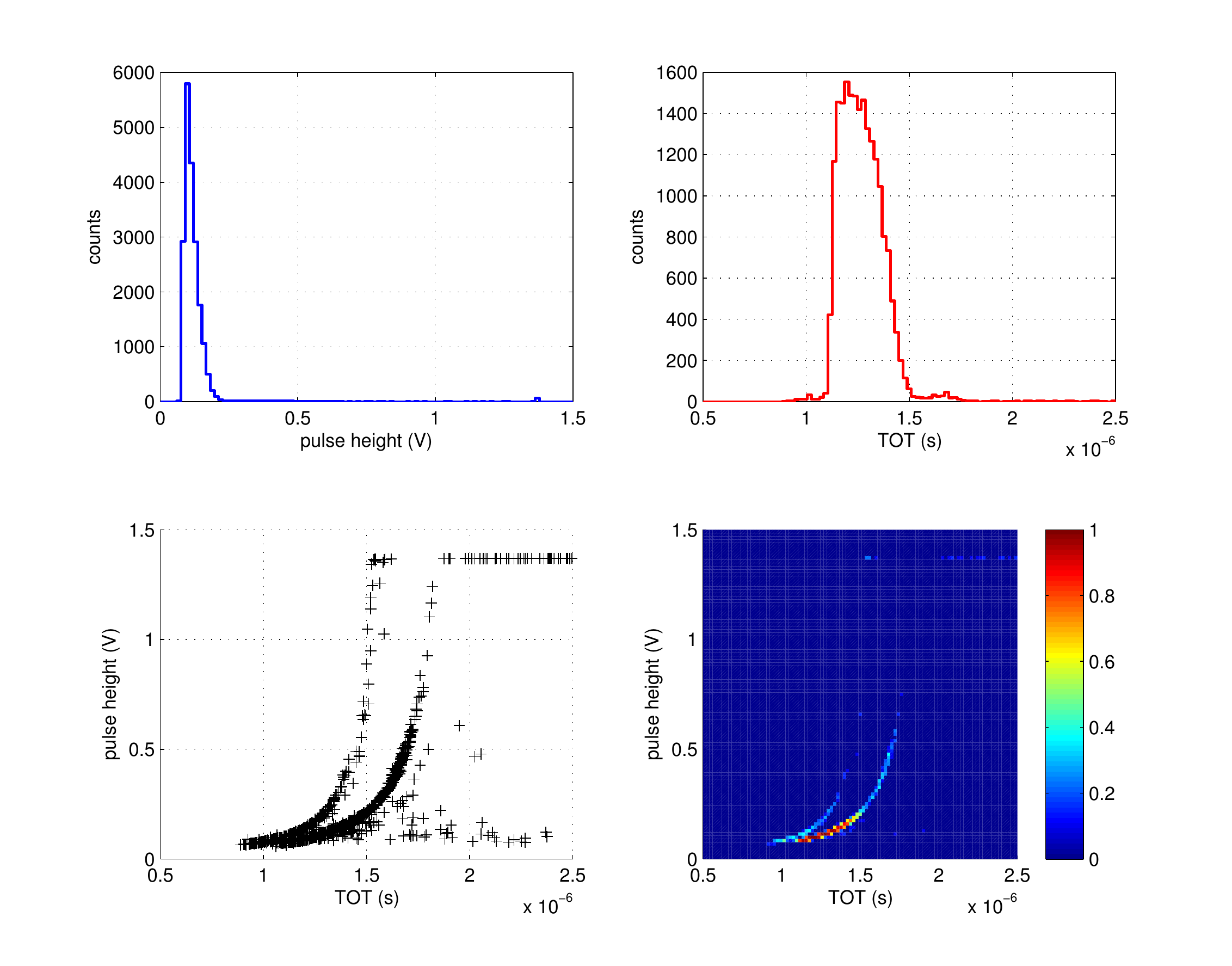}
\caption{\footnotesize  Measurement of $S2$ (AmBe source only) -
amplifier $1\, \mu s$. \label{1usS2}}
\end{figure}
\begin{figure}[!ht]
\centering
\includegraphics[width=12cm,angle=0,keepaspectratio]{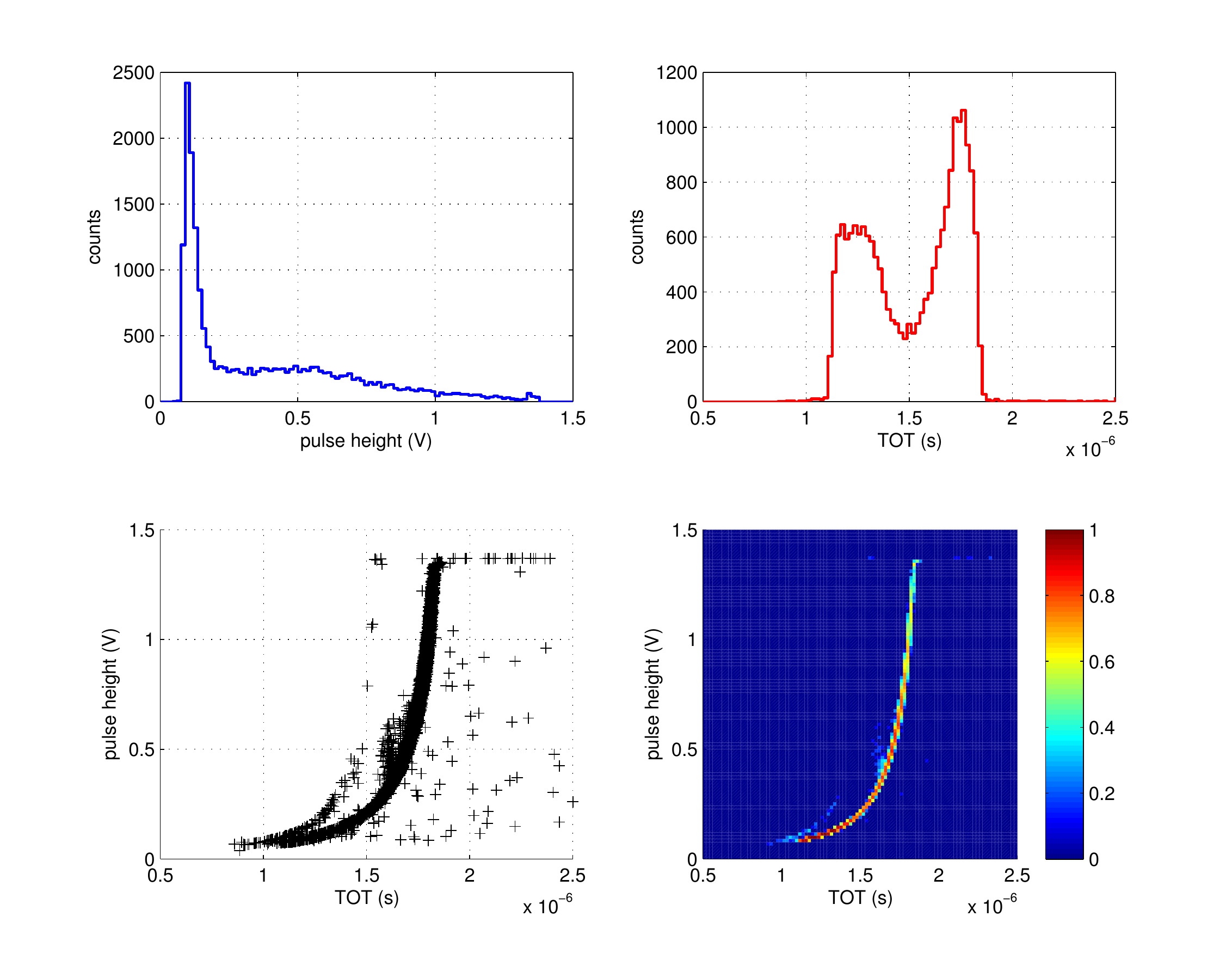}
\caption{\footnotesize Measurement of $S3$ (beam and AmBe source) -
amplifier $1\, \mu s$. \label{1usS3}}
\end{figure}
\\ By comparing Figure \ref{1usS1}
with Figures \ref{1usS2} and \ref{1usS3}; we note that in the
correlation plot TOT-PH the $\gamma$-ray contribution is mixed with
the neutron one as it is in the PHS. The low TOT events for neutrons
and $\gamma$-rays (below $1.5\,\mu s$) are just as
indistinguishable.
\\ We repeat the analysis for the fast amplifier. Figures \ref{3nsS1},
\ref{3nsS2} and \ref{3nsS3} show the signal analysis for the three
sets of measurements for the fast amplifier.
\begin{figure}[!ht]
\centering
\includegraphics[width=12cm,angle=0,keepaspectratio]{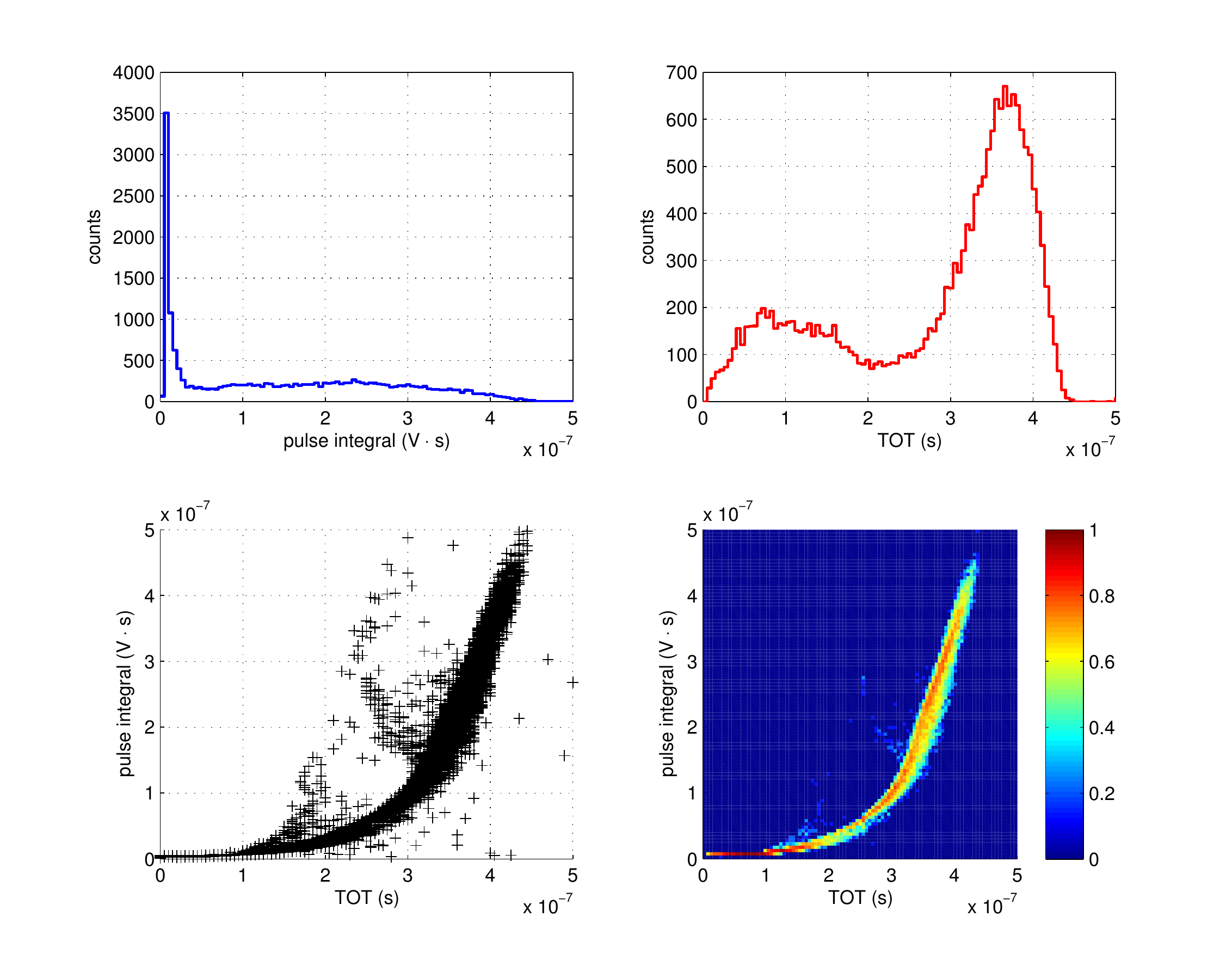}
\caption{\footnotesize Measurement of $S1$ (beam only) - amplifier
$3\,ns$. \label{3nsS1}}
\end{figure}
\begin{figure}[!ht]
\centering
\includegraphics[width=12cm,angle=0,keepaspectratio]{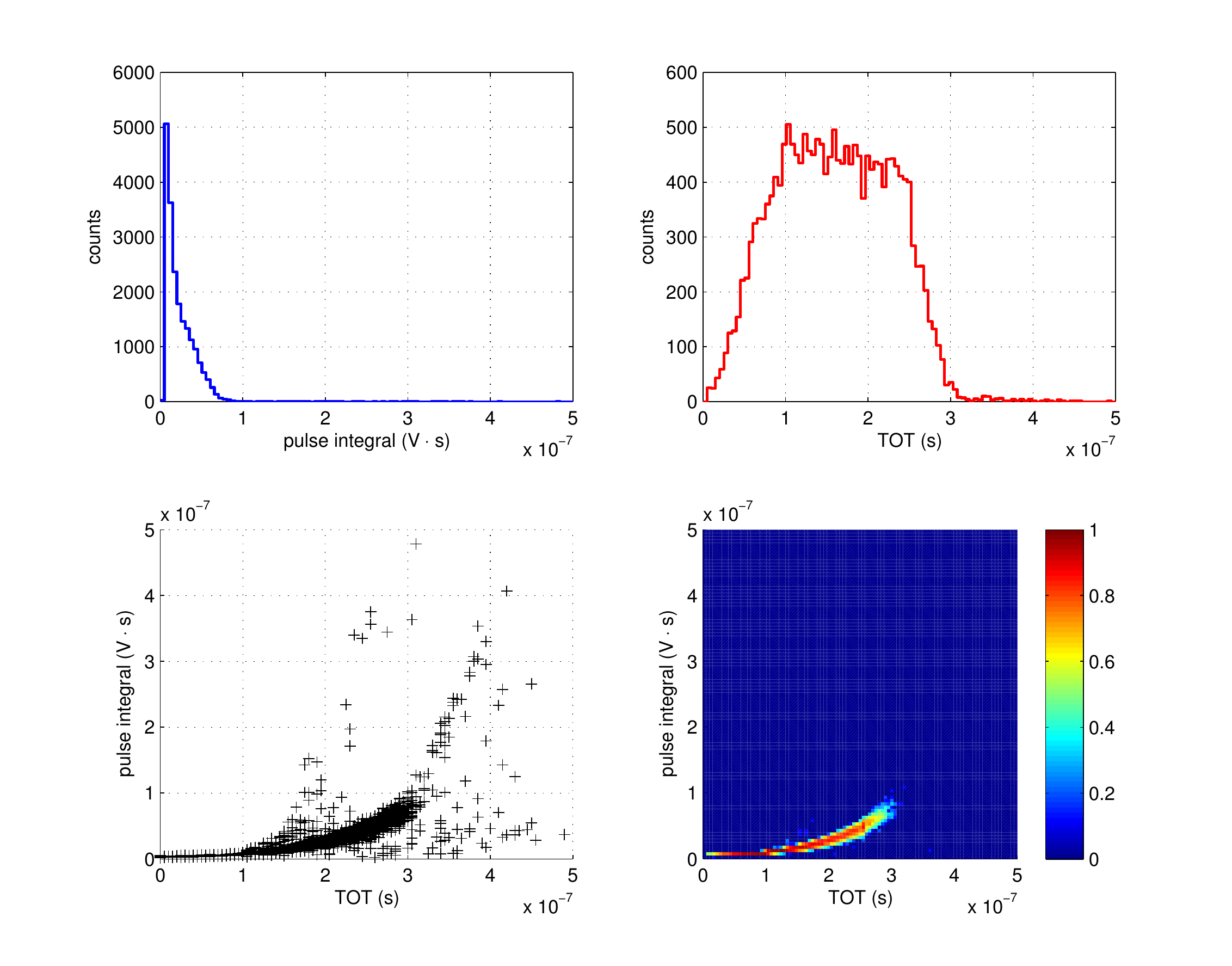}
\caption{\footnotesize Measurement of $S2$ (AmBe source only) -
amplifier $3\,ns$. \label{3nsS2}}
\end{figure}
\begin{figure}[!ht]
\centering
\includegraphics[width=12cm,angle=0,keepaspectratio]{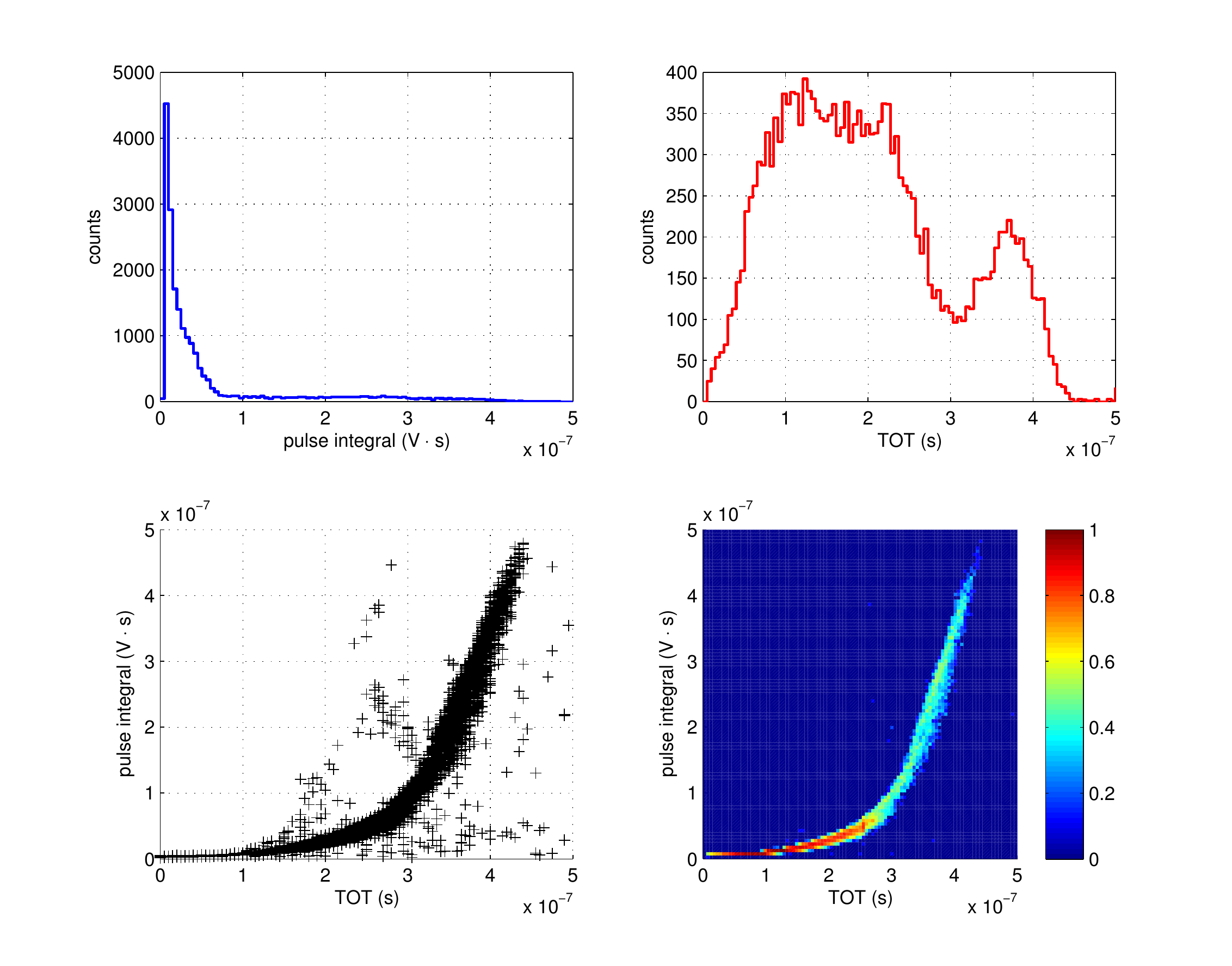}
\caption{\footnotesize Measurement of $S3$ (beam and AmBe source) -
amplifier $3\,ns$. \label{3nsS3}}
\end{figure}
\\In Figures \ref{3nsS1} the contribution is mostly given by
neutrons. In the TOT spectrum we notice two peaks corresponding to
the $\alpha$-particle and the $^7Li$-fragment as it is in the PHS.
\\ There is no physical difference in the signal shape using the TOT that can be
exploited to discriminate against background. The amplitude
discrimination remains the best way to reduce $\gamma$-ray
sensitivity of neutron detectors based on a solid converter.

\chapter{The Multi-Blade prototype}\label{MBprotochapt}
I have to thank my group head, Bruno Gu\'{e}rard, who pushed me a
lot for the prototype construction and believed in the success of
this prototype concept. I also want to thank Jean-Claude Buffet and
Sylvain Cuccaro for their important suggestions and the work that
made possible the construction of the Multi-Blade prototype.
\bigskip
\bigskip
\begin{figure}[!ht]
\centering
\includegraphics[width=10cm,angle=0,keepaspectratio]{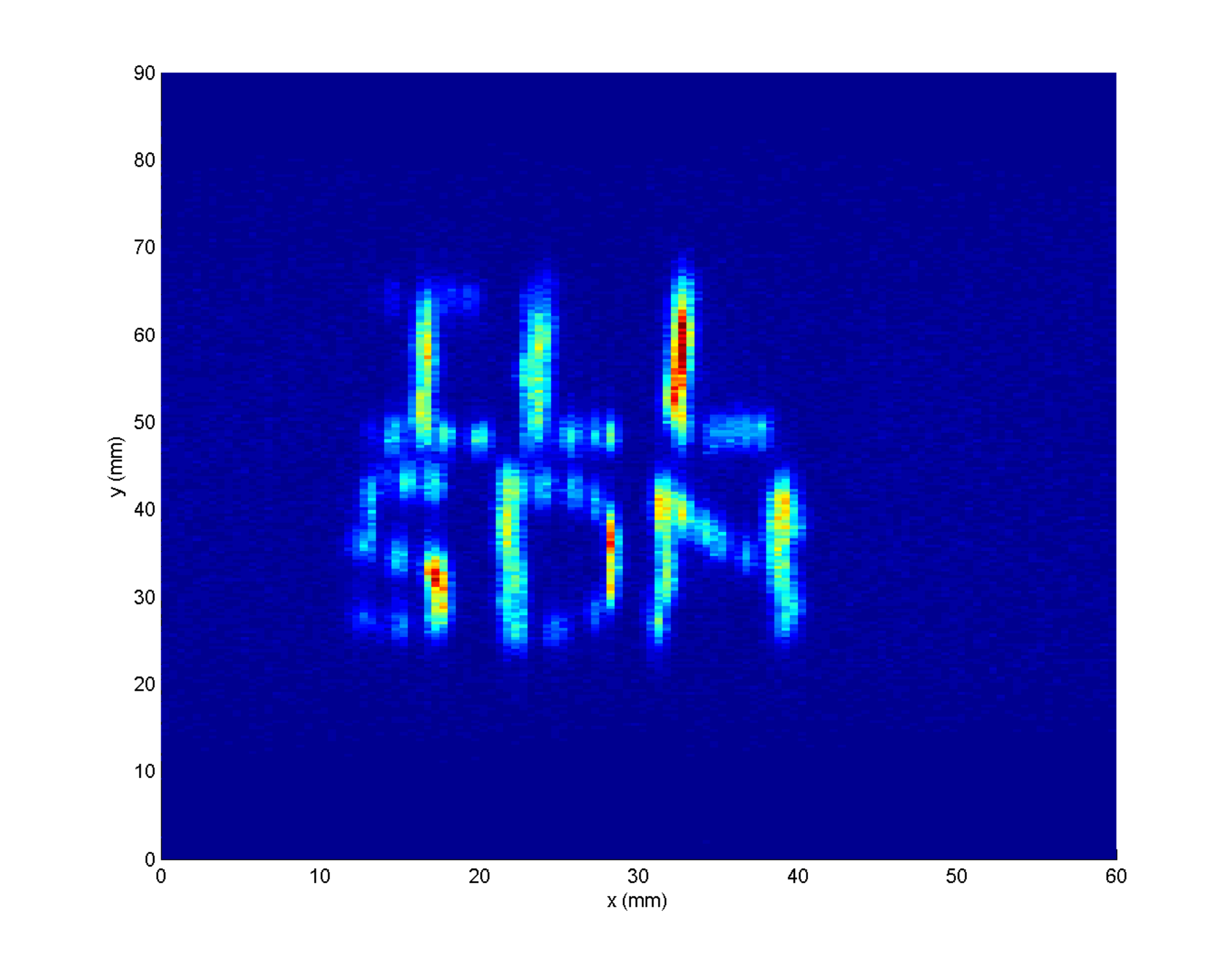}
\end{figure}

\newpage
\section{Detectors for reflectometry and Rainbow}
\begin{figure}[!ht]
\centering
\includegraphics[width=10cm,angle=0,keepaspectratio]{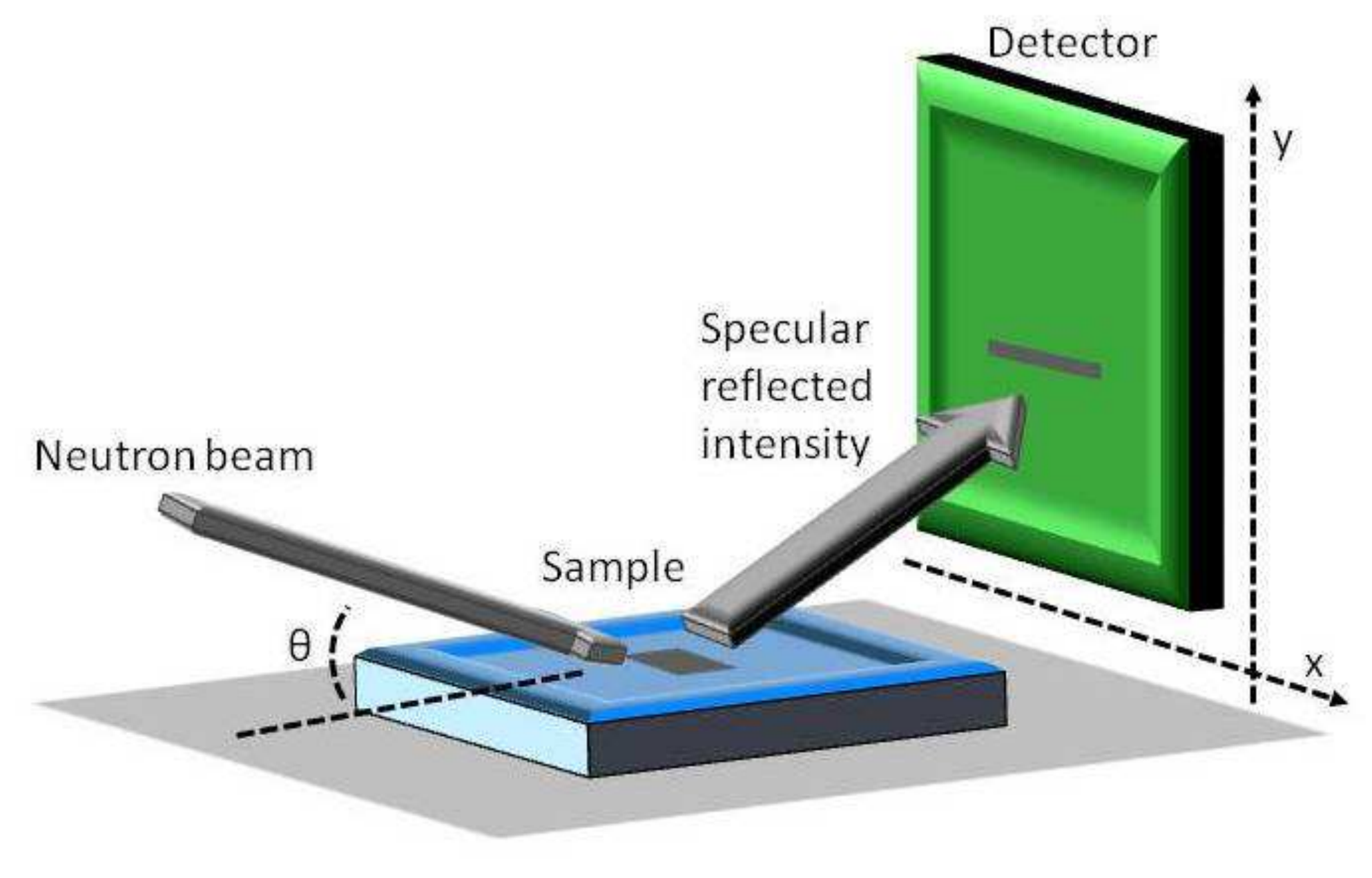}
\caption{\footnotesize A neutron reflectometry instrument
schematic.} \label{nrs5}
\end{figure}

\begin{figure}[!ht]
\centering
\includegraphics[width=9cm,angle=0,keepaspectratio]{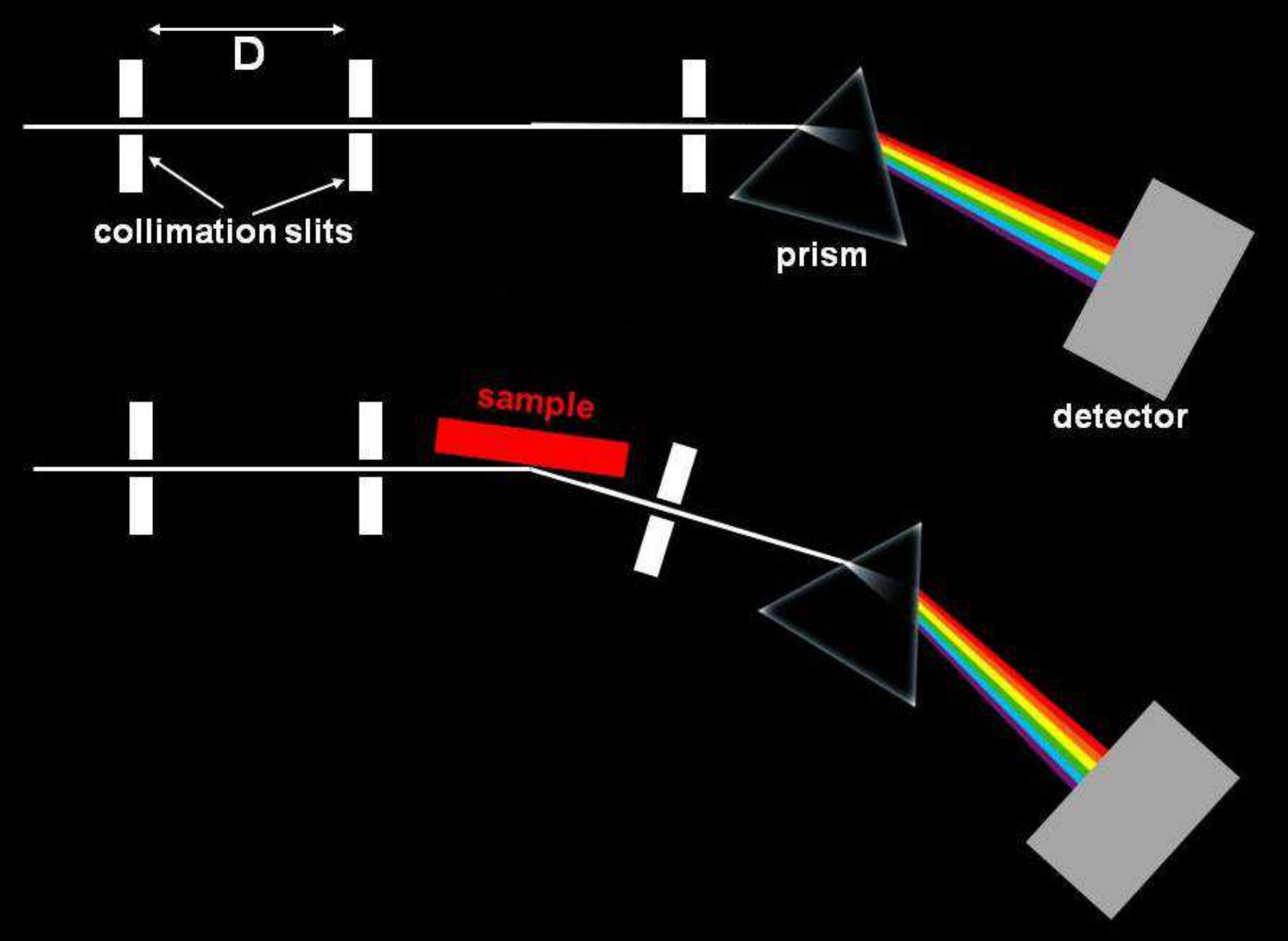}
\caption{\footnotesize Schematic of the reflectometer Rainbow
\cite{rainbow2} involving a prism to deduce the neutron wavelength
by refractive encoding.} \label{rain1}
\end{figure}

Because of its favorable properties, $^3He$ has been the main actor
in neutron detection for years. Nowadays its shortage pushes many
researchers to investigate alternative ways to efficiently detect
neutrons. For large area detectors, i.e. ToF spectrometers, several
squared meters in size, it is crucial to find an $^3He$ replacement.
At ILL efforts have been made to develop the Multi-Grid
\cite{jonisorma}. This is a large area neutron detector that
exploits up to $30$ $^{10}B_4C$-layers in a cascade configuration;
its optimization has already been explained in details in Chapter
\ref{Chapt1}. On the other hand, there are several neutron
instruments that can work with small detector size. A neutron
reflectometer, such as Figaro at ILL needs a detector surface of
$400\times250\,mm^2$. For these applications a limited amount of
$^3He$ is required and its shortage is not the main issue to be
addressed.
\\ Neutron scattering science is still growing its instruments' power
and together with that the neutron flux a detector must tolerate is
increasing. The peak brightness at ESS, the new European Spallation
Source, will be higher than that of any of the short pulse sources,
and will be more than one order of magnitude higher than that of the
World's leading continuous source. The time-integrated brightness at
ESS will also be one to two orders of magnitude larger than is the
case at today's leading pulsed sources \cite{esstdr},
\cite{gebauer1}.
\\ The Multi-Blade concept wants to address the counting rate
capability of $^3He$-based detectors for high flux applications. We
want to develop a detector suitable for neutron reflectometry
instruments.
\\ The main goal in neutron reflectometry instruments is to achieve a
high angular resolution at high counting rates.
\\ A neutron detector for a reflectometry instrument is in general compact in
size and the spatial resolution required is of the order of $1\,mm$
in order to achieve the needed angular resolution. Figure \ref{nrs5}
shows a reflectometry instrument schematic. Neutron reflection by a
sample is measured as a function of the momentum transfer $q$, as
shown in Section \ref{neutrefltheoint}, the value of $q$ can be
obtained from Equation \ref{eqaf16}
($q=\frac{4\pi}{\lambda}\sin(\theta)$). If $\theta$ is kept fixed
the reflectometer works in ToF-mode and the neutron beam is chopped
to get the ToF information that leads to $\lambda$. If, on the other
hand, the neutron beam is monochromatic and $\theta$ is scanned, the
reflectometer works in monochromatic-mode. A detector for a
reflectometer is characterized by a non-uniform spatial resolution.
Referring to Figure \ref{nrs5}, a high spatial resolution is only
needed for the $y$ direction. This is true because for a large
number of applications only the specular reflection is needed and
the other coordinate ($x$) is generally integrated over.
\\ A PSD (Position Sensitive Detector) is necessary when not only specular reflection
occurs but one wants to quantify more sample features, e.g.
off-specular reflection arising from the presence of in-plane
structures. It is more important to know the actual position of the
reflected intensity on the detector to determine $\theta$ for
off-specular studies.
\\ In order to give the specifications of a neutron reflectometer we
list some features of Figaro at ILL. The neutron wavelength range
explored is between $1$\AA \, and $30$\AA. The sample to detector
distance can be varied between $1.2\,m$ and $2.9\,m$. The white beam
flux before the instrument was recorded as
$1.4\cdot10^{10}\,neutrons/cm^2 s$; the flux on the sample using the
widest chopper pair with collimation slits openings of $0.8
\times40\,mm^2$ and $0.4\times30\,mm^2$ gives a neutron count rate
of $4\cdot10^{4}\,neutrons/cm^2 s$ on a sample area of $40cm^2$. The
Figaro detector is a PSD of $400\times250\,mm^2$ constructed from a
single block of aluminium, with 64 square ($7mm$ side), $25cm$ long
channels. Each tube is filled with 8 bars of $^3He$ and 2 bars of
$CF_4$ and contains a $15\,\mu m$ Stablohm wire of $250\,mm$ active
length, which detects neutrons by charge division. The vertical
position resolution is $2\,mm$ and the horizontal position
resolution is $8\,mm$ \cite{figaro}.
\\ In a $^3He$-based detector the counting rate is limited by the
space charge effect, as the ions created by each avalanche need more
time, compared to electrons, to be evacuated. At high rate they tend
to accumulate and consequently they decrease the actual electric
field in the gas volume, and as a result the detector loses
efficiency. If Figure \ref{nrs5} represents the Figaro's detector,
the $^3He$-tubes are placed vertically along $y$ in order to split
the reflected intensity over several tubes.
\\Morever, $^{3}He$ detectors of this type are limited in spatial resolution for
two main reasons. The first is that $^{3}He$-based detectors are
gaseous detectors which exploit anode wires for read-out; in one
direction wires can not be mounted with a $mm$ spacing because this
causes mechanical issues, on the other direction, along the wire,
the spatial resolution that can be achieved by a charge division
read-out is limited at about $2\,mm$ on a $30cm$ wire length, i.e.
$\sim0.7\%$. The second reason limiting the spatial resolution is
the gas pressure that can be reached. To reduce the particle traces,
and thereby increase the spatial resolution, those detectors are
operated at high gas pressure, resulting in mechanical constraints.
The higher the gas pressure, the more severe are the mechanical
problems arising in the detector vessel conception. There is a
reasonable limit in the resolution the can be reached with this
technique, around $1\,mm$ \cite{ott}.
\\ In many areas of soft and hard matter research science, the amount
of material to investigate is rather limited. Partly because the
fabrication of larger samples is too expensive or not feasible, yet,
partly because the interesting features depend on the size. The
development of a neutron reflectometer optimized for small samples
is under study \cite{rainbow1}. There is a great deal of interest in
expanding the technique of neutron reflectometry beyond static
structural measurements of layered structures to kinetic studies
\cite{cubitt2}. The time resolution for kinetic studies is limited
by the available neutron flux.
\\ For both working modes (monochromatic and ToF) a reflectometer is
operated, the actual neutron flux reaching the sample is a small
fraction of the incoming neutron beam. For monochromatic
instruments, this limitation arises from the monochromators used to
get a neutron beam of a defined energy. For instruments which use
the ToF technique the flux is limited by the choppers. E.g. the
reflectometer D17 \cite{cubittD17} at ILL has a transmission of only
$10^{-3}$ at about $2\,$\AA \, if a wavelength resolution $\Delta
\lambda/\lambda$ of $1\%$ is needed.
\\ The resolution in momentum transfer $q$ is related to the neutron
wavelength $\lambda$ and the angular resolution in $\theta$:
\begin{equation}\label{ecfaweqef54906}
\left(\frac{\Delta q}{q}\right)^2=\left(\frac{\Delta
\lambda}{\lambda}\right)^2+\left(\frac{\Delta
\theta}{\theta}\right)^2
\end{equation}
In ToF, $\Delta \lambda$ is determined by the chopper settings and
time resolution, and in a monochromatic approach, by the
monochromator resolution. When only considering specular reflection
the angular resolution $\Delta \theta$ is determined by collimation
and beam divergence.
\\ In \cite{cubitt2} and \cite{rainbow2} a new instrument layout is
presented: Rainbow. This instrument would involve a prism refraction
to deduce the wavelength in place of choppers thus providing a large
gain in useful neutron flux. Hence, it open the possibility of
sub-second kinetic studies. In this new approach, as we will
discuss, $\Delta \theta$ in Equation \ref{ecfaweqef54906} is now
determined also by the detector spatial resolution and its distance
to the sample.
\\ By using the reflectometer D17 \cite{cubittD17} at ILL, the actual measurement
time of $1\,s$ is only possible by loosening the $q$-resolution of
the instrument: between $4\%$ (at about $20\,$\AA) and $10\%$ (at
about $2\,$\AA). The advantage of using a prism would be to measure
faster and without the cost in resolution. The technique would be of
equal value for experiments with sample areas much smaller than can
be practically measured at present.
\\ Figure \ref{rain1} shows a schematic of Rainbow.
A white and continuous neutron beam is collimated before the sample.
A standard distance used between the collimation slits is about
$D=3\,m$. The prism is placed after the sample and is calibrated by
passing the direct beam through the prism alone to measure the
deflection of the beam due to refraction. The deflected angle
$\varphi$ by which a certain neutron wavelength is refracted depends
only on the prism angle $\alpha$ and the scattering length density
of the prism material. It is given by Snells law:
$\varphi=\arccos\left(n\,\cos\left(\alpha\right)\right)$. Where $n$
is refractive index in Equation \ref{eqaf7} which depends on the
neutron wavelength $\lambda$ and $\alpha$ is the angle of the beam
to the prism surface.
\\ With a sample in the direct beam, the prism and the detector need
to be rotated such that the sample reflection strikes the center of
the prism and is refracted. The intensity of the refracted spectrum
is measured as a function of the deflection. For specular
reflection, not only the incoming beam is well collimated but also
the reflected beam before refraction. It then reaches the prism
surface and each wavelength is refracted at unique angle $\varphi$.
\\ In a practical situation, we should consider a spread in the incoming
angle from the collimation, the imperfect flatness of the prism
surface, and the resolution of the detector.
\\ In \cite{rainbow2} it has been demonstrated that the
detector spatial resolution, in one direction, has to be of about
$0.3\,mm$ to reach $\Delta \lambda/\lambda=5\%$ resolution at
$2\,$\AA.
\\ On the actual D17 instrument \cite{cubittD17} one can implement the ''rainbow'' principle.
Its detector spatial resolution is $\Delta x=2\,mm$. In order to get
a sufficiently high angular resolution for the reflected beam, the
detector is positioned as far as possible from the sample, i.e.
about $3\,m$. Due to practical limits in $^3He$ detector resolution
and collimation, a resolution of $\Delta \lambda/\lambda<5\%$ at
short wavelengths is probably not practical. Therefore, the
development of an area detector with $\Delta x=0.2\,mm$ required in
one dimension only (the other dimension can be summed) is crucial
\cite{cubitt2}.
\\ Although $^{3}He$ shortage affects scientific research; this is not
the main issue for neutron reflectometry applications. A promising
alternative, to accomplish the high spatial resolution and the high
count rate capability, is to exploit solid $^{10}B$-films employed
in a proportional gas chamber. The challenge with this technique is
to attain a suitable detection efficiency which is about $63\%$ for
the Figaro detector at $2.5$\AA. This can be achieved by operating
the $^{10}B$ conversion layer at grazing angle relative to the
incoming neutron direction. The Multi-Blade design is based on this
operational principle and it is conceived to be modular in order to
be adaptable to different applications. A prototype has been
developed at ILL and the results obtained on our monochromatic test
beam line are presented here. A significant concern in a modular
design is the uniformity of detector response. Several effects might
contribute to degrade the uniformity and they have to be taken into
account in the detector concept: overlap between different
substrates, coating uniformity, substrate flatness, parallax errors,
etc.

\section{The Multi-Blade concept}
\begin{figure}[!ht]
\centering
\includegraphics[width=10cm,angle=0,keepaspectratio]{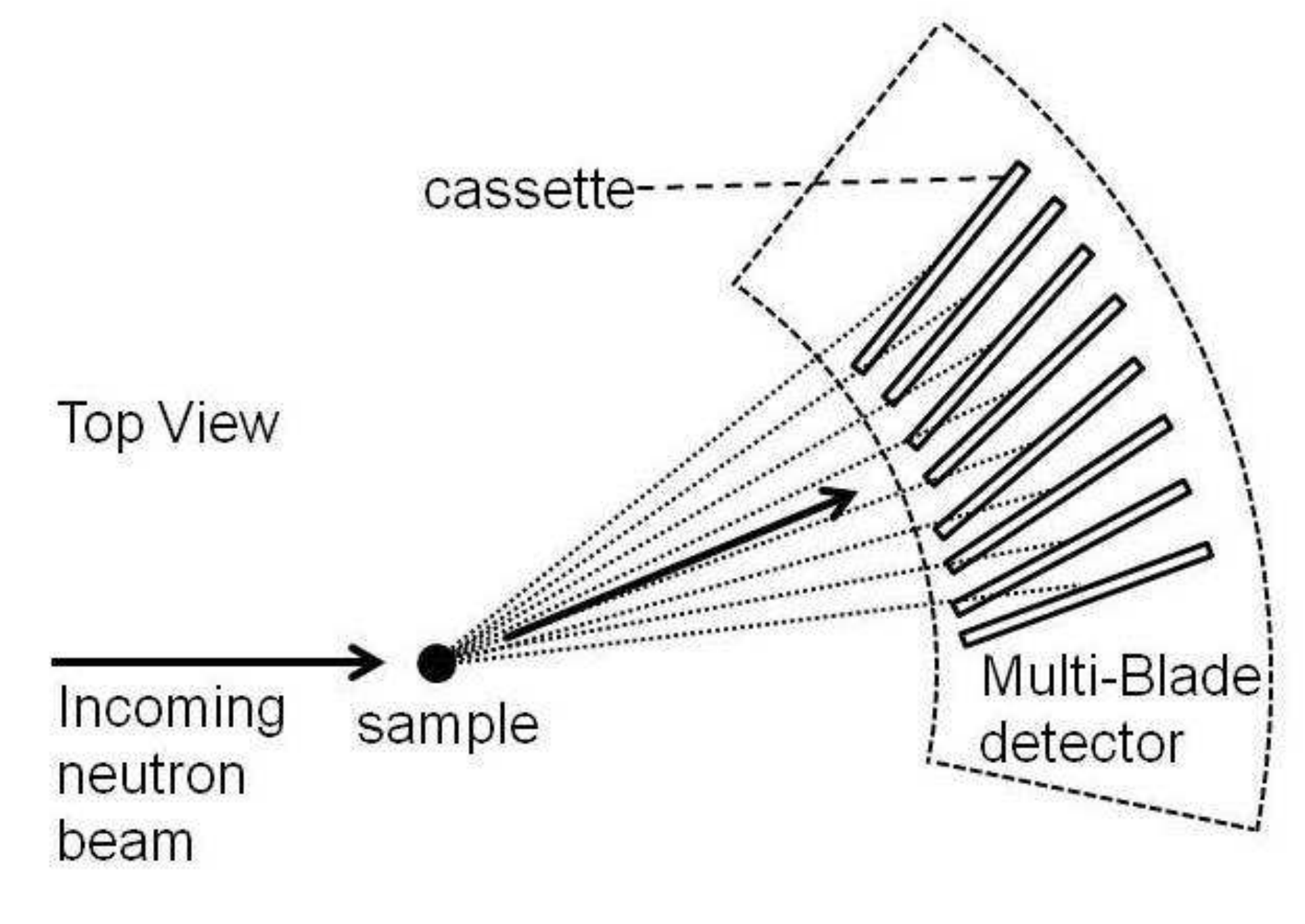}
\caption{\footnotesize The Multi-Blade detector sketch (top
view).}\label{schemMBlight6}
\end{figure}

\begin{figure}[!ht]
\centering
\includegraphics[width=13cm,angle=0,keepaspectratio]{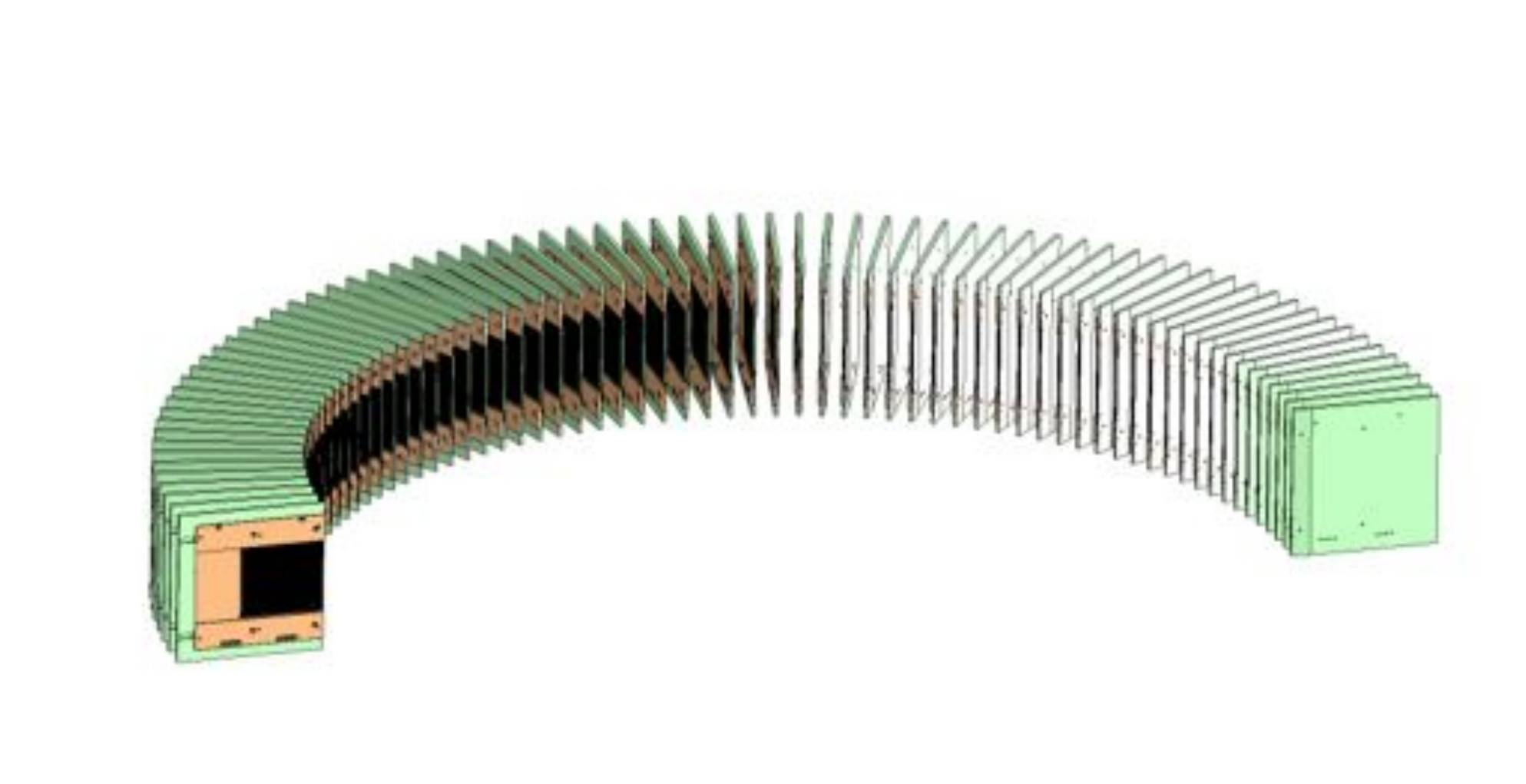}
\caption{\footnotesize Several \emph{cassettes} arranged in a
cylindrical configuration around the sample position.}
\label{schemMBlight645}
\end{figure}

\begin{figure}[!ht]
\centering
\includegraphics[width=5cm,angle=0,keepaspectratio]{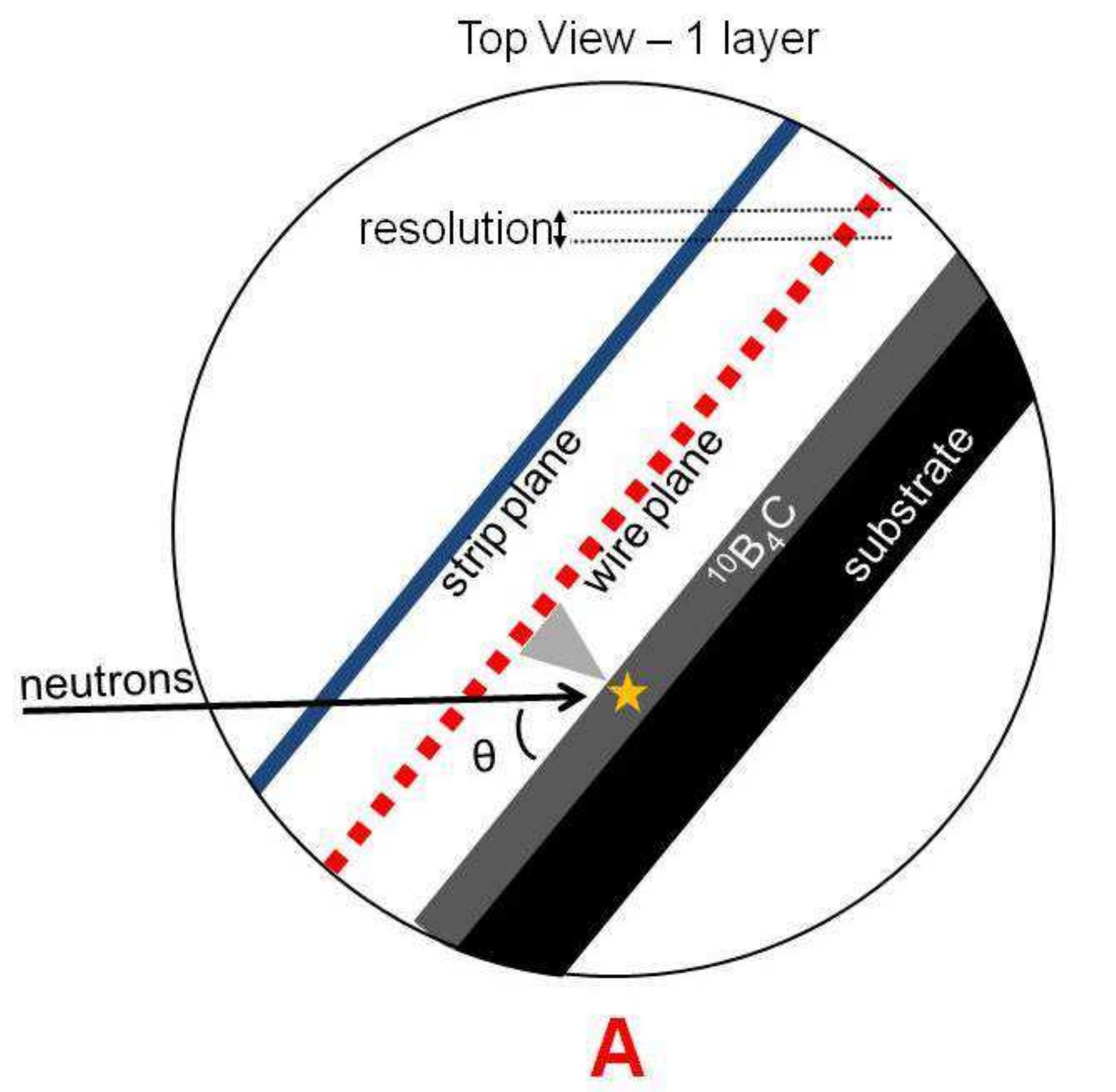}
\includegraphics[width=5cm,angle=0,keepaspectratio]{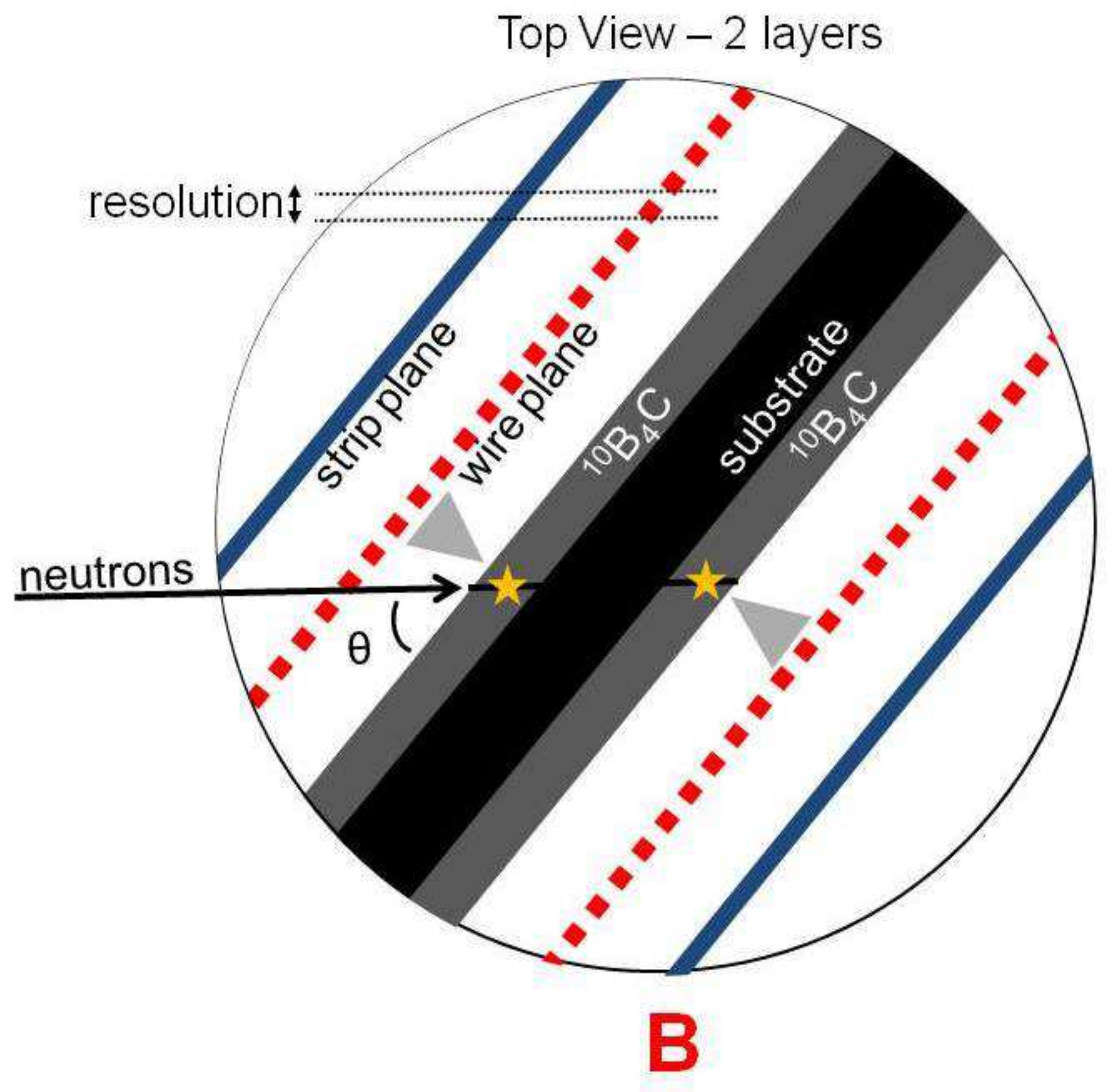}
\includegraphics[width=5.4cm,angle=0,keepaspectratio]{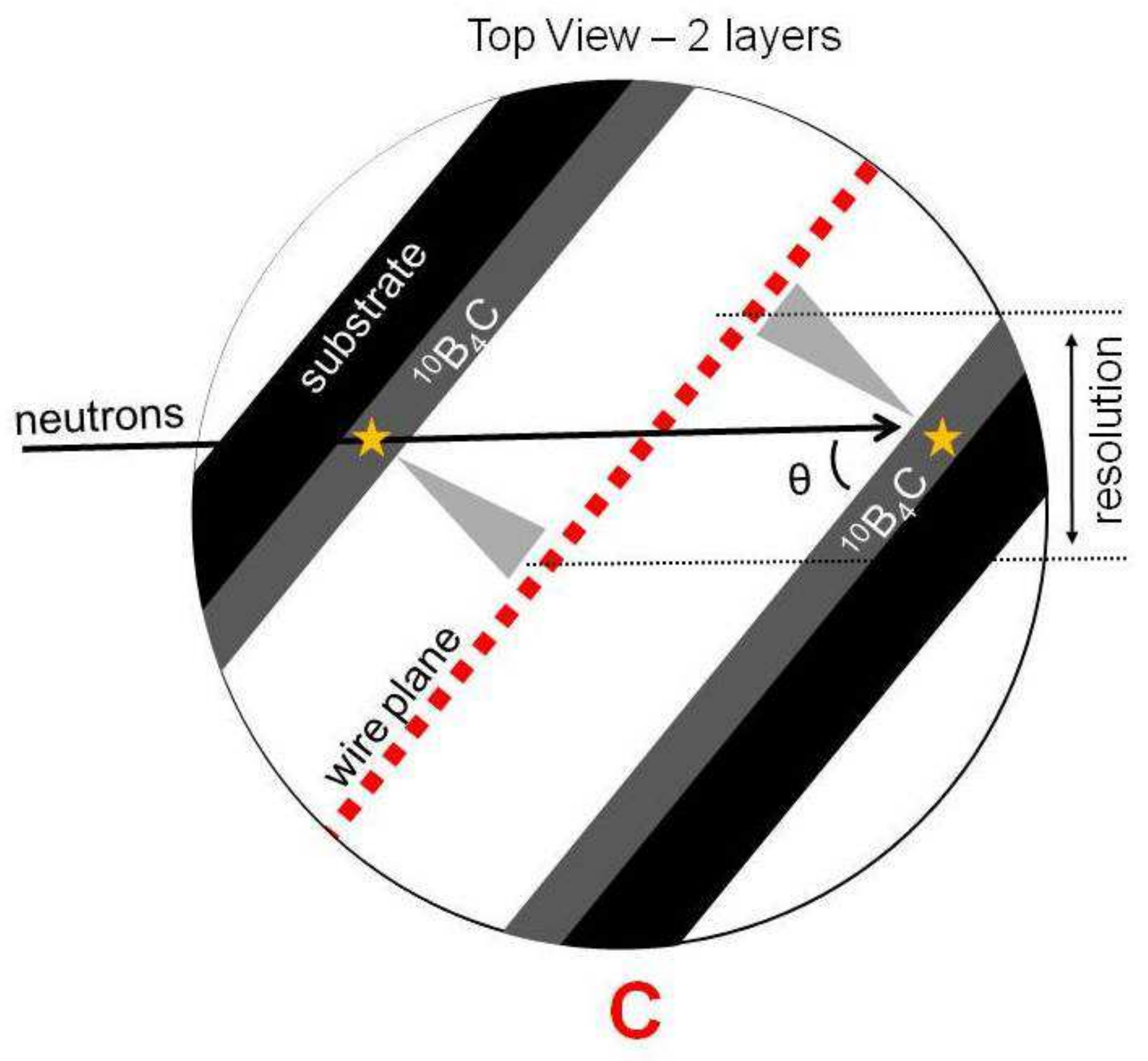}
\caption{\footnotesize Cross-section of one cassette. Three options
are shown: A, a single converter layer; B and C, with two
converters.}\label{figabc09}
\end{figure}

\begin{figure}[!ht]
\centering
\includegraphics[width=7.8cm,angle=0,keepaspectratio]{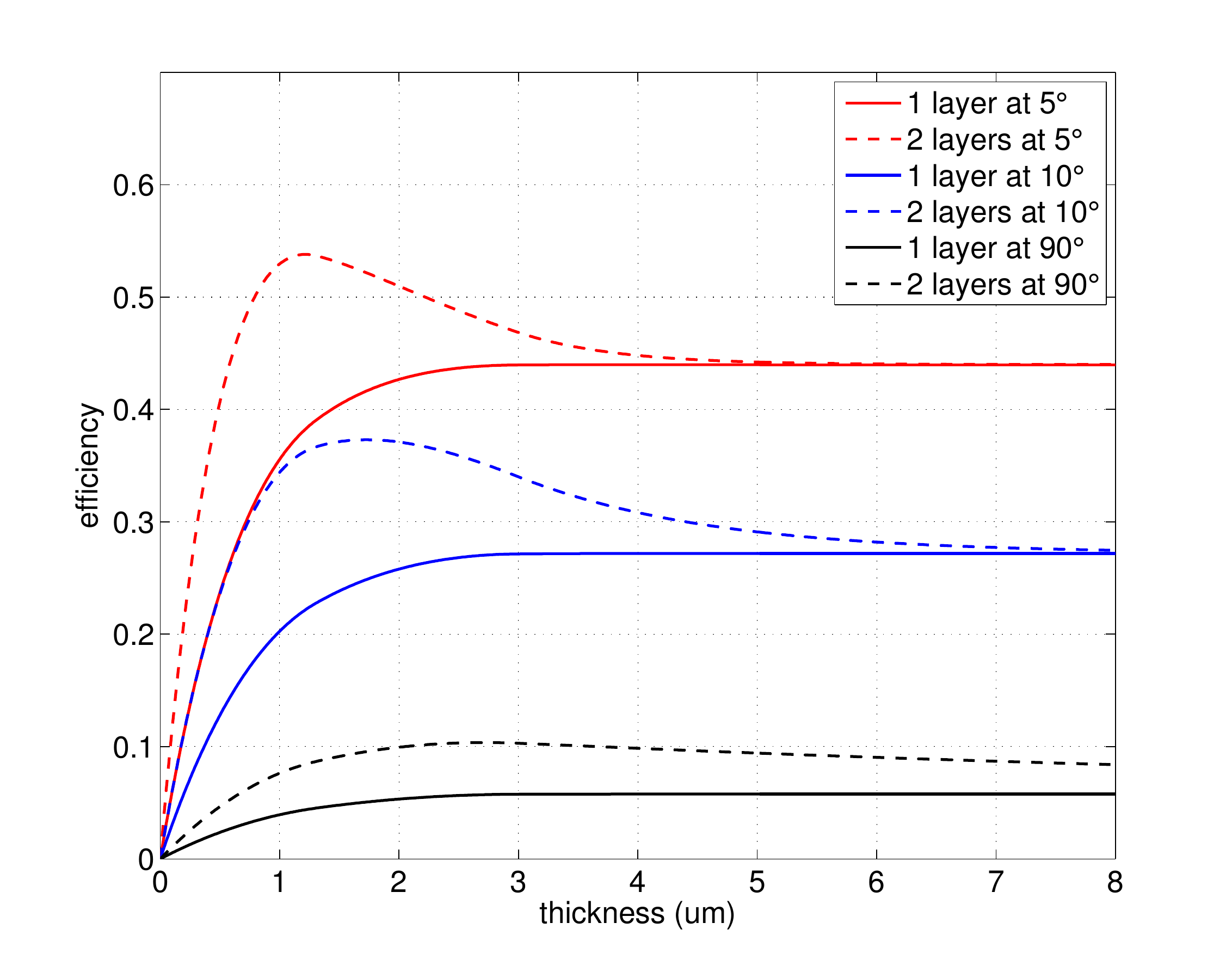}
\includegraphics[width=7.8cm,angle=0,keepaspectratio]{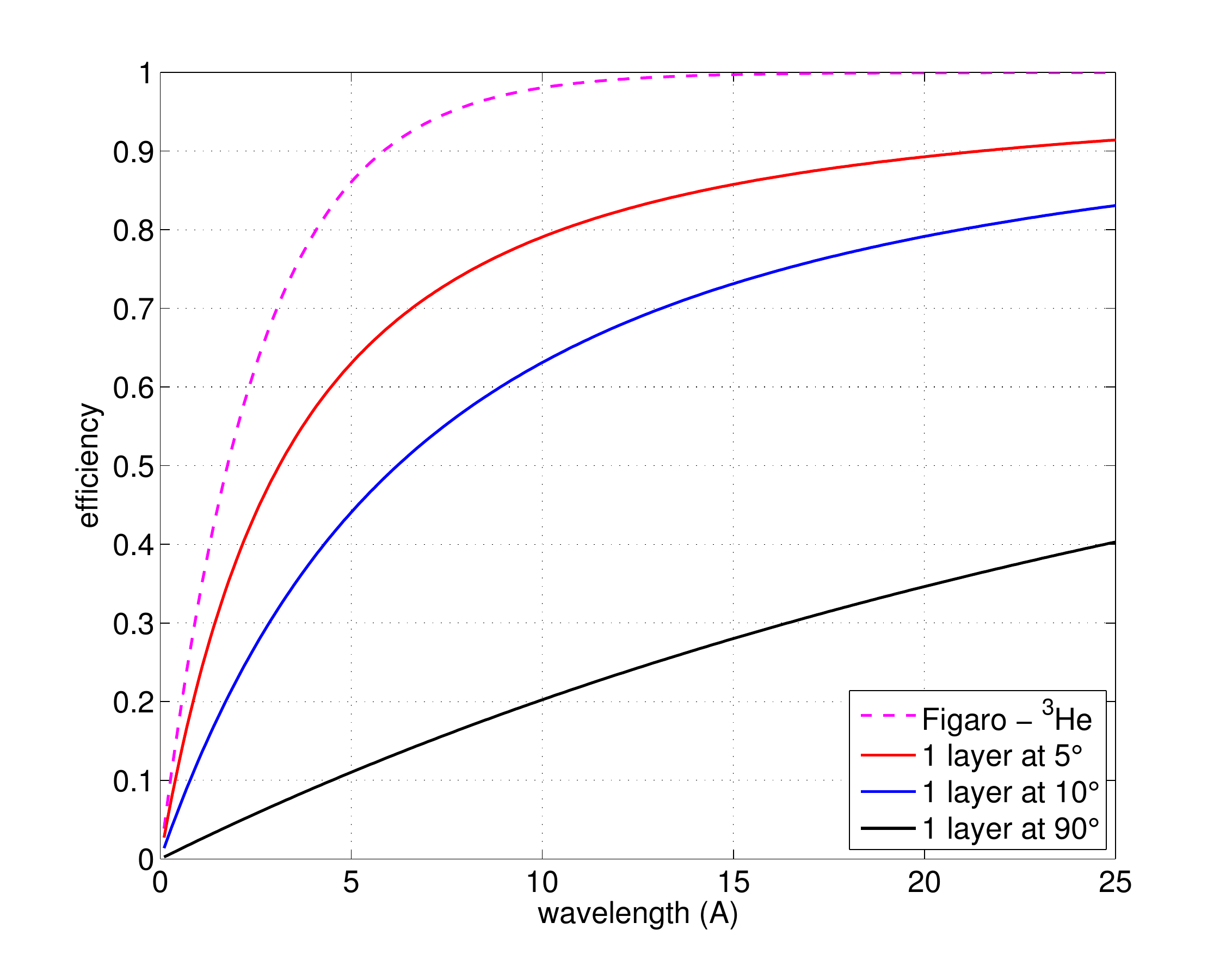}
\caption{\footnotesize $^{10}B_4C$ layers ($\rho=2.24\,g/cm^3$)
detection efficiency at $2.5$\AA \, as a function of the layer
thickness for the options A and B for three inclinations (left),
efficiency as a function of the neutron wavelength for three
inclinations of a single $3\,mu m$ layer (right). An energy
threshold of $100\,KeV$ is applied. The efficiency of the Figaro's
detector is shown as well ($6.9\,mm$ tubes filled with $8\,bars$ of
$^3He$).}\label{figgig345htyigv}
\end{figure}

\begin{figure}[!ht]
\centering
\includegraphics[width=8cm,angle=0,keepaspectratio]{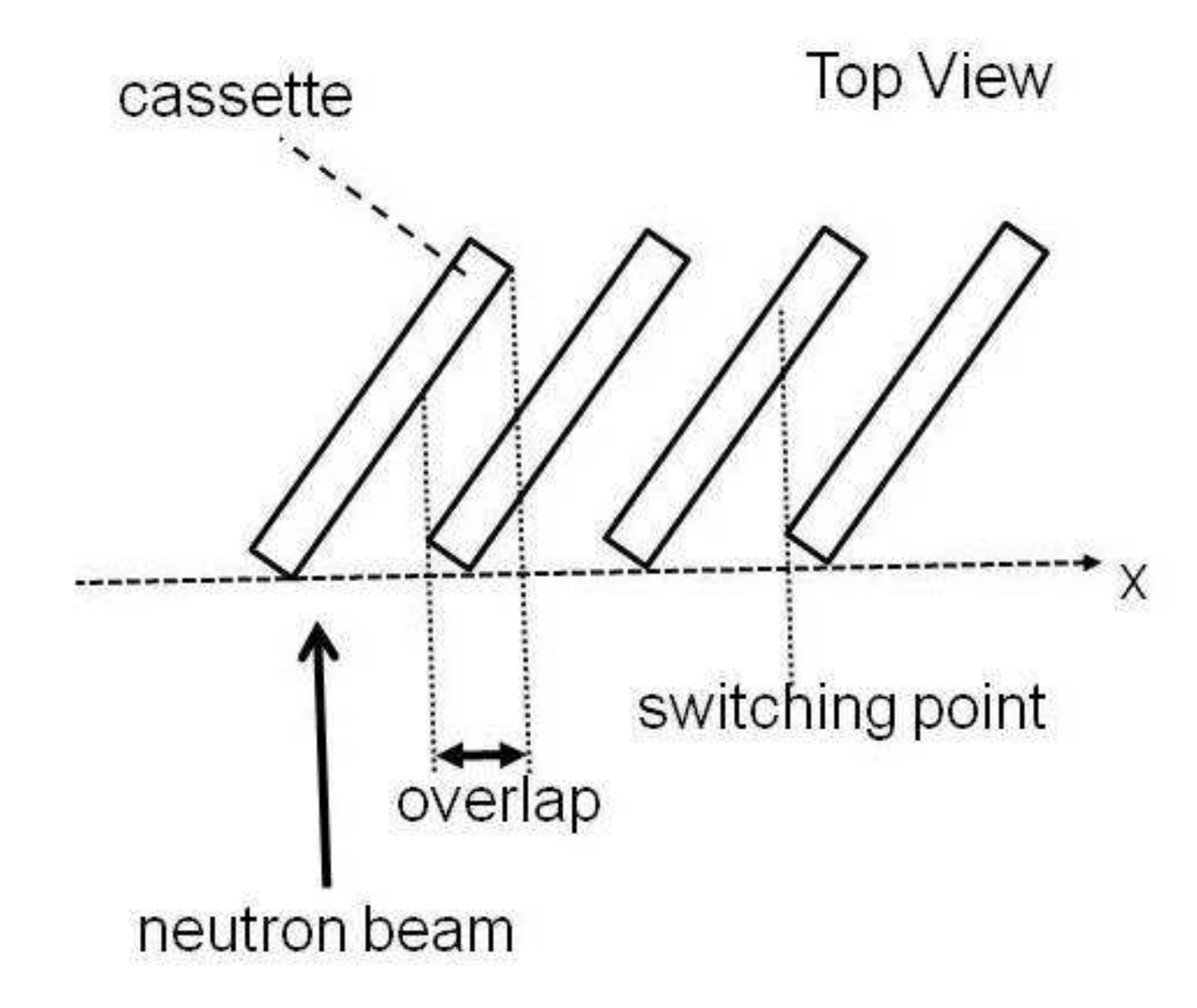}
\caption{\footnotesize Four cassettes disposed one after the other.
Their overlap and the switching region between one and an other is
an important aspect to be studied.}\label{problemsmb89dir84}
\end{figure}

The Multi-Blade concept was already introduced at ILL in 2005
\cite{buff1} and a first prototype was realized in 2012
\cite{buff3}. Its design is conceived to be modular in order to be
versatile to be applied in many applications on several instruments.
The Multi-Blade exploits solid $^{10}B$-films employed as a neutron
converter in a proportional gas chamber as in \cite{jonisorma}. The
challenge with this technique is to attain a suitable detection
efficiency. This latter can be achieved by operating the $^{10}B$
conversion layer at grazing angle relative to the incoming neutrons
direction. Moreover the inclined geometry leads to a gain in spatial
resolution and as well in counting rate capability compared to
$^{3}He$ detectors.
\\ Figures \ref{schemMBlight6} and \ref{schemMBlight645} show the Multi-Blade detector
schematic, it is made up of several identical units called
\emph{cassettes}. Each \emph{cassette} acts as an independent MWPC
(Multi Wire Proportional Chamber) which holds both the neutron
converter and the read-out system. The fully assembled detector is
composed of several cassettes inclined toward the sample position.
The angle subtended by each cassette looking at the sample position
is kept constant in order to maintain the spatial resolution and the
efficiency as uniform as possible.
\\ The cassettes must be arranged taking into account an overlap
between them in order to avoid dead space over the whole detector
surface. Moreover, once the instrument geometry changes the cassette
arrangement in the detector should also change; if the
sample-detector distance changes, the cassettes inclination should
change too, if we want to avoid dead spaces.
\\ Each cassette should contain one or more neutron converters, e.g.
$^{10}B_4C$ layers, and the read-out system that has to assure the
two-dimensional identification of the neutron event. Figure
\ref{figabc09} shows the cross-section of the cassette concept for
three different configurations.
\\ In the A and B solutions, in each cassette a single converter layer
is facing each read-out system. The read-out is a wire plane and a
strip plane placed orthogonally. The space between the strips and
the converter is filled with stopping gas at atmospheric pressure to
ensure the gas multiplication. The converter layer as well is
polarized and together with the strip plane acts as a cathode; the
wire plane, on the other hand, acts as an anode plane.
\\ In the C configuration a single wire plane performs the
two-dimensional read-out through charge division on resistive wires.
A single read-out system is facing two converter layers. The space
between the two converters is filled with stopping gas. The two
converter layers act as cathodes.
\\ Referring to configurations A and B, the identification of the
position of a neutron event is the coincidence of wire and strip
hits. The spatial resolution given by the strips does not depend on
the inclination of the cassette. The spatial resolution given by the
wire plane increases as the angle with the incoming neutron
direction decreases. E.g. if the resolution is given by the wire
pitch, the actual resolution is improved by a factor about $10$ at
$\theta=5^{\circ}$ ($1/\sin(\theta=5^{\circ})\sim 10$).
\\ Moreover, the actual neutron flux over the detector would be
divided by the same factor increasing its counting rate capability;
the same flux is shared by several wires.
\\ While the spatial resolution is improved by inclination in both options A and B in Figure
\ref{figabc09}, all the advantage of working at grazing angle is
lost in the C configuration. In \cite{kleinjalousie} can be found
the actual implementation of such a detector. In options A and B the
charge generated by neutron capture fragments in the gas gives a
signal on the facing wires and strips. In the solution C, for a
given incoming neutron direction there will be two regions on the
converters where neutrons are converted. The smaller the angle at
which we operate the detector, the larger is the distance between
those two regions. The uncertainty on the conversion point is then
given by this distance which is much bigger than the wire pitch. On
the other hand, option C has half the number of read-out channels as
compared to A and B. However, for us, high spatial resolution is
crucial.
\\ We decided to concentrate on the implementation of the option A
and B.
\\ In Chapter \ref{chaptreflectometry} we showed how the solid
converter layer efficiency increases as a function of its
inclination and how much neutron reflection affects that efficiency.
From reflectivity measurements we learnt for a common used neutron
wavelength range, e.g. from $1$\AA \, to $30$\AA, that all the
effects due to neutron reflection from $^{10}B_4C$ are negligible
down to grazing angles of $\theta=2^{\circ}$. Therefore reflectivity
is negligible for any kind of holding substrate that results into
different converter roughness. As a result we decided to operate the
Multi-Blade at either $\theta=10^{\circ}$ or $\theta=5^{\circ}$ in
order to maximize the detection efficiency without any reflection
concerns and keeping the mechanics simple.
\\ Figure \ref{figgig345htyigv} shows the detection efficiency for $^{10}B_4C$
layers ($\rho=2.24\,g/cm^3$) calculated according to the model
developed in Chapter \ref{Chapt1} and neglecting neutron reflection.
An energy threshold of $100\,KeV$ is applied. We considered two
possible configurations: options A and B in Figure \ref{figabc09},
with one converter or two. On the left we show the neutron detection
efficiency, at $2.5$\AA, as a function of the converter layer
thickness for the solutions A and B. While the efficiency shows a
maximum for the two layer option, it is saturated over $3\,\mu m$
for the single layer. The single converter option can attain a
maximum efficiency of $28\%$ at $10^{\circ}$ and $44\%$ at
$5^{\circ}$ ($2.5$\AA) to be compared with the double-layer
efficiency of $37\%$ and $54\%$ respectively. The addition of the
second layer, at $\theta=5^{\circ}$ leads to an increase of the
efficiency of about $10\%$ with respect to the solution A. The
advantage of having only one converter is that the coating can be of
any thickness above $3\,\mu m$ and the efficiency is not affected,
while for the two layer option its thickness should be well
calibrated. Moreover, in the two layer configuration the substrate
choice is also crucial because it should be kept as thin as possible
to avoid neutron scattering and this leads to possible mechanical
issues. On the solution A, the substrate choice can be more flexible
because it has not to be crossed by neutrons.
\\ On the right in Figure \ref{figgig345htyigv} we show the efficiency as a
function of the neutron wavelength for the the single layer of
thickness $3\,\mu m$ (configuration A). The Figaro's detector
efficiency \cite{figaro} efficiency is also plotted, it is a
$^3He$-based detector made up of $6.9\,mm$ tubes filled at
$8\,bars$. In the plots shown the detector gas vessel Aluminium
window is also taken into account as a neutron loss. For the
Figaro's detector we used a $5\,mm$ thick window, and, since the
Multi-Blade detector will be operated at atmospheric pressure, we
used a $2\,mm$ window. $^3He$-based detectors' efficiency can be
increased by increasing the $^3He$ pressure in the vessel; on the
other hand, for a solid converter based detector the gas acts only
as a stopping means, hence its pressure can be kept at atmospheric
values. Consequently the gas vessel construction has less
constraints.
\\ In each of the solutions proposed for the cassette concept, see Figure \ref{figabc09},
the read-out system has to be crossed by neutrons before reaching
the converter. The mechanical challenge in the read-out system
construction is to minimize the amount of material on the neutron
path to avoid scattering that can cause misaddressed events in the
detector.
\\ Figure \ref{problemsmb89dir84} shows four cassettes. They have to
overlap to avoid dead spaces and the event loss, due to the zone
where we switch the cassette, should be minimized. At the cassette
edge electric field distortions and structure holding materials can
cause a loss in the efficiency and consequently deteriorate the
detector uniformity.
\\ In the prototype realization all these problems have been taken into
account, their solutions will be explained in the next Section.
\section{Multi-Blade version V1}\label{sectv1mbg}
\subsection{Mechanical study}
The prototype was conceived
to clarify the advantages and disadvantages of the options A and B
shown in Figure \ref{figabc09}.
\\ The detector works as a standard MWPC operated at atmospheric pressure
stopping gas such as $Ar/CO_2$ $(90/10)$ or $CF_4$.
\\ Since we want to avoid neutrons to be scattered before reaching the
converter layer, we need to minimize the amount of matter that has
to be crossed by neutrons: the read-out system and the cassette
window.
\begin{figure}[!ht]
\centering
\includegraphics[width=12cm,angle=0,keepaspectratio]{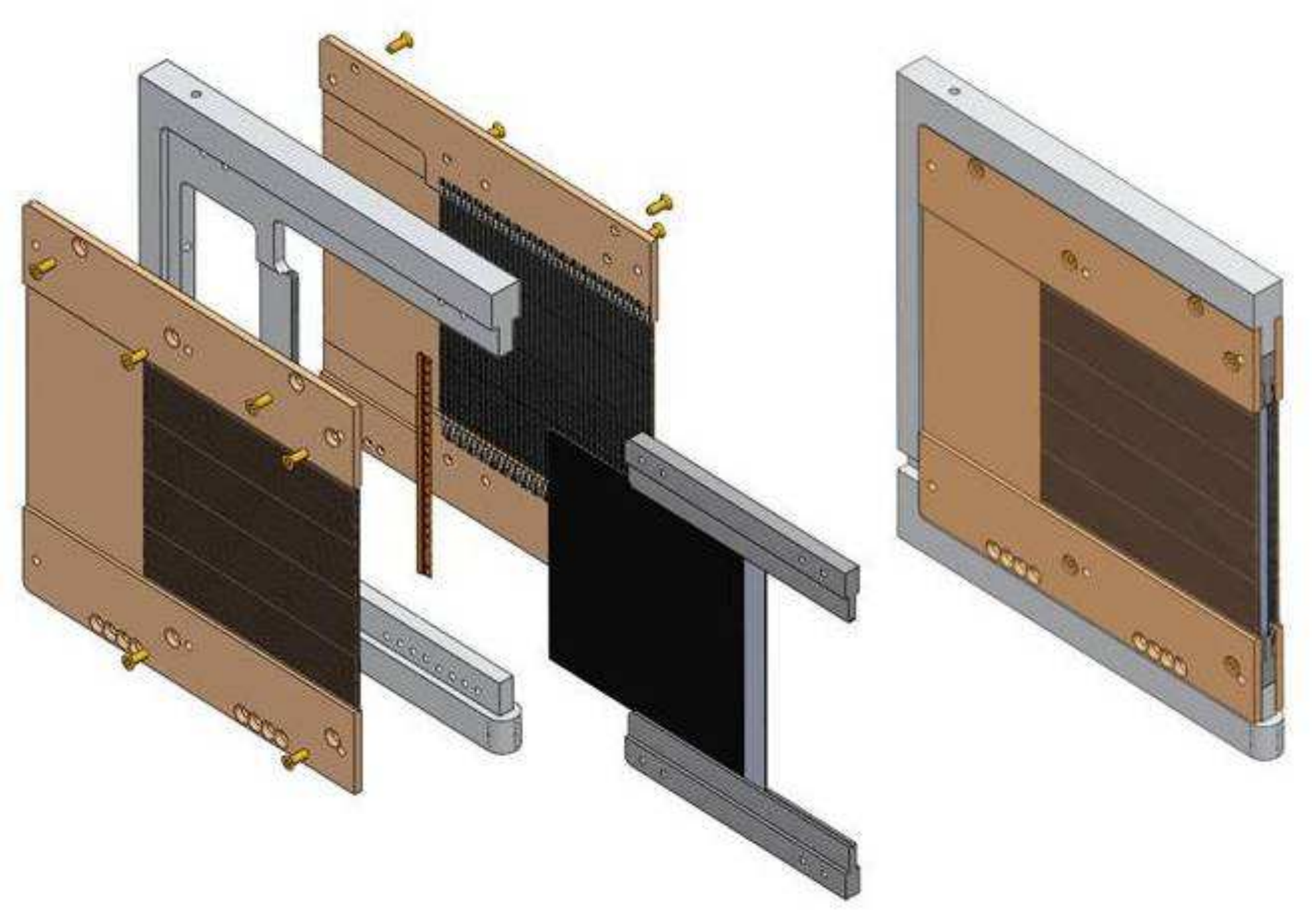}
\caption{\footnotesize Exploded and assembled view of a cassette.}
\label{viewfogi9fo}
\end{figure}
\\ Figure \ref{viewfogi9fo} shows a cassette drawing. An Aluminium
substrate of thickness $0.5\,mm$ is coated on both sides by a
$^{10}B_4C$-layer, i.e. a blade. One layer will work as a
back-scattering layer and the second as a transmission layer. The
converter is surrounded symmetrically by two polyimide PCBs. Each of
them holds a cathode strip plane and a anode wire plane. The
converter layer substrate is grounded and it acts as a cathode
plane. Therefore a half cassette is a complete MWPC containing one
neutron converter layer and a two-dimensional read-out system.
\\ The entire structure is supported by an Aluminium U-shaped holder. The latter shape was
conceived to remove any material that can scatter the incoming
neutrons. Moreover, each holder presents two gas inlets in order to
supply the stopping gas directly inside the gap between the
converter and the PCBs. The exhausted gas will flow out from the
frontal opening of the cassette inside the gas vessel.
\begin{figure}[!ht]
\centering
\includegraphics[width=7cm,angle=0,keepaspectratio]{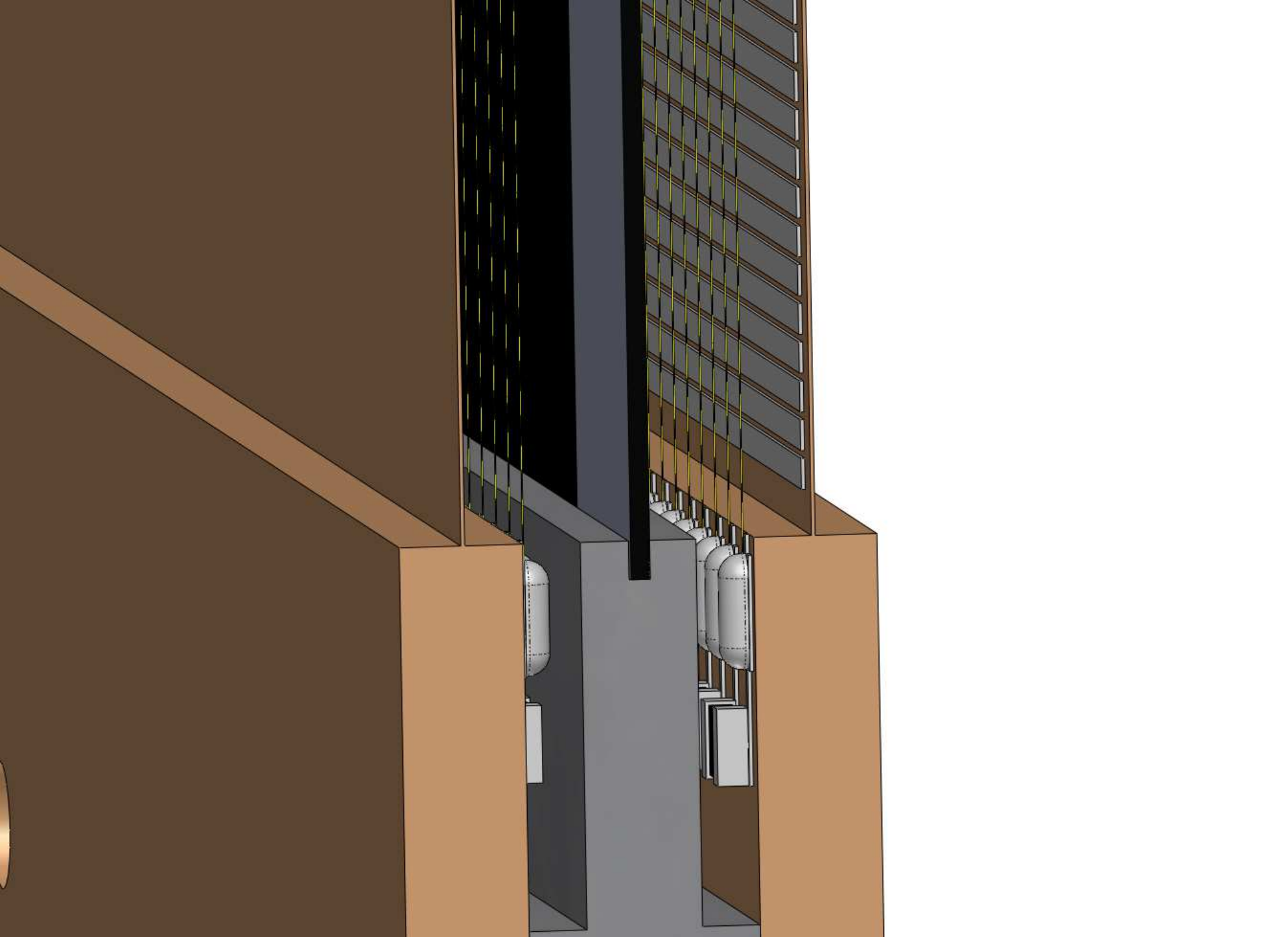}
\caption{\footnotesize Detail of a cassette: the two polyimide PCBs
surrounding the coated blade.} \label{viewfogi9forfwef4}
\end{figure}
\\ Figure \ref{viewfogi9forfwef4} shows the detail of an assembled
cassette. The polyimide PCBs have to be crossed before neutrons can
be converted, hence, in order to reduce the amount of material that
can induce neutron scattering, and thus misaddressed detected
events, those PCBs are as thin as possible according to the
mechanical constraints in their inner active region. The strips are
deposited on the polyimide and the anode wires are stretched
orthogonally over the strip plane. The copper strips are $0.8\, mm$
wide and spaced by $0.2\, mm$; tungsten wires are $15\, \mu m$ thick
and they are spaced by  $2.5\, mm$. The final electric signal is
obtained by gas amplification on the anode wires placed in the gas
volume. In order to decrease the number of read-out channels, anode
wires and cathode strips are grouped by resistive chain for charge
division read-out. Each full \emph{cassette} has then 4 anode
outputs and 4 cathodes outputs making 4 charge division read-out
chains. The resistors are placed on the PCBs surface.
\\ The polyimide PCBs are $60\, \mu m$ thick in the inner region:
$25\, \mu m$ is the polyimide thickness and $35\, \mu m$ is the
copper strips thickness.
\\ The sensitive area of each cassette is $10\times 9\,cm^2$ but, since it will be oriented at $10\,^{\circ}$ with
respect to the incoming neutron direction, the actual sensitive area
offered to the sample is given by $(10\,cm \cdot
\sin(10\,^{\circ}))\times 9\,cm = 1.7\times 9\, cm^2$. As a result,
the actual wire pitch, at $10\,^{\circ}$, is improved down to
$0.43\,mm$.
\\ The detector will be installed to have the
better resolution in the direction of the reflectometry instrument
collimation slits; i.e. the cassettes, which can be mounted either
horizontally or vertically, will be oriented with the wires parallel
to the instrument slits.
\\ Figure \ref{dgfag5465yb5hv} shows a drawing of 8 cassettes stacked one after
the other and placed in the gas vessel.
\\ As already mentioned, the main issue to be addressed in the final
detector is the uniformity, as soon as it is made of several units,
their arrangement is crucial to get a uniform response in
efficiency. A misalignment in one of the modules can give rise to a
drop in the efficiency or dead zones.
\\ The cassettes have to be arranged in order to overlap to avoid
dead zones. For this reason this detector is suitable for fixed
geometry reflectometry instruments, where the distance between
sample and detector is kept constant and the arrangement does not
change.
\begin{figure}[!ht]
\centering
\includegraphics[width=14cm,angle=0,keepaspectratio]{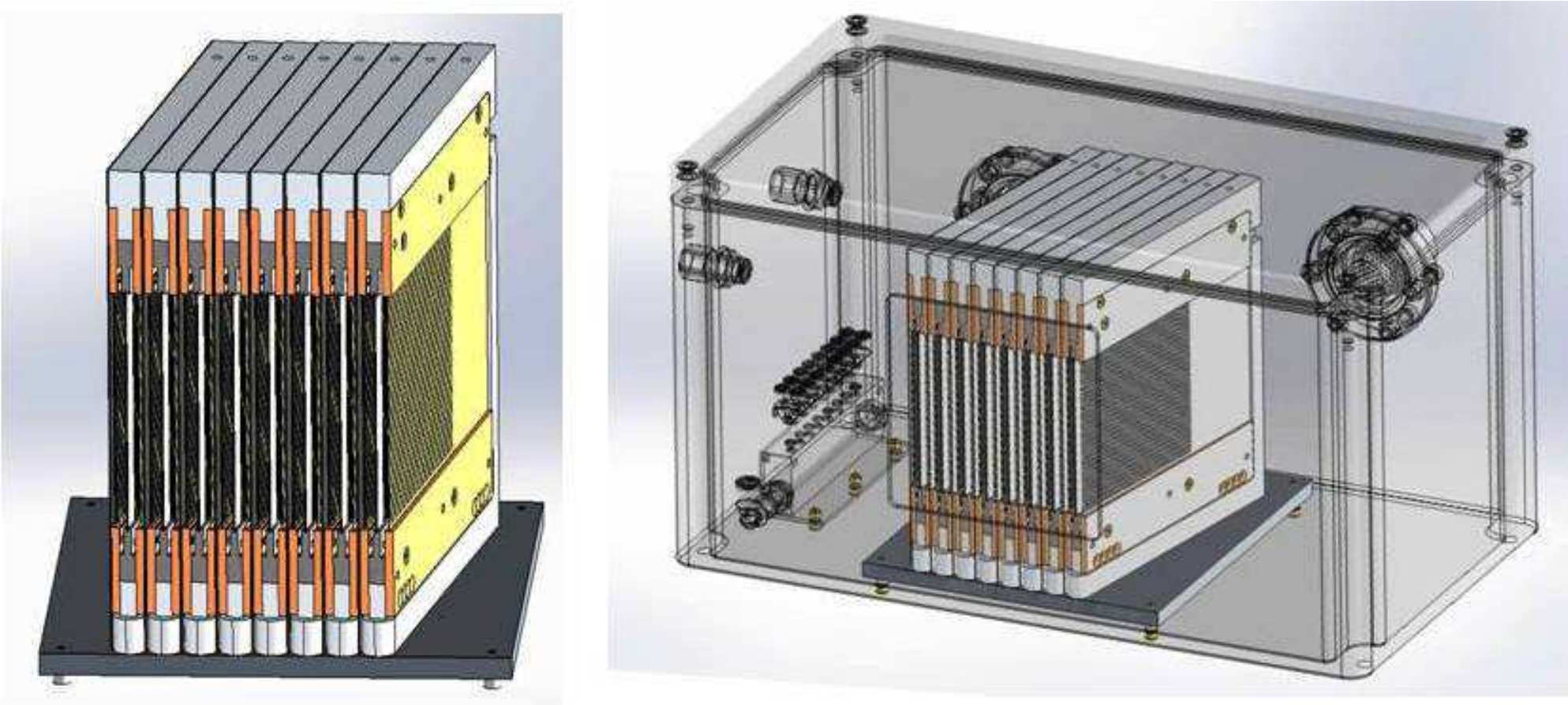}
\caption{\footnotesize A 8-cassettes Multi-Blade in its gas vessel.}
\label{dgfag5465yb5hv}
\end{figure}
\\ The final prototype will be mounted in a gas vessel, see Figure
\ref{dgfag5465yb5hv}, together with the gas distribution unit which
splits the inlet in the several cassettes, and the electronic
connections.
\\ Since the gas is flushed cost effective materials can be used
because their outgassing is not an issue.
\\ The presented prototype allows to study both the single layer and
the double converter solutions. Its realization, advantages and
mechanical issues, will be explained in the following Section.
\subsection{Mechanics}
\begin{figure}[!ht]
\centering
\includegraphics[width=12cm,angle=0,keepaspectratio]{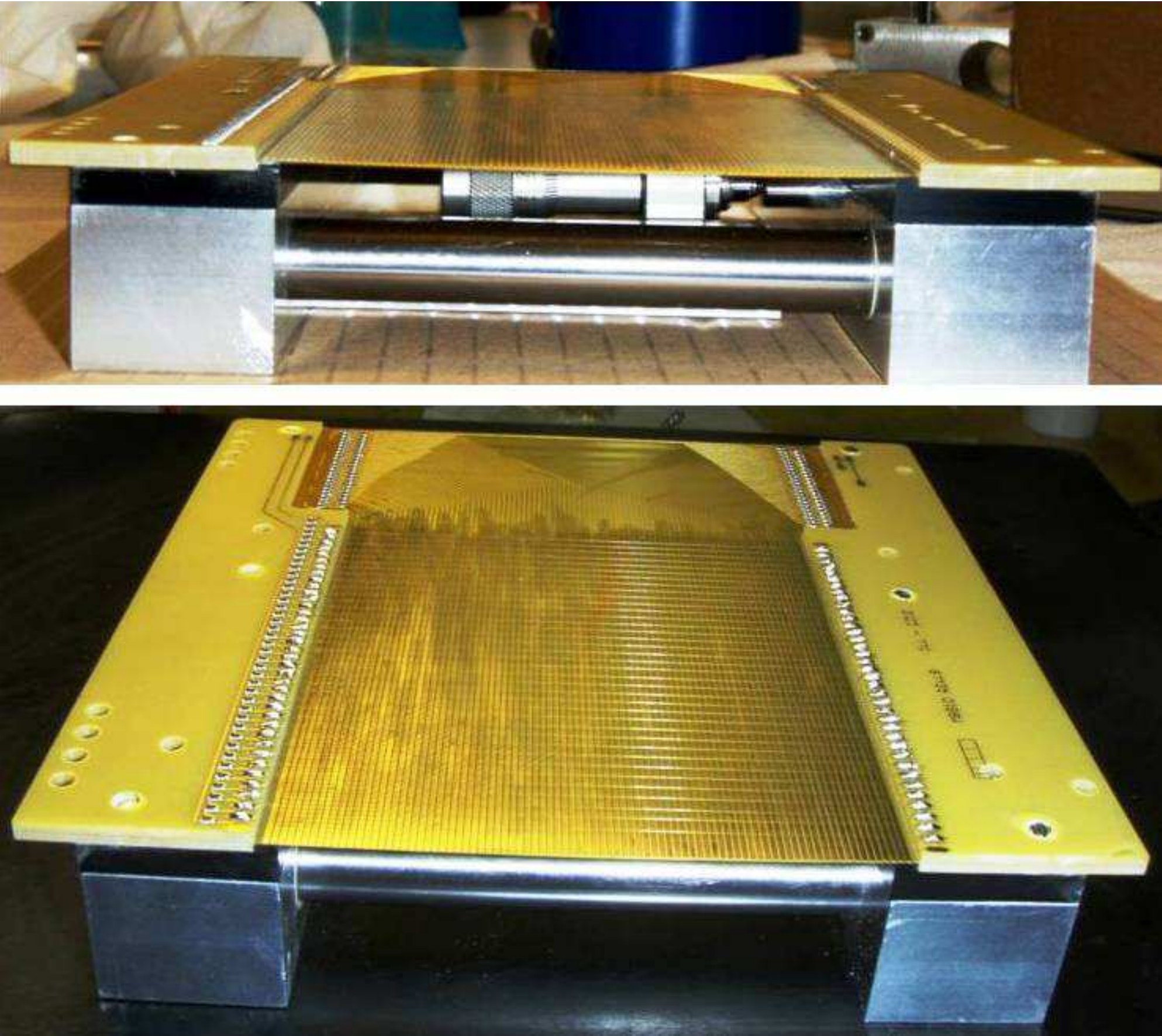}
\caption{\footnotesize A polyimide PCBs where anode wires are
mounted orthogonal to the cathode strips.} \label{figMBv1prima1}
\end{figure}

\begin{figure}[!ht]
\centering
\includegraphics[width=14cm,angle=0,keepaspectratio]{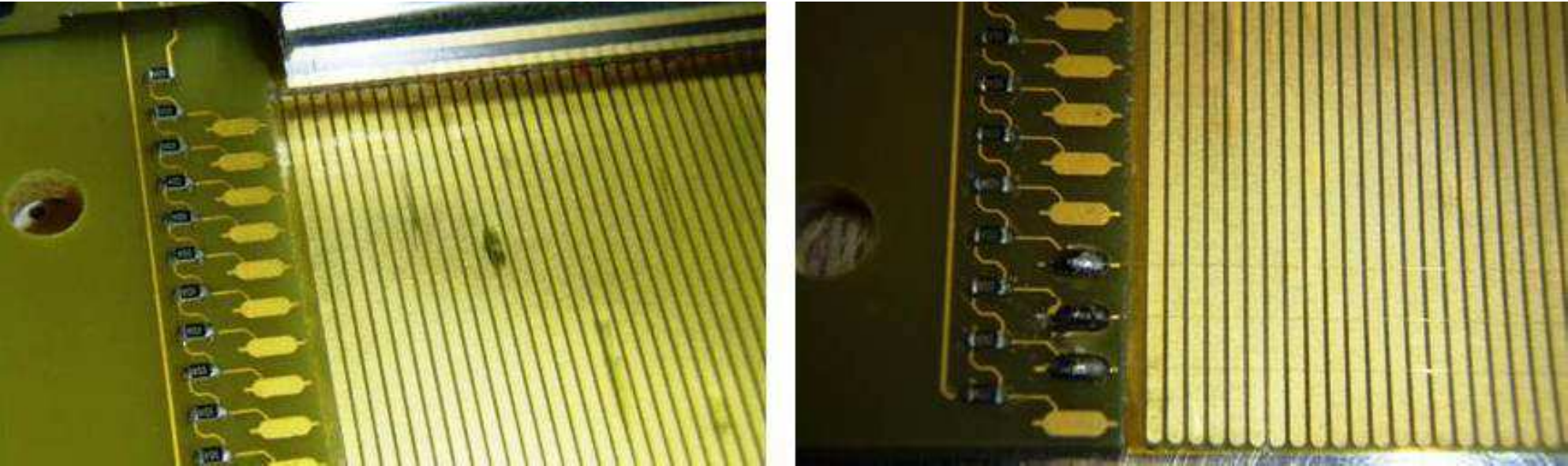}
\caption{\footnotesize Detail of a polyimide PCBs: resistors for
charge division link the wire pads where anodes are soldered.}
\label{figMBv1prima2}
\end{figure}

\begin{figure}[!ht]
\centering
\includegraphics[width=14cm,angle=0,keepaspectratio]{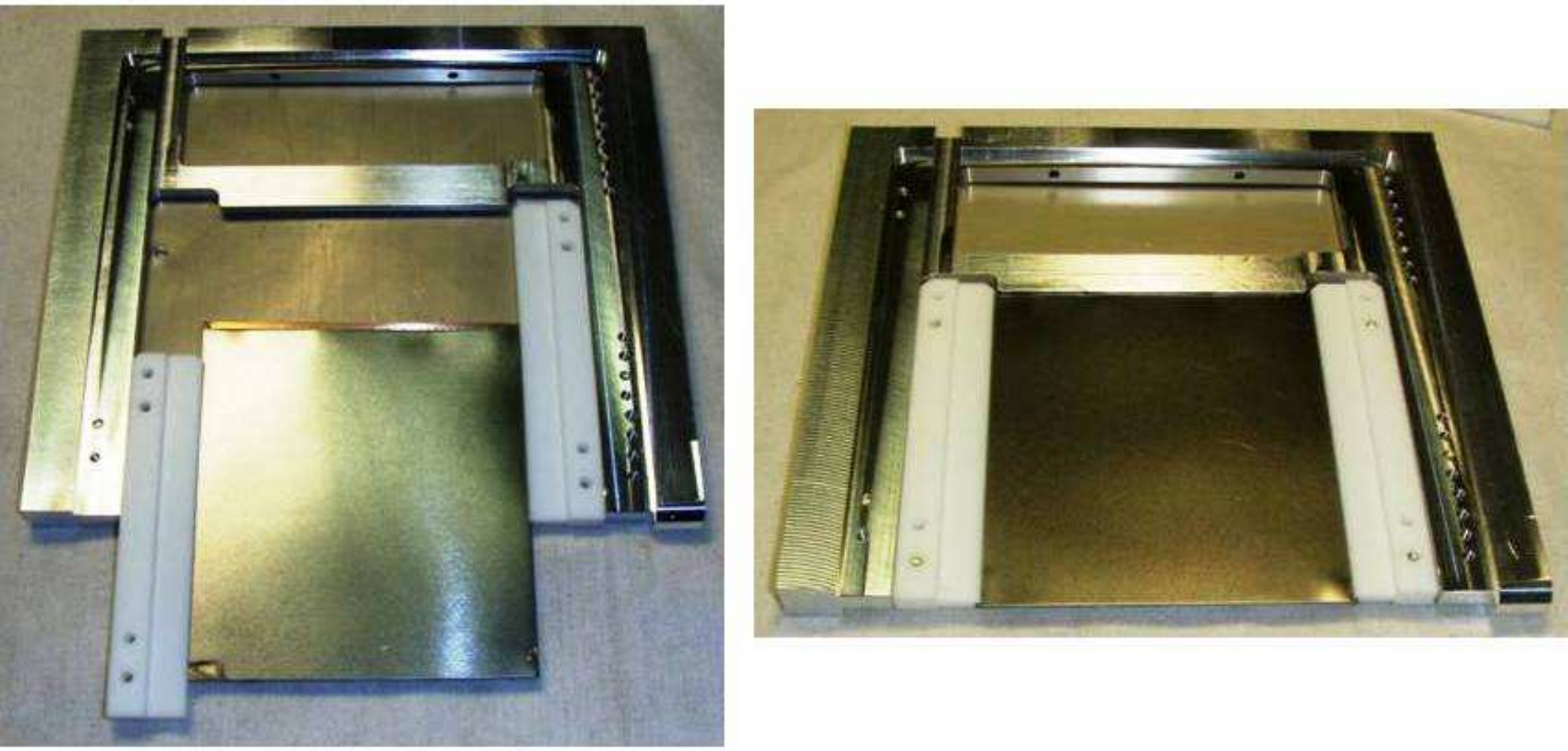}
\caption{\footnotesize The Aluminium holder and a blade (substrate
coated with $^{10}B_4C$ both sides) inserted.} \label{figMBv1prima3}
\end{figure}

\begin{figure}[!ht]
\centering
\includegraphics[width=10cm,angle=0,keepaspectratio]{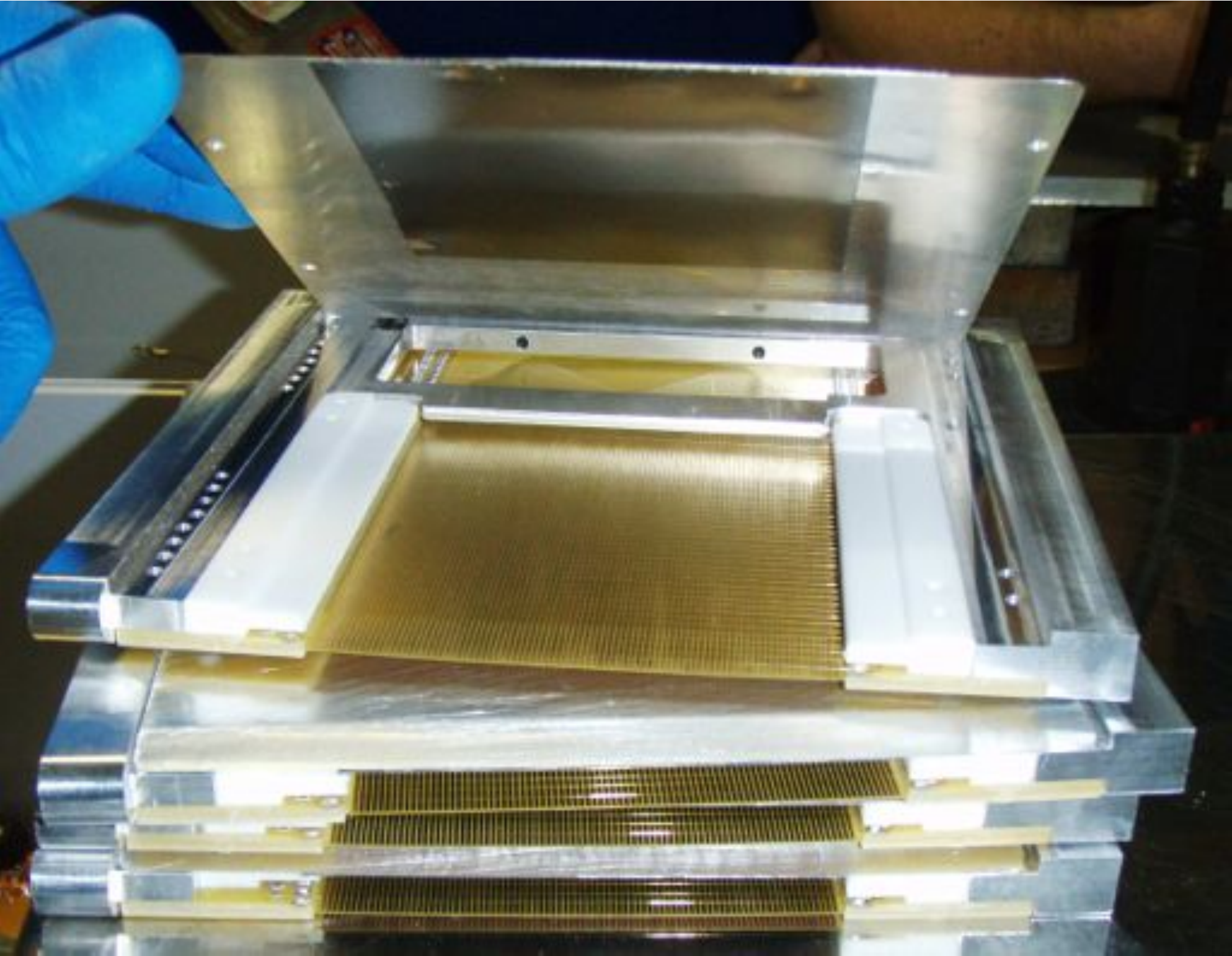}
\caption{\footnotesize Four cassettes assembled. The read-out PCB is
installed on the Aluminium holder.} \label{figMBv1prima4}
\end{figure}

\begin{figure}[!ht]
\centering
\includegraphics[width=14cm,angle=0,keepaspectratio]{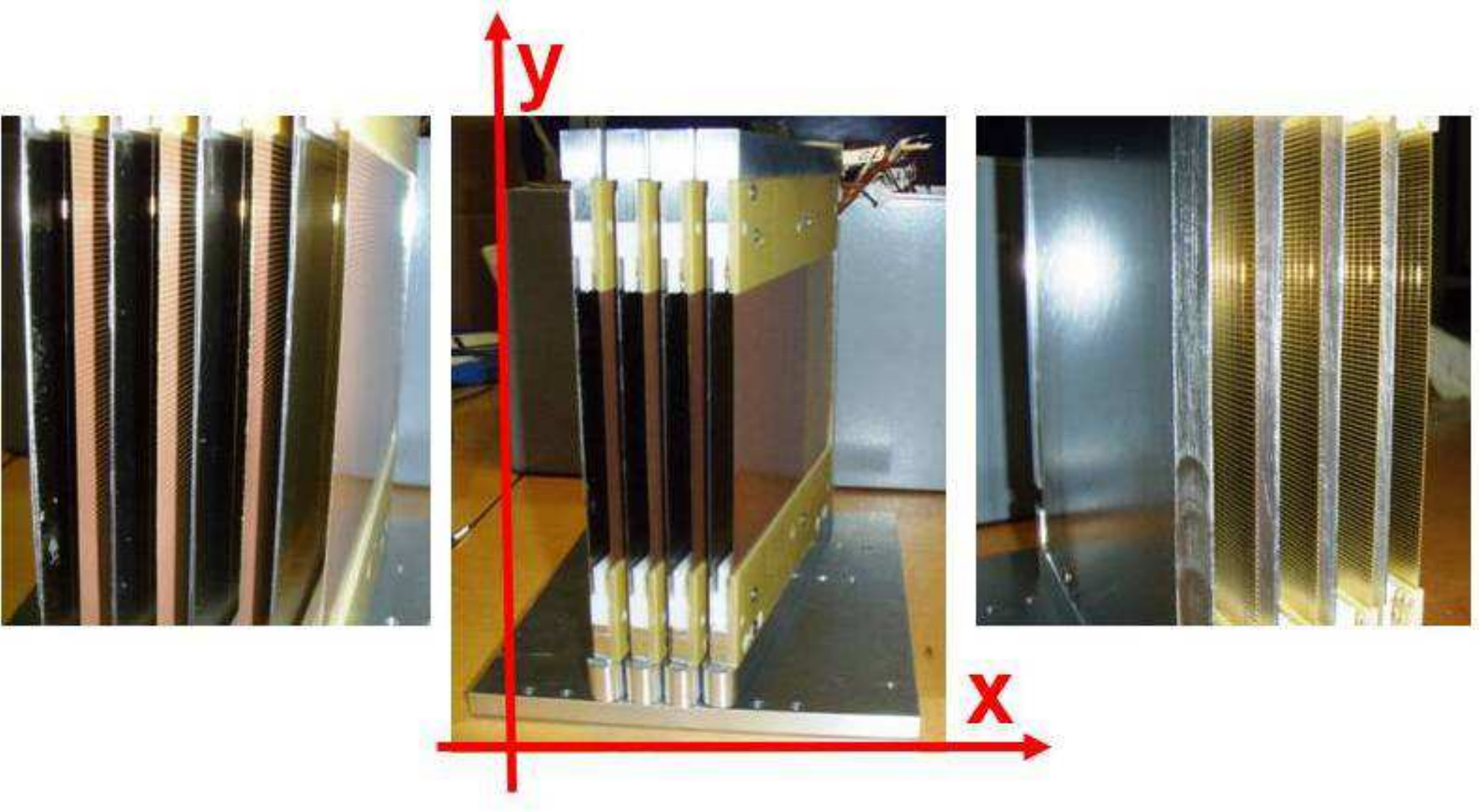}
\caption{\footnotesize Four cassettes stacked one after the other at
$10\,^{\circ}$ and parallel each other.} \label{figMBv1prima6}
\end{figure}

\begin{figure}[!ht]
\centering
\includegraphics[width=10cm,angle=0,keepaspectratio]{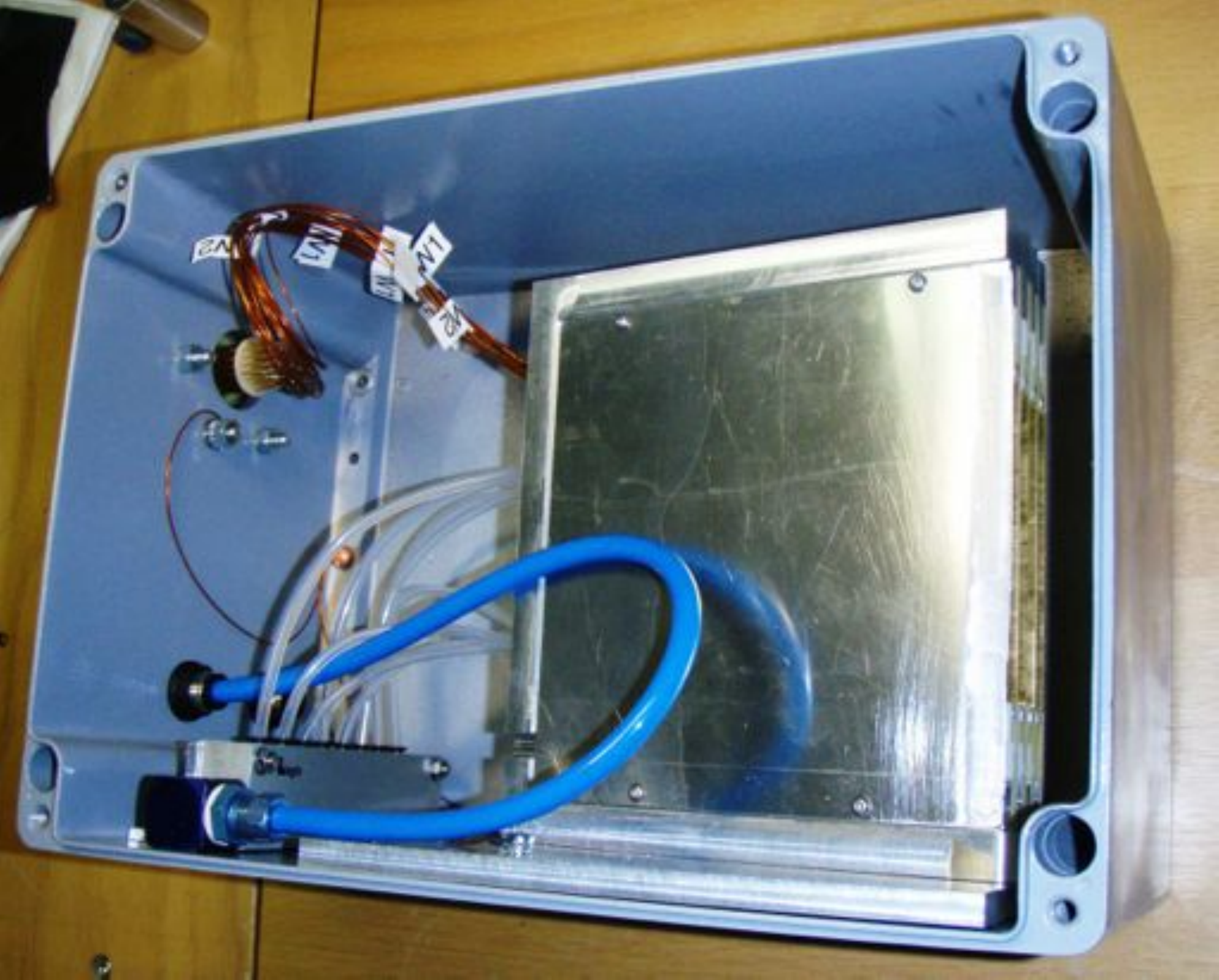}
\caption{\footnotesize The four cassette assembly in the gas vessel.
Each cassette is supplied by two gas inlets. The detector  entrance
window is the one on the right.} \label{figMBv1prima7}
\end{figure}

\begin{figure}[!ht]
\centering
\includegraphics[width=12cm,angle=0,keepaspectratio]{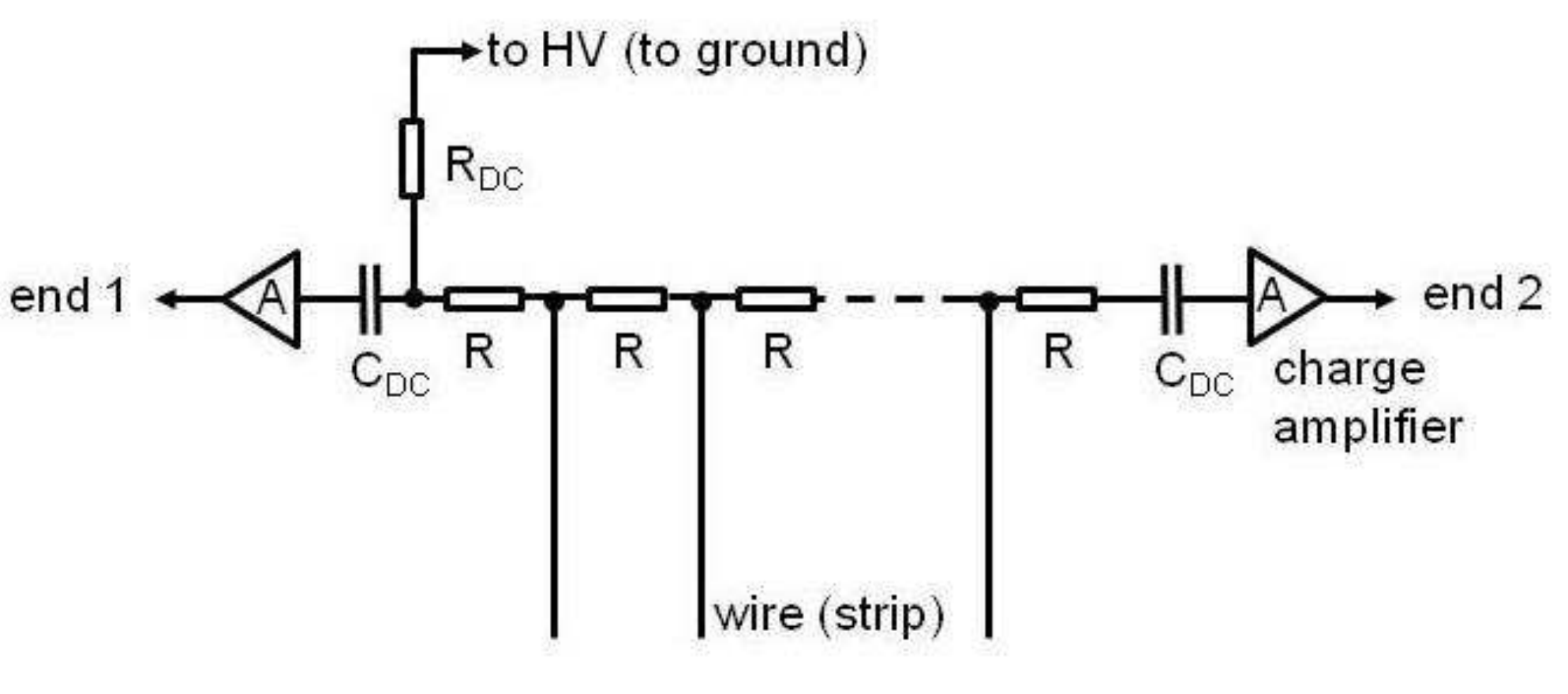}
\caption{\footnotesize The Multi-Blade front-end electronics
schematic.} \label{figMBv1prima8}
\end{figure}

The first prototype (V1) consists of four cassettes operated at
$10\,^{\circ}$. Given the cassette active region size, the prototype
active area, considering their overlap is about $6\times 9\,cm^2$.
\\ Figure \ref{figMBv1prima1} shows a polyimide PCB. The latter is
composed by a stack of three layers: two thick PCBs where in the
middle is fixed a $25\, \mu m$ polyimide foil. The inner part is
soft and the external part serves as a holder. 86 copper strips are
deposited on the surface of the thin region (see Figure
\ref{figMBv1prima2}).
\\ 39 anode wires (37 active wires and 2 guard wires)
are mounted and soldered on pads at $2\,mm$ distance from the
cathode plane. Both for anodes and for cathodes a resistive chain is
soldered on the rigid PCB. The total resistance is $6\,K\Omega$ for
the anode chain and $8\,K\Omega$ for the cathode chain. At the wire
plane edge a guard wire was installed to compensate the electric
field distortion, hence this wire will not produce any signal.
\\ The total gap between the converter and the cathode plane, i.e.
half cassette width, will be about $4\,mm$, thus any deformation of
either the substrate or the strip plane will produce a variation in
the local electric field produced between the anode plane and
cathodes. Consequently where the cathode is closer to the wire plane
the detector will manifest a higher gain. This effect mainly
degrades the uniformity over the cassette surface. The overall
uniformity on the whole detector surface is then degraded by the
single cassette uniformity and their arrangement in the space:
overlap and switching from one to another.
\\ It is crucial to control the flatness of both the substrate and
the PCB. The first manufactured PCB was composed of a thin polyimide
held on three sides by the rigid PCB. The provider was not able to
assure the polyimide flatness with this design. We changed the PCB
design in order to be able to pull on both sides and restore its
flatness. The polyimide is held by only two of its sides (see Figure
\ref{figMBv1prima1}). The PCB is held by a tool (see Figure
\ref{figMBv1prima1}) that allows to stretch the foil before being
mounted on the Aluminium holder. This tool allows also to mount
wires on the PCB keeping the system under tension. The $15\, \mu m$
tungsten wires are mounted on the PCB under a tension of $35\,g$.
Once wiring is over, the PCB can be installed on the holder.
\\ Figure \ref{figMBv1prima3} shows the Aluminium holder where the double side coated substrate with $^{10}B_4C$
\cite{carina} is inserted. In order to keep the wire tension, the
PCB, without removing the stretching tool, can be placed on the
holder. The holder and the PCBs present four different fixation
screw shifted by $0.25\,mm$ from each other. The PCB can be screwed
on the holder according to its actual size after stretching. This
ensures the right tension on the wires and the flatness of the
cathode plane.
\\ As for the read-out plane, the converter holding substrate must
be flat too. After sputtering, between the $Al$-substrate and the
$^{10}B_4C$ coating, a significant residual stress remains due to
the difference in the thermal expansion coefficient of $Al$
($\sim23.5\cdot10^{-6}\,1/K$) and $^{10}B_4C$
($\sim5.6\cdot10^{-6}\,1/K$). When they are cooled down to room
temperature the $Al$ contracts more than $^{10}B_4C$. Experiential
evidences of that have been observed: on single side coated
substrates, the un-coated side is shorter than the coated side,
resulting into a bending of the blade. When a double-side coated
blade has to be inserted into the holder (see Figure
\ref{figMBv1prima3}) the constraints on the sides makes the blade
bend and unstable. On both sides two PCBs have to be installed
resulting into two identical and symmetrical MWPC.
\\ We wanted to study both option A and B with this prototype but due
to the blade mechanical issue we convert the prototype in a single
layer detector.
\\ In order to keep the substrate with the converter flat enough to ensure a uniform
electric field, we mounted it on an Aluminium lid placed where a PCB
was removed (see Figure \ref{figMBv1prima4}). We used a $3\,\mu m$
thickness $^{10}B_4C$ coating instead. The gap between the wires and
the converter was increased up to $6\,mm$, while the gap between the
wires and the strips is about $2\,mm$. The MWPC is asymmetric.
Figure \ref{figMBv1prima4} shows the four cassettes equipped with
the read-out systems and the converters.
\\ The version V1 of the Multi-Blade detector allowed only to study
the single layer configuration A, as mechanical issues had made
impossible the initial configuration B realization.
\\ The number of read-out channels per cassette were reduced from 8 to 4: 2 anode and 2 cathode outputs.
\\ Four cassettes were stacked at $10\,^{\circ}$ with respect to the beam and parallel to each
other. Figure \ref{figMBv1prima6} shows the detail of the four
cassettes stacked from two points of view. We define as the
$x$-coordinate where the wire pitch is projected at $10\,^{\circ}$.
The $y$-coordinate is defined by the direction orthogonal to the strips orientation. \\
The four cassettes were then installed in the gas vessel, see Figure
\ref{figMBv1prima7}. Each cassette is supplied by two inlets to let
the gas to flow directly inside them. The entrance window of the
detector is the one on the right in Figure \ref{figMBv1prima7}.
\\ The front-end electronics of the prototype is connected outside the gas vessel
and consists of a decoupling circuit and charge amplifiers. A
schematic of the whole front-end electronic chain is shown in Figure
\ref{figMBv1prima8}. Both wires and strips are connected in the same
way by their resistive chain, the AC signal is decoupled by two
capacitors at both ends from the DC component used to polarize the
wires at the HV and the strips to the ground potential.
\\ The charge is amplified by charge amplifiers. We used inverting
amplifiers of $6\,V/pC$ and $1\,\mu s$ shaping time for anodes and
non-inverting amplifiers of $32\,V/pC$ and $2\,\mu s$ shaping time
for cathodes. \\ Each chain ends into two signal outputs that can be
either summed to get the energy information (PHS) or subtracted and
divided to get the positional information.
\subsubsection{Polyimide layers characterization}
\begin{figure}[!ht]
\centering
\includegraphics[width=5.2cm,angle=0,keepaspectratio]{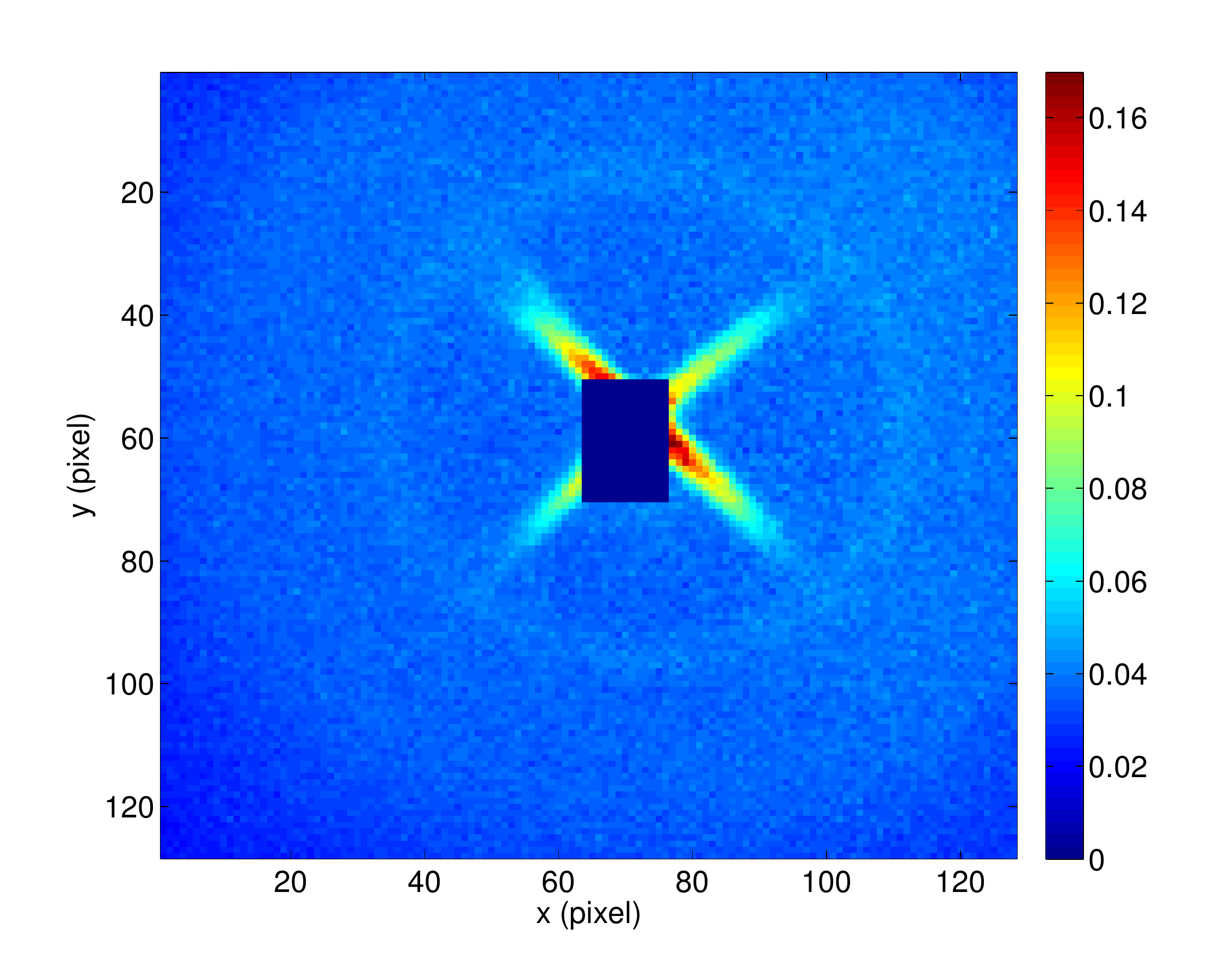}
\includegraphics[width=5.2cm,angle=0,keepaspectratio]{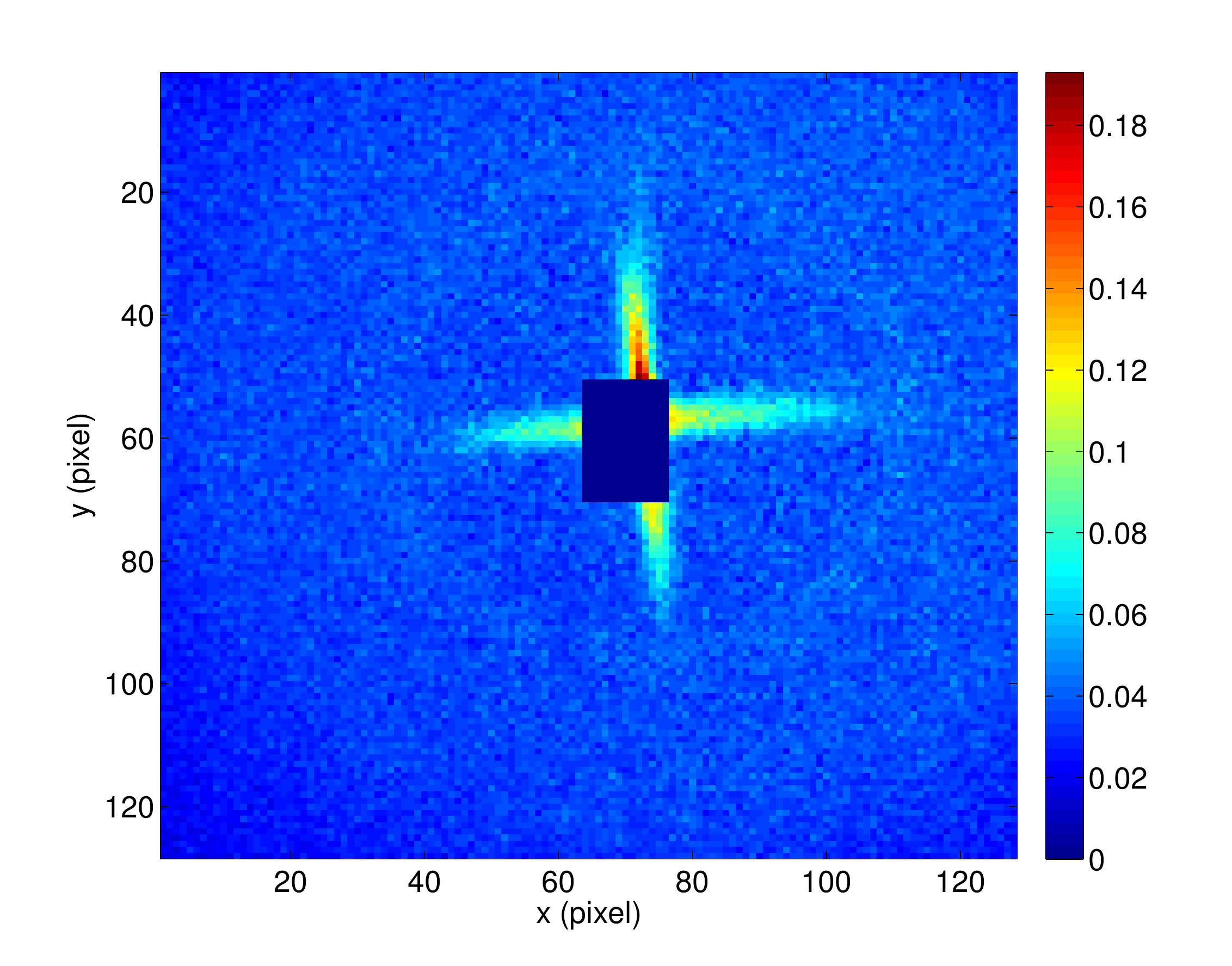}
\includegraphics[width=5.2cm,angle=0,keepaspectratio]{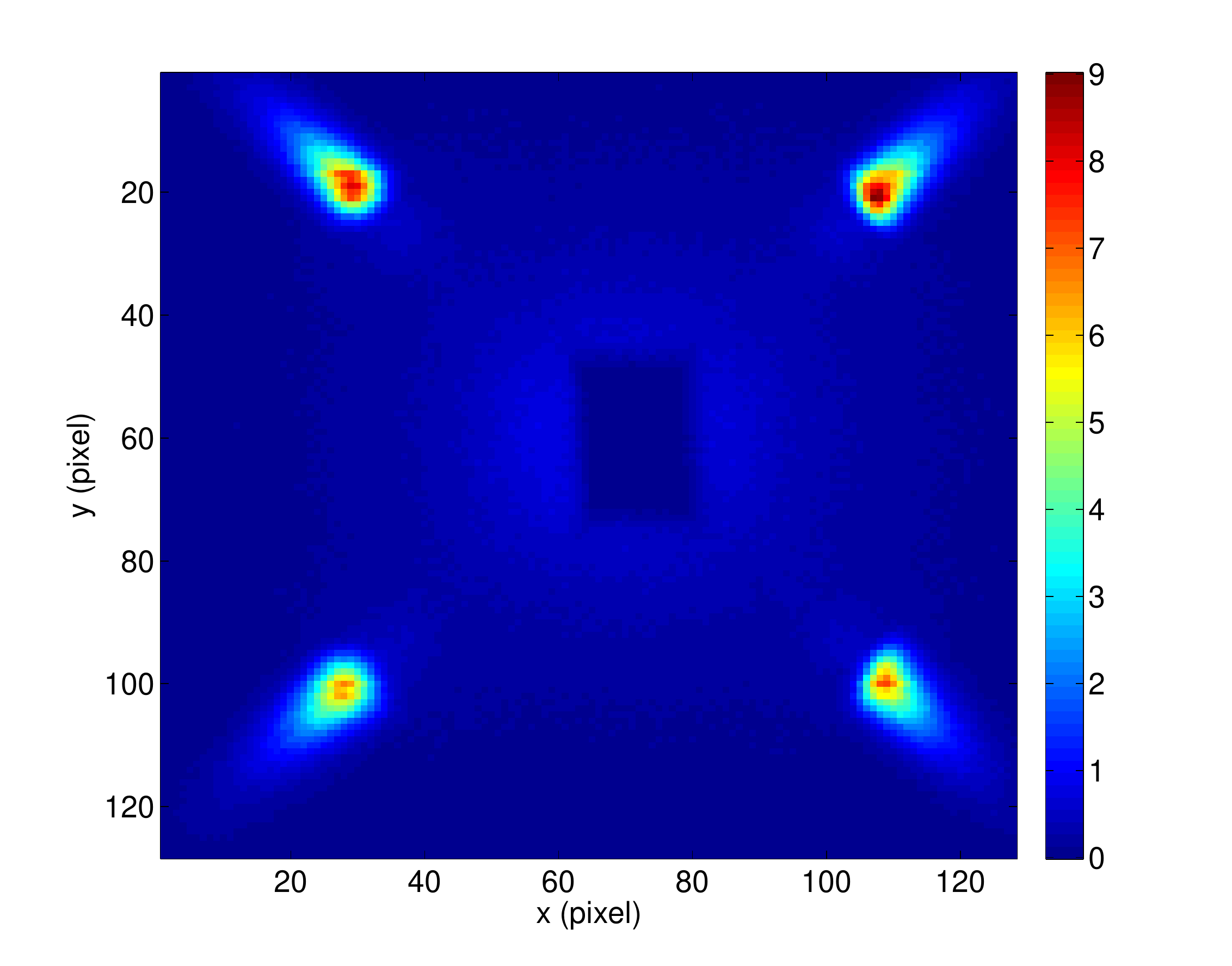}
\caption{\footnotesize Scattered neutrons ($2.5$\AA) by 12 polyimide
layers on the BIDIM detector ($26\times26\,cm^2$ active area) at
$23\,cm$ distance (left and center) and at $3\,cm$ distance. The
polyimide layers are rotated by $45^{\circ}$ between the left and
center plots. The color scale is in $Hz$.} \label{weg5e4}
\end{figure}

\begin{figure}[!ht]
\centering
\includegraphics[width=10cm,angle=0,keepaspectratio]{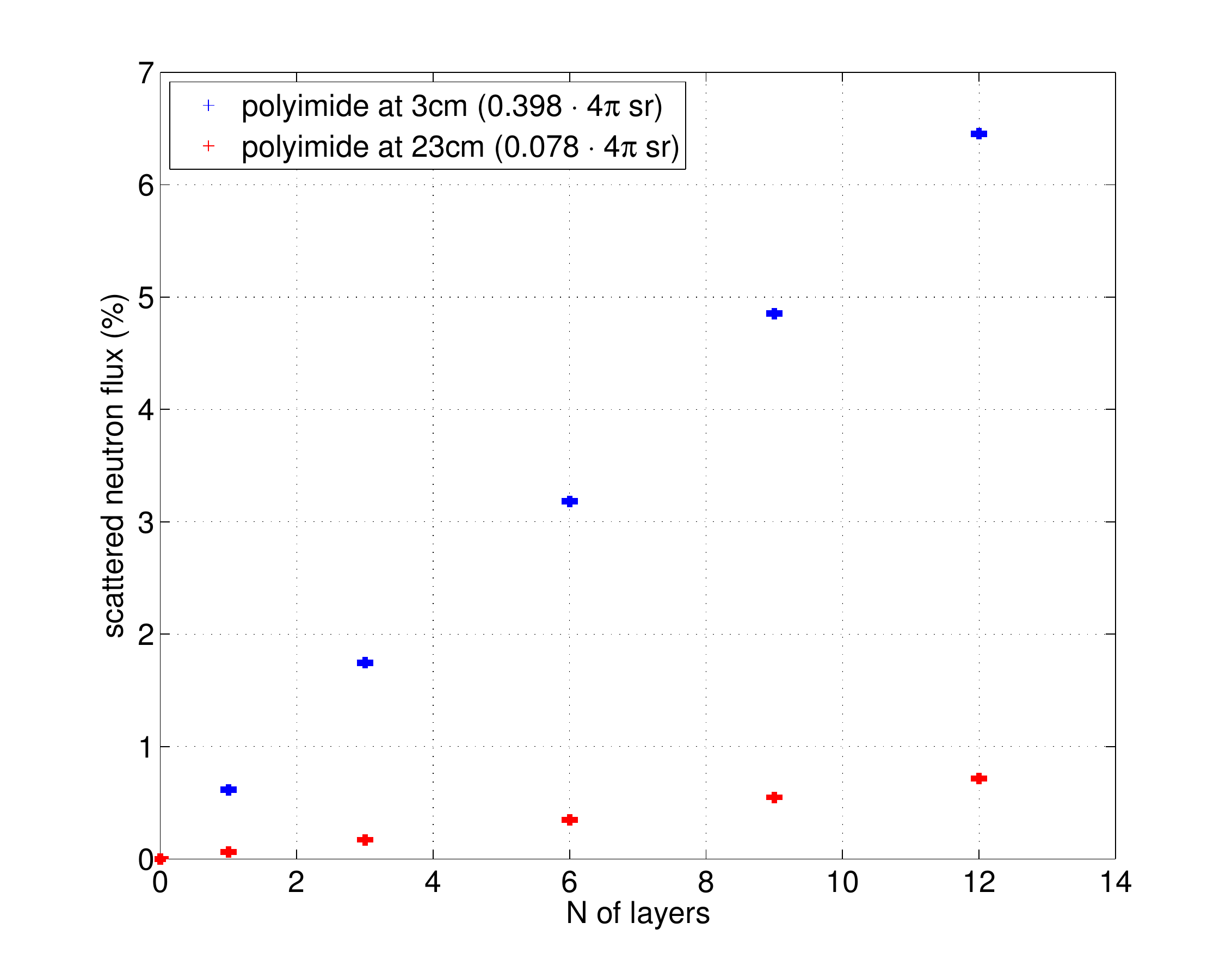}
\caption{\footnotesize Percentage of neutrons ($2.5$\AA) scattered
by polyimide as a function of number of layers. At $3\,cm$ distance
the detector covers $39.8\%$ of the full solid angle, at $23\,cm$
$7.8\%$.} \label{weg5e454}
\end{figure}

Transmission measurements on the polyimide layers used in the
Multi-Blade prototype have been performed on CT1 ($2.5$\AA\, neutron
beam) at ILL. The neutron beam was collimated and, as shown in
Appendix \ref{apphexmeasexplan}, it was calibrated to
$(122020\pm80)Hz$. The polyimide samples were placed after the
collimation slit in front of a a two-dimensional $^3He$-based
detector (BIDIM) placed at distance $D$. The detector has an
efficiency for $2.5$\AA\, neutrons of $70\%$, a spatial resolution
of $2\times2\,mm^2$ and $26\times26\,cm^2$ active area. A single
polyimide layer is composed of a stack of $25\, \mu m$ thick
polyimide and $35\, \mu m$ thick copper strips.
\\ We place different number of layers in front of the BIDIM in
order to simulate the increase of thickness to be crossed by
neutrons due to the inclination. An inclination of $10^{\circ}$
corresponds to about 6 layers and $5^{\circ}$ to about 12 layers.
\\ We repeat the measurement for two distances between detector and sample: $D=3\,cm$ and
$D=23\,cm$. Diffraction from the samples is not isotropic because of
the sample structure itself. Figure \ref{weg5e4} shows three
measurements on a 12 layer sample for the two distances normalized
to the incoming neutron flux (the color scale is in $Hz$). The cross
arising at $D=23\,cm$ is due to the diffraction through extended
fibers which is the polyimide molecule. Those fibers are placed
crossed in the layer manufacturing process.
\\ We calculate the scattered beam by the layers, for the two distances, as the
ratio of the counting rate on the whole detector surface over the
incoming neutron flux. The result as a function of number of layer
is shown in Figure \ref{weg5e454}. At $D=3\,cm$ the detector surface
covers a solid angle of $0.398\cdot4\pi\,sr$, and at $D=23\,cm$ of
$0.078\cdot4\pi\,sr$.
\\ At $10^{\circ}$ (that equals about 6 layers)
we expect to diffuse at most about $\frac{3\%}{0.398}\simeq7.5\%$
(at $2.5$\AA) of the beam assuming the same scattering in
$4\pi\,sr$.
\subsection{Results}
\subsubsection{Operational voltage}
\begin{figure}[!ht]
\centering
\includegraphics[width=7.8cm,angle=0,keepaspectratio]{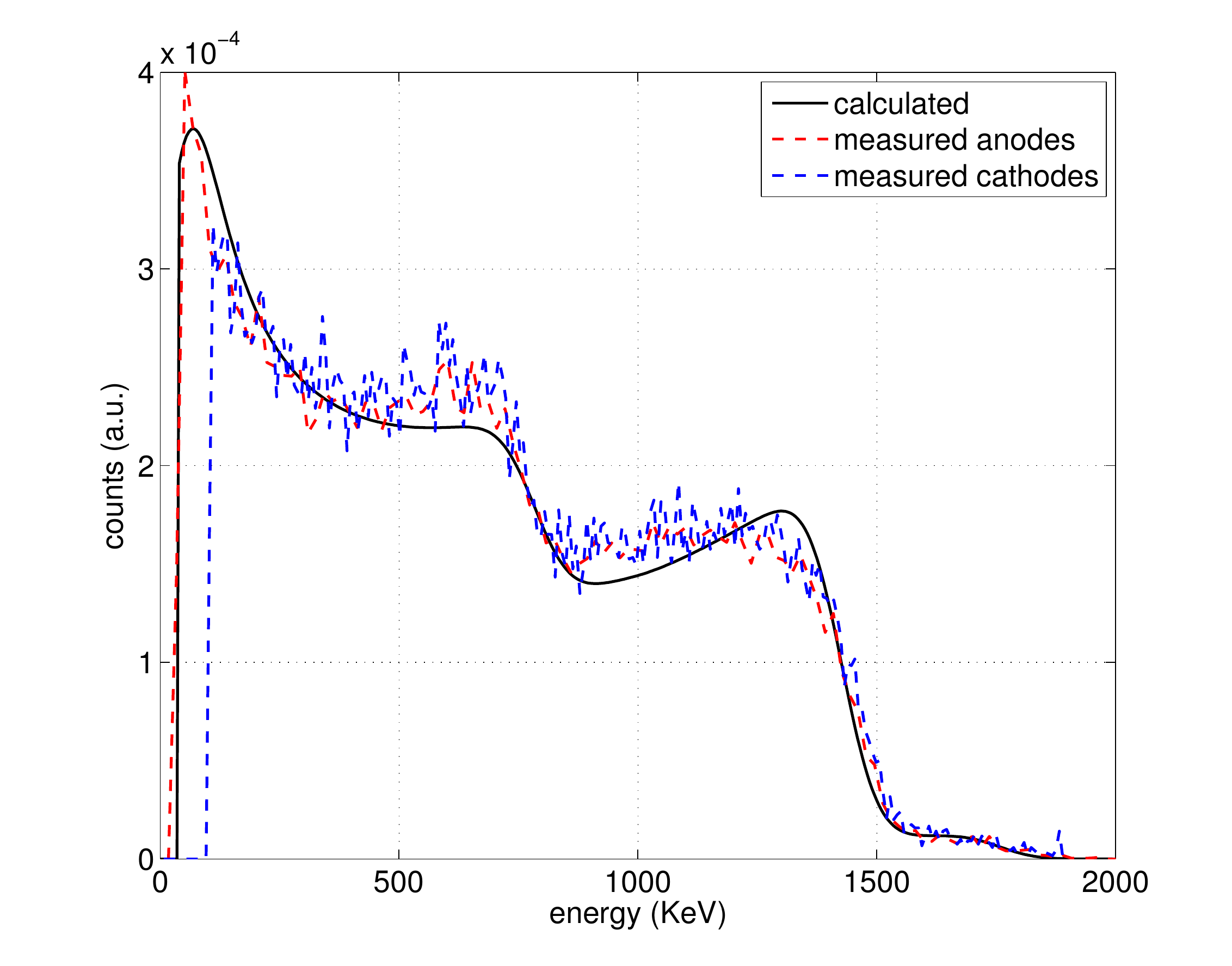}
\includegraphics[width=7.8cm,angle=0,keepaspectratio]{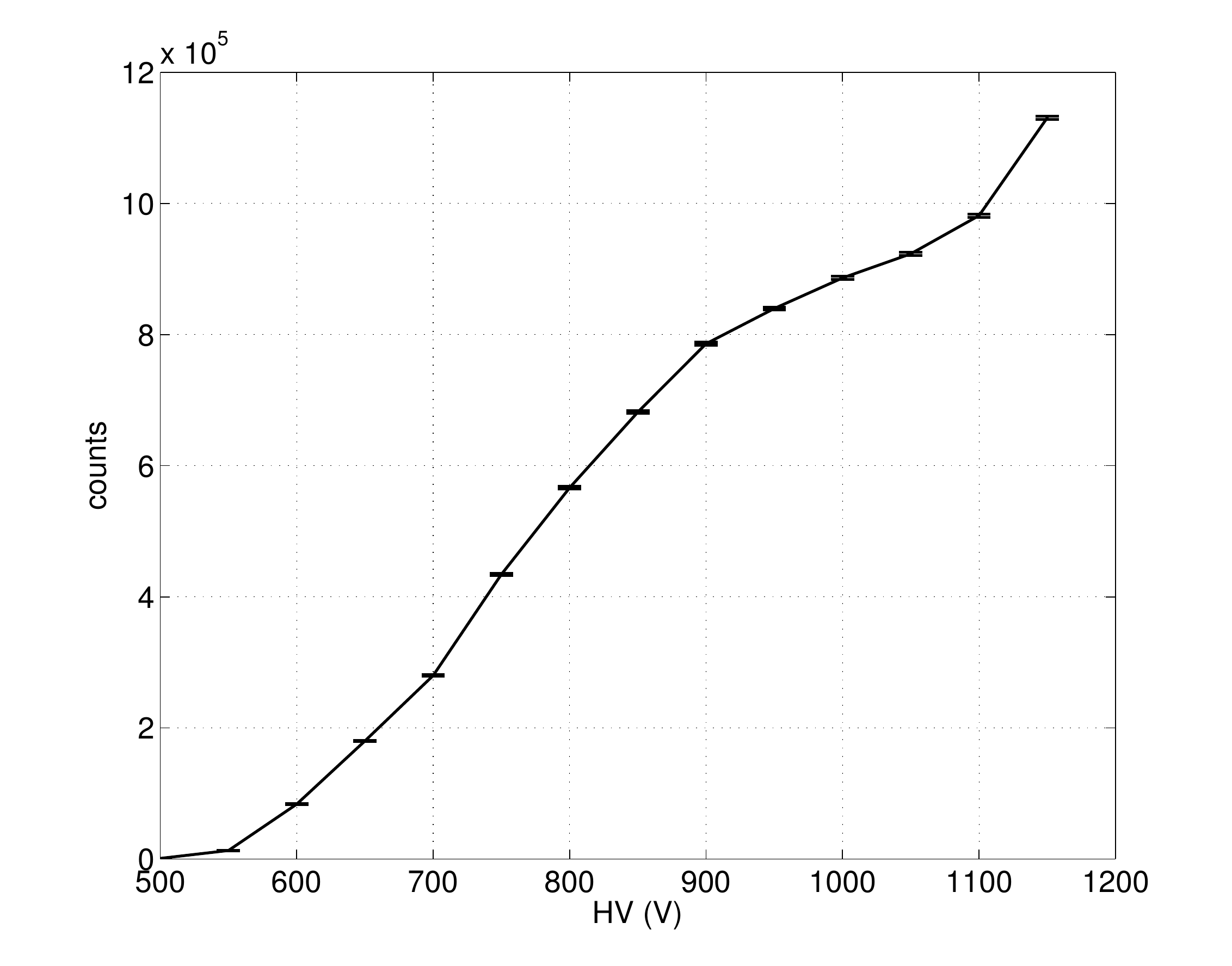}
\caption{\footnotesize PHS measured on strips and wires at $1000\,V$
and calculated PHS (left). The Multi-Blade detector plateau
(right).}\label{phspaltembv1456}
\end{figure}

A counting curve was measured in order to set the right bias voltage
to be applied to polarize the prototype. Each cassette output was
connected to get the energy information and then to measure the PHS.
Given the electronic noise, a $25\,mV$ threshold was used for the
anode amplifiers; for the cathode amplifiers we used $100\,mV$.
Figure \ref{phspaltembv1456} shows a PHS for both strips and wires,
compared with a PHS calculated according to Section
\ref{SectThPHSCalc} at $1000\,V$.
\\The working voltage chosen through the counting curve is $1000\,V$.
\subsubsection{Gain}
The Multi-Blade prototype is operated in proportional mode, its gain
has been measured on CT2 at ILL. The neutron flux of
$(15280\pm20)neutrons/s$ ($2.5$\AA) was quantified using the
Hexagonal detector (see Appendix \ref{apphexmeasexplan}).
\\ The prototype was polarized at $1000\,V$. Its $3\,\mu m$ $^{10}B_4C$ converter layer
was exposed to the beam orthogonally and
$\Phi_d=(904\pm2)neutrons/s$ were counted. It results into a
$\varepsilon=(5.92\pm4)\%$ detection efficiency. The current flowing
through the detector was measured and it is about
$I_{prop.}=180\,pA$.
\\ The detector operational voltage was set to $100\,V$.
The measurement of the current output was repeated operating the
detector in ionization mode, resulting into $I_{ion.}=3.1\,pA$.
\\ The average charge created for a detected neutron both proportional and ionization modes are:
\begin{equation}
Q_{prop.}=\frac{I_{prop.}}{\Phi_d}=199\,fC/neutron, \qquad
Q_{ion.}=\frac{I_{ion.}}{\Phi_d}=3.4\,fC/neutron
\end{equation}
This results into a gain of about $G=58$ at $1000\,V$.
\subsubsection{Efficiency}
Detection efficiency of the Multi-Blade prototype V1 has been
measured on CT2 at ILL by using a collimated and calibrated neutron
beam of wavelength $2.5$\AA.
\\ The neutron beam was calibrated using an $^3He$-based detector.
The procedure is explained in details in the Appendix
\ref{apphexmeasexplan}. After the calibration, the neutron flux the
Multi-Blade was exposed to is $\left(15280\pm20\right)\,neutrons/s$
over an area of $2\times 7\,mm^2$.
\\ The efficiency was measured for the following bias voltages
$950\,V$, $1000\,V$ and $1050\,V$. The efficiency was measured on
the four cassettes under the angle of $10^{\circ}$ and then
averaged. The results are listed in Table \ref{tabdeghjknflta99}.
\begin{table}[!ht]
\centering
\begin{tabular}{|c|c|c|c|}
\hline \hline $HV (V)$ &  $\varepsilon $(at$ \,\, 2.5$\AA) & Threshold $(KeV)$ & calculated $\varepsilon $(at$ \,\, 2.5$ \AA)\\
\hline
$950$     &  $\left( 24.8 \pm 0.2\right)\%$ & $180$ & $24.6\%$  \\
$1000$   &  $\left( 26.2 \pm 0.2\right)\%$  & $120$ & $26.3\%$ \\
$1050$    &  $\left( 27.5 \pm 0.2\right)\%$  & $90$  & $27.3\%$  \\
\hline \hline
\end{tabular}
\caption{\footnotesize Multi-Blade detection efficiency at $2.5$\AA
\, for three bias voltages and calculated efficiency for a given
threshold. } \label{tabdeghjknflta99}
\end{table}
\\ The result is in a good agreement with what can be calculated
from the theory (Chapter \ref{Chapt1}) by using an energy threshold
of $180\,KeV$, $120\,KeV$ and $90\,KeV$ for the three voltages from
$950\,V$ to $1050\,V$ respectively.
\subsubsection{Uniformity}
\begin{figure}[!ht]
\centering
\includegraphics[width=7.8cm,angle=0,keepaspectratio]{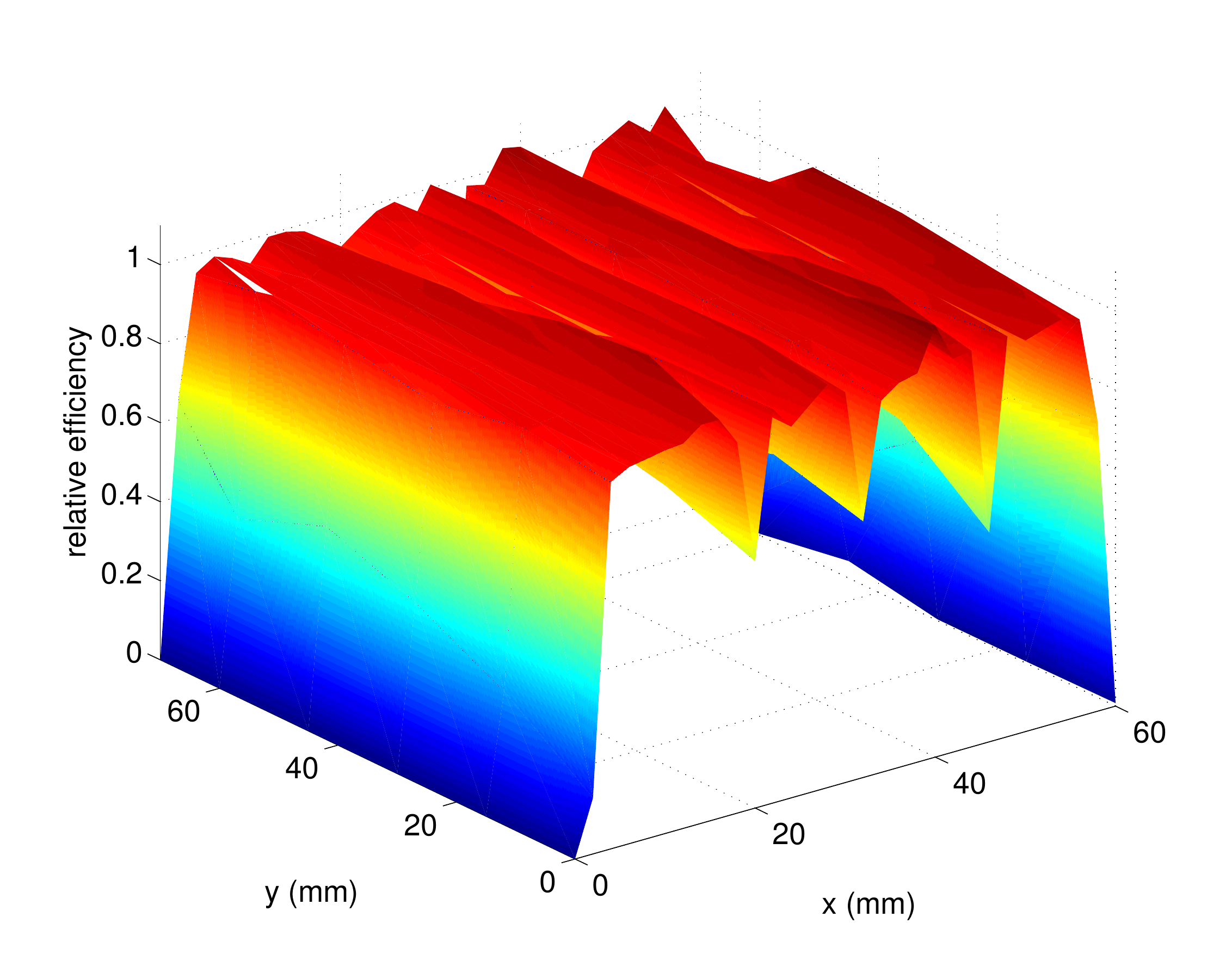}
\includegraphics[width=7.8cm,angle=0,keepaspectratio]{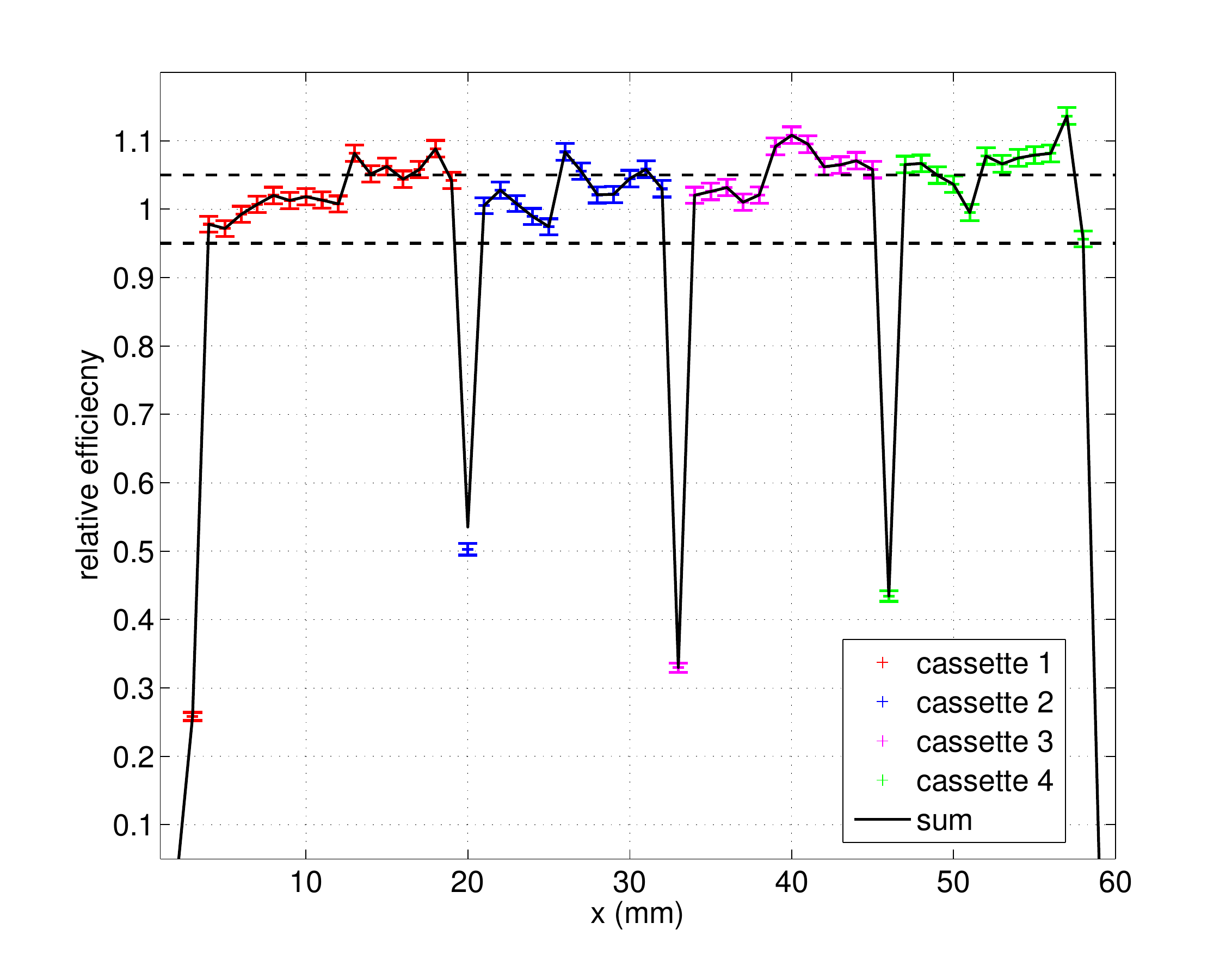}
\caption{\footnotesize Relative efficiency scan over the whole
detector surface.}\label{crgw5465u6u}
\end{figure}

The main issue in the Multi-Blade design is the uniformity over its
active surface. The cassettes overlap to avoid dead zones, and, in
the switching between one cassette to another, a loss in efficiency
can occur. There are mainly two reasons that cause the efficiency
drop: at the cassette edge the electric field is not uniform and
there is some material that scatters neutron on the way to the next
cassette. In order to reduce the dead zone at the cassette edge,
i.e. any material that can cause scattering, each cassette is
cropped (see Figure \ref{figMBv1prima6}) to be parallel to the
incoming neutron direction.
\\ Moreover, when a neutron is converted at the cassette edge, it
produces a fragment that half of the time travels toward outside of
the cassette and half time inward. We expect not to have generated
charge for about $50\%$ of the events. In addition to that the
electric field at the edge may not be uniform and the guard wire
contributes to enlarge the dead zone because it does not generate
charge amplification.
\\ We scan with a collimated neutron beam and we register a PHS over
the whole prototype surface by using a $1\,mm$ step for the
$x$-direction and a $10\,mm$ step for the $y$-direction. We
integrate the PHS for each position and we obtain a local counting.
We normalize it to 1 on the average efficiency. Figure
\ref{crgw5465u6u} shows the relative efficiency scan over the whole
detector. The scan along the cassettes in the position $y=40\,mm$ is
also shown.
\\ Each cassette shows a quite uniform response
along its strips ($y$-direction); the maximum efficiency relative
variation is below $2\%$. On the other hand, in the gap between two
cassettes the efficiency drops about $50\%$ in a region which is
$2\,mm$ wide.
\subsubsection{Spatial resolution}
\begin{figure}[!ht]
\centering
\includegraphics[width=8cm,angle=0,keepaspectratio]{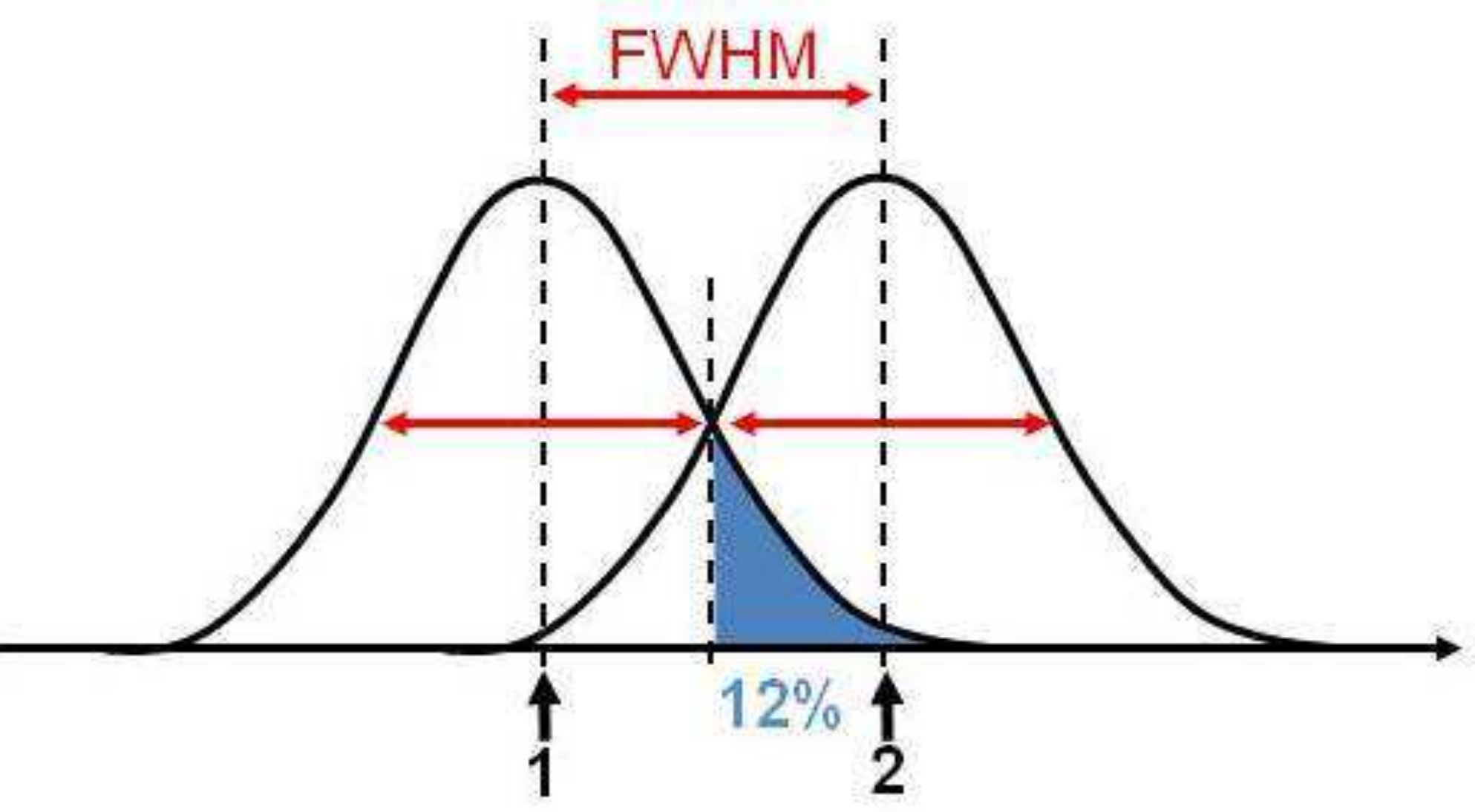}
\caption{\footnotesize Two neutrons hitting a detector at a FWHM
distance. We identify them with $12\%$ probability to be on the
wrong side.}\label{f5y22y4td}
\end{figure}

\begin{figure}[!ht]
\centering
\includegraphics[width=7.8cm,angle=0,keepaspectratio]{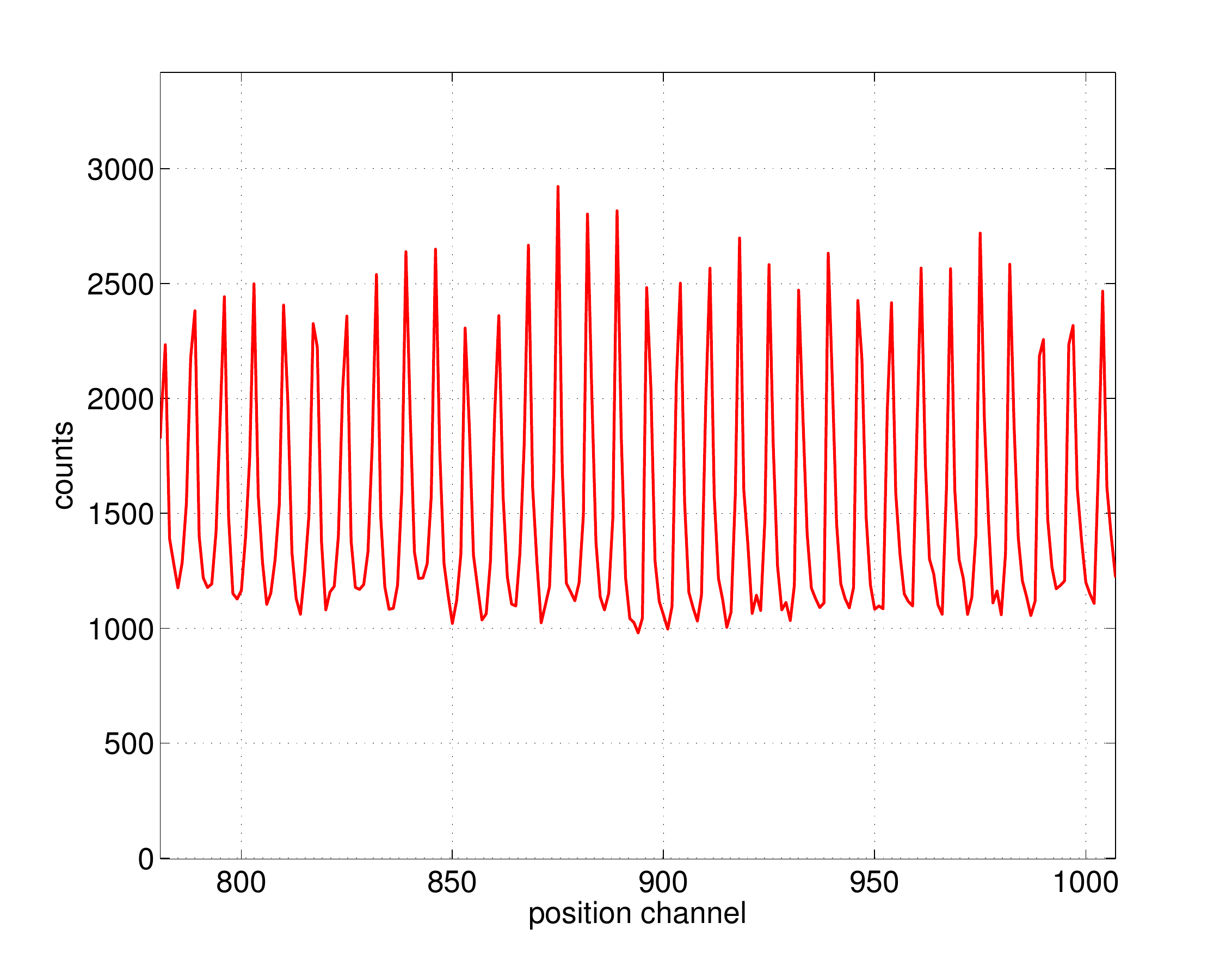}
\includegraphics[width=7.8cm,angle=0,keepaspectratio]{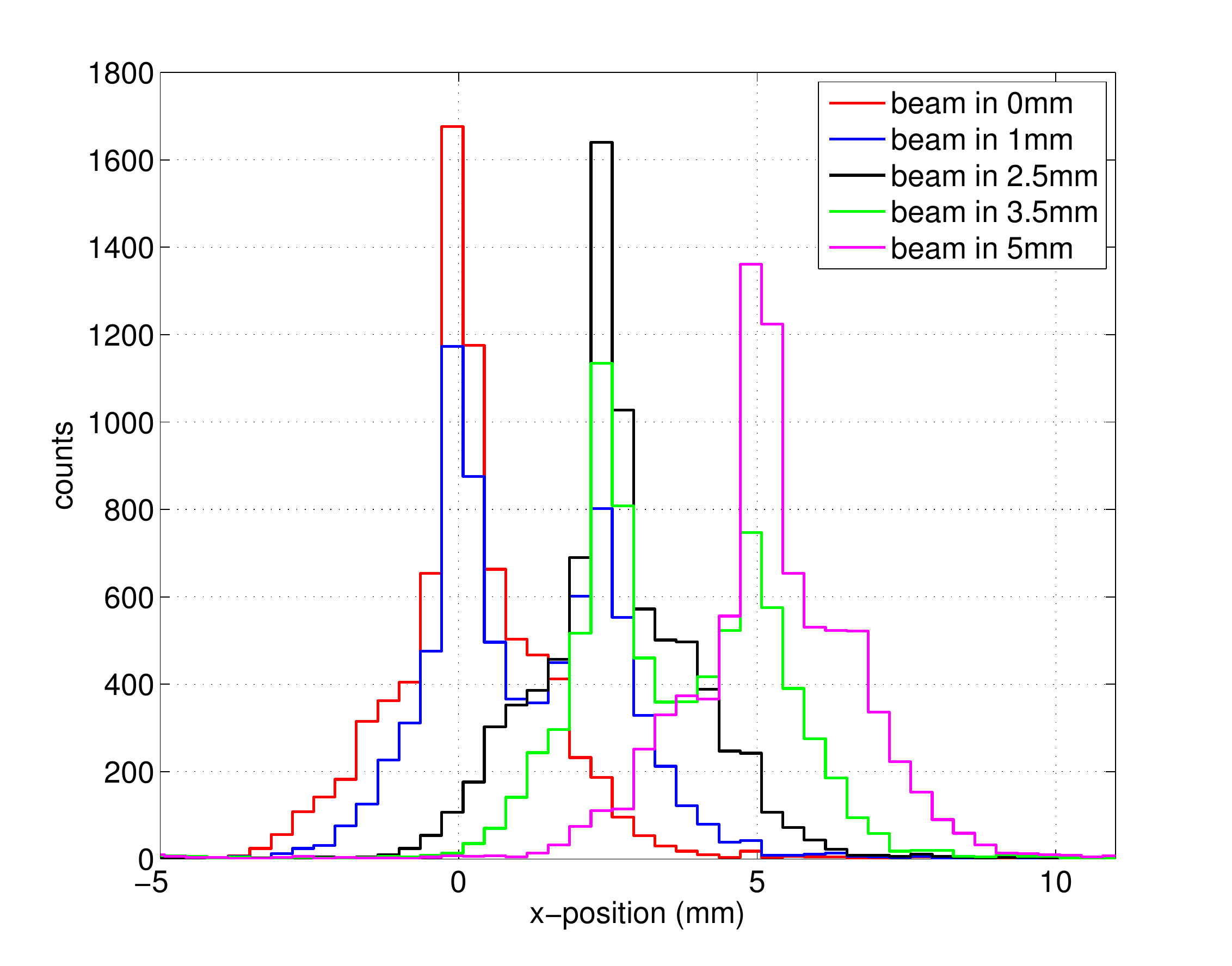}
\caption{\footnotesize Diffuse beam wire response in charge division
(left), collimated neutron beam response as a function of the beam
position (right).}\label{figurewirerspsv1}
\end{figure}

\begin{figure}[!ht]
\centering
\includegraphics[width=10cm,angle=0,keepaspectratio]{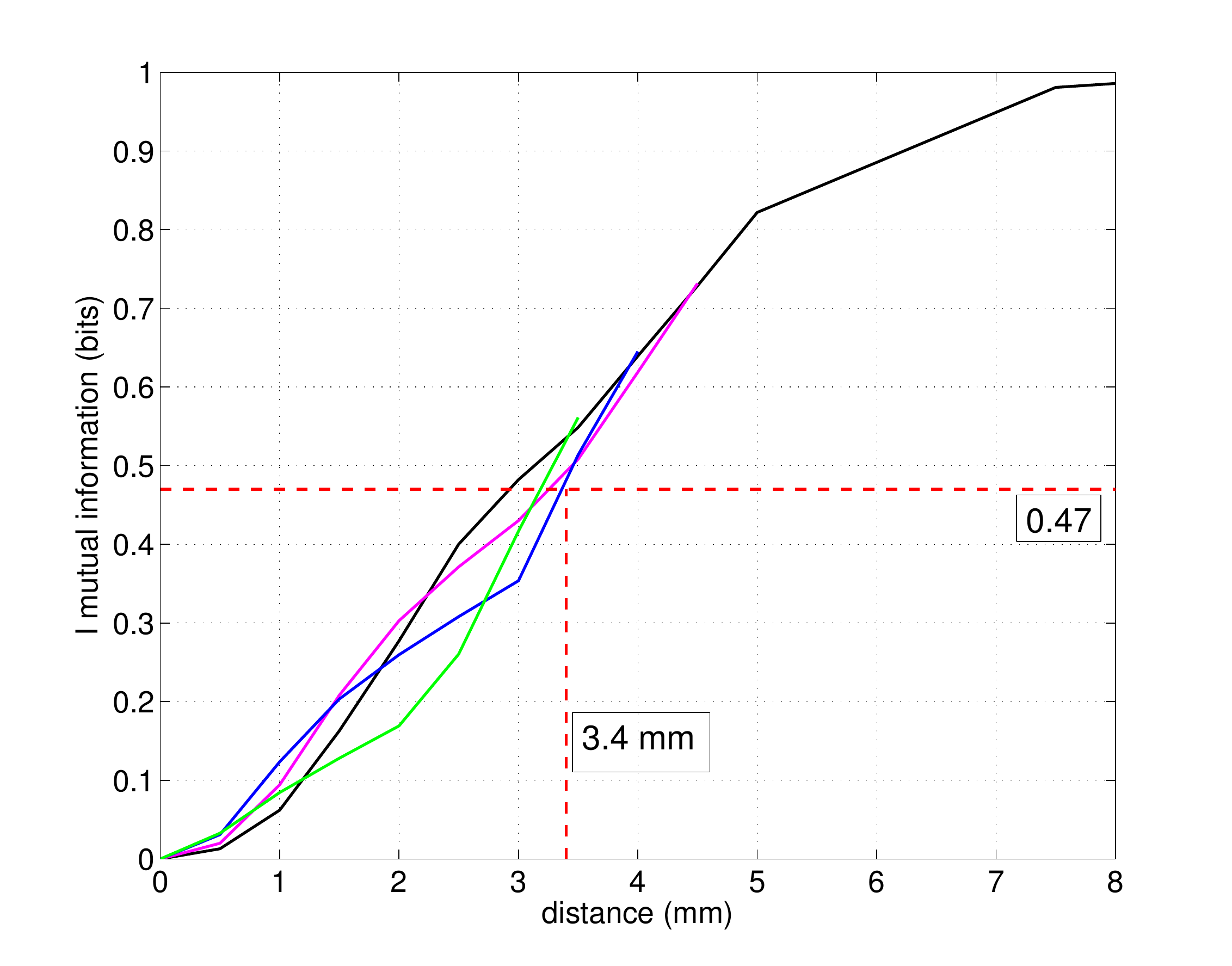}
\caption{\footnotesize Mutual information as a function of the
distance between the response distributions of the neutron detector.
The horizontal line defines an information of $0.47$ bits that
corresponds to a $3.4\,mm$ spatial resolution (before projection) in
the worse case.}\label{spatfigu57}
\end{figure}

\begin{figure}[!ht]
\centering
\includegraphics[width=10cm,angle=0,keepaspectratio]{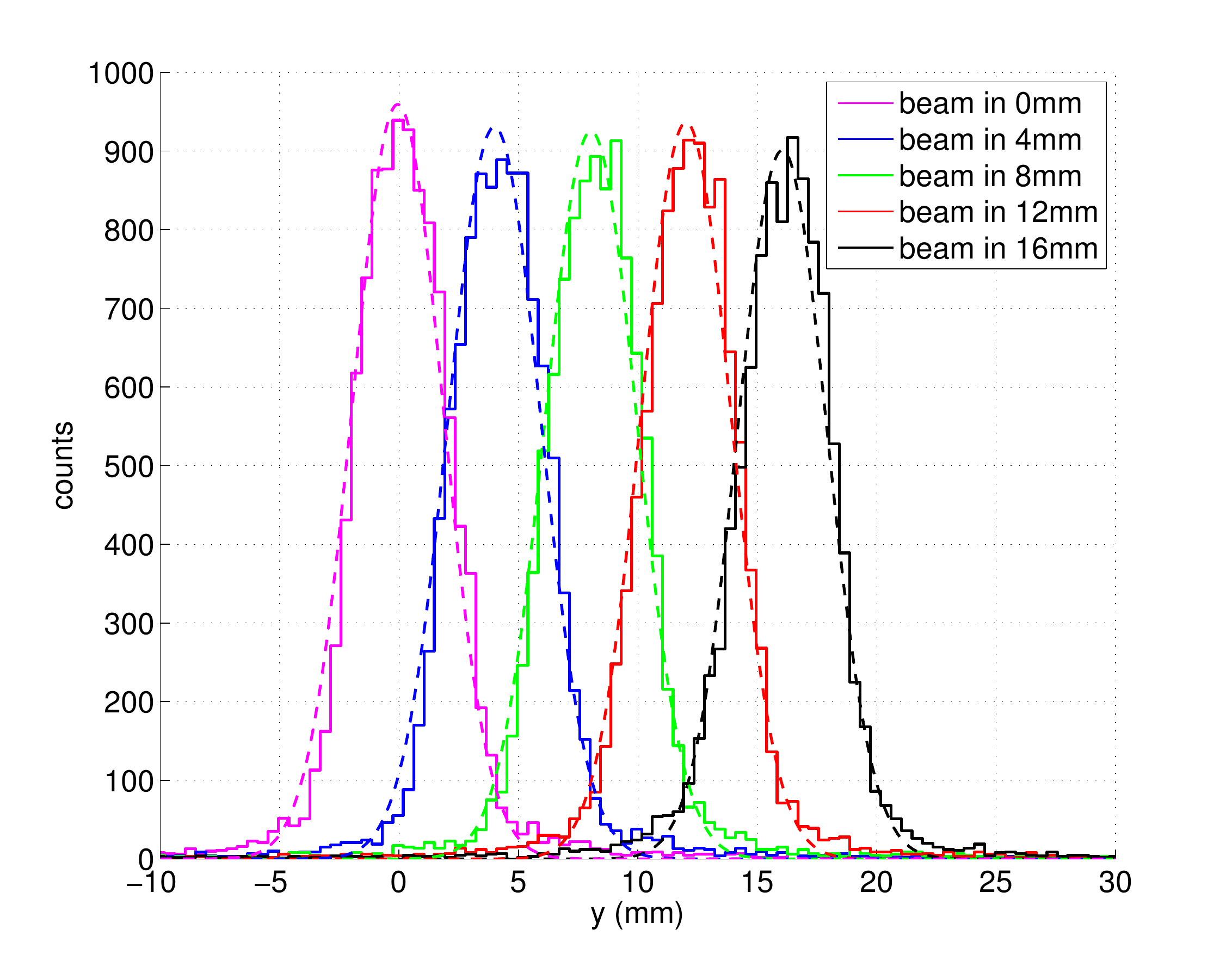}
\caption{\footnotesize Fine beam neutron scan along the strip
cathodes. The spatial resolution is given by the FWHM and
corresponds to $4.4\,mm$.}\label{figrespyresolmb4}
\end{figure}

When we want to calculate a detector spatial resolution one should
be careful as to which definition has to be adopted in order to give
a meaningful result. If the detector response is a continuous
function or discrete the problem should be tackled in a different
way.
\\ The spatial resolution is defined
as the ability to distinguish between two events as a function of
their distance. We have a continuous detector response if a large
amount of events that occur at a certain position on the detector
will generate a continuous distribution in space. A widely used
criterion is to define the spatial resolution as the FWHM (Full
Width Half Maximum) of a distribution of those events. In the
particular case the events distribution in space ends up to be
gaussian the FWHM is equivalent to $2.35\cdot\sigma$, with $\sigma$
its standard deviation. In $88\%$ of the cases, for two neutrons
hitting the detector at a distance of a FWHM (see Figure
\ref{f5y22y4td}), we will identify them on ''the right side''. We
will identify them with $12\%$ probability to be on the wrong side.
\\  When the detector response can be considered continuous, the spatial
resolution as the FWHM represents a suitable definition. This
happens when the pixel size is much smaller than the resolution. On
the other hand it is not possible to define the spatial resolution
as a parameter of a distribution if the events are discretely
distributed over a few pixels. That is when the granularity of the
detector is comparable with the spatial resolution.
\\ As example, let's imagine a detector read-out system
made up of strips. Imagine that their width is a few $mm$. A gas
volume faces the strip plane. If the ionizing particle ranges in gas
is order of a few $cm$, the induced charge on the strips results in
a continuous distribution over the strips with charge centroid in
the center of the resulting gaussian distribution. \\ On the
contrary, if the particle ranges in the gas were much shorter,
$\sim1\,mm$, the strip response would be not any more gaussian, but
almost an individual strip will participate to the induction
process.
\\ We consider a MWPC of wires spaced by a few
$mm$ and particle tracks comparable to the wire pitch. The detector
response to a single event is not continuous. As most of the charge
is originated in the multiplication region thanks to the avalanche
process close to the wire, each wire acts as an independent
detector. Each wire acts as a electromagnetic lens focusing the
charge for a certain detector segment. Thus, the detector is
segmented and, thus, discrete. If the primary charge is only
generated in the influence cell of a single wire, only this wire
will generate a signal; on the other hand, if a ionization process
covers two segments two wires will be involved. In this case the
spatial resolution is not given by the wire pitch because there can
be situations where two wires reacts to the single event. Moreover,
spatial resolution is neither two wire pitch because there are cases
when only one wire reacts. In fact, by putting forward that the
detector has a two wire pitch resolution we were asking to be able
to discriminate between events with a precision of $100\%$. That is
not incongruent with out FWHM resolution definition to get a $88\%$
precision.
\\ We should reformulate the spatial resolution definition for a more general case.
\\ In order to do that, one can refer to the Shannon information
theory \cite{patrickinforis}. If we consider two streams of neutrons
hitting the detector in two different positions we will get,
according to its resolution, two resulting response distributions
that can either be completely separate, can overlap partially or can
be indistinguishable. The resolution definition is based on the
probability to correctly assign an event to its real neutron
original stream, i.e. to the right distribution.
\\ This problem can be treated as a lossy communication channel.
In a binary communication channel the bit 0 or 1 can be sent and, in
presence of noise, it can be received correctly or flipped.
\\That is the same as determining the probability to assign a
neutron that belonged to one of two distributions to the right one.
\\ We define as the mutual information $I(X;Y)$
the information that the receiver get in a noisy channel, i.e. the
detection process when 1 bit of information was sent. $X$ is the
received message if $Y$ was sent. If the mutual information is 1, no
loss in the channel is present. \\ In the detector case four
possibilities are possible: a 0 was sent and 0 is received, a 0 was
sent and a 1 is received, and similar for a 1 is sent.
\\ The mutual information is operationally defined as:
\begin{equation}\label{eqrp1}
I(X;Y) =
\frac{1}{2}\sum_{x=0}^{1}\sum_{y=0}^{1}p(x|y)\log_2\frac{p(x|y)}{\frac{1}{2}(p(x|0)+p(x|1))}
\end{equation}
where $p(x|y)$ is the conditional probability to send y and get x.
\\ $I$ represents the information we have obtained from a single
neutron impact about the point source from which it originates. By
resolution we can then understand the distance between two neutron
hits needed in order to reach a given threshold of information $I$.
\\ We can define, in the case of a symmetric distribution, as $a$
the probability to make the right reception of the message; i.e. we
send 0 and we get 0 or we send 1 and we get 1. This probability is
naturally symmetric for the neutron labels 0 or 1. The resulting
probability $1-a$ would be the probability to make the wrong
reception; i.e. to assign a neutron to the wrong stream. In the
gaussian continuous case, at one FWHM distance, those probabilities
correspond to $a=88\%$ and $1-a=12\%$. Hence, if we calculate the
mutual information for this case we get $0.47$ bits of information.
Therefore, the probability of $88\%$ to make the right reception
corresponds to an information of $0.47$ bits (in the symmetric and
continuous case).
\\ As a result, a way to define the resolution which is independent
from any spatial distribution is to calculate the mutual information
and asking to get a minimum threshold value of $0.47$ bits. This
definition is suitable as well for discrete response, when the
detector granularity is comparable to the resolution we want to
calculate as for the continuous case where it will become the FWHM
definition if the distribution is gaussian. As a matter of fact, a
scan over several neutron streams has to be performed to determine
the lowest mutual information obtained.
\\ Going back to the MWPC example, in the case the
detector response involves either one or two wires according to the
particle track position, the mutual information will be higher for
higher spacing between distributions because their overlap is
smaller.
\\ The value to take as the detector resolution
is the one that gives the spacing to achieve a mutual information of
at least $0.47$ bits.
\paragraph{Spatial resolution: x}
The version V1 of the Multi-Blade prototype is operated at
$10^{\circ}$ between the neutron incoming direction and the detector
converter layer. We recall that the wire plane is projected on the
neutron incoming direction. An improvement by a factor
$\sin(10^{\circ})\sim0.17$ is achieved on the horizontal spatial
resolution with respect to an orthogonal incidence. E.g. if the
spatial resolution, before projection, were a wire pitch (in our
prototype $2.5\,mm$) this results in an actual resolution of about
$0.45\,mm$.
\\ In Figure \ref{figurewirerspsv1} is shown the charge division
response of the wire plane by using either a diffuse beam or a
collimated beam down to $1\,mm$ footprint ($2.5$\AA).
\\ The charge division method is able to identify each wire anode
position; when a collimated beam is in one position we can either
get a single wire reacting or two. This effect is due to the fact
that the wire plane splits the gas volume into almost independent
cells, thus the charge generated by primary ionization in one cell
makes its associated wire react. In our case, since we are using a
mixture of $Ar/Co_2$ ($90/10$) at atmospheric pressure the $^{10}B$
neutron capture reaction fragment ranges make a few $mm$. Therefore,
if the track is contained in one single wire cell, only a single
wire reacts; on the other hand if the track travels across two cells
we get a two wire response. Since the wire plane is read-out in
charge division, if two wires react, the hit will be identified to
be in between the two wires, corresponding to the charge centroid.
The response distribution, for a given hitting position, will have
tails corresponding to these events. Figure \ref{figurewirerspsv1}
shows the resulting distribution as the neutron beam moves along the
detector. One can wonder now in what is the actual spatial
resolution in this situation. In order to quantify it is necessary
to apply the informational-theoretical approach explained above.
\\ We calculate the mutual information between the distribution in
Figure \ref{figurewirerspsv1} for all the possible combinations and
we will take as resolution the worse result at an information
threshold level of $0.47$ bits.
\\ Figure \ref{spatfigu57} shows the mutual information as a
function of the distance of the neutron distribution response of our
detector. We notice that in the worst case we end up with $3.4\,mm$;
which translates into a spatial resolution of $0.6\,mm$ at
$10^{\circ}$.
\\Note that the spatial resolution lies in between
a single wire pitch ($2.5\,mm$) and two; because we are asking the
detector to be able to discriminate between two neutrons that hit
the detector at one resolution distance with a confidence level of
$88\%$.
\paragraph{Spatial resolution: y}
As already mentioned, particles tracks in gas make few $mm$. Since
the cathodes read-out strips are $0.8\,mm$ wide and they are spaced
by $0.2\,mm$ and the read-out is performed by a charge division
chain, there are several strips that are involved in the induction
process per each event. The charge division makes the charge
centroid along the $y$-direction in the detector.
\\ For the cathodes the response can be considered continuous and the
FWHM method is suitable.
\\ Figure \ref{figrespyresolmb4} shows the strip response as a
function of the position of the collimated beam hitting the
detector. By performing a gaussian fit we obtain a spatial
resolution (FWHM) for the vertical direction $y$ of about $4.4\,mm$.
\subsubsection{Images}
\begin{figure}[!ht]
\centering
\includegraphics[width=14cm,angle=0,keepaspectratio]{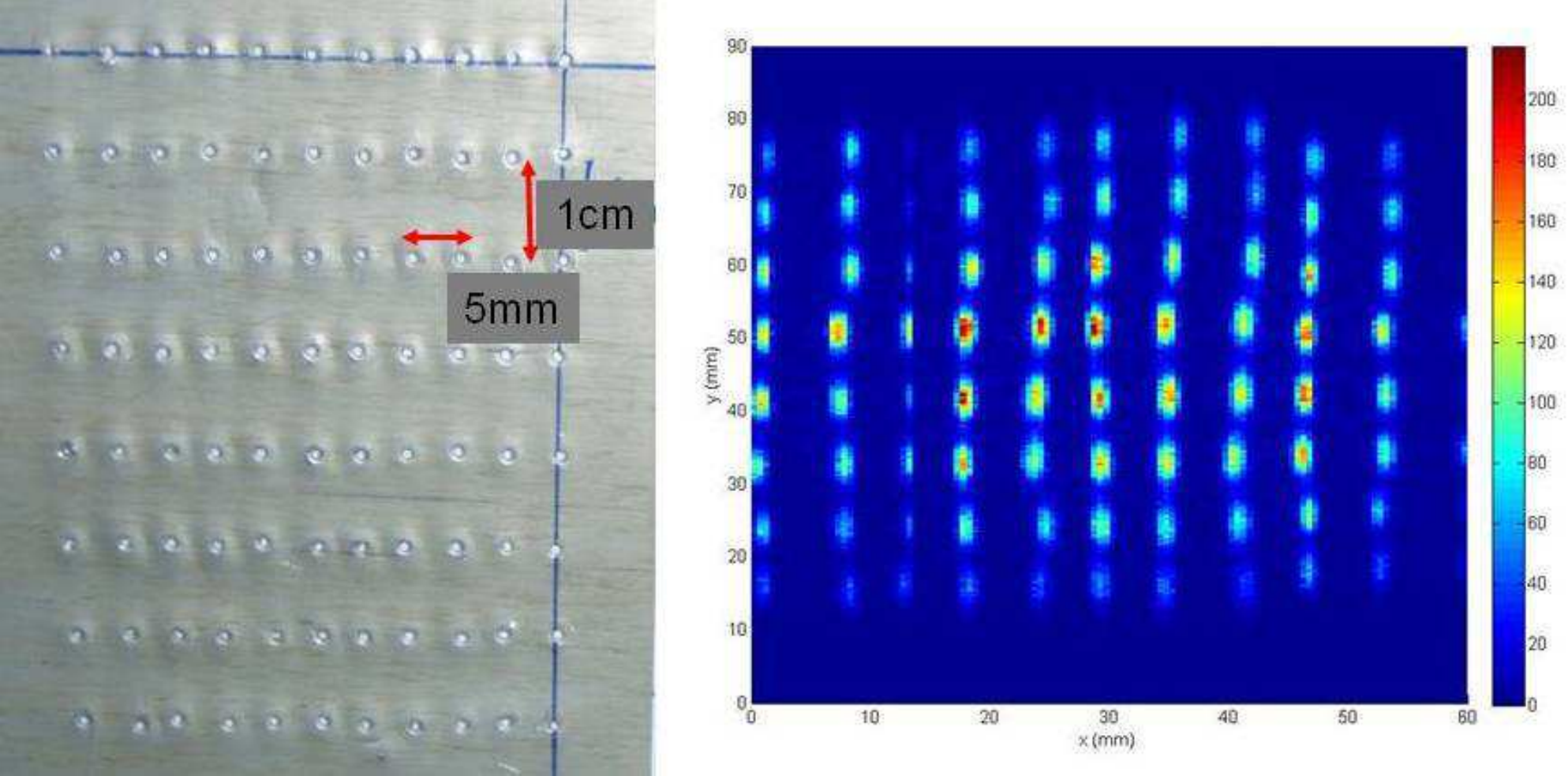}
\caption{\footnotesize The Cd mask (left) used to generate the image
on the right.}\label{figrmcwtcwb45}
\end{figure}

To validate the results we want to generate an image with our
prototype. In order to do that we acquire both anode and cathode
signals and we reconstruct an event using their time coincidence.
\\ Figure \ref{figrmcwtcwb45} shows an image reconstructed with the Multi-Blade
obtained by placing a Cd mask in order to get a pattern. The mask
consists of 80 holes of $1\,mm$ size spaced by $5\,mm$ along $x$ and
by $1\,cm$ along $y$. \\ The number of bins on the image is set to
be equal to the number of wires (37) for the $x$-direction and is
256 bins for the $y$-direction.
 \\ Due to the neutron beam divergence the spots
on the image appear much wider than the detector resolution.
\\ In the reconstructed image we observe the intensity that drops at the cassettes edges.
\section{Multi-Blade version V2}\label{sectv2mbg}
\subsection{Mechanical study}
\begin{figure}[!ht]
\centering
\includegraphics[width=5.5cm,angle=0,keepaspectratio]{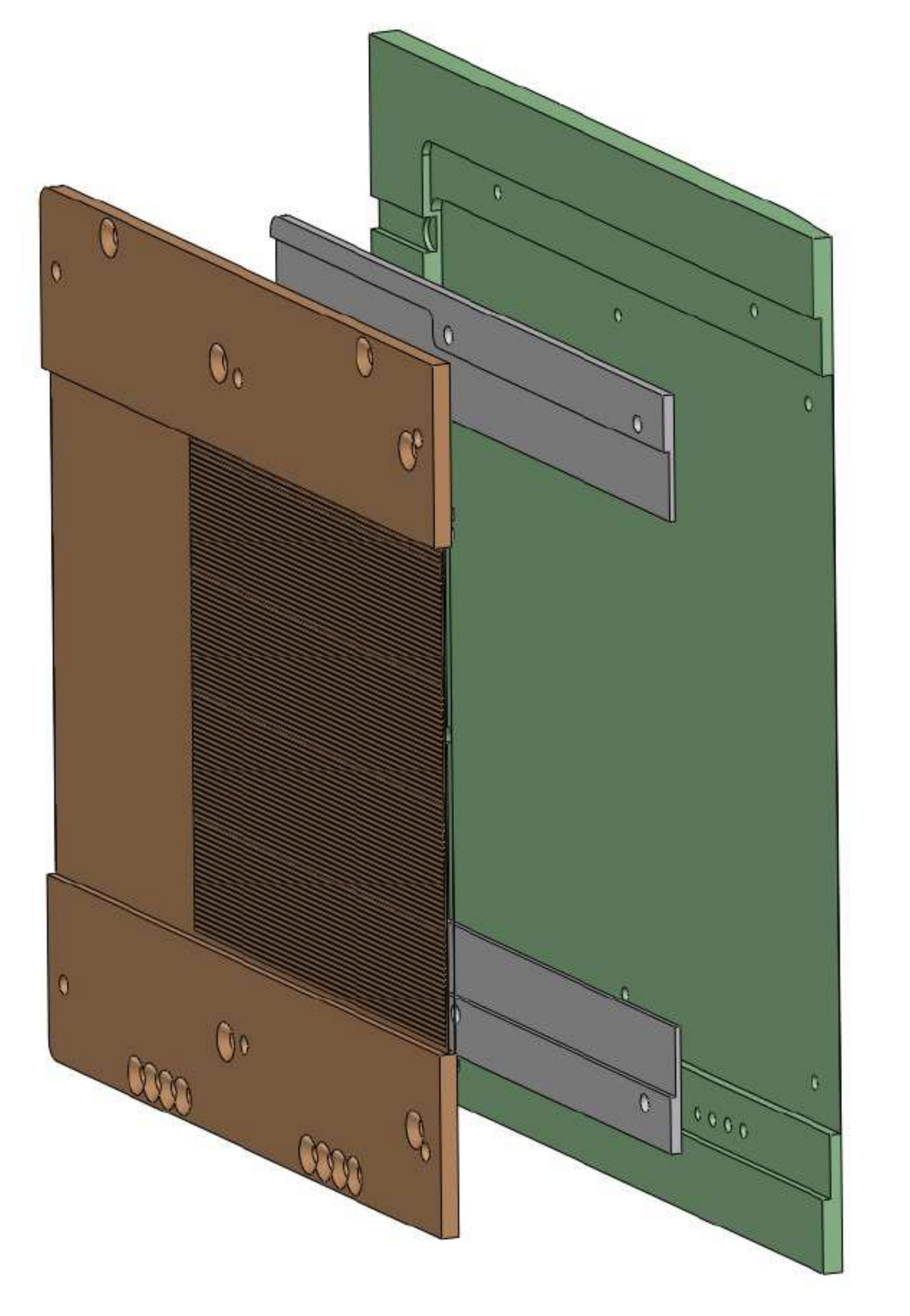}
\includegraphics[width=9.2cm,angle=0,keepaspectratio]{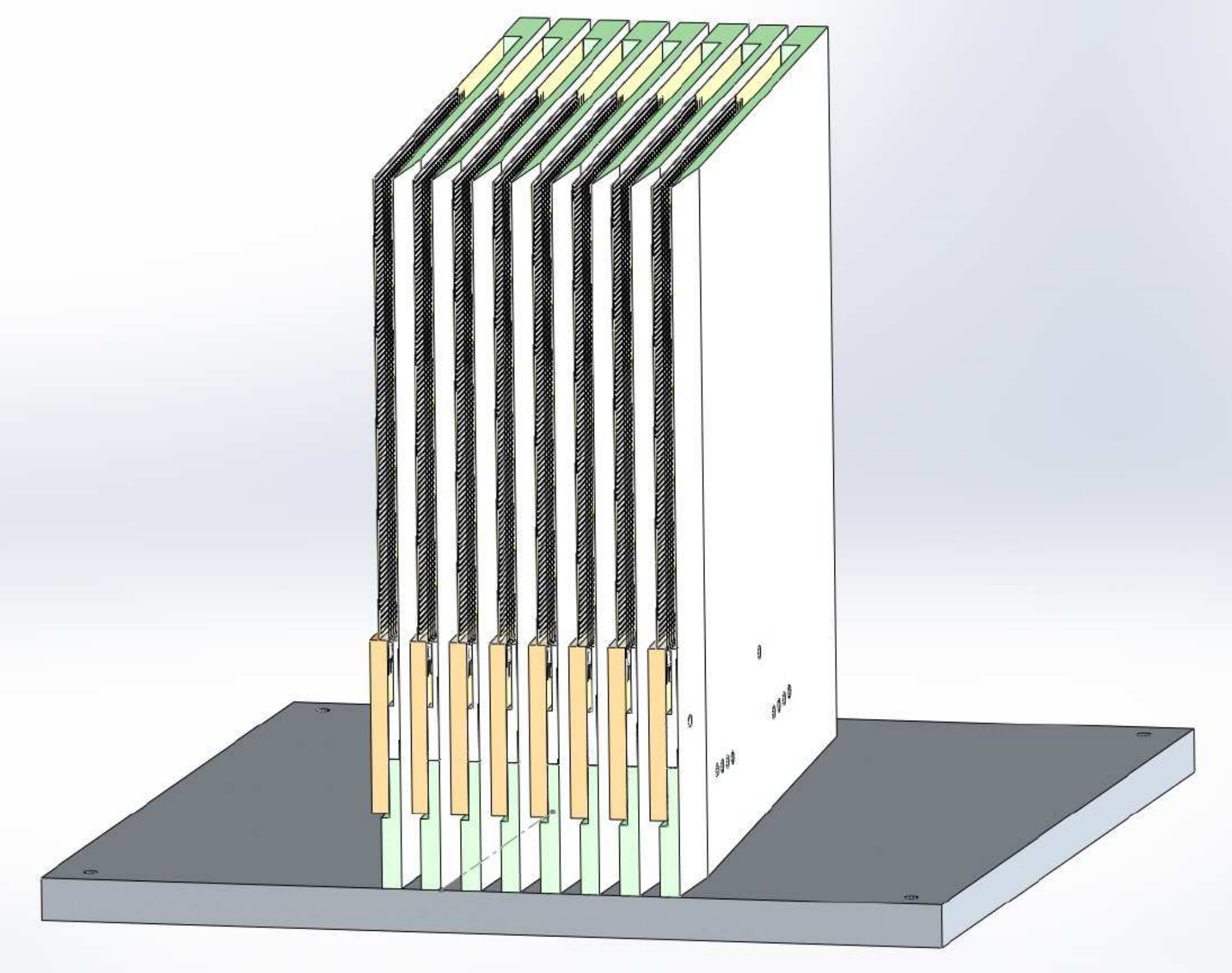}
\caption{\footnotesize A cassette conceived to hold one converter
layer (left) and a stack of several cassette (right).}
\label{mbv2scehm543ergg}
\end{figure}

We learned from the Multi-Blade version V1 that the single layer
configuration (option A) presents less mechanical constraints.
Moreover, the substrate holding the converter has not to be crossed
by neutrons that makes its manufacture easier.
\\ The converter layer can be thick because the efficiency is
saturated above $3\,\mu m$; the substrate can be thick also because
it has not to be crossed by neutrons to hit a second converter.
Hence, the substrate can be an integrated part of the cassette
holder, the converter layer can be directly deposited over its
surface. The read-out system used is the same as in version V1.
Neutrons have still to cross the PCBs before being converted. Figure
\ref{mbv2scehm543ergg} shows a cassette and a stack of them
conceived for the single layer option.
\\ The cassettes are oriented at $5\,^{\circ}$ with respect to the incoming neutron
direction. The sensitive area of each cassette is $10\times 9\,cm^2$
but, the actual sensitive area offered to the sample is given by
$(10\,cm \cdot \sin(5\,^{\circ}))\times 9\,cm = 0.9\times 9\, cm^2$.
As a result, the actual projected wire pitch is improved down to
$0.22\,mm$.
\subsection{Mechanics}
\begin{figure}[!ht]
\centering
\includegraphics[width=14cm,angle=0,keepaspectratio]{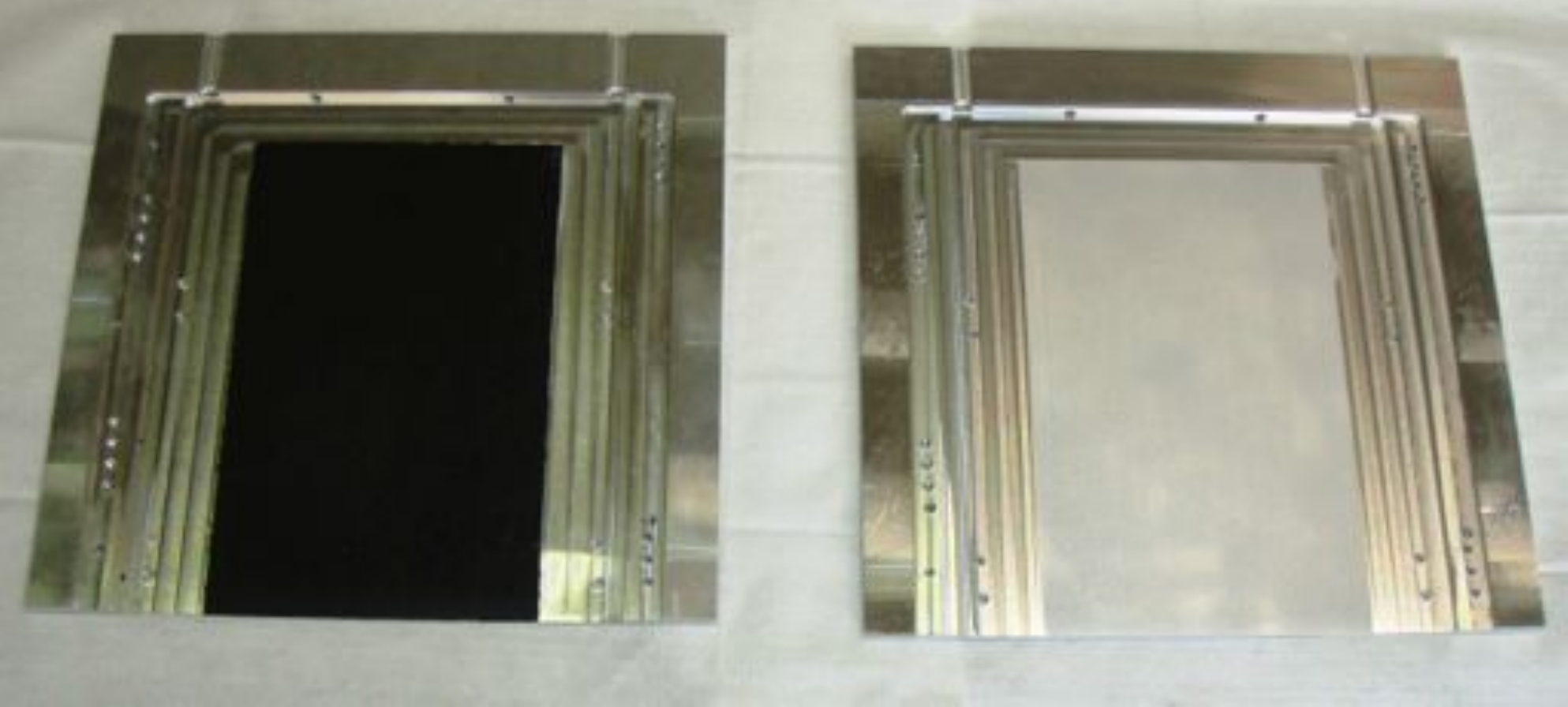}
\caption{\footnotesize A cassette V2 coated and un-coated with
$^{10}B$ painting.} \label{asd56}
\end{figure}

\begin{figure}[!ht]
\centering
\includegraphics[width=10cm,angle=0,keepaspectratio]{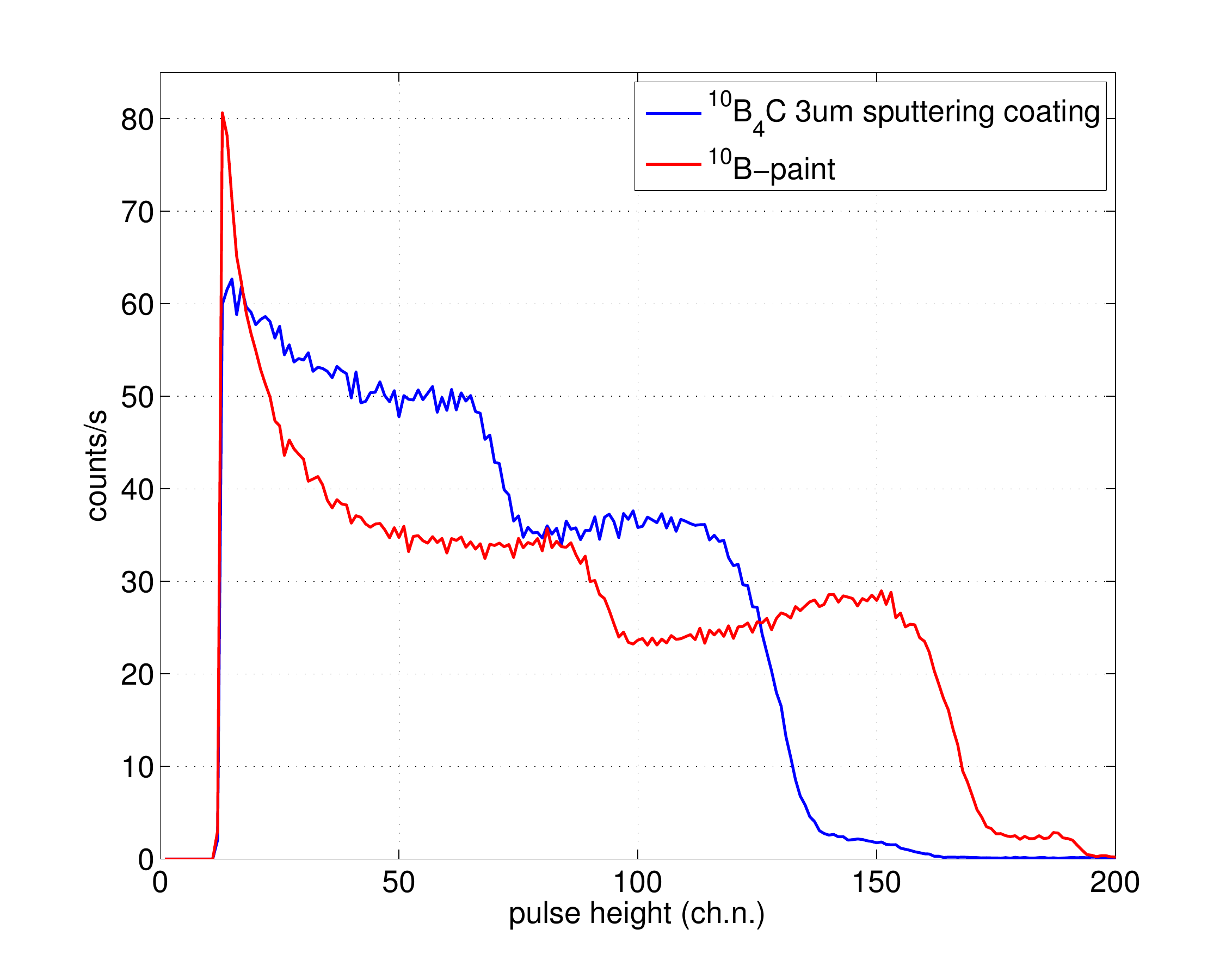}
\caption{\footnotesize Comparison between the PHS of the sputtered
coated layers \cite{carina} and the $^{10}B$ painting.}
\label{eretvrtr57457}
\end{figure}

\begin{figure}[!ht]
\centering
\includegraphics[width=10cm,angle=0,keepaspectratio]{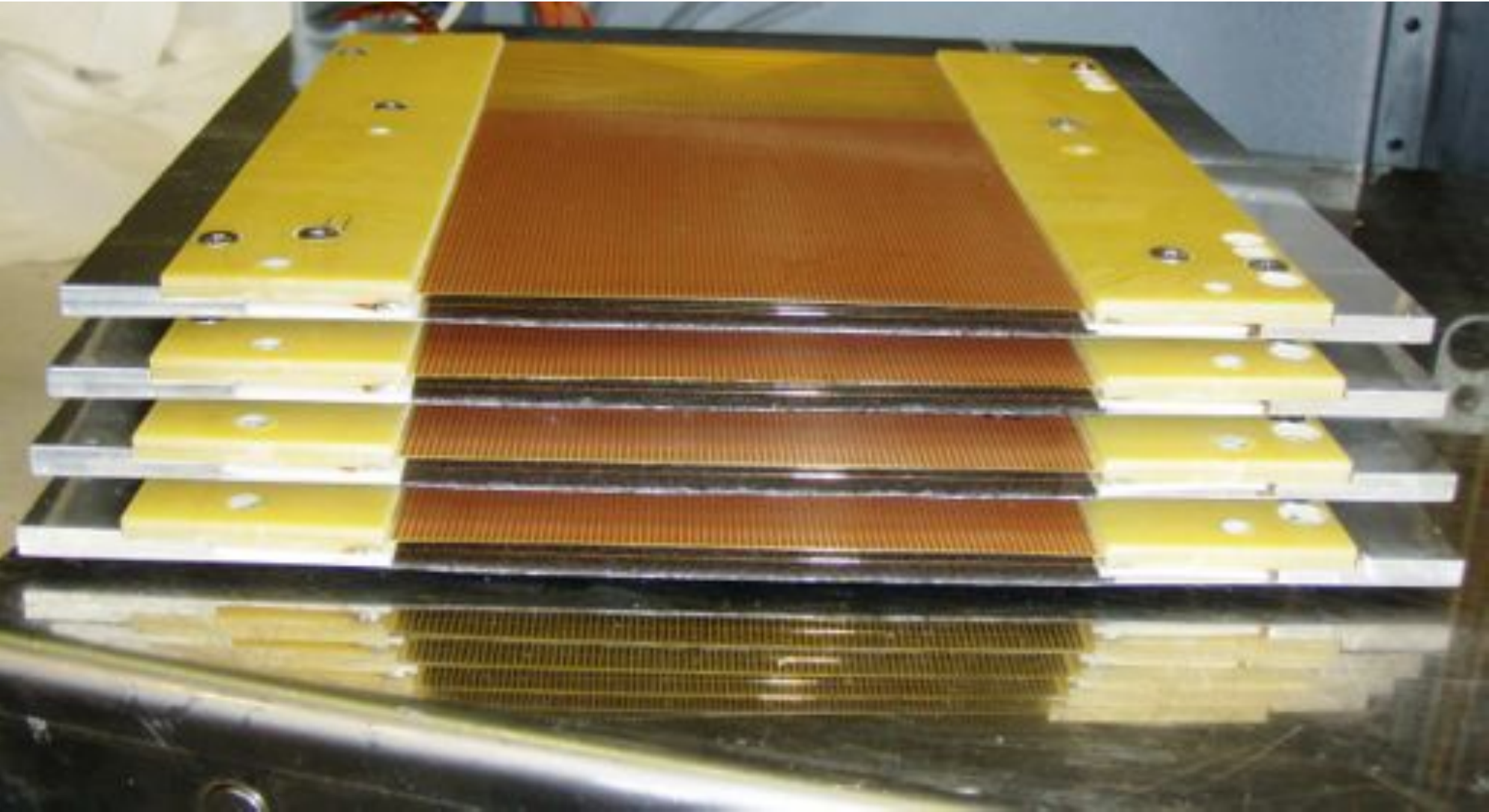}
\caption{\footnotesize Four fully assembled cassettes for the
Multi-Blade version V2.} \label{fulcassv2546}
\end{figure}

\begin{figure}[!ht]
\centering
\includegraphics[width=9cm,angle=0,keepaspectratio]{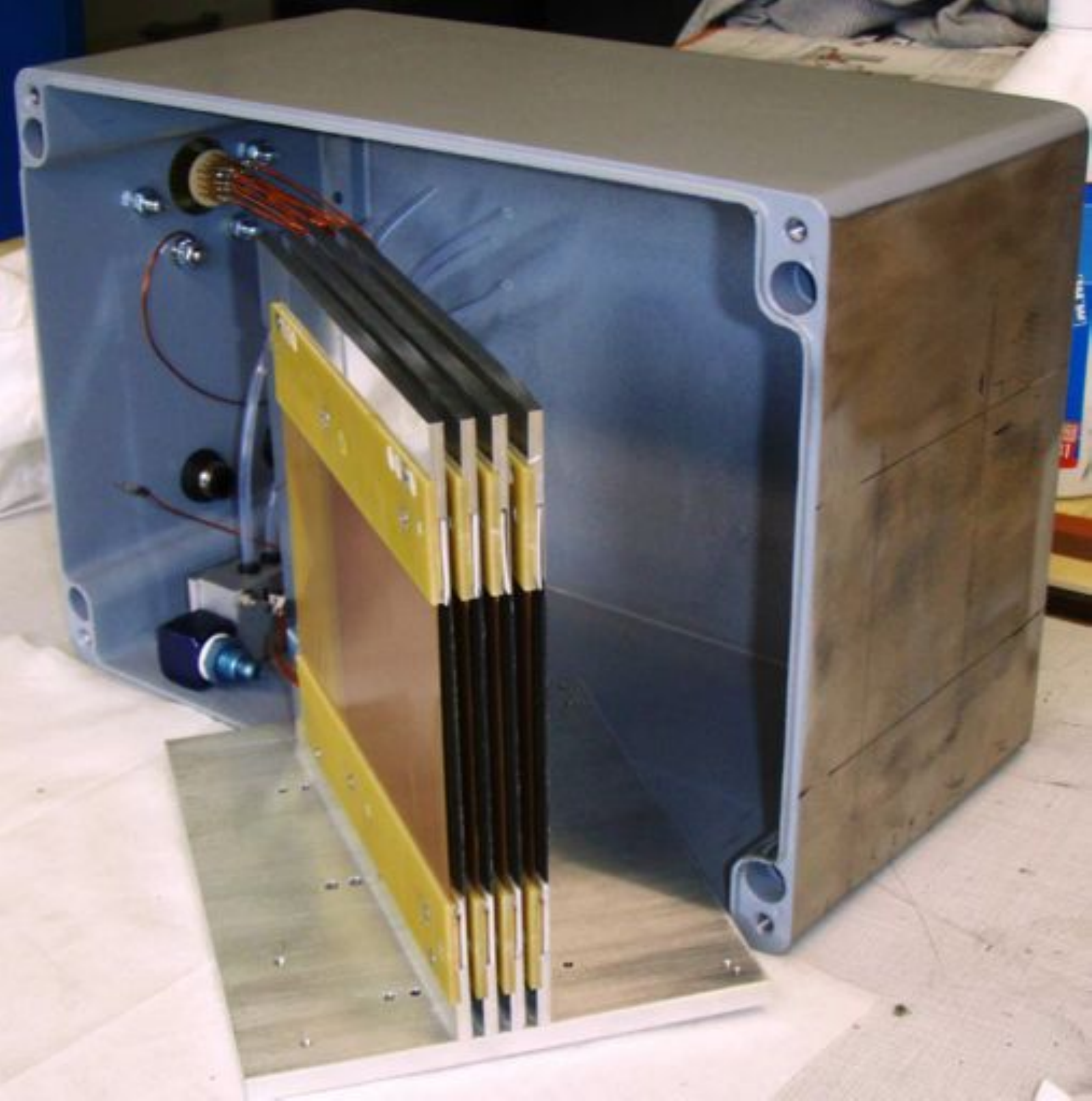}
\caption{\footnotesize The four cassettes installed at $5\,^{\circ}$
with respect to the detector window being installed in the gas
vessel.} \label{fullv246tf}
\end{figure}

The second prototype (V2) consists also of four cassettes but we
operate them at $5\,^{\circ}$. At this inclination the expected
efficiency at $2.5$\AA \, is about $43\%$ if we employ the sputtered
coating of the version V1 \cite{carina}. The cassettes are the ones
shown in Figure \ref{mbv2scehm543ergg}, conceived to study the
single converter layer option. A rigid substrate is directly coated
with the converter material. The cassette width in the version V1
was about $12\,mm$, in the version V2 we reduce their actual size to
$6\,mm$. Consequently the MWPC gap, between the converter and the
cathodes, is $4\,mm$. With respect to the version V1 the wire plane
is closer to the converter.
\\The prototype active area, considering the cassette overlap, is
about $3.2\times 9\,cm^2$.
\\ The read-out PCBs are those used in the version V1.
\\ Figure \ref{asd56} shows a cassette substrate both coated and
un-coated.
\\ Since the efficiency is saturated as the thickness of the layer
exceeds $3\,\mu m$, we study the possibility to use different
converters. We can deposit a painting containing $^{10}B$ grains and
make a coating a few hundreds of microns thick. A thick layer also
functions as an integrated collimator. Any neutron that comes from
the sides of the detector has less probability to be detected and is
more likely absorbed in the outer layers. Hence, only neutrons which
impinge the detector from the front have a serious chance to
generate a signal. Neutron background is then decreased.
\\ The uniformity of the coating, even in the single layer configuration, is an
important aspect to guarantee the converter flatness. The latter has
to ensure the precision of the neutron incidence angle, in fact if
it varies slightly the efficiency changes widely. Furthermore, a
deviation from the converter flatness also induces the variation of
the electric field and then the local gain of the detector changes.
\\ The roughness of the converter should be below the
neutron capture fragment ranges, which is of the order of a few $\mu
m$ for $^{10}B$. In fact, the gain in efficiency due to an
inclination comes from the fact that the neutron path travels close
to the surface. If the surface is irregular (on the $\mu m$ scale or
more), that can be seen as equivalent for a neutron to hit a surface
perpendicularly, there is not much gain in efficiency. It is crucial
that the size of our grains, in the painting, is less than the
particles ranges, i.e. their size should be below the micron scale
for $^{10}B$.
\\ The conductivity of the painting can be an issue.
If the resistivity is too large the charge evacuation is not
guaranteed and consequently the actual electric field is affected.
We mix a glue with $^{10}B$ grains of sizes $<10 \mu m$, the layer
resistivity was measured to be about $50\, M\Omega \cdot m$ and we
make a $0.5\,mm$ layer. As the grain size is not smaller than the
fragment ranges we know that there can be an efficiency issue. We
did not have access to a finer-grained $^{10}B$ powder: our grinding
technique resulted in $\sim10\,\mu m$ grain size. We used this
powder.
\\ Figure \ref{eretvrtr57457} shows two PHS: one is taken with a $3\,\mu m$ thick $^{10}B_4C$ layer \cite{carina}
and the other with the $^{10}B$ painting both installed in a MWPC.
Both at normal incidence and with a $2.5$\AA \, neutron beam. The
variation in gain on the two spectra is due to the difference in the
gas gap between the wire plane and the converter, since the painting
is a few $mm$ closer to the wires than the sputtered layer.
\\ The painting efficiency is $1.5\%$ lower than the sputtered
coating.
\\ In order to investigate if the resistivity of the painting is not to large to avoid the evacuation
of the charges, we place the painting layer on a very intense beam
of $560\,KHz$ and we measure the counting rate as a function of
time. There are no losses after several hours. The resistivity of
the $^{10}B$ painting seems to be acceptable.
\\ The painting is suitable for single layer application supposed that we can control the flatness
of the layer and to use smaller grains. We do not guarantee the
sputtered layers efficiency under an angle because of the size of
the grains we used in the painting.
\\ The converter painting was not optimized, hence we expect some
problems due to its not perfect regularity. This effect will be more
evident at the edge of each cassette where the flatness affect to a
greater extent the electric field glitches.
\\ We mount the prototype using the painting. Four cassettes were assembled.
Figure \ref{fulcassv2546} and \ref{fullv246tf} show the cassettes
and the installation in the gas vessel for testing.
\\ The electronics used is the same as in the version V1 of the
Multi-Blade.
\subsection{Results}
\subsubsection{Operational voltage}
The measure of the counting curve gives the bias voltage of
$800\,V$.
\\ The operational voltage is lower than the one used for the first prototype
because the gap between anodes and cathodes was reduced in the new
design. For this reason the PHS is degraded also because the maximum
path in $Ar/CO_2$ of an $\alpha$-particle makes almost $9\,mm$ it is
more likely in the version V2 to hit the opposite cathode before
depositing its entire energy in the gas volume.
\begin{figure}[!ht]
\centering
\includegraphics[width=7.8cm,angle=0,keepaspectratio]{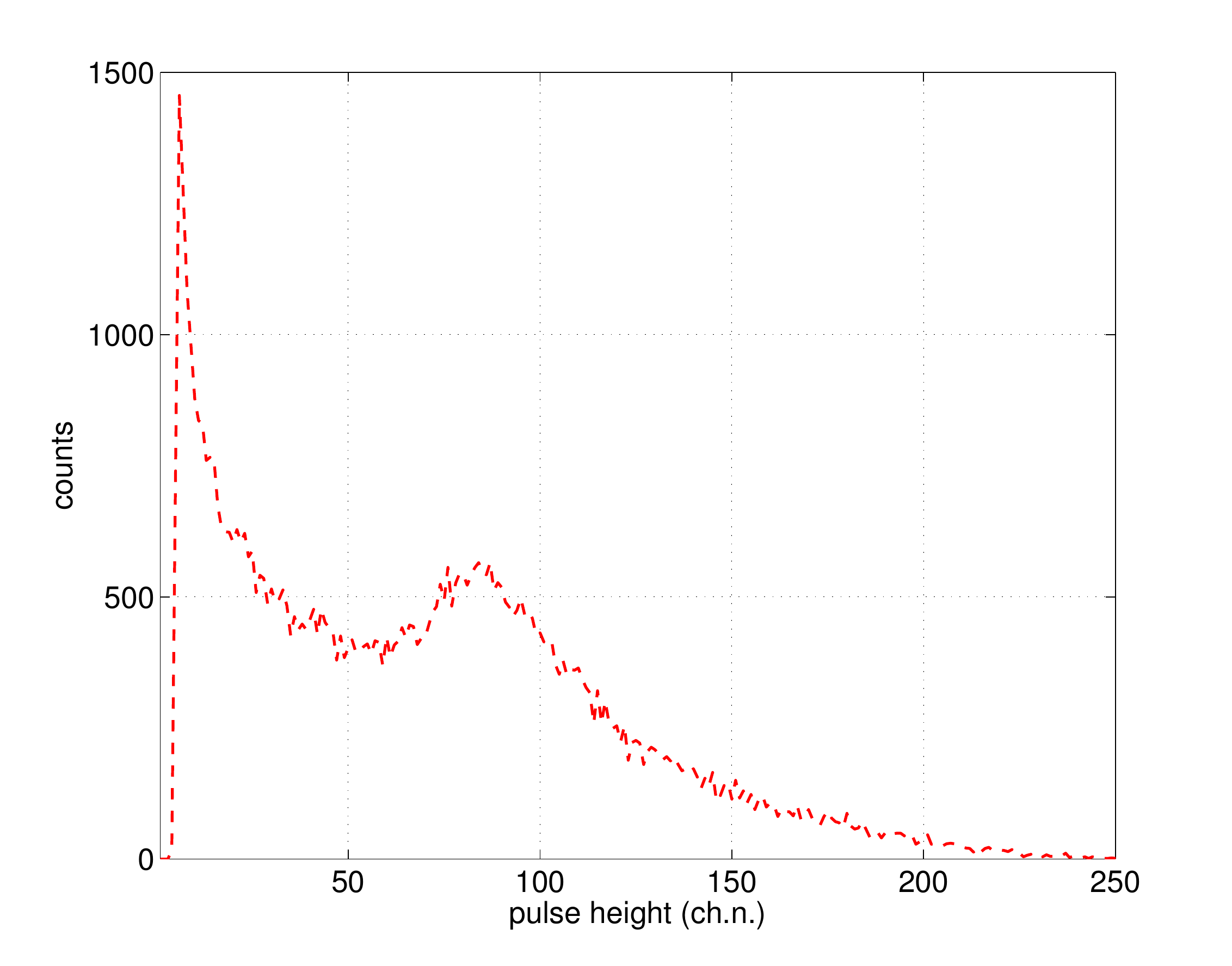}
\includegraphics[width=7.8cm,angle=0,keepaspectratio]{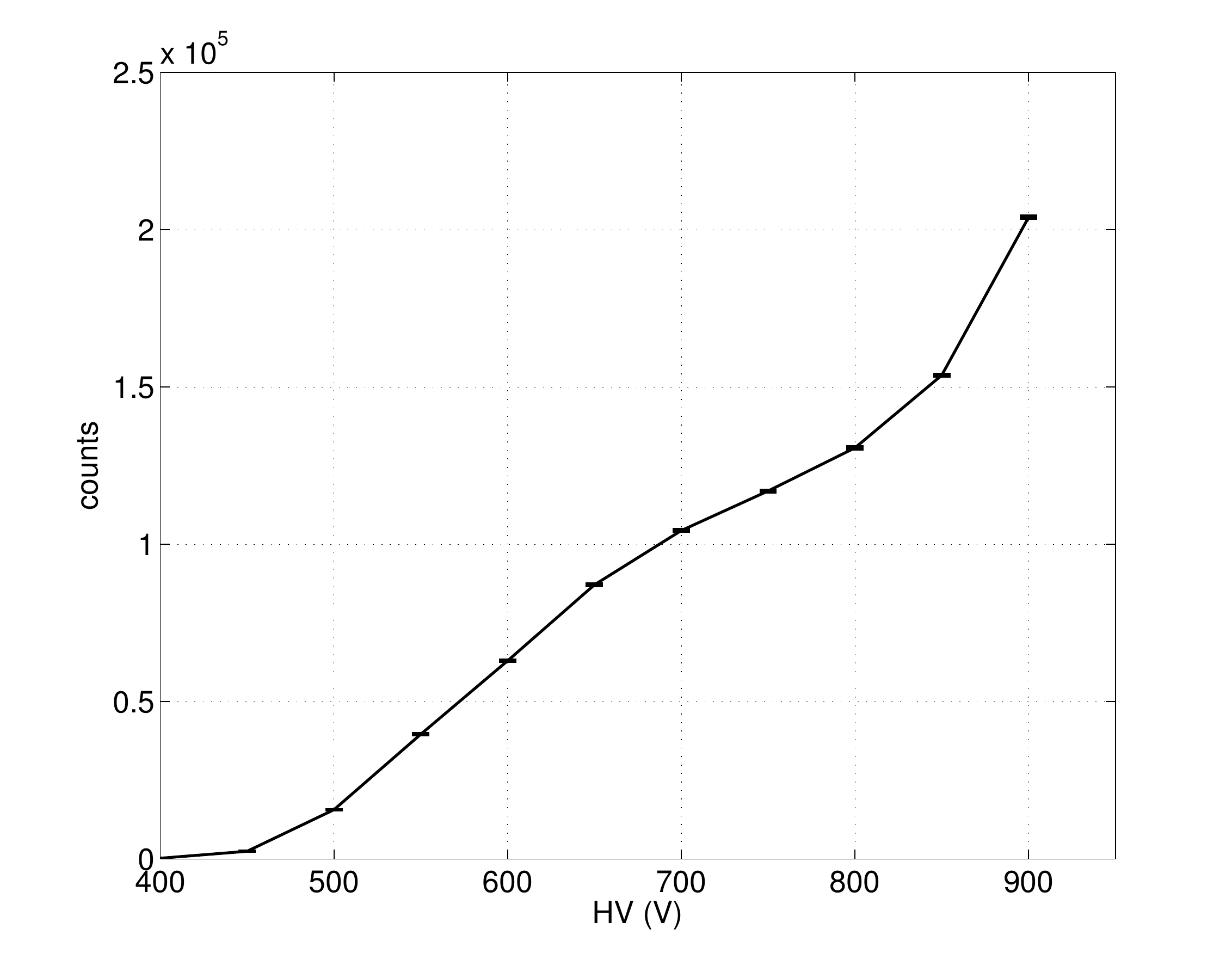}
\caption{\footnotesize PHS measured on wires at $800\,V$ (left). The
Multi-Blade detector plateau (right).}\label{phspaltembv1456ewc5}
\end{figure}
\subsubsection{Efficiency}
Detection efficiency of the Multi-Blade prototype V2 has been
measured on CT2 at ILL by using a collimated and calibrated neutron
beam of wavelength $2.5$\AA using the procedure explained in details
in the Appendix \ref{apphexmeasexplan}.
\\ The neutron flux used is $\left(12010\pm20\right)\,neutrons/s$
over an area of $2\times 6\,mm^2$.
\\ The efficiency was measured for the operational voltage $800\,V$.
The efficiency was measured on the four cassettes under the angle of
$5^{\circ}$ and then averaged. The results is:
\begin{equation}
\varepsilon\left( \mbox{at }2.5\mbox{\AA}\right)=\left( 8.32 \pm
0.05\right)\%
\end{equation}
The expected efficiency for a sputtered layer of which the roughness
is widely below the $\mu m$ scale is about $43\%$ (at $2.5$\AA).
Having an angle only increases the efficiency by a few percent
because of the grains size. The surface irregularity makes the
inclination effect vanish. Neutrons only impinge almost
perpendicular on the microscopic grain structure.
\\ At $5^{\circ}$ neutron reflection by the surface is
negligible. Hence, by using a sputtered layer or, if one can better
control the painting flatness and have smaller grain size, there
should be no reason not to get the calculated efficiency.
\subsubsection{Uniformity}
\begin{figure}[!ht]
\centering
\includegraphics[width=10cm,angle=0,keepaspectratio]{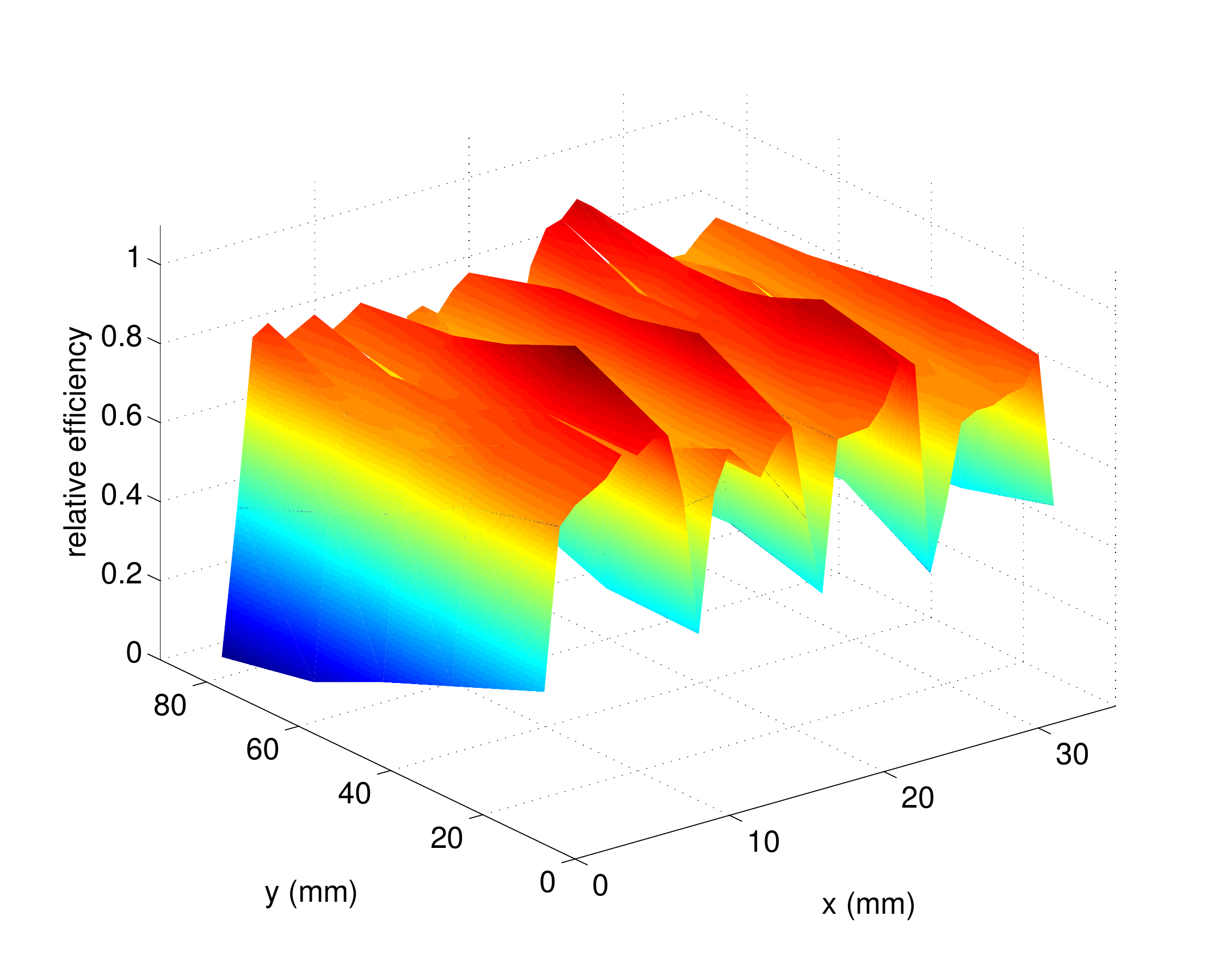}
\caption{\footnotesize Relative efficiency scan over the whole
detector surface.} \label{unifgasf945ntgww54}
\end{figure}

In the version V2 of the Multi-Blade the mechanics is more compact
in order to avoid dead zones in the overlap between the cassettes.
It has been discussed that the two issues which degrade the
uniformity are those dead zones and the electric field at each
cassette edge.
\\ Even though the mechanics design is more efficient in the version V2, the electric field
issue remains.
\\Moreover, the coating done by the painting was slightly irregular
mostly at the edges of each cassette. This diminishes the precision
by which we switch between one and the following. A large amount of
converter material at the edge will absorb most of the neutrons
without generating any signal.
\\ Figure \ref{unifgasf945ntgww54} shows the relative efficiency scan over the
detector surface. Compared with the version V1 uniformity is worse.
The gap between the converter and the wire plane is $2\,mm$, hence
any irregularity on the layer surface will affect the local gain of
the detector. Even along each cassette ($y$-direction) the gain
varies by about $10\%$ while in the sputtered version it only varied
of about $2\%$.
\\ Moreover, now looking at the $x$-direction, at the cassettes edge
the efficiency now drops more than $50\%$. This is due to the amount
of converter at the edge that does not generate signal but only
absorbs neutrons.
\\ The version V2 is in principle more compact and if the coating
would be precise the uniformity was expected to be better than in
the version V1.

\subsubsection{Spatial resolution}
\begin{figure}[!ht]
\centering
\includegraphics[width=14cm,angle=0,keepaspectratio]{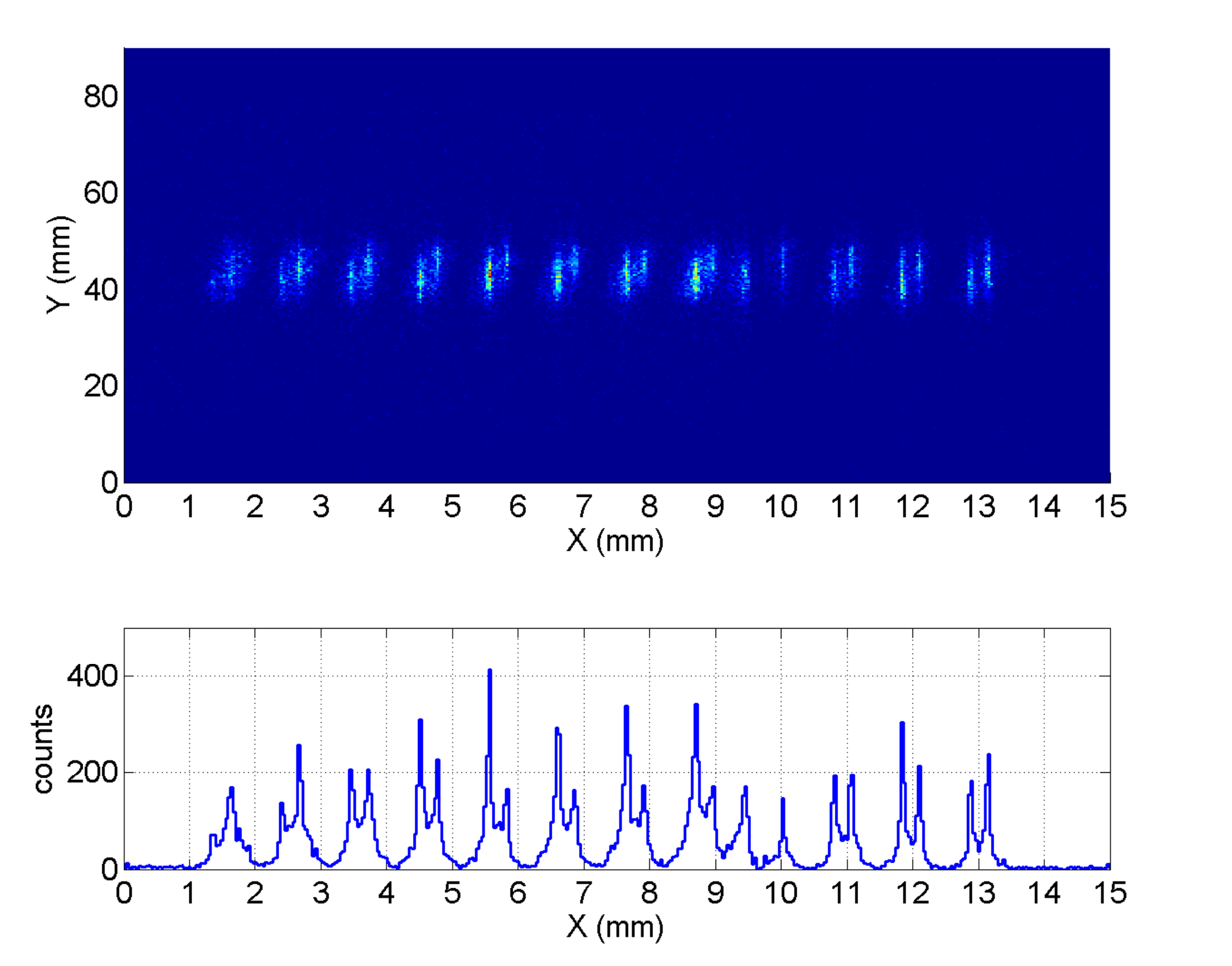}
\caption{\footnotesize An image and its projection on the $x$ axis
taken with the prototype. Each slit is $0.2\,mm\times10\,mm$ large
and it is spaced by $0.9\,mm$.} \label{figexp0n74jdfiwhud}
\end{figure}

\begin{figure}[!ht]
\centering
\includegraphics[width=10cm,angle=0,keepaspectratio]{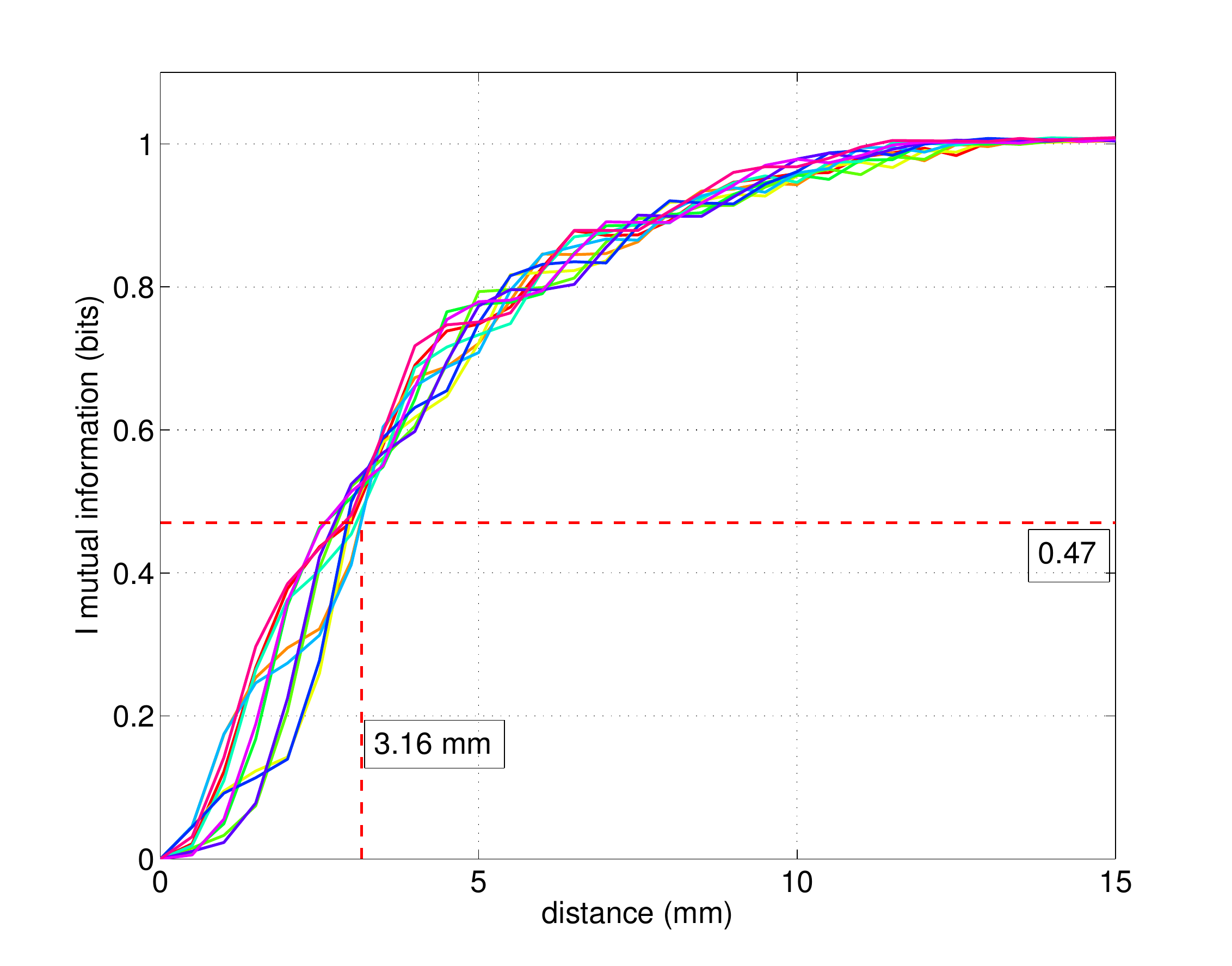}
\caption{\footnotesize Mutual information as a function of the
distance between the response distributions of the neutron detector.
The horizontal line defines an information of $0.47$ bits that
corresponds to a $3.16\,mm$ spatial resolution (before projection)
in the worse case.} \label{mutinforfgjk830nsbbcd}
\end{figure}

\begin{figure}[!ht]
\centering
\includegraphics[width=10cm,angle=0,keepaspectratio]{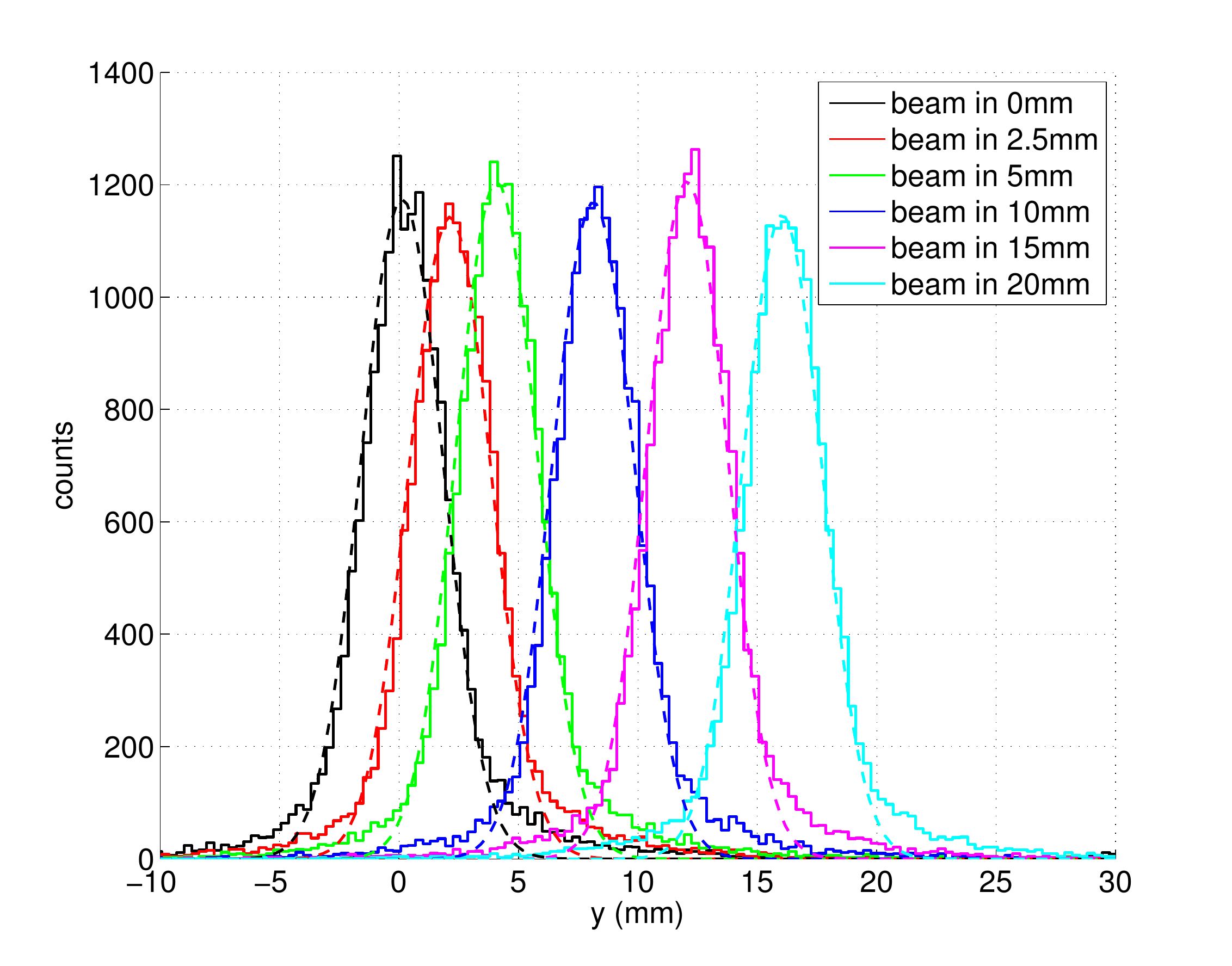}
\caption{\footnotesize Fine beam neutron scan along $y$. The spatial
resolution is given by the FWHM and corresponds to $4\,mm$.}
\label{spaty8jndh}
\end{figure}

The spatial resolution was calculated as already shown for the
Multi-Blade version V1. By using a very collimated beam, of about
$0.2\times10\,mm^2$, we scan one cassette and a half of the
detector. Each step is $0.9\,mm$ along the $x$-direction. The
cassette 1 is from $x=0\,mm$ to $x=10\,mm$, the cassette 2 starts at
$x=10\,mm$. Figure \ref{figexp0n74jdfiwhud} shows the reconstructed
image and its projection on the $x$-direction obtained by adding
together all the images taken in the scan.
\\ For each step either a wire or two are firing. A the switching
point between the two cassette we observe the drops in the counts.
\paragraph{Spatial resolution: x}
We quantify the spatial resolution along the $x$-direction in the
same way as for the version V1.
\\ We scan the detector surface to obtain the events distribution to
calculate the mutual information which is shown in Figure
\ref{mutinforfgjk830nsbbcd}. We use the threshold of $0.47\,bits$
which corresponds to the standard FWHM resolution definition and we
obtain a value of $3.16\,mm$.
\\ This value is slightly better of the one found for the version V1
because in the version V2 the gap between wires and converter is
diminished. It is the first part of the ionization path, on average,
that gives more signal and it is closer to the fragment emission
point, this improves the spatial resolution.
\\ Since the detector is inclined at
$5^{\circ}$ the actual spatial resolution is given by the
projection: $3.16\,mm\cdot\sin(5^{\circ})=0.275\,mm$.
\paragraph{Spatial resolution: y}
The spatial resolution given by the strips is also enhanced thanks
to the narrower gas gap.
\\ Figure \ref{spaty8jndh} shows a scan performed along $y$. The
spatial resolution is given by the FWHM and is $4\,mm$.
\subsubsection{Dead time}
\begin{figure}[!ht]
\centering
\includegraphics[width=10cm,angle=0,keepaspectratio]{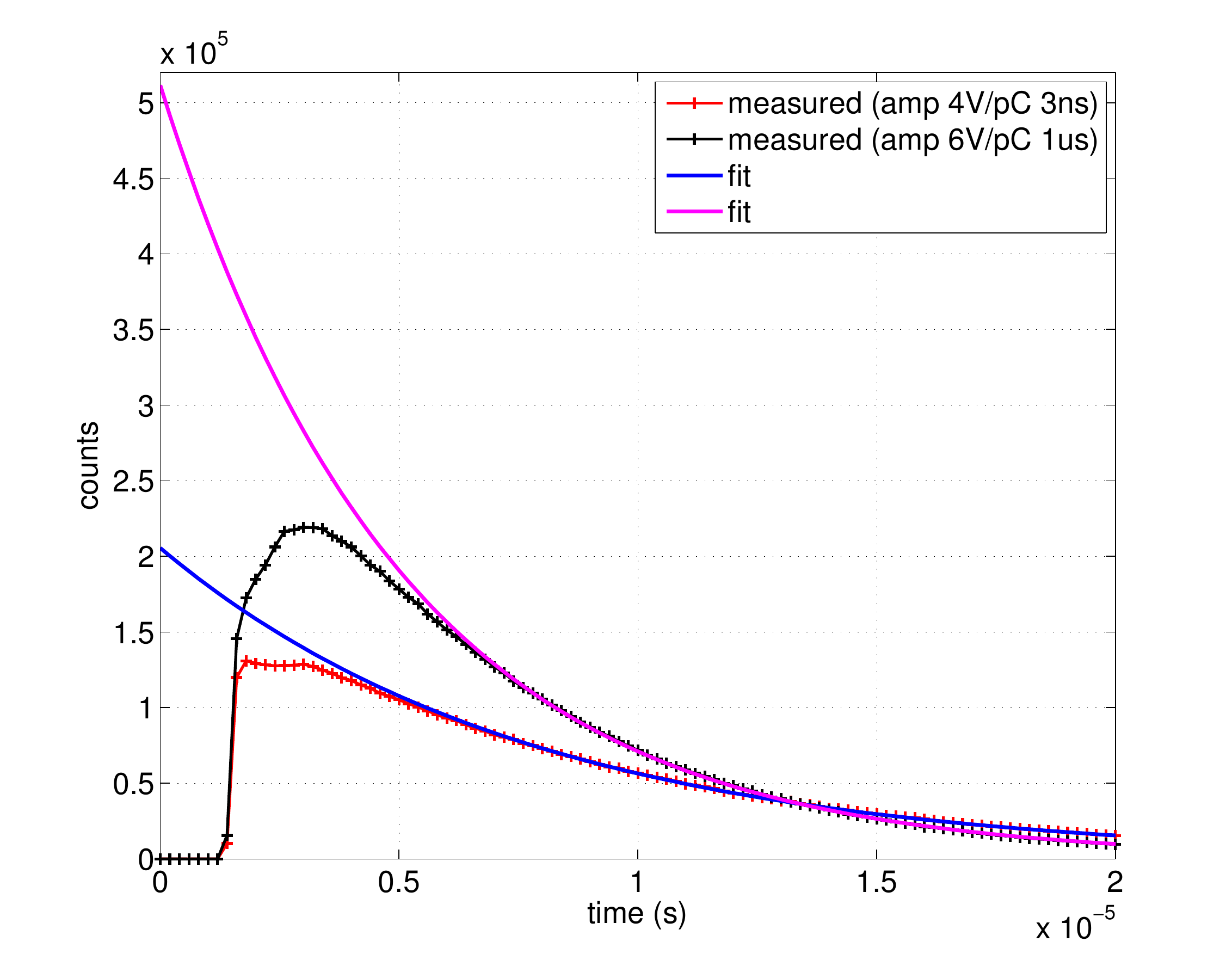}
\caption{\footnotesize Time distribution of neutron events recorded
with the Multi-Blade, the fit shows the theoretical behavior. Two
anode amplifiers have been used.} \label{dtmbv6scfa}
\end{figure}

The intrinsic dead time of a detector is due to its physical
characteristics; here we measure the entire dead time from the
detector to the end of the whole electronic chain. It is the
detector plus electronics dead time we measure.
\\ Neutrons arrive at the detector according to an exponential distribution
assuming the process to be Poissonian. A way to measure dead time is
to record the difference in the arrival time between two successive
neutrons on the detector; in principle their distribution should
follow an exponential:
\begin{equation}
f(t)=\frac1{\tau}e^{-t/\tau}
\end{equation}
where the time $\tau$ represents the average time is in between two
events; $\nu=1/\tau$ is the counting rate.
\\ In practice the detector is characterized by a dead time $t_D$
which is the minimum time interval that separates two correctly
recorded events. As a result the distribution measured with the
detector should move away from the theoretical behavior near and
below $t_D$. Moreover, if the detector is ideally non-paralyzable,
the distribution has to show a sharp cutoff at $t_D$ because the
probability of measuring an event between $t=0$ and $t=t_D$ is zero.
\\ In a paralyzable case the passage is smoother.
\\ Figure \ref{dtmbv6scfa} shows the measured times between
neutrons on the detector. The two anode outputs of a single cassette
where added and the resulting signal discriminated. The time between
every couple of discriminated events was recorded for $T=300\,s$.
\\ The measurement was performed by using two kind of amplifiers. One is
the standard Multi-Blade amplifier of $1\,\mu s$ shaping time and
the second is a fast amplifier with $3\,n s$ shaping time.
\\ The fits in Figure \ref{dtmbv6scfa} represent the theoretical
behavior. \\ The value for $\tau$, for both the measured
distributions, was obtained by the calculation of the maximum
likelihood estimator for the exponential distribution. It is:
\begin{equation}
\tau = \frac{\sum_i n_i (t_i-t_s)}{\sum_j n_j}
\end{equation}
where $t_s$ is the minimum time for which we consider the measured
distribution to behave as expected. We can assume that a time $t_s$
exists above which the measured distribution follows the exponential
behavior. We assume $t_s=7\,\mu s$.
\\ By knowing $\tau$ and the measurement duration $T=300\,s$, the
total number of neutrons that have generated a signal in the
detector but, due to dead time, have not all been recorded, is given
by:
\begin{equation}
N_0 = \frac{T}{\tau}
\end{equation}
If we integrate instead the measured distribution we obtain the
number of events recorded $N_m$.
\\ The dead time is simply given by:
\begin{equation}
t_D = \frac{N_0-N_m}{N_0}\tau = \frac{\frac{T}{\tau}-N_m}{T}\tau^2
\end{equation}
For the $1\mu s$ amplifier we obtain $t_D =
(1.58\pm0.08)\cdot10^{-6}\,s$, the fast amplifier gives $t_D =
(1.5\pm0.1)\cdot10^{-6}\,s$.

\subsubsection{Images}
\begin{figure}[!ht]
\centering
\includegraphics[width=14cm,angle=0,keepaspectratio]{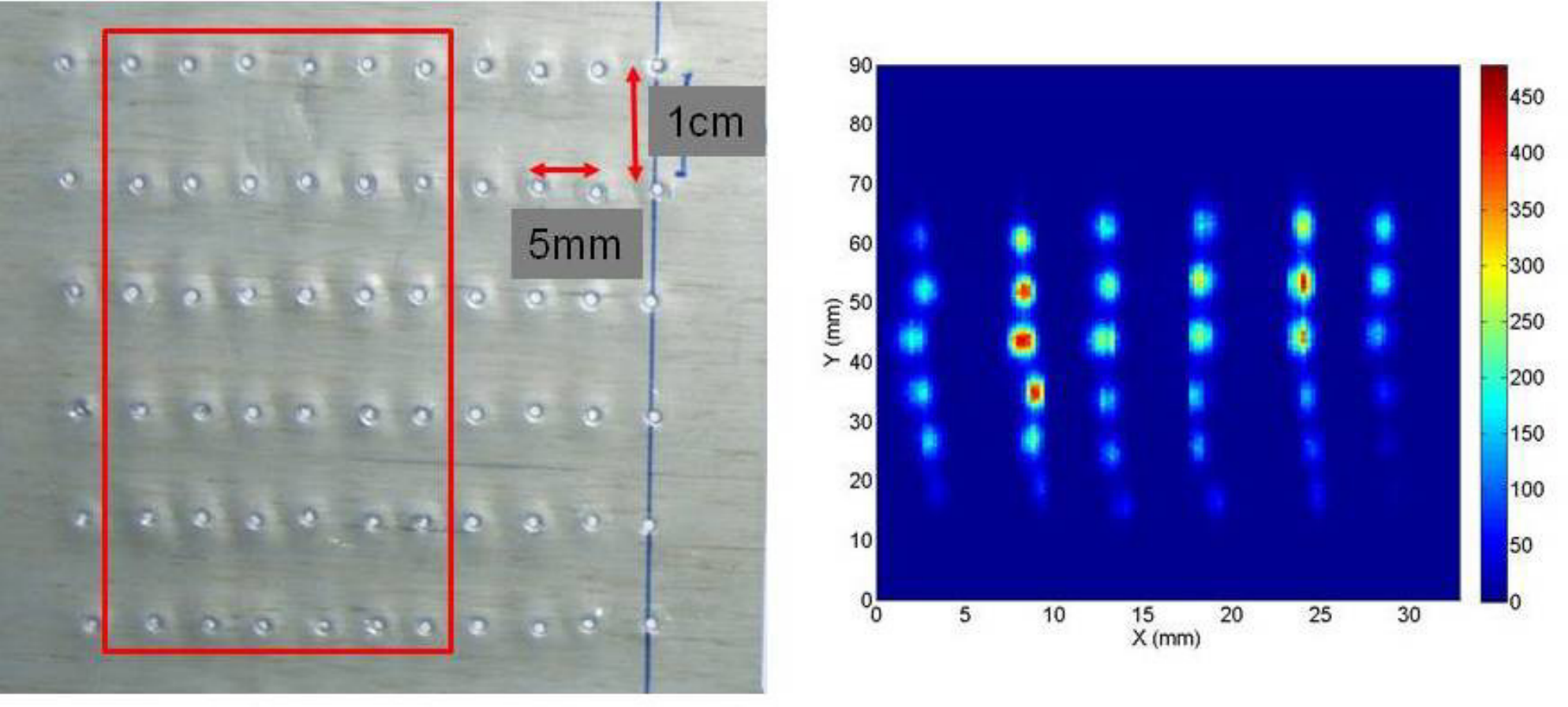}
\caption{\footnotesize The Cd mask (left) used to generate the image
on the right.} \label{ecfwercgqwetgq}
\end{figure}

Figure \ref{ecfwercgqwetgq} shows an image reconstructed with the
Multi-Blade version V2 obtained by placing the Cd mask with holes in
front of the detector. We repeat the mask consists of 80 holes of
$1\,mm$ size spaced by $5\,mm$ along $x$ and by $1\,cm$ along $y$.
\\ The number of bins on the image is set to equal the number of
wires (37) for the $x$-direction and is 256 bins for the
$y$-direction.
 \\ Due to the neutron beam divergence the spots
on the image appear much wider than the detector resolution.
\\ In the reconstructed image we observe the intensity that drops according
to the cassettes edges. We observe that any deviation of an hole
position on the mask is perfectly reproduced on the image.

\chapter*{Conclusions}
\addcontentsline{toc}{chapter}{Conclusions}

Although $^3He$ has been the main actor in thermal neutron
detection, the World is now experiencing a shortage of $^3He$. The
main issue to be addressed for large area neutron detectors (several
square meters) is to find an alternative technology to detect
neutrons because this rare isotope of Helium is not available
anymore in large quantities. This is not the main concern for small
area detectors ($\sim1\,m^2$) where the main effort is focused on
the performances. There is a great interest in expanding the
detector performances because for instance $^3He$ detectors are
limited in spatial resolution and counting rate capability.
\\ The detectors developed and implemented at ILL are based on $^{10}B$ layers used
as neutron converter in a gas proportional chamber. In particular we
used $^{10}B_4C$-layers deposited by magnetron sputtering technique
on holding substrates.
\\ The Multi-Grid is a large area detector that
has been developed at ILL to face the $^3He$ shortage problem. It
employs up to $30$ $^{10}B_4C$-layers in a cascade configuration.
\\ The concept of the Multi-Blade was introduced at ILL in 2005
but it has never been implemented until 2012. The Multi-Blade
prototype is a small size detector for application in neutron
reflectometry instruments based on single $^{10}B_4C$ layers. The
goal of the Multi-Blade is to go beyond the limits of $^3He$-based
detectors in terms of spatial resolution (which is about $1\,mm$ for
$^3He$ gaseous detectors) and counting rate capability.
\\ For both applications there are several aspects that we
investigated in order to validate this alternative technology.
\\ Although the physical process behind the neutron conversion
through solid converter is well known, the theoretical modeling for
these new detectors can open further developments. We elaborated
analytical expressions and equations to help the detector design.

\bigskip

In a neutron facility a detector is always exposed to other kinds of
radiation that we consider as a background to be suppressed. The
detection of a background event (mostly $\gamma$-rays) can give rise
to misaddressed events in a neutron detector. Generally the
$\gamma$-ray background can be a few orders of magnitude more
intense than the neutron signal. This has been proved by the
measurement of the typical background in a Multi-Grid prototype
detector installed on the time-of-flight spectrometer IN6 at ILL.
The low $\gamma$-ray sensitivity of a neutron detector is then a key
feature and it must be determined to validate the $^{10}B$
technology. While for $^3He$ detectors there is a clear energy
separation between neutron and $\gamma$-ray events, this is not the
case for solid converter based detectors. The neutron Pulse Height
Spectrum (PHS) for a $^{10}B$ detector is extended in a continuum
down to zero energy. In addition to that a $^{10}B$-based detector
should in principle show a higher sensitivity to $\gamma$-rays
because of the larger amount of material it is composed of with
respect to a standard $^3He$ detector.
\\ We quantified the $\gamma$-ray
sensitivity of $^{10}B$ and $^3He$ detectors by using a set of
calibrated $\gamma$-ray sources. For $^{10}B$-based detectors we
confirm that the energy spectrum given by $\gamma$-rays mostly
involves the low energy region of the spectrum. A suitable
$\gamma$-ray rejection, below $10^{-6}$, can be achieved by using a
discrimination on the energy level. The method consists of an energy
threshold, as is the case for $^3He$ detectors. We demonstrated that
a limited efficiency price of only about $0.5\%$ has to be paid in
neutron detection if a strong $\gamma$-ray rejection is necessary,
below $10^{-6}$.
\\ However, contrary to $^3He$ detectors, there is no clear
separation between the neutron Pulse Height Spectrum (PHS) and the
$\gamma$-ray PHS. Therefore, we investigate another method to
separate neutron from photon events trying to improve the
discrimination; but since it turned out that there is no physical
difference in the signal shape using the Time Over Threshold method
(TOT), this can not be exploited to discriminate against background.
The amplitude discrimination remains the best and simpler way to
reduce $\gamma$-ray sensitivity of neutron detectors based on a
solid converter.
\\ The Multi-Grid detector appears a promising replacement of
$^3He$-based large area detectors. We can conclude that such
detectors' $\gamma$-ray sensitivity is comparable to that of $^3He$
detectors.

\bigskip

In a multi-layer detector the arrangement of those layers is crucial
to optimize the performances. We studied solid converter films based
detectors from the theoretical point of view. With the magnetron
sputtering deposition method both sides of a substrate are coated
with the same thickness of converter. The suite of equations we
developed demonstrates that this method is also suited to make
optimized blades. In fact, for a single blade the same converter
thickness for both sides of a substrate has to be chosen in order to
maximize the efficiency. We demonstrated this result to be valid in
the case the optimization is done for a single neutron wavelength
and in the more general case when we deal with a distribution of
wavelengths. If the absorption of the substrate is not negligible we
also calculated the deviation from an ideal transparent substrate.
\\ We showed that also in a multi-layer detector all the blades have
to hold two layers of the same thickness in order to maximize the
detection efficiency. The demonstration is valid for a single
neutron wavelength and for whichever distribution of wavelengths. On
the other hand, the thicknesses of different blades can be distinct
and they can be optimized. The optimization procedure to be
implemented has been developed for both the monochromatic case and
for the case of a distribution of wavelengths.
\\ The blade-by-blade optimization in the case of a multi-layer detector for a
single neutron wavelength can achieve a few percent more efficiency
over the best detector with identical blades but this can lead to
several blades less in the detector. In the case of a distribution
of wavelengths, the optimization does not give important
improvements in the overall efficiency compared with a monochromatic
optimization done for the barycenter of that distribution. On the
other hand, the optimization of the efficiency for a neutron
wavelength distribution is often more balanced between short and
long wavelengths than the barycenter optimization.
\\ Since the $\gamma$-ray discrimination is an important feature of a neutron detector
and it is strictly related to its PHS shape, we calculated the PHS
analytical expression based on very simple assumptions and we showed
it is in a good agreement with measurements. Thanks to this model,
we understood the overall shape of the PHS which can be an important
tool if one wants to improve the $\gamma$-ray to neutron
discrimination in neutron detectors.
\\ In a solid converter-based detector both the PHS shape, for $\gamma$-ray sensitivity, and
the efficiency have to be taken into account in its optimization.
Both of them are influenced by the choice of the layer thicknesses.

\bigskip

Theoretical modeling was a useful tool to develop the Multi-Blade
detector of which we constructed two prototypes in order to
demonstrate its feasibility. The Multi-Blade employs $^{10}B_4C$
layers operated at grazing angle with respect to the neutron
incoming direction. The read-out is performed by a standard gas
amplification process implemented by a plane of wires. There are at
least three advantages in operating the detector at grazing angle.
The detection efficiency increases, the spatial resolution and the
counting rate capability are improved. Thanks to the inclination the
detector active surface is projected over the incoming neutron beam.
The effective wire pitch is then smaller than the pitch of the
finest possible wire mounting and the same local neutron flux is
shared among several wires.

\bigskip

In order to get a suitable detection efficiency, compared with
$^3He$, the converter layers must be operated at an angle of
$10^{\circ}$ or below. By approaching lower and lower angles,
neutrons can be reflected by the barrier potential of the converter
surface and are then lost for detection. We studied the reflection
process by strongly absorbent surfaces and we developed a model that
we have compared with our experiments carried out at ILL. This model
indicates that even for a strong neutron absorber, the reflectivity
measurement does not depend on the technique used. A monochromatic
scan over the angle or a Time of Flight (ToF) measurement lead to
the same result. This is confirmed by experiments.
\\ The theoretical equations derived to optimize the solid
converter-based detectors have been corrected for neutron reflection
for $^{10}B_4C$ in the neutron wavelength range from $1$ to $30$
\AA. Above $2^{\circ}$ there is no need to correct for neutron
reflection even if the layer is smooth at the $nm$ scale. We
observed that a too small converter roughness drastically increases
neutron reflection below $2^{\circ}$. A suitable roughness can help
to diminish neutron reflection. Since the ranges of the neutron
capture fragments in $^{10}B_4C$ are about a few $\mu m$, the layer
irregularity should never exceed the $\mu m$ scale. A larger
roughness cancels the gain in efficiency in operating the layer at
grazing angle. Our theoretical model about neutron reflection on
strong absorbers allowed us to fit the measured profiles that led to
the determination of the scattering length density of the used
$^{10}B_4C$ layers, it is about
$(2.5-1.1\,i)\cdot10^{-6}\,$\AA$^{-2}$.

\bigskip

We studied two approaches to be used in the Multi-Blade
implementation: either with one or with two converters. The latter
has more technical issues that makes its realization more difficult.
The single layer detector is finally the choice to make to keep the
mechanics reasonably simple. The extra advantage of having only one
converter is that the coating can be of any thickness above $3\,\mu
m$ without affecting the efficiency, while for the two layer option
its thickness should be chosen carefully. Moreover, in the two layer
configuration the substrate choice is also crucial because it should
be kept as thin as possible to avoid neutron scattering and this
leads to mechanical issues. In a single-layer detector it can be
integrated in the holding structure.
\\ We conceived a detector to be modular in order to be
versatile: it is composed of modules called \emph{cassettes}. We
operated the two Multi-Blade prototypes at either
$\theta=10^{\circ}$ and $\theta=5^{\circ}$. In each of the solutions
proposed for the cassette concept the read-out system has to be
crossed by neutrons before reaching the converter. The mechanical
challenge in the read-out system construction is to minimize the
amount of material on the neutron path to avoid scattering that can
cause misaddressed events in the detector. The choice fell on
polyimide substrates; they induce a few percent scattering of the
incoming neutrons. It can be eventually replaced with more suitable
materials.
\\ The detector is operated at atmospheric pressure. This makes it
suitable to be operated in vacuum. Moreover, cost effective
materials can be used inside the detector because outgassing is not
an issue.
\\ Since the detector is modular the main issue is its uniformity.
In the presented prototype we got a $50\%$ drop in efficiency in the
overlap region between cassettes.
\\ The presented Multi-Blade showed a very high spatial resolution,
it was measured to be about $0.3\,mm$ in one direction and about
$4\,mm$ in the other one.
\\ We measured the neutron detection efficiency for both prototypes
at $2.5$\AA \, neutron wavelength. The first prototype has an
efficiency of about $28\%$ employing sputtered $^{10}B_4C$-layers
inclined at $10^{\circ}$. This result is in a perfect agreement with
the expected efficiency we can calculate by using our theoretical
model we developed for solid neutron converters. Since the
efficiency, in the single layer option does not depend on the
converter thickness above $3\,\mu m$, in the second prototype we
investigated a different deposition method: a $^{10}B$ glue-based
painting. This thick painted layer functions also as an integrated
collimator inside the detector. The resistivity of the $^{10}B$
painting is larger than the sputtered $^{10}B_4C$-layers but it
seems to be acceptable and does not cause issues to the charge
evacuation. We measured the efficiency of the second prototype
operated at $5^{\circ}$ and we only got about $8\%$. The coarse
granularity of the painting makes the inclination effect vanish. The
expected efficiency for a sputtered layer of which the roughness is
widely below the $\mu m$ scale is about $43\%$ (at $2.5$\AA). There
is no reason not to get the calculated efficiency.
\\ We measured the detector dead time, including the read-out electronics, to
be about $1.5\,\mu s$.
\\ The single layer option represents a good candidate to go beyond
the performances of $^3He$ detectors. Further studies need to
address the uniformity problems. If a simple coating technique is
found, e.g. painting containing grains of a suitable size that
assures a uniform layer, the Multi-Blade could be a cost-effective
and high performance alternative. Its production in series is easy
to be implemented.

\bigskip

We have reached a profound insight in the principles of solid
converters in neutron detection by successfully confronting our
theoretical investigations with experiment.
\\ We investigated several technological aspects of $^{10}B$-based detectors
by prototyping the Multi-Blade idea. Some explored avenues are very
promising while others indicate technological difficulties which
need to be resolved. Nevertheless, the results are sufficiently
encouraging to foresee a relatively simple construction.

\appendix

\chapter{The stopping power law}\label{classstoppow5}
\section{Classical derivation}
\begin{figure}[ht!]
\centering
\includegraphics[width=7cm,angle=0,keepaspectratio]{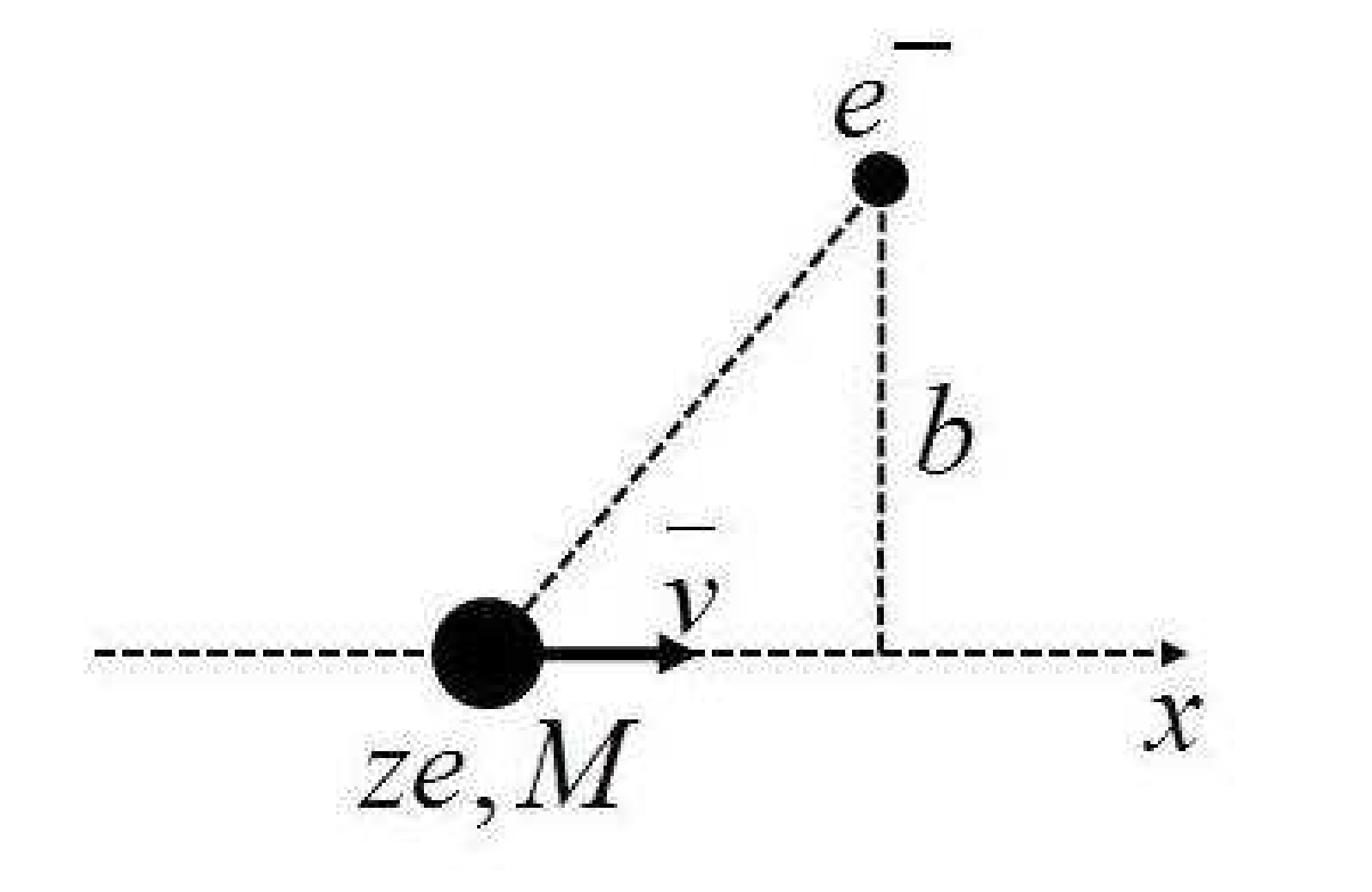}
\caption{\footnotesize Collision of a heavy particle with an atomic
electron.}\label{bet1}
\end{figure}

Consider a heavy particle with charge $ze$, mass $M$ and velocity
$v$ passing through a medium and suppose there is an atomic electron
at some distance $b$ from the particle trajectory (see Figure
\ref{bet1}). We assume that the electron is free and at rest, and
that it only moves slightly during the interaction so that the
position where the electric field acts in the collision can be
considered constant.
\\ Let's calculate the energy gained by the electron by finding the
momentum impulse it receives from the collision with the heavy
particle:
\begin{equation}\label{eqha14}
\Delta p = \int F \, dt = e \int E_{\bot}\frac{dt}{dx}dx= e \int
E_{\bot}\frac{dx}{v}
\end{equation}
where $E_{\bot}$ is the normal component of the electric field to
the particle trajectory. By using the Gauss' law on a infinitely
long cylinder of radius $b$ and centered on the particle trajectory:
\begin{equation}\label{}
\Phi(E_{\bot})=\int E_{\bot} 2\pi b \,dx = 4\pi ze \qquad
\Longrightarrow \qquad \int E_{\bot} dx = \frac{2 z e}{b}
\end{equation}
hence:
\begin{equation}\label{}
\Delta p = e \int E_{\bot}\frac{dx}{v} = \frac{2ze^2}{bv}
\end{equation}
The energy gained by the electron results to be:
\begin{equation}\label{eqha15}
\Delta E(b) = \frac{(\Delta p)^2}{2m_e}= \frac{2z^2e^4}{m_e b^2v^2}
\end{equation}
If we consider $n_e$ the number density of electrons, the energy
lost to all the electrons located at a distance between $b$ and
$b+db$ in a thickness $dx$ is:
\begin{equation}\label{eqha13}
-dE(b) = \Delta E(b)\, n_e\, dV = \frac{2z^2e^4}{m_e b^2v^2} \,n_e
\,(2\pi b \,db \,dx)=\frac{4 \pi z^2e^4}{m_e v^2} n_e \frac{db}{b}\,
dx
\end{equation}
The total energy loss in $dx$ is obtained by integration over $db$
of the Equation \ref{eqha13} in the interval $[b_{min}, b_{max}]$.
It has to be pointed out that if we integrate between $0$ and
$+\infty$ this is contrary to our original assumptions. E.g.
collisions at very large $b$ would not take place over a short
period of time, thus the impulse calculation in Equation
\ref{eqha14} would not be valid. On the other hand, for $b=0$,
Equation \ref{eqha15} gives an infinite energy transfer. Thus:
\begin{equation}\label{eqha16}
-\frac{dE}{dx} = \frac{4 \pi z^2e^4}{m_e v^2} n_e \,
\ln\left(\frac{b_{max}}{b_{min}}\right)
\end{equation}
To estimate $b_{min}$ and $b_{max}$ we should advance some physical
arguments. The maximum energy transferable, classically, is in a
head-on collision where the electron obtains an energy of $\frac 1 2
m_e (2v)^2$. If we take relativity into account this becomes
$2\gamma^2 m_e v^2$, where $\gamma = (1-\beta^2)^{-1/2}$ and $\beta
= \frac v c$. From Equation \ref{eqha15} we find:
\begin{equation}\label{eqha17}
\frac{2z^2e^4}{m_e b_{min}^2v^2}=2\gamma^2 m_e v^2 \qquad
\Longrightarrow \qquad b_{min} = \frac{ze^2}{\gamma m_e v^2}
\end{equation}
For $b_{max}$, we should recall that the electrons are not free but
bound to atoms with some orbital frequency $\nu$. In order for the
electron to absorb energy, the perturbation caused by the passing
particle must take place in a time short compared to the period
$\tau=\frac 1 \nu$ of the bound electron, otherwise the perturbation
is adiabatic and no energy is transferred. The typical interaction
time is $t\simeq \frac b v$, which relativistically becomes
$t'=\frac t \gamma = \frac b {v \gamma}$, thus:
\begin{equation}\label{eqha18}
\frac b {v \gamma}\leq \tau = \frac 1 \nu \qquad \Longrightarrow
\qquad b_{max} = \frac{\gamma v}{\nu}
\end{equation}
Where $\nu$ should be considered as the mean frequency averaged over
all bound states because there are several bound electron states
with different frequencies.
\\ Finally, by substituting \ref{eqha17} and \ref{eqha18} in Equation \ref{eqha16},
we obtain:
\begin{equation}\label{eqha19}
-\frac{dE}{dx} = \frac{4 \pi z^2e^4}{m_e v^2} n_e \,
\ln\left(\frac{\gamma^2 m v^3}{z e^2 \nu}\right)
\end{equation}
This is the Bohr's classical formula and it gives a reasonable
description of the energy loss for very heavy particles.
\section{The Bethe-Bloch formula}
The correct quantum-mechanical calculation leads to the Bethe-Bloch
formula:
\begin{equation}\label{eqha20}
-\frac{dE}{dx} =2\pi N_A r_e^2 m_e c^2 \rho \frac Z A
\frac{z^2}{\beta^2}\left( \ln \left( \frac{2m_e\gamma^2 v_2
W_{max}}{I^2}\right)-2\beta^2-\delta-2\frac C Z\right)
\end{equation}
where:
$$
\centering
\begin{tabular}{llll}
\hline
$r_e$ &  classical electron radius & $\rho$ & absorbing material density\\
$m_e$ &  electron mass             & $z$ & charge of the incident particle \\
$N_A$ &  Avogadro's number         & $\beta$ & $=v/c$ incoming particle velocity\\
$I$ &  mean excitation potential & $\gamma$ &  $= (1-\beta^2)^{-1/2}$\\
$Z$ &  atomic number of absorbing material & $\delta$ & density correction\\
$A$ &  atomic weight of absorbing material & $C$ & shell correction\\
$W_{max}$ &  max energy transferred in a single collision &  & \\
\hline
\end{tabular}
$$
\\ The maximum energy transfer is that produced by a head-on
collision; for an incident particle of mass $M$:
\begin{equation}\label{}
W_{max}
=\frac{2m_ec_2\beta^2\gamma^2}{1+2\frac{m_e}{M}\sqrt{1+\beta^2\gamma^2+\frac{m_e^2}{M^2}}}
\end{equation}
The mean excitation potential $I$ is directly related to the average
orbital frequency $\nu$.
\\ The quantities $\delta$ and $C$ are correction at high and low
energies respectively. The density factor, $\delta$, arises from the
fact that the electric field of the incoming particle also tends to
polarize the atoms along its path. The shell correction, $C$,
accounts the effect which arise when the velocity of the incident
particle is comparable, or smaller, than the orbital velocity of the
bound electrons.
\\ A qualitative behavior of the Bethe-Bloch formula is shown in
Figure \ref{figstopremmip} as a function of both particle energy $E$
and track length $x$, respectively. The point where $\frac{dE}{dx}$
has a minimum in energy is known as \emph{minimum ionizing}.

\chapter{Connection with Formulae in \cite{gregor}}\label{connfgreg}
The relations between the formulae in \cite{gregor} and the
expression used in Chapter \ref{Chapt1}, Section \ref{secttheoeff}
are the following:
\begin{itemize}
    \item the particle ranges $R$ are denoted by $L$;
    \item the branching ratios of the $^{10}B$
reaction (expressed by $F_p$) are $F_1=0.94$ and $F_2=0.06$;
\item the thickness of the layer is $d=D_F$.
\end{itemize}
\noindent Hence, the relation between the expressions \ref{T21},
\ref{T22}, \ref{T23} and the formulae in \cite{gregor} is:
\begin{equation}\label{expla78}
\begin{aligned}
\varepsilon_T(d_T)&=0.94\cdot\varepsilon_{T}(R^{94\%}_1,R^{94\%}_2)+0.06\cdot\varepsilon_{T}(R^{6\%}_1,R^{6\%}_2)\\
&=S_1(D_F,L_1,0.94)+S_1(D_F,L_2,0.94)+S_2(D_F,L_1,0.06)+S_2(D_F,L_2,0.06)
\end{aligned}
\end{equation}
\\ Valid for both equations ($18a$) ($D_F \leq L_i$) and ($18b$) ($D_F
> L_i$) in Section $4.2$ of \cite{gregor}. In the case of the back-scattering
mode, equations ($25a$) and ($25b$) in \cite{gregor}, we consider
one layer of converter and we replace $\varepsilon_{T}(d_T)$ into
$\varepsilon_{BS}(d_{BS})$ in the expression \ref{expla78}.
\\ \noindent In a different way, for both equations ($18a$) and ($18b$), we can also write:
\begin{equation}
F_p \cdot \varepsilon_{T}=S_p(D_F,L_1,F_p)+S_p(D_F,L_2,F_p)
\end{equation}

\chapter{Connection with Formulae in \cite{salvat}}\label{connfsalvat}
The relations between the formulae in \cite{salvat} and the
expression used in Chapter \ref{Chapt1}, Section \ref{SectThPHSCalc}
are the following:
\begin{itemize}
    \item the macroscopic cross-section ($\Sigma$) is expressed in terms of mean free path $\Sigma=\frac 1
    l$;
    \item the variable $u$ is denoted by its cosine $u=\cos(\theta)$;
\end{itemize}
The formulae $(4)$ in \cite{salvat} corresponds to the Equation
\ref{eqae1}, except for a factor $\frac 1 2$, where $l=\frac 1
\Sigma$.
\\ The formulae $(3)$ in \cite{salvat} corresponds to the Equation
\ref{eqae3}.

\chapter{Highly absorbing layer neutron reflection model}\label{modelreflect}
The model used to fit the reflectivity profile for a strong neutron
absorber, such as $^{10}B_4C$, is the one shown in the Equations
\ref{eqaf22} in Chapter \ref{chaptreflectometry}. $R$ and $A$ are
the reflectivity and the absorption of an absorbing layer deposited
on a substrate. These quantities depend on $q$ by
$q=\frac{4\pi}{\lambda}\sin(\theta)$.
\\ In order to fit the reflectivity and absorption profiles we
calculate the reduced chi-square:
\begin{equation}\label{wctgwwercgw5wd54667yqc4yhw45}
\chi^2=\frac{1}{(2N-p)}\sum_{n=1}^N
\left[\left(R_{meas.}(q_n)-R_c(q_n)\right)^2+\left(A_{meas.}(q_n)-A_c(q_n)\right)^2\right]
\end{equation}
where $N$ is the number of points in each set of measurements and
$p$ is the number of parameters we used in the fit. In our specific
case they are $p=8$: $\theta_{over}$, $\theta_{shift}$, $^{10}B_4C$
scattering length density (real and imaginary parts), layer
roughness $\sigma_r$, layer thickness $d$, Ge-detector efficiency
$\varepsilon_{Ge}$ and an overall normalization $M$.
\\ Before the sample the neutron beam is collimated by two slits.
Even if the beam divergence is known it is not easy to calculate the
exact beam footprint at the sample position. In addition to that,
the beam can be not uniform, even after collimation, and the samples
do not have sharp edges. $\theta_{over}$ corrects for these effects
in the horizontal direction. However, the remaining effect in the
vertical direction will introduce a modification of the
normalization. This is why we introduce the normalization factor
$M$. Generally it has values close to $1$.
\\ To model the layer roughness we used the Equation
\ref{eqaf15} applied to the Fresnel coefficients (Equation
\ref{equaf21}) in the case of absorption.
\\ In Equation \ref{wctgwwercgw5wd54667yqc4yhw45},
$R_{meas.}$ and $A_{meas.}$ are the measured raw curves; $R_{c}$ and
$A_{c}$ are the reflectivity and the absorption as a function of $q$
predicted by our theoretical model after applying the instrumental
corrections.
\\ The overillumination correction is a factor $I_{over}$ that has to be
applied to the data to consider the finite extension of the sample.
This correction is $I_{over}=1$ for all the $\theta$ for which the
sample projection is larger than the beam size (the full beam hits
the sample), and it is less than 1 for the values of $\theta$ for
which the sample projection is smaller (only a part of the beam hits
the sample). This function has also to be smoothed using a smoothing
filter (a 3 points moving average filter) to include the non uniform
beam intensity and the instrumental resolution. It is:
\begin{equation}
I_{over}= \begin{cases} \frac{\sin(\theta)}{\sin(\theta_{over})} &\mbox{if } \theta<\theta_{over} \\
1 & \mbox{otherwise} \end{cases}
\end{equation}
The misalignment of the instrument is included in the angle
$\theta_{shift}$. This affects the actual value for the angle used
to calculate the reflectivity and absorption:
\begin{equation}
\theta_c=\theta+\theta_{shift}
\end{equation}
The values for the profiles once applied the instrumental
corrections can be calculated by:
\begin{equation}
R_c=M\cdot I_{over} \cdot R(\theta_c), \qquad
A_c=M\cdot \varepsilon_{Ge} \cdot I_{over} \cdot A(\theta_c)\\
\end{equation}
We recall the quantities $R$ and $A$ depend on the fitting parameter
$d$, the layer scattering length density, the reflection angle
$\theta_c$ and the layer roughness $\sigma_r$.

\chapter{Neutron flux measurement}\label{apphexmeasexplan}
\begin{figure}[h]
\centering
{\includegraphics[width=0.31\linewidth,angle=0,keepaspectratio]{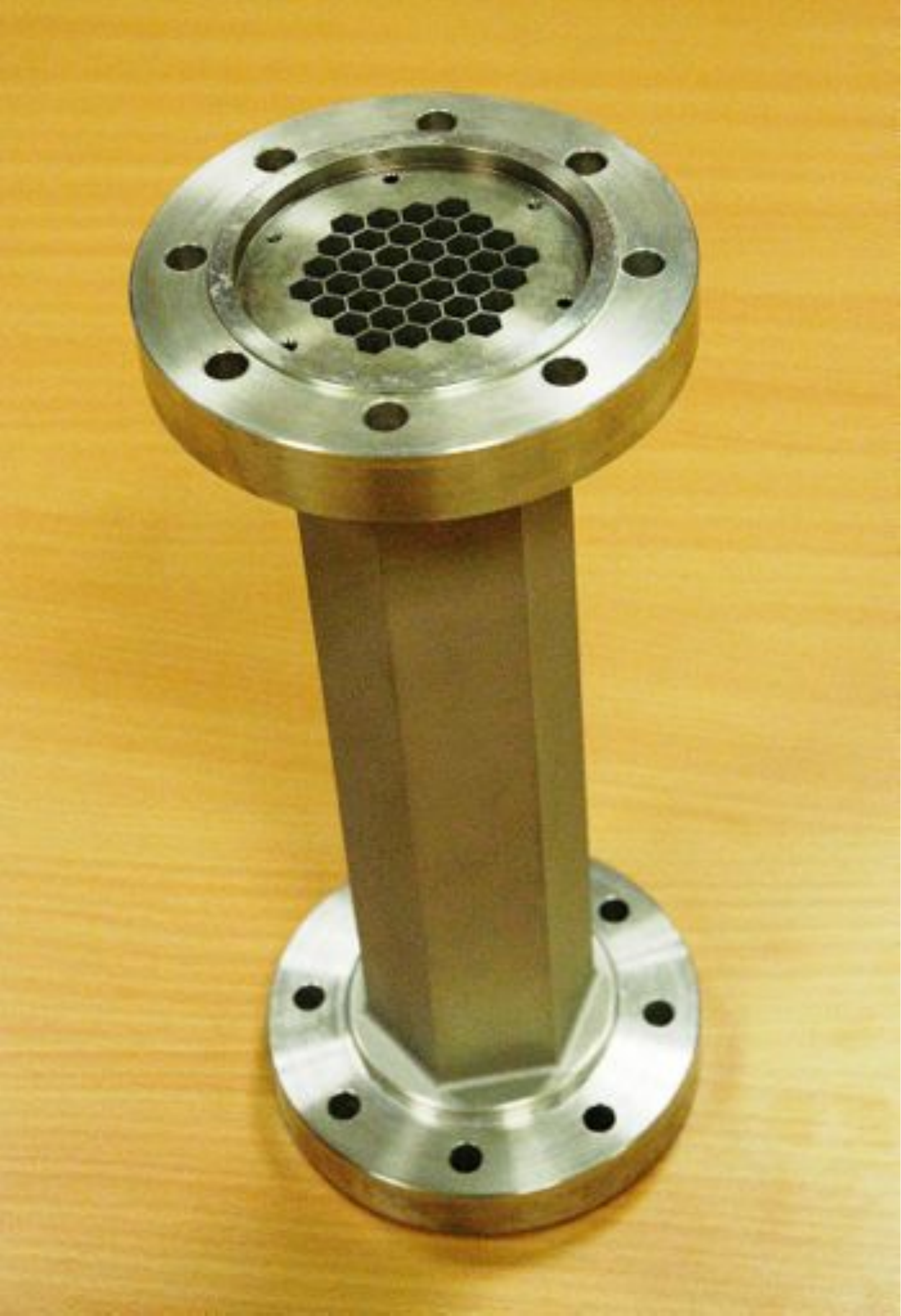}}
{\includegraphics[width=0.61\linewidth,angle=0,keepaspectratio]{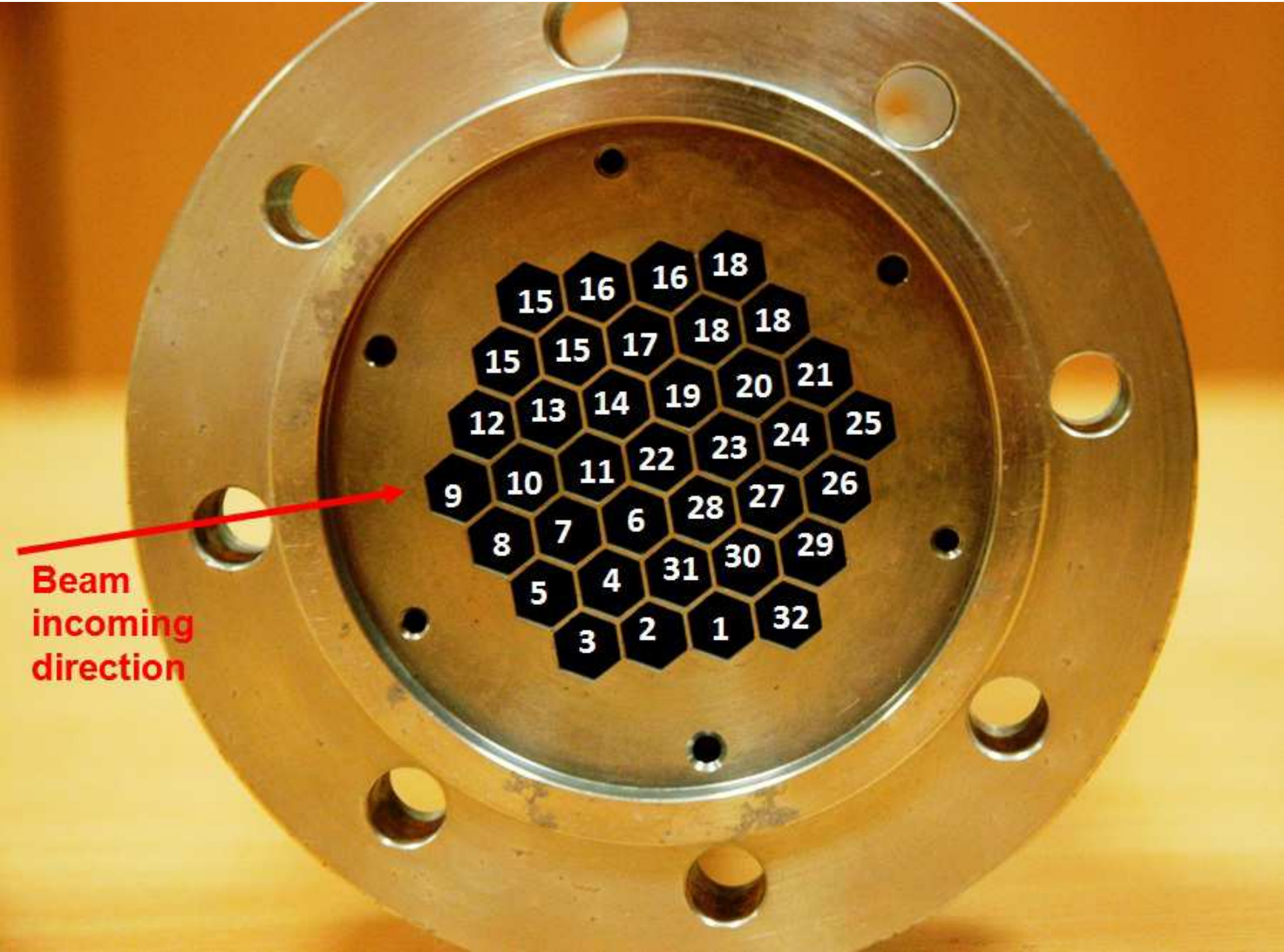}}
\caption{\footnotesize Hexagonal $^3He$-tubular detector used to
measure the actual neutron flux (the red arrow shows the neutron
beam direction and the number are the channel number of each tube,
some tubes are connected together). \label{hexpictu1}}
\end{figure}

\begin{figure}[h!]
\centering
{\includegraphics[width=0.49\linewidth,angle=0,keepaspectratio]{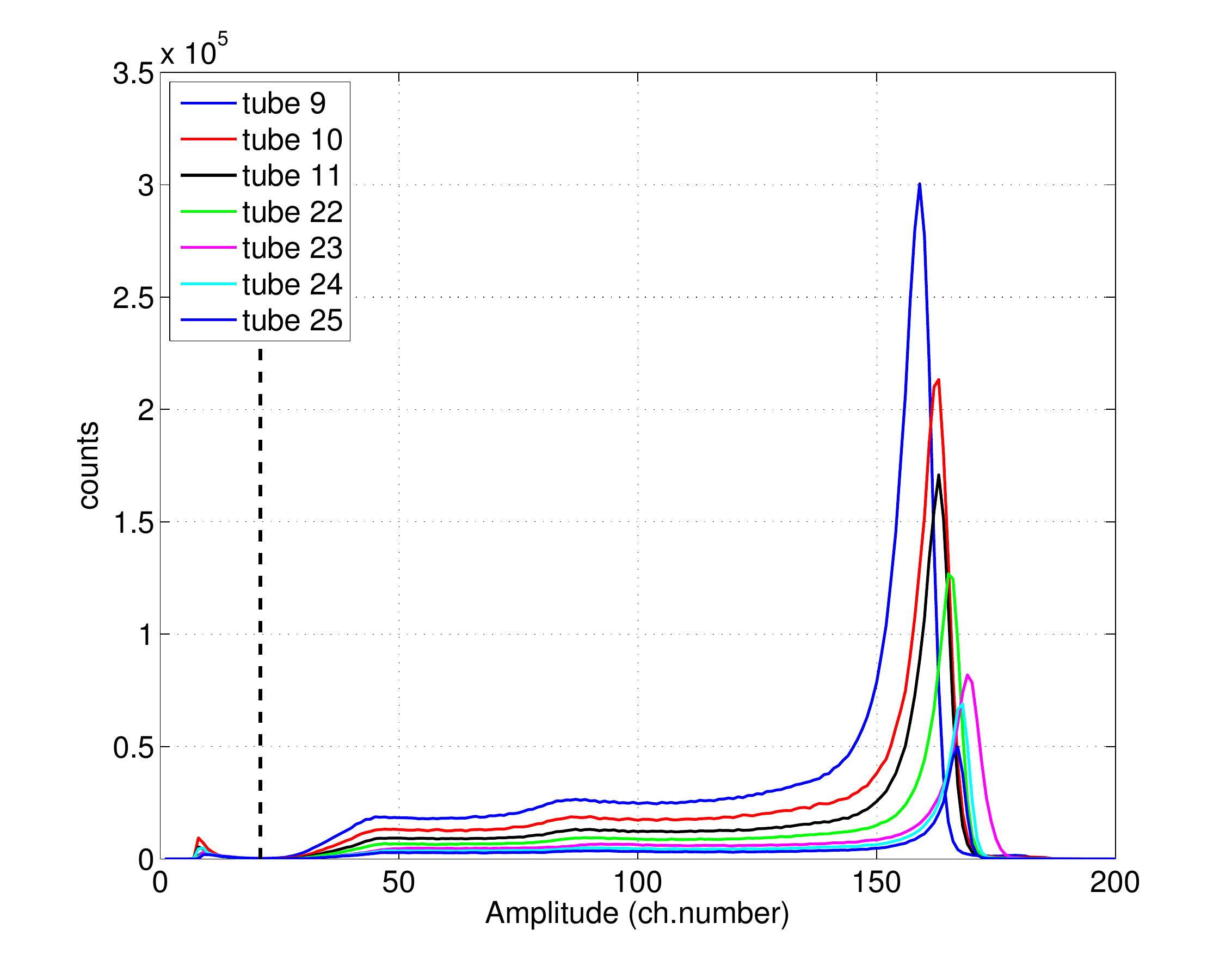}}
{\includegraphics[width=0.49\linewidth,angle=0,keepaspectratio]{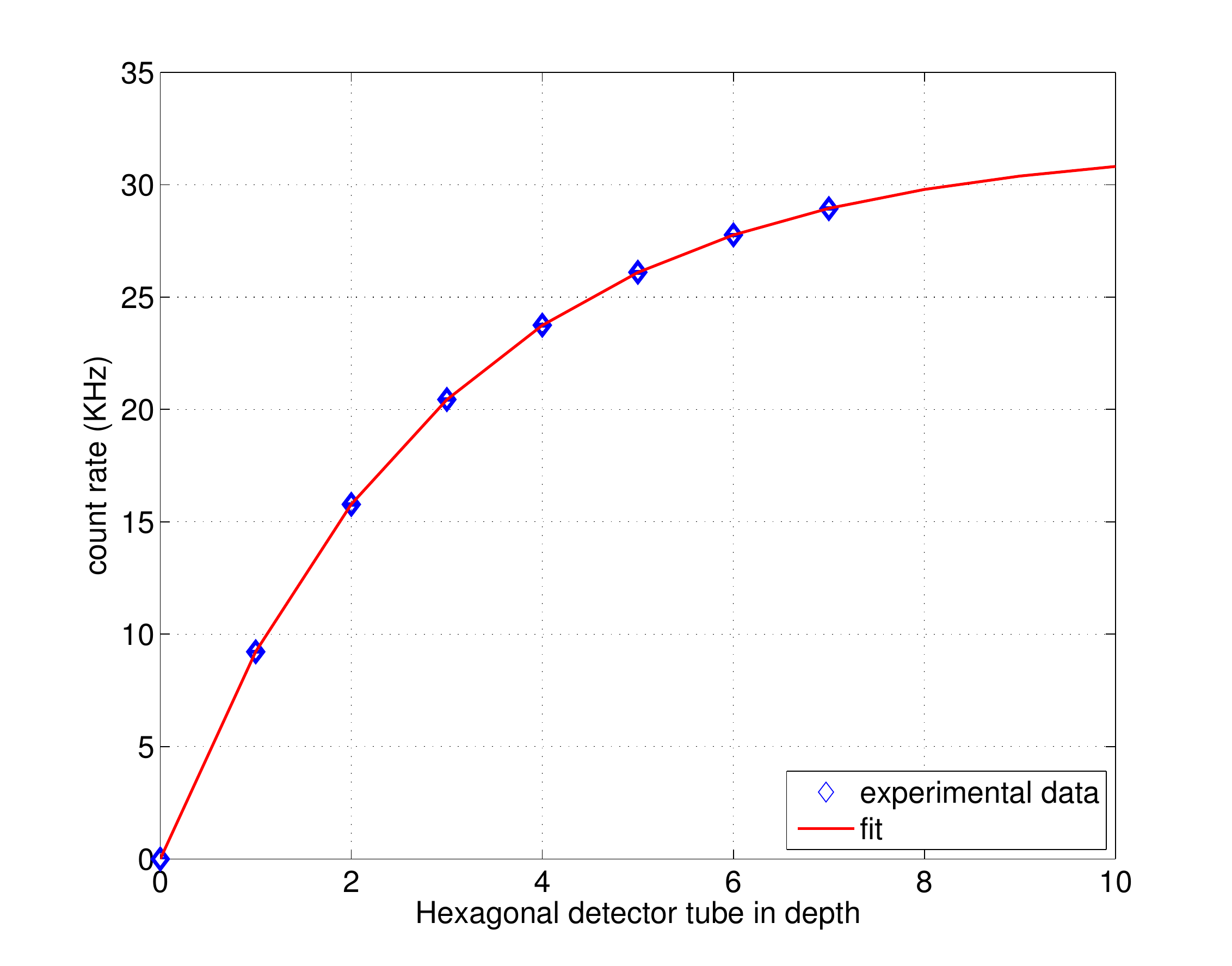}}
\caption{\footnotesize Hexagonal detector PHS (left) and neutron
cumulative flux across the 7 central tubes of the detector (right).
\label{paramhex}}
\end{figure}

It is commonly useful to know the actual neutron flux of a beam
line, e.g. CT1 at ILL. By knowing the neutron flux, for example, the
efficiency of a prototype can be estimated. In order to perform an
accurate neutron flux measurement, the so-called \emph{Hexagonal
detector} is used. The latter, shown in Figure \ref{hexpictu1}, is a
multi-tube $^3He$-based gaseous detector composed of $37$
$7\,mm$-diameter tubes. Each of them acts as a proportional chamber;
the drift voltage is common to the $37$ hexagonal shaped tubes and
they are read-out individually. Thus, the neutron detected in each
tube can be quantified.
\\ We start by setting each channel (tube) threshold above the noise
level; it is not problematic to have some gain difference between
the channels, because further adjustment can be done later by
software.
\\ The beam is collimated at a size which is smaller
than the single tube diameter. By counting the detected neutrons per
tube, we scan the detector in order to find the position where the
central seven tubes are aligned on the neutron trajectory (see
Figure \ref{hexpictu1}). The additional tubes on the two sides allow
to help in the alignment.
\\ Once alignment is over, we record a PHS for each tube in a given
time $T$ which is long enough to get a small statistic error
(relative errors below $1\%$ can be archived in few minutes).
\\ We repeat the measure in the same conditions but without the beam
in order to quantify the background that has to be subtracted from
the measurement with the beam on.
\\ In Figure \ref{paramhex}, on the left, is shown the seven $^3He$-PHS relative to the
central tubes and the dashed line represent the threshold, added by
software, used to calibrate the detector efficiency.
\\ Hence, we integrate, from the threshold to the end, the PHS and
we obtain the counting per tube that has to be divided by the
measurement time $T$ to get the counting rate. We cumulate the rates
of the seven central tubes in order to get the plot shown in Figure
\ref{paramhex} on the right. This represents the number of neutron
detected per second in a detector made up of $1$, $2$, $\dots$, $N$
tubes. 
\\ One can now imagine to arrange an infinite number of $^3He$-tubes on the
beam path, ideally each neutron is counted, if it is not lost before
by scattering out of the beam direction.
\\ Therefore, by performing a fit of the plot in Figure
\ref{paramhex}, the asymptotic value gives the actual neutron flux
at the neutron wavelength it was measured.


\end{document}